\documentclass[11pt,a4paper]{article}
\pdfoutput=1

\usepackage[colorlinks=true, linkcolor=black!50!blue, urlcolor=blue, citecolor=blue, anchorcolor=blue]{hyperref}
\usepackage[font=small,labelfont=bf,margin=0mm,labelsep=period,tableposition=top]{caption}
\usepackage[a4paper,top=1.5cm,bottom=2cm,left=1.5cm,right=1.5cm,bindingoffset=0mm]{geometry}

\usepackage{placeins,setup,cite}
\usepackage{graphicx}
\usepackage{float}
\usepackage{afterpage}
\usepackage{epsfig}
\usepackage{amssymb}
\usepackage{amsmath}
\usepackage{bm}
\usepackage{multirow}
\usepackage{url}
\usepackage{xcolor}
\usepackage{ulem}
\usepackage{url}
\usepackage{booktabs,multirow}
\usepackage[T1]{fontenc}

\makeatletter
\def\thickhline{%
             \noalign{\ifnum0 =`}\fi\hrule \@height \thickarrayrulewidth \futurelet
             \reserved@a\@xthickhline}
\def\@xthickhline{\ifx\reserved@a\thickhline
                \vskip\doublerulesep
                \vskip -\thickarrayrulewidth
                \fi
                \ifnum0 =`{\fi}}
\makeatother
\newlength{\thickarrayrulewidth}
\usepackage{tikz}
\usetikzlibrary{positioning}
\usetikzlibrary{arrows.meta}
\usepackage{varwidth}
\usepackage{xcolor}
\definecolor{mtplotlib1}{HTML}{1f77b4}
\definecolor{mtplotlib2}{HTML}{ff7f0e}
\definecolor{mtplotlib3}{HTML}{2ca02c}
\definecolor{mtplotlib4}{HTML}{d62728}

\tikzset{%
  >={Latex[width=2mm,length=2mm]},
            base/.style = {rectangle, rounded corners, draw=black,
                           minimum width=4cm, minimum height=1cm,
                           text centered}, 
            mystyle/.style={rectangle, rounded corners, draw=black,
            minimum width=12cm, minimum height=1cm,
            text centered}, 
    col0/.style = {base, fill=white!30},
    col1/.style = {base, fill=mtplotlib1!30},
    col11/.style = {mystyle, fill=mtplotlib1!30},
    col2/.style = {base, fill=mtplotlib2!30},
    col3/.style = {base, fill=mtplotlib3!30},
    col4/.style = {base, minimum width=2.5cm, fill=mtplotlib4!15,}
}

\newcommand{\be}{\begin{equation}}
\newcommand{\ee}{\end{equation}}
\newcommand{\bea}{\begin{eqnarray}}
\newcommand{\eea}{\end{eqnarray}}
\newcommand{\bi}{\begin{itemize}}
\newcommand{\ei}{\end{itemize}}
\newcommand{\ben}{\begin{enumerate}}
\newcommand{\een}{\end{enumerate}}
\newcommand{\la}{\left\langle}
\newcommand{\ra}{\right\rangle}
\newcommand{\lc}{\left[}
\newcommand{\rc}{\right]}
\newcommand{\lp}{\left(}
\newcommand{\rp}{\right)}

\def\frac#1#2{{{#1}\over {#2}}}
\def\gsim{\mathrel{\rlap{\lower4pt\hbox{\hskip1pt$\sim$}}
    \raise1pt\hbox{$>$}}}       
\def\lsim{\mathrel{\rlap{\lower4pt\hbox{\hskip1pt$\sim$}}
    \raise1pt\hbox{$<$}}}

\newcommand{\rep}{\mathrm{rep}}

\newcommand{\draft}[1]{}

\def\beq{\begin{equation}}
\def\eeq{\end{equation}}

\numberwithin{equation}{section}
\numberwithin{figure}{section}
\numberwithin{table}{section}

\usepackage{tabularx}
\newcolumntype{C}[1]{>{\centering\arraybackslash}p{#1}}

\def\CTprime{CT18${}^\prime$\xspace}
\newcommand{\NNprime}{NNPDF3.1${}^\prime$\xspace}

\definecolor{darkblue}{rgb}{0.0,0,0.5}
\definecolor{darkgreen}{rgb}{0.0,0.3,0.0}
\definecolor{redish}{rgb}{0.675,0,0.2}
\definecolor{red}{rgb}{0.8,0,0}
\definecolor{green}{rgb}{0,0.6,0}
\definecolor{bluish}{rgb}{0.2,0.2,0.675}
\definecolor{neonpink}{rgb}{0.98,0.27,0.8}
\definecolor{sepia}{rgb}{0.44, 0.26, 0.08}

\begin{document}
\vspace{-2.0cm}
\begin{flushright}
Edinburgh 2021/31\\
FERMILAB-PUB-22-121-QIS-SCD-T \\
MSUHEP-22-010\\
Nikhef 2021-033 \\
SMU-HEP-22-01\\
\end{flushright}
\vspace{0.1cm}

\begin{center}
{\Large \bf The PDF4LHC21 combination of global PDF fits for the LHC Run III}\\
  \vspace{0.7cm}
  {\small
    The {\bf PDF4LHC Working Group:}\\[0.2cm] Richard~D.~Ball$^1$, 
    Jon~Butterworth$^2$,
    Amanda~M.~Cooper-Sarkar$^3$,
    Aurore~Courtoy$^{4}$,
    Thomas~Cridge$^2$,
    Albert~De~Roeck$^5$,
    Joel~Feltesse$^6$,
    Stefano~Forte$^7$,
    Francesco~Giuli$^5$,
    Claire~Gwenlan$^2$,
    Lucian~A.~Harland-Lang$^8$,
    T.~J.~Hobbs$^{9,10}$,
    Tie-Jiun Hou$^{11}$,
    Joey~Huston$^{12}$,
    Ronan~McNulty$^{13}$, 
    Pavel~M.~Nadolsky$^{14}$,
    Emanuele~R.~Nocera$^1$,
    Tanjona~R.~Rabemananjara$^{15,16}$,
    Juan~Rojo$^{15,16}$,
    Robert~S.~Thorne$^2$,
    Keping~Xie$^{17}$, 
    and
    C.-P.~Yuan$^{12}$
  }\\
  
\vspace{0.4cm}

{\it \small
  ~$^1$ Higgs Centre, University of Edinburgh, JCMB, KB, Mayfield Rd, Edinburgh EH9 3JZ, Scotland\\[0.1cm]
  ~$^2$ Department of Physics and Astronomy, University College London, London, UK\\[0.1cm]
  ~$^3$ Department of Physics, University of Oxford, UK\\[0.1cm]
  ~$^{4}$ Instituto de F\'isica, Universidad Nacional Aut\'onoma de M\'exico\\Apartado Postal 20-364, 01000 Ciudad de M\'exico, Mexico\\[0.1cm]
  ~$^5$CERN, CH-1211 Geneva, Switzerland\\[0.1cm] 
  ~$^6$ IRFU, CEA, Université Paris-Saclay, 91191 Gif-sur-Yvette, France\\[0.1cm]
  ~$^7$ Tif Lab, Dipartimento di Fisica, Universit\`a di Milano and
INFN, Sezione di Milano, Via Celoria 16, I-20133 Milano, Italy\\[0.1cm]
   ~$^8$ Rudolf Peierls Centre, Beecroft Building, Parks Road, Oxford, UK \\[0.1cm]
   ~$^{9}$ Fermi National Accelerator Laboratory, Batavia, IL 60510, USA\\[0.1cm]
  ~$^{10}$ Department of Physics, Illinois Institute of Technology, Chicago, IL 60616, USA\\[0.1cm]
  ~$^{11}$ Department of Physics, College of Sciences, Northeastern University, Shenyang 110819, China \\[0.1cm]
   ~$^{12}$ Department of Physics and Astronomy, Michigan State University, East Lansing, MI 48824 USA \\[0.1cm]
   ~$^{13}$ School of Physics, University College Dublin, Dublin 4, Ireland.\\[0.1cm]
   ~$^{14}$ Department of Physics, Southern Methodist University, Dallas, TX 75275-0181, USA \\[0.1cm]
    ~$^{15}$ Department of Physics and Astronomy, Vrije Universiteit, NL-1081 HV Amsterdam\\[0.1cm]
~$^{16}$ Nikhef Theory Group, Science Park 105, 1098 XG Amsterdam, The Netherlands\\[0.1cm]
  ~$^{17}$ Department of Physics and Astronomy, University of Pittsburgh, Pittsburgh, PA 15260, USA\\[0.1cm]
 }

\vspace{0.7cm}

{\bf \large Abstract}

\end{center}

A precise knowledge of the quark and gluon structure of the proton,
encoded by the parton distribution functions (PDFs), is of paramount
importance for the interpretation of high-energy processes at present and future lepton-hadron
and hadron-hadron colliders.
Motivated by recent progress in the PDF determinations
carried out by the CT, MSHT, and NNPDF groups,
we present an updated combination of global PDF fits: PDF4LHC21.
It is based on the Monte Carlo combination of the CT18, MSHT20, and NNPDF3.1
sets followed by either its Hessian reduction or 
its replica compression.
Extensive  benchmark studies are carried out
in order to disentangle the origin of the differences between the three global PDF sets.
In particular, dedicated fits based on almost identical theory settings
and input datasets are performed by the three groups, highlighting
the role played by the respective fitting methodologies.
We compare the new PDF4LHC21 combination with its predecessor, PDF4LHC15, demonstrating
their good overall consistency and a 
modest reduction of PDF uncertainties for key LHC processes
such as electroweak gauge boson production
and Higgs boson production in gluon fusion.
We study the phenomenological implications of PDF4LHC21 for a representative selection of inclusive, fiducial, and differential cross sections at the LHC.
The PDF4LHC21 combination is made available via the {\sc\small LHAPDF} library and
provides a robust, user-friendly, and efficient method to
 estimate the PDF uncertainties associated
to theoretical calculations for the upcoming Run III of the LHC and beyond.

{\it We dedicate this paper to the memory of James Stirling,  Dick Roberts and Jon Pumplin, all of whom sadly died in the past few years. All three were instrumental in the development and evolution of PDF fitting, Jon for the CTEQ-TEA collaboration and James and Dick for the MSHT collaboration.}

\tableofcontents
\section{Introduction and motivation}
\label{sec:introduction}

Uncertainties associated with our limited knowledge of the quark and gluon structure of the proton, encoded by its collinear unpolarised parton distribution functions (PDFs), represent one of the most significant limiting factors in the theoretical interpretation of crucial processes at the LHC. These include the extraction of fundamental Standard Model (SM) parameters such as Higgs boson coupling measurements~\cite{LHCHiggsCrossSectionWorkingGroup:2016ypw}, the $W$-boson mass~\cite{ATLAS:2017rzl,LHCb:2021bjt,Bozzi:2011ww} and the strong coupling constant $\alpha_s(m_Z)$, as well as direct BSM searches for heavy resonances~\cite{Beenakker:2015rna} and indirect BSM searches via effective field theory~\cite{Carrazza:2019sec,Greljo:2021kvv}. Despite encouraging progress from first-principles lattice QCD calculations~\cite{Lin:2017snn,Constantinou:2020hdm}, the dominant paradigm for PDF determinations remains their phenomenological extraction from a global QCD analysis~\cite{Gao:2017yyd,Kovarik:2019xvh,Ethier:2020way,Rojo:2019uip} from a wide range of hard-scattering processes, see~\cite{Hou:2019efy,Bailey:2020ooq,Alekhin:2017kpj,NNPDF:2017mvq,Ball:2021leu,Moffat:2021dji,ATLAS:2021vod} for recent analyses.

The determination of the parton distributions of the proton by means of a global analysis is, however, a rather challenging endeavour, which requires the resolution of delicate issues that can otherwise compromise the reliability of the results obtained. An incomplete list of topics that need to be dealt with in a global PDF fit include  limitations of fixed-order theory calculations, internal or external inconsistencies of the experimental measurements, incomplete correlation models, choice of techniques for PDF error estimation and propagation, choice of PDF parametrisation, implementation of theoretical constraints on the PDF shape (positivity, integrability, counting rules, or Regge behaviour), treatment of heavy-quark PDFs~\footnote{See also Sect.~22 of~\cite{SM:2010nsa} for a benchmarking of the general-mass heavy quark schemes used in PDF fits. The differences between the schemes were found to be modest and well understood, especially at NNLO.}, and the choice of SM parameters, such as $\alpha_s(m_Z)$, heavy-quark masses, and CKM matrix elements. In-depth  understanding of the differences and similarities between global PDF determinations requires dedicated benchmark exercises involving close collaboration between the different PDF fitting groups, and with the experimental groups that published the fitted data. 

With this motivation, the PDF4LHC Working Group was established in 2008~\cite{DeRoeck:2009zz} in order to coordinate, facilitate, and promote scientific discussions and collaborative projects within the PDF theory and experimental LHC communities. The first PDF4LHC benchmarking exercise was performed in 2010~\cite{Alekhin:2011sk}, resulting in an initial set of recommendations~\cite{Botje:2011sn} for PDF usage at Run I of the LHC. Subsequently, several dedicated studies and benchmark exercises were carried out~\cite{Ball:2012wy,Rojo:2015acz,Andersen:2016qtm,Andersen:2014efa}, often in the collaborative context of the Les Houches workshops. In 2015, following a year-long study, the PDF4LHC15 combined sets were released~\cite{Butterworth:2015oua} together with an updated set of recommendations for PDF usage and uncertainty estimation at the LHC Run II. PDF4LHC15 was based on the combination of the CT14~\cite{Dulat:2015mca}, MMHT2014~\cite{Harland-Lang:2014zoa}, and NNPDF3.0~\cite{NNPDF:2014otw} global analyses and benefited from a number of technical developments regarding the transformation of Hessian PDF sets into their MC representation~\cite{Watt:2012tq,Hou:2016sho} and vice versa~\cite{Gao:2013bia, Carrazza:2015aoa,Carrazza:2016htc} and the replica compression of MC sets~\cite{Carrazza:2015hva}. The PDF4LHC15 combined PDF sets were made available via the {\sc\small LHAPDF} interface~\cite{Buckley:2014ana} and have been extensively used by the theoretical and experimental LHC communities.

Since the release of PDF4LHC15, several important developments have taken place in subjects of direct relevance to global PDF determinations. First of all, a large number of new datasets have been measured at the LHC, providing valuable information on the proton PDFs over a wide kinematic range and for many complementary flavour combinations. Secondly, a number of landmark NNLO QCD calculations~\cite{Heinrich:2020ybq} have been completed for processes of key relevance for global PDF fits, specifically inclusive jet~\cite{Currie:2016bfm} and dijet~\cite{Currie:2017eqf} production, direct photon production~\cite{Campbell:2016lzl}, differential top quark pair production~\cite{Czakon:2015owf}, and charged-current (CC) deep-inelastic scattering (DIS) with heavy-quark mass effects~\cite{Gao:2017kkx}. Thirdly, recent years have witnessed steady progress in the development of novel fitting methodologies, improved parametrisation strategies, techniques for error estimation, and machine learning algorithms. An update of the PDF4LHC15 combination is thus both timely and relevant, especially given the upcoming restart of data-taking at the LHC during Run III and the subsequent  high-luminosity era~\cite{Cepeda:2019klc,Azzi:2019yne}.

The goal of this work is to present PDF4LHC21, a combination of three recent global PDF analyses, CT18~\cite{Hou:2019efy}, MSHT20~\cite{Bailey:2020ooq}, and NNPDF3.1~\cite{NNPDF:2017mvq}, and to study its implications for the phenomenology program of the LHC Run III. A prerequisite for this combination has been an extensive set of benchmark studies aiming to understand better the origin of the differences between the three global PDF fits in terms of their input data, theory settings, and fitting methodology. Special attention has been paid to the assumptions underlying the experimental correlation models in the interpretation of high-precision LHC measurements. These  are often limited by systematic uncertainties, see e.g.~\cite{Harland-Lang:2017ytb,Bailey:2019yze,Thorne:2019mpt,Boughezal:2017nla,AbdulKhalek:2020jut,Amat:2019upj,Hou:2019gfw,Aad:2015mbv,ATLAS:2018owm,Amoroso:2020lgh,ATLAS:2021vod} and references therein. The new NNPDF4.0 PDF set~\cite{Ball:2021leu,NNPDF:2021uiq} was only released after the benchmarking exercise leading to PDF4LHC21 was completed, and hence will not be included. Comparisons between NNPDF4.0 and PDF4LHC21 are presented in App.~\ref{app:nnpdf40}. Furthermore, the present study has benefited from the lessons provided by independent PDF studies carried out by the ATLAS~\cite{ATLAS:2021qnl,ATLAS:2021vod} and CMS~\cite{CMS:2021yzl} Collaborations, while not explicitly including them in the combination.

One goal of the present study is to disentangle the effects of the fitted data and the settings of the theory calculations from those associated with the respective fitting methodologies, including the choice of PDF parametrisation, the implementation of theoretical constraints, or the error estimation techniques adopted. To achieve this, it is useful for the three groups to perform fits on a common dataset, with common parameter settings, for the purposes of the benchmarking only (full global PDF sets are then used later in forming the PDF4LHC21 combination). This common {reduced} dataset, representing an intersection of the sets fitted by each group, has a smaller total number of data points than the groups' default datasets. A central result of this study has thus been the production and comparison of variants of CT18, MSHT20, and NNPDF3.1 each based on this common reduced dataset, with the settings of the underlying theory calculations also homogenised as far as possible. 
As will be shown, while the agreement between the three groups is greatly improved for these fits to the common reduced dataset, there remain differences that should therefore be attributed to the methodological choices made by each group.

By and large, the results of the benchmark studies presented in this work demonstrate that the differences observed between the three global PDF sets can be explained by genuinely valid choices related to the input dataset, theory settings, and fitting methodology adopted in each case. It is not our goal here to resolve these differences by imposing a single choice of ``optimal'' settings, but rather to consider the resulting spread of results as a genuine contribution to a conservative estimate of PDF uncertainties in LHC processes~\cite{Ball:2021dab}. We thus proceed with the PDF4LHC21 combination by using the same procedure as in PDF4LHC15, namely we combine an equally large number of Monte Carlo replicas from the global fits of each group and then either compress the resulting ensemble or construct a Hessian representation. The PDF4LHC21 combination obtained in this manner is found to be consistent with the previous PDF4LHC15 combination and exhibits a modest reduction in PDF uncertainties in some critical LHC processes, notably for electroweak gauge boson and Higgs boson production measurements. This finding is confirmed by an extensive series of comparisons  between PDF4LHC21 and PDF4LHC15, as well as with individual PDF sets, at the level of parton distributions at low and high energy scales, partonic luminosities, and representative inclusive, fiducial, and differential LHC cross sections.

The outline of this paper is as follows. First of all, Sect.~\ref{sec:combination_inputs} summarises the main features of the three global sets that enter the PDF4LHC21 combination and compares them at the level of PDFs and of partonic luminosities. Sect.~\ref{sec:benchmarking} presents the outcome of the benchmark studies carried out between these three PDF sets, in particular the fits based on an identical reduced dataset. Following this preparatory analysis, the PDF4LHC21 combination is constructed and described in Sect.~\ref{sec:pdf4lhc21}, the optimal number of compressed replicas and Hessian eigenvectors is determined, and the PDF4LHC21 and PDF4LHC15 combinations are compared. The implications of the PDF4LHC21 combination for LHC processes are assessed in Sect.~\ref{sec:phenomenology}. In Sect.~\ref{sec:usage} we list the PDF4LHC21 sets that are released, provide prescriptions to evaluate uncertainties in LHC processes, and give usage recommendations for specific applications. Finally in Sect.~\ref{sec:conclusions}  we summarise our main results and consider briefly possible future developments. 

A number of technical discussions are collected in the appendices: a review of the tools relevant for Monte Carlo combination, compression, and Hessian reduction of PDF fits (App.~\ref{app:tools}); a study of the interplay between NNPDF4.0 and PDF4LHC21 (App.~\ref{app:nnpdf40}); further dedicated studies utilising the reduced fits to investigate differences of interest in the global fits (App.~\ref{app:specifics}), and a summary of the $L_2$-sensitivity studies (App.~\ref{app:l2_sensitivity}).
\section{Inputs to the PDF4LHC21 combination}
\label{sec:combination_inputs}

In this section, we describe the parton sets that are used as input to the PDF4LHC21 combination: CT18, MSHT20, and NNPDF3.1.  Variants of  CT18 (\CTprime) and NNPDF3.1 (\NNprime) that involve changing the heavy quark masses to a common value and a small variation of input data sets for NNPDF3.1, will also be discussed (and compared to their parent PDFs). It is these variants, plus the default MSHT20 PDFs, that will ultimately be used for the combination in PDF4LHC21. 

\subsection{CT18}
\label{subsec:ct18}

The CT18 PDFs~\cite{Hou:2019efy} are the newest general-purpose PDF release from the CTEQ-TEA (CT) collaboration, and fit to a wide range of high-energy data, including high-precision LHC experiments at 7 and 8 TeV, the HERA I+II combined data set~\cite{H1:2015ubc}, as well as the default sets included in the CT14 analysis \cite{Dulat:2015mca}. The CT14 PDFs were included in the PDF4LHC15 combination; CT14HERA2, as its name implies, came out after the CT14 PDFs and included in addition the HERAI+II data sets~\cite{Hou:2016sho}.  CT18 NNLO includes a total of 3681 data points from over 39 different experiments, with almost 700 data points from LHC experiments. These data were selected out of about two dozen of candidate LHC data sets examined in pre-fit studies. Compared to the previous fits to less precise data, the CT18 analysis elevated stringency of goodness-of-fit criteria according to the general approach laid out in~\cite{Kovarik:2019xvh}, as summarised below. The experimental data sets were selected to accommodate these criteria.

\begin{table}[!t]
  \small
  \renewcommand{\arraystretch}{1.2}
\begin{center}
\small
\hskip-0.7cm
\begin{tabular}{|lr|c|c|c|}
\hline
 \textbf{Experimental data set $E$} &  & $N_{\mathrm{pt}}$   & $\chi^{2}/N_{\mathrm{pt}}$  & $S$ \tabularnewline
\hline
  LHCb 7 TeV 1.0 fb$^{-1}$ $W/Z$ forward rapidity                           & \cite{LHCb:2015okr}        &  33   &  1.63 (  1.21) &  2.3 ( 0.9) \tabularnewline 
  LHCb 8 TeV  2.0 fb$^{-1}$ $Z\rightarrow e^{-} e^{+}$ forward rapidity     & \cite{LHCb:2015kwa}        &  17   &  1.04 (  1.06) &  0.2 ( 0.3) \tabularnewline
  ATLAS 7 TeV 4.6 fb$^{-1}$, $W/Z$ combined$^{\ddagger}$                    & \cite{ATLAS:2016nqi}      &  34   &  8.45 (  2.61) & 16 ( 5.1) \tabularnewline
 CMS 8 TeV 18.8 fb$^{-1}$ muon charge asymmetry $A_{ch}$                                & \cite{CMS:2016qqr} &  11   &  1.04 (  1.10) &  0.2 ( 0.3) \tabularnewline
  LHCb 8 TeV 2.0 fb$^{-1}$ $W/Z$ cross sec.                                            & \cite{LHCb:2015mad}        &  34   &  2.17 (  1.75) &  4.0 ( 2.7) \tabularnewline 
  ATLAS 8 TeV 20.3 fb$^{-1}$, $Z$ $p_T$ cross sec.                                    & \cite{ATLAS:2015iiu}         &  27   &  1.12 (  1.05) &  0.5 ( 0.2) \tabularnewline
  CMS 7 TeV 5 fb$^{-1}$, single incl. jets, $R=0.7$          & \cite{CMS:2014nvq}  & 158   &  1.23 (  1.19) &  2.0 ( 1.7) \tabularnewline
  ATLAS 7 TeV  4.5 fb$^{-1}$, single incl. jets, $R=0.6$                    & \cite{ATLAS:2014riz}         & 140   &  1.45 (  1.45) &  3.4 ( 3.4) \tabularnewline
  CMS 8 TeV 19.7 fb$^{-1}$, single incl. jets, $R=0.7$, (extended)     & \cite{CMS:2016lna} & 185   &  1.14 (  1.12) &  1.3 ( 1.2) \tabularnewline
  CMS 8 TeV 19.7 fb$^{-1}$, $t\bar{t}$ norm. double-diff. top $p_T$ and $y$  & \cite{CMS:2017iqf}    &  16   &  1.18 (  1.19) &  0.6 ( 0.6) \tabularnewline
  ATLAS 8 TeV 20.3 fb$^{-1}$, $t\bar{t}$ $p_{T}^{t}$ and $m_{t\bar{t}}$ abs. spectrum    & \cite{ATLAS:2015lsn}         &  15   &  0.63 (  0.71) & -1.1 (-0.8) \tabularnewline
\hline
\end{tabular}
\end{center}
\caption{Numbers of points, $\chi^2/N_{pt}$, and the effective Gaussian variables for the newly added LHC measurements in the CT18 and CT18Z NNLO fits.
The ATLAS 7 TeV $W/Z$ data (4.6 fb$^{-1}$), labelled by $\ddagger$, are included in the CT18A and CT18Z global fits, but not in CT18 and CT18X. 
\label{tab:CT18_LHC} }
\end{table}

The LHC data in the CT18 NNLO fit -- the default of four fits that also include CT18Z, A, and X -- are listed in Table~\ref{tab:CT18_LHC}. The non-LHC data sets can be found in the CT18 paper. Shown in the table for each data set $E$ are the number of data points, $N_{\mathrm{pt}, E}$, the $\chi^2$ values for those data, and $S_E$, an equivalent Gaussian variable \cite{Lai:2010vv,Dulat:2013hea,Kovarik:2019xvh} that quantifies the level of agreement with $E$ as the difference of $\chi^2_E$ from its global best-fit value in units of the standard deviation for $E$, equal to $\sqrt{2 N_{{\rm pt},E}}$ for large enough $N_{{\rm pt},E}$.  
In this paper, we adopt 
\begin{equation}
S_E=\sqrt{2\chi^2_E}-\sqrt{2N_{{\rm pt},E}-1} 
\label{eq:S}
\end{equation}
in accord with the definition in \cite{Fisher:1925,Lai:2010vv}. (The subscripts ``$E$'' are omitted in the tables.) Positive values of $S_E$ above two units indicate that the data set is not described well by the PDF fit. Large negative values may indicate overfitting or overestimated experimental errors. The philosophy in CT18 has been to include all points in a particular data set in order to cover as wide a kinematic range as possible (for example, the full rapidity range for the LHC jet data). Such kinematic coverage makes use of the full constraining power of the data set and also reveals instances where there may be conflicts, {\it e.g.}, between different rapidity intervals. Where relevant, statistical correlations are taken into account to allow for the use of multiple observables for a given process, such as for the ATLAS top-quark pair production data.  

To better dissect the potential sensitivity and PDF impact of candidate experimental data sets, two numerical packages were developed for fast preliminary analysis: \texttt{PDFSense}~\cite{Wang:2018heo,Hobbs:2019gob} and \texttt{ePump}~\cite{Schmidt:2018hvu,Hou:2019gfw}. \texttt{PDFSense} can predict which data sets will have the largest impact on the global PDF fit, and \texttt{ePump} applies Hessian probability to quickly estimate  the impact  of the data on the fit before the actual fit is carried out. These programs help to select the new data sets that will have the greatest impact on the PDFs. 

The CTEQ global PDF fitting code itself was parallelized in order to allow a fast turn-around time when running on high performance clusters. For much of the data, \texttt{APPLgrid}~\cite{Carli:2010rw}/\texttt{fastNLO}~\cite{Wobisch:2011ij,Britzger:2012bs} tables were computed (to be multiplied by point-by-point NNLO/NLO K-factors). Also, \texttt{fastNNLO}~\cite{Czakon:2017dip} tables were used for computing NNLO $t\bar{t}$ cross sections. 

As with CT14HERA2, the SACOT-$\chi$ heavy quark scheme at NNLO \cite{Guzzi:2011ew} was used, with a charm pole mass of 1.3 GeV and a bottom pole mass of 4.75 GeV. CT18 places kinematic cuts on the data used in the global fit that reduces the possible impact of any deuteron corrections. DIS cross sections on iron (CCFR, CDHSW, and NuTeV) and proton-copper Drell-Yan (E605) data are corrected to the corresponding cross sections on deuterium using a phenomenological parametrization of the nuclear-to-deuteron cross section ratios. This model is acceptable with the present accuracy of data \cite{Accardi:2021ysh}.

The $x$ dependence of the PDFs is parametrized by Bernstein polynomials, multiplied by the standard $x^a$ and $(1-x)^b$ factors that control the small-$x$ and large-$x$ behaviors, respectively. There are 5-8 independent fitting parameters for each parton flavour, while the strangeness PDF has four fitting parameters. Some parameters may be determined by sum rules or fixed to physically reasonable values. For example, the CT18 parameters are such that at large $x$ the PDFs are non-negative (to avoid having negative differential cross sections that are common with sign-indefinite PDFs) and compatible with quark counting rules in accord with common models of the nonperturbative nucleon structure \cite{Courtoy:2020fex,Courtoy:2021xcc}. Dependence on the parametrization was tested by redoing the fits using a large number of parametrization forms. The fits improved (with stable results) for up to around 30 parameters. Even more parameters tended to destabilize the fits. 

The final PDF ensemble, based on 29 parameters, is distributed in the form of the central PDF and 58 Hessian error PDFs (for positive/negative eigenvector directions) for estimation of the PDF uncertainties. While this ensemble utilises one out of many acceptable PDF functional forms, given in Appendix C in \cite{Hou:2019efy}, and, similarly, it employs one possible prescription for estimating the systematic errors, the published PDF uncertainties by construction cover central PDFs obtained in the candidate fits with alternative settings. Many such alternative choices have been explored: in particular, the final PDF uncertainties seen in Fig.~6 of \cite{Hou:2019efy} cover the central PDFs obtained with more than 250 alternative functional forms as well as by varying either the QCD scales in some experiments or the prescription for estimation of experimental correlated systematic errors.\footnote{In the default CT18 approach, the relative systematic errors provided by the experiments are converted into the absolute ones by the multiplicative ``extended $T$'' prescription developed in the previous CT papers \cite{Stump:2003yu,Lai:2010vv,Ball:2012wy,Gao:2013xoa,Dulat:2015mca,Hou:2016nqm}.}

\begin{figure}[!b]
\includegraphics[width=1\textwidth]{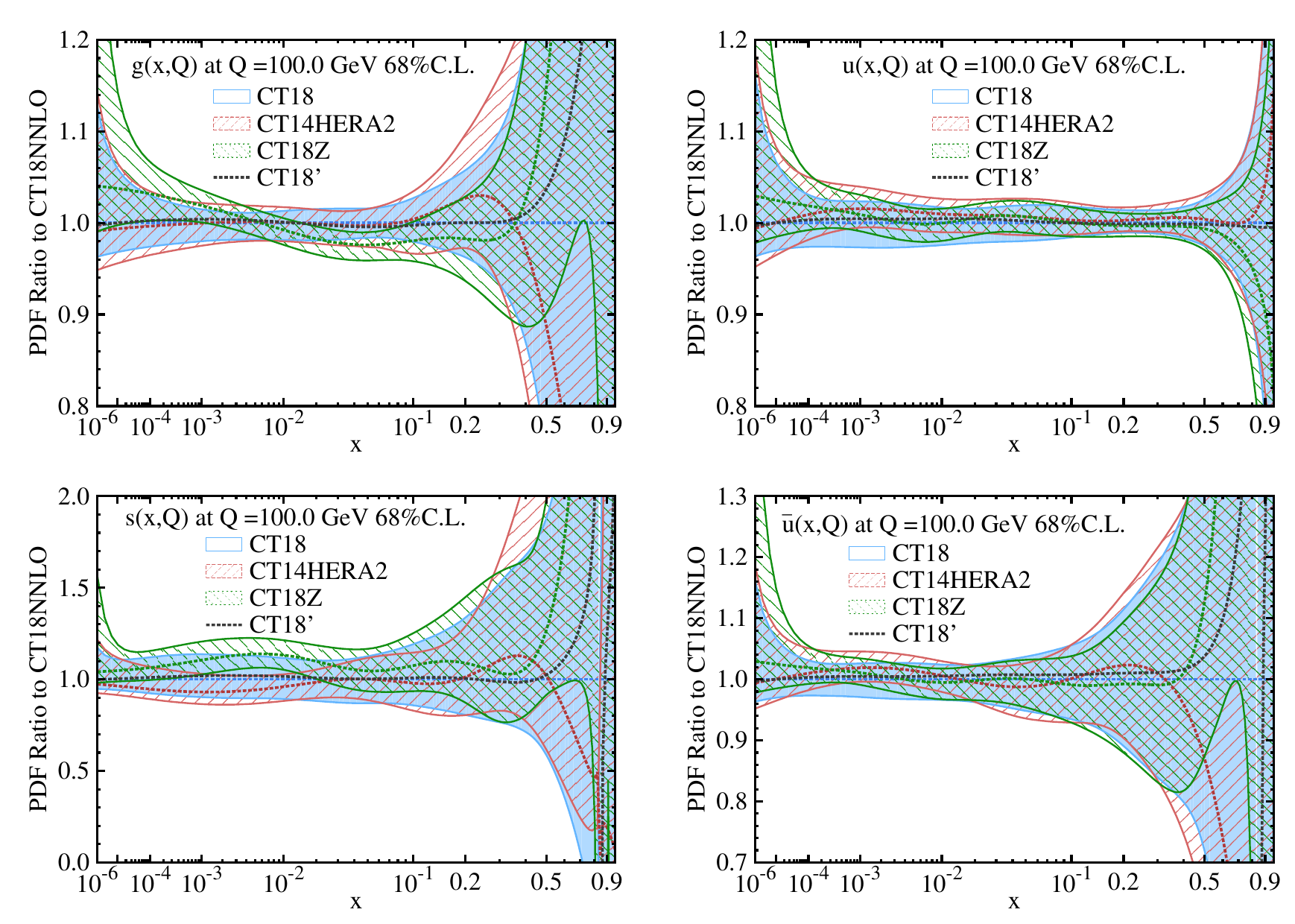}
\caption{A comparison of the CT18${}^\prime$, CT18, CT18Z, and CT14HERA2 parton distribution functions and their uncertainties at NNLO  for
	$g, u, s,$ and  $\bar{u}$ at $Q=100$ GeV. }
\label{fig:ct18pdf}
\end{figure}

The key outcome of this detailed analysis is that the spread of PDF solutions is considerably augmented due to the variety of methodological choices that can be made, as well as due to some inconsistencies between the fitted experiments that impose competing pulls on the PDFs in some regions. The final uncertainty must reflect this spread, and hence the CT18 uncertainties are enlarged comparatively to those according to the $\Delta\chi^2=1$ criterion or based on a single functional form. The CT18 uncertainty balances between two competing demands of precision and robustness. While the CT18 analysis can obtain a smaller estimated uncertainty that is close to the MSHT20 one by using the dynamic ``tier-2" tolerance, such estimate is less robust under the explored variations of the underlying assumptions, as those may produce excursions outside of the nominal uncertainty bands. 

The size of the CT18 uncertainty also reflects considerations about the quality of the fits. 
The CT18 analysis performs minimisation of the global log-likelihood function $\chi^2$ as the key statistic quantifying the overall agreement of theory and data.
By definition, the central PDF corresponds to the minimum $\chi^2$ solution for all experiments. For each fitted experiment, $\chi^2$  takes into account the statistical errors and (correlated and uncorrelated) systematic errors of a data set.  
By this conventional measure, employed in particular to gauge the quality of the fits throughout this article, the CT18 fit obtains good overall agreement with the full data set. It takes a further step and applies {\it strong} goodness-of-fit tests \cite{Kovarik:2019xvh} to examine internal consistency of the resulting fits. 
For a number of key data sets or even their parts, including several LHC experiments with $S_E > 2$ in Table~\ref{tab:CT18_LHC}, the CT analysis finds enhanced values of $\chi^2_E$  that are improbable if differences between data and theory are purely random. These enhancements also exist in the fits by the other groups, as seen in Tables~\ref{tab:MSHT20_LHC} and \ref{tab:NNPDF31_LHC}. With such enhanced $\chi^2_E$ values, the PDFs in future fits are likely to show more variability, and hence the CT18 uncertainty must account for these inconsistencies, or ``tensions". 

\begin{figure}[t]
	\begin{center}
		\includegraphics[width=0.65\textwidth]{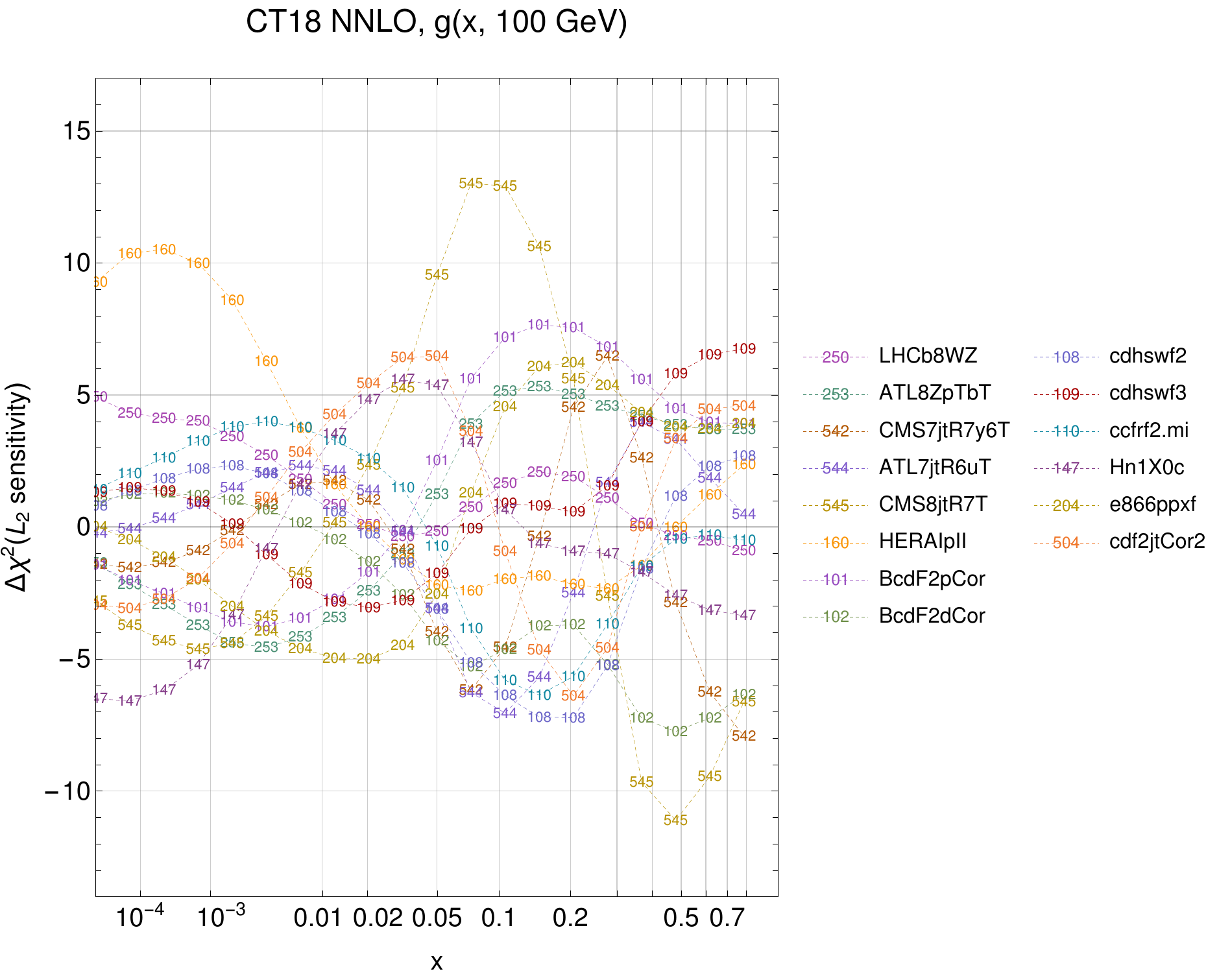}
	\end{center}
	\vspace{-2ex}
	\caption{
		The $x$-dependent $L_2$ sensitivity of the CT18 data sets with strongest
		pull upon the gluon PDF, $g(x,Q\!=\!100\,\mathrm{GeV})$. A number of tensions
		among the leading data sets are revealed by examining those regions of $x$
		where $S_{f, L2}(E)$ peaks in the `positive
		direction' for certain experiments, while $S_{f, L2}(E)$ is sharply negative for others.
	}
\label{fig:L2glu}
\end{figure}

For those applications in which the full span of PDF solutions is essential, this analysis provides a supplemental PDF set called CT18Z that is compatible with CT18 within their respective 90\% confidence level (CL) uncertainties. The CT18Z ensemble has an elevated strangeness PDF at $x=0.02-0.1$ as a result of inclusion of the ATLAS 7 TeV W/Z data set \cite{ATLAS:2016nqi}. These data were left out of CT18 due to the apparent tension with the other data sets, cf. Table~\ref{tab:CT18_LHC}, that weakens the fit according to the strong goodness-of-fit criteria, a tension observed by the other fitting groups as well, if not so severely. The other prominent distinction of CT18Z is its enhanced low-$x$ gluon distribution, caused by using a new $x_B$-dependent choice of the factorisation scale for the DIS data that mimics the effects of small-$x$ resummation and results in a reduction of $\chi^2_E$ of over 50 units for the combined HERA data set. 

The CT18Z ensemble achieves a $\chi^2/N_\mathrm{pt}$=1.19, for a total of 3493 data points. (The numbers for the $\chi^2$  and $S_E$ values for CT18Z for the LHC experiments are also given in Table~\ref{tab:CT18_LHC}.) For comparison, $\chi^2/N_\mathrm{pt}$=1.17 for the CT18 fit at NNLO. The default CT18 ensemble is sufficient for most applications; together, CT18 and CT18Z provide a more complete map of the PDF solutions. 

Figure~\ref{fig:ct18pdf} compares the central values and uncertainties for the gluon, up quark, strange quark and $\bar{u}$ antiquark for  CT18, CT14HERA2 and CT18Z NNLO PDFs at a scale of $100$ GeV. The CT18 gluon distribution is similar to that of CT14HERA2, with a reduction in uncertainty and with the gluon being somewhat softer at very high $x$, primarily due the influence of the LHC jet data. 

In the same figure, the black dashed curves indicate the central PDFs of a modified version of the CT18 Hessian PDFs, designated as ``\CTprime'', that enters the PDF4LHC21 combination discussed in this article. The ``\CTprime'' PDFs assume the charm pole mass  $m_c^{\mbox{\tiny pole}}=1.4\mbox{ GeV}$, which leads primarily to the shown changes in the central PDFs, without tangibly modifying the uncertainties. For the purpose of the PDF4LHC21 combination, the Hessian \CTprime NNLO ensemble shown in Fig.~\ref{fig:ct18pdf} is approximated by an ensemble of 300 Monte-Carlo replicas. The Hessian and Monte Carlo ensembles are equivalent within the accuracy of the input ensemble, their minor numerical differences reflect stochastic fluctuations during the replica generation \cite{Hou:2016sho}.

While not recommended for the general use, CT18A and CT18X are two auxiliary fits that lie between CT18 and CT18Z and include only the ATLAS 7 TeV $W/Z$ data and only the $x$-dependent DIS scale, respectively. 
Hessian eigenvector sets are provided for all these PDF ensembles at both NLO and NNLO (a LO PDF set, although not recommended, is in progress). Recently, a CT18QED NNLO analysis with two realisations of the \texttt{LUX} model \cite{Manohar:2016nzj,Manohar:2017eqh} for the photon PDF was also released \cite{Xie:2021equ}. 

\begin{figure}[tb]
 \includegraphics[width=0.50\textwidth]{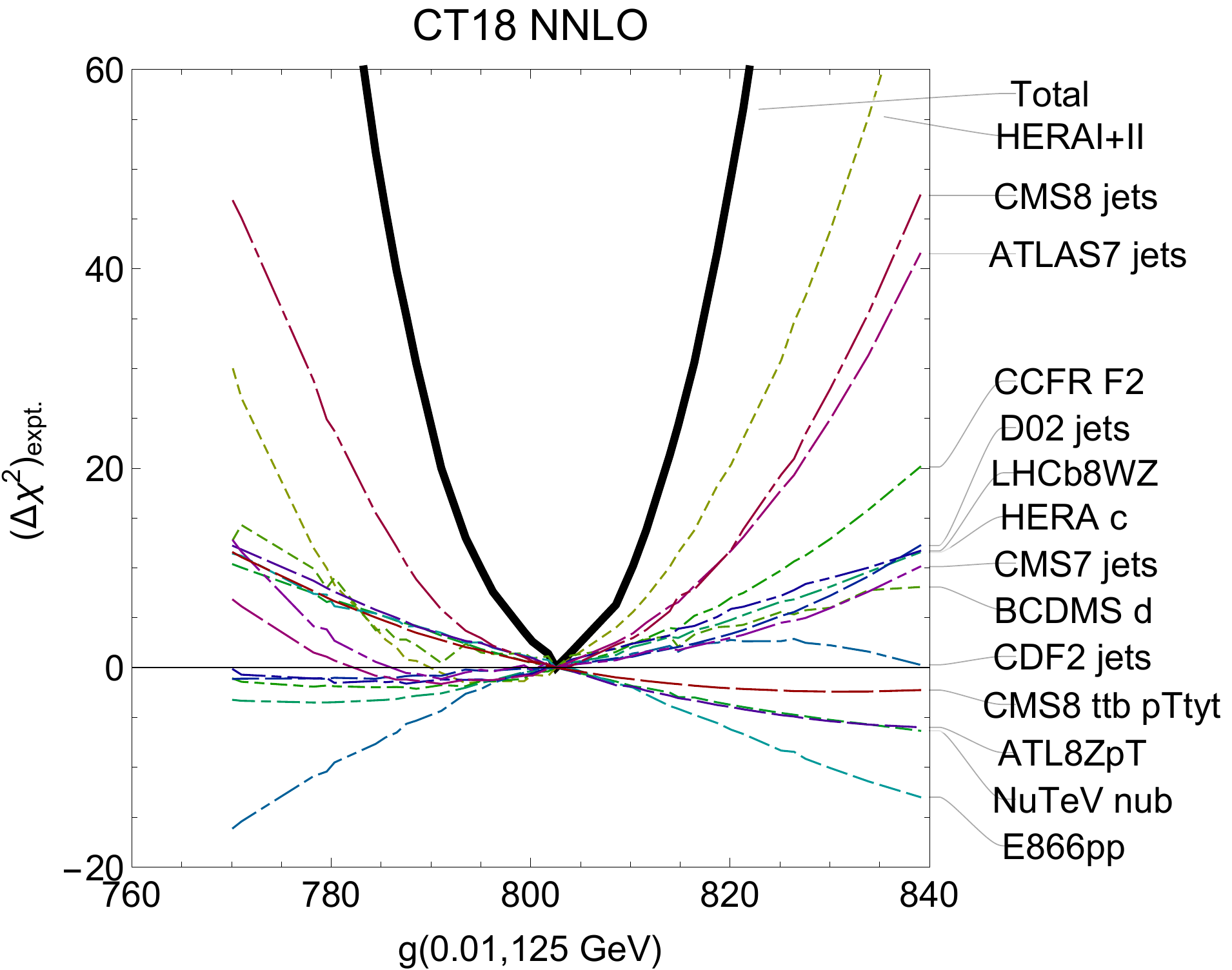}\quad
  \includegraphics[width=0.50\textwidth]{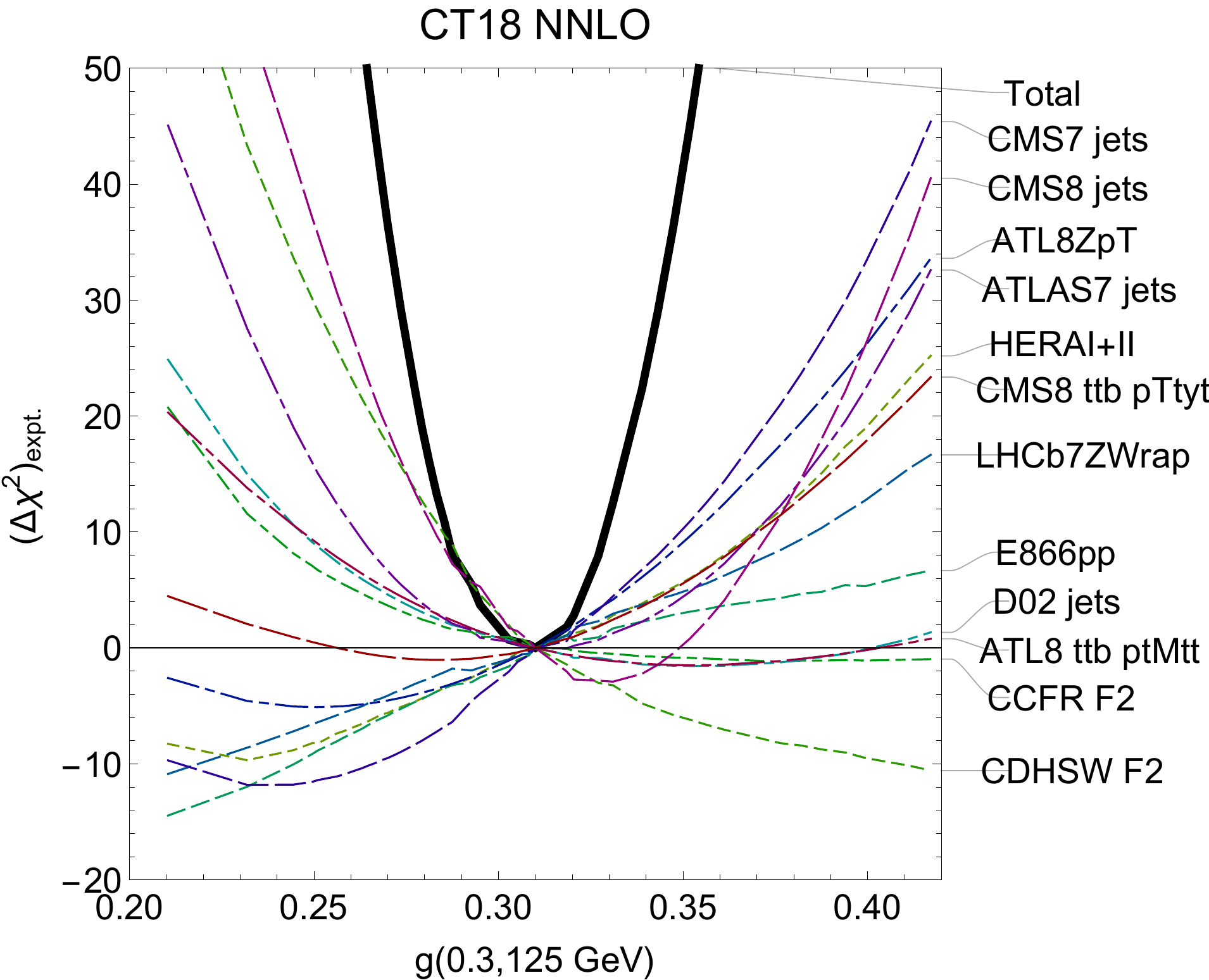}
	\caption{LM scans for the gluon PDF at $Q=125$ GeV and $x=0.01$ (left) and $x=0.3$ (right), based upon the CT18 NNLO fits.
		\label{fig:LMg18}}
\end{figure}

We conclude this section by mentioning two elucidating techniques to explore interplay of constraints from individual experiments directly within the fits. 
The $L_2$ sensitivity technique \cite{Hobbs:2019gob} is summarized in App.~\ref{app:l2_sensitivity}. It quantifies the degree to which each data set influences the global PDF fit (for a particular parton, or for a particular parton-parton luminosity) as a function of a given kinematic variable (parton $x$, parton-parton mass, etc). It also indicates the tensions that exist among the data sets.   Plotting $S_{f, L2}(E)$
against $x$ yields useful information regarding the pulls of the CT18(Z) data sets 
upon the PDFs or PDF combinations. 
This also permits
rapid visualization of possible tensions directly within a given fit, observed when a PDF variation of some parton density is correlated with  variations of $\chi^2_E$ for some experiments ({\it i.e.}, $S_{f, L2}(E) > 0$), while it is 
anti-correlated with other experiments ($S_{f, L2}(E) < 0$) at the same values of $(x, Q)$.
An example is given in Fig.~\ref{fig:L2glu}, the $L_2$ sensitivity for the gluon distribution at a $Q$ value of 100 GeV. At an $x$ value of 0.01, relevant for Higgs production through gluon-gluon fusion, experiments such as the HERA I+II data want to decrease the magnitude of the gluon, while experiments such as E866 and the ATLAS 8 TeV $Z$ $p_T$ data prefer a larger gluon at this $x$ value. This clearly demonstrates the discriminating power of the $L_2$ sensitivity for exploring the mutual agreement of the data sets included in a global fit.  

While the $L_2$ sensitivity approximately estimates the experimental sensitivities over a wide range of parton $x$, the Lagrange Multiplier (LM) scan technique~\cite{Stump:2001gu} allows a detailed examination of the constraining power of a data set at a particular value of $x$. 
The LM scans and $L_2$ sensitivity are especially consistent with one another in identifying the leading experiments with the strongest pulls on the PDFs in the kinematic region under consideration. This can be seen in Fig.~\ref{fig:LMg18} for the gluon at $x=0.01, ~Q= 125$ GeV (left) and at $x=0.3$ (right). Additional cross-checks of the final fits include an examination of the distributions of the best-fit nuisance parameters and of the systematic error shifts required, as well as comparisons of the shifted data points to the unshifted data points for each experiment. All are needed for a complete understanding of the resultant PDFs.

\subsection{MSHT20}
\label{subsec:msht20}

The MMHT14 PDFs \cite{Harland-Lang:2014zoa} have recently been superseded by the MSHT20 \cite{Bailey:2020ooq} sets. The acronym MSHT is now intended to be a permanent naming convention and 
stands for Mass Scheme Hessian Tolerance, i.e. it incorporates some of the
central and enduring features of the approach. 
The analysis includes new theoretical developments, and an extended parameterisation, in particular for $\bar d/\bar u$ and the strange quark and more eigenvector sets. 
There is much new, largely LHC data, but also final HERA and Tevatron data sets and very
nearly all cross sections are included at NNLO in QCD perturbation theory. 
The fit quality is generally very good, but there are some 
problems with correlated uncertainties and tensions for some data sets.
It is found that NLO QCD is clearly no longer sufficient 
for real precision. 

As in the MMHT14~\cite{Harland-Lang:2014zoa} analysis, heavy flavour in DIS is obtained using 
a general mass variable flavour scheme based on 
the TR scheme~\cite{Thorne:1997ga,Thorne:2006qt}, using the ``optimal'' choice~\cite{Thorne:2012az}
for smoothness near threshold. Deuteron and heavy nuclear corrections are applied, the 
former being fit using a $4$ parameter model, 
as in MMHT14 and the latter use the same corrections~\cite{deFlorian:2011fp} as MMHT14 with 
the fit allowing an additional penalty-free freedom of order $1\%$. 
Data are fit using systematic uncertainties using either nuisance parameters if possible 
(the preferred method) or with
the correlation matrix provided, and statistical correlations are also applied whenever these are 
available.
(Some old data sets which are dominated  by uncorrelated uncertainties and/or where 
there is a limited 
understanding of correlations have errors added in quadrature.) 
In general there is a fit to absolute cross sections in preference to normalised in order to avoid loss of information from normalisations. 

The analysis includes many new NNLO corrections compared to MMHT14.
Use is made of the NNLO calculations for dimuon 
production~\cite{Berger:2016inr}, where the correction is negative, and larger in size at 
lower $x$, allowing the strange quark to be larger in the fit to 
the dimuon data and helping to relieve tension between the dimuon 
data~\cite{NuTeV:2001dfo} and LHC $W,Z$ data~\cite{ATLAS:2016nqi,ATLAS:2017rue,ATLAS:2019fgb}
which prefers a larger strange quark. Nearly all other data have the theoretical 
calculations at full NNLO precision, in particular NNLO cross-section calculations~\cite{Currie:2016bfm} 
for all LHC jet data included, i.e.  
inclusive jet production at 2.76, 7 and 8~TeV, using the larger available 
jet radius, e.g. $R=0.6,0.7$ and scales $\mu_{R,F}=p_{T,jet}$. (Older Tevatron jet  data 
are still included but with the threshold approximation for NNLO~\cite{Kidonakis:2000gi} - which is a better approximation 
for these data which also carry little weight.) 
CMS $7$~TeV$~W+c$ data~\cite{CMS:2013wql} have only NLO theory available for the specific 
measurement, but the few data points carry little weight.
The $Z p_T$ distribution and all top quark cross sections used are included at full NNLO.
Electroweak  corrections are included where possible, if these are not already 
subtracted from the data supplied. 

There has been a very significant extension of the parameterisation. 
In MMHT14 the parameterisation used for PDFs was $A(1-x)^{\eta} x^{\delta}
(1+\sum_{i=1}^n a_i T_i(1-2x^{1/2}))$, where $T_i(1-2x^{1/2}))$ are Chebyshev polynomials.
In \cite{Martin:2012da} it was demonstrated how the achieved precision possible could improve with 
increasing $n$ using a fit to pseudo-data. In MMHT14 $n=4$ was deemed sufficient, 
but using $n=6$ will lead to much better than $1\%$ precision. Hence, MSHT now  
extend the parameters of different flavour PDFs using 
$n=6$ and also now parameterise $(\bar d/\bar u)$ instead of $(\bar d - \bar u)$, with 
the sole constraint on the ratio being that $(\bar d/\bar u)\to $ constant  as $x\to 0$. 
This leads to significant improvements in the global fit; mainly from 
changing to $(\bar d /\bar u)(x,Q_0^2)$, and extending $d_V(x,Q_0^2)$, $g(x,Q_0^2)$  
and $s^+(x,Q_0^2)$,  with overall $\Delta \chi^2 = -73$. 
Overall there is an improvement in the fit to high-$x$ fixed-target data, 
a reduction in tension between E866 DY ratio data and LHC data, 
and an improvement in the description of the LHC lepton asymmetry data.  
Using $n=6$ for the parameterisations except for $s - \bar s$, means an increase 
to 52 parton parameters. As for MMHT14 the default $\alpha_S(M_Z^2)=0.118$.

The first new data set to be updated compared to the MMHT14 PDFs was the final HERA total 
cross section data~\cite{H1:2015ubc}. This was studied in \cite{Harland-Lang:2016yfn} and found to  have a limited effect on the PDFs, 
but there was some trouble fitting the lower $Q^2, x$ data. 
Also included is the  final combined $\tilde{\sigma}^{\bar c c}$ and 
$\tilde{\sigma}^{\bar b b}$ data~\cite{H1:2018flt} where the fit at low $Q^2$ is not optimal, but similar results are seen in other PDF studies~\cite{H1:2018flt}.
Another important additional new data set is D0 electron/$W$ asymmetry~\cite{D0:2013lql}.
The $W^{\pm}$ boson is produced preferentially 
in the proton/antiproton direction, but 
the $V\!-\!A$ structure of the lepton decay means $e^{\pm}$
is emitted preferentially opposite to $W^{\pm}$ -- 
leptons at particular $\eta^e$ come from a range of 
$\eta^W$ values and dilute the direct constraint on PDFs at given $x$. 
Mapping the lepton to $W$ asymmetry requires 
PDF-dependent modelling, with a small uncertainty and 
this gives a more direct constraint from $W$ asymmetry data. 
MSHT20 see a reduced uncertainty on $d/u$ compared to using the $e$ asymmetry, particularly at
very high $x$, where $d_V$ is reduced.

The MSHT20 analysis contains a large amount of new LHC data. This includes
extremely high precision data on $W,Z$ at 7~TeV from ATLAS,
and high precision $W^{\pm}$ data and double differential $Z$ data at 8~TeV;
CMS 8~TeV precise data on the $W^{+,-}$ rapidity distribution;
LHCb data at $7$ and $8$~TeV on $W,Z$ rapidity
distributions at higher rapidity; $W+c \,\,{\rm jets}$ data at $7$~TeV from CMS;
ATLAS high mass Drell Yan data at 8~TeV;
ATLAS data on $W^{\pm}~+$  jets at 8~TeV; $Z\,\,p_T$ distributions at 8~TeV;
new data on $\sigma_{t \bar t}$ at 8~TeV plus ATLAS
single differential distributions in $p_{T,t}, M_{t \bar t}, y_t, y_{t \bar t}$
and CMS double differential distributions in $p_{T,t},y_t$ both at 8~TeV;
inclusive jet data from ATLAS at 7~TeV and CMS at 2.76, 7 and 8~TeV.
All these recent LHC data updates are included in the fit at NNLO (except for $W+c$).  
The fit quality is generally good, as seen in Table \ref{tab:MSHT20_LHC}. There are 
relatively poor $\chi^2$ values for some sets,
seemingly observed by other groups.

\begin{table}[!t]
  \small
   \renewcommand{\arraystretch}{1.2}
\begin{center}
\begin{tabular}{|l|c|c|c|c|}
\hline
{\bf Experimental data set}          & $N_{pt}$    &   $\chi^2/N_{\rm pt}$     & $S$  \\
\hline
D0 $W$ asymmetry \cite{D0:2013lql}                                  & 14    & 0.86                    &  -0.3  \\
$\sigma_{t \bar t}$ Tevatron +CMS+ATLAS $7,8$~TeV  \cite{CDF:2013hmv}-\cite{CMS:2013hon}&  17   & 0.85&  -0.4    \\
LHCb 7$+$8 TeV $W+Z$ \cite{LHCb:2015okr,LHCb:2015kwa}                              &  67   & 1.48     &  2.6   \\
LHCb 8 TeV $e$ \cite{LHCb:2015mad}                                    &   17  & 1.54                  &  1.5    \\
CMS 8 TeV  $W$ \cite{CMS:2016qqr}                                    &   22  & 0.58                   &  -1.5   \\
ATLAS 7 TeV jets $R=0.6$ \cite{ATLAS:2014riz}                          &  140  & 1.59                 &  4.4     \\
CMS 7 TeV   $W+c$    \cite{CMS:2013wql}                               & 10    & 0.86                  &  -0.2   \\
ATLAS 7 TeV $W,Z$   \cite{ATLAS:2016nqi}                               &  61   & 1.91                 &  4.3     \\
CMS 7 TeV jets $R=0.7$  \cite{CMS:2014nvq}                          & 158   & 1.11                    &  1.0     \\
ATLAS 8 TeV $Z p_T$   \cite{ATLAS:2015iiu}                             & 104   & 1.81                 &  5.0     \\
CMS 8 TeV jets     \cite{CMS:2016lna}                                & 174   & 1.50                   &  4.2      \\
ATLAS 8 TeV $t \bar t \to l + {\rm j}$ single-diff \cite{ATLAS:2015lsn} & 25  & 1.02                  &  0.1      \\
ATLAS 8 TeV $t \bar t \to l^+l^-$ single-diff   \cite{ATLAS:2016pal}   & 5   & 0.68                   &  -0.4      \\
ATLAS 8 TeV high-mass Drell-Yan    \cite{ATLAS:2016gic}                 & 48  & 1.18                  &  0.9       \\
ATLAS 8 TeV $W^{+,-} + {\rm jet}$   \cite{ATLAS:2017irc}               & 32  & 0.60                   &  -1.7        \\
CMS 8 TeV $(d \sigma_{t \bar t}/d p_{T,t} dy_t)/\sigma_{t \bar t}$ \cite{CMS:2017iqf}    & 15  & 1.50 &  1.3            \\
ATLAS 8 TeV $W^+,W^-$   \cite{ATLAS:2019fgb}                           & 22  & 2.61                   &  4.2         \\
CMS 2.76 TeV jets      \cite{CMS:2015jdl}     & 81  & 1.27                                            &  1.7        \\
CMS 8 TeV $t \bar t$ $y_t$ distribution   \cite{CMS:2015rld}           & 9   & 1.47                   &  1.0         \\
ATLAS 8 TeV double differential $Z$ \cite{ATLAS:2017rue}                & 59  & 1.45                  &  2.3        \\
\hline
    \end{tabular}
\vspace{-0.0cm}
\vspace{+0.2cm}
\caption{
Numbers of points, fit qualities $\chi^2/N_{\rm pt}$ and $S$ values for new collider data added to the NNLO MSHT20 fit.
}
\label{tab:MSHT20_LHC}
\end{center}
\end{table}

The main effect of the new LHC data on the MSHT20 PDFs is on the details of flavour, i.e. the $d_V$ 
shape, an increase in the strange quark 
for $0.001<x<0.3$ and the $\bar d,\bar u$ details, 
though some of these are also partially from the parameterisation 
change.  There is a slight decrease in the high-$x$ gluon. 
Generally the fit is good, but the most straightforward approach gives a 
distinctly poor fit quality to some data sets due to
tensions between different kinematic regions (e.g. rapidity bins) or 
different differential distributions of the same data. Often, this is 
clearly related to modelling-type systematic 
uncertainties, particularly for jet and $t \bar t$ data, as illustrated 
in detail in \cite{Harland-Lang:2017ytb,Bailey:2019yze}, and for some data a smooth 
decorrelation, similar to that advocated for 8~TeV ATLAS inclusive jet data~\cite{ATLAS:2017kux}, is used.

MSHT20 goes from 25 eigenvector pairs to 32 - there is one extra parameter for each PDF and two for $s + \bar s$. 
The mean tolerance is $T\sim 3-4$. About half the constraints are primarily provided by precision 
electroweak collider data,
largely D0 $W$ asymmetry, $7~$TeV  and $8$~TeV ATLAS $W,Z$ and CMS $W$ data.  
8-10 eigenvectors are mainly constrained by the E866 Drell-Yan ratio which is 
vital for the $\bar d /\bar u$ constraint, 
$\sim$ 10 eigenvectors are constrained by fixed target DIS data (i.e. BCDMS, NMC, NuTeV, 
CCFR) and these data sets still mainly constrain high-$x$ quarks,
$\sim 10$ eigenvectors are constrained by CCFR, NuTeV dimuon data, i.e. 
this is still the main constraint on the strange quark and its asymmetry. 
Hence, a fully global fit is found to be necessary for a full constraint on all PDFs without 
use of assumptions and/or models. 
HERA data provides good constraints on the widest variety of PDF parameters, 
mainly the gluon and light sea, but now it is very rarely the best. However the HERA data 
are a very strong constraint on the best fit PDFs, and  central values and uncertainties 
at small $x$ are still strongly constrained 
by HERA data.

\begin{figure}[!t]
\centering
\includegraphics[width=0.48\textwidth,trim= 0.0cm 0.1cm 0.0cm 0.2cm,clip]{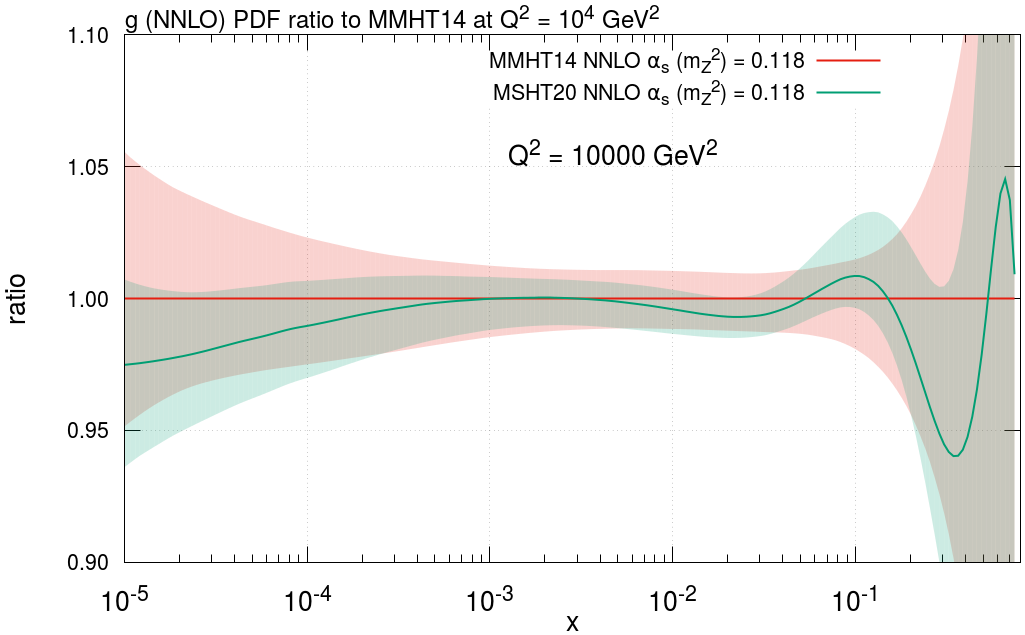} 
\includegraphics[width=0.48\textwidth,trim= 0.0cm 0.1cm 0.0cm 0.2cm,clip]{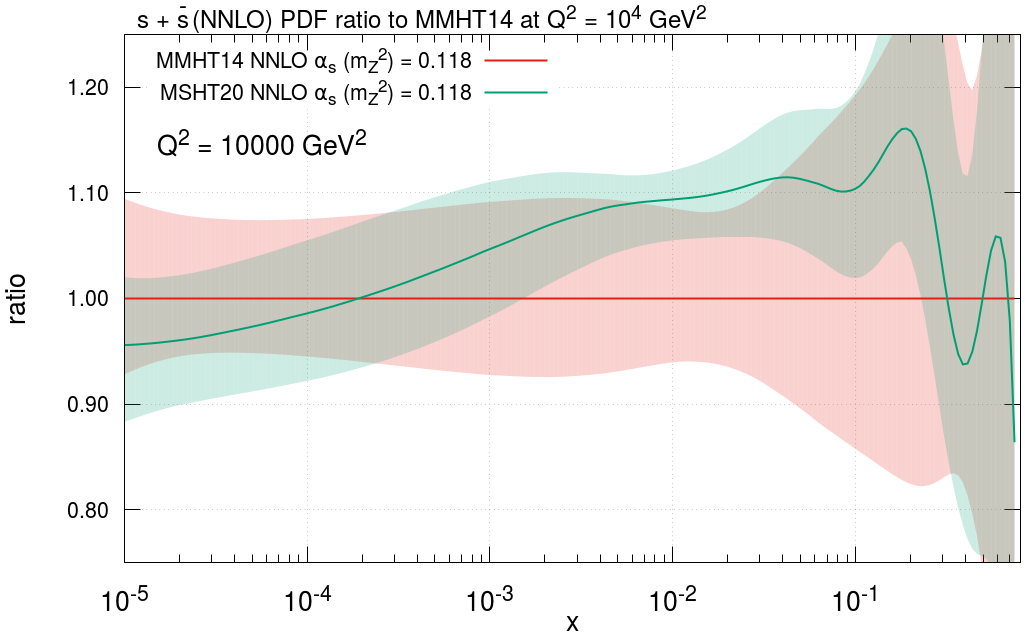}\\
\phantom{xxx}\\
\includegraphics[width=0.48\textwidth,trim= 0.0cm 0.1cm 0.0cm 0.2cm,clip]{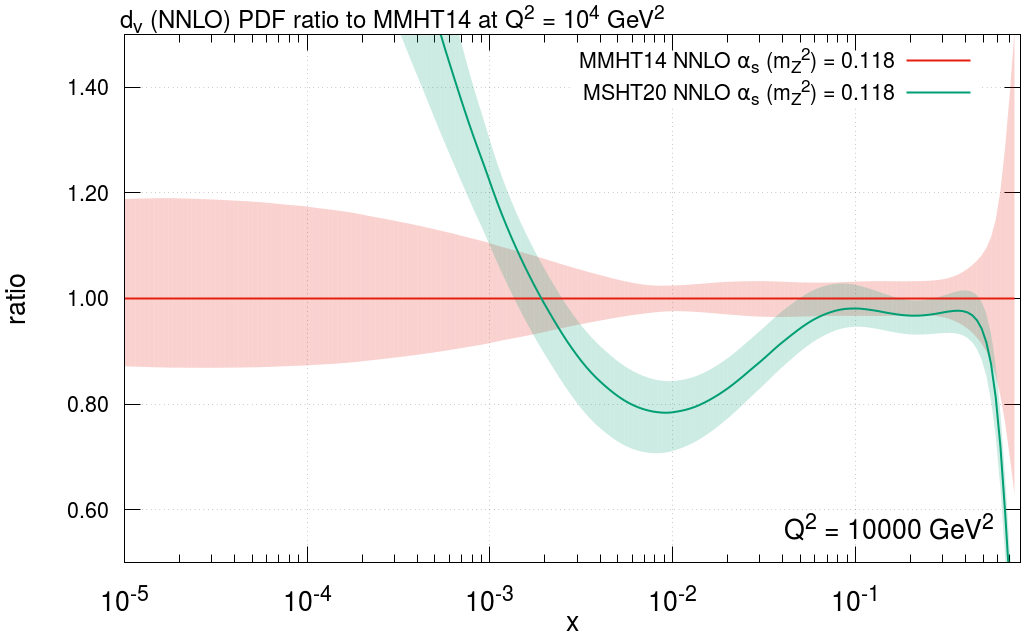} 
\includegraphics[width=0.48\textwidth,trim= 0.0cm 0.1cm 0.0cm 0.2cm,clip]{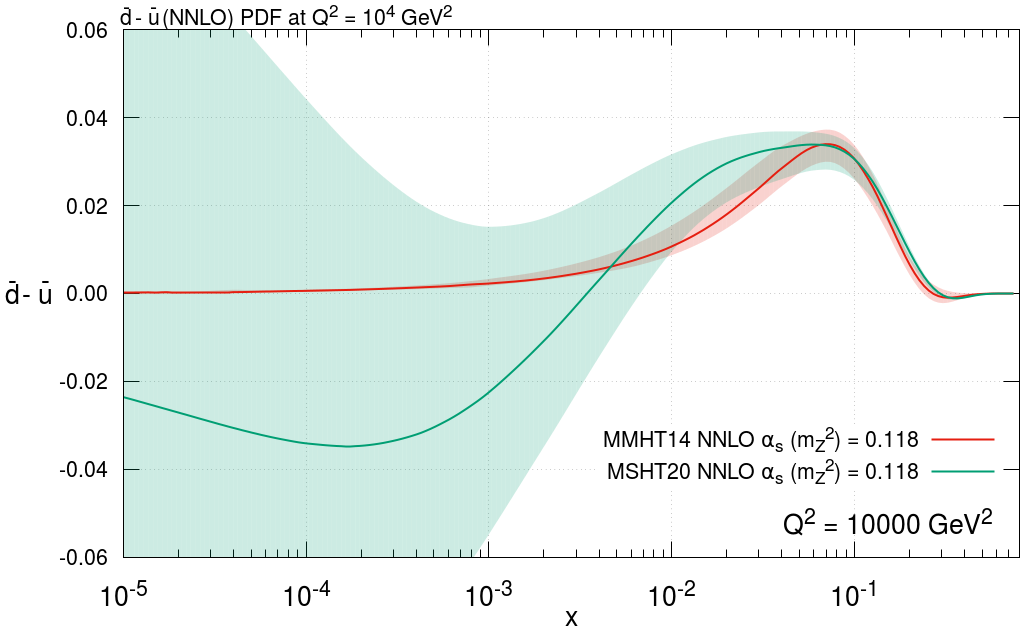}
\caption{The MSHT20 gluon (top left), strange quark sum (top right), down valence (bottom left) and $\bar d - \bar u$ (bottom right) compared to MMHT14.}
\label{fig:MSHTvsMMHT}
\end{figure}

We compare the new MSHT20 PDFs compared to those of MMHT14. 
First we show the gluon distribution, Fig. \ref{fig:MSHTvsMMHT} (top left), 
where there is no significant change 
in the central value, though the uncertainty is reduced. The details in shape 
at high $x$ depend on the LHC jet, $Z \,p_T$ and
differential $t \bar t$ data. The $Z \,p_T$ data pull the gluon up and differential 
$t \bar t$ data pulls the gluon down, each also affecting the lower $x$ normalisation 
via the momentum sum rule. 
Not all jet data pull in the same direction though the total effect is 
slightly downwards.
More significant changes in the PDFs include an increase in the strange quark below $x=0.1$, 
Fig. \ref{fig:MSHTvsMMHT} (top right), due to 
ATLAS 7, 8~TeV $W$ and $Z$ data which influence PDFs similarly. There is also a 
significant change in the shape in valence quarks, most notably 
$d_V$, due to LHC data on $W,Z$ and the improved parameterisation flexibility,
Fig. \ref{fig:MSHTvsMMHT} (bottom left).
The strange asymmetry is similar to MMHT14, but now is non-zero outside uncertainties. 
There is also a change in the details of light antiquarks at high-$x$ where 
constraints are weak, and a slight decrease at low $x$ due to compensation for 
the increase in the strange quark.
The details of the $\bar u, \bar d$ difference, shown in Fig. \ref{fig:MSHTvsMMHT} (bottom right) 
are completely changed due to 
the new type of parameterisation. 
There is a huge increase in uncertainty at small $x$, and a slight tendency 
for negative $\bar d -\bar u$. 
However, a different impression is formed by considering
$\bar d /\bar u$ which has small low-$x$ uncertainty 
and notably the ratio $\to 1$ as $x \to 0$ to a good accuracy even without this being 
a constraint. 

MSHT20 also includes PDFs at NLO (and even still at LO, where the fit is very 
poor). However, there is significant 
deterioration in fit quality for some of the precision LHC data,  and NNLO is now 
very much preferred. The strong coupling value obtained from the analysis is 
$\alpha_S(M_Z^2)=0.1174\pm{0.0013}$~\cite{Cridge:2021qfd}.
There are constraints from a variety of new LHC data, but in different 
directions. In general jet data prefer slightly lower, while $W,Z$ data prefer slightly higher $\alpha_S(M_Z^2)$, and  
no single new set constrains $\alpha_S(M_Z^2)$ more strongly than a number of older data sets. 
For quark masses, unlike previous results~\cite{Harland-Lang:2015qea} 
which preferred lower values 
($m_c^{{\rm pole}}\sim 1.25$~GeV), the default choice of $m_c^{{\rm pole}}=1.4$~GeV 
is close to optimal. 
There is no strong pull from the default choice  
$m_b^{{\rm pole}}=4.75$~GeV. The PDFs have also been presented with an inclusion of relevant electroweak corrections, in particular a photon parton distribution \cite{Cridge:2021pxm} using a very similar formalism to that in \cite{Harland-Lang:2019pla}. 

There are no direct constraints applied to the MSHT20 PDFS other than those imposed by data. Since there is a
finite flexibility in the parameterisation, and Chebyshev polynomials oscillate between values of $\pm 1$, it is possible for PDFs to become negative in principle. Indeed, the parameterisation of the gluon, which involves two separate terms, is designed to allow this possibility at input at very small $x$. In practice the gluon does become negative at some very low $x$ and $Q^2$ values, as will always happen eventually if performing backwards evolution, but this feature quickly disappears as evolution to higher scales takes place. At high values of $x$ the limitations in fluctuations allowed by a smooth parameterisation and data constraints, which make all PDFs positive in regions where data are constraining, result in any negative PDF values being at extremely high $x$ and with PDFs values of $10^{-4}$ or less. At small $x$ the three light quarks have a 
common power-like behaviour as $x \to 0$, but their normalisations are allowed to differ. For the up and down sea quarks this is a new feature for MSHT20 but, as mentioned above, the normalisations turn out to be very similar and the ratio has a small uncertainty.    

For clarity, unlike the case for the other groups, the MSHT input to PDF4LHC21 is unchanged from the MSHT20 published PDF set.

\subsection{NNPDF3.1}
\label{subsec:nnpdf31}

Input to the PDF4LHC21 combination discussed in this work is a variant of the
NNPDF3.1 analysis~\cite{NNPDF:2017mvq} called NNPDF3.1.1. This PDF set is made available through the NNPDF web site\footnote{See \href{https://nnpdf.mi.infn.it/nnpdf3-1-1/}{https://nnpdf.mi.infn.it/nnpdf3-1-1/}}, and, analogously to \CTprime, will be denoted as \NNprime henceforth. This supersedes the NNPDF3.0 parton set~\cite{NNPDF:2014otw} included in the previous PDF4LHC15 combination~\cite{Butterworth:2015oua}. As noted in Sect.~\ref{sec:introduction}, the NNPDF collaboration has recently delivered a more updated analysis, NNPDF4.0~\cite{Ball:2021leu}. Because this became available only while the PDF4LHC21 combination was already at an advanced stage, it is not considered here. A description of the NNPDF4.0 baseline PDF set and its comparison to the PDF4LHC21 combination is nevertheless presented in App.~\ref{app:nnpdf40}.

In comparison with NNPDF3.0, NNPDF3.1 incorporates legacy measurements for completed experiments and a significant number of new LHC measurements. In the first category there are the combined HERA measurements of inclusive NC and CC DIS cross-sections~\cite{H1:2015ubc}, the H1~\cite{H1:2009uwa} and ZEUS~\cite{ZEUS:2014wft} measurements of the bottom quark structure function $F_2^b(x,Q^2)$, and the Tevatron-D0 measurements of the $W$ asymmetry in the electron~\cite{D0:2014kma} and muon~\cite{D0:2013xqc} channels. In the second category there are various measurements performed by ATLAS, CMS and LHCb. For ATLAS, NNPDF3.1 includes: the inclusive $W^\pm$ and $Z$ distributions, differential in rapidity, measured at 7~TeV~\cite{ATLAS:2016nqi} (albeit only the subset corresponding to the central rapidity region); the low-mass DY distribution, differential in rapidity and invariant mass, measured at 7~TeV~\cite{ATLAS:2014ape}; the $Z$ boson distributions, double differential in the $Z$-boson transverse momentum and either in the $Z$-boson rapidity or in the invariant mass of the lepton pair, measured at 8~TeV~\cite{ATLAS:2015iiu}; the top quark pair production distribution, differential in the rapidity of the top quark and normalised to the total top quark pair production cross-section, measured at 8~TeV~\cite{ATLAS:2015lsn}; the total cross-sections for top quark pair production at 7, 8 and 13~TeV~\cite{ATLAS:2014nxi,ATLAS:2016zet}; and the single-inclusive jet production distributions, differential in rapidity and transverse momentum of the jet, measured at 7~TeV~\cite{ATLAS:2014riz}. For CMS, NNPDF3.1 includes: the $W^\pm$ distributions, differential in rapidity, measured at 8~TeV~\cite{CMS:2016qqr}; the $Z$ boson distribution, double differential in the $Z$-boson transverse momentum and rapidity, measured at 8~TeV~\cite{CMS:2015hyl}; the top quark pair production distributions, differential in the rapidity of the top quark pair and normalised to the total top quark pair production cross-section, measured at 8~TeV~\cite{CMS:2015rld}; the total cross-sections for top quark pair production at 7, 8 and 13~TeV~\cite{CMS:2016yys,CMS:2015yky}; and the single-inclusive jet distributions, differential in rapidity and transverse momentum of the jet, measured at 2.76~TeV~\cite{CMS:2015jdl}. For LHCb, NNPDF3.1 includes the complete set of inclusive $W$ and $Z$ production distributions, differential in rapidity, in the muon channel measured at 7 and 8~TeV~\cite{LHCb:2015okr,LHCb:2015mad}.

Theoretical predictions for nearly all the measurements included in NNPDF3.1 are performed at NNLO in the strong coupling $\alpha_s$, the exceptions being massive charm neutrino-DIS dimuon production and single-inclusive jet production, for which NNLO corrections were not available when the original NNPDF3.1 analysis was released. In these two cases, PDF evolution accurate to NNLO was combined with matrix elements accurate only to NLO. For very precise single-inclusive jet measurements, an additional fully correlated theoretical systematic uncertainty, estimated from scale variation of the NLO calculation, was incorporated in the total covariance matrix to account for missing higher-order corrections. For all DIS measurements, NNLO corrections are included exactly,  while for all other measurements NNLO corrections are implemented by means of $K$-factors, that is hadron-level bin-by-bin ratios of the NNLO to NLO predictions computed with a pre-defined PDF set, and applied to the NLO computation. The PDF dependence of the $K$-factors is much smaller than all other relevant uncertainties. Fast-interpolation tables, combining PDF and $\alpha_s$ evolution accurate up to NNLO with weighted grids for matrix elements accurate to NLO (obtained from various Monte Carlo generators depending on the process), are pre-computed with {\sc APFELgrid}~\cite{Bertone:2013vaa,Bertone:2016lga}. No electroweak corrections are applied upon checking that they never exceed experimental uncertainties. Likewise, no nuclear corrections are applied to DIS measurements involving deuterium or heavy nuclei as targets. 

In contrast to CT18 and MSHT20, in NNPDF3.1 the charm PDF is parametrised~\cite{Ball:2016neh} on the same footing as the light quark PDFs. The FONLL matched general-mass variable flavour number scheme~\cite{Forte:2010ta} is extended for this purpose~\cite{Ball:2015tna,Ball:2015dpa}. Within this formalism, a massive correction to the charm-initiated contribution is included alongside the contribution of fitted charm as a non-vanishing boundary condition to PDF evolution. At NNLO this correction requires knowledge of massive charm-initiated contributions to the DIS coefficient functions up to $\mathcal{O}(\alpha_s^2)$. Because these were known only to $\mathcal{O}(\alpha_s)$ when NNPDF3.1 was released~\cite{Kretzer:1998ju}, the NLO expression for this correction is used: this corresponds to setting the unknown $\mathcal{O}(\alpha_s^2)$ contribution to the massive charm-initiated term to zero. Parametrising charm leads to improvements in fit quality without an increase in PDF uncertainty, and it stabilizes the dependence of PDFs on the charm mass, all but removing it in the light quark PDFs. The values of the charm and bottom quark pole masses are set according to the Higgs cross section working group recommendation~\cite{LHCHiggsCrossSectionWorkingGroup:2016ypw}, namely $m_c=1.51$~GeV and $m_b=4.92$~GeV. The value of the strong coupling at the mass of the $Z$ boson is fixed to $\alpha_s(M_Z)=0.118$.

The NNPDF3.1 analysis has been used as baseline in several complementary studies of some of its theoretical aspects. First, a simultaneous determination of the strong coupling $\alpha_s$ and of PDFs, including correlations between the two, was carried out in~\cite{Ball:2018iqk}. All relevant sources of experimental, methodological and theoretical uncertainty were studied in detail, finding $\alpha_s(M_Z)=0.1185\pm 0.0005^{\rm (exp)}\pm 0.0001^{\rm meth}\pm 0.0011^{\rm th}$, in good agreement with the PDG average~\cite{ParticleDataGroup:2020ssz}. Second, the photon PDF was determined in a dedicated variant of the NNPDF3.1 analysis~\cite{Bertone:2017bme} by means of the LUXqed formalism~\cite{Manohar:2016nzj,Manohar:2017eqh}. A few percent uncertainty was found on the photon PDF, with photons carrying up to $\sim 0.5\%$ of the proton's momentum; corrections up to $\simeq 10\%$ ($\simeq 20\%$) due to photon-induced contributions were found for high-mass DY ($W^+W^-$) production. Third, in a variant of the NNPDF3.1 analysis~\cite{Ball:2017otu}, NLO and NNLO fixed-order PDF evolution and DIS structure functions were supplemented with NLO+NLL$x$ and NNLO+NLL$x$ small-$x$ resummation. A quantitative improvement in the perturbative description of the HERA inclusive and charm-production reduced cross-sections was observed. Finally, a general methodology to incorporate theoretical uncertainties in PDF determinations~\cite{Ball:2018lag} was used to study the impact of missing higher order uncertainty (MHOU) in the fixed-order QCD calculations~\cite{NNPDF:2019vjt,NNPDF:2019ubu} and of nuclear uncertainties~\cite{Ball:2018twp,Ball:2020xqw} in datasets involving nuclear targets. Results showed that, in both cases, PDF accuracy improves while PDF precision reduces only moderately.

The NNPDF3.1 analysis has been also incrementally extended to incorporate additional measurements, typically for LHC processes not previously used for PDF determination, in dedicated studies. Prompt photon production was addressed in~\cite{Campbell:2018wfu}; single top quark production in~\cite{Nocera:2019wyk}; di-jet production in~\cite{AbdulKhalek:2020jut}; and $W$+jet production in~\cite{Faura:2020oom}. As part of two of these analyses, the theoretical details entering the computation of the observables have been revisited: NNLO corrections were included systematically in the analysis of single-inclusive jet and dijet production~\cite{AbdulKhalek:2020jut}, as were NNLO massive corrections in the analysis of neutrino-DIS dimuon production~\cite{Faura:2020oom}. In the case of prompt photon and single top quark production, it was found that the data has little or no impact in the global fit, given the rather large uncertainty of the corresponding measurements; some impact was found in the case of $W$+jet production, which remains consistent with the rest of the NNPDF3.1 dataset; and a very significant impact, depending on the dataset analysed, together with hints of tension with other measurements in the NNPDF3.1 dataset (particularly top quark pair production), were found in the case of di-jet production.

\begin{table}[tb]
  \small
  \renewcommand{\arraystretch}{1.2}
\begin{center}
\begin{tabular}{|l|c|c|c|c|c|c|c}
\hline
& \multicolumn{3}{c|}{NNPDF3.1~\cite{NNPDF:2017mvq}}
& \multicolumn{3}{c|}{\NNprime}\\
\hline
{\bf Experimental data set} & $N_{\rm pt}$   & $\chi^2/N_{\rm pt}$ & $S$
        & $N_{\rm pt}$   & $\chi^2/N_{\rm pt}$  & $S$ \\
\hline
D0 $W$ electron asymmetry~\cite{D0:2014kma}
&  8 & 2.70 & $+2.70$ & 11 & 3.07 & $+3.64$ \\
D0 $W$ muon asymmetry~\cite{D0:2013xqc}
&  9 & 1.56 & $+1.18$ &  9 & 1.58 & $+1.21$ \\
ATLAS low-mass DY 7~TeV~\cite{ATLAS:2014ape}
&  6 & 0.90 & $-0.03$ &  6 & 0.89 & $-0.05$\\
ATLAS $W$,$Z$ 7~TeV~\cite{ATLAS:2016nqi}
& 34 & 2.14 & $+3.88$ & 61 & 1.99 & $+4.58$ \\
ATLAS $Z$ $p_T$ 8~TeV ($p_T$, $m_{\ell\ell}$)~\cite{ATLAS:2015iiu}
& 44 & 0.93 & $-0.28$ & 44 & 0.94 & $-0.23$ \\
ATLAS $Z$ $p_T$ 8~TeV ($p_T$, $y_Z$)~\cite{ATLAS:2015iiu}
& 48 & 0.94 & $-0.25$ & 48 & 0.95 & $-0.20$\\
ATLAS single-inclusive jets 7~TeV ($R=0.6$)~\cite{ATLAS:2014riz}
& 31 & 1.07 & $+0.33$ &140 & 1.25 & $+2.00$\\
ATLAS $\sigma_{t\bar t}^{\rm tot}$ 7, 8, 13~TeV~\cite{ATLAS:2014nxi,ATLAS:2016zet}
&  3 & 0.86 & $+0.04$ &  3 & 0.95 & $+0.15$\\
ATLAS $t\bar t$ $\ell$+jets 8~TeV (1/$\sigma$ $d\sigma/dy_t$)~\cite{ATLAS:2015lsn}
&  9 & 1.45 & $+0.99$ &  4 & 3.56 & $+2.69$\\
CMS $W$ rapidity 8~TeV~\cite{CMS:2016qqr}
& 22 & 1.01 & $+0.11$ & 22 & 1.03 & $+0.17$\\
CMS $Z$ $p_T$ 8~TeV~\cite{CMS:2015hyl}
& 28 & 1.32 & $+1.18$ & 28 & 1.34 & $+1.25$\\
CMS single-inclusive jets 2.76~TeV~\cite{CMS:2015jdl}
& 81 & 1.03 & $+0.23$ & --- & --- & --- \\
CMS single-inclusive jets 8~TeV~\cite{CMS:2016lna}
& --- & --- & --- & 185 & 1.30 & $+2.72$\\
CMS $\sigma_{t\bar t}^{\rm tot}$ 7, 8, 13~TeV~\cite{CMS:2016yys,CMS:2015yky}
&  3 & 0.20 & $-1.14$  & 3 & 0.18 & $-1.20$\\
CMS $t\bar t$ $\ell$+jets 8~TeV
(1/$\sigma$ $d\sigma/dy_{t\bar t}$)~\cite{CMS:2015rld}
&  9 & 0.94 & $-0.01$  & 9 & 1.67 & $+1.36$\\
CMS $t\bar t$ 2D 2$\ell$ 8~TeV
(1/$\sigma$ d$\sigma$/d$y_tdm_{t\bar t}$)~\cite{CMS:2017iqf}
& --- & --- &  --- & 16 & 0.81 & $-0.48$\\
LHCb $W,Z\to\mu$~7~TeV~\cite{LHCb:2015okr}
& 29 & 1.76 & $+1.55$ &  29 & 1.96 & $+3.11$\\
LHCb $W,Z\to\mu$~8~TeV~\cite{LHCb:2015mad}
& 30 & 1.37 & $+1.39$ &  30 & 1.36 & $+1.35$\\
\hline
\end{tabular}
\vspace{+0.3cm}
\caption{The numbers of points, $\chi^2/N_{\rm pt}$ and $S$ values for new collider data in the NNPDF3.1 fit~\cite{NNPDF:2017mvq} and in  the \NNprime fit variant adopted in the present combination. The Tevatron and LHC data sets already included in NNPDF3.0 are kept in NNPDF3.1, but not necessarily in \NNprime. These are not indicated in the table. Note that, despite the number of LHC data points being larger in \NNprime than in NNPDF3.1, the total number of data points in the two analyses is similar, mainly because the Tevatron single-inclusive jet measurements (not indicated in the table) are no longer included in \NNprime. See text for details. }
\label{tab:NNPDF31_LHC}
\end{center}
\end{table}

The variant of the NNPDF3.1 analysis used in this work, \NNprime, benefits from some of the studies outlined above. In terms of data, we replace the HERA measurements of charm~\cite{H1:2012xnw} and bottom~\cite{H1:2009uwa,ZEUS:2014wft} structure functions with their legacy counterparts~\cite{H1:2018flt}. An incorrect kinematic cut in the analysis of the D0 electron $W$ asymmetry~\cite{D0:2014kma} is amended. Likewise we correct a small bug affecting the CDF $Z$ rapidity distribution~\cite{CDF:2010vek}, whereby the last two bins had not been merged consistently with the published measurement. The ATLAS measurements of the $W$ and $Z$ cross-sections at 7~TeV, differential in rapidity~\cite{ATLAS:2016nqi}, are extended to include also the forward rapidity region. The implementation of the ATLAS normalised distribution for top quark pair production at 8~TeV, differential in the rapidity of the top quark~\cite{ATLAS:2015lsn}, is revisited by taking into account a new piece of information on statistical correlations, as discussed in~\cite{Amoroso:2020lgh}; the single-inclusive jet measurements from ATLAS~\cite{ATLAS:2013pbc} and CMS~\cite{CMS:2015jdl} at $\sqrt{s}=2.76$~TeV and from ATLAS~\cite{ATLAS:2011qdp} at $\sqrt{s}=7$~TeV are no longer included because NNLO QCD corrections are not available for these measurements, either at all or for the scale choice used for other jet data~\cite{Currie:2018xkj}. For similar reasons the CDF single-inclusive jet data~\cite{CDF:2007bvv} are also not included. These datasets were already removed in the NNPDF3.1-related studies mentioned above~\cite{Nocera:2019wyk,Faura:2020oom,Ball:2018iqk,NNPDF:2019vjt,NNPDF:2019ubu,Ball:2020xqw}. Finally, in order to make the \NNprime dataset more similar to the CT18 and MSHT20 ones, we include all the rapidity bins of the ATLAS single-inclusive jet measurements at 7~TeV~\cite{ATLAS:2014riz} while decorrelating systematic uncertainties across different bins (only the central rapidity bin was included in the original NNPDF3.1 analysis). For the same reason we also include the CMS single-inclusive jet production measurement at 8~TeV~\cite{CMS:2016lna} and the CMS normalised distribution for top quark pair production at 8~TeV, double differential in the rapidity of the top quark and in the invariant mass of the top quark pair~\cite{CMS:2017iqf}.

In terms of theoretical treatment the changes are the following. For DIS we correct a bug in the {\sc APFEL} computation of the NLO CC structure functions, that mostly affects the large-$x$ region; and we re-analyse the NuTeV dimuon cross-section data by including the NNLO charm-quark massive corrections~\cite{Gao:2017kkx,Berger:2016inr}, as explained in~\cite{Faura:2020oom}, and by updating the value of the branching ratio of charmed hadrons into muons to the PDG value~\cite{ParticleDataGroup:2020ssz}, as explained in~\cite{Ball:2018twp}. For fixed-target DY, we include the NNLO QCD corrections for the E866 measurement~\cite{Towell:2001nh} of the proton-deuteron to proton-proton cross-section ratio: these corrections had been inadvertently overlooked in NNPDF3.1. For single-inclusive jets, we update the  theoretical treatment of the ATLAS and CMS measurements at $\sqrt{s}=7$~TeV~\cite{ATLAS:2014riz,CMS:2012ftr}, by systematically including NNLO corrections with $K$-factors. Moreover NLO and NNLO theoretical predictions are computed with factorisation and renormalisation scales equal to the optimal scale choice advocated in~\cite{Currie:2018xkj}, namely, the scalar sum of the transverse momenta of all partons in the event,  see~\cite{AbdulKhalek:2020jut}. The same treatment is also adopted in the analysis of the newly added CMS single-inclusive jet measurements at 8~TeV~\cite{CMS:2016lna}. We finally choose the same values of the charm and bottom quark masses as in the MSHT20 analysis, namely $m_c=1.4$~GeV and $m_b=4.75$~GeV. All the other values of the physical parameters are as in the original NNPDF3.1 analysis.

A summary of the new measurements included in the NNPDF3.1 analysis and in its variant used in the PDF4LHC21 combination are provided in Table~\ref{tab:NNPDF31_LHC}. For each dataset we also indicate the number of data points included in the NNLO fits, the $\chi^2$ per number of data points and the value of the $S$ metric, as in Table~\ref{tab:CT18_LHC}. The overall fit quality deteriorates in comparison with the original NNPDF3.1 analysis. The deterioration, which brings the total $\chi^2$ per datapoint close to that of the MSHT20 analysis, see Table~\ref{tab:MSHT20_LHC}, is driven by a significant deterioration in the fit to the D0 $W$ electron asymmetry and the ATLAS and CMS top quark pair measurements. This pattern has been also observed in the recent NNPDF4.0 analysis~\cite{Ball:2021leu} and, in the case of top quark pair production, was traced back to tension with the additional jet data.

\begin{figure}[!t]
  \centering
  \includegraphics[width=0.49\textwidth]{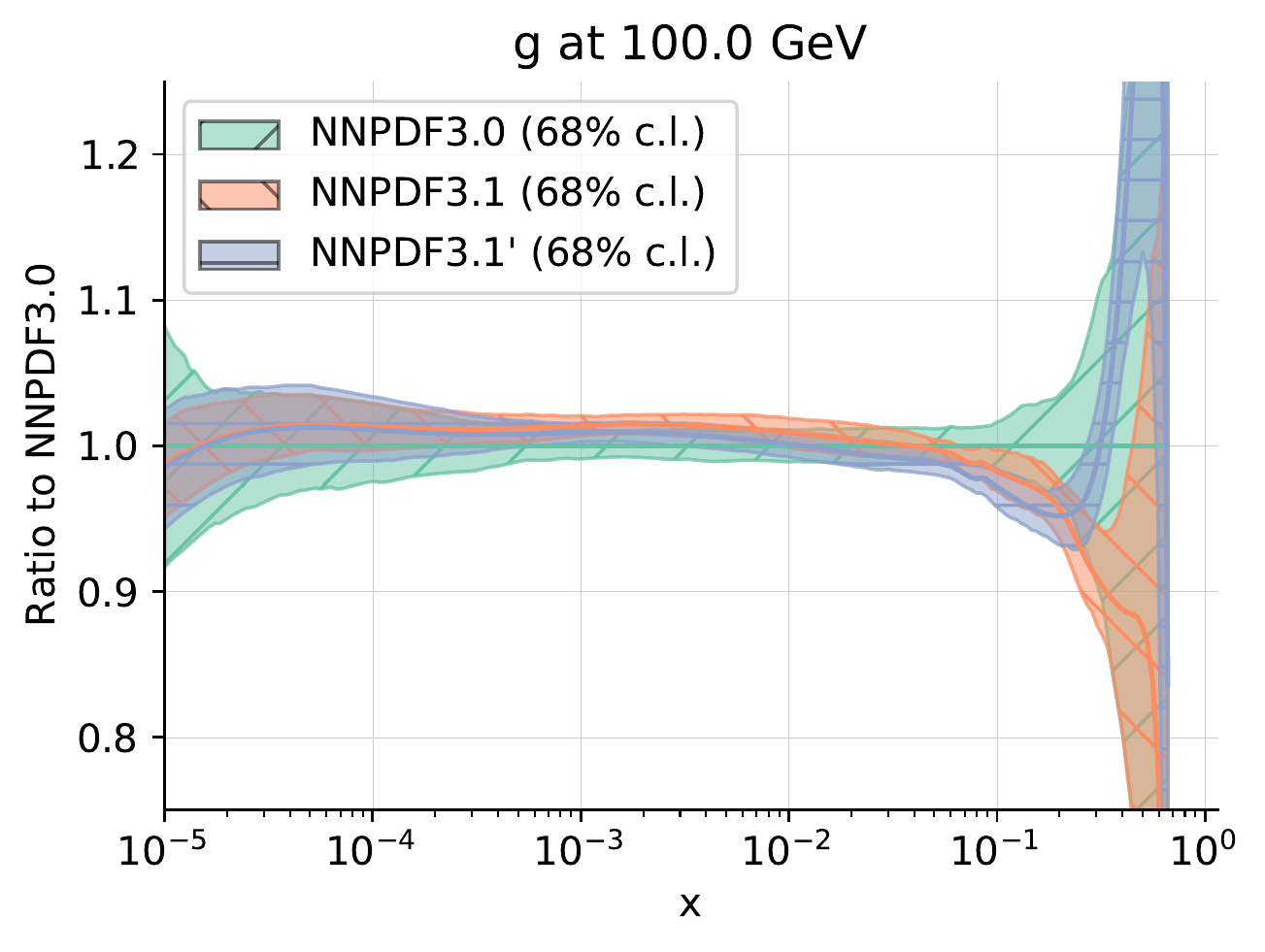}
  \includegraphics[width=0.49\textwidth]{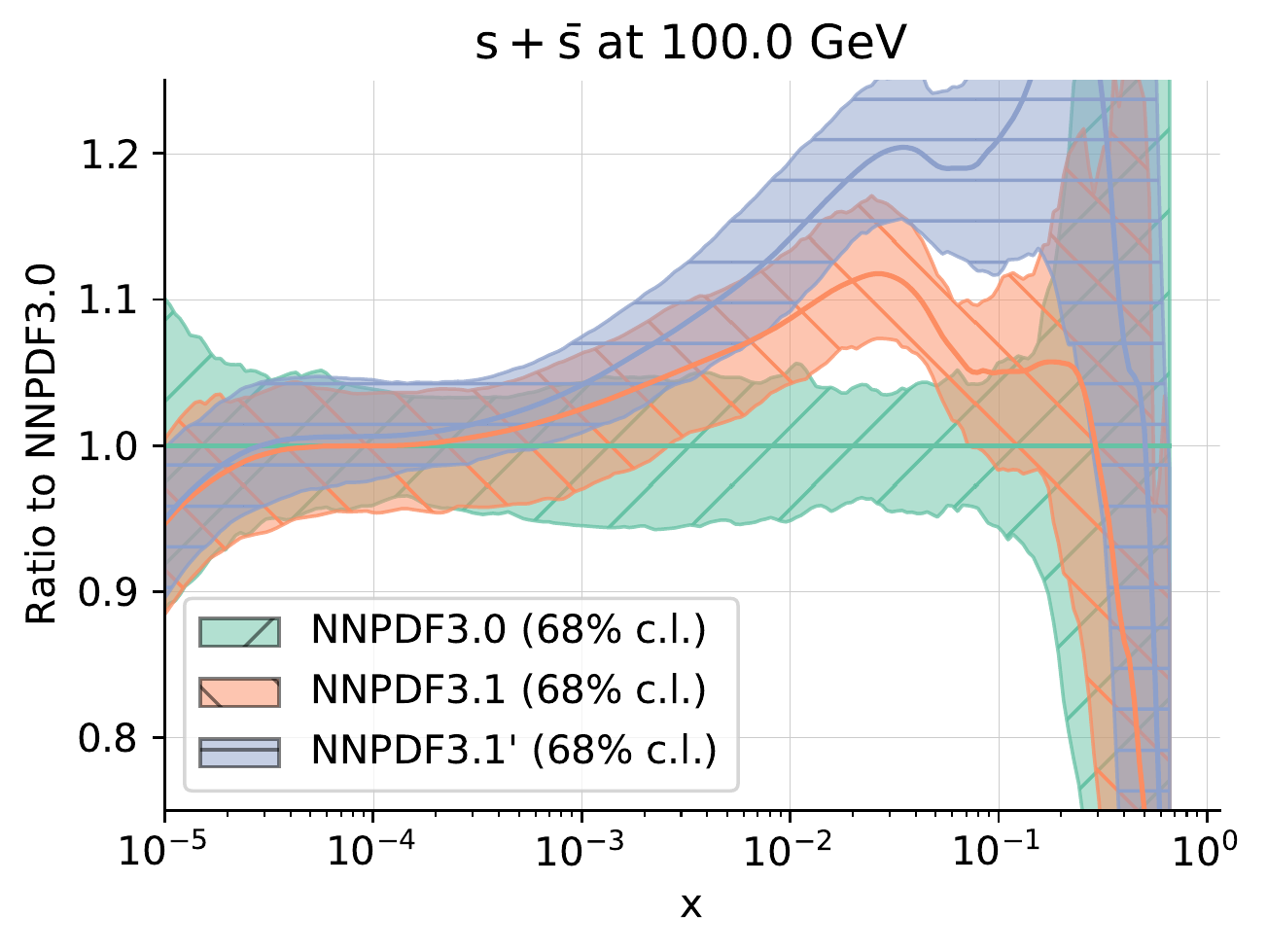}\\
  \includegraphics[width=0.49\textwidth]{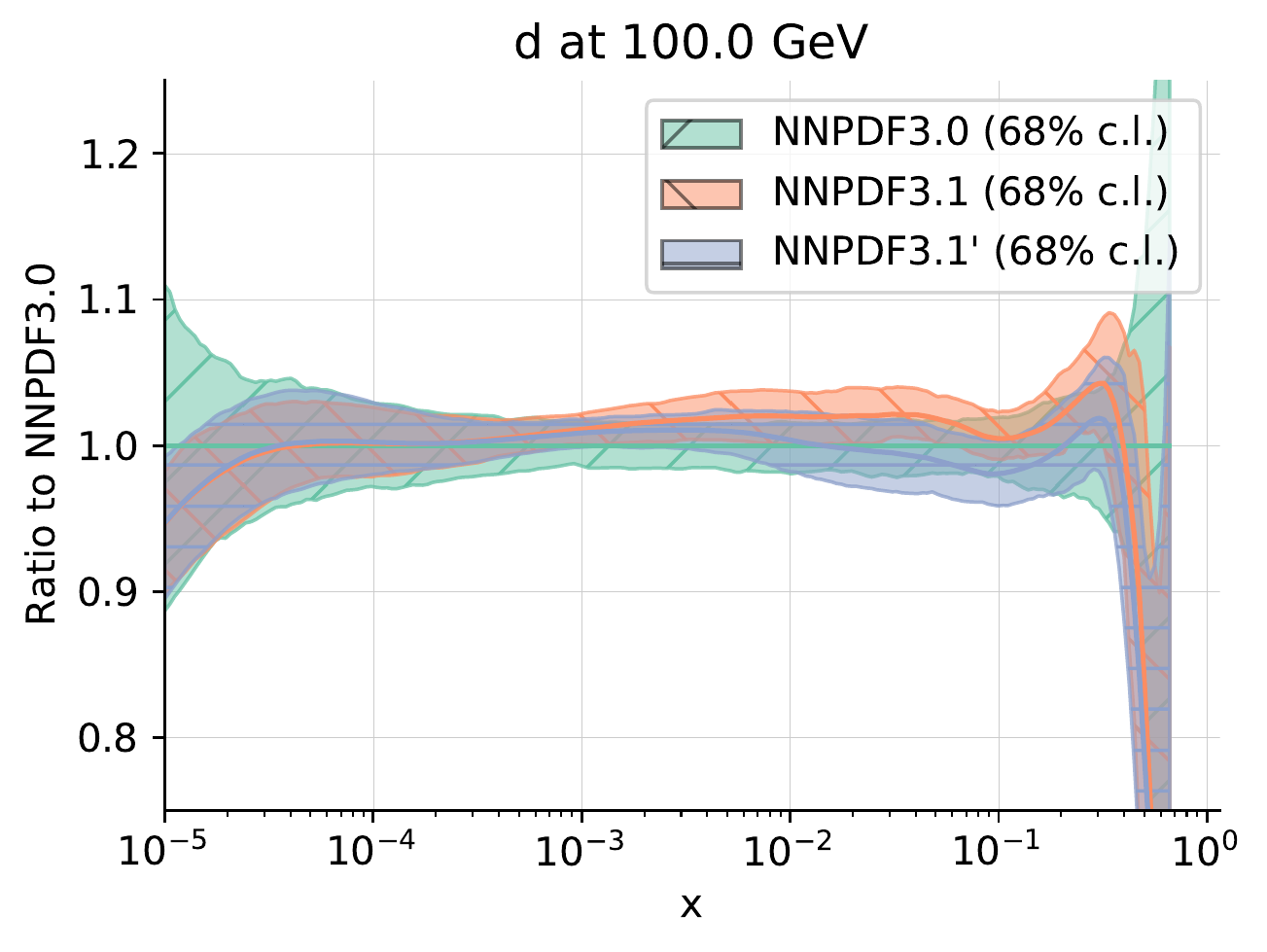}
  \includegraphics[width=0.49\textwidth]{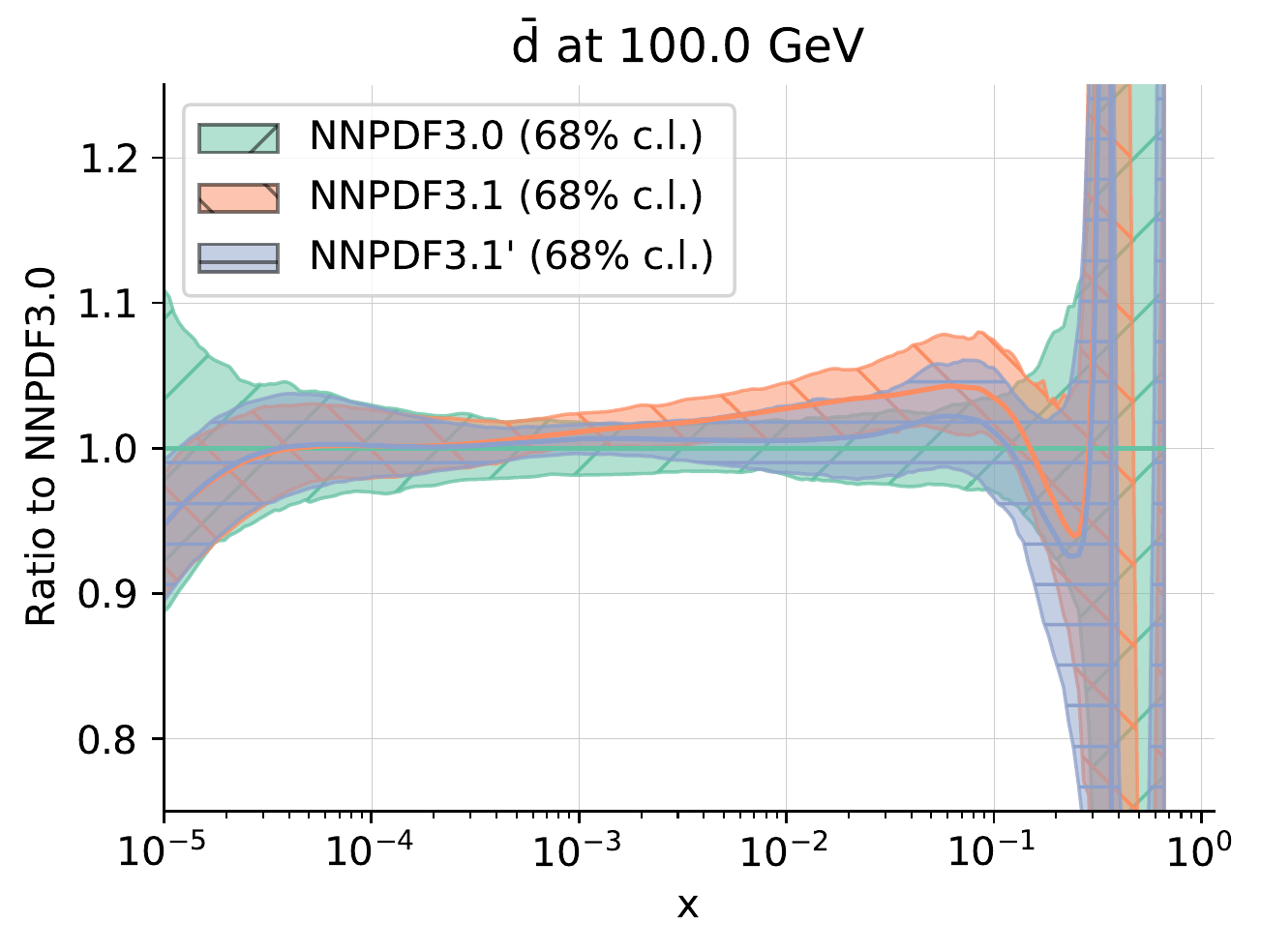}\\
  \caption{A comparison of the gluon (top left), total strange (top right),
    down (bottom right) and anti-down quark (bottom right) PDFs between NNPDF3.0~\cite{NNPDF:2014otw}
    NNPDF3.1~\cite{NNPDF:2017mvq} and the NNPDF3.1 variant used in this work, \NNprime.
    Results are shown at $Q=100$~GeV and are normalised to the central value
    of NNPDF3.0. Error bars correspond to 68\% confidence level intervals}
  \label{fig:NNPDF31variant}
\end{figure}

Selected PDF flavours from the NNPDF3.0, NNPDF3.1 and \NNprime  sets are compared  in Fig.~\ref{fig:NNPDF31variant}. We display the gluon, strange quark sum, down and anti-down distributions at $Q=100$~GeV, normalised to the NNPDF3.0 result. As expected, the three PDF sets are generally consistent, with the PDF central values of each set being almost always included in the PDF uncertainties of the other across the entire range of $x$. Some differences are nevertheless seen. These are the largest for the total strangeness. The difference between NNPDF3.1 and NNPDF3.0 is due to the partial inclusion of  ATLAS 2011 $W,Z$
differential measurements~\cite{ATLAS:2016nqi}. The difference between NNPDF3.1 and the variant  used in this work is explained by the improved treatment of the NuTeV data: \NNprime incorporates NNLO massive QCD corrections to the dimuon cross-sections, which were not available at the time the original NNPDF3.1 set was produced, and an update of the value for the branching ratio of charmed hadrons into muons. The combined effect of these two updates is an enhancement of the total strangeness in comparison to the original NNPDF3.1 analysis, as already reported in~\cite{Faura:2020oom}. To compensate for this effect, the down quark and antiquark PDFs are correspondingly slightly suppressed. Differences in the gluon PDF are possibly due to the different treatment of single-inclusive jet data: Tevatron and 2.76~TeV ATLAS and CMS measurements are no longer included in \NNprime, and NNLO $K$-factors are incorporated for the remaining 7~TeV ATLAS and CMS measurements (no NNLO $K$-factors were used in NNPDF3.1, as they were not yet available). The precision of the PDFs in the NNPDF3.1 and \NNprime  parton sets is very similar; both are more precise than NNPDF3.0.
\subsection{Comparison between input global fits}
\label{subsec:inputs_comparison}
Here we present a comparison between the 
CT18, MSHT20, and NNPDF3.1 global analyses, specifically
between the variants that will
enter the PDF4LHC21 combination discussed in Sect.~\ref{sec:pdf4lhc21}
and that have been described earlier in this section. 
These variants differ from the  nominal releases by a number of
(in general) small differences, such as the values
of the heavy-quark masses, which are set to $m_c^{\rm pole} = 1.4~{\rm GeV}$ and $m_b^{\rm pole} = 4.75~{\rm GeV}$ for all three groups,
and to some variation in the input dataset. We denote the variants by \CTprime and \NNprime. 
Note that there is no MSHT20${}^\prime$, as the version entering the PDF4LHC21 combination is the MSHT20 PDF set with no changes.
The comparisons presented in this section should be contrasted with the
comparisons derived from the fits based on the common reduced dataset
presented in Sect.~\ref{sec:reducedvsglobal} below.

\begin{figure}[!t]
\centering
\includegraphics[width=0.49\textwidth]{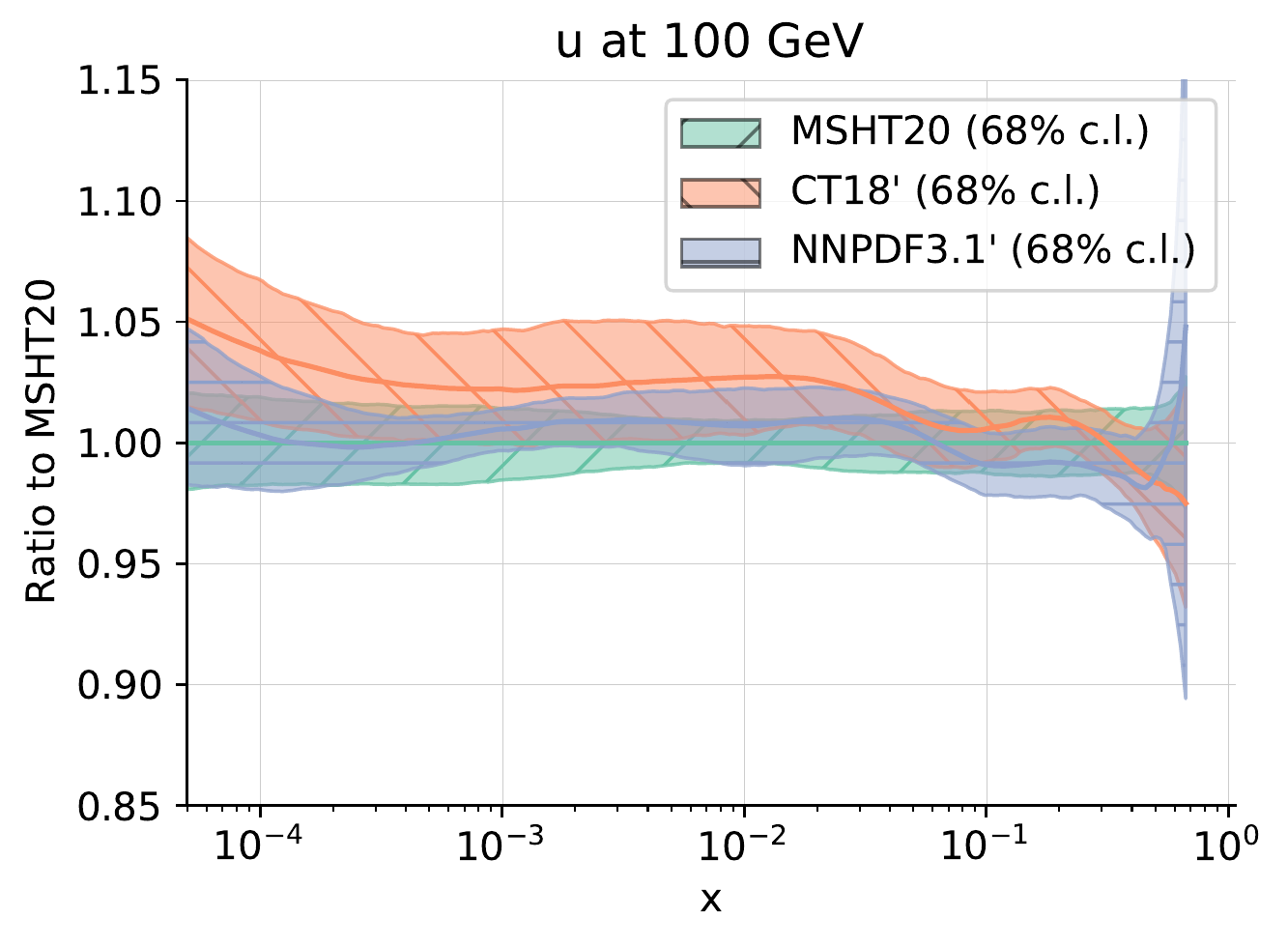}
\includegraphics[width=0.49\textwidth]{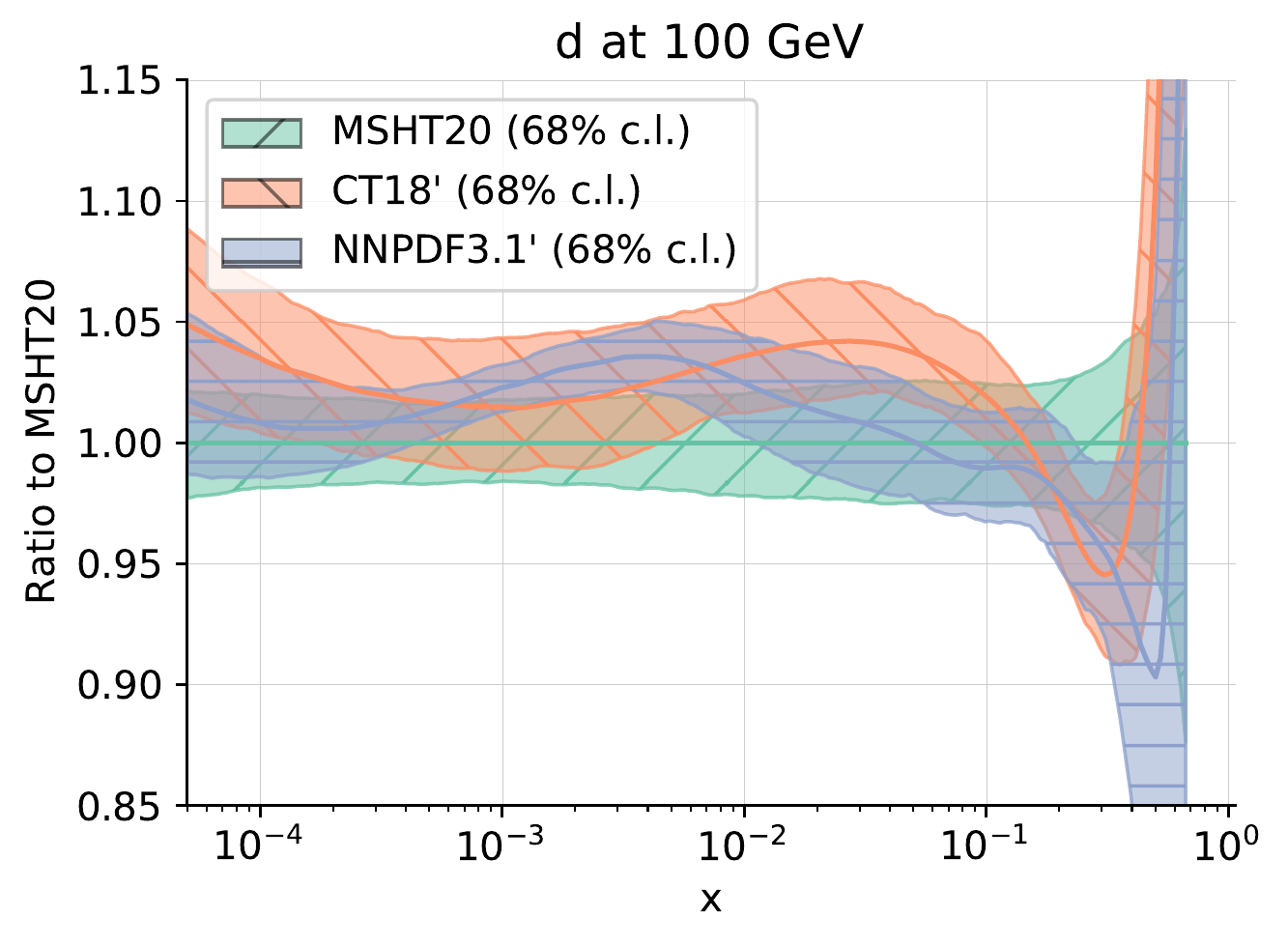}\\
\includegraphics[width=0.49\textwidth]{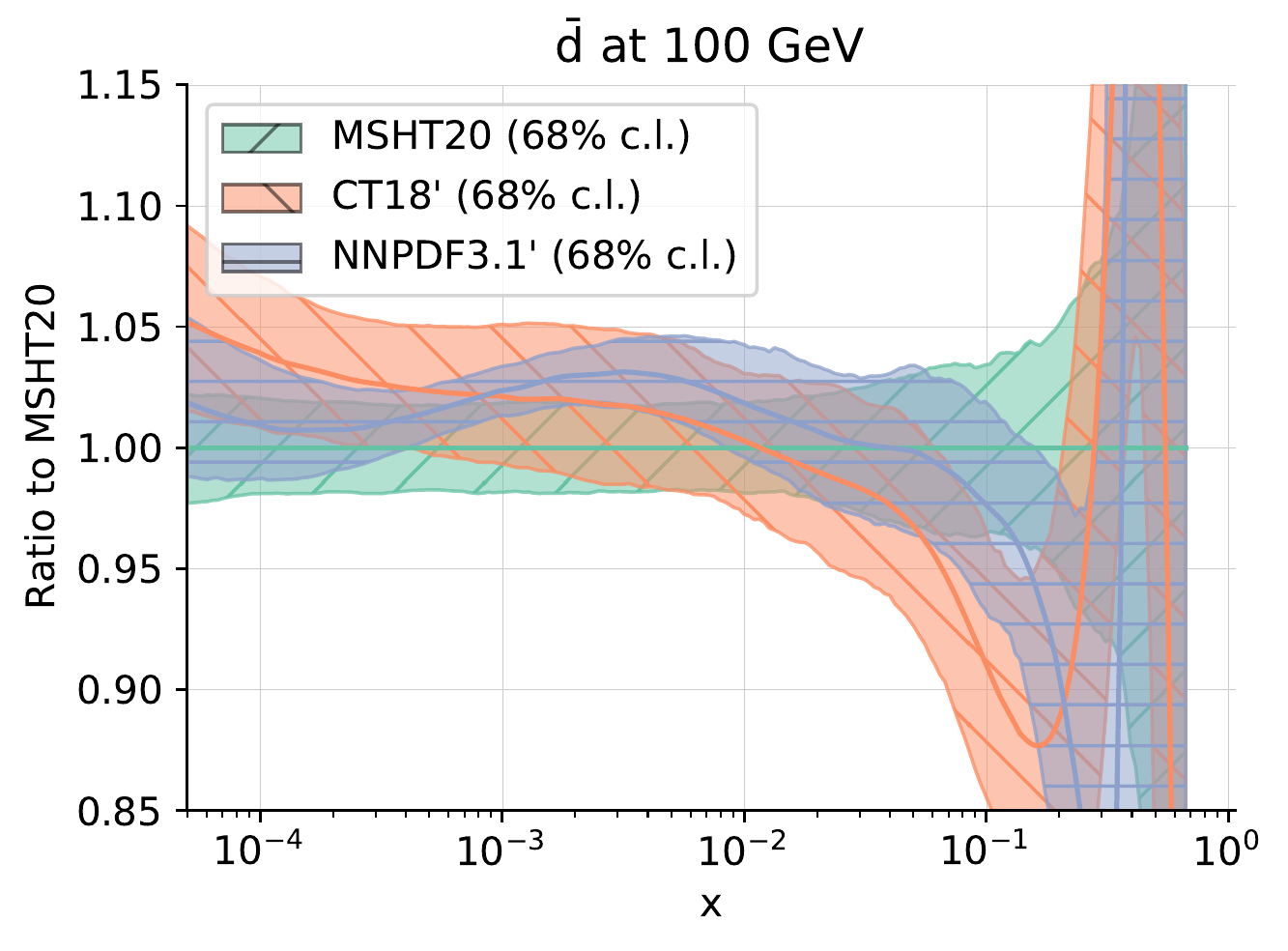}
\includegraphics[width=0.49\textwidth]{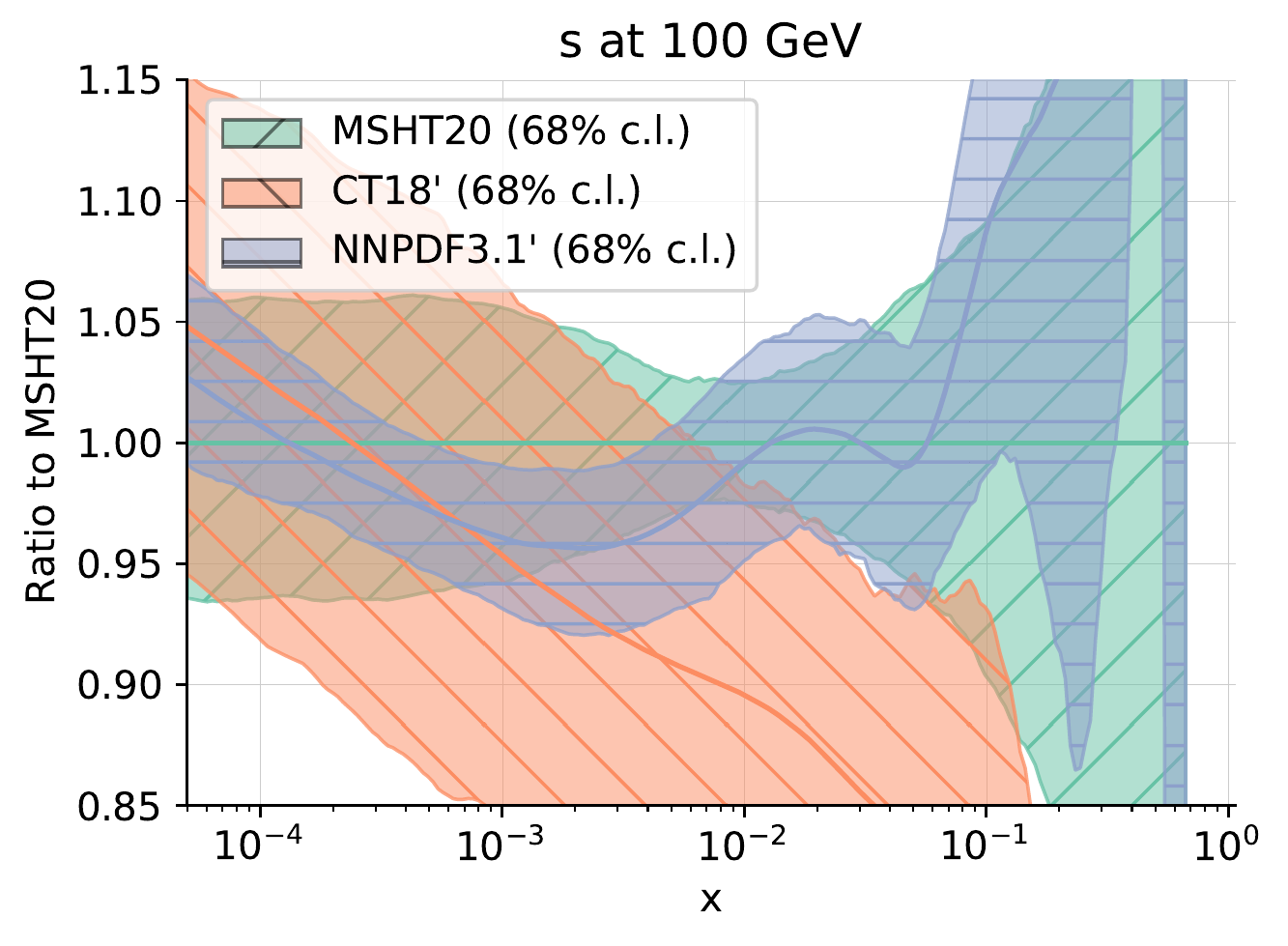}\\
\includegraphics[width=0.49\textwidth]{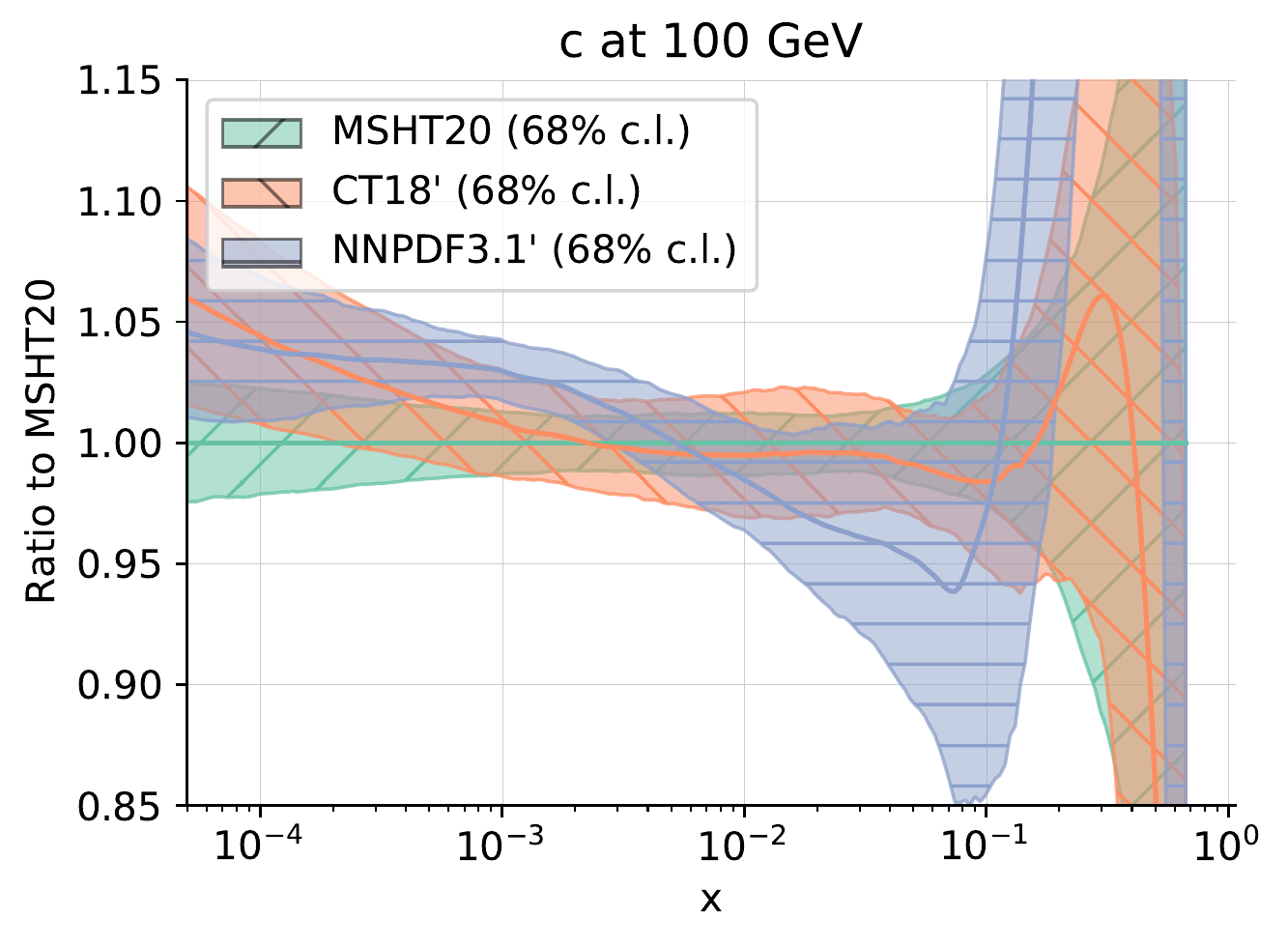}
\includegraphics[width=0.49\textwidth]{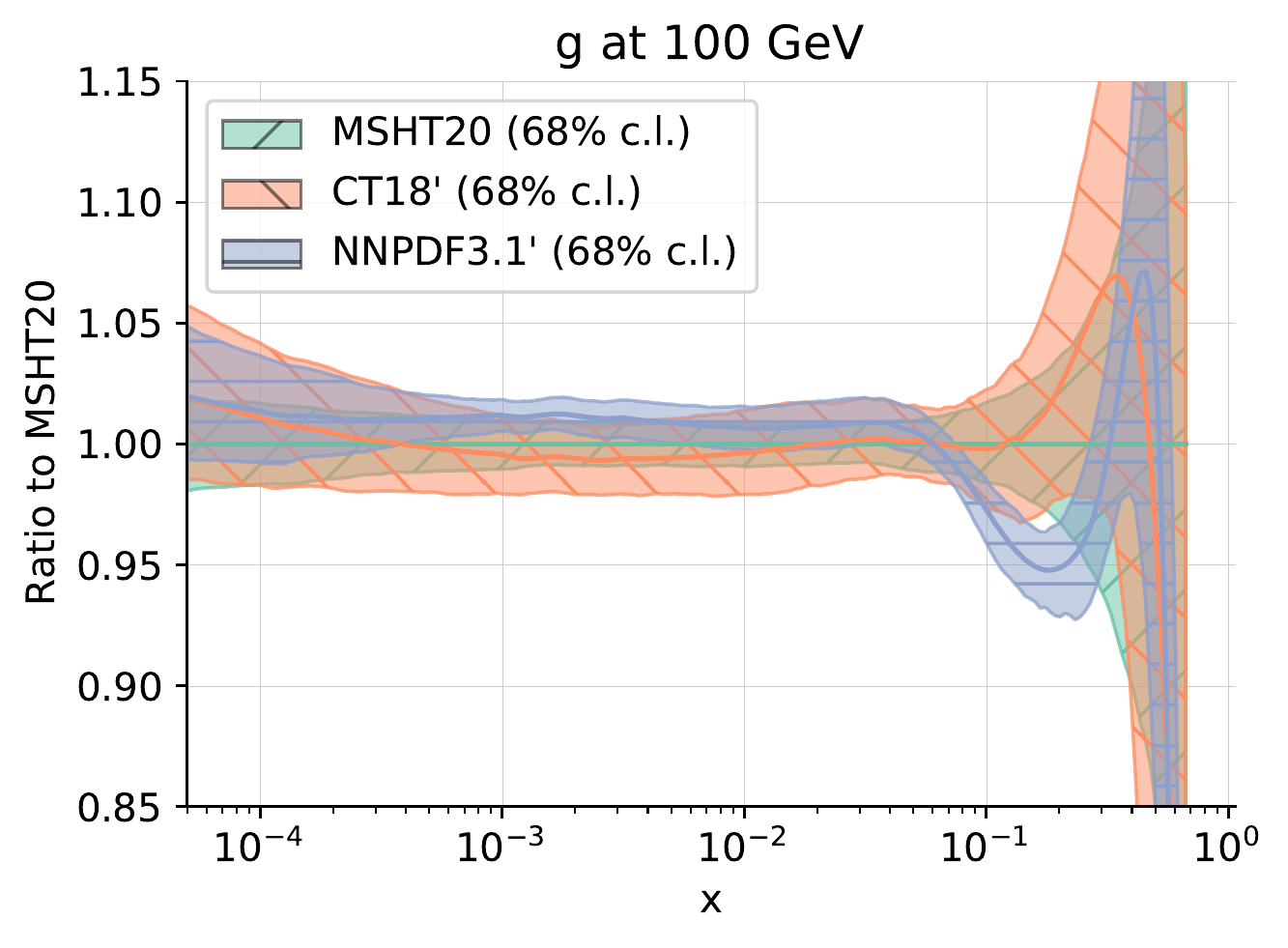}\\
\caption{\small Comparison of the \CTprime, MSHT20, and \NNprime global sets, normalised to the central
value of MSHT20, as a function of $x$ at $Q=100$ GeV. We show the results for the up, down, anti-down, strange, charm quark and gluon PDFs. Note that we consider the variants of the three global fits that are used as input for the combination, and hence in the cases of \CTprime and MSHT20 we are displaying the Monte Carlo representations of the original Hessian sets. Error bands correspond to 68\% CL uncertainties.
}
\label{fig:global_fits_comparison}
\end{figure}

First of all, we compare the three global fits at the level of the $x$-dependent PDFs they produce.
Then we display the comparison of the partonic luminosities, which will be a further subject of Sect.~\ref{sec:phenomenology} once we consider the implications of the PDF4LHC21 combination for LHC phenomenology.

Figure~\ref{fig:global_fits_comparison} displays a comparison of the \CTprime, MSHT20, and \NNprime global sets normalised to the central value of MSHT20 as a function of $x$ at $Q=100$ GeV. We show the results for the gluon and the up, down, anti-down, strange, and charm quark PDFs\footnote{Note that for CT18$'$ and MSHT20 we are displaying the Monte Carlo representations of the original Hessian sets.}.

Several interesting observations can be derived from Fig.~\ref{fig:global_fits_comparison}. In the case of the gluon PDF, the three sets are in good agreement for most of the $x$ range except for $x\simeq 0.2$, where \NNprime undershoots MSHT20 by a few percent, though the differences are barely outside the respective 68\% CL bands.
The dataset dependence of the gluon PDF in the three global fits will
be scrutinised in Apps.~\ref{sec:highxgluon} and \ref{app:l2_sensitivity}.
For the well-constrained up and down quarks, the three global
fits agree within uncertainties in the entire range of $x$
considered.
The same is true for the down antiquark PDF, which is affected
by larger uncertainties especially in the large-$x$ region.
Some more marked differences are observed for the cases of the strange and charm quark
PDFs.

\begin{figure}[!t]
\centering
\includegraphics[width=0.49\textwidth]{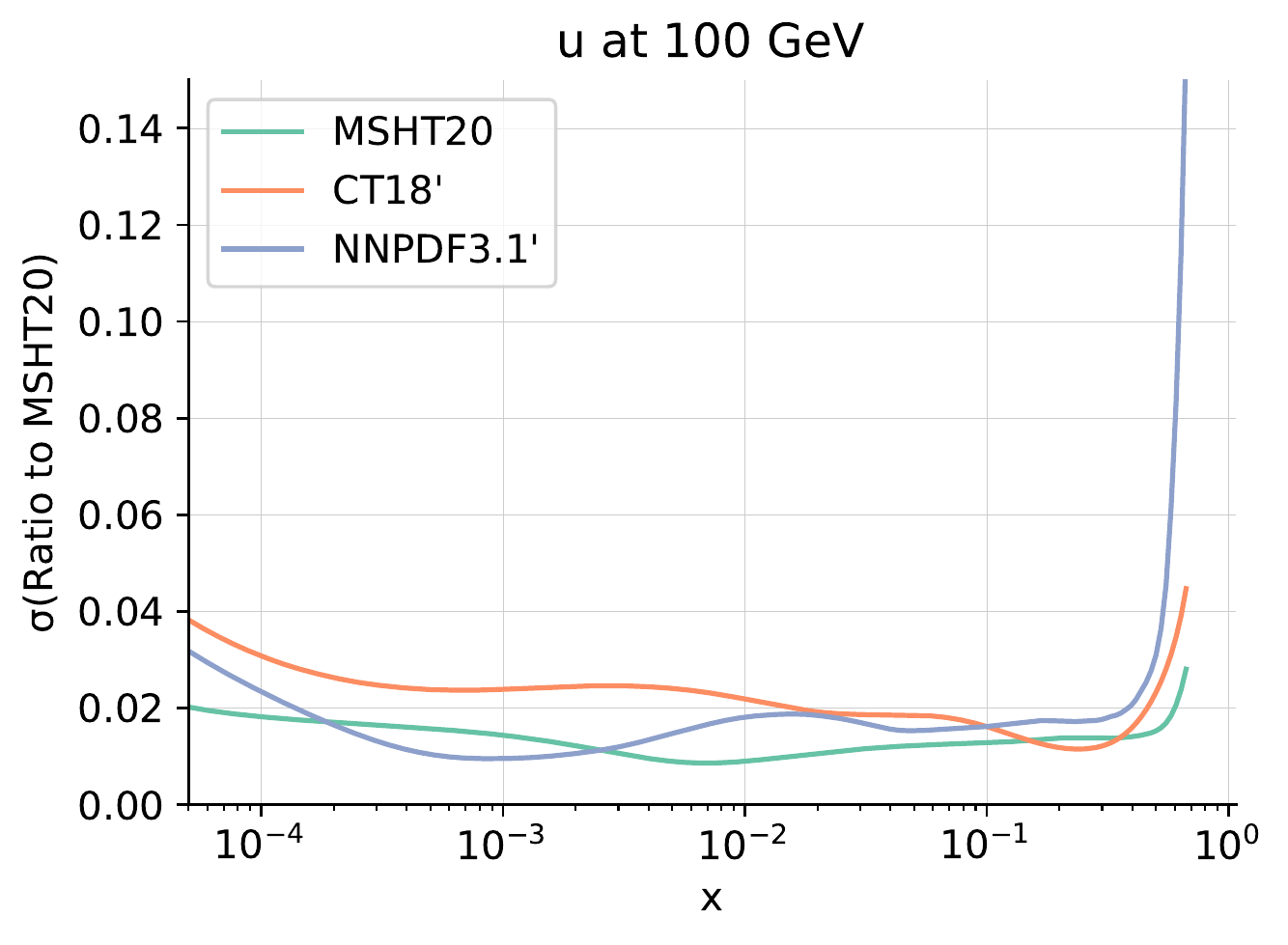}
\includegraphics[width=0.49\textwidth]{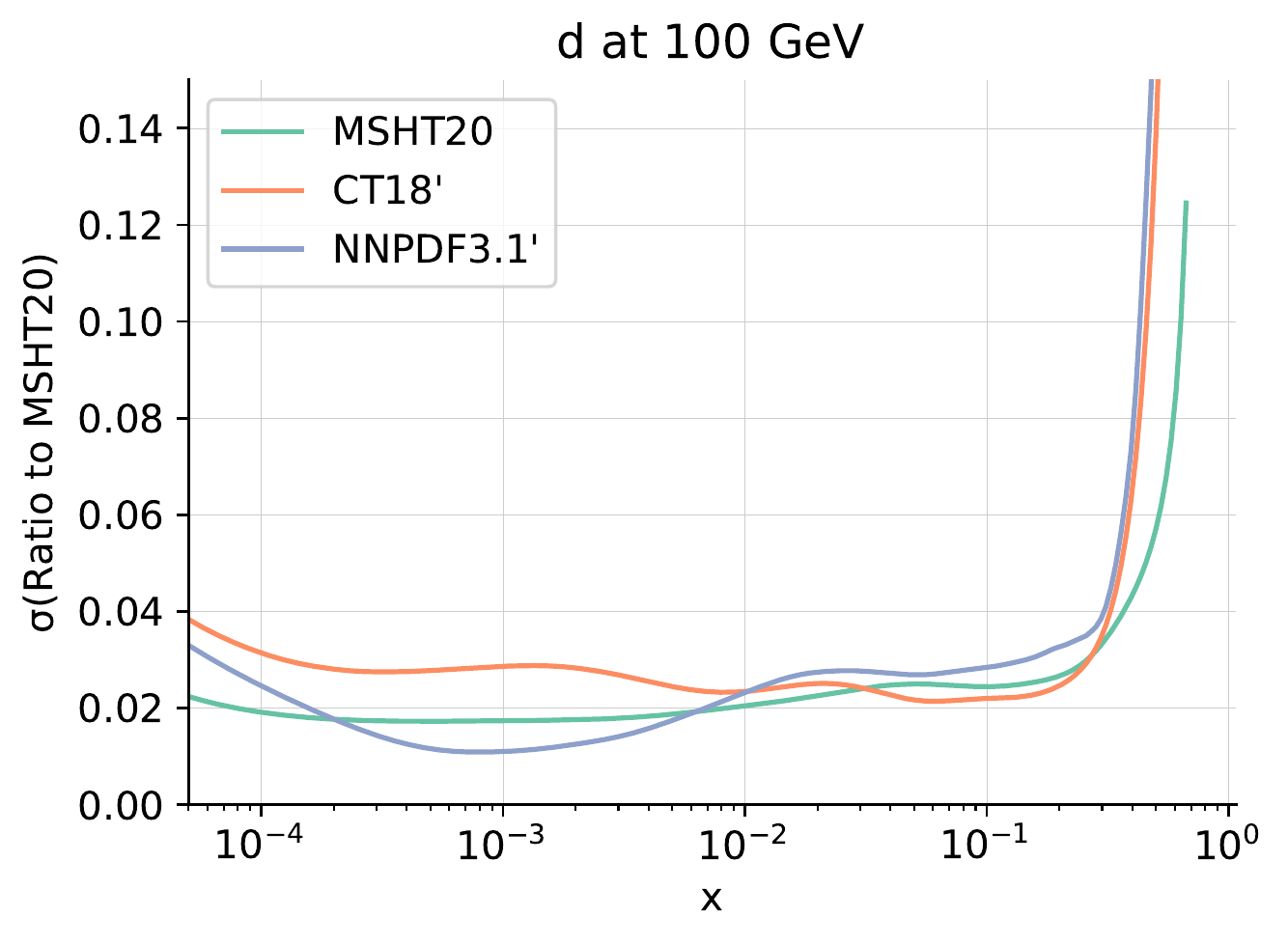}\\
\includegraphics[width=0.49\textwidth]{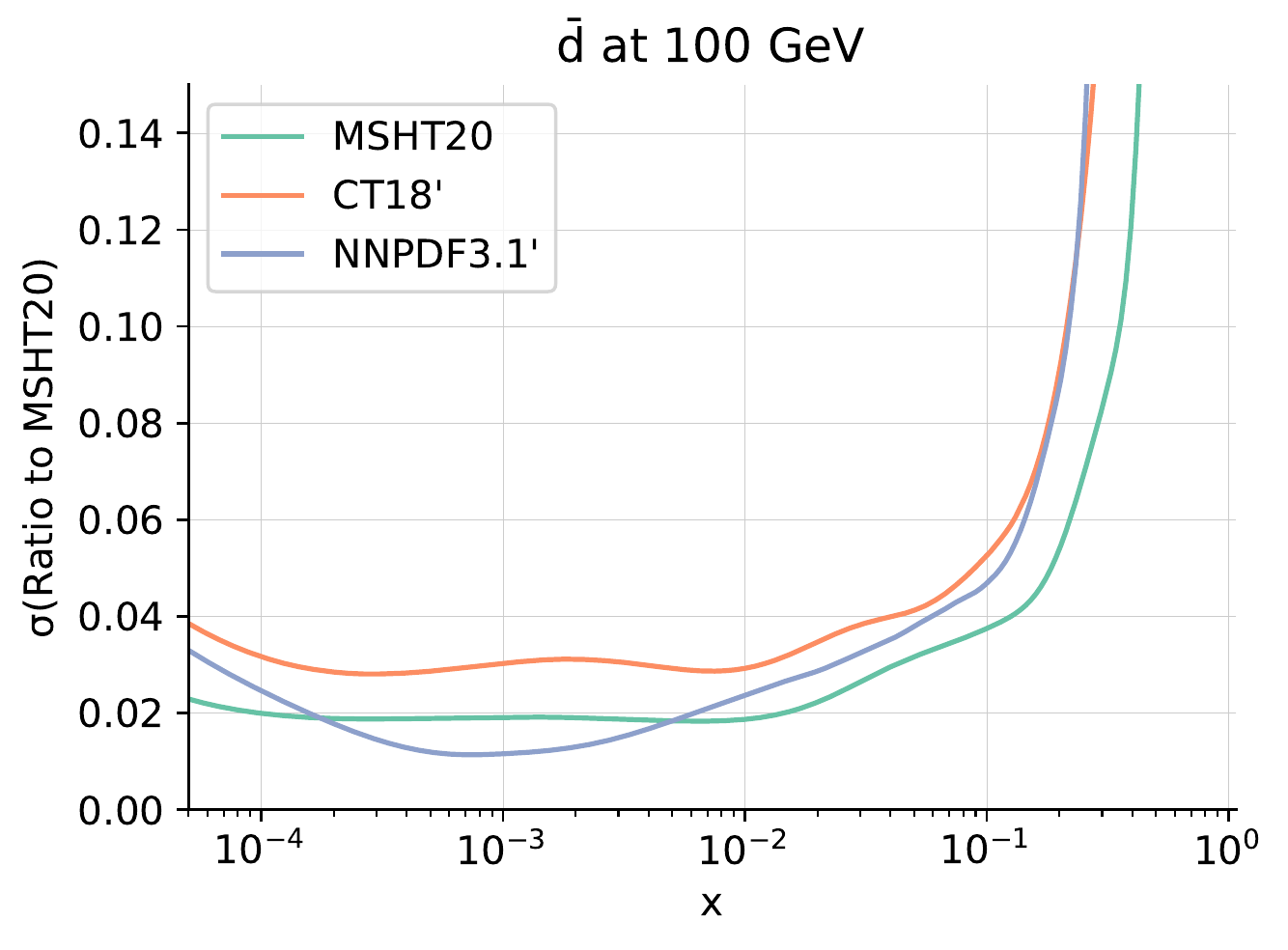}
\includegraphics[width=0.49\textwidth]{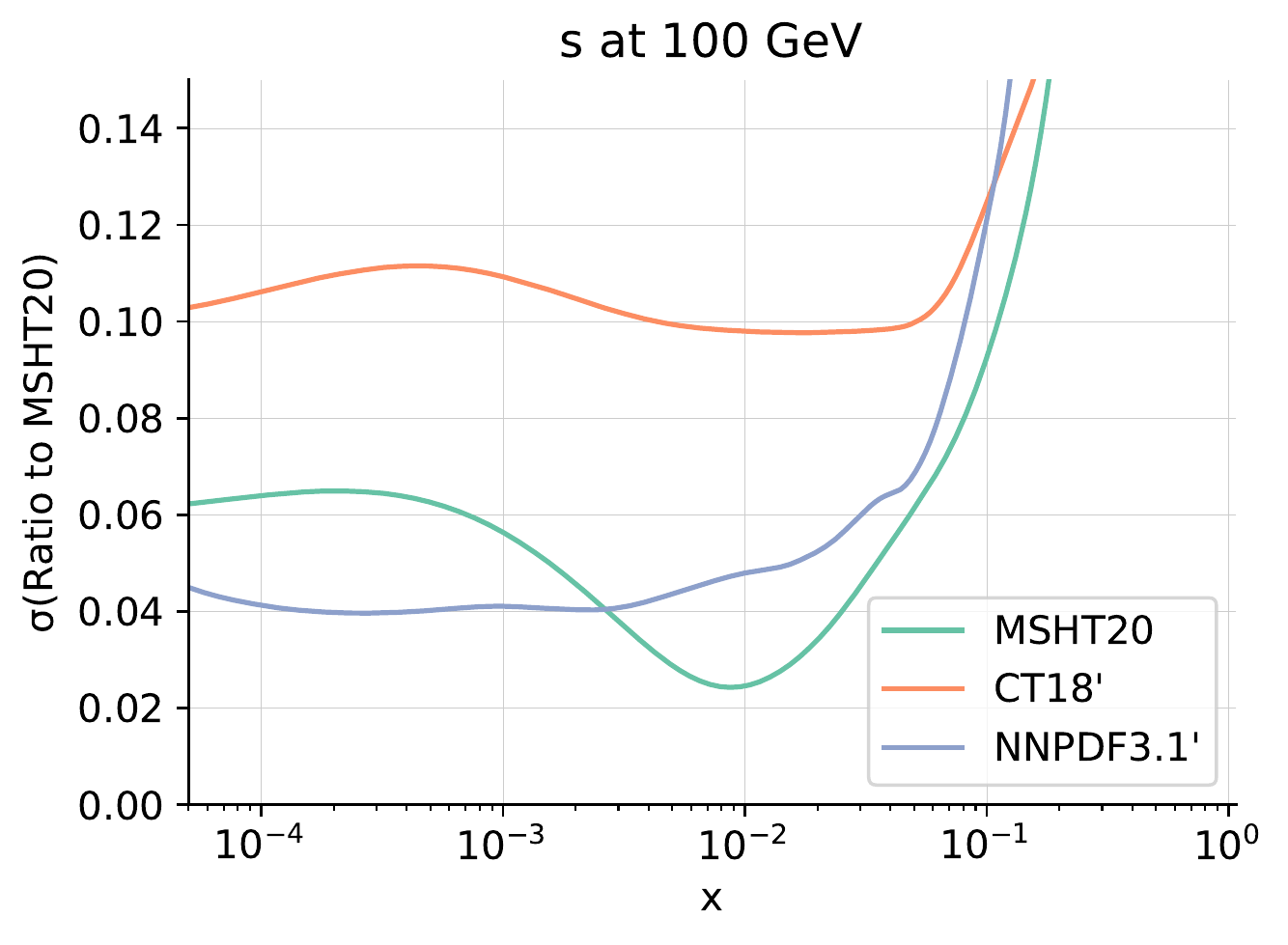}\\
\includegraphics[width=0.49\textwidth]{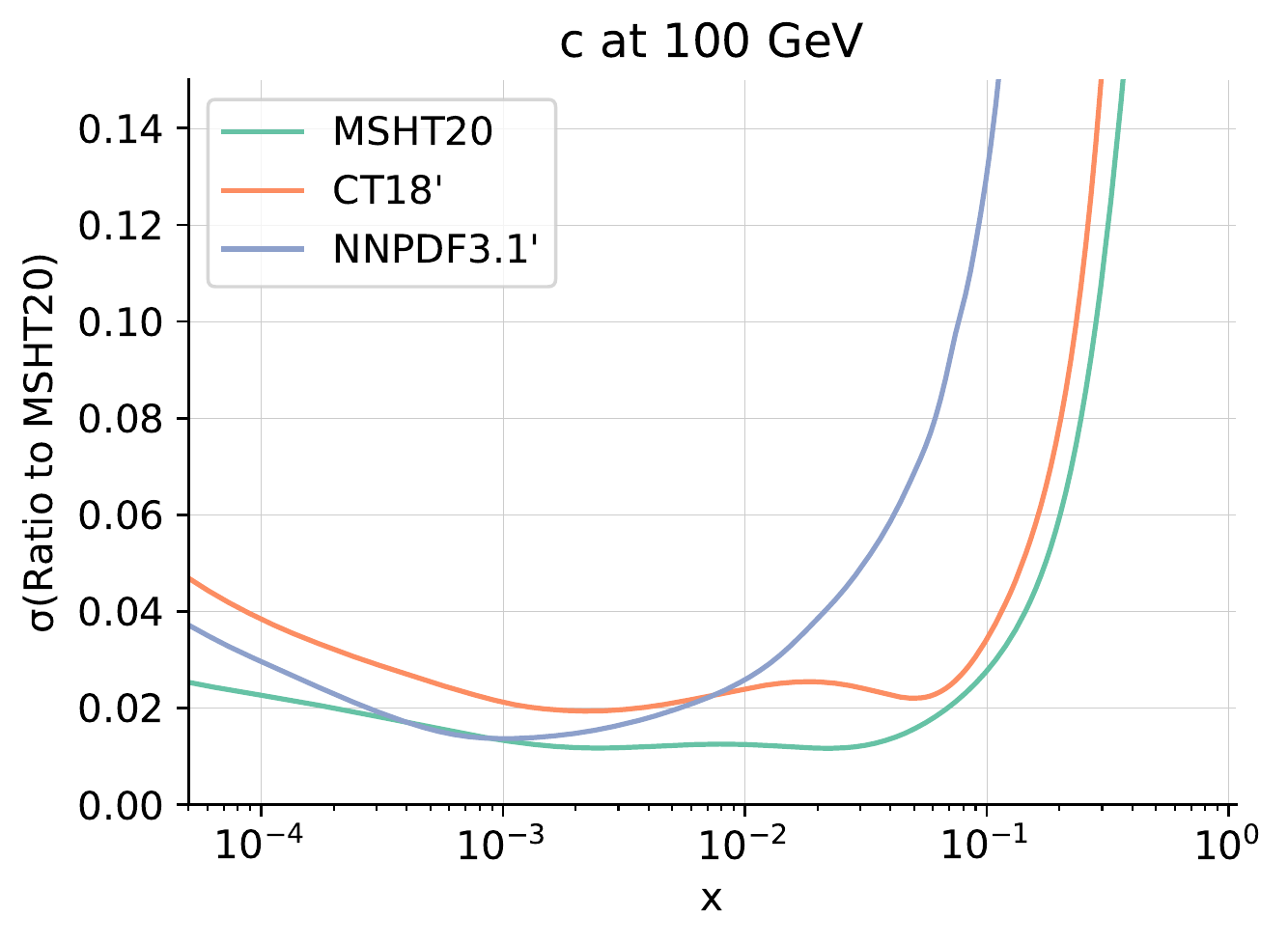}
\includegraphics[width=0.49\textwidth]{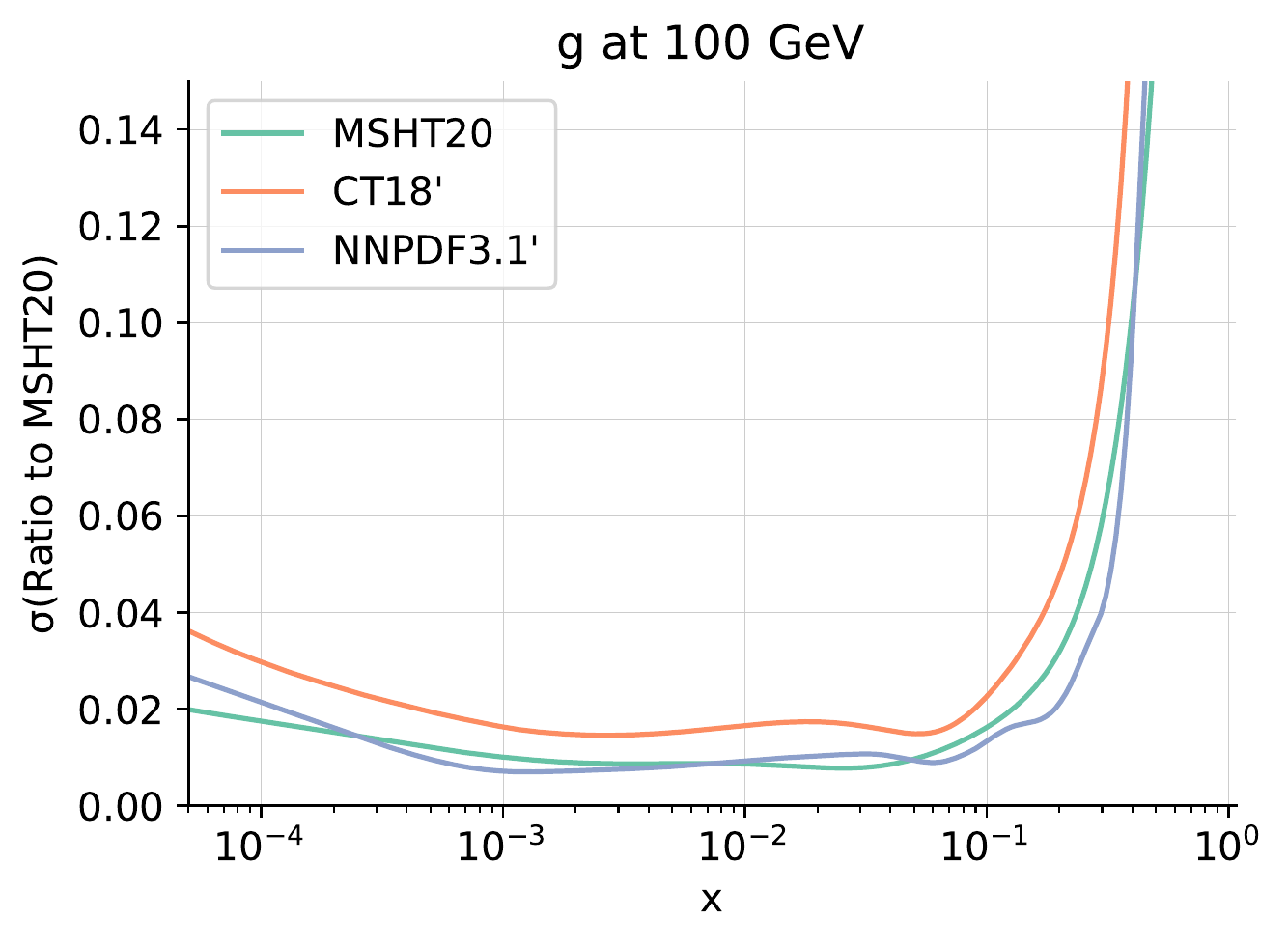}\\
\caption{\small Same as Fig.~\ref{fig:global_fits_comparison},
  now comparing the one-sigma PDF uncertainties associated to the
  three global fits.
}
\label{fig:global_fits_comparison_uncertainties}
\end{figure}

Concerning the strangeness content of the proton, the three groups only agree within uncertainties at low $x$, and there are appreciable
differences in the central values, with \CTprime being suppressed for
$x\gsim 10^{-3}$ and \NNprime being enhanced for $x\simeq 0.1$
as compared to MSHT20.
The smaller strangeness in \CTprime can be traced back in part to the exclusion
of the ATLAS $W,Z$ 2016 dataset from their baseline (affecting $x\approx 0.02$), combined
with a different dimuon branching ratio and a small missing NNLO QCD massive correction to dimuon production, in contrast with the choices adopted in \NNprime and MSHT20. 
A dedicated investigation of the impact of modelling choices in the NuTeV
cross-sections on strangeness is presented in App.~\ref{sec:strangeness}; see also App.~\ref{app:l2_sensitivity} for an illustration of the pulls of the various data sets on $s(x,Q)$ in various $x$ regions.
In addition, there are slight differences between \NNprime and MSHT20 for $10^{-3} \lesssim x \lesssim 10^{-2}$, with \NNprime reduced relative to MSHT20. This may reflect the inclusion of the ATLAS 8~TeV $W$ and $Z$ data in the latter which has been observed to further raise the strangeness in this region \cite{Bailey:2020ooq}.

Some expected differences in the charm PDF are also observed.
While for $x\lsim 0.2$ the charm PDF from the three groups
is consistent within uncertainties, for $x\gsim 0.2$ it is significantly larger
in \NNprime.
The reason is that in the latter case the charm PDF is fitted rather than generated perturbatively,
which results into a sizable enhancement in the large-$x$ region which persists to high scales.

In Fig.~\ref{fig:global_fits_comparison_uncertainties} we display
a similar comparison to that of Fig.~\ref{fig:global_fits_comparison},
  now comparing the one-sigma PDF uncertainties associated to the
  three global fits.
  While in general there is reasonable agreement at the qualitative level
  between the three fits in most cases, one can also appreciate
  some significant differences.
   For the well-constrained up and down quarks in the valence region,
  the uncertainties in the three groups are essentially identical.
  Concerning the gluon PDF, very similar uncertainties are obtained in the MSHT20
  and \NNprime analyses, while those of \CTprime can be somewhat larger, by a factor of
  $\sim\!1.5\!-\!2$, for select regions of $x$.
  In the case of the strange PDF, the uncertainties in \CTprime are larger than those of either MSHT20 or \NNprime for $x\lesssim 10^{-1}$.
  Also, the inclusion of fitted charm in \NNprime leads to a marked
  increase in uncertainties for $x\gsim 10^{-2}$ compared to charm which is entirely perturbatively generated; and the $x$ dependence of the charm-PDF uncertainty obtained
  by CT and MSHT largely reflects that of the gluon.

\begin{figure}[!t]
  \centering
  \includegraphics[width=0.49\textwidth]{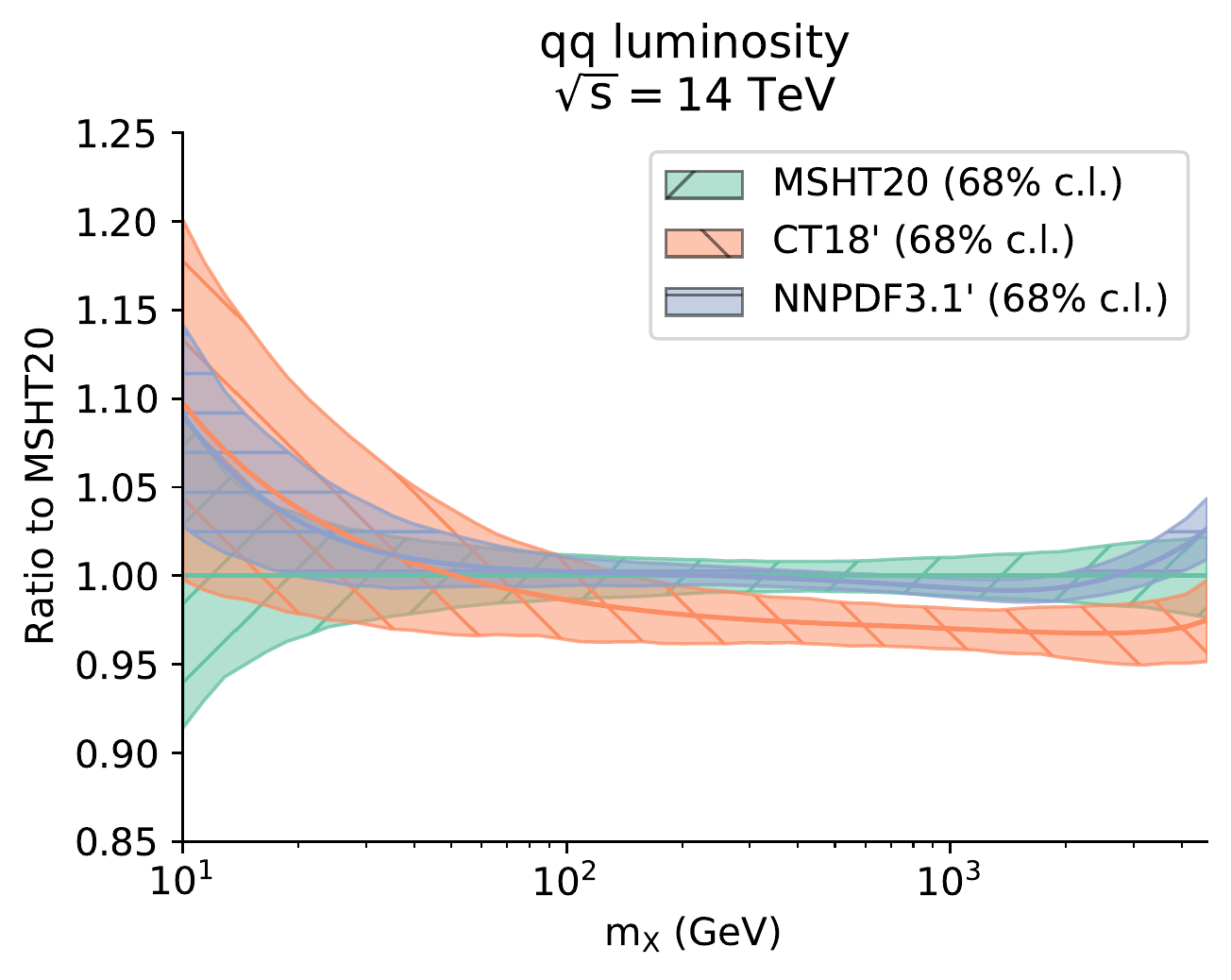}
  \includegraphics[width=0.49\textwidth]{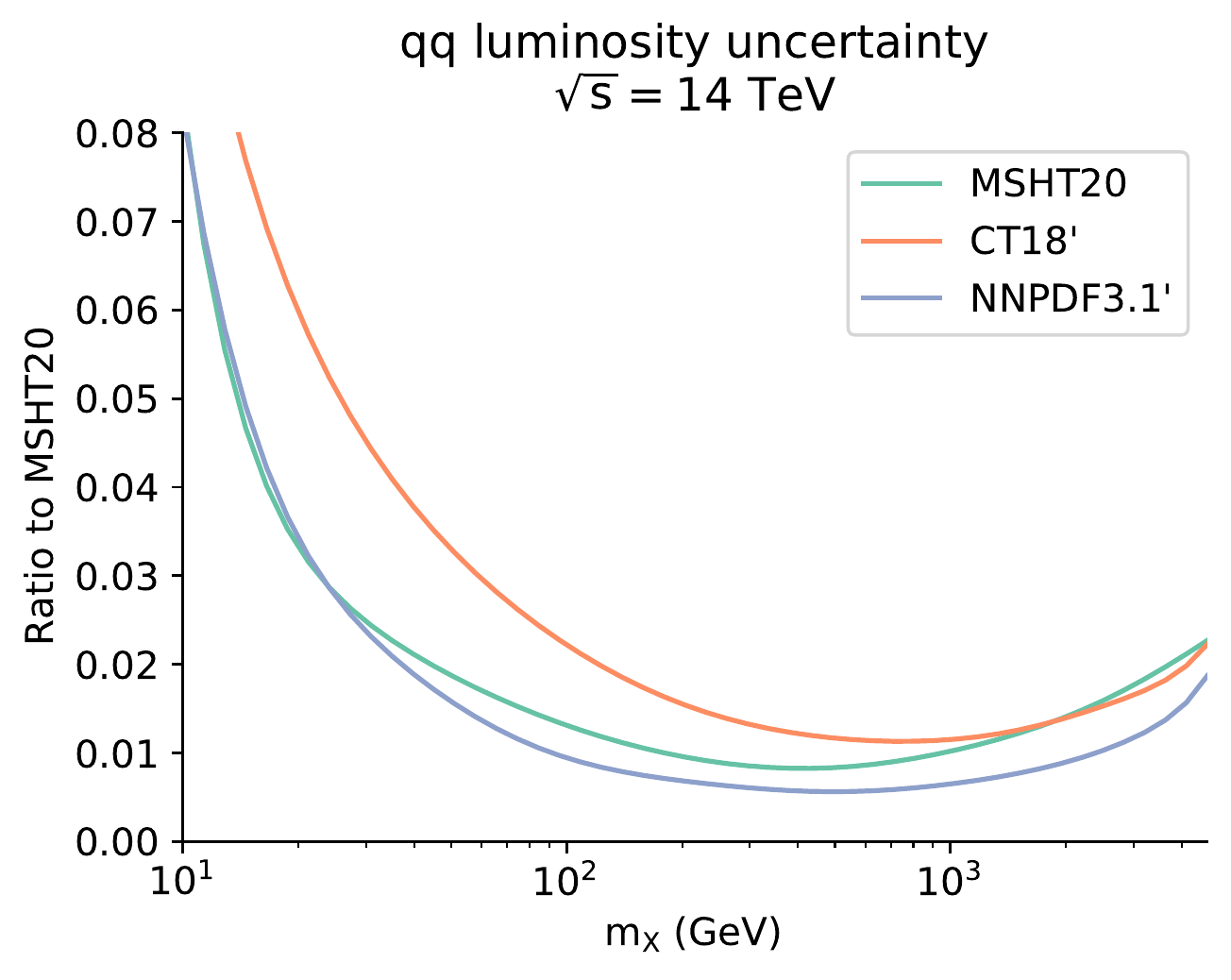}\\
  \includegraphics[width=0.49\textwidth]{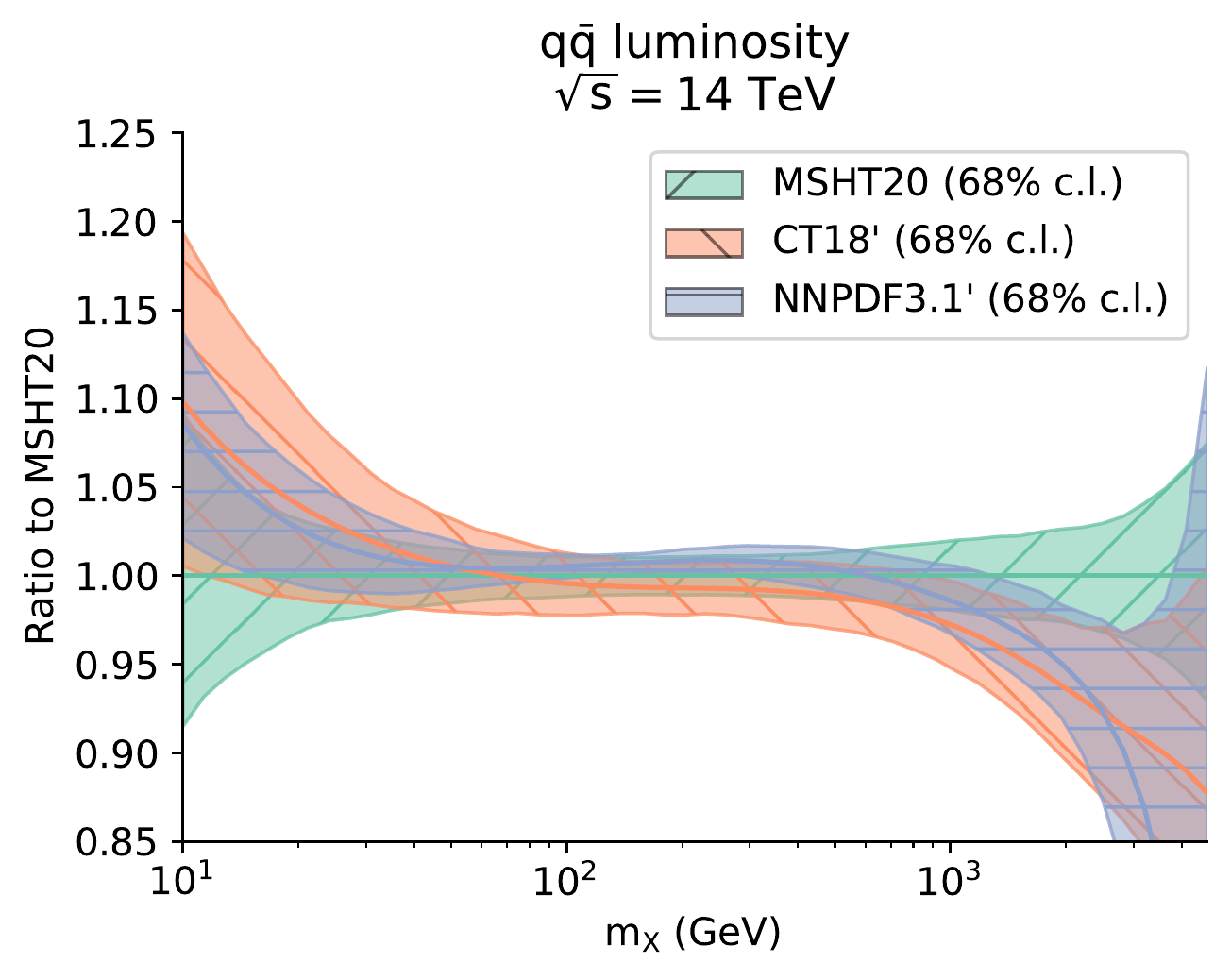}
  \includegraphics[width=0.49\textwidth]{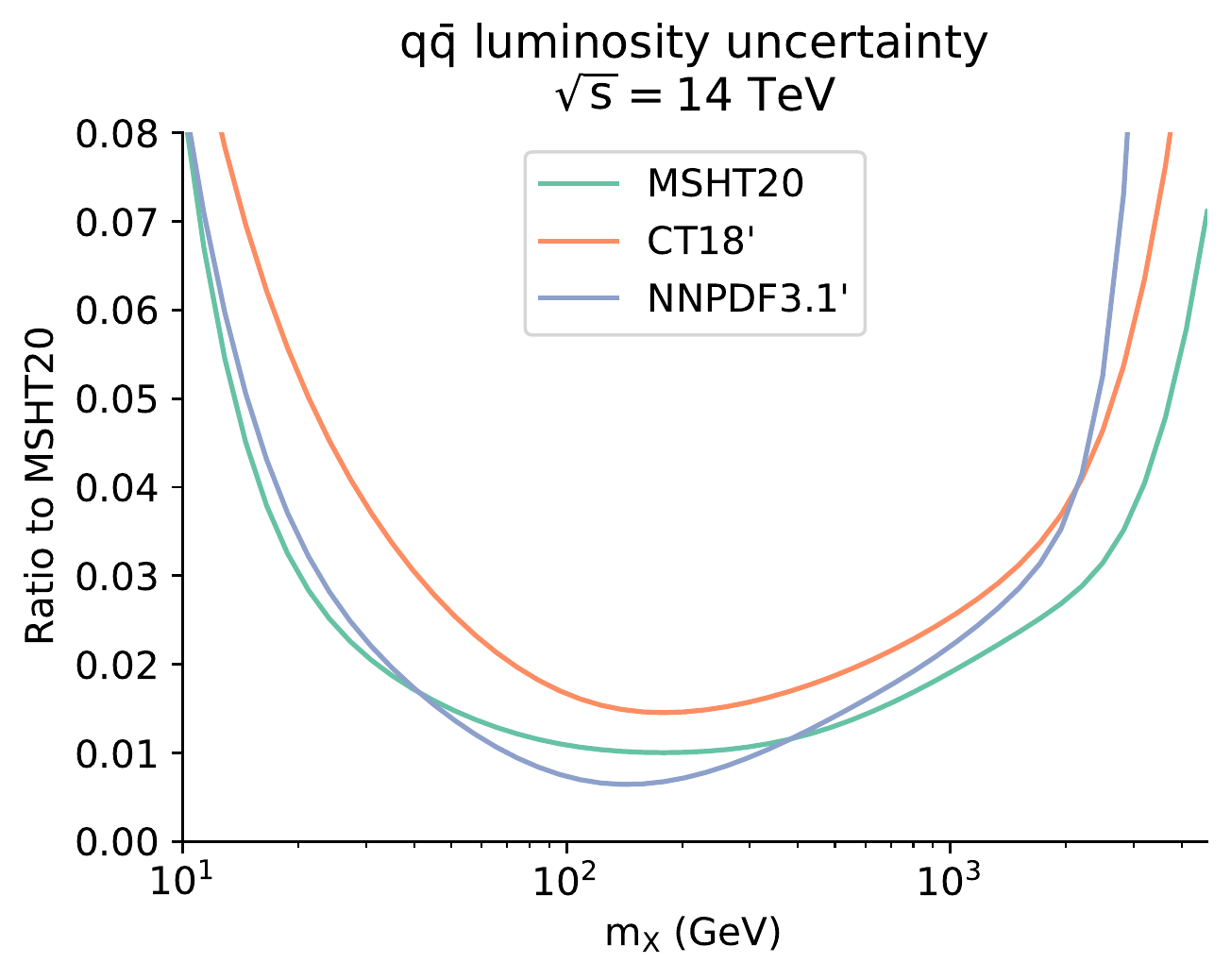}\\ 
  \includegraphics[width=0.49\textwidth]{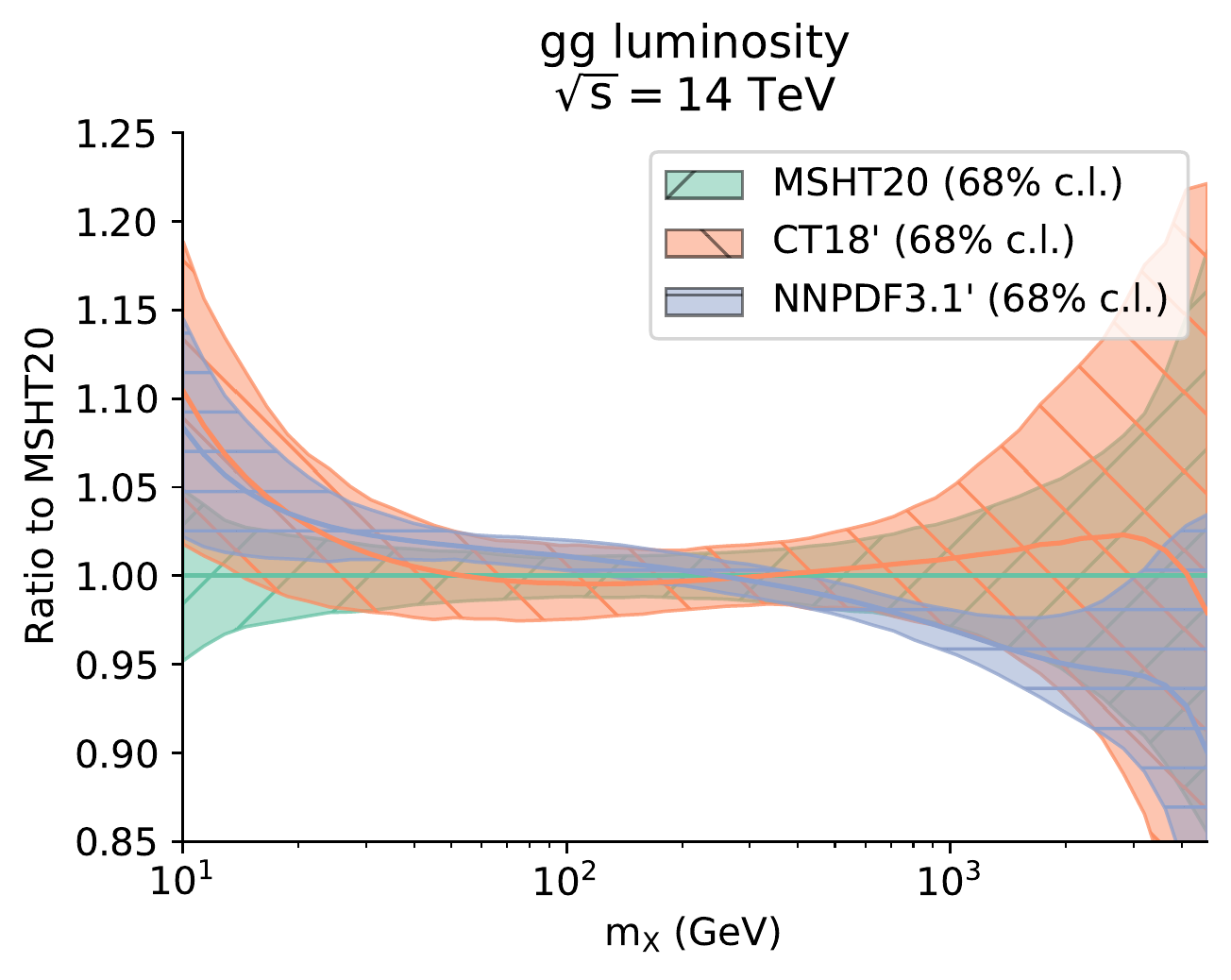}
  \includegraphics[width=0.49\textwidth]{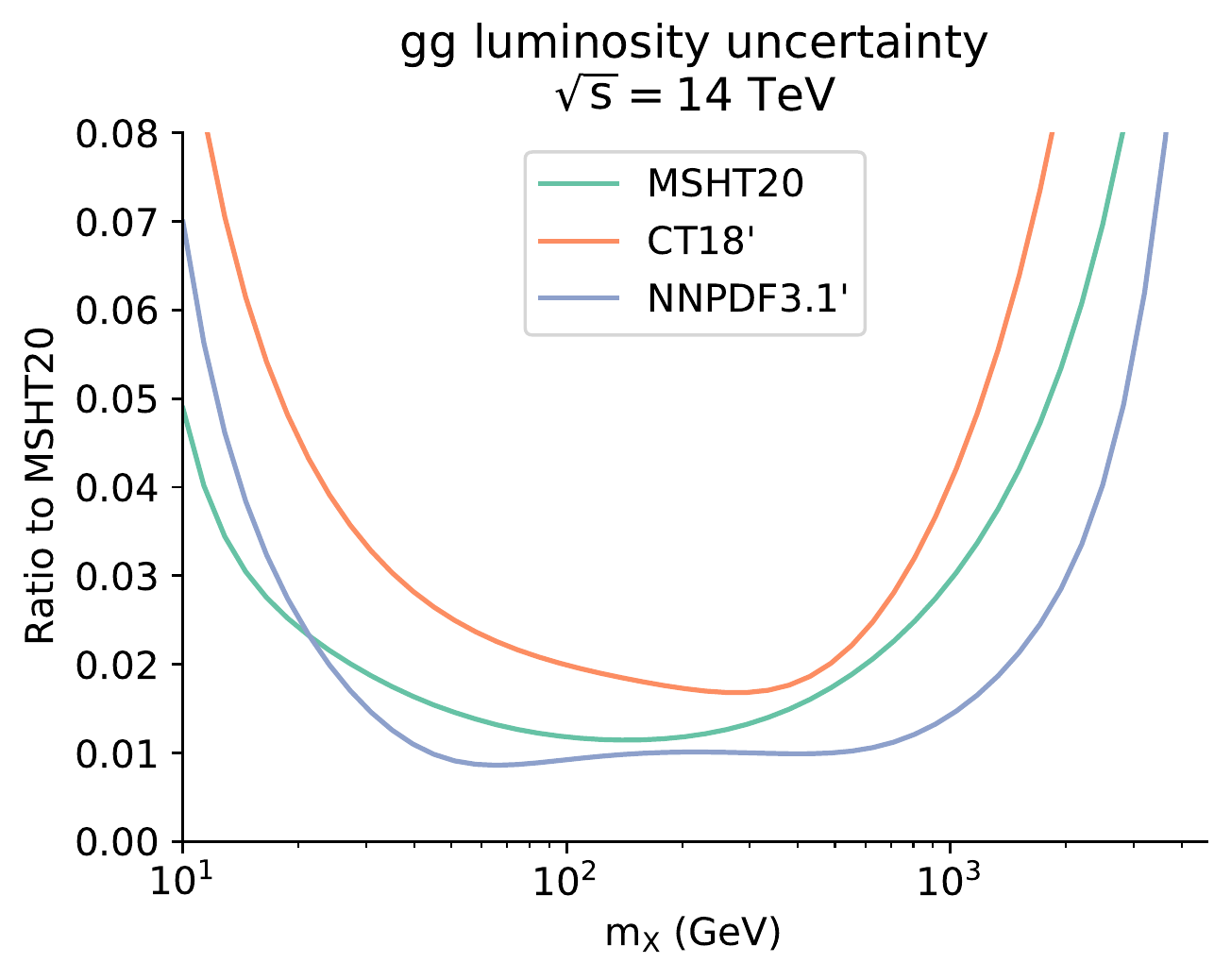}\\ 
\caption{\small Comparison of the partonic luminosities, evaluated at $\sqrt{s}=14$ TeV  as a function
    of the final state invariant mass $m_X$, between \CTprime, MSHT20, and \NNprime. Results are shown for the quark-quark, quark-antiquark, and gluon-gluon luminosities normalised to the
    central value of the MSHT20 prediction (left panels) as well as for the corresponding 68\% CL relative uncertainties (right panels).}
\label{fig:global_fits_comparison_luminosities}
\end{figure}

We consider finally the comparison at the level of parton luminosities.
Fig.~\ref{fig:global_fits_comparison_luminosities} displays
the partonic luminosities evaluated 
at $\sqrt{s}=14$ TeV  according to the definition in \cite{Campbell:2006wx} as functions of
the final state invariant mass $m_X$, for the versions of \CTprime, MSHT20, and \NNprime to be included in the combination.
No cut in the rapidity of the produced final state, $y_X$, has been applied.
Results are shown for the quark-quark, quark-antiquark,
and gluon-gluon  luminosities normalised to the
central value of the MSHT20 prediction 
as well as for the corresponding 68\% CL relative PDF uncertainties.
From this comparison of the partonic luminosities between the three global PDF
sets that enter the present combination one sees that
for the gluon-gluon luminosity there is good agreement within uncertainties
for the full range of $m_X$ values, though the central value of \NNprime
is lower than that of \CTprime and MSHT20 in the region
$m_X\gsim 1$ TeV. For the quark-antiquark luminosity, \NNprime and \CTprime are very close
to each other in the whole $m_X$ range, with MSHT20 a bit higher at large $m_X$ but
also in agreement at intermediate invariant mass values.
For the quark-quark luminosity, \NNprime and MSHT are very close
across the whole $m_X$ range, with \CTprime being lower by a few percent for $m_X\gsim 100$
GeV.
With this exception, the partonic luminosities from the three groups
are found to agree within uncertainties over the whole kinematic range
relevant for LHC phenomenology.

Concerning the magnitude of the relative luminosity uncertainties themselves, we see that for the gluon-gluon luminosity \CTprime has the largest uncertainty, while \NNprime has a smaller uncertainty then either MSHT20 or \CTprime at high invariant mass, $m_X\gsim 500$ GeV .
For the quark-quark and quark-antiquark luminosities, at low invariant mass \CTprime again has the highest uncertainties, while at very high invariant mass $m_X\gsim 2$ TeV \NNprime has larger uncertainties than either MSHT20 or \CTprime for the quark-antiquark luminosity.

All in all, from these comparisons between the three global fits, 
we see that, while they are sufficiently in agreement for the PDF4LHC21 combination to be meaningful, they also show non-negligible differences which will result in a more conservative result in the combination than might be obtained from using the sets individually.

\section{Benchmarking of global fits}
\label{sec:benchmarking}

In this section we present the outcome of a dedicated benchmarking exercise
carried out among the three global fits considered in the combination
and whose settings are described in Sect.~\ref{sec:combination_inputs}.
In this benchmarking we have strived to homogenise as much as possible the input
datasets and theoretical settings, such that any residual differences can be attributed
to the effect of methodological choices adopted by each of the three groups.

First of all, we describe the rationale for the choice of reduced dataset and of the common settings
for the benchmark comparison.
Then we compare each of the reduced PDF fits with its global fit counterparts.
We assess the outcome of the three reduced fits both at the level of the PDFs and of the $\chi^2$ dataset-by-dataset,
and finally we carry out the corresponding comparisons at the level of partonic luminosities at a center-of-mass energy of 14 TeV, relevant 
for the LHC. This represents both a more detailed description and an update on that presented in \cite{Cridge:2021qjj}.

\subsection{Choice of the data and theory settings}
\label{sec:reduced_fits_settings}

In order to establish any differences in the global PDF fits, and then to pinpoint their origin, we adopt baseline
settings for this benchmarking comparison, removing as many differences as possible in input dataset,
methodological choices, and theory settings.
We therefore choose a common input dataset, common settings for the theoretical  calculations,
and we set the strangeness asymmetry to zero at the input scale.
In the following we denote the outcome of PDF fits based on these uniform settings
as ``reduced fits'', offering ease of comparison at the expense of the full breadth typically offered in global fits.
We emphasise that the goal of these reduced fits is not to produce the most precise and accurate
PDF determination possible, but rather to disentangle the impact
of the fitting methodology adopted by each group from possible differences in the dataset implementation
or in the corresponding theoretical calculations.

We begin with the choice of input data, which
is chosen as the largest subset of data fit by all three groups in an (almost) identical manner.
Furthermore, we adopt the most conservative kinematic cuts made by any group, i.e. $Q^2 > 4~{\rm GeV}^2$ and $W^2 > 15~{\rm GeV}^2$.
Given the numerous subtle differences between groups, the final list of common data
that enters the reduced PDF fit is rather restricted, and summarised in Table~\ref{tab:datasets}.
This reduced fit dataset also satisfies the competing requirement of being sufficiently large and varied so as to provide some constraints on all the relevant PDF combinations and their uncertainties. We expect this common choice to reduce any differences between the PDFs due to data selection, and hence illuminate
the origin of differences related to the underlying methodological procedures adopted by the three groups.

\begin{table}[!t]
\begin{center}
\fontsize{9}{11}\selectfont 
\renewcommand\arraystretch{1.4}
\begin{tabular}{|c|c|c|c|}
\hline Dataset & Reference & Dataset & Reference \\ \hline
BCDMS proton, deuteron DIS & \cite{BCDMS:1989qop,BCDMS:1989ggw} & LHCb 8~TeV $Z \rightarrow ee$ & \cite{LHCb:2015kwa}\\ 
NMC deuteron to proton ratio DIS & \cite{NewMuon:1996fwh} & ATLAS 7~TeV high precision $W,Z$ (2016) & \cite{ATLAS:2016nqi} \\  
NuTeV $\nu N$ dimuon & \cite{Mason:2006qa} & D0 $Z$ rapidity & \cite{D0:2007djv} \\ 
HERA I+II inclusive DIS & \cite{H1:2015ubc} & CMS 7~TeV electron asymmetry & \cite{CMS:2012ivw}\\
E866 Drell-Yan ratio $pd/pp$ DIS & \cite{NuSea:2001idv} & ATLAS 7~TeV $W,Z$ rapidity (2011) & \cite{ATLAS:2011qdp} \\ 
LHCb 7, 8~TeV $W,Z$ rapidity & \cite{LHCb:2015okr,LHCb:2015mad} & CMS 8~TeV inclusive jet & \cite{CMS:2016lna} \\ 
\hline
    \end{tabular}
\end{center}
\caption{The measurements included in the initial round of PDF fits to a reduced dataset, together
with the corresponding publication reference.
This dataset is chosen as the largest subset of data fit by CT18, MSHT20, and NNPDF3.1
in an (almost) identical manner.
}
\label{tab:datasets}   
\end{table}

From Table~\ref{tab:datasets} one sees how the reduced PDF fits considered here still fit data from older fixed target DIS experiments, such as BCDMS and NMC, the crucial full HERA combined dataset is also included, whilst the NuTeV dimuon data is included to constrain the strangeness. Then, newer LHC data on Drell-Yan, including the important high precision ATLAS 7~TeV $W,Z$ data (2016), is included, whilst the CMS 8~TeV inclusive jet data constrains the gluon at high $x$. The constraints placed by this reduced fit dataset will necessarily be significantly more limited than in the usual full global fits, but this set-up provides a more straightforward baseline for comparison. Additional datasets and further complexities can then be added to move to the full global fits,
providing a well-defined and robust starting point for subsequent extensions of the benchmarking
exercise, some of which are presented in App.~\ref{app:specifics}.

With differences in input data now removed (or at least significantly minimised), we must also make the theoretical and methodological settings as uniform as possible in order to avoid other potential sources of differences. Specifically,
we adopt the following common choices among the three groups:

\begin{itemize}
\item Same heavy-quark masses: $m_c^{\rm pole} = 1.4~{\rm GeV}$ and $m_b^{\rm pole} = 4.75~{\rm GeV}$.
\item Same value of the strong coupling: $\alpha_s(M_Z^2) = 0.118$.
\item No strangeness asymmetry at the input scale, i.e. $(s-\bar{s})(Q_0) = 0$. Note that NNLO QCD evolution
generates nevertheless a non-zero strangeness asymmetry for $Q > Q_0$.
\item The charm PDF is entirely generated by perturbative evolution: that is, any 3FNS nonperturbative ({\it i.e.} {\it intrinsic}) charm PDF is assumed to be zero.
\item Positive-definite quark distributions.
\item No deuteron or nuclear corrections/uncertainties.
\item A common branching ratio for charm hadrons to muons $B(D\rightarrow \mu)$, which is taken as fixed and not allowed to float during the fit.
\item NNLO corrections for the heavy quark structure functions relevant for the description of the neutrino dimuon process.
\end{itemize}

While these common choices help to eliminate known sources of differences between the three global
fits, each group still uses their own version of the theoretical calculation at the appropriate order in QCD.
However, as further discussed in App.~\ref{app:specifics}, no major difference has been observed
in these calculations except for the known case of the HERA structure functions, due to the different heavy quark
general-mass variable-flavour number (GM-VFN) schemes adopted by each group.
As mentioned above, these choices would not necessarily be justified if the aim were to achieve
the most accurate and precise PDF fit possible, but the aim in this section is instead to understand
better the origin of the differences between the three groups.

Imposing the same values of the heavy-quark masses and of the strong
coupling constant removes an obvious source of potential difference between fits.
The lack of a strangeness asymmetry and requirement of purely perturbatively-generated charm remove further differences in approach taken by individual global fitting groups (irrespectively of whether or not they are physically justified), whilst the requirement of positive-definite quark distributions can be important when dealing with such reduced datasets due to the limited constraints, particularly on the poorly-known anti-quark PDFs at large $x$.
No deuteron or nuclear corrections are applied, as all three groups apply them, or not, in different ways, see the corresponding
discussions in Sect.~\ref{sec:combination_inputs}. Finally, the last two requirements of a fixed branching ratio for charm hadrons to muons and NNLO corrections for the dimuon data are relevant
for the description of the NuTeV dimuon data and relate to specific differences
in the strange PDF, which are discussed in more detail in App.~\ref{sec:strangeness}.

At this point it is worth stressing once again that neither the dataset nor common theory settings of the reduced fit correspond to the baseline data or theory settings adopted by any group.
Rather they represent a compromise to the least common denominator in each case and should not be regarded as the best choices for a global PDF fit.
Indeed, some of the choices made are known to be suboptimal. This setup therefore applies to the benchmarking exercise only.
Furthermore, even with these differences in the input data and theory removed, methodological differences remain,
such as those related to the choice of GM-VFN
scheme~\cite{SM:2010nsa,Guzzi:2011ew,Andersen:2014efa}, the definition
and treatment of the PDF uncertainties, and the overall fitting methodology.

\subsection{Reduced fits versus global fits}
\label{sec:reducedvsglobal}

\begin{figure}[!t]
\centering
\includegraphics[width=0.49\textwidth,trim= 0.5cm 0.5cm 0.3cm 0.4cm,clip]{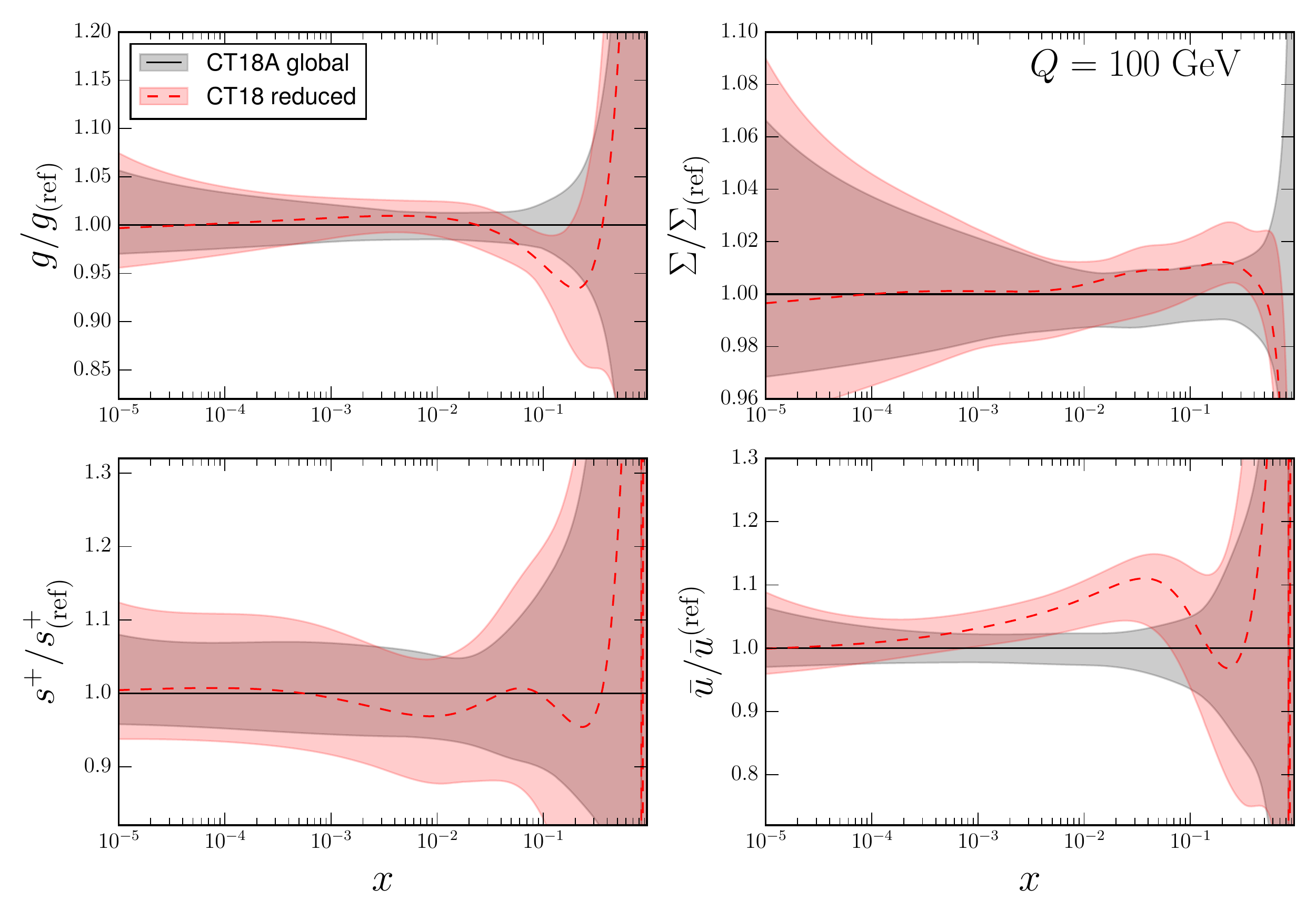}
\includegraphics[width=0.49\textwidth,trim= 0.5cm 0.5cm 0.3cm 0.4cm,clip]{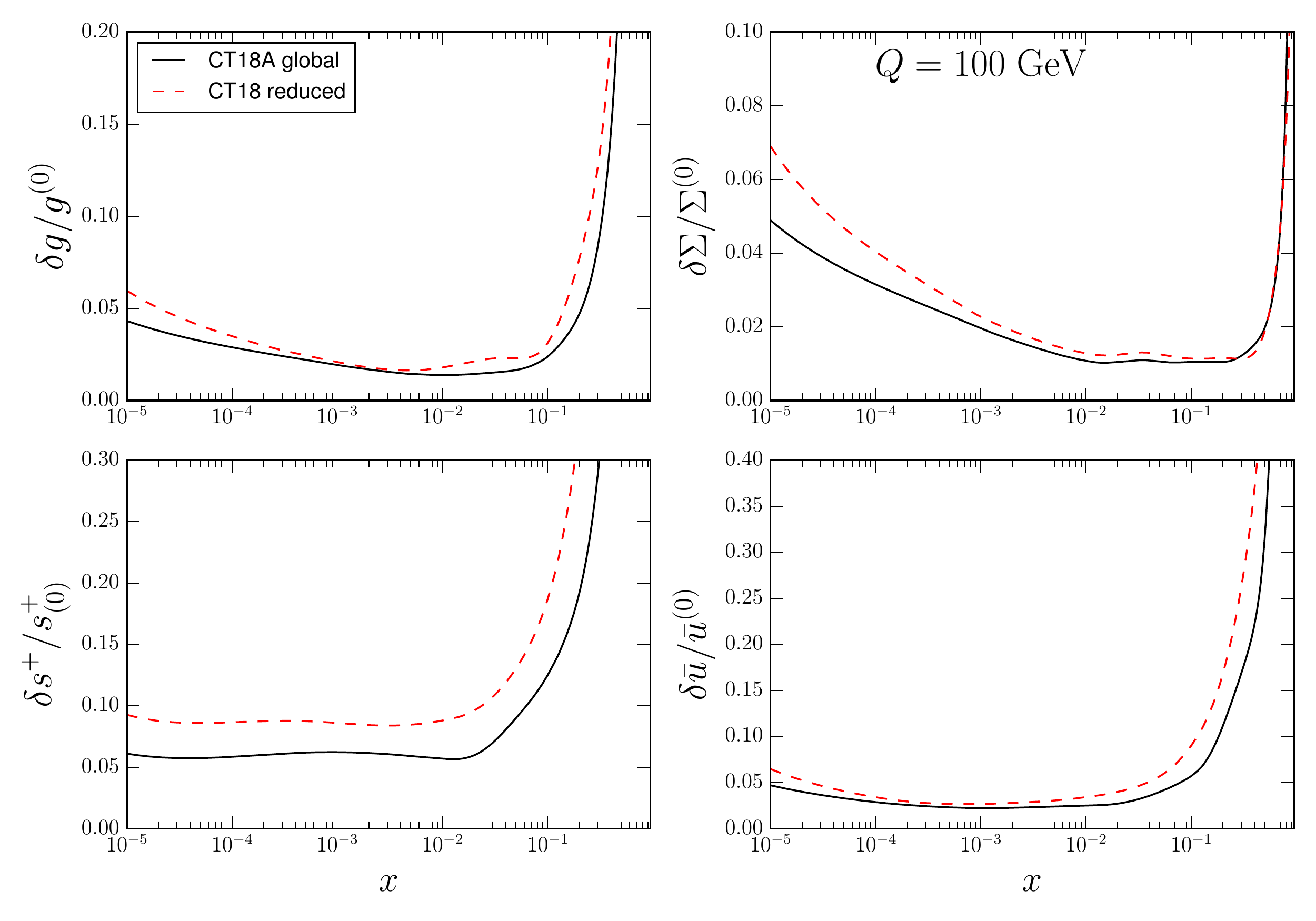}
\caption{\small Comparison of the CT18 reduced dataset PDF fit relative to the CT18A global analysis (left)
and of the corresponding $1\sigma$ PDF uncertainties (right panels).
The gluon, singlet, total strangeness and anti-up quark PDFs are displayed at $Q=100$ GeV.
The CT18A global fit is used in this comparison
as it includes the ATLAS 7, 8~TeV $W,Z$ data which are also included in the reduced fit.}
\label{fig:CTreducedvsglobal}
\end{figure}

\begin{figure}[!t]
\centering
\includegraphics[width=0.49\textwidth,trim= 0.5cm 0.5cm 0.3cm 0.4cm,clip]{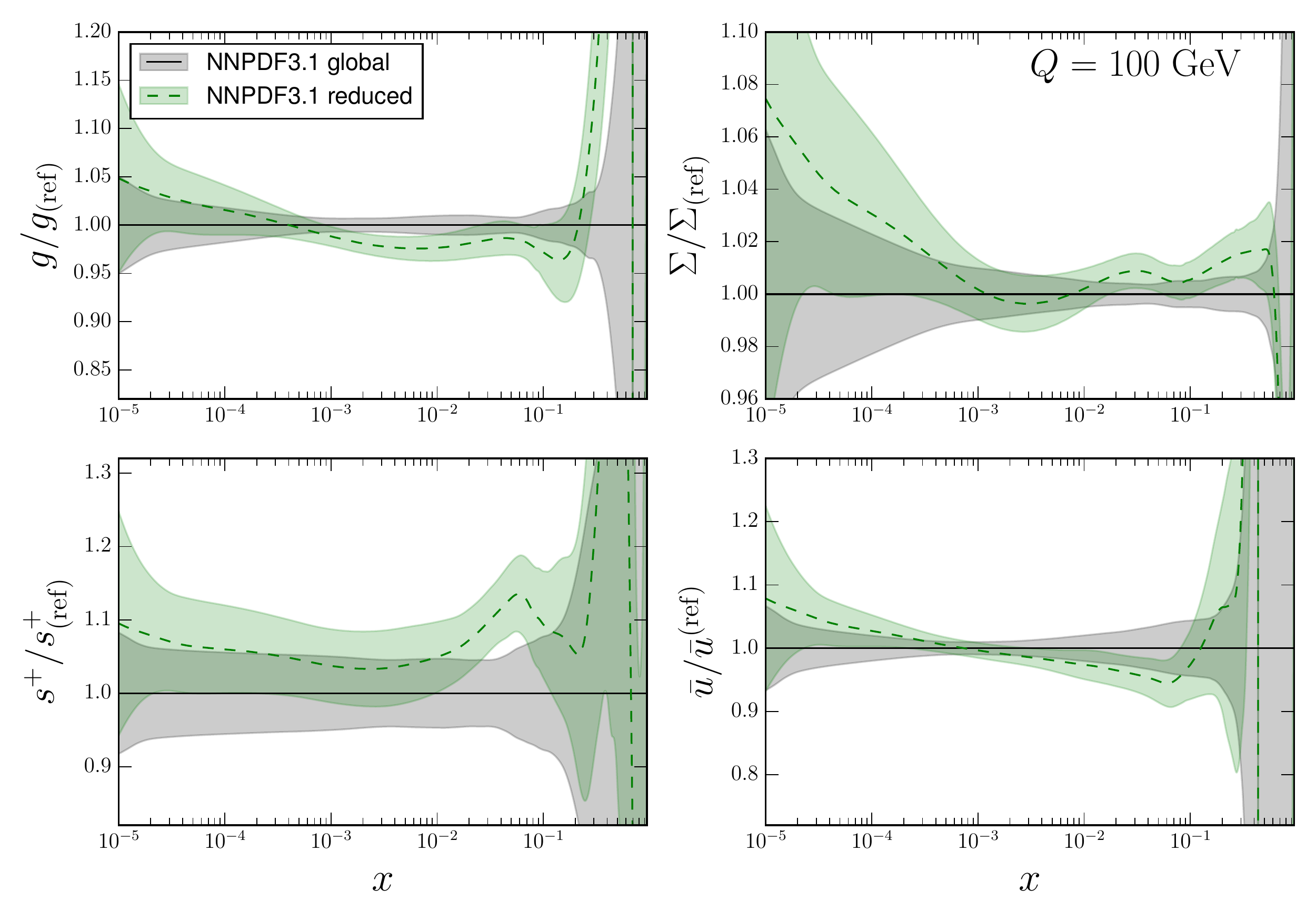}
\includegraphics[width=0.49\textwidth,trim= 0.5cm 0.5cm 0.3cm 0.4cm,clip]{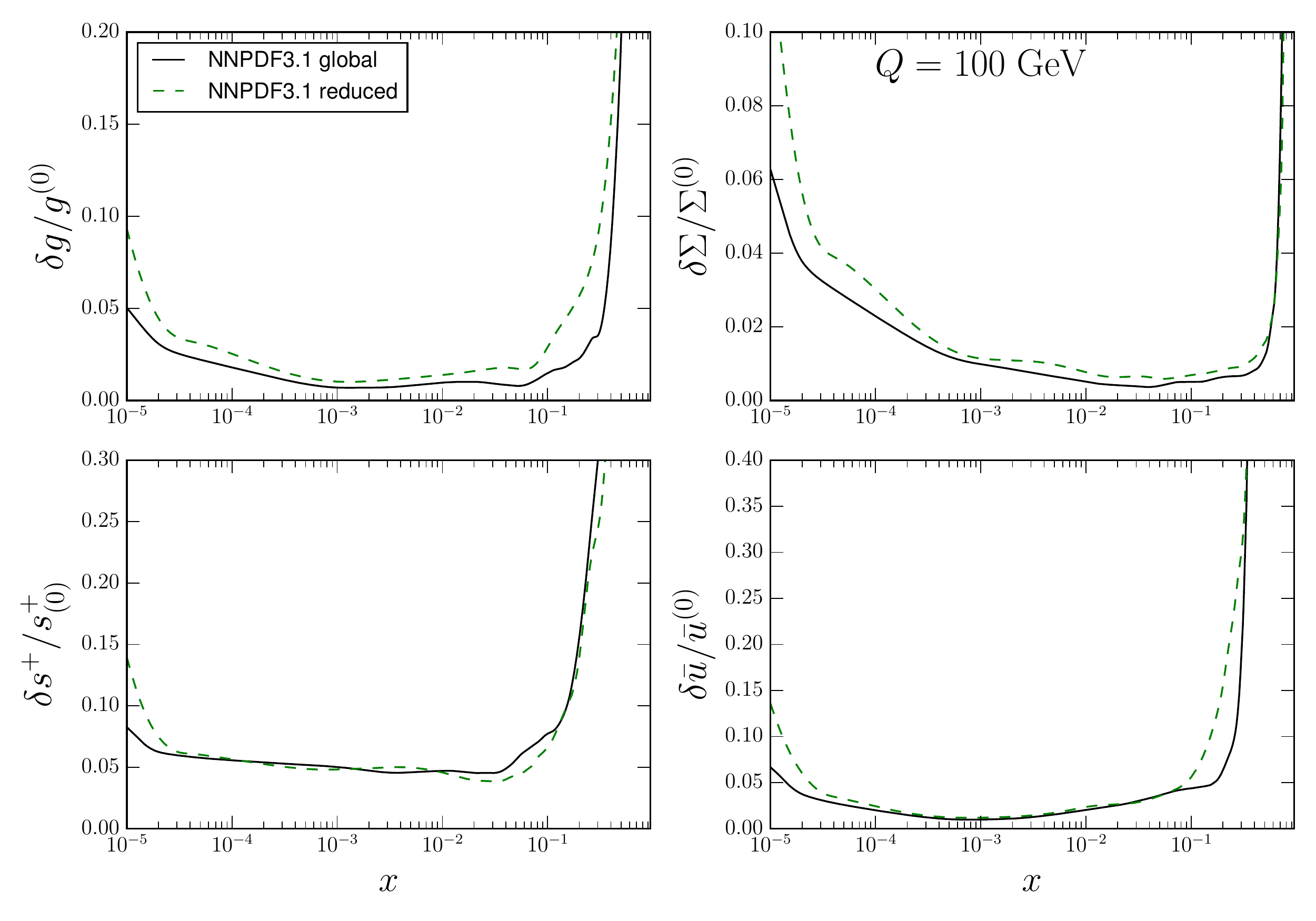}
\caption{Same as Fig.~\ref{fig:CTreducedvsglobal} for NNPDF3.1.}
\label{fig:NNPDFreducedvsglobal}
\end{figure}

\begin{figure}[!t]
\centering
\includegraphics[width=0.49\textwidth,trim= 0.5cm 0.5cm 0.3cm 0.4cm,clip]{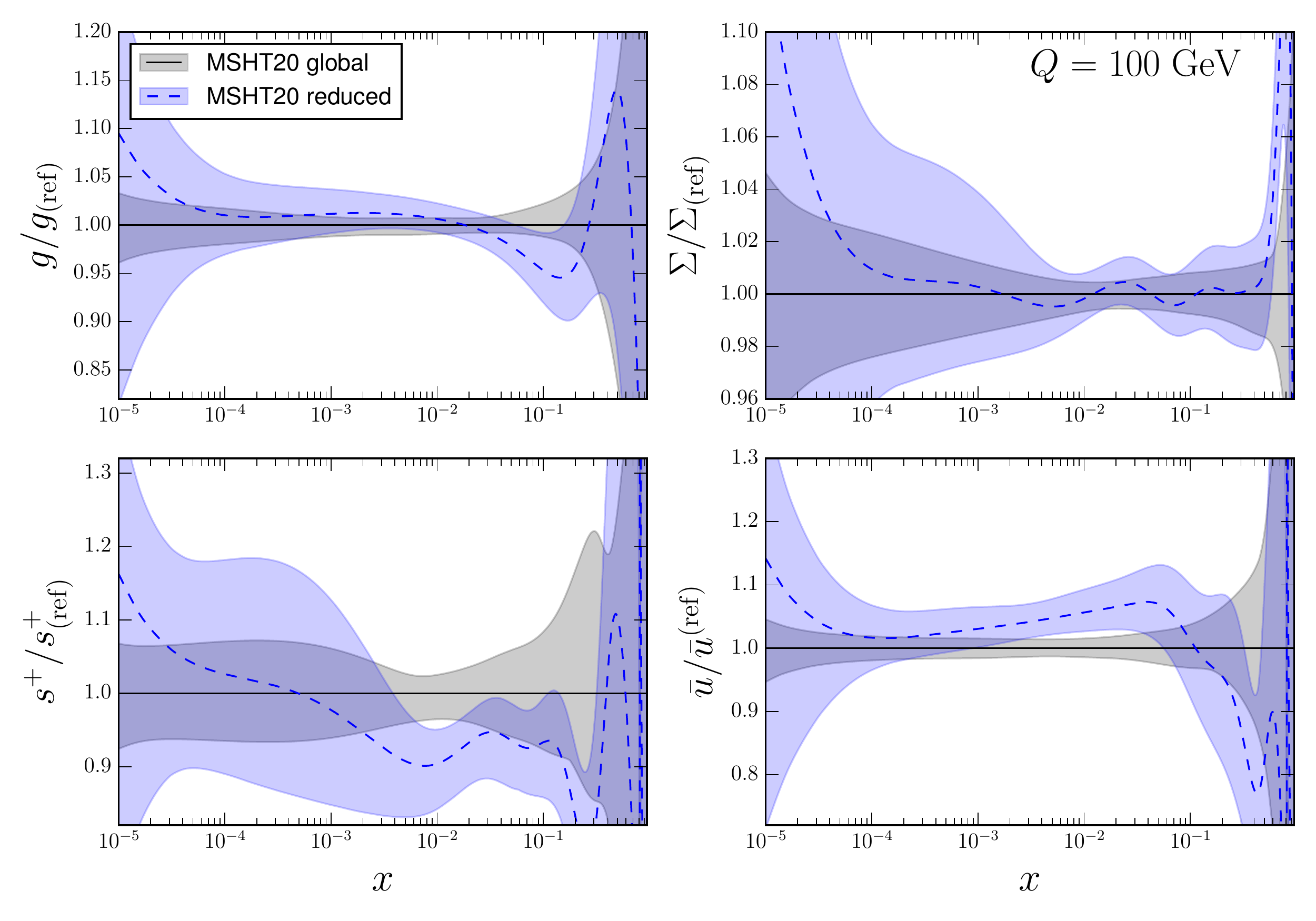}
\includegraphics[width=0.49\textwidth,trim= 0.5cm 0.5cm 0.3cm 0.4cm,clip]{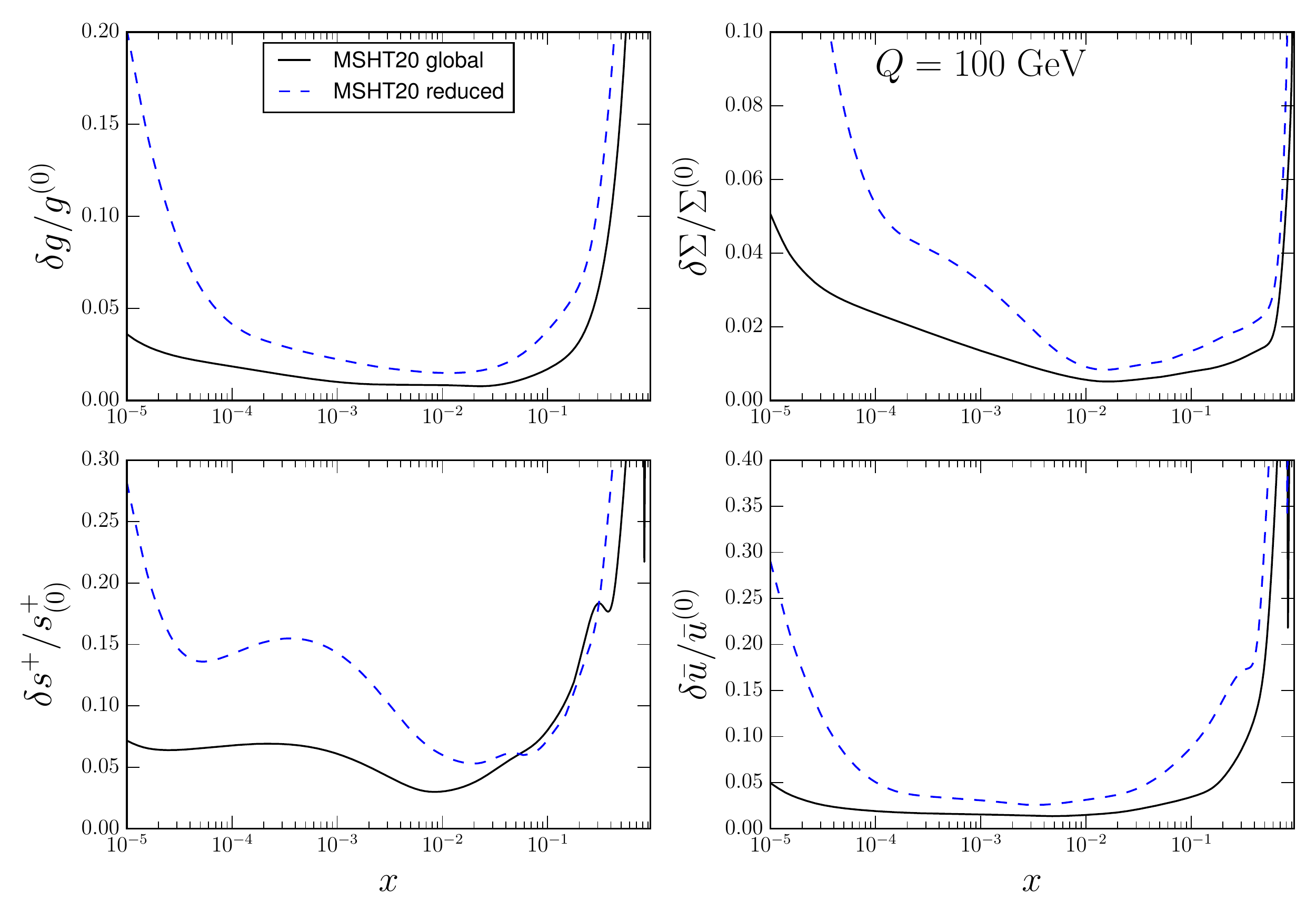}
\caption{Same as Fig.~\ref{fig:CTreducedvsglobal} for MSHT20.}
\label{fig:MSHTreducedvsglobal}
\end{figure}

At this stage, before each group performs the benchmarking of the reduced fits, it is useful to compare the reduced fit produced by each group with their published global fit.
In Figs.~\ref{fig:CTreducedvsglobal},~\ref{fig:NNPDFreducedvsglobal}, and~\ref{fig:MSHTreducedvsglobal}
we compare the PDFs from the reduced fits and those from the corresponding global analyses by CT18, NNPDF3.1, and MSHT20 respectively.
In the left panels, we compare the two fits as ratios to the central value
of the global fit, while in the right panels we display the associated
$1\sigma$ PDF uncertainties.
For illustration purposes, we display the gluon, singlet, total strangeness and anti-up quark PDFs at $Q=100$ GeV,
though similar considerations apply to the other flavour combinations.
The CT18A global fit is used in this comparison,
as it includes the ATLAS 7~TeV 
$W,Z$ data, which are also included in the reduced fit.
The NNPDF reduced fit is compared to the published NNPDF3.1 fit, not the variant 
\NNprime (or NNPDF3.1.1) that enters the combination (as described in Sect.~\ref{subsec:nnpdf31}).

We focus mostly on the comparison for CT18 for brevity, though similar qualitative considerations
apply to the NNPDF3.1 and MSHT20 results.
Overall, good compatibility is observed between the central values of the CT18 reduced fit and the CT18A global fit according to the plotted ratios, with changes in the high-$x$ gluon shape resulting from the diminished number of jet and other measurements relevant in this region in the reduced fit.
The singlet and strangeness PDFs are both compatible between the reduced and global fits within the uncertainties, whilst there is an increase in the anti-up $\bar{u}$ PDF at intermediate $x$, which signals a change in the flavour decomposition in the reduced fit.
Such changes are not unexpected, given the significant curtailment in the total size of the reduced dataset.

In the comparisons of magnitudes of PDF uncertainty bands in the right panel, clearly there is some increase in the nominal uncertainties of the reduced fit, particularly at low $x$ for the singlet, at large $x$ for the gluon
and the up antiquark, and across the whole range of $x$ for strangeness.
As with the changes to the central PDFs observed above, increases in the PDF uncertainties (as evaluated in an identical manner to the global fit) are generally to be expected, given the significantly reduced amount of data constraining the PDFs.
This said, this increase is not so large to make the reduced fits unreliable, and actually
the resulting PDF uncertainties turn out to be rather competitive as compared to what could
have been expected from the limited dataset listed in Table~\ref{tab:datasets}.
Hence, we conclude that the use of a reduced dataset should not undermine the main conclusions
derived from the present benchmarking exercise.

Similar differences in the central values and PDF uncertainties are observed
in comparisons of the MSHT20 and NNPDF3.1 reduced fits with their respective global fits.
Again, both groups report differences in the high $x$ gluon and some differences in the flavour decomposition.
In the NNPDF3.1 reduced analysis, one finds an increased strangeness PDF relatively to the NNPDF3.1 global fit, as
is further examined in App.~\ref{sec:strangeness}.
MSHT20, on the other hand, sees a reduced strangeness relative to their global fit due, in part perhaps, to the exclusion of the ATLAS 8~TeV $W,Z$ data~\cite{ATLAS:2016nqi} from the reduced fit, these data have been shown to increase the strangeness in the regions of $x$ where there is a deficit in Fig.~\ref{fig:MSHTreducedvsglobal} \cite{Bailey:2020ooq}. In addition, the requirement to fix the charm hadrons to muons branching ratio for the dimuon data, rather than allowing it to float in the fit, also lowers the strangeness in this region. 
Both groups see increased uncertainties in their reduced fits, as expected, particularly in the less constrained regions of the PDFs at low and high $x$: this is particularly true for the MSHT reduced fit at low $x$.

\subsection{Benchmarking of reduced PDF fits}
\label{sec:benchmarking_reduced}

Now that the previous subsection has assessed the main differences between the reduced and global fits, we begin with the benchmarking of the reduced fits, by comparing the outcomes obtained by the three
groups.
As discussed in Sect.~\ref{sec:reduced_fits_settings}, the use of a common dataset
and of similar fit settings should improve the agreement between the three PDF sets
as compared to the baseline fits reported in Sect.~\ref{subsec:inputs_comparison}.

Several approaches can be taken to perform this benchmarking comparison. Firstly, we can compare the central values and uncertainties among the three reduced fit PDFs themselves.
Secondly, we then seek to identify specific datasets causing observed differences by comparing the reduced fits at the level of the dataset-by-dataset individual $\chi^2$. 
To separate the effects of differences in theory predictions from other sources, the $\chi^2$ values for each common experiment of the three fits can be compared using a fixed PDF parametrisation, specifically by adopting the PDF4LHC15 NNLO set as the common input
PDF set.
Where such differences were seen, data and theory predictions themselves were directly compared to identify the origin of the differences.

\begin{figure}[!t]
\centering
\includegraphics[width=0.49\textwidth,trim= 0.5cm 0.5cm 0.3cm 0.4cm,clip]{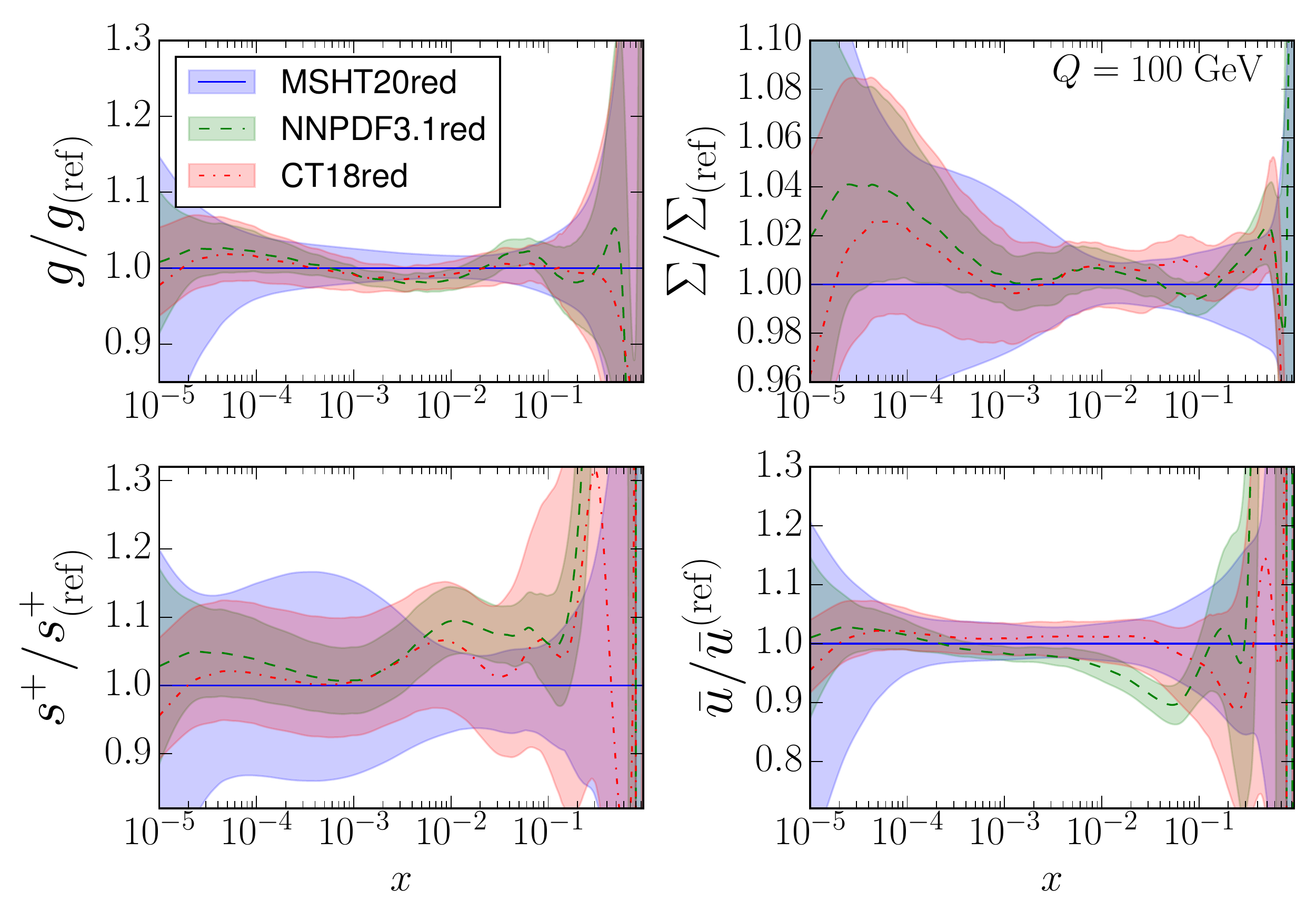}
\includegraphics[width=0.49\textwidth,trim= 0.5cm 0.5cm 0.3cm 0.4cm,clip]{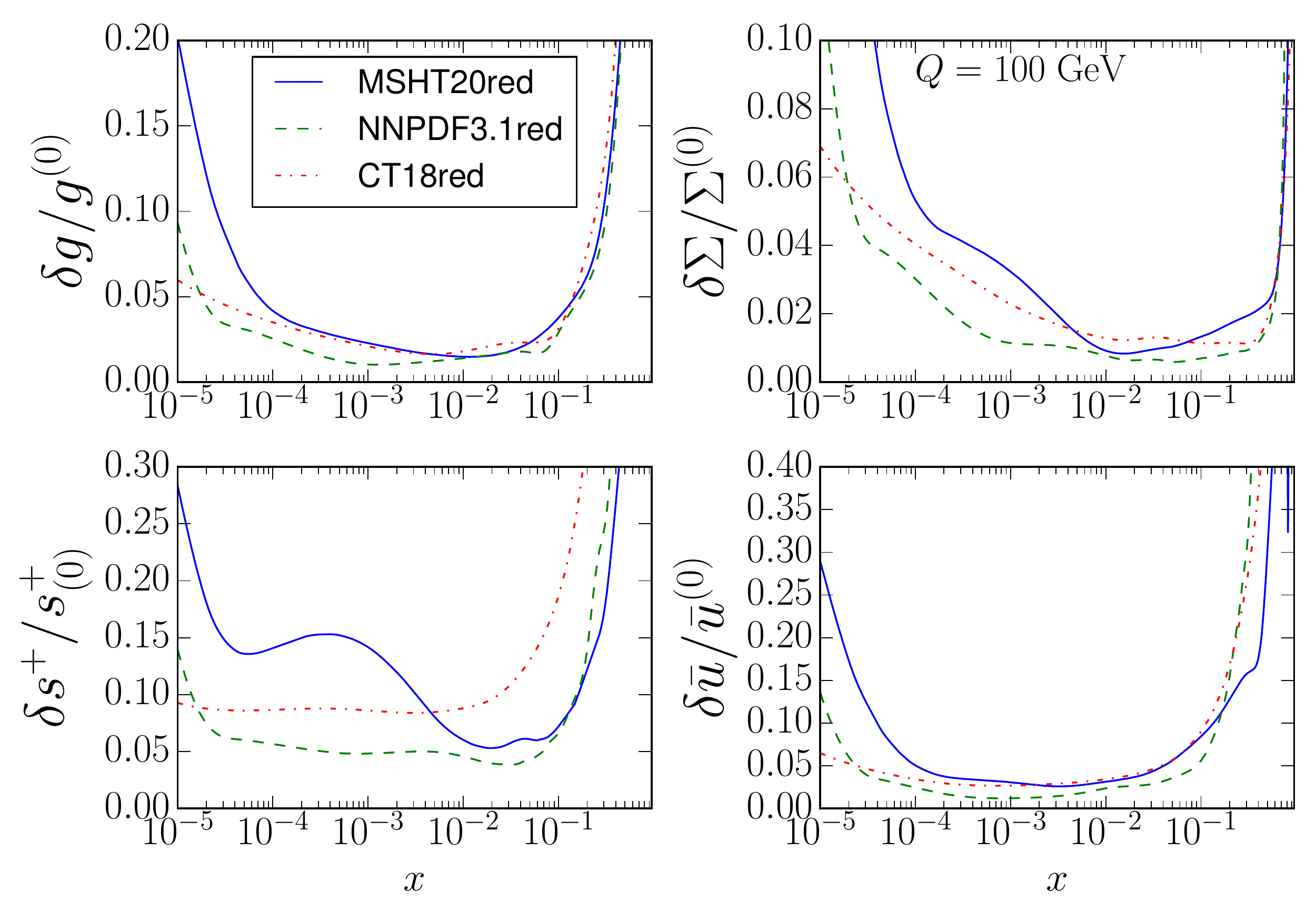}
\caption{Comparison between the reduced PDF fits from the three groups, in the same format
as in Fig.~\ref{fig:CTreducedvsglobal}. For the three groups, PDF errors correspond
to $1\sigma$ intervals.
In the left panels, PDFs are displayed normalised to the central value of the MSHT20 reduced PDF set.
}
\label{fig:reducedvsglobal}
\end{figure}

We therefore begin by comparing the PDFs and uncertainties from three reduced fits in Fig.~\ref{fig:reducedvsglobal} using 
the same format
as in Fig.~\ref{fig:CTreducedvsglobal}. 
In the left panels, PDFs are displayed normalised to the central value of the MSHT20 reduced PDF set.
The main message from this comparison is that
there is good general agreement between the three reduced fits, with the error bands of most flavours overlapping over the wide $x$ range. Starting with the gluon, we note that all three groups agree within uncertainties over the entirety of the $x$ range.
This finding strongly suggests that differences in the high-$x$ gluon shape between the global fits and relative to the reduced fits are driven by the datasets included. This region is investigated further in App.~\ref{sec:highxgluon}, 
and a further independent analysis is performed in App.~\ref{app:l2_sensitivity} by examining the $\chi^2$ pulls of individual experiments 
using the $L_2$ sensitivity. 
The three singlet PDFs are also in very good agreement for all $x$.
The strangeness is also largely consistent, albeit the NNPDF3.1 central reduced fit is notably high around $10^{-2} \lesssim x \lesssim 10^{-1}$, though this difference is within the overlap of the respective
PDF uncertainties. The origin of the different trends
in the strangeness PDF is further scrutinised in App.~\ref{sec:strangeness} and App.~\ref{app:l2_sensitivity}.
The up antiquark PDF is in good agreement between the MSHT and CT reduced fits over all $x$, the NNPDF reduced fit $\bar{u}$, however, is lower than both MSHT and CT in the $10^{-2} \lesssim x \lesssim 10^{-1}$ region, signalling a difference in the high-$x$ flavour decomposition.

The relative $1\sigma$ PDF uncertainties of the three reduced fits, displayed in the rightmost panels of
Fig.~\ref{fig:reducedvsglobal}, turn out to be similar in size in regions with good data constraints.
The agreement between the PDF uncertainties for the gluon in $x\gsim 10^{-2}$ among the three groups
is particularly remarkable. For lower $x$ values, the NNPDF3.1 gluon uncertainty is  smaller. This has an impact on the $gg$ PDF luminosity, as will be discussed later. 
The MSHT20 reduced fit displays larger uncertainties outside of these regions, i.e. where constraints are lacking in the reduced fit --- particularly at low $x$.
A further examination of the uncertainties of the reduced and global fits
is ongoing and will be reported in the future.

In order to further identify any differences in the reduced fits, we examine their goodness-of-fit values for each individual dataset, as given by $\chi^2/N_{\rm pt}$.
Before calculating these for the PDFs from the reduced fits themselves, it is useful to compare the agreement between theory and data with a fixed PDF4LHC15 NNLO parametrisation as the common input, {\it i.e.}, in lieu of fitting.
%
Table~\ref{tab:reducedfit_chisqs_PDF4LHC15in} indicates the values of the $\chi^2/N_{\rm pt}$ for the measurements that enter the fits to
the reduced datasets  and listed in  Table~\ref{tab:datasets}.
The results are obtained using the codes from each of the three groups, for the common theory
settings listed in Sect.~\ref{sec:reduced_fits_settings}.
Hence this comparison 
is sensitive only to differences in the implementation of the various datasets or to differences in the 
theoretical calculations performed by each group.
In addition to the presented $\chi^2$ values, theoretical predictions for individual data points using the same PDF4LHC15 set were compared, which allowed us to track the differences among the theoretical computations implemented in the three fits. Note that, in some cases, a poor $\chi^2$ was expected, for example for
the ATLAS 7~TeV $W$,$Z$~(2016) data set, which had not been included in fits for the PDFs in the PDF4LHC15 combination.

\begin{table}[!t]
\centering
\footnotesize
  \renewcommand\arraystretch{1.30} 
\begin{tabular}{|l|c|c|c|c|}
\hline
 Dataset                              & $N_{\rm pt}$ & \multicolumn{3}{c|}{  $\chi^2/N_{\rm pt}$}     \\
 \hline
                                     &   &  $\qquad$CT18$\qquad$ & MSHT20 &  NNPDF3.1   \\
\hline
BCDMS $F_2^p$                       &  329/163$^{\dagger\dagger}$/325$^\dagger$  &   1.35    &   1.20    &   1.51    \\
 BCDMS $F_2^d$                       &  246/151$^{\dagger\dagger}$/244$^\dagger$  &   0.97    &   1.27   &   1.24    \\
 NMC $F_2^d/F_2^p$                   &  118/117$^\dagger$                         &   0.92    &   0.93   &   0.94    \\
 NuTeV dimuon $\nu+\bar{\nu}$ &  38+33                                     &   0.75    &   0.73   &   0.84    \\
HERAI+II                            &   1120                                     &   1.27    &   1.24   &   1.74    \\
E866 $\sigma_{pd}/(2\sigma_{pp})$   &     15                                     &   0.45    &   0.54   &   0.59    \\
 LHCb 7~TeV \& 8TeV  $W$,$Z$           &  29+30                                     &   1.5     &   1.34   &   1.76    \\
LHCb 8~TeV $Z \rightarrow ee$        &     17                                     &   1.35    &   1.65   &   1.25    \\
ATLAS 7~TeV $W$,$Z$~(2016)            &     34                                     &   6.71    &   7.46   &   6.51    \\
 D0 Z rapidity                       &     28                                     &   0.61    &   0.58   &   0.61    \\
 CMS 7~TeV electron $A_{\rm ch}$           &     11                                     &   0.45    &   0.5    &   0.73    \\
ATLAS 7~TeV $W$,$Z$~(2011)            &     30                                     &   1.21     &   1.23   &  1.31    \\
 CMS 8~TeV incl. jet                  &  185/174$^{\dagger\dagger}$                &   1.53    &   1.89   &   1.78    \\
\hline
Total    $N_{\rm pt}$                     &    ---                                     &   2263    &    1991       &   2256   \\
\hline
Total    $\chi^2/N_{\rm pt}$              &    ---                                     &   1.31    &    1.36       &   1.62 \\
\hline
\end{tabular}
\vspace{0.3cm}
        \caption{\small
The values of the $\chi^2/N_{\rm pt}$ for the measurements that enter the fits to
the reduced datasets  and listed in  Table~\ref{tab:datasets}.
The results are obtained using the codes from each of the three groups, for the common theory
settings listed in Sect.~\ref{sec:reduced_fits_settings} and in all cases
using the PDF4LHC15 PDF as input, i.e., with fixed PDFs. 
%
%
$^{\dagger\dagger}$MSHT\ \ $^{\dagger}$NNPDF.}
\label{tab:reducedfit_chisqs_PDF4LHC15in}                
\end{table}

This exercise confirmed that there is good overall agreement both between the datasets and theoretical computations implemented in the reduced fits of three groups.
Note that, as indicated in the table, different groups may have slightly different numbers of points for some datasets, in particular for the BCDMS data, with MSHT20 having substantially fewer points due to their use of a different version of the same data, which encodes essentially the same information and constraints.
However, there are some differences in the $\chi^2$ results among the three groups, the most significant of these being in the fit quality of the combined HERA dataset of NNPDF3.1, which has a $\chi^2$ value per point which is 0.5 worse than CT18 and MSHT20. Since this dataset comprises half of the overall reduced fit dataset, this difference is also visible in the overall fit quality.
This is, however, a known difference indicative of a mismatch of the heavy-flavour schemes between the PDF4LHC15 combination and
those used by the three groups, which is most prevalent in NNPDF3.1 and in the low $Q^2$ region probed by the HERA dataset~\cite{SM:2010nsa,Guzzi:2011ew,Andersen:2014efa}. In particular, this difference disappears
in the corresponding comparison after fitting (Table~\ref{tab:reducedfit_chisqs_fit}), showing that it is well understood.

The theoretical predictions 
for the reduced dataset
based on the MSHT and CT codes are generally in a very good mutual agreement, with both reporting very similar overall fit qualities of 1.31 and 1.36 over the whole reduced fit dataset.
MSHT does have a higher $\chi^2/N_{\rm pt}$ than CT and NNPDF for the CMS 8~TeV inclusive jet data, but this is expected due to the inclusion of statistical correlations in MSHT (which also is the origin of the reduced number of points for this dataset in MSHT).
As mentioned above, all three groups describe poorly the ATLAS 7~TeV $W,Z$ (2016) data using the PDF4LHC15 PDFs with its lower strangeness distribution, which is now inconsistent with that favoured by this dataset.

On the other hand, after the three groups fit their PDFs to the reduced dataset, the agreement between theory and data improves
markedly across the board, see the corresponding $\chi^2/N_{\rm pt}$ values in Table~\ref{tab:reducedfit_chisqs_fit}.
Very good total figure-of-merit values of  $\chi^2/N_{\rm pt}$=1.14, 1.15, 1.20 are achieved for the CT, MSHT and NNPDF reduced fits respectively. Moreover, this global agreement is also seen in the dataset-by-dataset comparisons, with the majority showing good agreement. There are, nonetheless, some differences, notably for the NuTeV dimuon data, described with $\chi^2/N_{\rm pt} \approx 0.8$ by CT and MSHT, and 1.2 by NNPDF.
This is consistent with the increased strangeness noted earlier in the NNPDF reduced fit in the region $10^{-2} \lesssim x \lesssim 10^{-1}$, given the dimuon data favour lower strangeness here. 
However, this occurs without improvement in the fit quality of the ATLAS 7~TeV $W,Z$ data which pulls the strangeness upwards in this $x$ region. The origin of this difference in the strangeness is investigated in App.~\ref{sec:strangeness}.
%
An additional difference is observed in the quality of the NNPDF fit to smaller datasets, such as the E866 Drell-Yan ratio data or the CMS 7~TeV electron asymmetry. For the former, there is a known difference between CT and MSHT due to the parameterisation of the $\bar{d}-\bar{u}$ 
asymmetry in the relevant high-$x$ region~\cite{Hou:2019efy,Bailey:2020ooq}.
For the latter, CT and MSHT both obtain 1.5 per point, whereas NNPDF achieves half of this at 0.76, again this will be analysed briefly later in the context of App.~\ref{sec:highxgluon}.

\begin{table}[!t]
\centering
\footnotesize
  \renewcommand\arraystretch{1.30} 
\begin{tabular}{|l|c|c|c|c|}
\hline
 Dataset                              & $N_{\rm pt}$ & \multicolumn{3}{c|}{  $\chi^2/N_{\rm pt}$}     \\
 \hline
                                     &   &  $\qquad$CT18$\qquad$ & MSHT20 &  NNPDF3.1   \\
\hline
  BCDMS $F_2^p$                       &  329/163$^{\dagger\dagger}$/325$^\dagger$   &    1.06   &   1.00   &   1.21  \\
 BCDMS $F_2^d$                       &  246/151$^{\dagger\dagger}$/244$^\dagger$   &    1.06   &   0.88   &   1.10  \\
   NMC $F_2^d/F_2^p$                   &  118/117$^\dagger$                          &    0.93   &   0.93   &   0.90  \\
  NuTeV dimuon $\nu + \bar{\nu}$ &  38+33                                      &    0.79   &   0.83   &   1.22  \\
   HERAI+II                            &   1120                                      &    1.23   &   1.20   &   1.22  \\
  E866 $\sigma_{pd}/(2\sigma_{pp})$   &     15                                      &    1.24   &   0.80   &   0.43  \\
   LHCb 7~TeV \& 8TeV  $W$,$Z$           &  29+30                                      &    1.15   &   1.17   &   1.44  \\
   LHCb 8~TeV $Z \rightarrow ee$        &     17                                      &    1.35   &   1.43   &   1.57  \\
  ATLAS 7~TeV $W$,$Z$~(2016)            &     34                                      &    1.96   &   1.79   &   2.33  \\
   D0 Z rapidity                       &     28                                      &    0.56   &   0.58   &   0.62  \\
  CMS 7~TeV electron $A_{\rm ch}$           &     11                                      &    1.47   &   1.52   &   0.76  \\
   ATLAS 7~TeV $W$,$Z$(2011)            &     30                                      &    1.03   &   0.93   &   1.01  \\
  CMS 8TeV incl. jet                  &  185/174$^{\dagger\dagger}$                 &    1.03   &   1.39   &   1.30  \\
  \hline
  Total    $N_{\rm pt}$                     &    ---    &   2263       &    1991       &   2256   \\     
  \hline                                                                                    
  Total    $\chi^2/N_{\rm pt}$              &    ---    &   1.14       &    1.15       &   1.20 \\
  \hline
  \end{tabular}
  \vspace{0.3cm}
          \caption{Same as Table~\ref{tab:reducedfit_chisqs_PDF4LHC15in}, now
          displaying the results obtained after each group has carried out
          the corresponding fits to this reduced dataset.
That is, the input PDF is now the best-fit value obtained for each group to the reduced
dataset rather than the common PDF4LHC15 PDF input used in Table~\ref{tab:reducedfit_chisqs_PDF4LHC15in}.
$^{\dagger\dagger}$MSHT\ \ $^{\dagger}$NNPDF.}
\label{tab:reducedfit_chisqs_fit}                                
\end{table}

\subsection{Partonic luminosities in the reduced fits}
\label{sec:reducedfit_lumis}

%
%
Next, we present a comparison of partonic luminosities
for the three reduced fits, 
both with and without a cut on the rapidity of the massive final state. 

\begin{figure}[t]
\centering
\includegraphics[width=0.49\textwidth,trim= 0.2cm 0.2cm 0.1cm 0.2cm,clip]{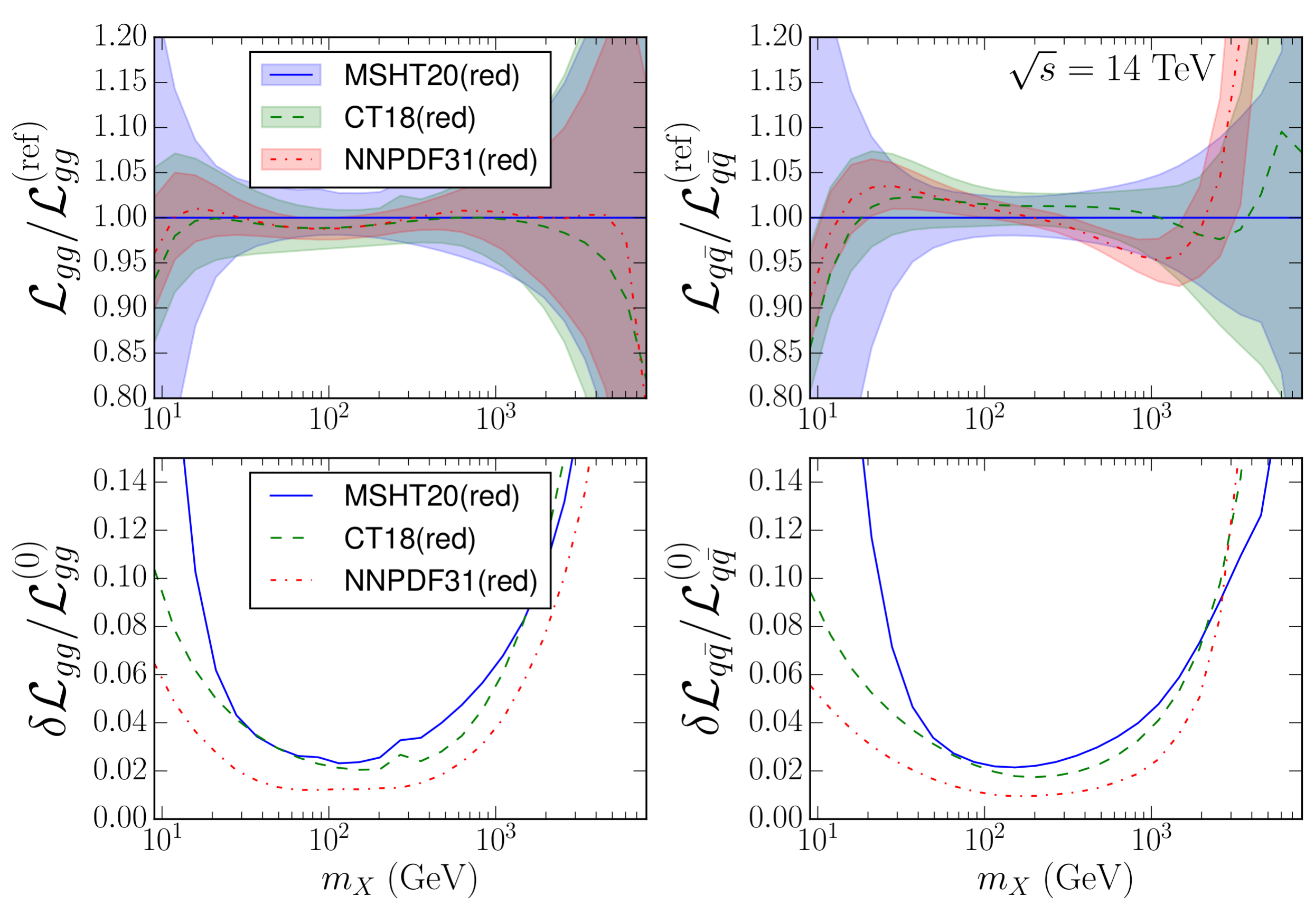}
\includegraphics[width=0.49\textwidth,trim= 0.2cm 0.2cm 0.1cm 0.2cm,clip]{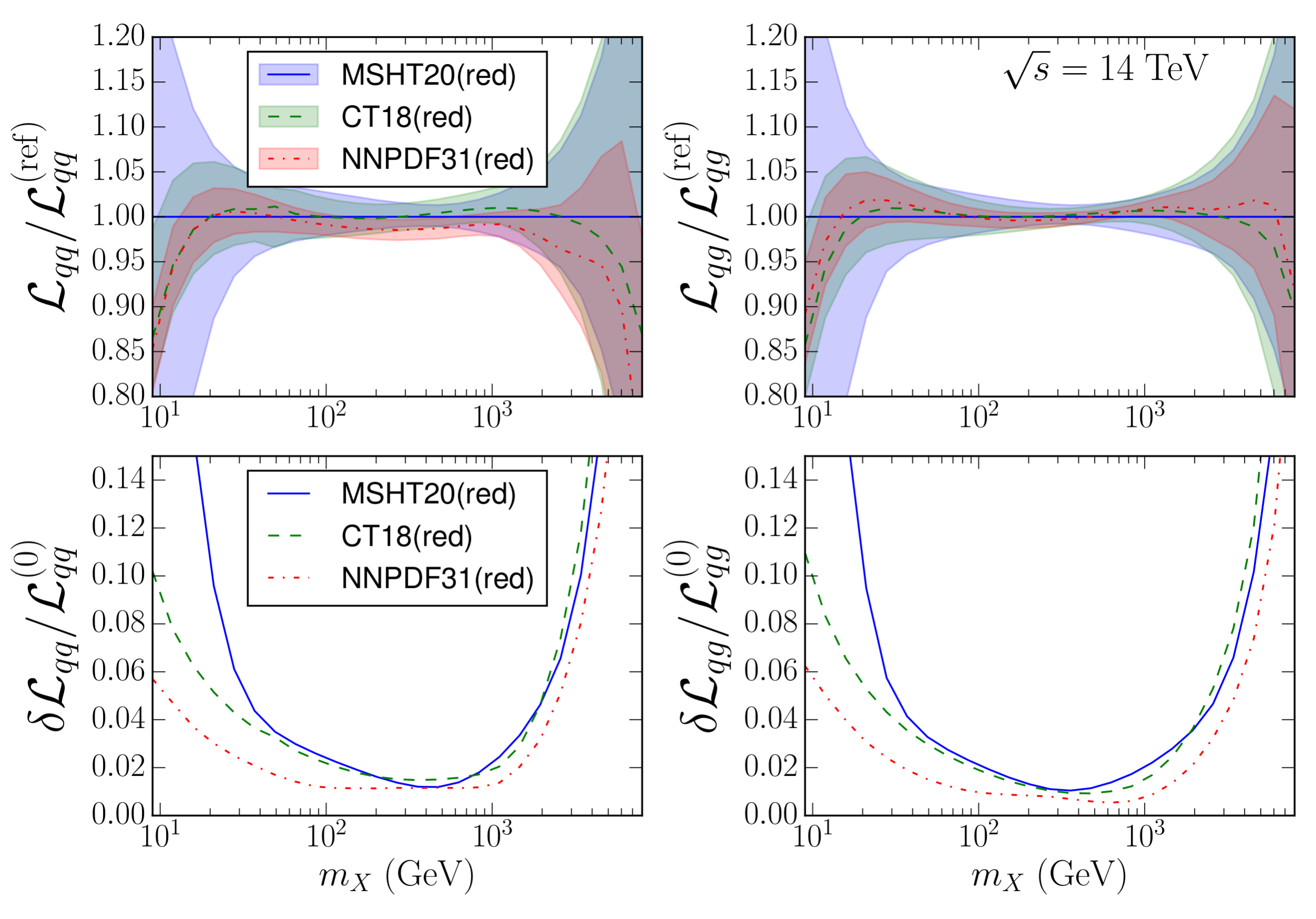}
\includegraphics[width=0.49\textwidth,trim= 0.2cm 0.2cm 0.1cm 0.2cm,clip]{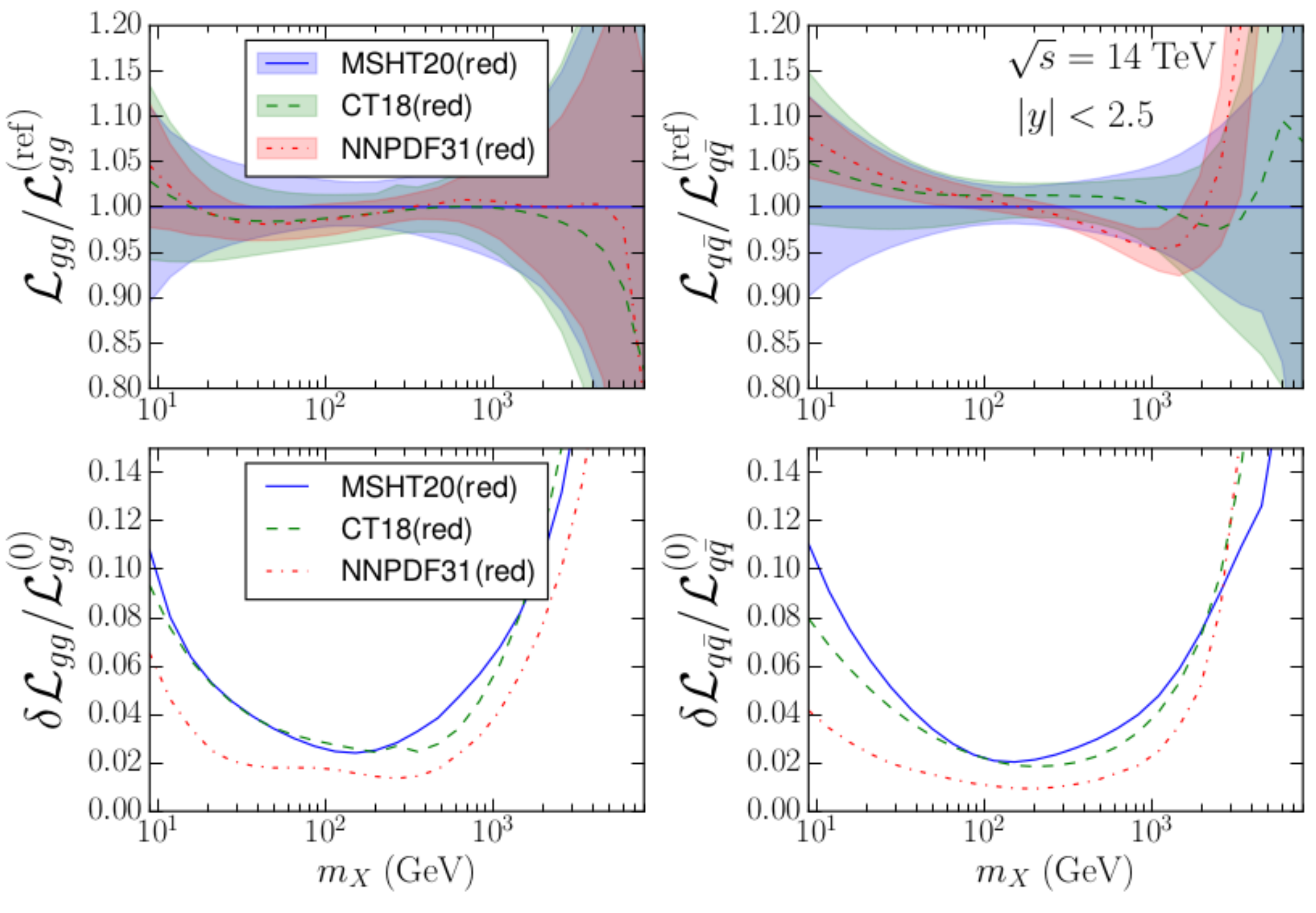}
\includegraphics[width=0.49\textwidth,trim= 0.2cm 0.2cm 0.1cm 0.2cm,clip]{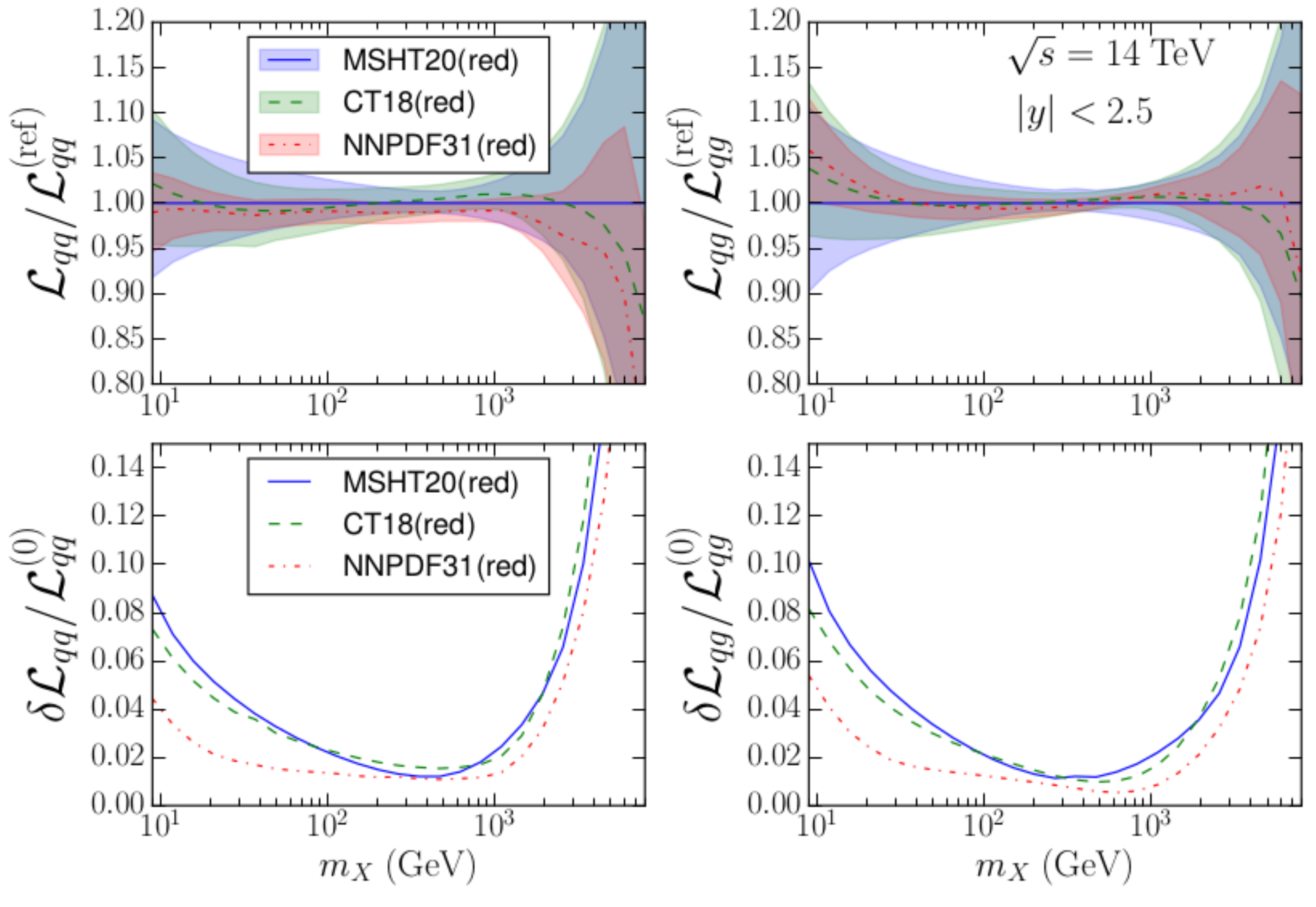}
\caption{Comparison of the  partonic luminosities between the CT18, MSHT20, and NNPDF3.1 reduced
fits at $\sqrt{s}=14$ TeV as a function of the invariant mass of the produced final state $m_X$.
Upper two rows: $gg$, $q\bar q$, $qq$, and $qg$ luminosities evaluated without any restriction
on the final-state rapidity $y_X$. Bottom two rows: same, for a rapidity  cut of $\lvert y_X \rvert < 2.5$.
}
\label{fig:Reducedfit_lumi_comp}
\end{figure}

Figure~\ref{fig:Reducedfit_lumi_comp} displays the comparison of these partonic luminosities
between the CT18, MSHT20, and NNPDF3.1 reduced
fits at $\sqrt{s}=14$ TeV as a function of the invariant mass of the produced final state $m_X$.
From left to right we show the  gluon-gluon, quark-antiquark, quark-quark and quark-gluon luminosities,
normalised to the central value of the MSHT20 prediction, together with the associated $1\sigma$
relative PDF uncertainties.
The upper two rows display the luminosities evaluated without any restriction
on the final-state rapidity $y_X$, while the bottom two rows
instead account for a rapidity  cut of $\lvert y_X \rvert < 2.5$,
which restricts the produced final state to lie within the ATLAS/CMS central acceptance region.

There is in general a very good agreement between the central values of the reduced fits at the level of the parton-parton luminosities. The important gluon-gluon luminosities agree within uncertainties across the entire $m_X$ range, and similarly the quark-quark and quark-gluon luminosities. The agreement of the quark-antiquark luminosity is also good, however the NNPDF3.1 reduced fit luminosity is on the edge of the uncertainty bands for a small portion of the $m_X$ range, perhaps due to the remaining difference in the quark flavour decomposition.
Some statistically insignificant differences seen among the central values of the three groups at very low and very high $m_X$ might be expected, as  these regions are the least constrained in general and particularly in the reduced fits.

The uncertainties of the 
$qq$ and $qg$ luminosities from the three groups are similar in the most constrained $m_X$ intervals.
At low $m_X$, the magnitudes of uncertainties differ among the groups for all four luminosities, with the MSHT reduced fit having significantly larger uncertainties here, as also seen in the reduced fit PDFs at low $x$, again more noticeably reflecting the lack of constraints in this region. 

Meanwhile, the $gg$ and $q\bar q$ luminosity uncertainties are similar in the CT and MSHT reduced fits in the central $m_X$ region, and these both are notably larger than the corresponding uncertainty of the NNPDF reduced fit. For example, the uncertainty on the $gg$ luminosity at the Higgs mass is 
2.3\% for the MSHT reduced fit, 
2.1\% for the CT reduced fit, and only
1.2\% for the NNPDF reduced fit.
This is despite the similarly sized uncertainties of the three gluon PDFs in Fig.~\ref{fig:reducedvsglobal} in the central $x$ region ($x\simeq 0.01$), most relevant for Higgs production in gluon fusion. 
However, the gluon uncertainties differ more in the extreme regions of $x$ (especially for $x$ less than 0.01). These differences may propagate into the $gg$ luminosity uncertainties in the second row of Fig.~\ref{fig:Reducedfit_lumi_comp}, when no rapidity cut is applied to the final state.

It is therefore informative to apply a rapidity cut in order to reduce the effect of extreme kinematic regions, particularly from low $x$.
The bottom panels of Fig.~\ref{fig:Reducedfit_lumi_comp} provide the same luminosity comparison as in the
upper panels, but with a rapidity cut of $\lvert y_X \rvert < 2.5$ applied.
This also allows for a more realistic comparison in the kinematic region used for
central-rapidity measurements by ATLAS and CMS.

As expected, the cut affects the low $m_X$ range, where the momentum fractions in the colliding partons are very unbalanced in many events. At low $m_X$, the cut significantly improves the agreement between the central values of the luminosities, whilst the uncertainties for all four parton-parton luminosities are reduced significantly and are brought into a better mutual agreement among the three groups.
Again however, the uncertainties on the gluon-gluon and quark-antiquark luminosities remain noticeably different even at central $m_X$ between the MSHT and CT reduced fits, on the one hand, and the NNPDF's reduced fit, on the other.

It is particularly interesting to note that the uncertainties in the $gg$ luminosity at the Higgs mass actually increase slightly for all three groups after applying the rapidity cut, with the MSHT, CT, and NNPDF reduced fits achieving the uncertainties of 
2.4\%, 
2.7\%, and  
1.7\%, respectively. The three uncertainties are now in a closer accord than without the cut; the smaller the uncertainty without the rapidity cut was, the larger the change after the cut has occurred,
consistent with the expectation that the uncertainties in the low-$x$ PDFs cause some of the observed differences. The increase in the $gg$ luminosity uncertainty at $m_X =125~{\rm GeV}$ for the NNPDF reduced fit may reflect the removal of the $x\sim 10^{-3}$ region of the gluon, which is better constrained in their reduced fit than the $x\sim 0.01$ region. On the other hand, naively one would expect the $gg$ luminosity uncertainty to be reduced for MSHT and CT after applying the rapidity cut, as their gluons' uncertainties are the smallest around $x\sim 0.01$. The increase (albeit only a very slight one in MSHT) therefore implies some anti-correlation between the contributions with one high-$x$ and one low-$x$ parton, which are now cut, and those with reasonably similar $x$, which remain. 

\subsection{Summary of the benchmarking with reduced fits}
To summarise the main findings of this benchmark comparison, once a common dataset and similar theory
and methodology settings are adopted by the three groups, the agreement among the reduced fits is indeed better than it was in the full global fits, both at the level of the PDFs and of the dataset-by-dataset $\chi^2$.
The satisfactory consistency of the three reduced fits at the PDF level is further confirmed by the  comparisons of the partonic luminosities in Fig.~\ref{fig:Reducedfit_lumi_comp}.
Another reassuring result is based on the comparisons of $\chi^2$ values and theoretical predictions using the common PDF4LHC15 set, which give us confidence that there are no ``trivial'' explanations of the differences observed among the global fits, such as an incorrectly implemented dataset or a buggy
theoretical calculation.

This said, the fact that the residual differences remain after the PDFs are refitted in this exercise, such as in the magnitude of the PDF uncertainties,
indicates that the methodological choices adopted by each group remain important even
when fitting to the same dataset (albeit a reduced one in these benchmark fits) with very similar theory settings. In some cases, methodological uncertainties, such as those associated with the functional form, fitting methodology, or the definition of the PDF errors can be as large or even larger than the PDF uncertainties associated with the fitted data.
The presence of such “irreducible” differences justifies the adoption of a PDF4LHC15-like strategy for the combination of the three global sets in the next section.
\section{The PDF4LHC21 combination}
\label{sec:pdf4lhc21}

In this section we present the outcome of the PDF4LHC21 combination, based on the variants
of the CT18, MSHT20, and NNPDF3.1 global PDF analyses - \CTprime, MSHT20 and \NNprime - described in
Sect.~\ref{sec:combination_inputs}.
First of all, we describe the generation of the Monte Carlo
replicas and the main features of the resulting
combined distribution, including a comparison with
the three constituent PDF fits.
Second, we present the results of the Monte Carlo
compression and of the Hessian  reduction of PDF4LHC21,
which lead to the {\sc\small LHAPDF} grids released
and recommended for phenomenological applications.
Third, we compare 
PDF4LHC21 with its predecessor PDF4LHC15 both
at the level of PDFs and of partonic luminosities.
Finally, we assess the behaviour of the PDF4LHC21 combination at large-$x$,
and provide a prescription to deal with cross-sections which may become negative in this region.
The corresponding comparisons at the level
of inclusive and differential LHC cross-sections
are then presented in Sect.~\ref{sec:phenomenology}.

\subsection{Generation and combination of Monte Carlo replicas}
\label{sec:generation_mc_replicas}

The first step in the PDF4LHC21 combination procedure is the construction of
a Monte Carlo sampling
of the probability distributions associated to each of the three inputs
to the combination.
This requires in particular transforming the native Hessian sets (\CTprime and MSHT20)
into a Monte Carlo representation using one of the available methods,
see~\cite{Watt:2012tq,Hou:2016sho} and references therein.
As an example of this transformation, in the Watt-Thorne method \cite{Watt:2012tq}, there exist two options to construct a Monte Carlo representation of an asymmetric Hessian PDF set, such as MSHT20.
First, one could use the following expression
\be
\label{eq:mc_replica_generation_1}
\mathcal{F}^{(k)} = \mathcal{F}\lp S_0\rp + \sum_{j=1}^{N_{\rm eig}}\lc
\mathcal{F}\lp S_i^{(\pm)}\rp  - \mathcal{F}\lp S_0\rp \rc \left|R_j^{(k)}\right| \,, \qquad k=1\,\ldots,N_{\rm rep} \, ,
\ee
where $\mathcal{F}$ indicates a general PDF-dependent quantity (in particular, $\mathcal{F}$
can be a PDF itself), $N_{\rm rep}$ is the number of replicas to be generated,
$ S_0$ indicates the central (best-fit) PDF set, and $S_i^{(\pm)}$ corresponds to the
$i$-th eigenvector (out of a total of $N_{\rm eig}$ eigenvectors) along the positive
or negative direction.
In Eq.~(\ref{eq:mc_replica_generation_1})
the $R_j^{(k)}$ are independent Gaussian random numbers, and the sign of 
 $S^\pm_j$ is chosen depending on the
sign of the  $R_j^{(k)}$.
The expression in Eq.~(\ref{eq:mc_replica_generation_1})
has the disadvantage that, in general, the average over the replicas
does not coincide with the best-fit predictions, that is,
\be
\la \mathcal{F}\ra_{\rm rep} = \frac{1}{N_{\rm rep}}\sum_{k=1}^{N_{\rm rep}} \mathcal{F}^{(k)}  \ne \mathcal{F}\lp S_0\rp \, .
\ee
To avoid this limitation, it is preferred to adopt a symmetrised version of
Eq.~(\ref{eq:mc_replica_generation_1}) taking the form
\be
\label{eq:mc_replica_generation}
\mathcal{F}^{(k)} = \mathcal{F}\lp S_0\rp + \frac{1}{2}\sum_{j=1}^{N_{\rm eig}}\lc
\mathcal{F}\lp S_i^{(+)}\rp  - \mathcal{F}\lp S^{(-)}_i\rp \rc R_j^{(k)} \,, \qquad k=1\,\ldots,N_{\rm rep}
\, ,
\ee
for which instead the property $\la \mathcal{F}\ra_{\rm rep} =  \mathcal{F}\lp S_0\rp$
is satisfied.

In the case of \CTprime, the Monte-Carlo replicas were generated by using a log-normal sampling procedure (the Gaussian sampling of the logarithm $X\equiv \ln \mathcal{F}$ of PDFs) and asymmetric errors~\cite{Hou:2016sho},
\begin{equation}
X^{(k)}= X(S_0) + \sum_{i=1}^{N_{\rm eig}}\left(\frac{X\lp S_i^{(+)}\rp-X\lp S_i^{(-)}\rp}{2}R_{i}^{(k)}+\frac{X\lp S_i^{(+)}\rp +X \lp S_i^{(-)}\rp-2X \lp S_0\rp}{2}\left(R_{i}^{(k)}\right)^{2}\right) + \Delta.\label{dXkAsym}
\end{equation}
A small shift $\Delta$ was applied to all \CTprime MC replicas in accord with \cite{Hou:2016sho} so that the central value of the resulting Monte Carlo set reproduces then the central value of the input \CTprime Hessian set.

By means of Eqs.~(\ref{eq:mc_replica_generation}) and (\ref{dXkAsym}), we have constructed Monte Carlo
representations of the \CTprime and MSHT20 variants described in
Sect.~\ref{sec:combination_inputs} and composed of $N_{\rm rep}=300$ replicas each.
We have verified that $N_{\rm rep}=300$ for each set is a sufficiently large number of replicas to ensure
that all the features of the native Hessian representation are appropriately reproduced.

The \CTprime and MSHT20 replicas are then combined together with $N_{\rm rep}=300$ native replicas of
\NNprime based on the fit settings described on Sect.~\ref{subsec:nnpdf31},
adding up to a total of  $N_{\rm rep}=900$ replicas which define the
PDF4LHC21 combination. 

\begin{figure}[!t]
\centering
\includegraphics[width=0.49\textwidth]{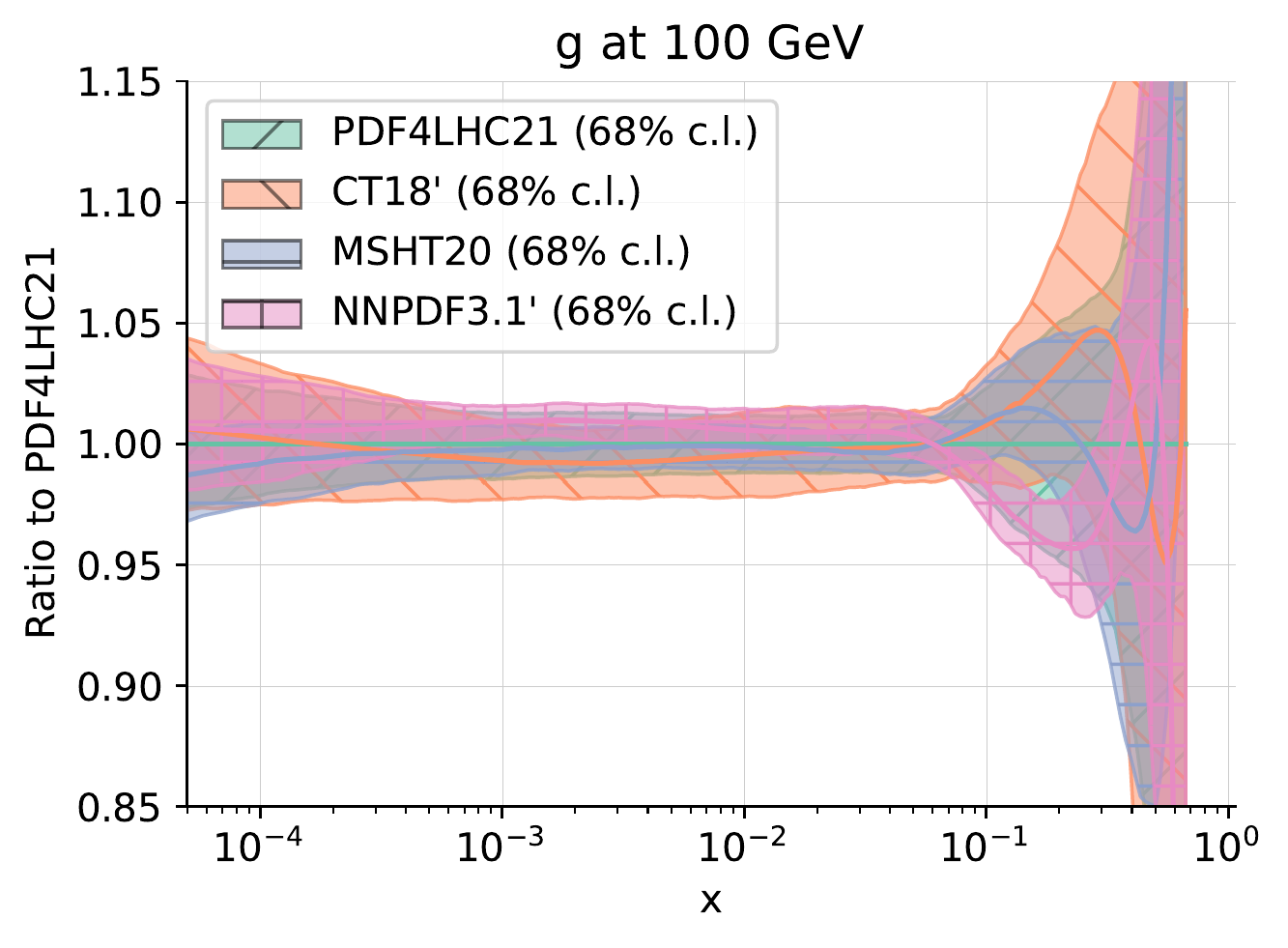}
\includegraphics[width=0.49\textwidth]{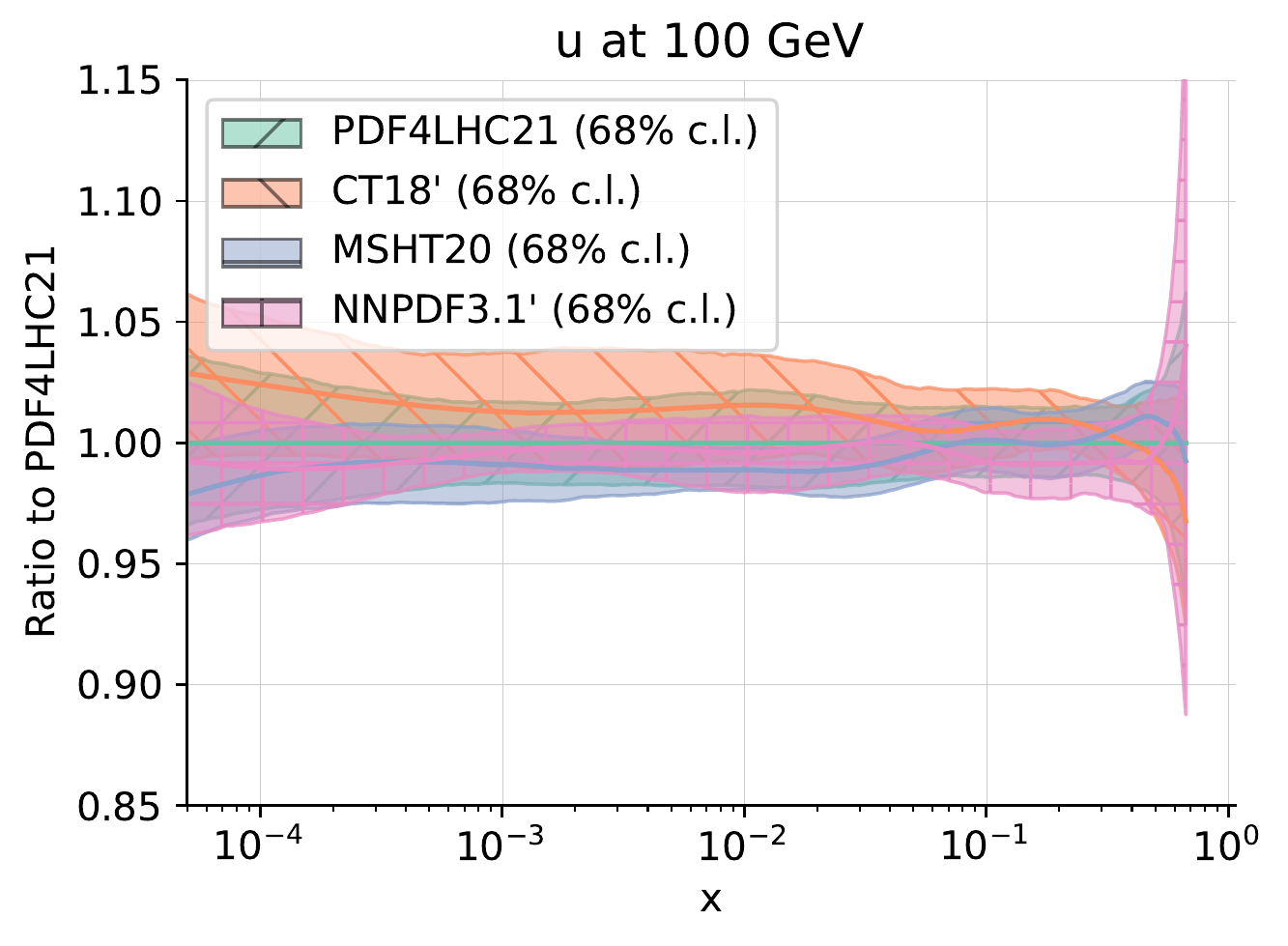}\\
\includegraphics[width=0.49\textwidth]{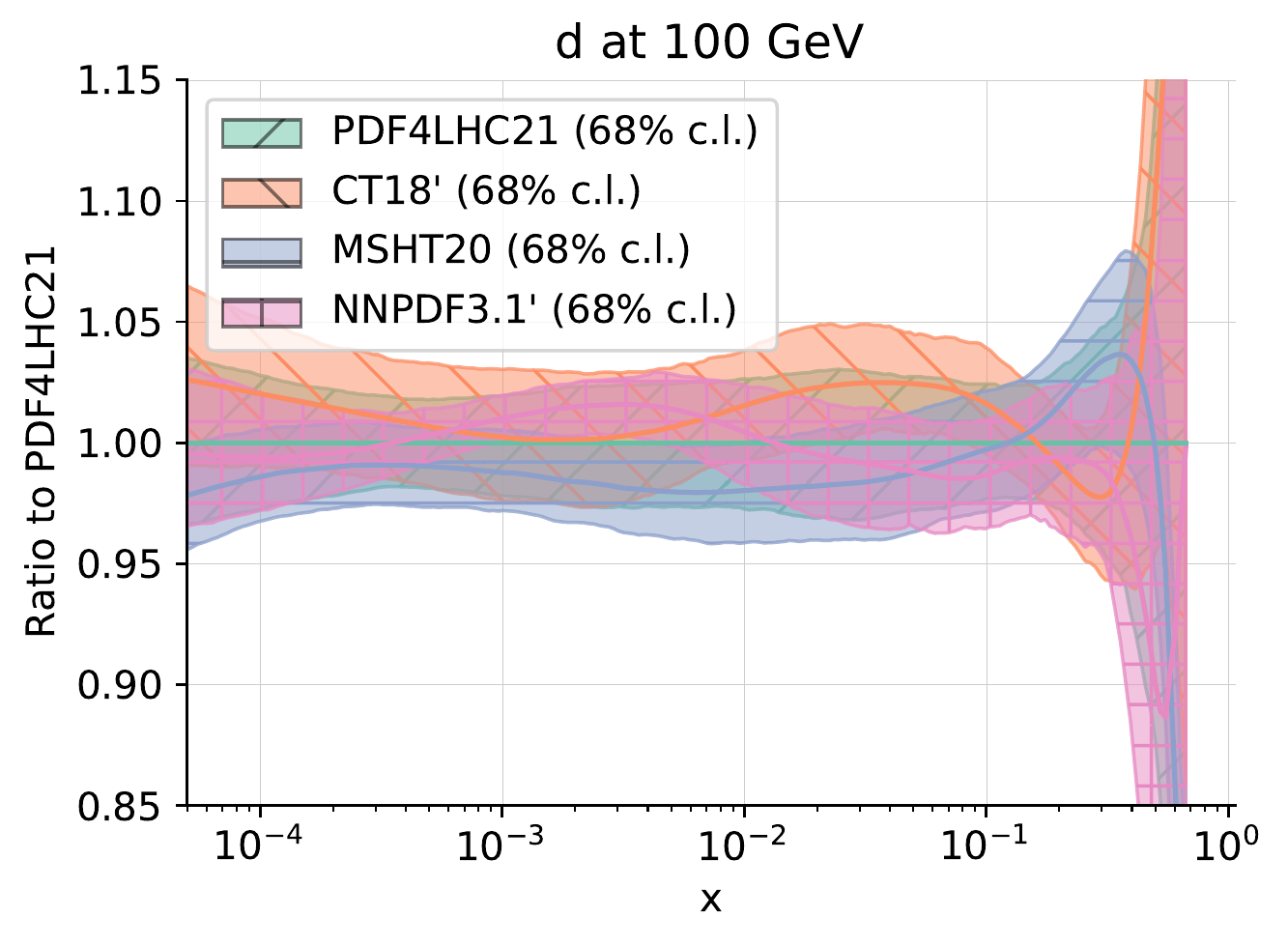}
\includegraphics[width=0.49\textwidth]{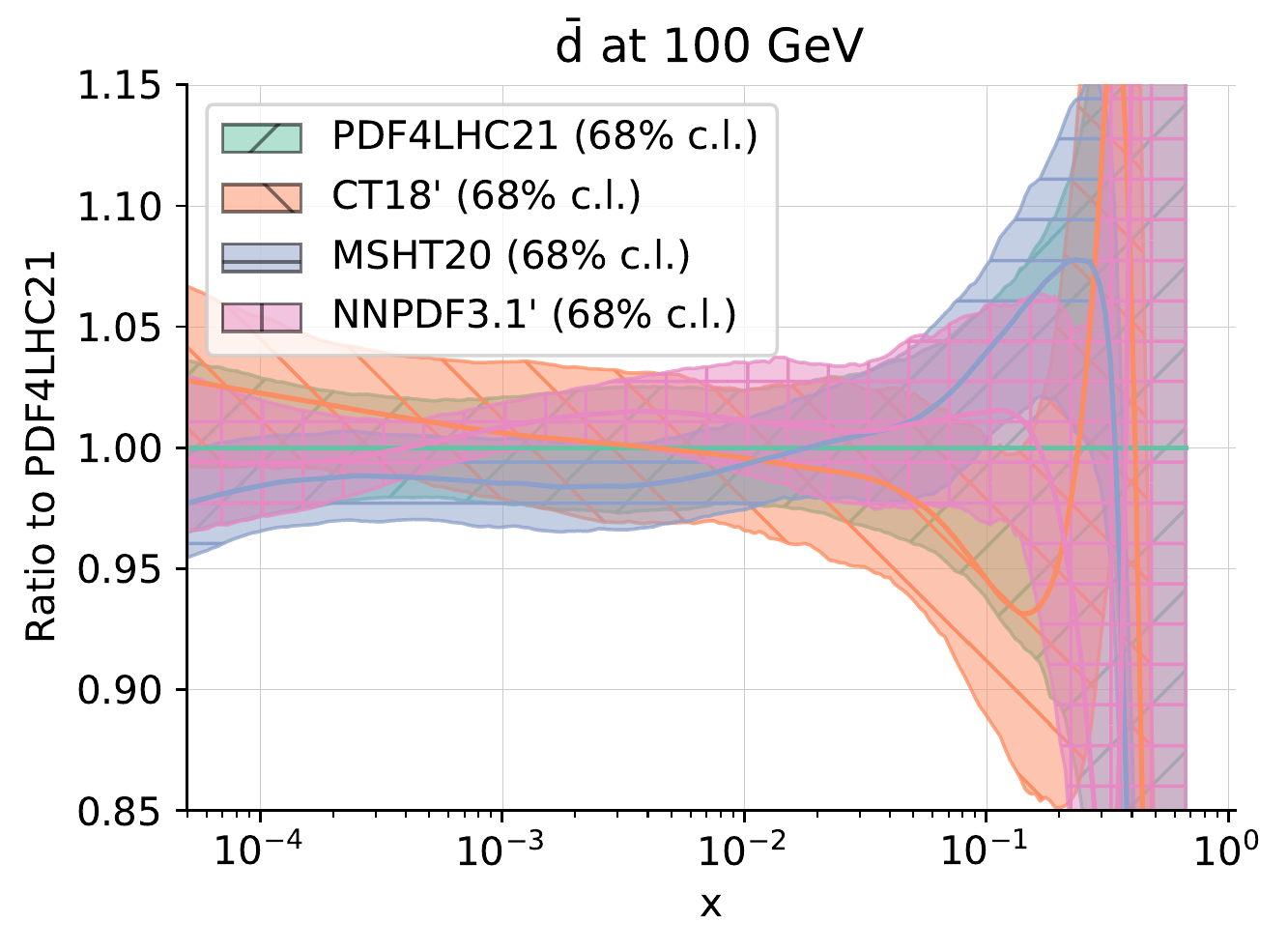}\\
\includegraphics[width=0.49\textwidth]{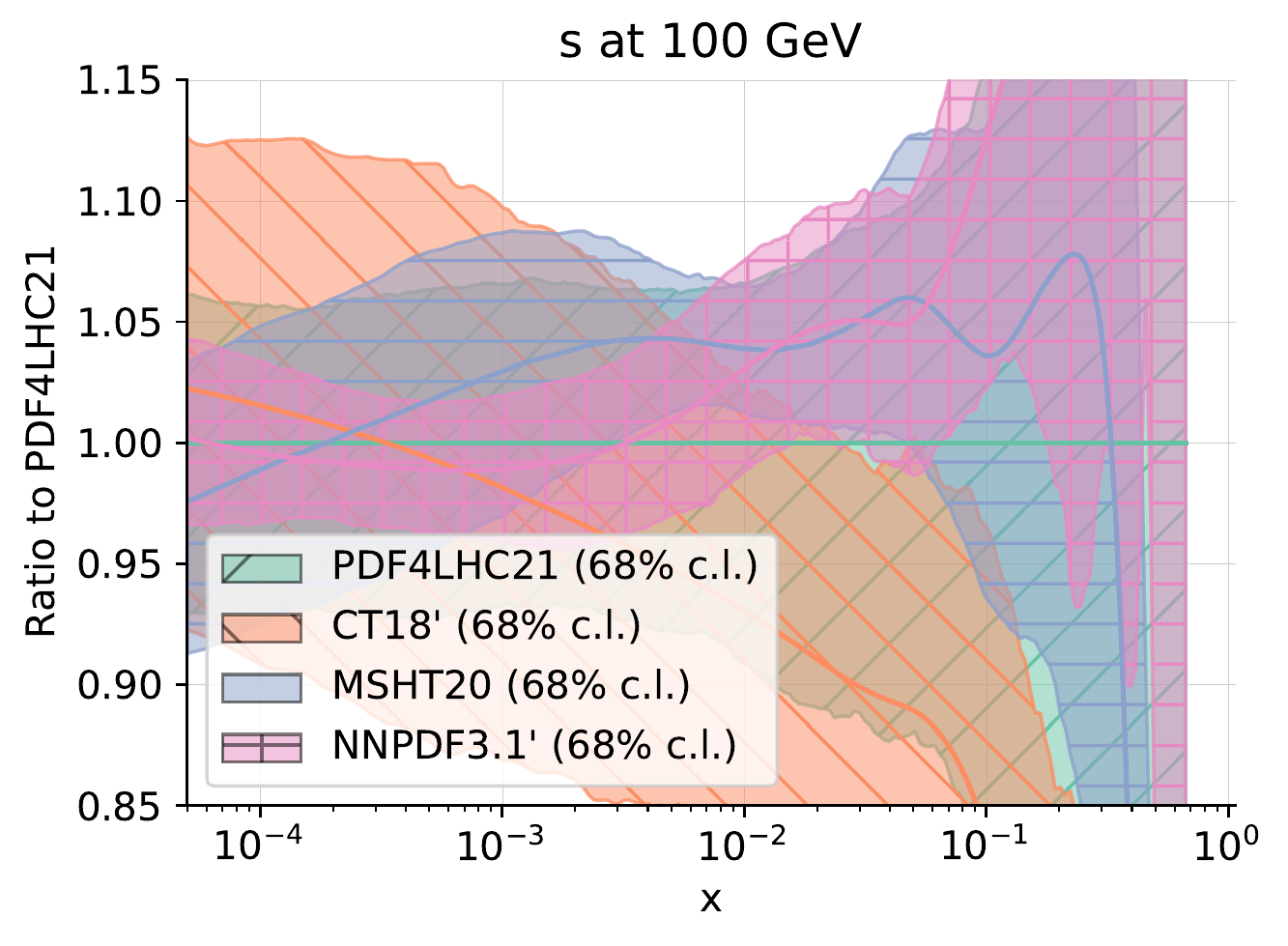}
\includegraphics[width=0.49\textwidth]{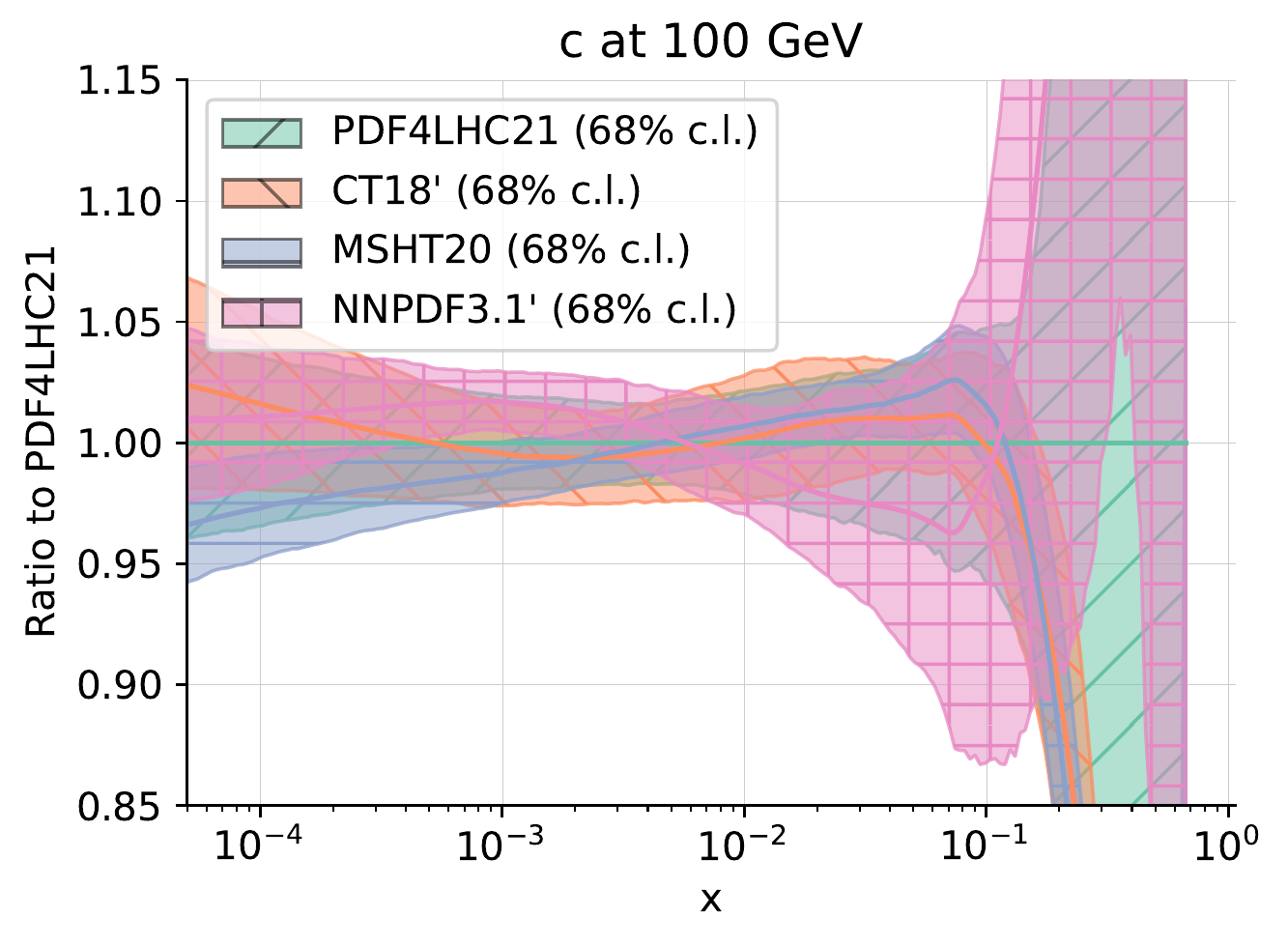}\\
\caption{\small  Comparison of the PDF4LHC21 combination (composed of $N_{\rm rep}=900$ replicas)
  with the three constituent sets at $Q=100$ GeV, normalised to the central value of the
  former and with their respective 68\%CL uncertainty bands.
  In the case of the Hessian sets (\CTprime and MSHT20) we display their
  Monte Carlo representation composed of $N_{\rm rep}=300$ replicas generated
  according to Eq.~(\ref{eq:mc_replica_generation}).
  The \NNprime band is also constituted by $N_{\rm rep}=300$ (native) replicas.
}
\label{fig:pdf4lhc21_vs_inputs_pdfs}
\end{figure}

\begin{figure}[!t]
\centering
\includegraphics[width=0.49\textwidth]{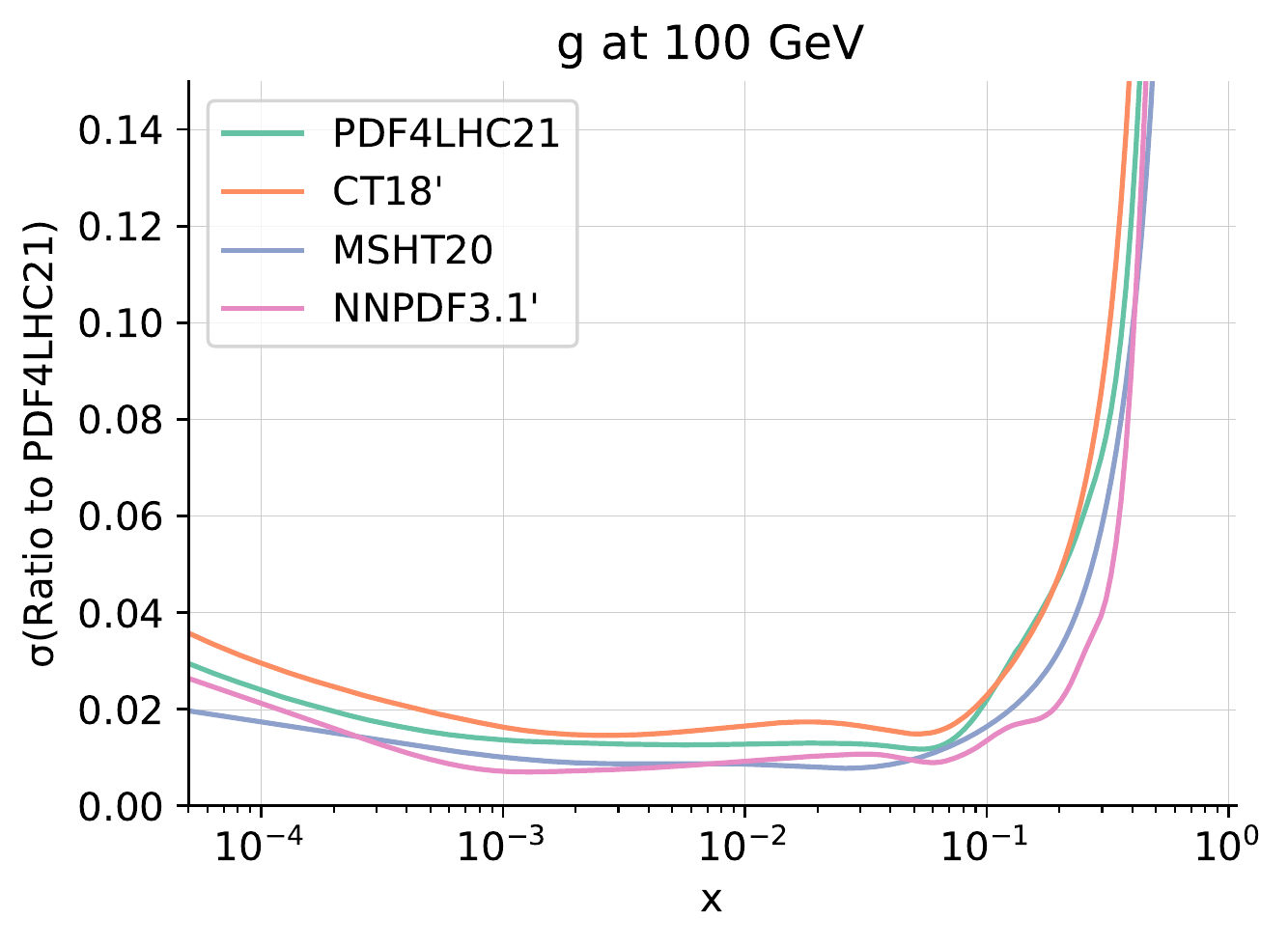}
\includegraphics[width=0.49\textwidth]{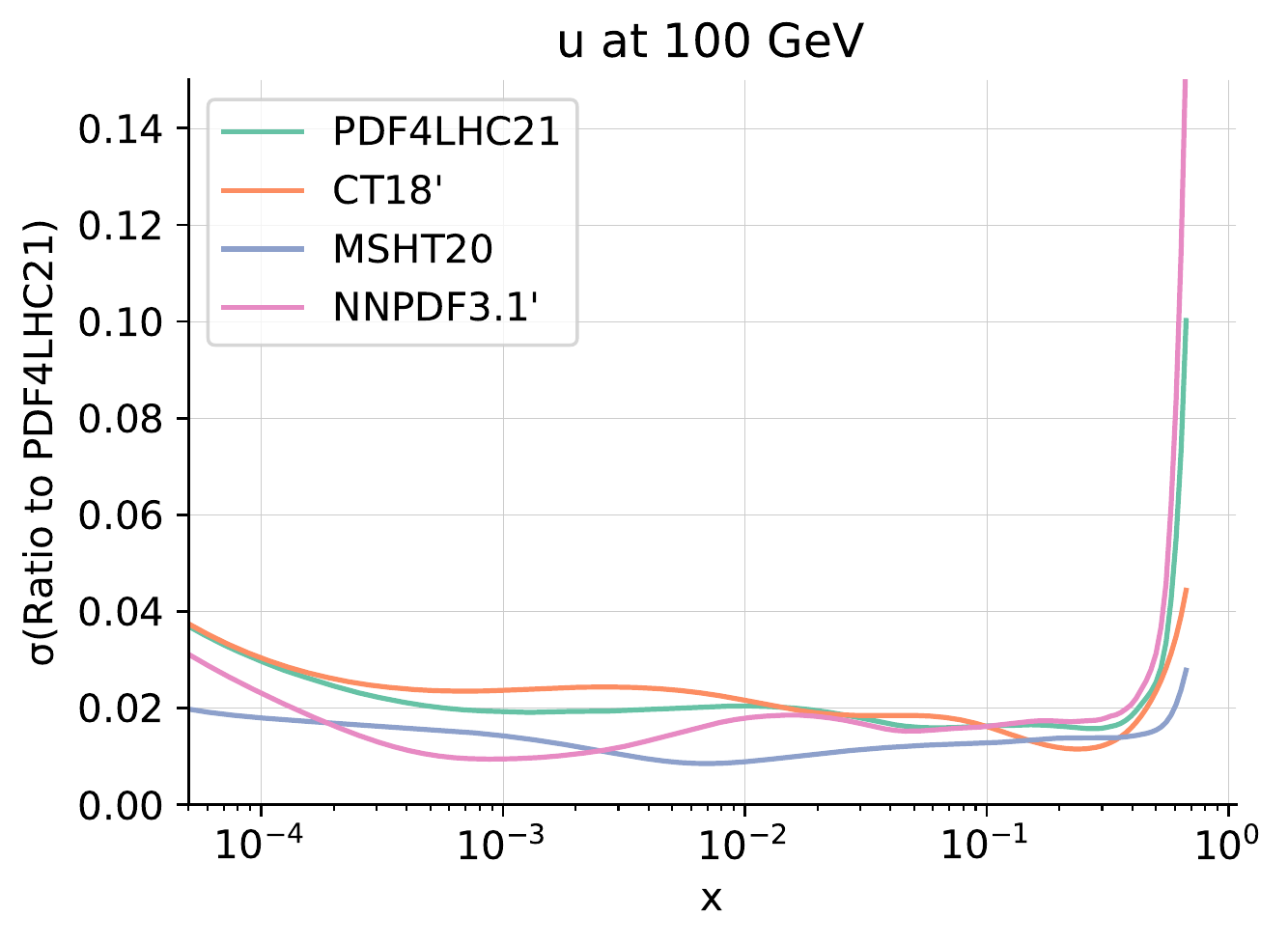}\\
\includegraphics[width=0.49\textwidth]{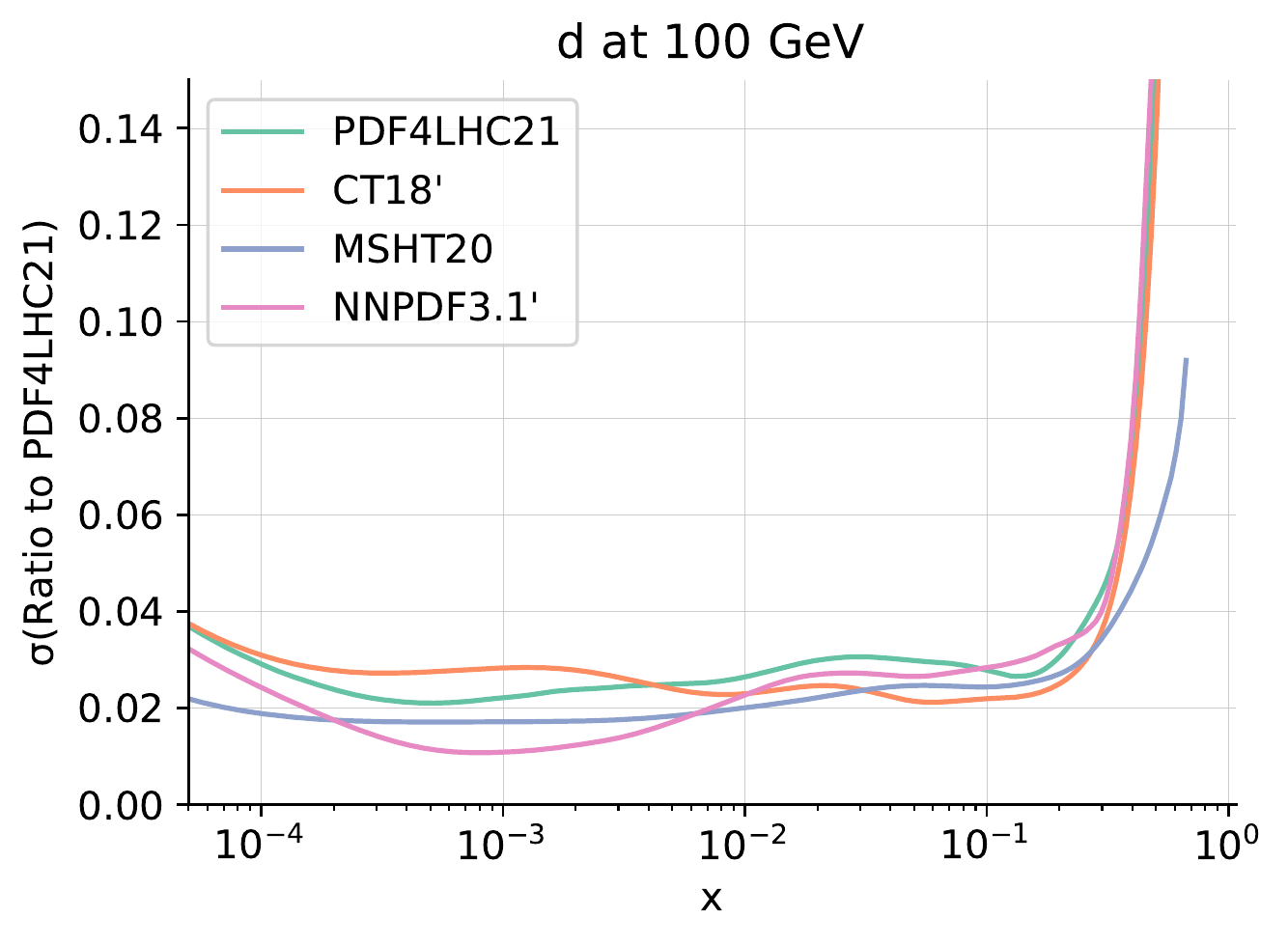}
\includegraphics[width=0.49\textwidth]{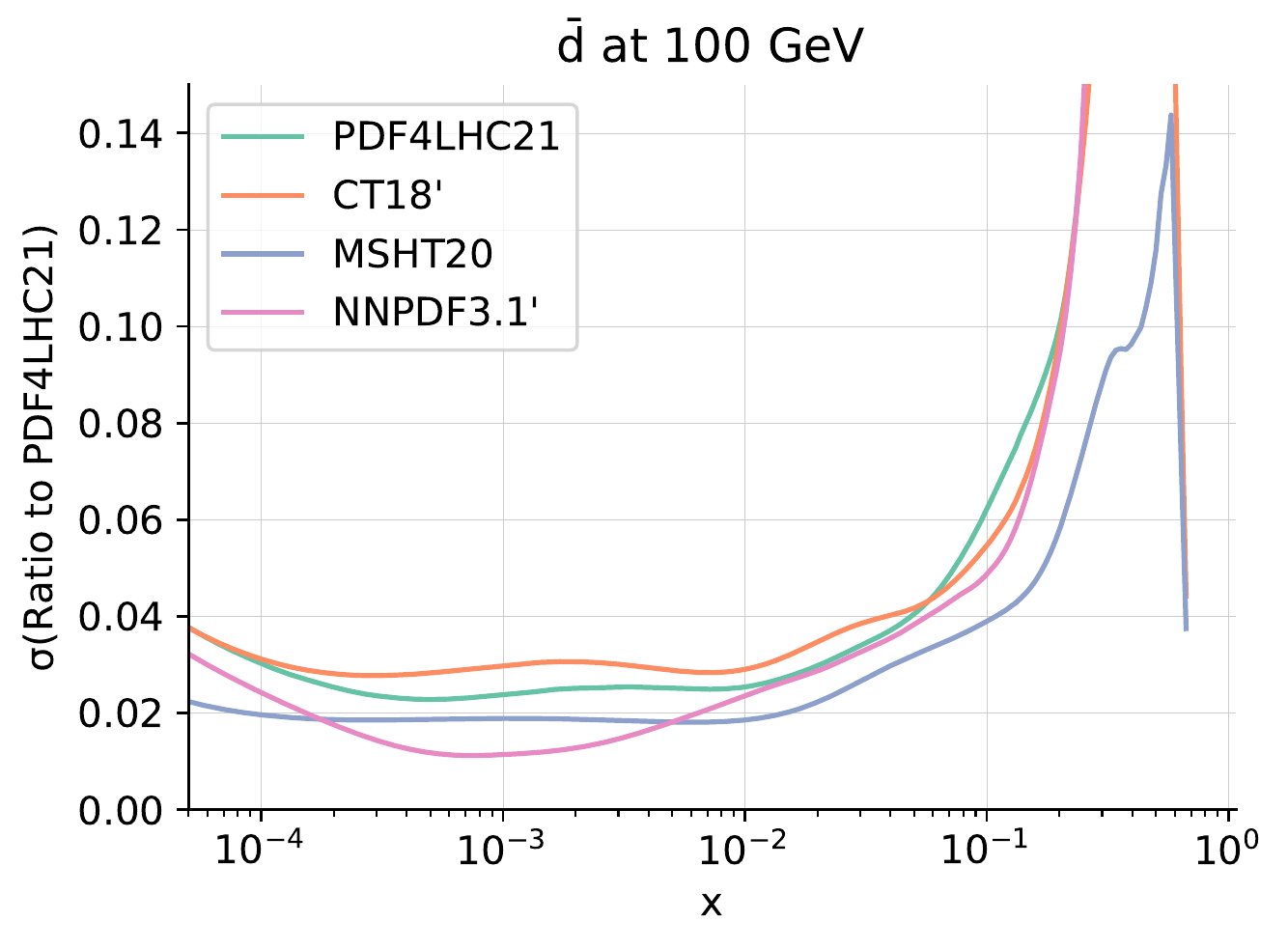}\\
\includegraphics[width=0.49\textwidth]{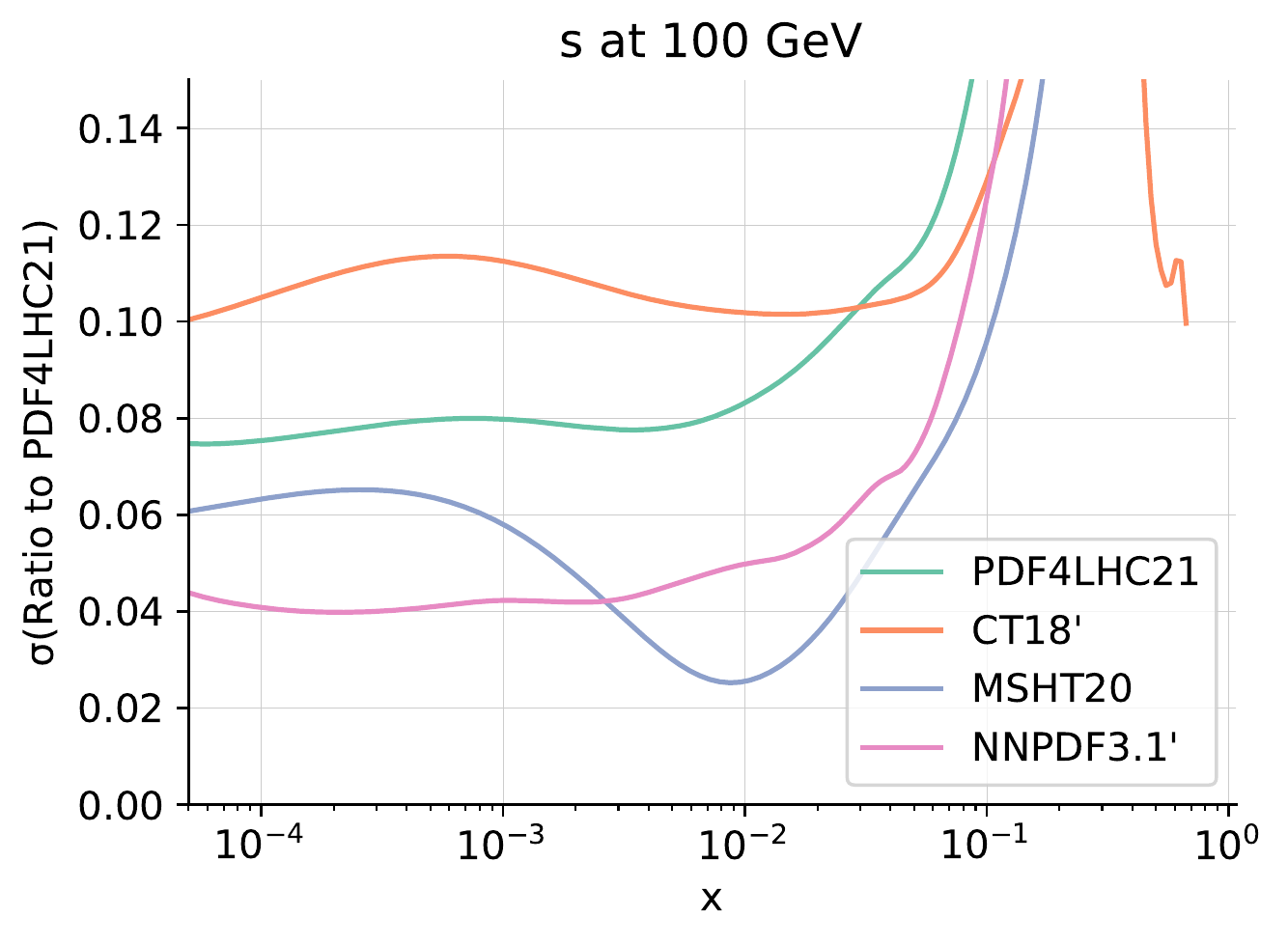}
\includegraphics[width=0.49\textwidth]{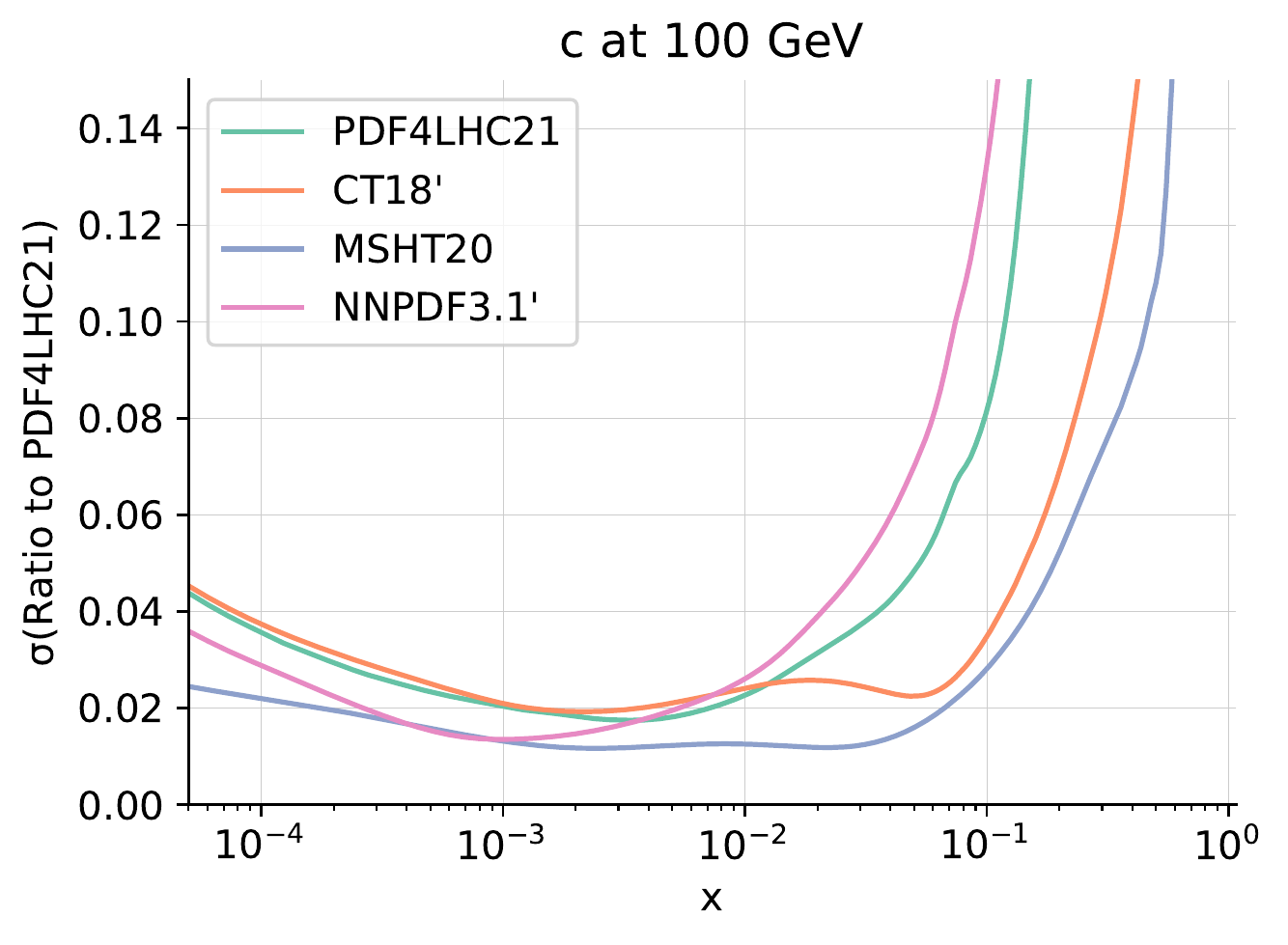}\\
\caption{\small  Same as Fig.~\ref{fig:pdf4lhc21_vs_inputs_pdfs}, now
  showing the relative PDF 68\% CL uncertainties (shown as fractions of the PDF4LHC21 central value)
  of the four PDF sets.}
\label{fig:pdf4lhc21_vs_inputs_uncs}
\end{figure}

The combined set with $N_{\rm rep}=900$ replicas will be called the {\it baseline} PDF4LHC21 set.
Let us discuss first the main features of the baseline set and its
comparison with the three constituent inputs.
Figure~\ref{fig:pdf4lhc21_vs_inputs_pdfs} displays this comparison
at $Q=100$ GeV, normalised to the central value of the baseline set, and showing
the 68\% CL uncertainty bands.
Then Fig.~\ref{fig:pdf4lhc21_vs_inputs_uncs} displays a similar comparison, now
for the relative PDF 68\% CL uncertainties (shown as fractions of the PDF4LHC21 central value) of the four PDF sets.

Several interesting observations can be derived from Figs.~\ref{fig:pdf4lhc21_vs_inputs_pdfs}
and~\ref{fig:pdf4lhc21_vs_inputs_uncs}.
All in all, the main qualitative features of the PDF4LHC21 combination
are consistent with the expectations derived from the comparison
between the three input sets presented in Sect.~\ref{subsec:inputs_comparison}, and the generic properties of the PDF4LHC combination prescription.
The PDF4LHC21 combination overlaps with the three constituent sets at the 68\% CL,
implying that there is no region where the PDF error bands of the combination
and of the individual sets do not touch.
In terms of the relative PDF uncertainties,
one does observe how the relation between those of PDF4LHC21 and those of the input sets
depends on both the flavour and on the range of $x$.
We recall that, by construction, the uncertainties of the PDF4LHC21 set are expected to be bracketed
by those of the constituents sets in the cases where there is good overall consistency
(since then one is applying effectively an average of the three sets), while for
those regions in $x$ and flavour where one has discrepancies between the three inputs,
the PDF4LHC21 uncertainties may be larger than those of any individual constituent set, since they include the spread in the central predictions \cite{Ball:2021dab}.
For example, for strangeness at $x\gsim 0.05$ the PDF4LHC21 uncertainties are larger
than those of the three inputs, due to their disagreements in this region, while the uncertainties in \CTprime are larger than those of either MSHT20 or \NNprime for $x\lesssim 10^{-1}$.
For the charm PDF the uncertainty for $x\gsim 10^{-2}$ is dominated by the fitted charm
contribution of \NNprime, which leads to a marked
  increase in uncertainties for $x\gsim 10^{-2}$ compared to charm which is entirely perturbatively generated.
 Concerning the gluon PDF, very similar uncertainties are obtained in the MSHT20
  and \NNprime analyses while those of \CTprime can be somewhat larger, by a factor of
  $\sim\!1.5\!-\!2$, for select regions of $x$.

\begin{figure}[!t]
\centering
\includegraphics[width=0.49\textwidth]{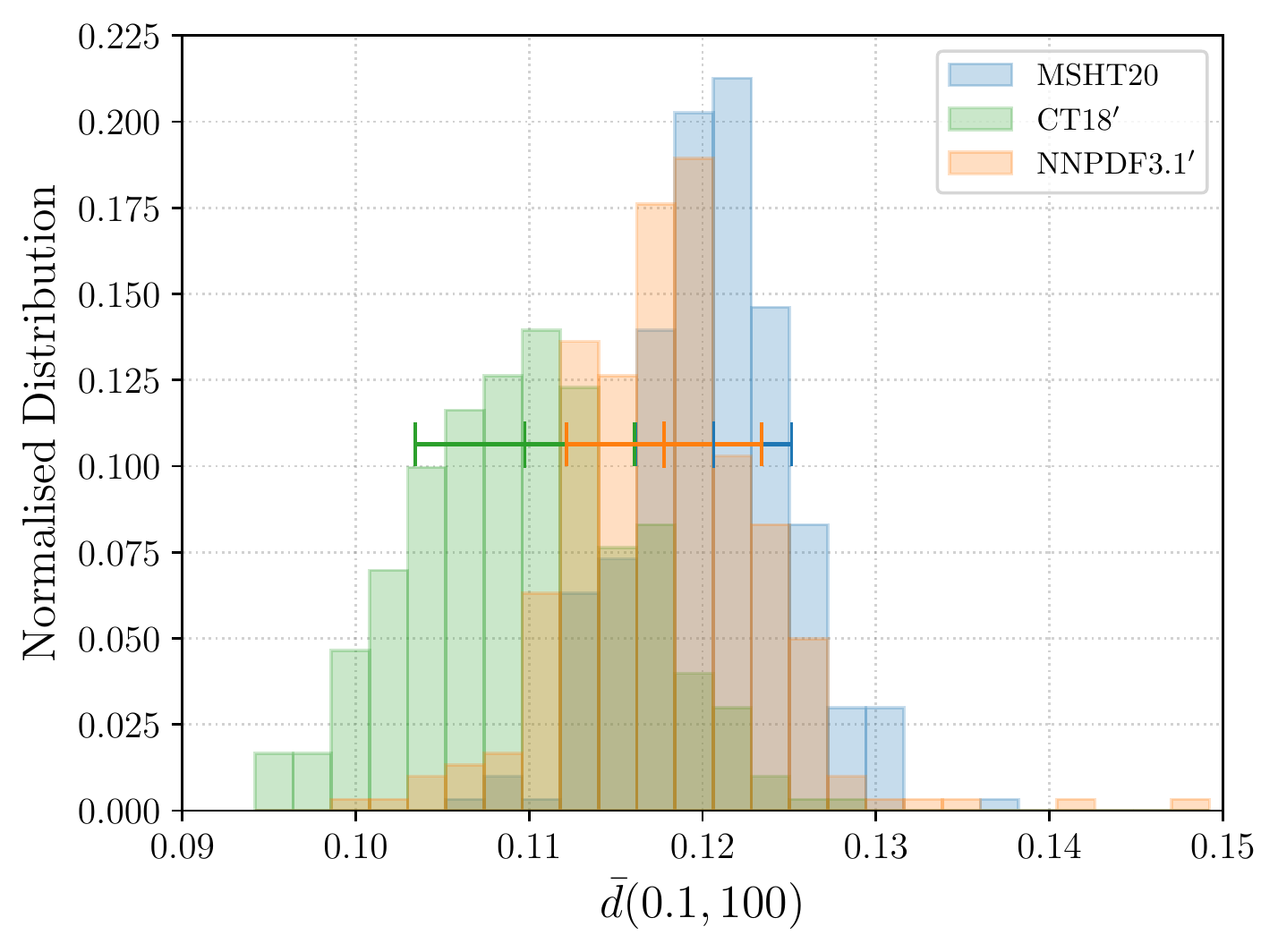}
\includegraphics[width=0.49\textwidth]{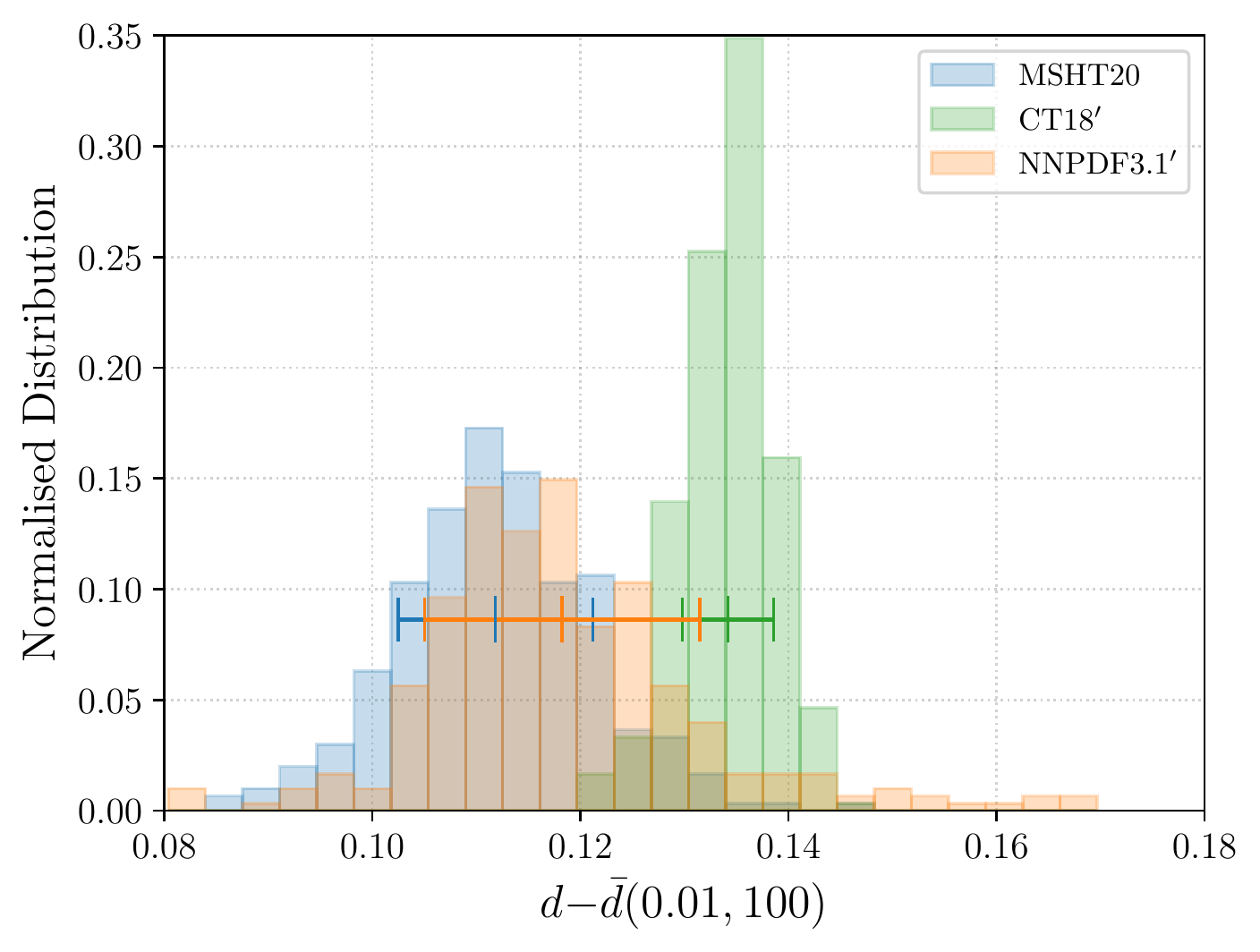}\\
\includegraphics[width=0.49\textwidth]{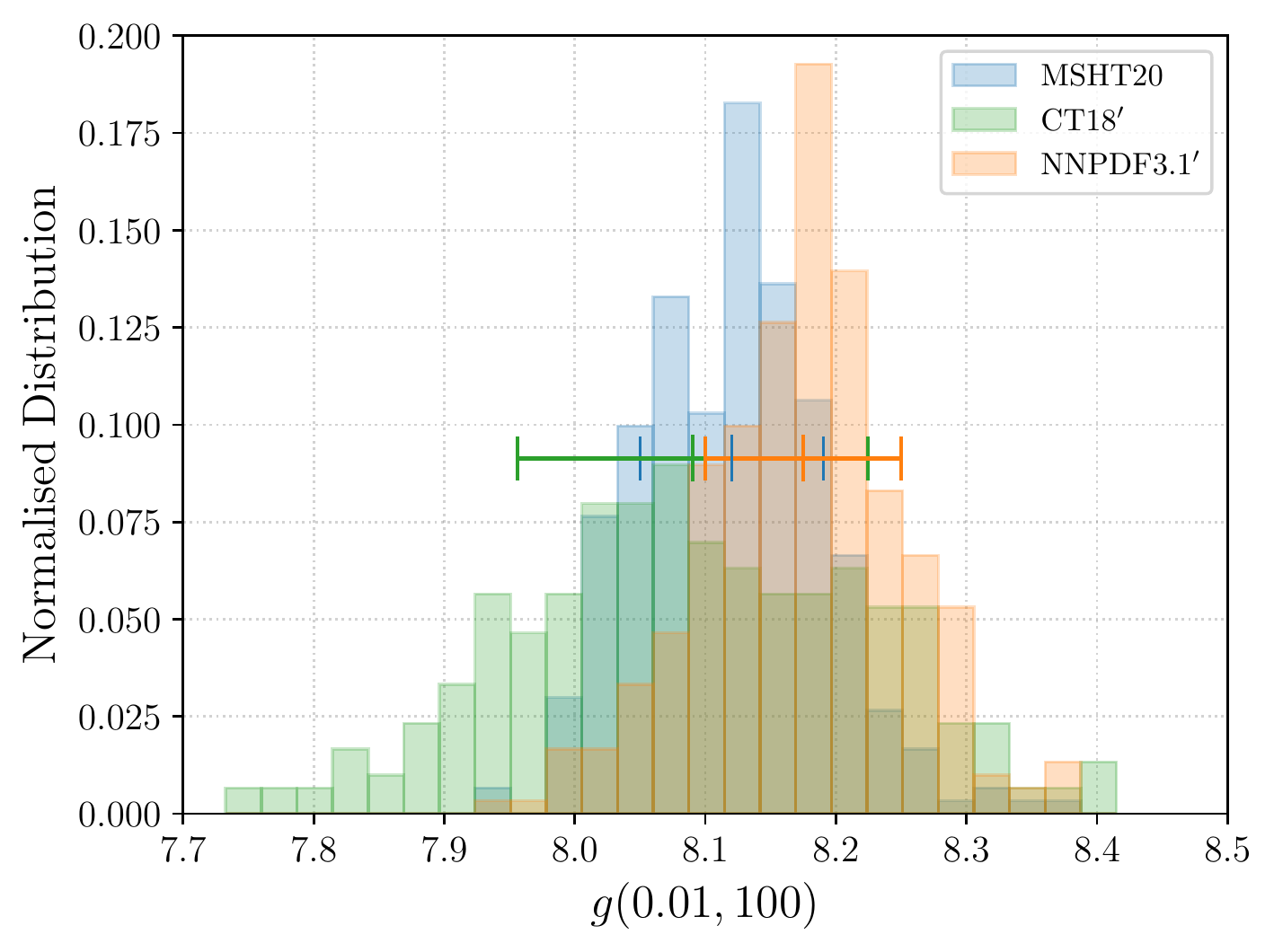}
\includegraphics[width=0.49\textwidth]{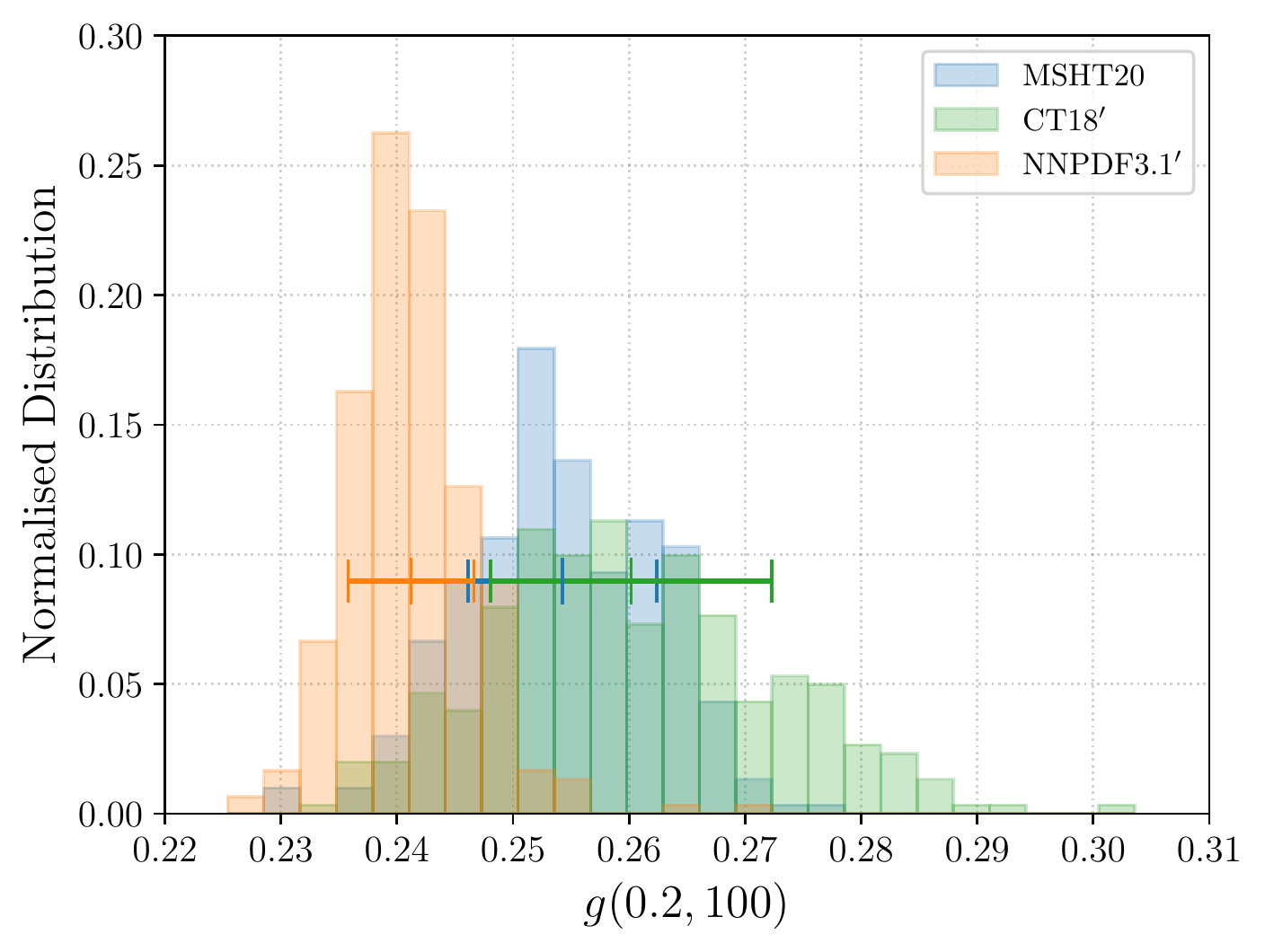}\\
\includegraphics[width=0.49\textwidth]{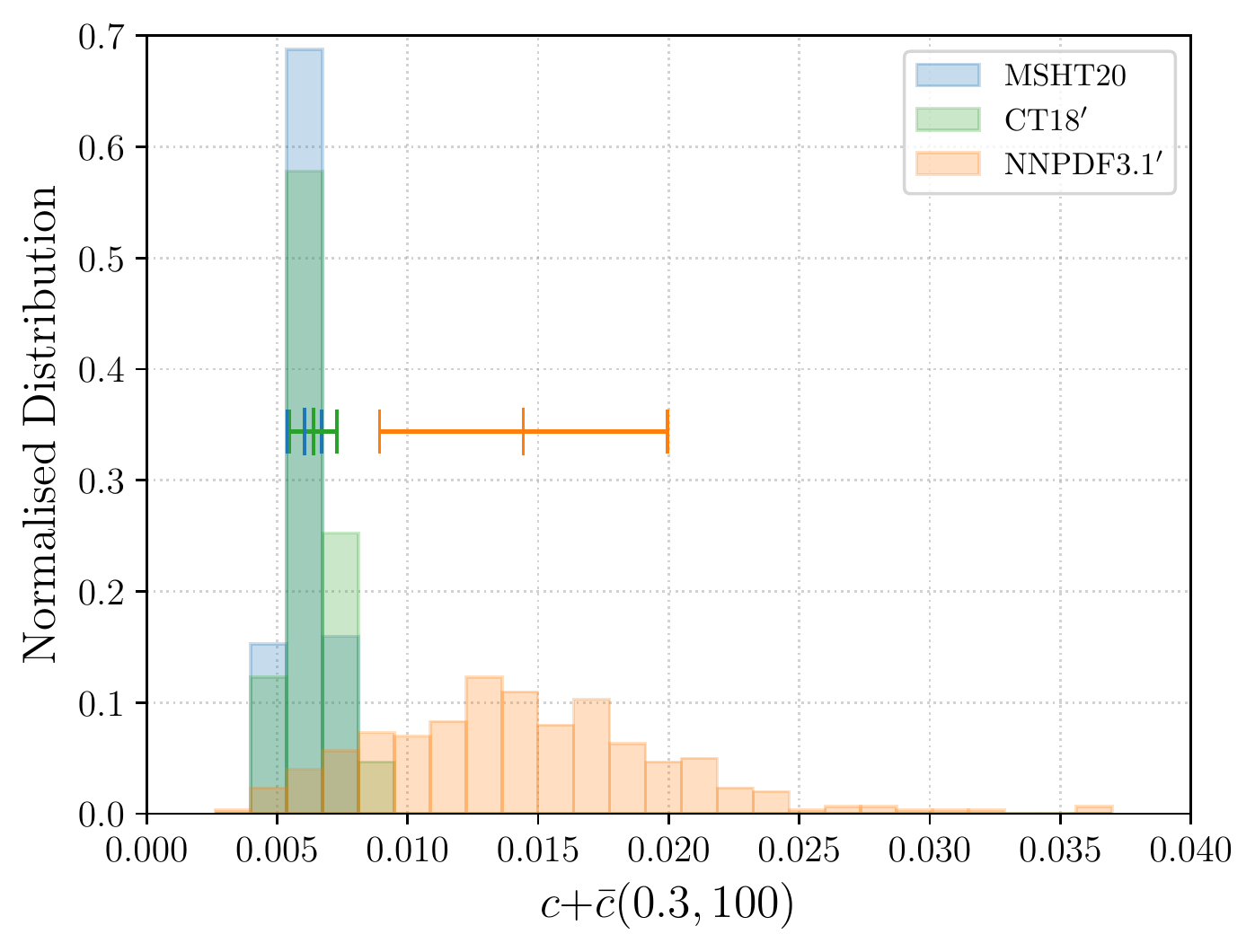}
\includegraphics[width=0.49\textwidth]{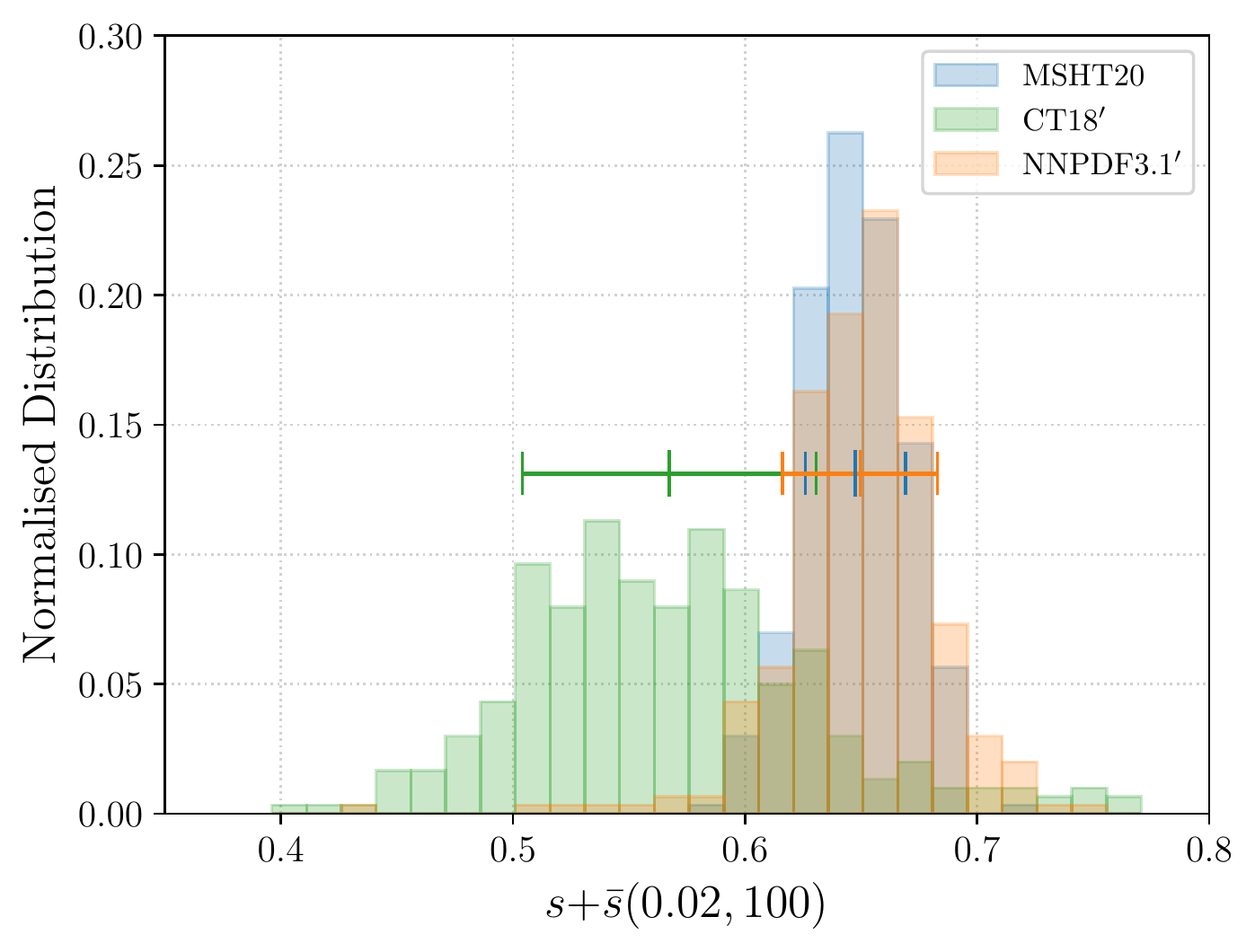}\\
\caption{\small The Monte Carlo representations of the probability
densities associated to the $N_{\rep}=300$ of \CTprime, MSHT20, and \NNprime.
Results are displayed at $Q=100$ GeV for $\bar{d}(x=0.1)$,  $(d-\bar{d})(x=0.01)=d_v(x=0.01)$, $s^+(x=0.02)$, $c^+(z=0.3)$, $g(x=0.01)$, and $g(x=0.2)$.
Note that the same binning is applied to the all distributions shown.
The horizontal lines indicate the median and 68\% CL intervals associated with each distribution (their vertical position is arbitrary).
}
\label{fig:prior-xDistr}
\end{figure}

To further compare the three input Monte Carlo sets to the PDF4LHC21 combination, we show, in Fig.~\ref{fig:prior-xDistr},
the
 Monte Carlo representations of the probability
 densities associated with \CTprime, MSHT20, and \NNprime, where
 $N_{\rep}=300$ replicas from each group are used.
 We display six points in $(x,Q^2,{\rm flavour})$ space: namely, for $\bar{d}(x=0.1)$,  $d_v(x=0.01)$, $g(x=0.01)$,
$g(x=0.2)$, $c^+(x=0.3)$, and $s^+(x=0.02)$, all taken at $Q=100$ GeV.
These points in the $(x,Q^2,{\rm flavour})$ space have been selected as representative
of those regions for which the differences among the three PDF sets displayed in Fig.~\ref{fig:global_fits_comparison} are most pronounced.
The reason is that these are precisely
the regions for which the combined probability distributions
might be expected to display non-Gaussian features.
Note that the same binning is applied to each of the distributions shown.
The median and 68\% CL intervals associated with each distribution are indicated by the horizontal lines.
Inspection of the probability distributions displayed
in Fig.~\ref{fig:prior-xDistr} reveals that indeed, for several
regions in the $(x,Q^2,{\rm flavour})$ space, the combination
of the three Monte Carlo probability distributions will result
in a combined distribution with clear non-Gaussian features.
In particular, the $d_v$ distribution at $x=10^{-2}$ exhibits a double-peaked distribution. The latter is understood from the substantial difference
between the distribution of the $N_{\rep}=300$ \CTprime replicas and those of   \NNprime and MSHT20. Similarly, the $c^+$ distribution at $x=0.3$ is  wider for \NNprime, resulting
in a fat tail to the right of the combined distribution. Lastly,
concerning the total strangeness $s^+$ at $x=0.02$, differences in the central values of three error sets lead to the observed bimodal structure. 
While these $(x, Q^2,{\rm flavour})$ points were selected to show the largest differences observed between the three input Monte Carlo sets, in the other regions, a general agreement between the constituent distributions is found. 
Hence the combination of the three groups will result in distributions
which are generally Gaussian to a good approximation, as will be explicitly
demonstrated in Sect.~\ref{sec:hessian_reduction}.

We conclude that the combination procedure exhibits the desired and expected
properties. Due to its size, the baseline PDF4LHC21 set is impractical for phenomenological applications. It can be approximated to good accuracy with smaller PDF ensembles obtained with the Monte Carlo replica compression and Hessian reduction techniques, described in Sects.~\ref{sec:mc_compression} and~\ref{sec:hessian_reduction}, respectively.

\subsection{Monte Carlo compression}
\label{sec:mc_compression}

The Monte Carlo replica compression algorithm presented
in~\cite{Carrazza:2015hva,Carrazza:2021hny} reduces a large number of replicas of the PDF4LHC21 combination ($N_{\rm rep}=900$) 
while preserving key statistical properties in a controlled way.

The main aim of compressing a Monte Carlo PDF set is to extract a subset of the replicas 
that most faithfully reproduces the statistical properties of the baseline PDF distribution.
The compression
methodology relies on two main ingredients: a proper definition of a distance metric
that quantifies the distinguishability between the baseline and the compressed
probability distributions,
and an appropriate minimisation algorithm that explores the space of possible combinations
of PDF replicas which leads to such minima.
More details about the Monte Carlo compression strategy
adopted in this work can be found in App.~\ref{subsec:mccompression},
with technical information about the algorithmic settings following~\cite{Carrazza:2021hny}.

We have explored in detail the dependence of the compression efficiency
with respect to the number of replicas.
We emphasise that a successful compression should be able to reproduce
not only the first and second moments of the baseline
distribution (mean, variance, and correlations) but also all higher-order
moments (skewness, kurtosis, etc).
In particular, we have noted in Sect.~\ref{subsec:inputs_comparison} that
in some cases the PDF4LHC21 combination exhibits marked non-Gaussian features,
and these should be reproduced by the compression algorithm.
We find that $N_{\rm rep}=100$ represents a good compromise between reconstructing the
main statistical features of the baseline distribution and ensuring a reasonably low
number of replicas as required in phenomenology applications.
This value has been obtained by inspection of not only PDFs and luminosities,
but also of representative LHC cross-sections and distributions.
We point the reader to App.~\ref{subsec:mccompression} for more details
about the justification of $N_{\rm rep}=100$  for the compressed PDF4LHC21
Monte Carlo set.

\begin{figure}[!t]
\centering
\includegraphics[width=0.49\textwidth]{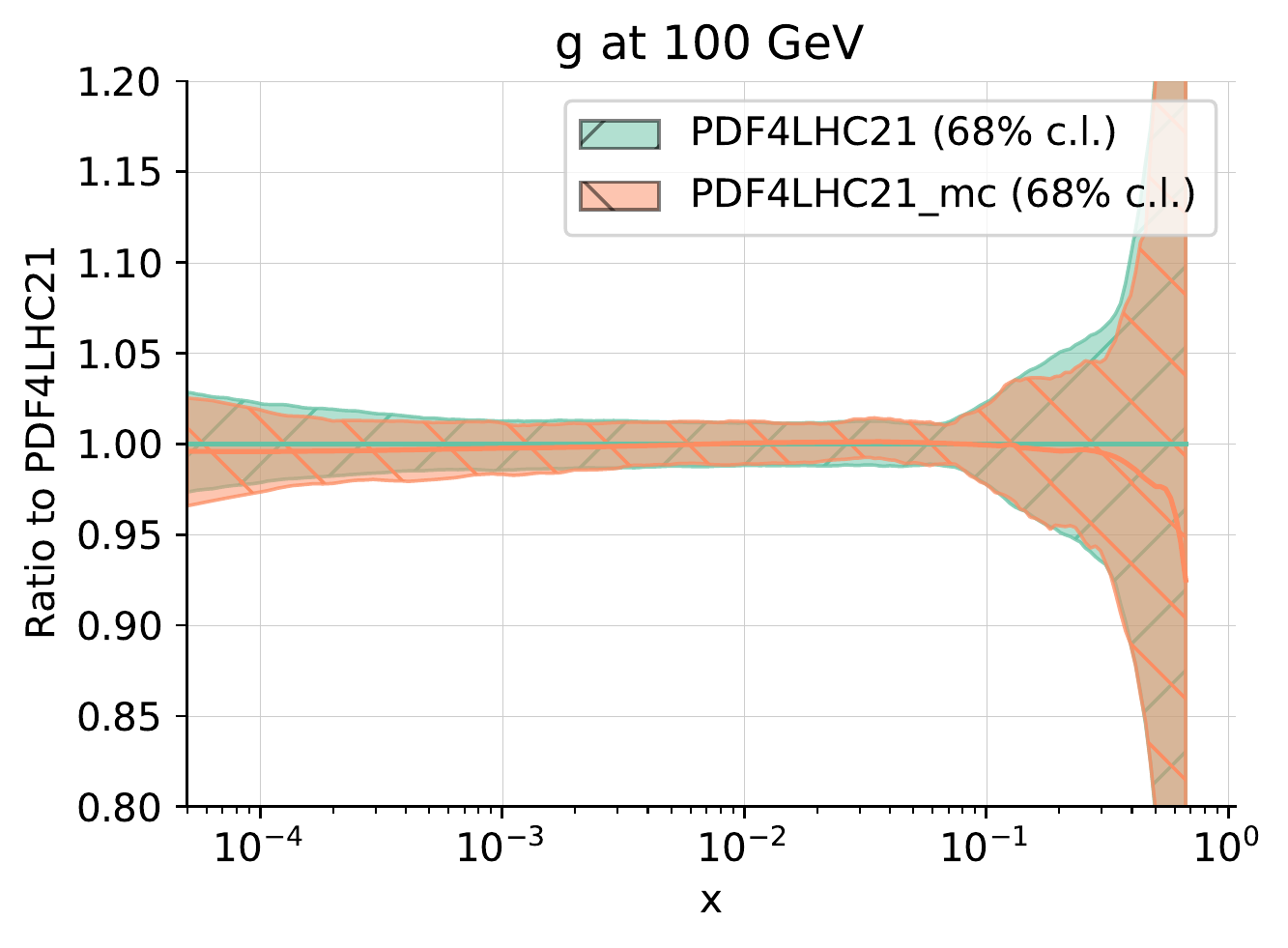}
\includegraphics[width=0.49\textwidth]{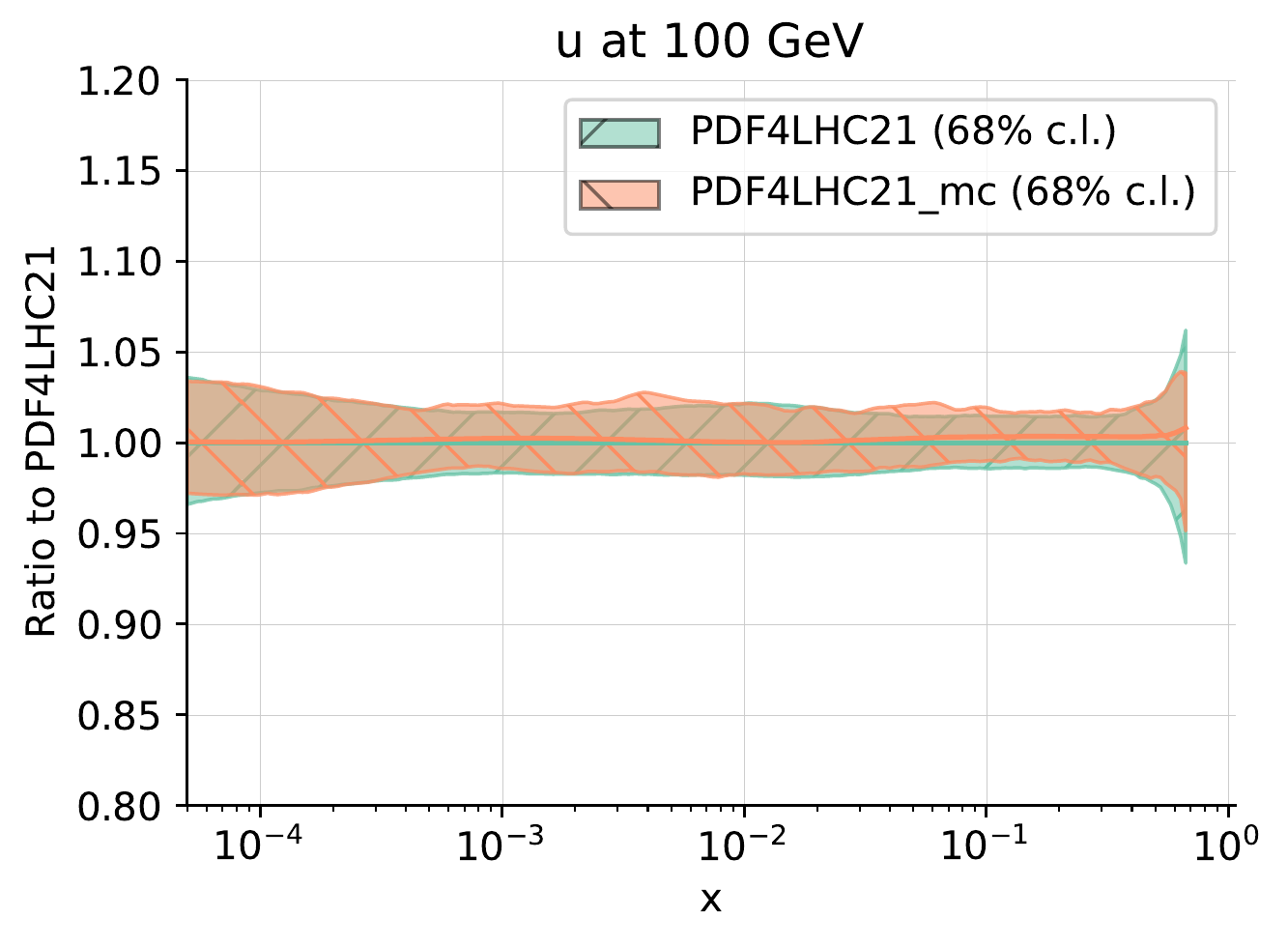}\\
\includegraphics[width=0.49\textwidth]{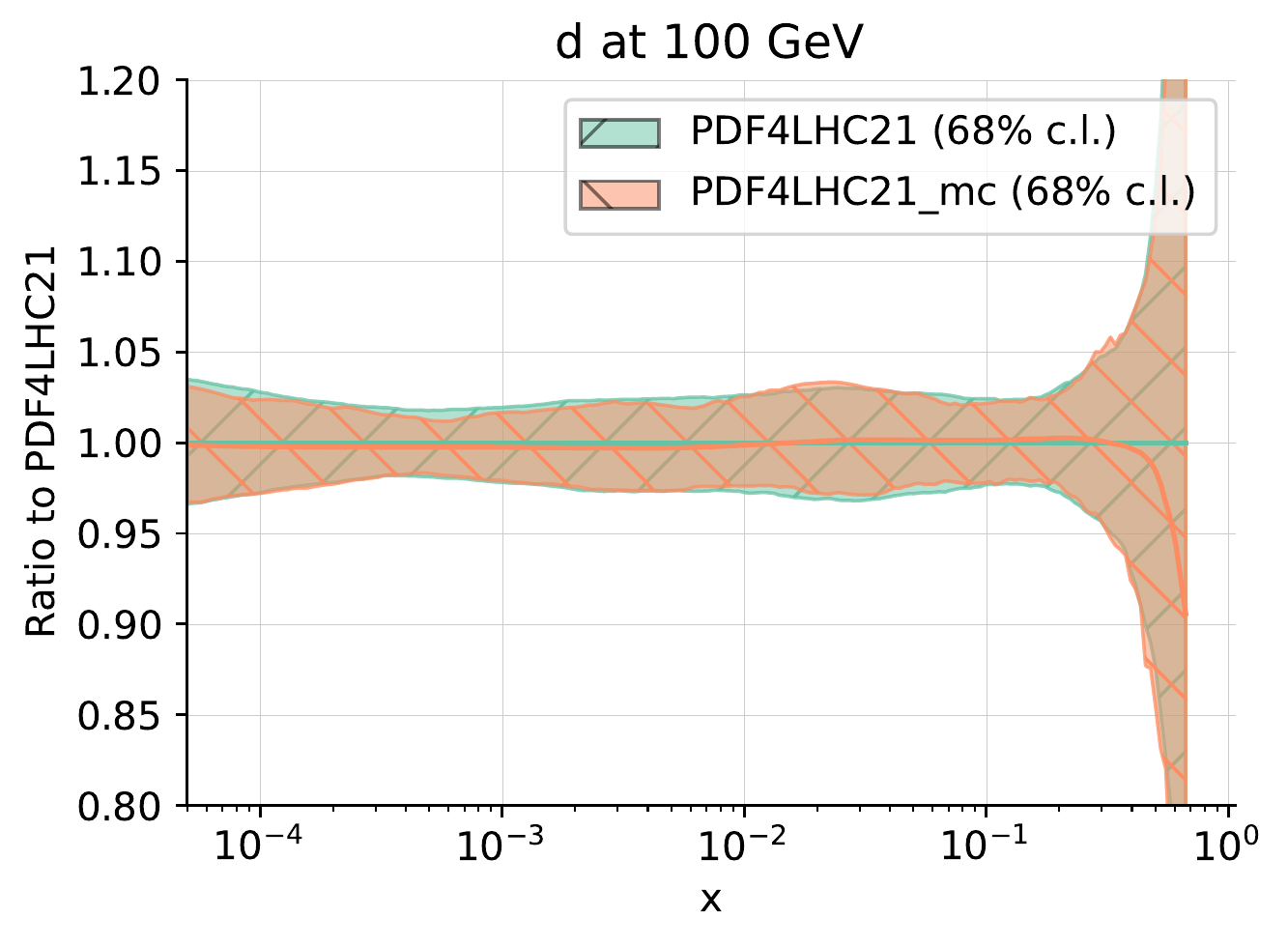}
\includegraphics[width=0.49\textwidth]{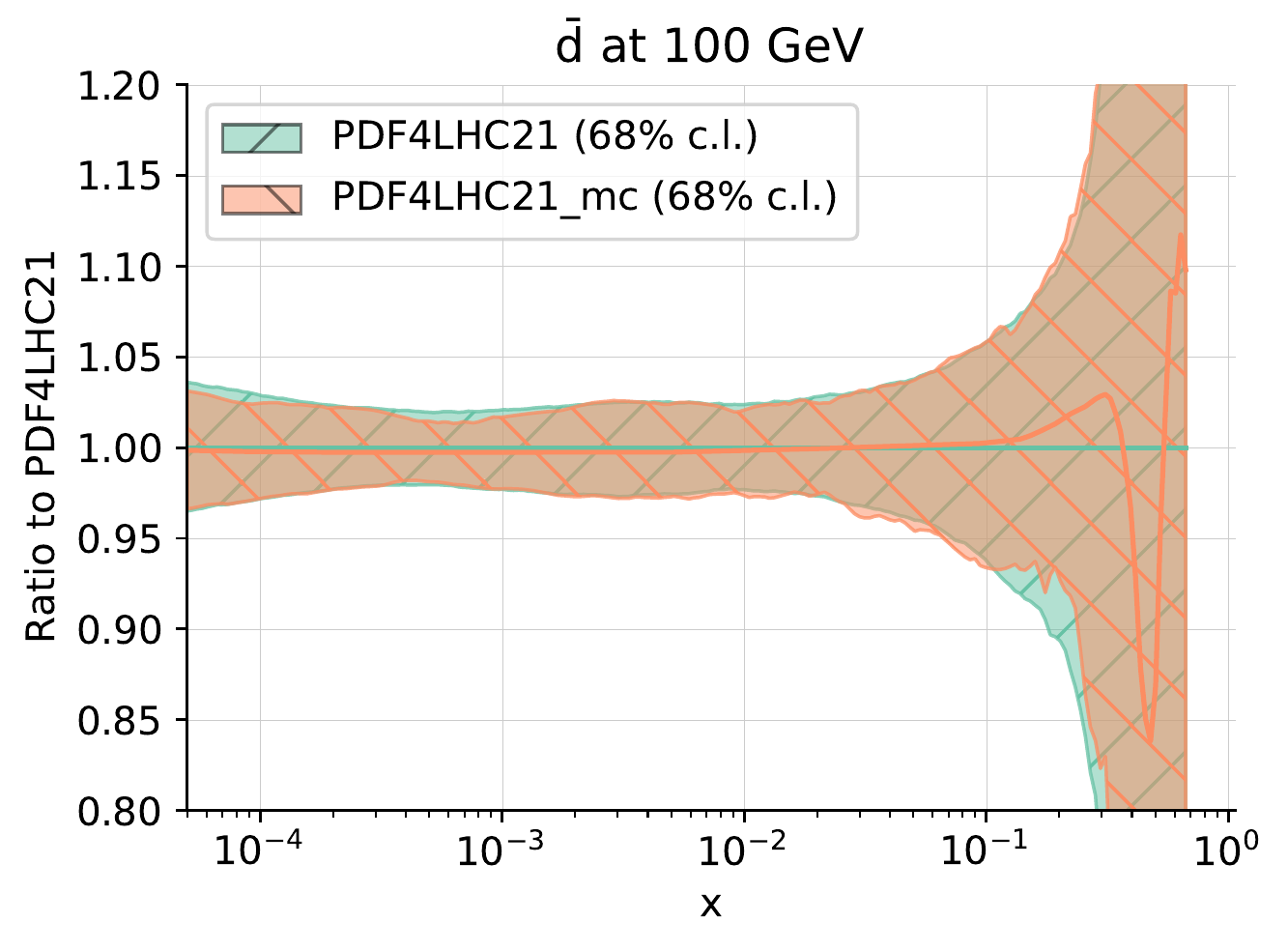}\\
\includegraphics[width=0.49\textwidth]{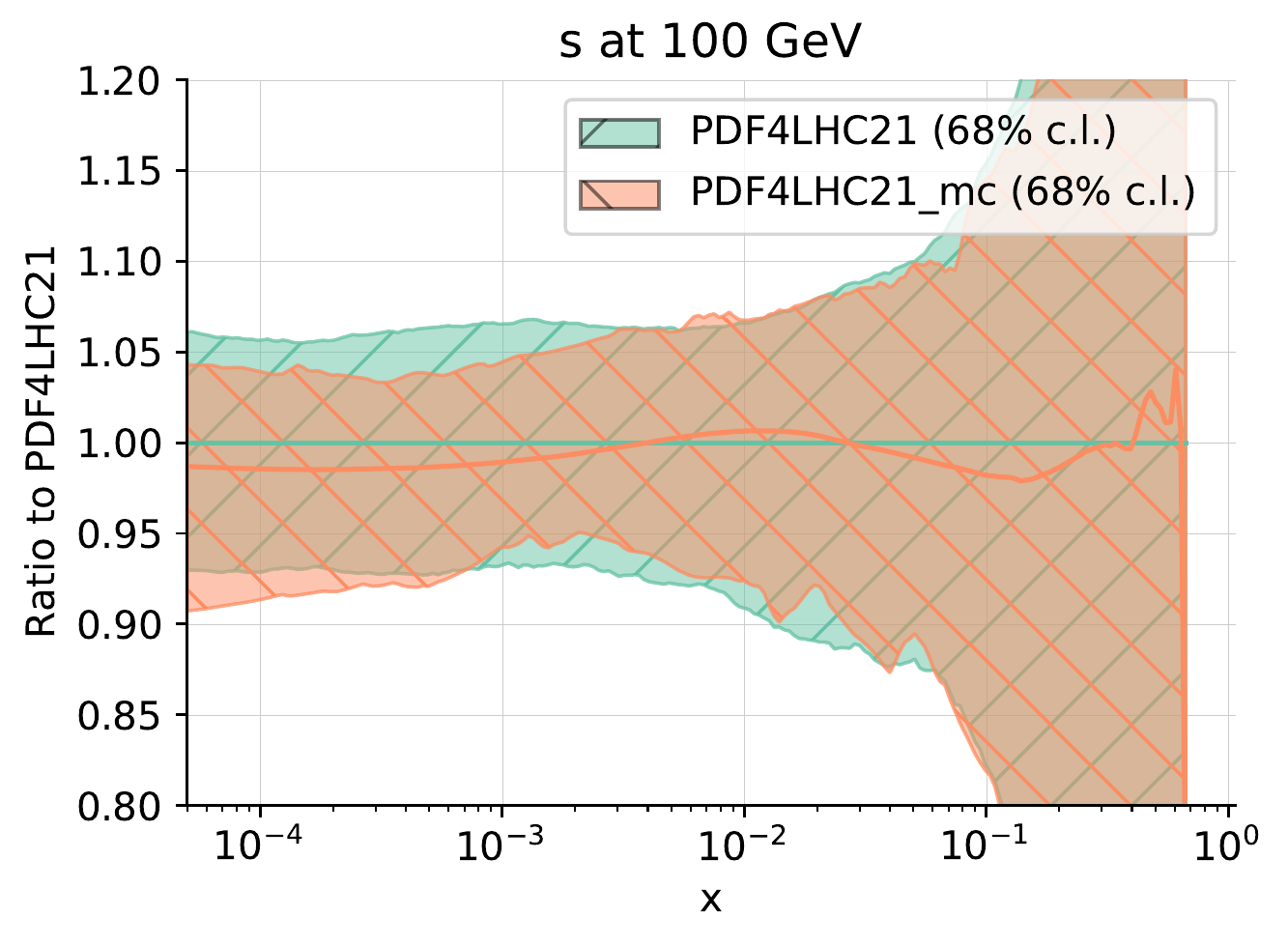}
\includegraphics[width=0.49\textwidth]{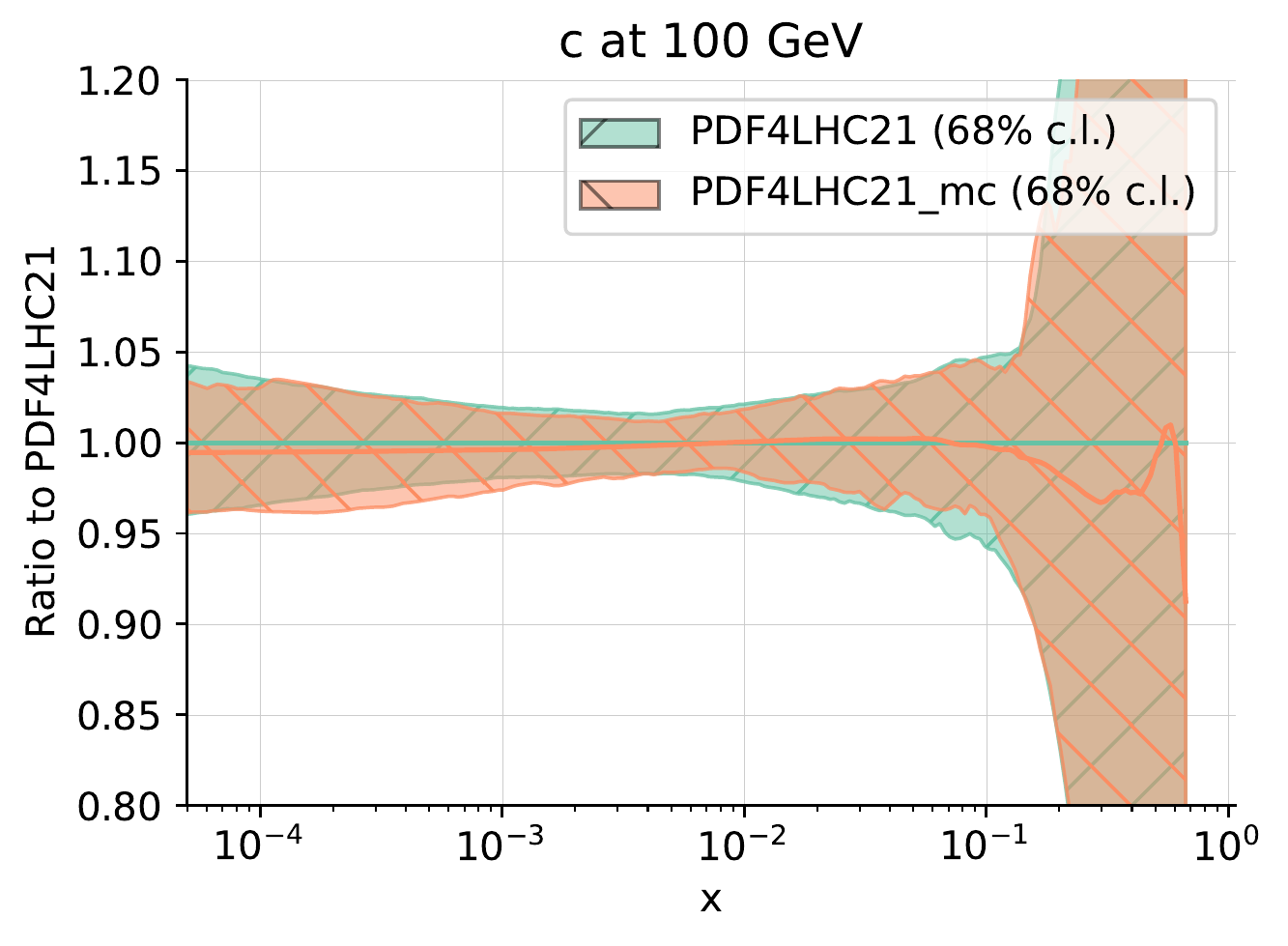}\\
\caption{\small Comparison of the baseline PDF4LHC21 combination, composed of $N_{\rm rep}=900$
replicas, with its compressed set composed of $N_{\rm rep}=100$ replicas.
Results are presented at $Q=100$ GeV and normalised to the central value
of the PDF4LHC21 combination.
}
\label{fig:cmc100-pdfs}
\end{figure}

Fig.~\ref{fig:cmc100-pdfs} displays the
comparison of the baseline PDF4LHC21 combination, with $N_{\rm rep}=900$
replicas, with its compressed set, labeled {\sc \small PDF4LHC21\_mc} from now on, with only $N_{\rm rep}=100$ replicas.
Results are presented at $Q=100$ GeV and normalised to the central value
of the PDF4LHC21 combination.
One can observe the good agreement between the baseline and the compressed distributions.
In terms of central values, any residual difference between the baseline
and compressed sets is much smaller than the PDF uncertainties themselves.
This comparison highlights how the compressed set
reproduces appropriately the baseline combination.

\begin{figure}[!t]
\centering
\includegraphics[width=0.49\textwidth]{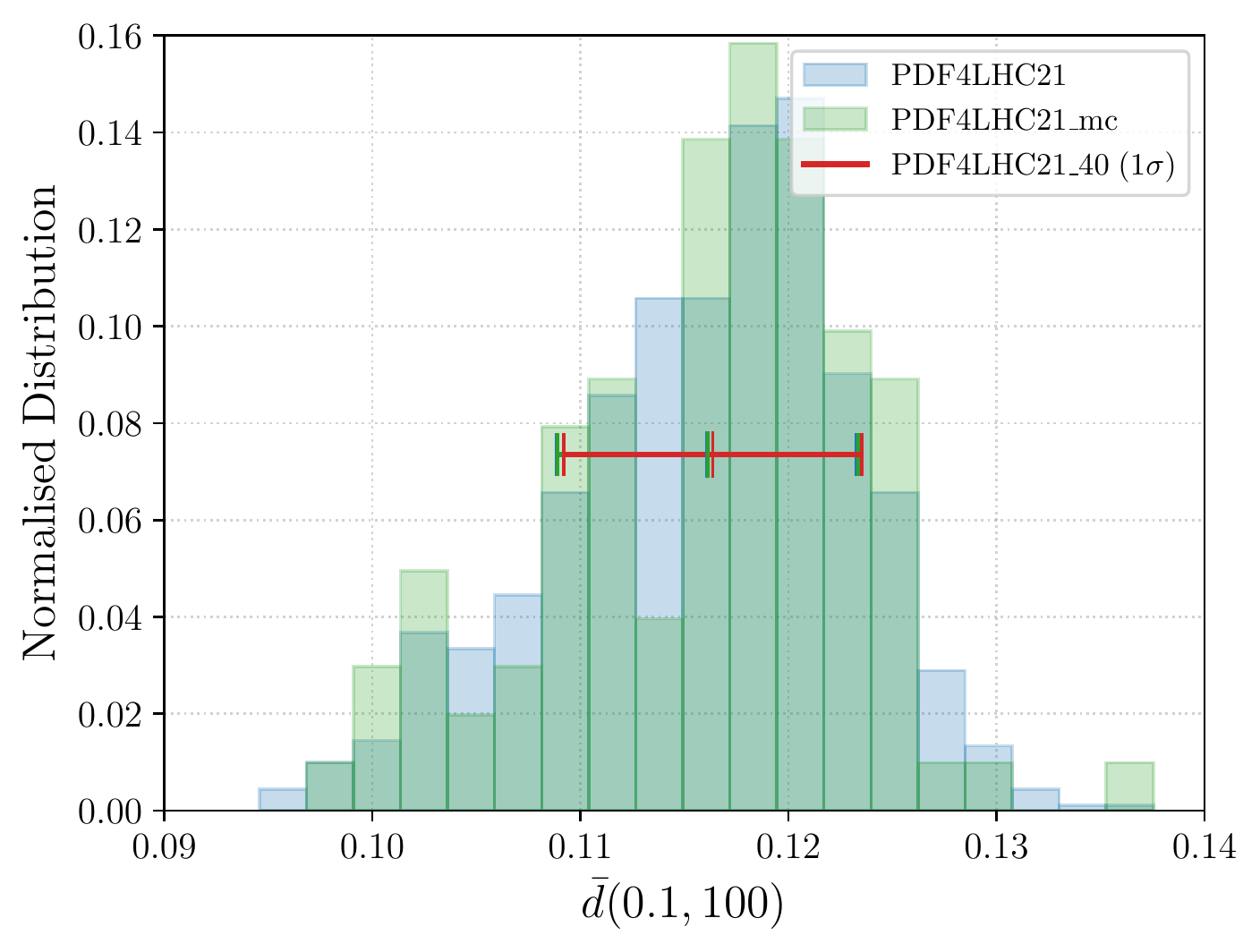}
\includegraphics[width=0.49\textwidth]{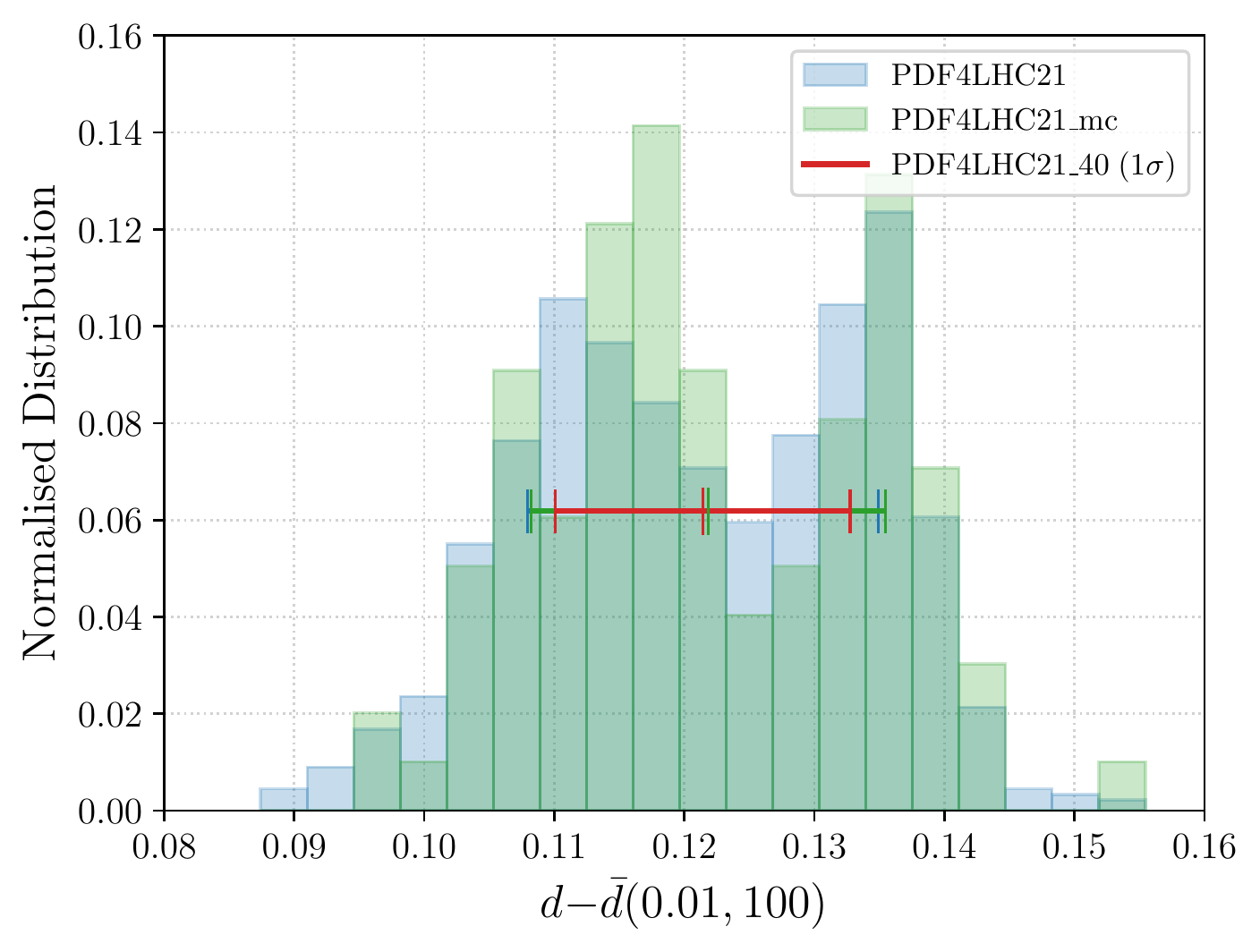}\\
\includegraphics[width=0.49\textwidth]{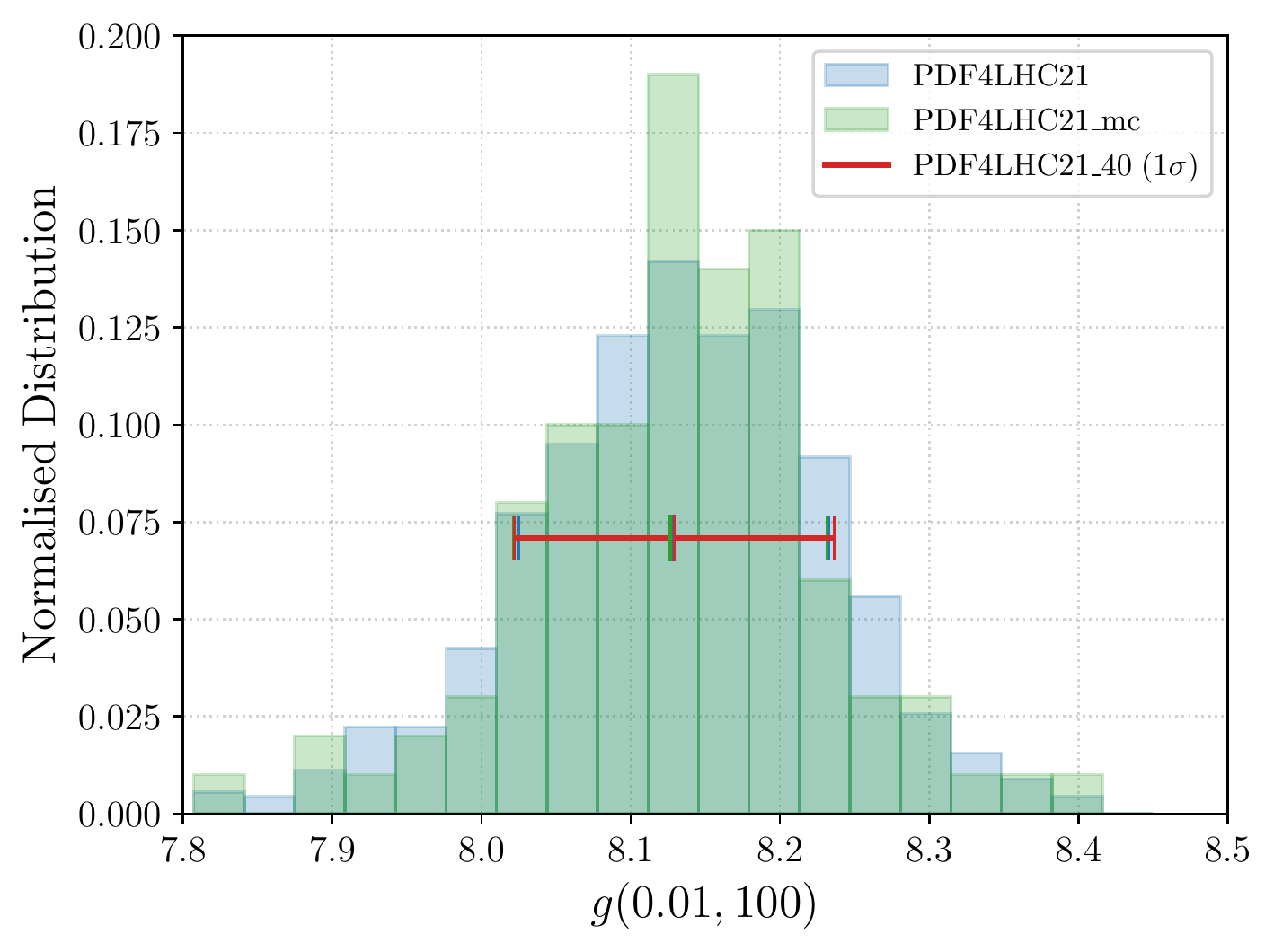}
\includegraphics[width=0.49\textwidth]{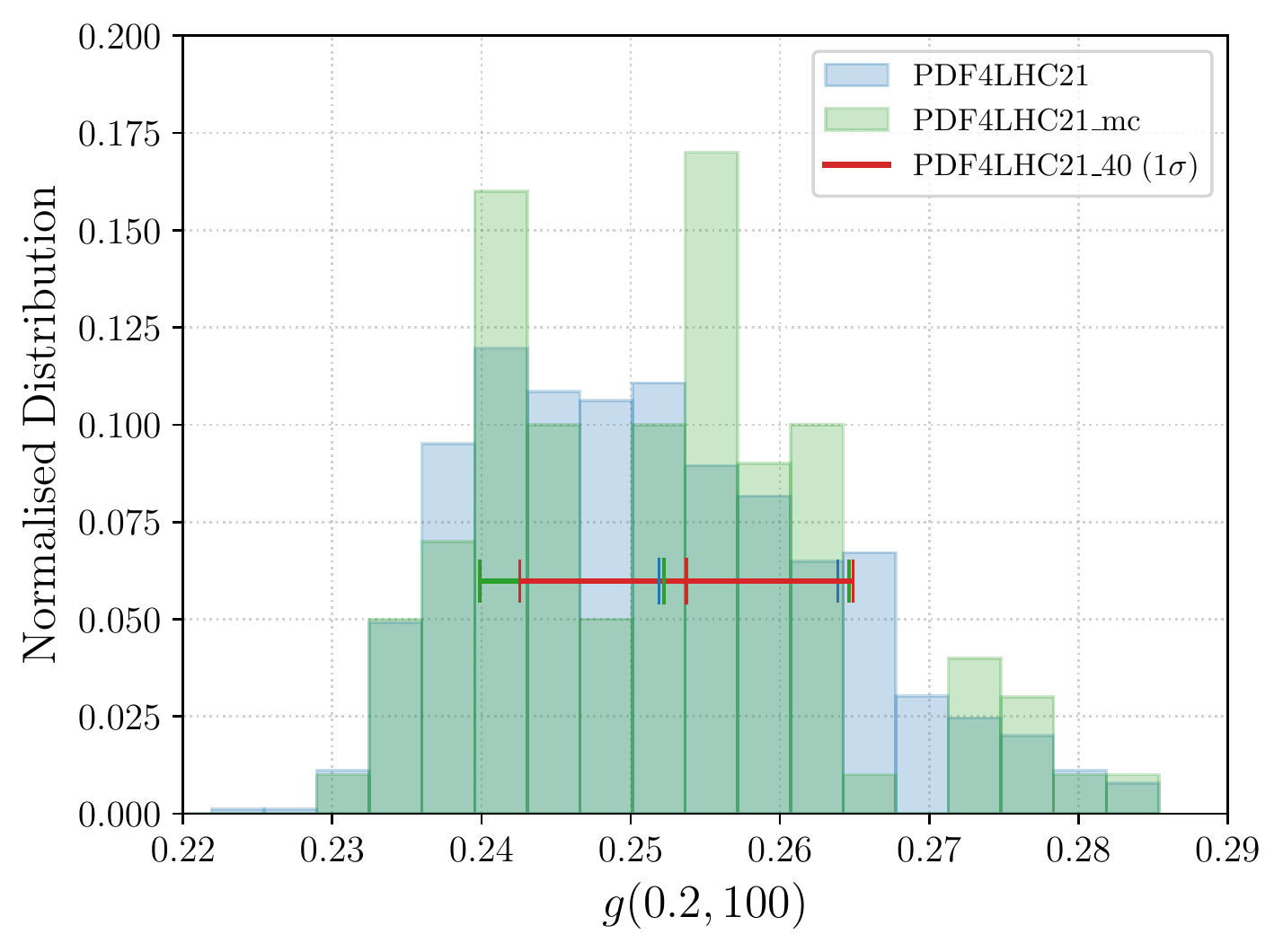}\\
\includegraphics[width=0.49\textwidth]{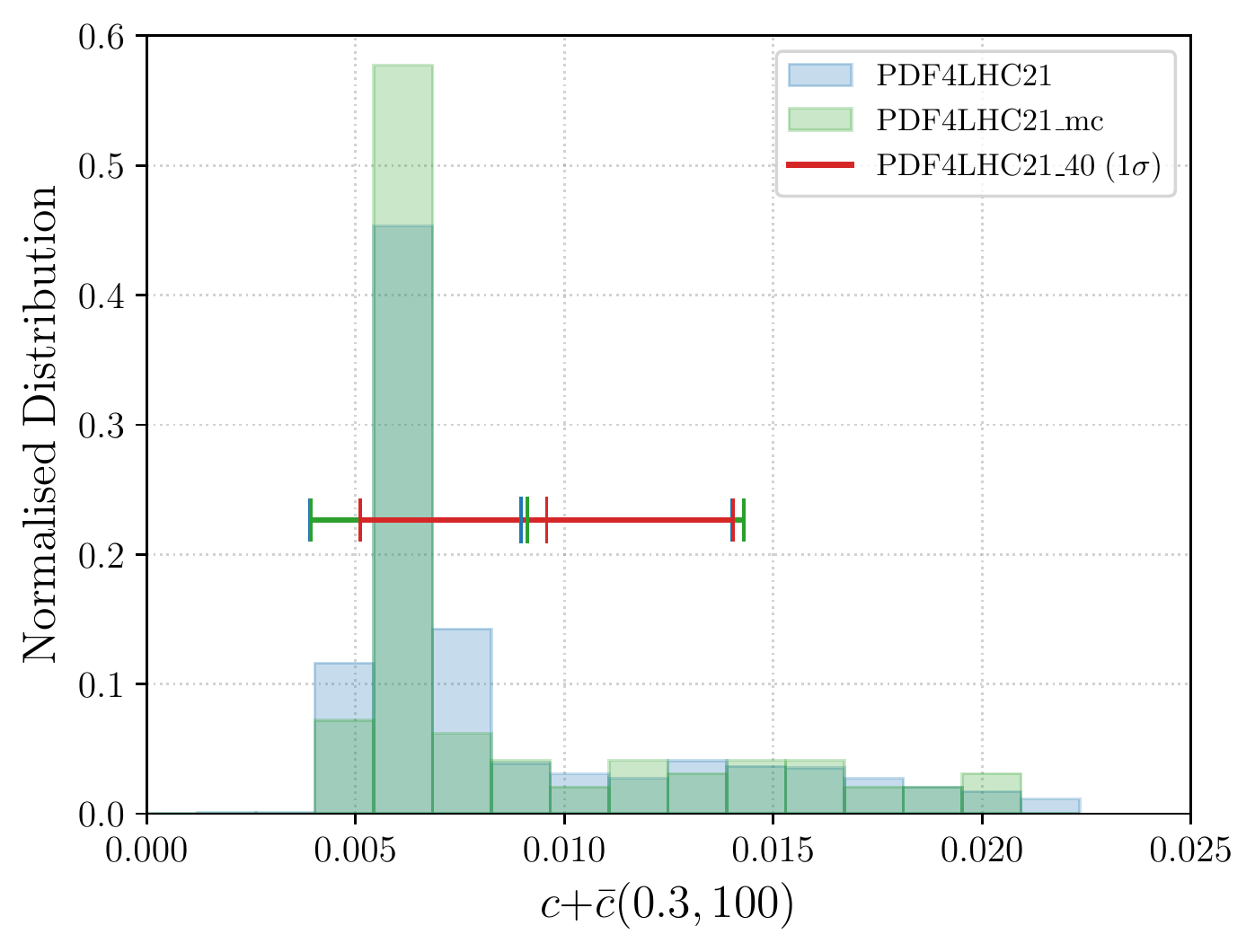}
\includegraphics[width=0.49\textwidth]{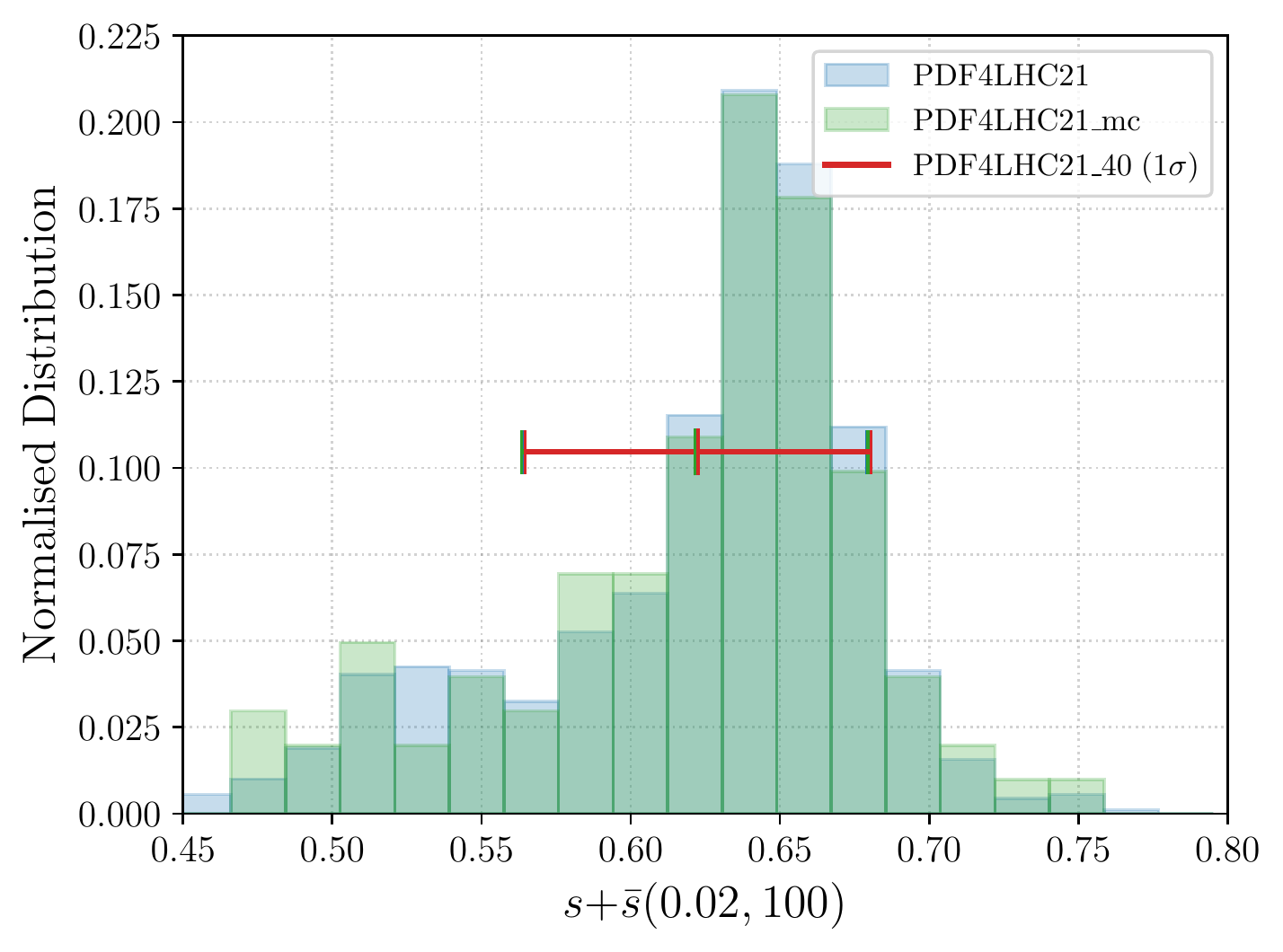}\\
\caption{\small Same as in Fig.~\ref{fig:prior-xDistr}, now with the
  probability distribution associated to the PDF4LHC21 
  900-replica combination (PDF4LHC21)
  set compared to the that of {\sc \small PDF4LHC21\_mc}. 
  Results are displayed at $Q=100$ GeV for $\bar{d}(x=0.1)$,  $(d-\bar{d})(x=0.01)=d_v(x,0.01)$, $g(x=0.01)$,
$g(x=0.2)$, $c^+(x=0.3)$, and $s^+(x=0.02)$.
The same binning is applied to the two distributions.
The horizontal lines indicate the median and 68\% CL intervals, also including those from the 40-member Hessian PDF set, {\sc \small PDF4LHC21\_40}, which will be described in Sec.~\ref{sec:hessian_reduction}.
}
\label{fig:cmc100-xDistr}
\end{figure}

In particular, for those PDF flavours and regions of $x$ where the one-sigma and
68\% CL intervals do not coincide (indicating a non-Gaussian distribution), the
compression algorithm manages to also reproduce these differences.
To demonstrate that this is indeed the case, Fig.~\ref{fig:cmc100-xDistr} displays
the Monte Carlo representations of the probability
densities associated to the $N_{\rep}=900$ replicas of the baseline PDF4LHC21 combination
compared to the compressed version with $N_{\rep}=100$ replicas, the {\sc \small PDF4LHC21\_mc} set.
Results are displayed at $Q=100$ GeV for $\bar{d}(x=0.1)$,  $d_v(x,0.01)$, $g(x=0.01)$,
$g(x=0.2)$, $c^+(z=0.3)$, and $s^+(x=0.02)$.
The choice is the same as in  Fig.~\ref{fig:prior-xDistr}, which was motivated
by the fact that these are the
flavours and regions of $x$ where there are discrepancies between the
three combination inputs and hence where non-Gaussian features
are more marked.
Note that the same binning is applied to the two distributions.
The horizontal lines indicate the median and 68\% CL intervals associated to the two distributions.

From Fig.~\ref{fig:cmc100-xDistr} one observes that indeed the compression algorithm
manages to reproduce the main features of the baseline distributions.
In all cases the median and 68\% CL intervals are reproduced, which is worth
pointing out
given that these distributions exhibit marked non-Gaussian features.
In particular, we find that the compressed set also contains the double-peak
structure in $d_v(x=10^{-2})$, the fat tails in $c^+(x=0.3)$ and $s^+(x=0.02)$,
as well as the skewness of the $\bar{d}(x=0.1)$ and $g(x=10^{-2})$ distributions.
Hence we conclude that the  performance of the compression method is satisfactory
and that it captures the non-trivial statistical features
of the baseline probability distribution.

\begin{figure}[t]
\centering
\includegraphics[width=0.49\textwidth]{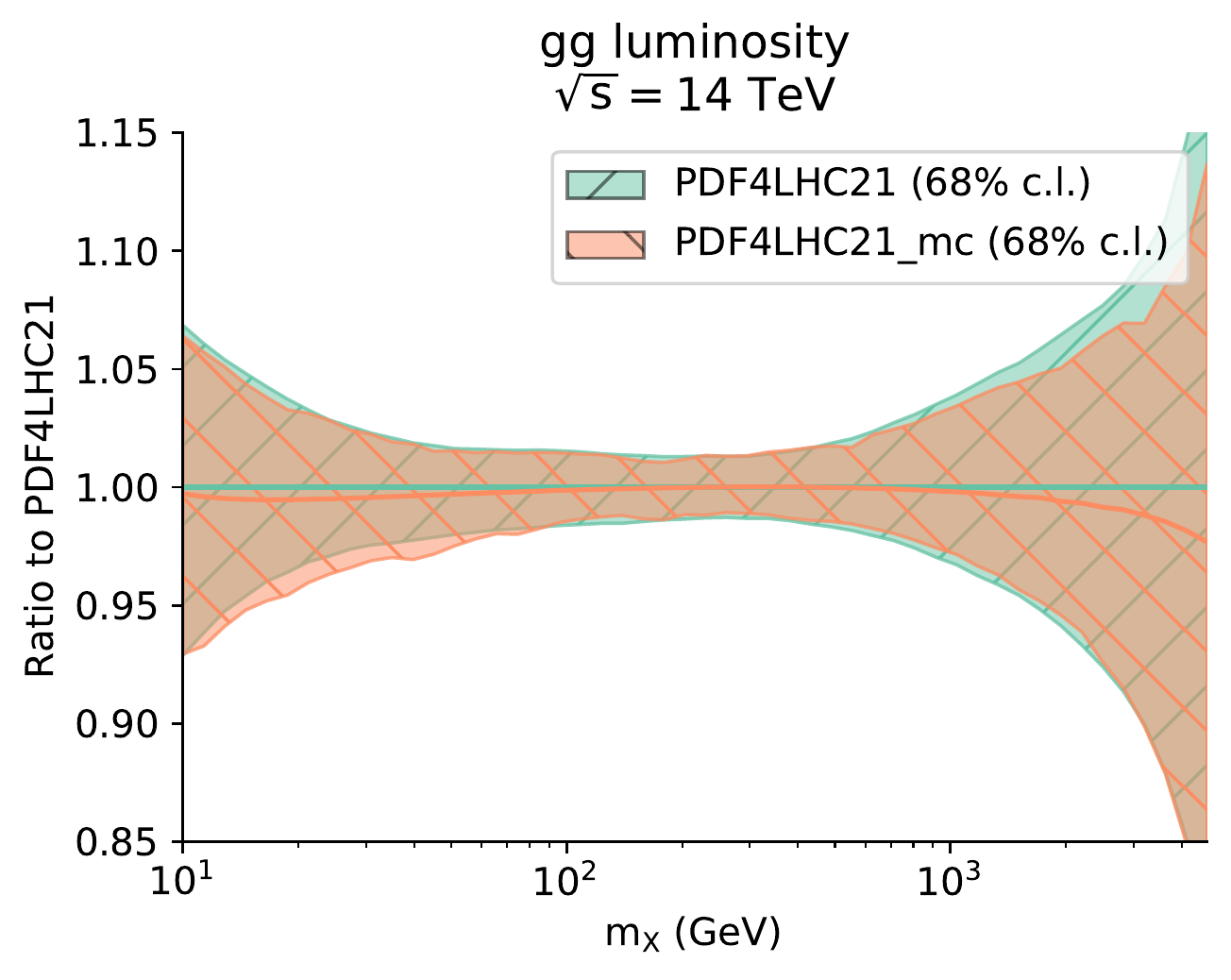}
\includegraphics[width=0.49\textwidth]{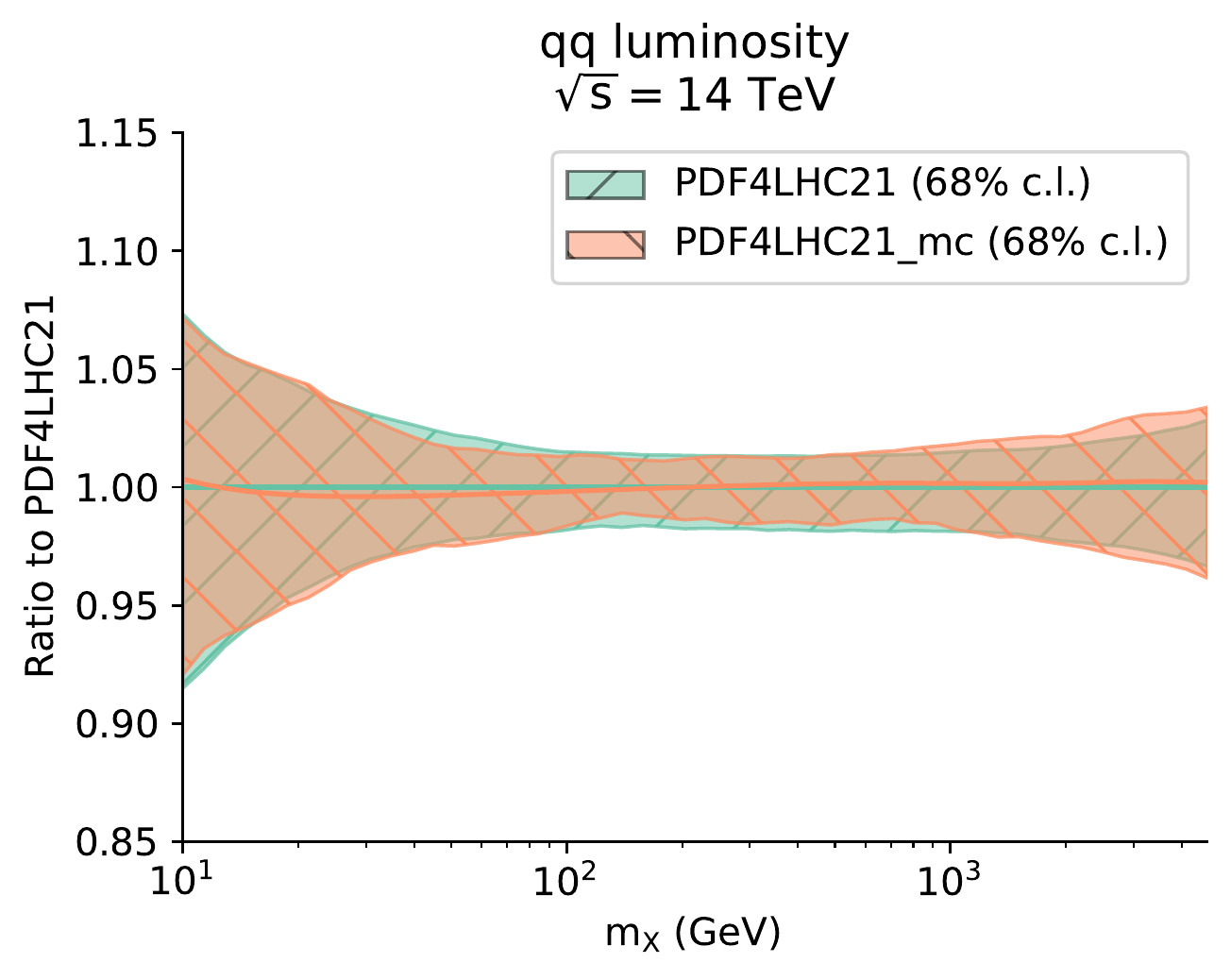}\\
\includegraphics[width=0.49\textwidth]{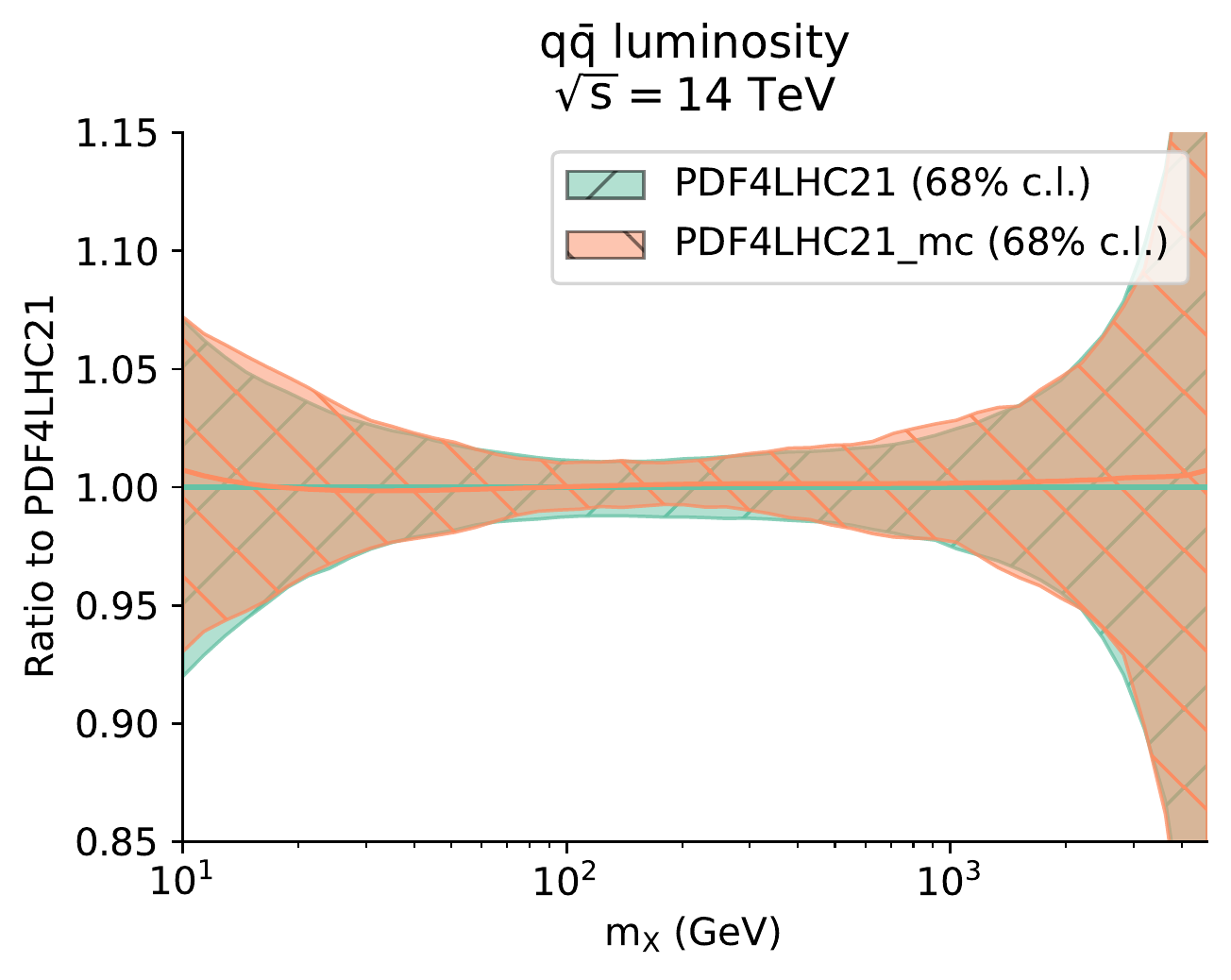}
\includegraphics[width=0.49\textwidth]{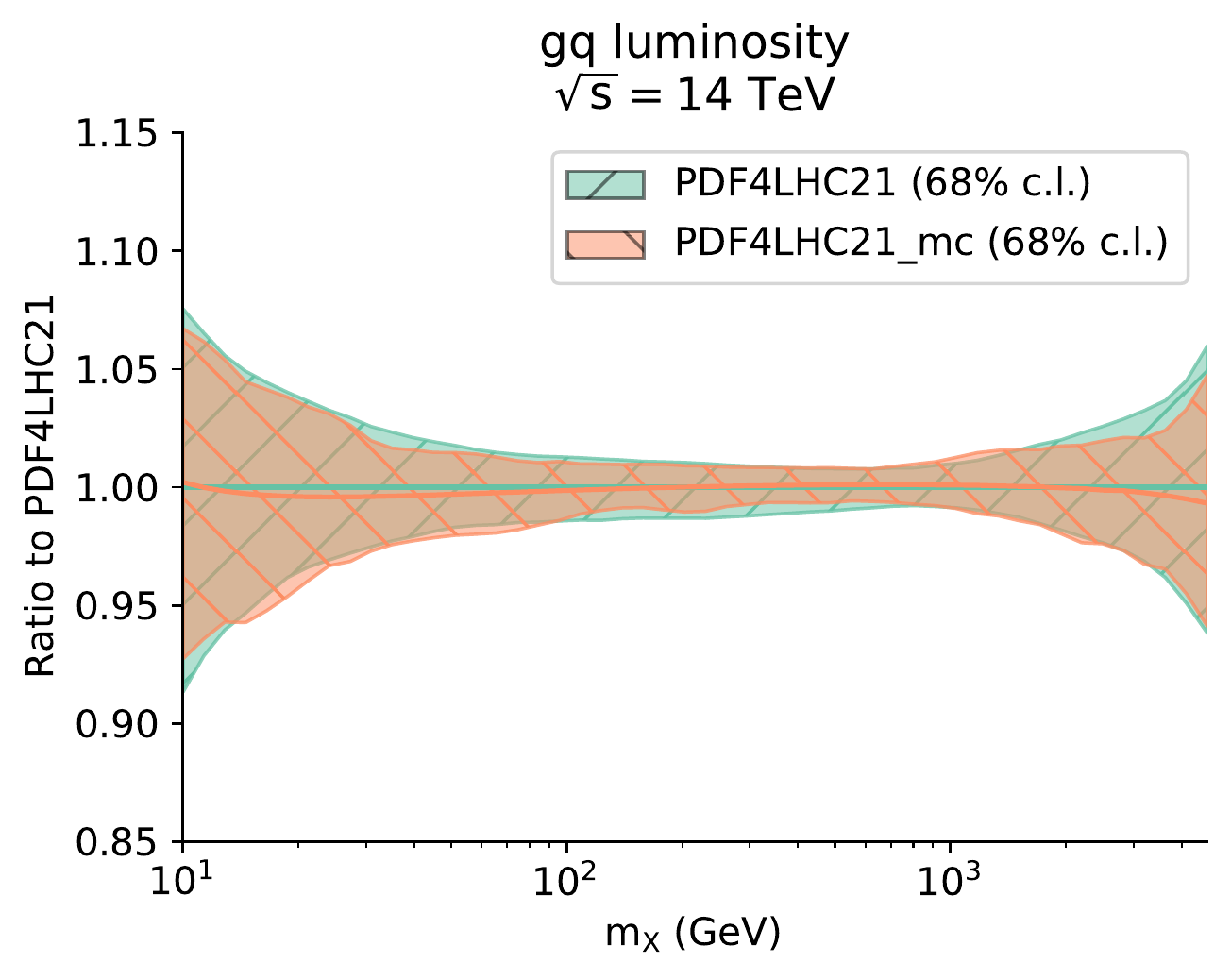}\\
\caption{\small Comparison of the gluon-gluon, quark-quark, quark-antiquark, and gluon-quark PDF
luminosities at the LHC 14 TeV as a function of the invariant mass $m_X$ between the PDF4LHC21
combination, with $N_{\rm rep}=900$ replicas, and its version compressed to $N_{\rm rep}=100$ replicas.
Results are displayed normalised to the
central value of the PDF4LHC21 combination.
}
\label{fig:cmc100-lumis}
\end{figure}

Further validation of the compression strategy is provided by
Fig.~\ref{fig:cmc100-lumis}, displaying the
comparison of the gluon-gluon, quark-quark, quark-antiquark, and gluon-quark
partonic
luminosities at the LHC 14 TeV as a function of the invariant mass
$m_X$ between the PDF4LHC21 combination with $N_{\rm rep}=900$ replicas, 
and the compressed {\sc \small PDF4LHC21\_mc} set, with only $N_{\rm rep}=100$ replicas.
Results are displayed normalised to the central value of the PDF4LHC21 combination.
As for the case of the PDFs, good agreement between the baseline and the compressed
sets is found, and in particular for those partonic combinations and $m_X$ regions
where the $1\sigma$ and the 68\% CL intervals differ, such as the large-$m_X$
gluon-gluon luminosity, or the small-$m_X$ quark luminosities, the compressed
set manages to capture appropriately the non-Gaussian features of the baseline
distribution.

To summarise, we have demonstrated that a compressed set with $N_{\rm rep}=100$
replicas reproduces satisfactorily the PDF4LHC21 combination
composed of $N_{\rm rep}=900$ replicas.
This compressed set can hence be reliably deployed for LHC phenomenology,
as will be shown in Sect.~\ref{sec:phenomenology}.
%

\subsection{Hessian reduction}
\label{sec:hessian_reduction}

A Hessian reduction was the first technique introduced for the combination of individual PDF sets, first 
using the {\sc \small META} approach in~\cite{Gao:2013bia}
and soon developed in an independent {\sc\small mc2hessian} approach in~\cite{Carrazza:2015aoa,Carrazza:2016htc}. 
Such a Hessian
representation  reproduces the Gaussian features (central value, variances, and correlations) 
of the baseline probability distribution, which can be approximately described by a multi-Gaussian distribution in a wide range of $x$ and $Q$ as discussed at the end of Sec.~\ref{sec:generation_mc_replicas}.

As was already the case in the PDF4LHC15 analysis~\cite{Butterworth:2015oua}, two different Hessian
reduction strategies have been investigated for this work. The output Hessian set of the present analysis, designated {\sc\small PDF4LHC21\_40}, is based on $N_{\rm eig}=40$ eigenvectors and constructed by means of the {\sc\small META-PDF} method, reviewed in technical detail in App.~\ref{subsec:HessianConv}. The second Hessian reduction method is the {\sc\small mc2hessian} algorithm, also discussed in App.~\ref{subsec:HessianConv}. This appendix compares performance of the two methods and justifies the choice adopted for the {\sc\small PDF4LHC21\_40} set.

\begin{figure}[t]
\centering
\includegraphics[width=1\textwidth]{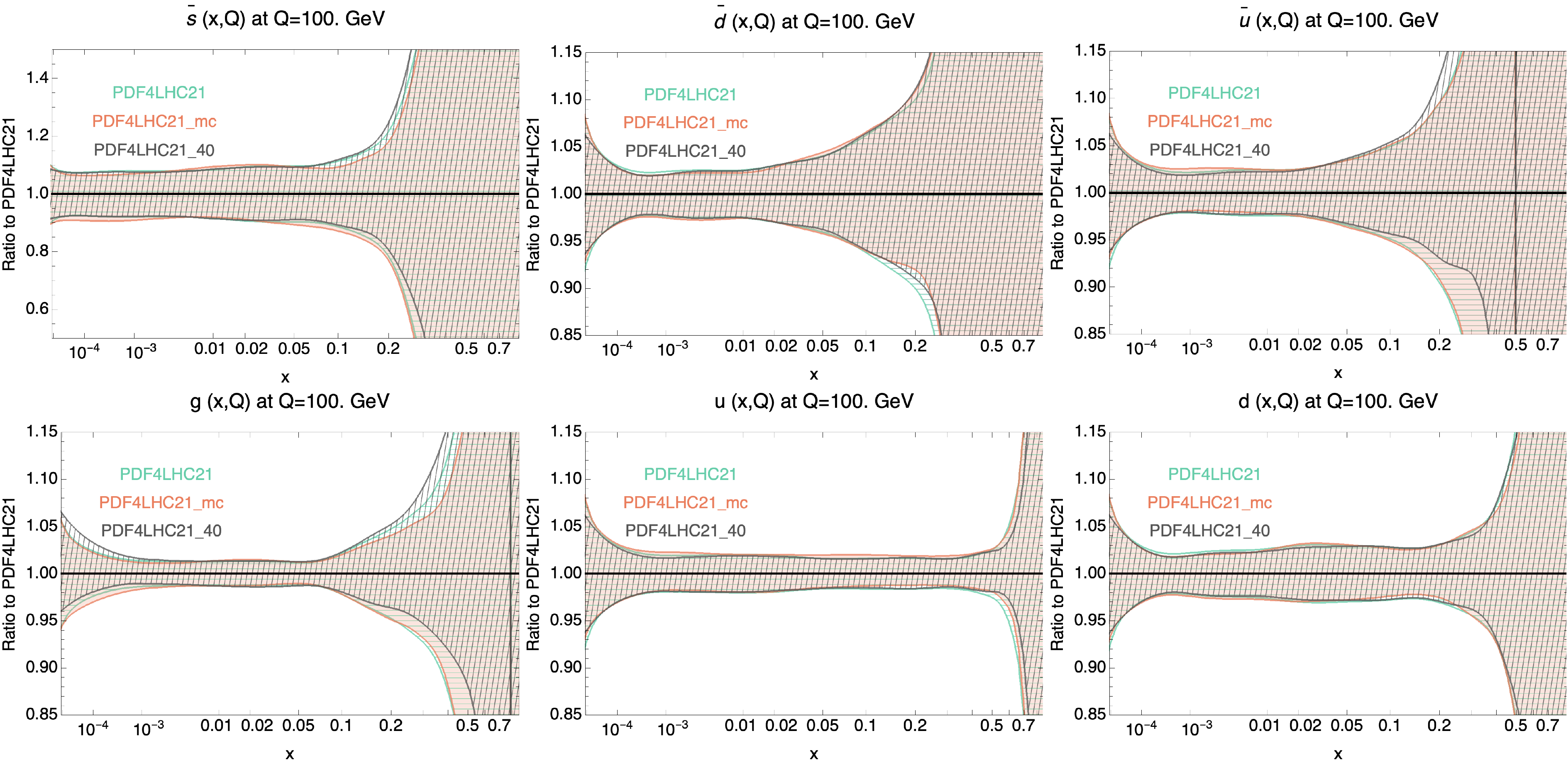}
\caption{\small Comparison of the baseline  PDF4LHC21 set
  with the corresponding MC compressed set, {\sc \small PDF4LHC21\_mc},
  and its Hessian representation with  $N_{\rm eig}=40$ eigenvectors, {\sc \small PDF4LHC21\_40}.
We show the results for the anti-strange, anti-down, anti-up,  gluon and the up, down
quark PDFs  at $Q=100$~GeV. 
}
\label{fig:pdf4lhc21_900priors_meta40}
\label{fig:pdf4lhc21_vs_pdf4lhc15_pdfs_hessian50}
\end{figure}

The {\sc\small META-PDF} approach is implemented in the {\sc\small MP4LHC} package. 
This method  constructs
a common meta-parametrization of the MC replicas from the baseline PDF ensemble
using Bernstein polynomials. Then, dimensionality reduction is performed in the space of meta-parameters to describe the probability manifold in the meta-parameter space with fewer parameters. 
In the current realisation, dimensionality reduction is performed by Principal Component Analysis (PCA). 

In more detail, all baseline replicas end up being fitted with the same parametric form,
or meta-parametrisation, which depends on a reference PDF and a  Bernstein polynomial of a certain degree. 
The spread of the solutions of the meta-fit can then be quantified by a Hessian or covariance matrix
with dimensions $N_{\rm par}\approx 140$, the number of parameters of the fitted functional form,
which hence provides the sought-for multi-Gaussian representation
of the baseline distribution.  To build the reduced Hessian META-PDF set for the current analysis, the eigenvector directions corresponding to the largest (less constrained) $N_{\rm eig}=40$ eigenvalues of the Hessian matrix are selected.

For the PDF4LHC21 combination, the reference PDF is chosen to be positive in the whole $x$ range and, as such, it slightly differs from the average replica of the baseline set at $x\gtrsim 0.5$ at $Q=2$~GeV. 
Sect.~\ref{sec:largex_pdf4lhc21} elaborates further on this large-$x$ behaviour. 
The positivity feature is an improvement with respect to the 30-member META set {\sc \small PDF4LHC15\_30} distributed by the PDF4LHC15 study. With the positivity of the reference PDF implemented, the {\sc\small PDF4LHC21\_40} ensemble reproduces the trend of the 68\% CL uncertainties of the baseline fit more faithfully than the PDF4LHC15\_30 set in
a manner comparable to the {\sc\small PDF4LHC21\_mc} ensemble. 
However, being a Hessian set, {\sc\small PDF4LHC21\_40} does not provide additional information about the baseline probability distribution that can be obtained with {\sc\small PDF4LHC21\_mc}. 
For example, in Fig.~\ref{fig:cmc100-xDistr}, the {\sc\small PDF4LHC21\_40} ensemble predicts only the 68\% CL intervals indicated by the red bars, which agree with the baseline and {\sc\small PDF4LHC21\_mc} ensembles. 
It does not predict the shape of the histograms in the figure, including their non-Gaussian features.  

Fig.~\ref{fig:pdf4lhc21_vs_pdf4lhc15_pdfs_hessian50} displays the comparison
between the baseline PDF4LHC21 set, the {\sc\small PDF4LHC21\_mc} Monte Carlo,  and the {\sc\small PDF4LHC21\_40}
Hessian representations.
Only the PDF uncertainty bands (not the central values) are shown for the two compressed sets. The error bands are normalised to the central PDF of the corresponding flavor of the baseline set.
We show the results for the gluon and the up, down, anti-down, anti-strange, and anti-up 
quark PDFs at $Q=100$ GeV.

In general there is good agreement between the baseline and compressed sets both in terms of central values and PDF uncertainties. Mild differences can be seen for {\sc\small PDF4LHC21\_40} at very large $x$, where the respective error bands for $\bar{s}, \bar{u}, \bar{d}, g$ and $s$ are marginally shifted upwards due to the shifts in the central values. At $Q=100$ GeV, those shifts are appreciable only for  $x\gsim 0.2$.
A similar level of agreement can be found at the level of parton luminosities. 
Sect.~\ref{sec:phenomenology}
shows how predictions based on these PDFs compare for representative LHC processes.

\begin{figure}
\centering
\includegraphics[width=0.45\textwidth]{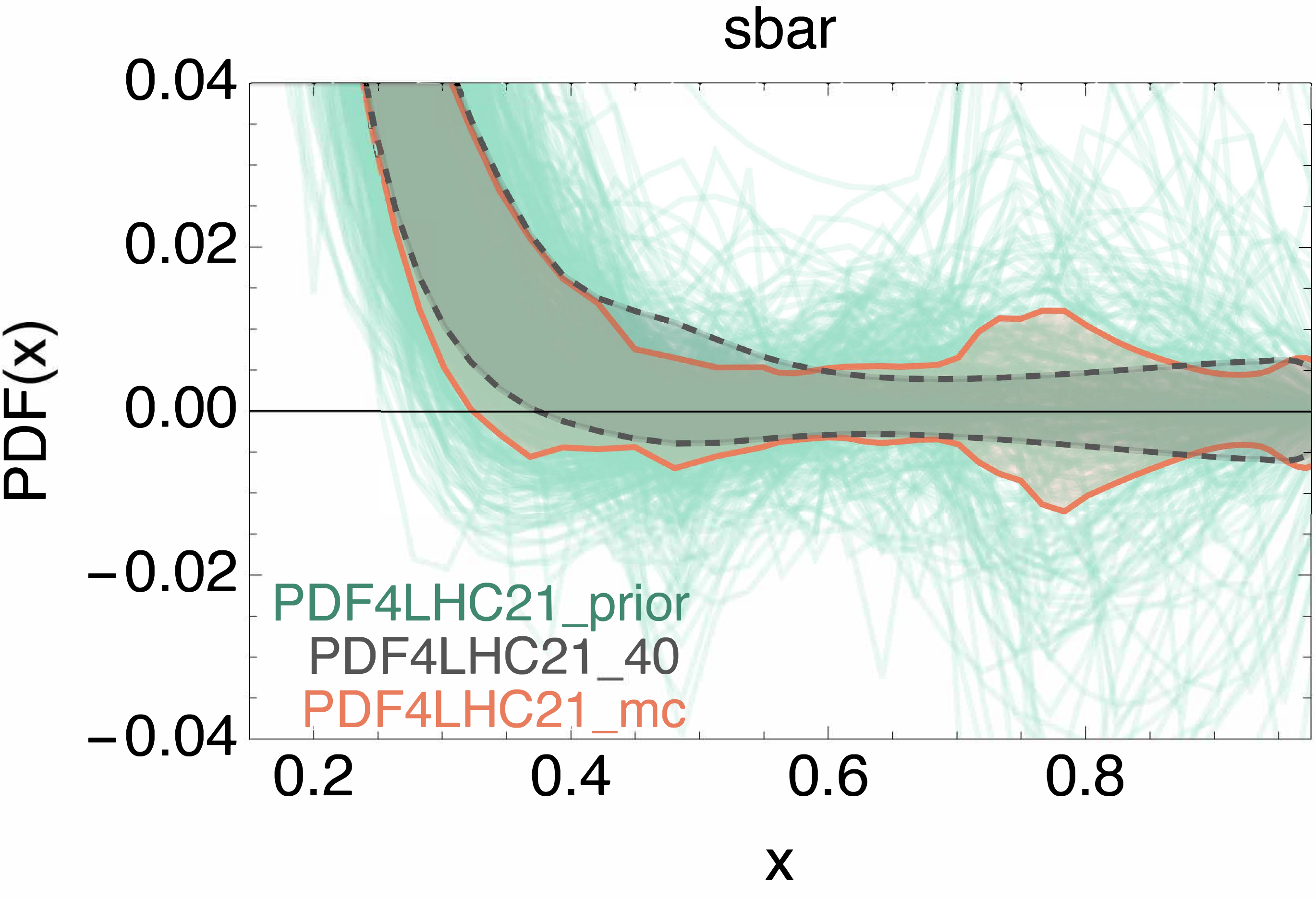}
\includegraphics[width=0.45\textwidth]{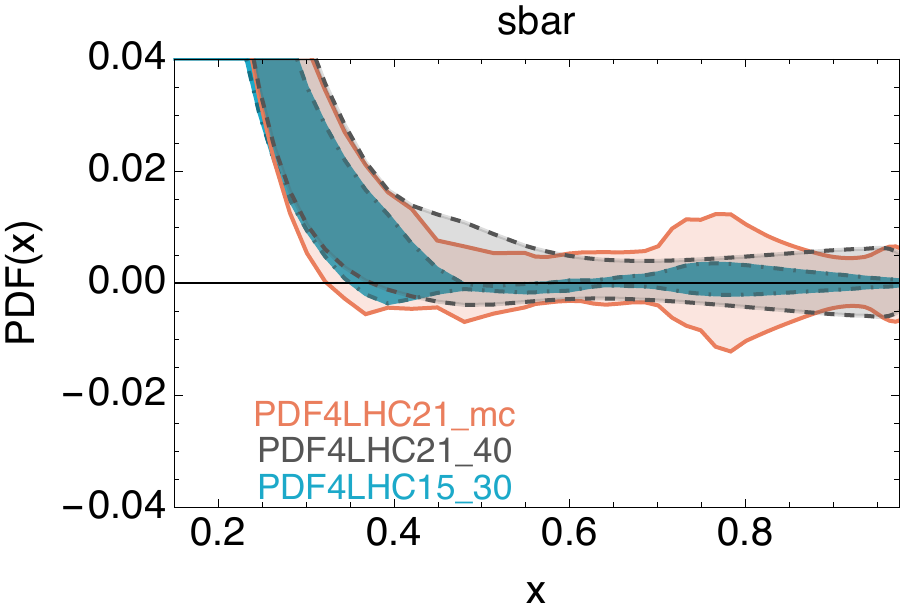}
\caption{\small Left: $\bar{s}(x,Q)$  now on an absolute scale and focusing on the large-$x$ region
at $Q=100$ GeV.
 The 68\% CL error bands of the {\sc\small PDF4LHC21\_40} and {\sc\small PDF4LHC21\_mc} ensembles are superimposed on the individual 900 replicas that compose  the {\sc \small PDF4LHC21} set (thin green lines).
  Right: the PDF4LHC21 compressed error bands from the left panel, without the $N_{\rm rep}=900$ set and overlaying the {\rm PDF4LHC15\_30} Hessian set, again based on the {\sc\small META-PDF} method.
}
\label{fig:pdf4lhc21_meta40_largex_ex}
\end{figure}

While more detailed comparisons of the two compression methods are reported in App.~\ref{app:tools}, Fig.~\ref{fig:pdf4lhc21_meta40_largex_ex} illustrates a typical comparison of the error bands at large $x$, where the differences of 
{\sc\small PDF4LHC21\_40} and {\sc\small PDF4LHC21\_mc} tend to be more pronounced. 
The left panel shows a comparison similar to those in Fig.~\ref{fig:pdf4lhc21_900priors_meta40}, now only for $\bar{s}(x,Q)$ on an absolute scale and focusing on $x>0.2$. Instead of the baseline 68\% CL error band, the panel shows the individual 900 replicas -- those demonstrate quite irregular behaviour that is approximated on average by both compressed error bands. The two error bands agree especially well at small- and moderate-$x$, while at $x>0.5$ the MC uncertainty shows somewhat more variability that traces variations of the baseline replicas. 

The right panel of Fig.~\ref{fig:pdf4lhc21_meta40_largex_ex} demonstrates that the {\sc\small PDF4LHC21\_40} ensemble more accurately reproduces the baseline uncertainty at very large $x$ than its {\sc\small PDF4LHC15\_30} counterpart: the uncertainty of the latter was artificially suppressed at $x>0.4$, the point where the reference PDF of the {\sc\small PDF4LHC15\_30} ensemble (set equal to the average of three input central PDFs) crossed zero. The META algorithm is more stable when the reference PDF is positive, and hence the {\sc\small PDF4LHC21\_40} uses positive-definite reference PDFs at the combination scale $Q=2$ GeV to obtain trustworthy uncertainties. 
The central PDFs of the PDF4LHC21 baseline are negative in some large-$x$ intervals, see Figs.~\ref{fig:pdf4lhc21_900_prior_scaled} and~\ref{fig-app:META40_CMC100}. 
The reference PDFs agree with the baseline central PDFs within the uncertainties, but remain positive by construction, as discussed below. Other sea (anti)quark distributions show similar improvements in their large-$x$ behaviour due the imposed positivity, as
will be shown also in Fig.~\ref{fig-app:META40_CMC100}. 
The consequences of positivity on phenomenology will be discussed in Sects.~\ref{sec:phenomenology}; the large-$x$ behaviour of the PDF4LHC21 combination is further scrutinised in Sect.~\ref{sec:largex_pdf4lhc21}. 

From this comparison, consistent with the previous ones, one finds that
 the Hessian reduced set provides
a satisfactory description of the baseline at high energies.\footnote{A comparison at the scale of the combination, $Q=2$ GeV, can be found in Fig.~\ref{fig-app:META40_CMC100}. }

\subsection{Comparison with PDF4LHC15}
\label{subsec:comp_pdf4lhc15}

In the previous sections, we have discussed how to construct the PDF4LHC21 Monte Carlo combination from its three input sets and subsequently how to compress the number of replicas or to produce a Hessian representation. We have also discussed the main features of this PDF4LHC21 combination, tracing them back  to the differences and similarities between the three constituent sets. We can now assess to which extent PDF4LHC21 differs from its predecessor, PDF4LHC15, both in terms of the central values and PDF uncertainties. In this subsection we compare the baseline PDF4LHC15 and PDF4LHC21 combinations, each made up of $N_{\rm rep}=900$ Monte Carlo replicas. The qualitative conclusions, however, remain the same if the compressed or Hessian versions are used.

\begin{figure}[!t]
\centering
\includegraphics[width=0.49\textwidth]{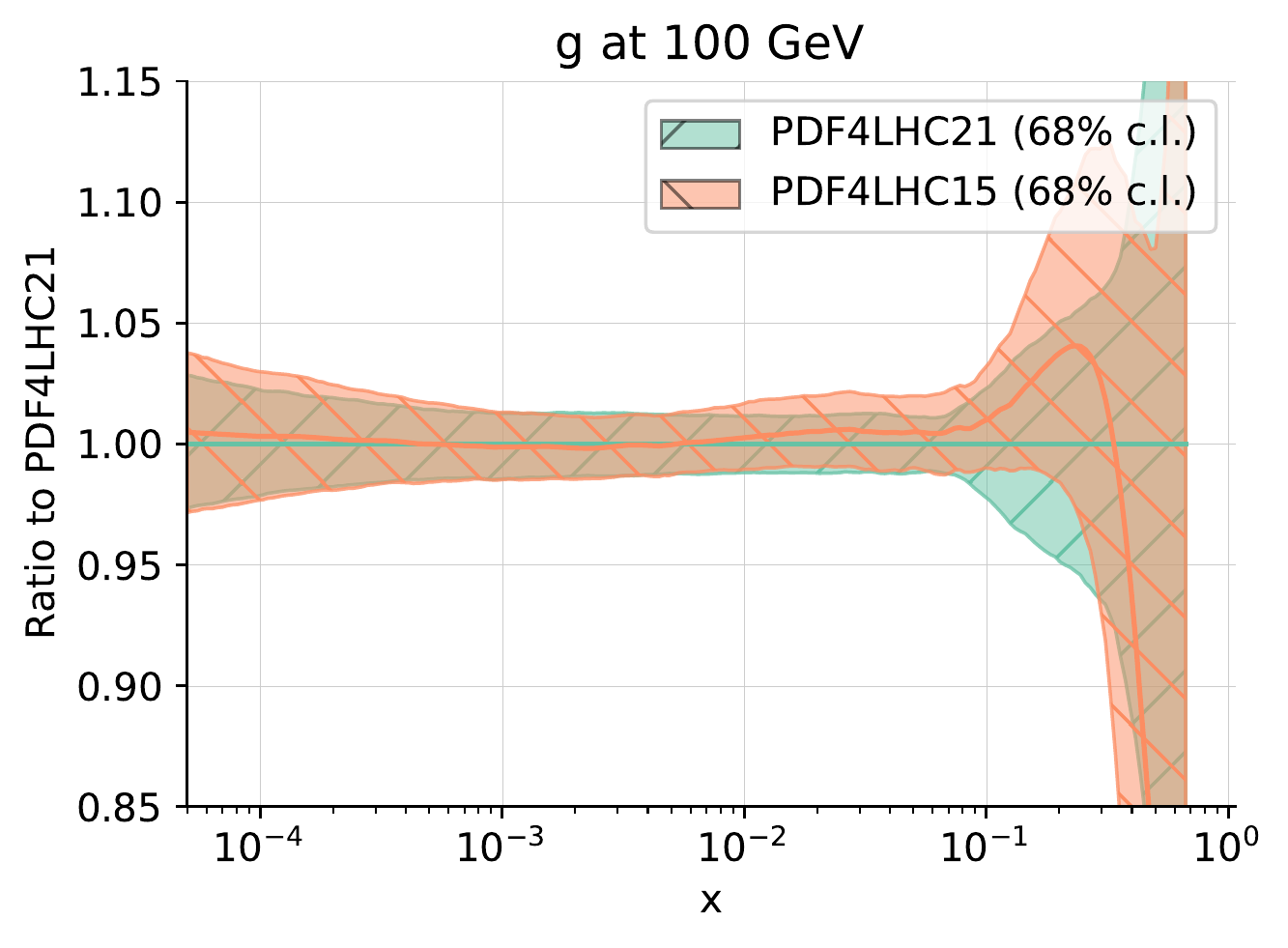}
\includegraphics[width=0.49\textwidth]{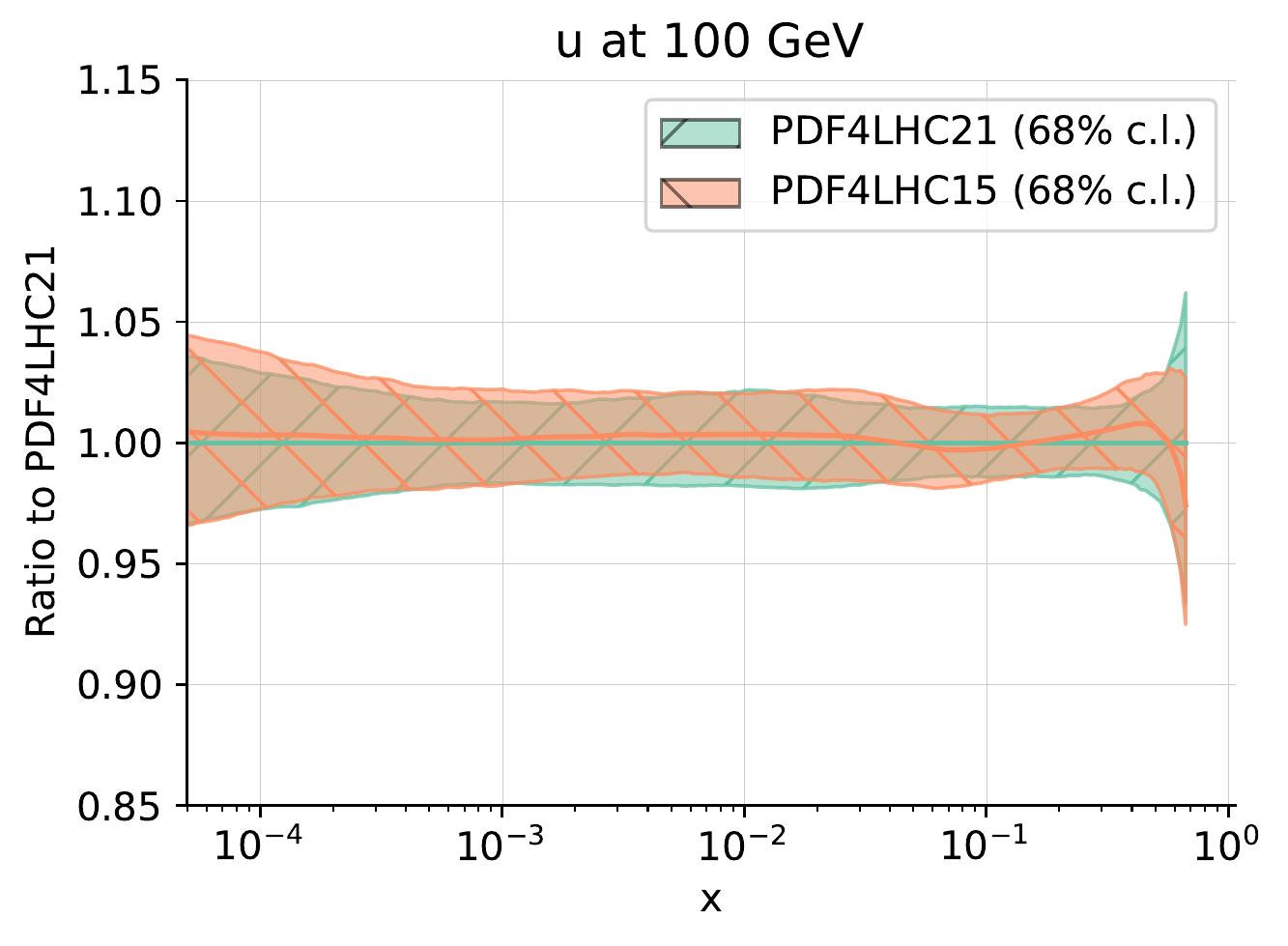}\\
\includegraphics[width=0.49\textwidth]{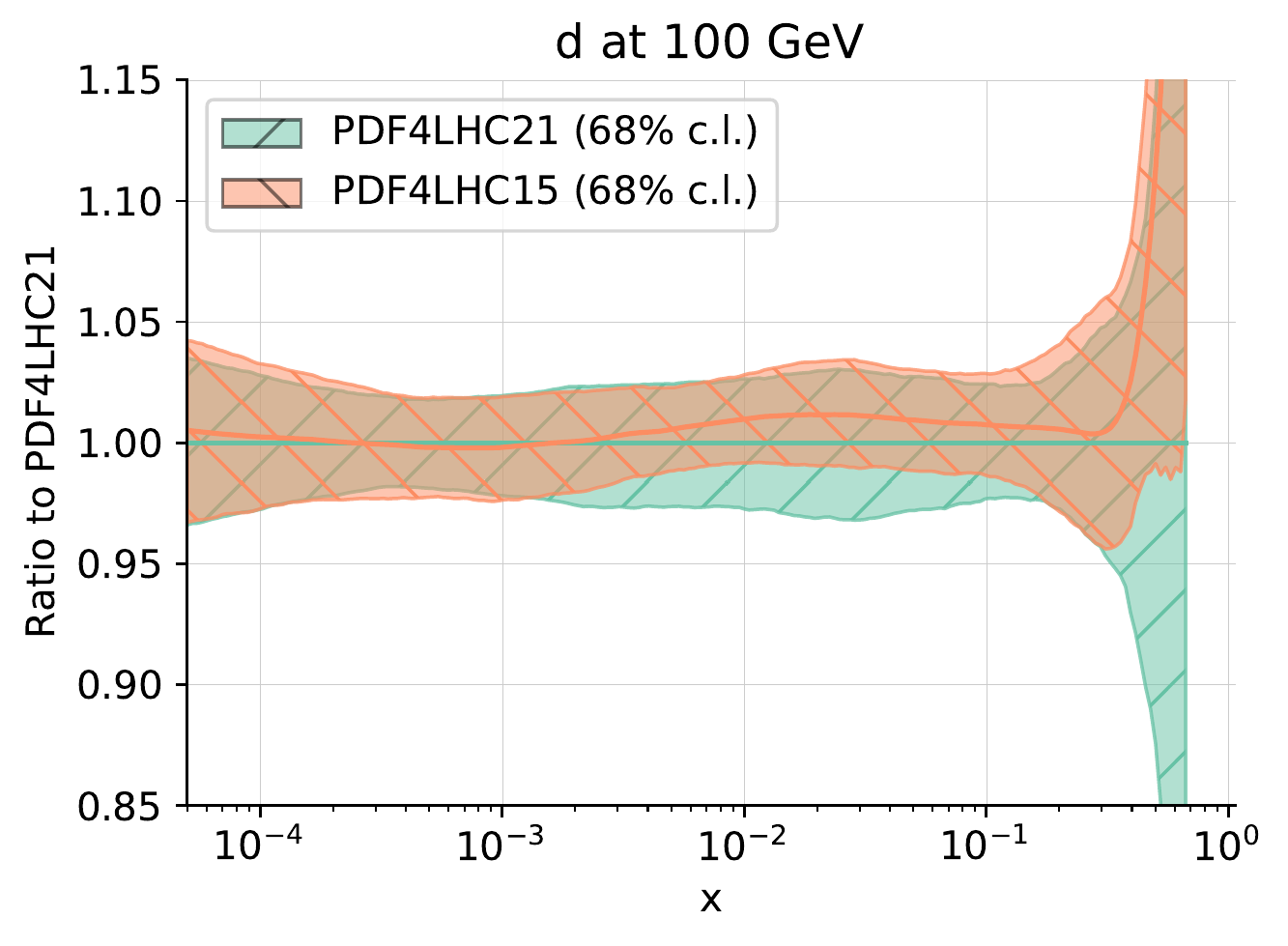}
\includegraphics[width=0.49\textwidth]{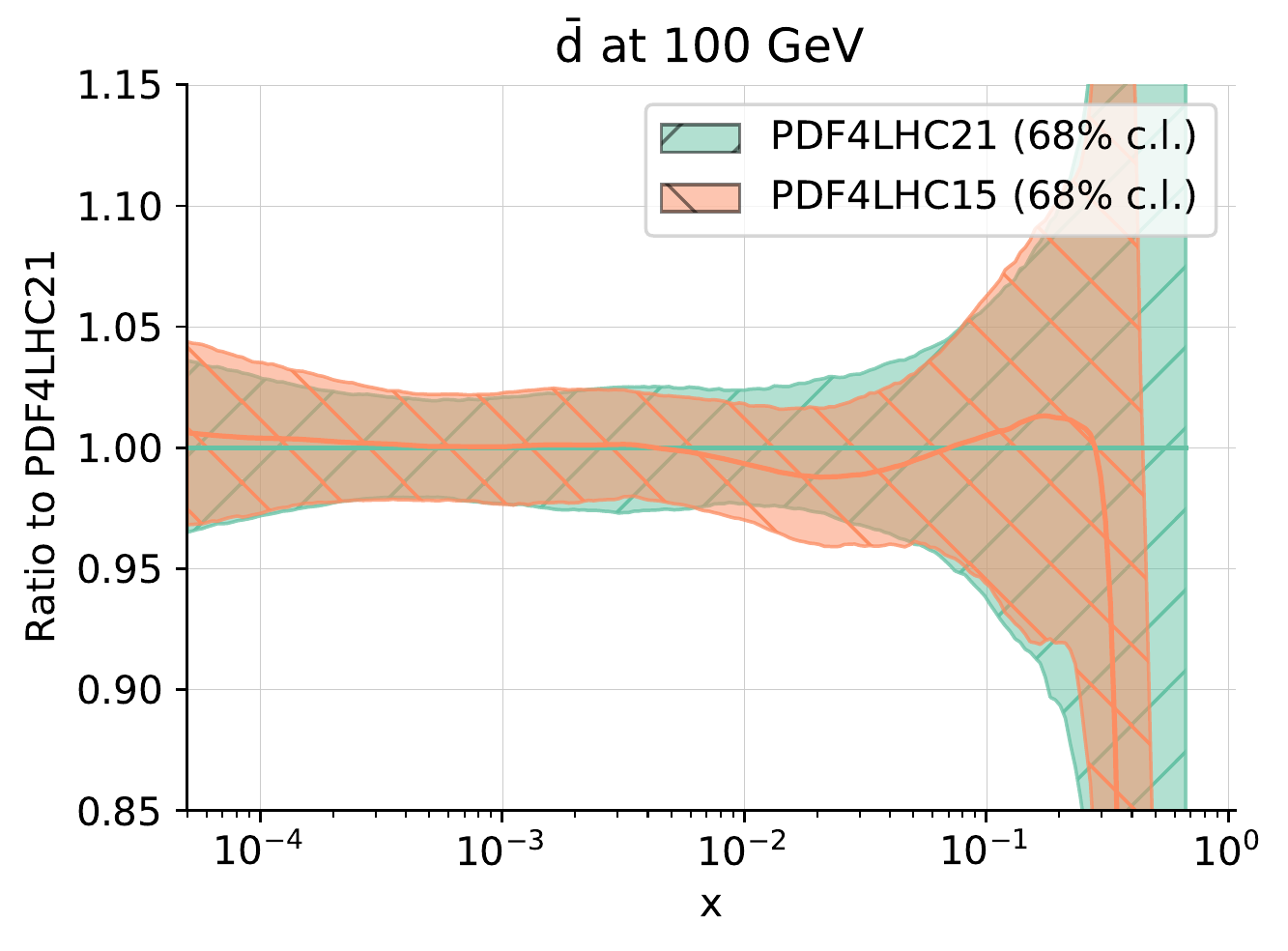}\\
\includegraphics[width=0.49\textwidth]{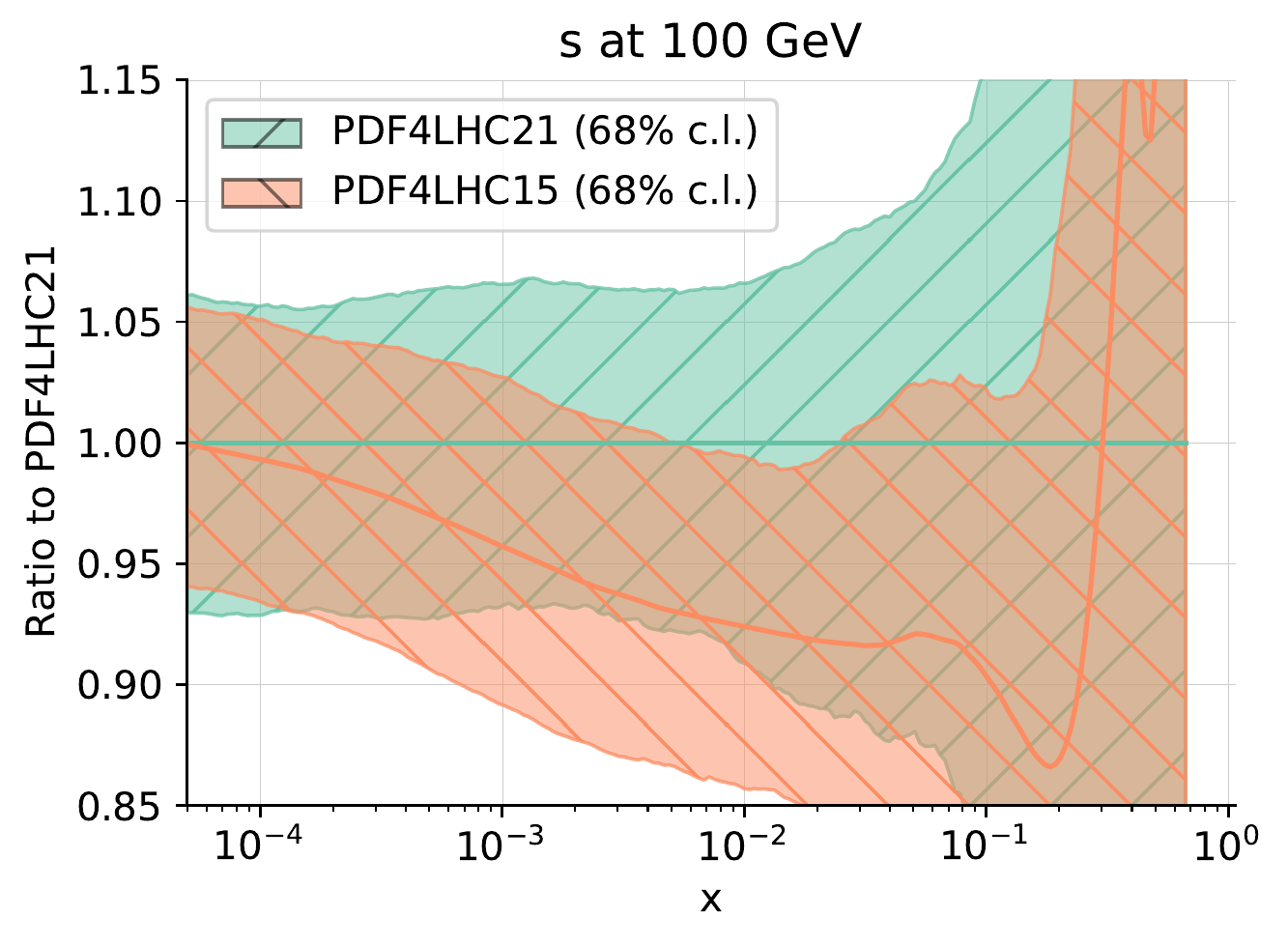}
\includegraphics[width=0.49\textwidth]{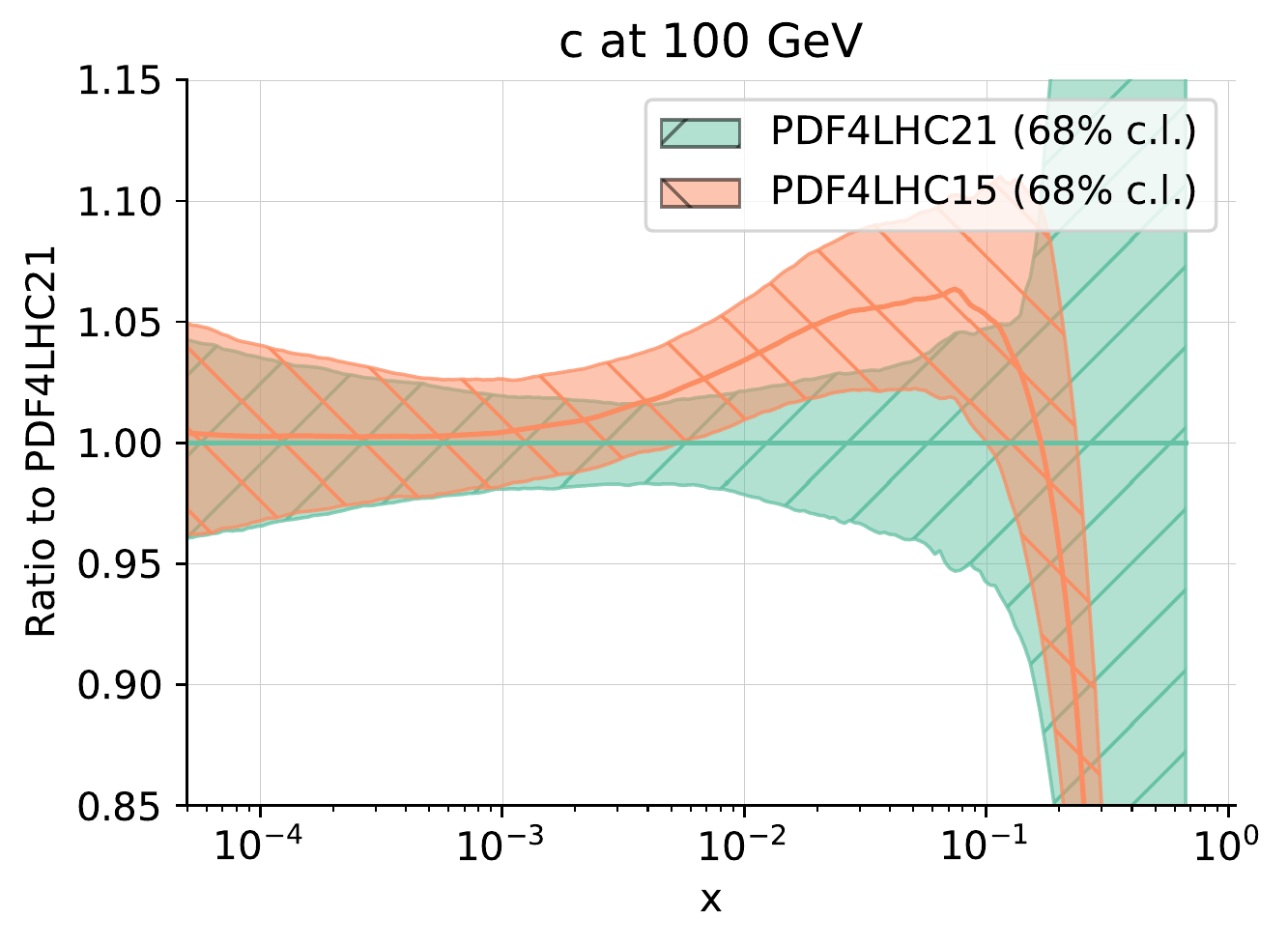}\\
\caption{\small Comparison between the baseline PDF4LHC15 and PDF4LHC21 ensembles, normalised to the central
  value of the latter, at $Q=100$ GeV. We show the results for the gluon and the up, down, anti-down, strange, and charm
  quark PDFs.
}
\label{fig:pdf4lhc21_vs_pdf4lhc15_pdfs}
\end{figure}

\begin{figure}[!t]
\centering
\includegraphics[width=0.49\textwidth]{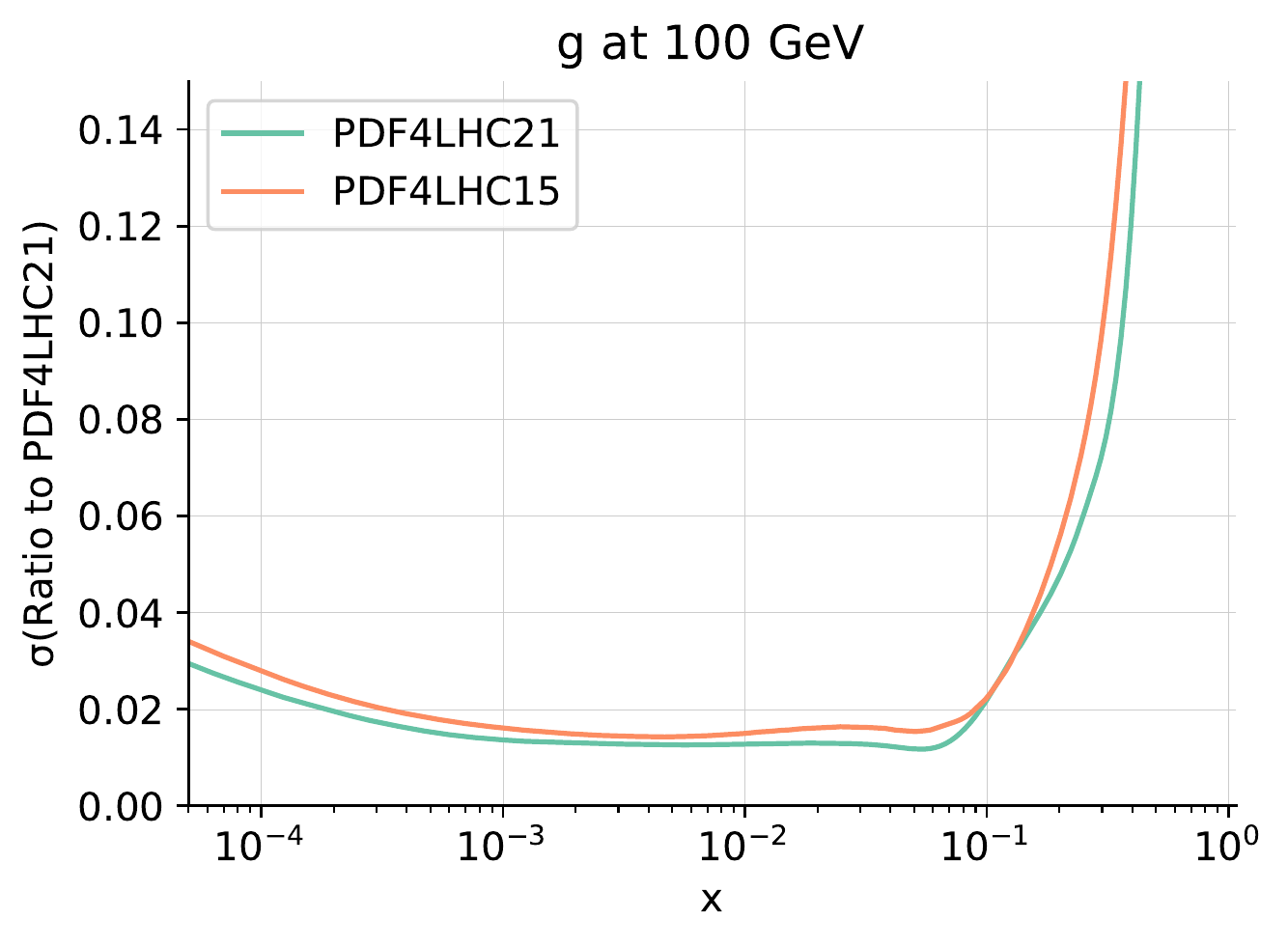}
\includegraphics[width=0.49\textwidth]{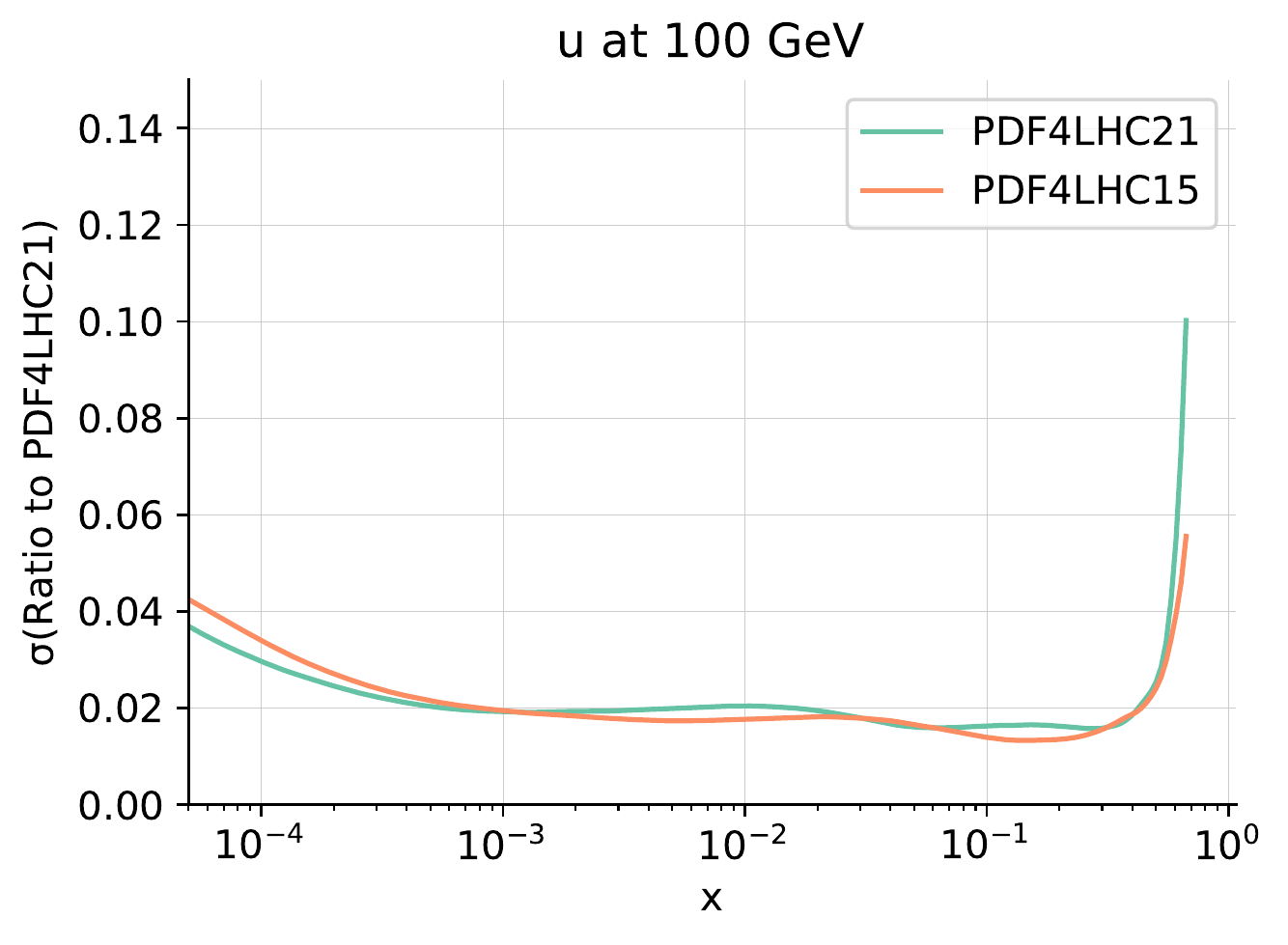}\\
\includegraphics[width=0.49\textwidth]{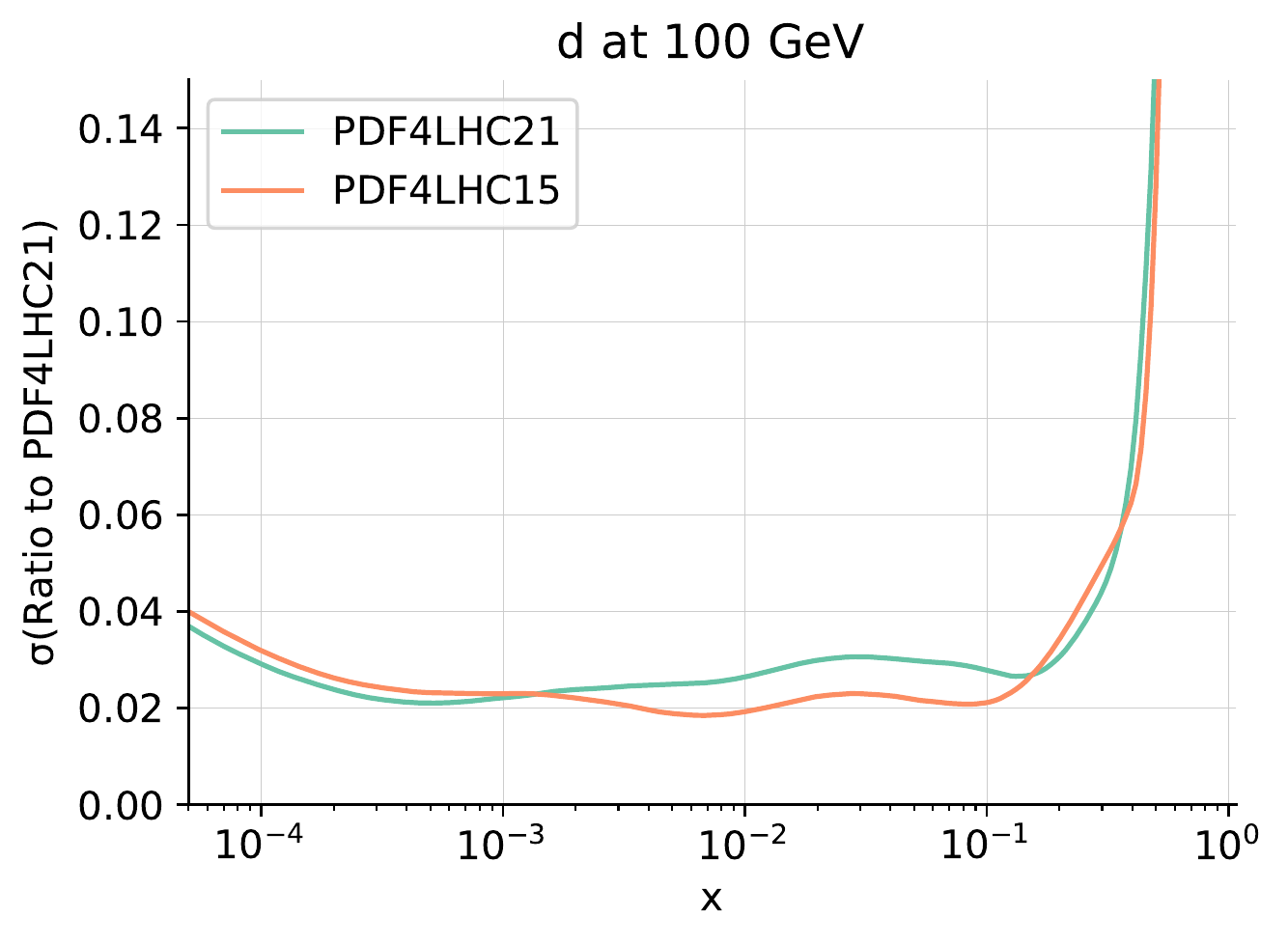}
\includegraphics[width=0.49\textwidth]{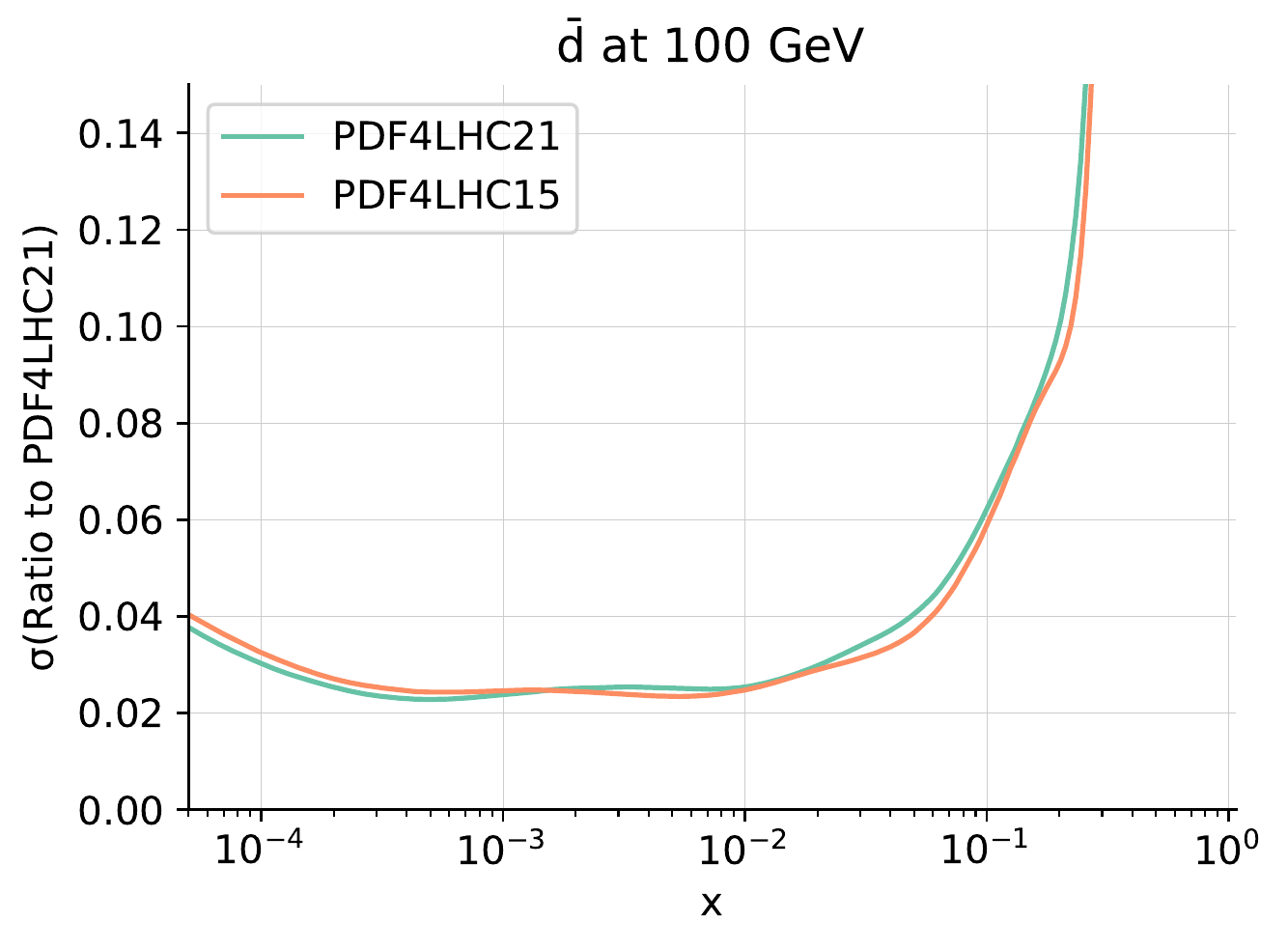}\\
\includegraphics[width=0.49\textwidth]{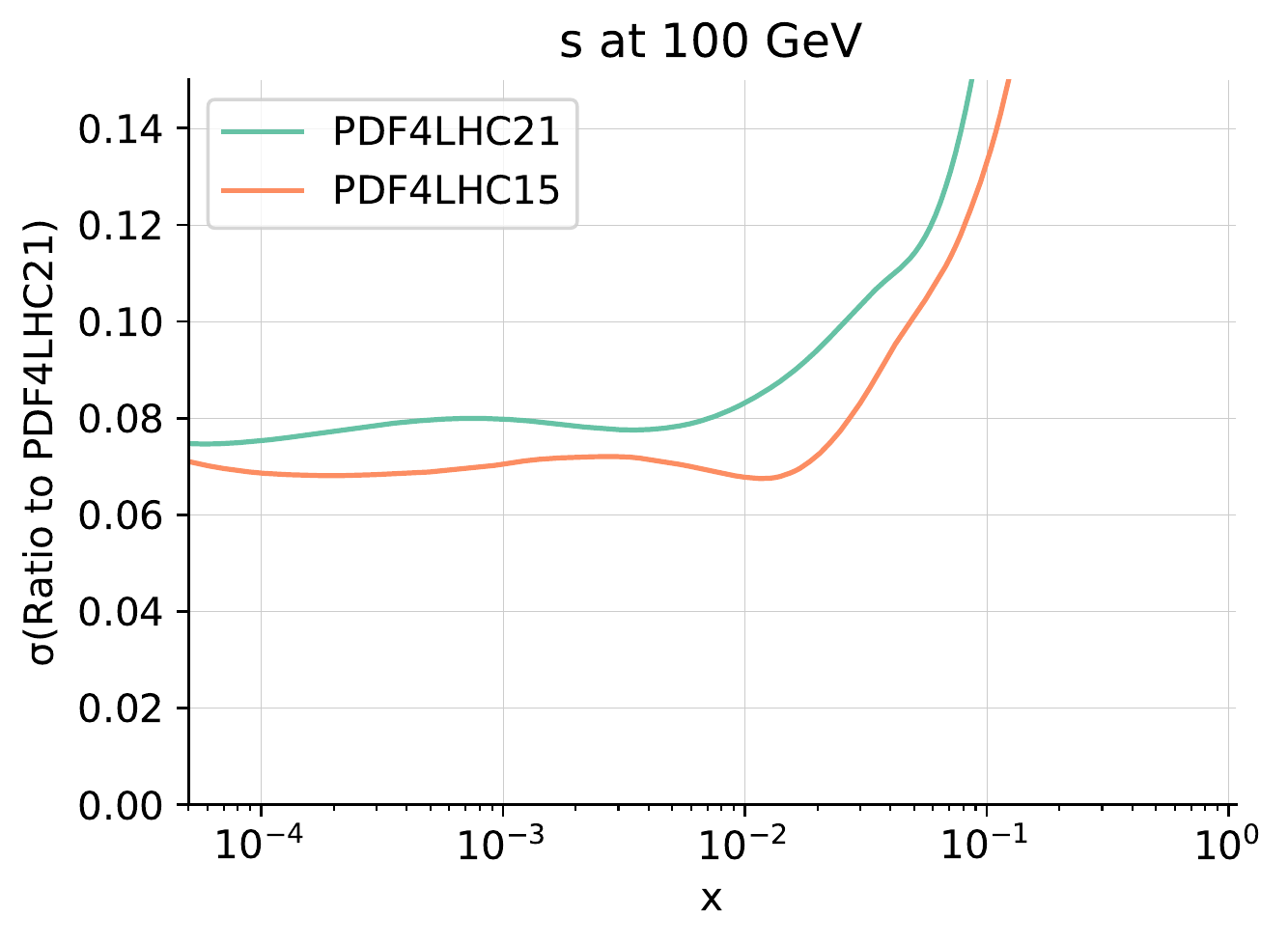}
\includegraphics[width=0.49\textwidth]{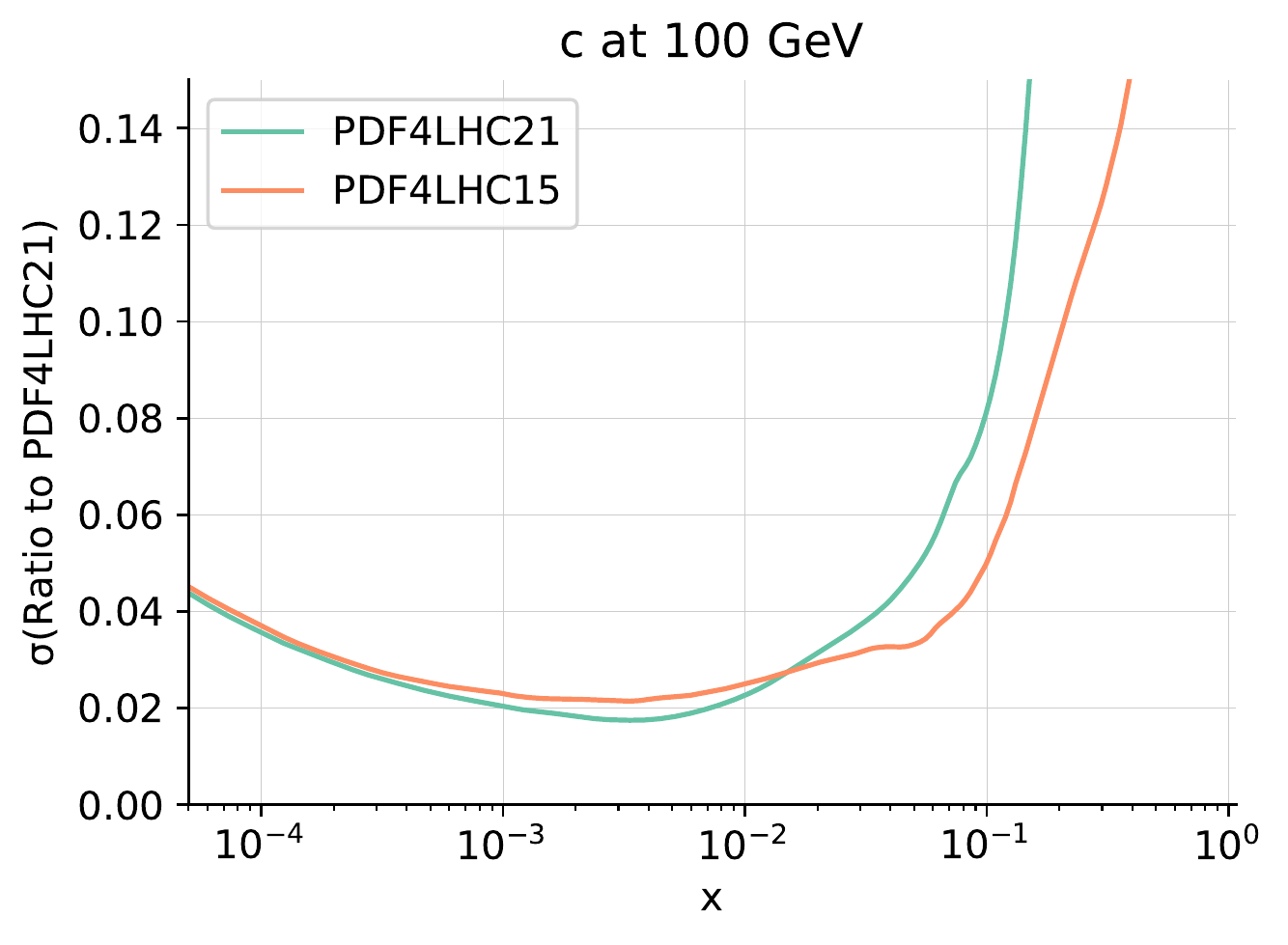}\\
\caption{\small Same as Fig.~\ref{fig:pdf4lhc21_vs_pdf4lhc15_pdfs} now displaying
  the corresponding 68\% CL relative PDF uncertainties.
}
\label{fig:pdf4lhc21_vs_pdf4lhc15_errors}
\end{figure}

First of all, Fig.~\ref{fig:pdf4lhc21_vs_pdf4lhc15_pdfs} displays a comparison between PDF4LHC15 and PDF4LHC21, normalised to the central value of the latter, at $Q=100$ GeV. We show the results for the gluon and the up, down, anti-down, strange, and charm quark PDFs.
Fig.~\ref{fig:pdf4lhc21_vs_pdf4lhc15_errors} displays the corresponding results for the relative 68\% CL PDF uncertainties.

Inspection of Figs.~\ref{fig:pdf4lhc21_vs_pdf4lhc15_pdfs} and~\ref{fig:pdf4lhc21_vs_pdf4lhc15_errors} reveals that the PDF4LHC15 and PDF4LHC21 ensembles are consistent for all flavours and values of $x$, since their 68\% CL uncertainty bands always overlap.
The agreement between the two combined sets is especially remarkable for the up and down quark, the down antiquark, and for the gluon  PDF for $x\lsim 0.1$, where the previous and the current combinations are extremely similar to each other. Some differences are observed in the large-$x$ region, but in all cases the shift in the central value is contained within the error band of PDF4LHC15. Given that, as discussed in Sect.~\ref{sec:combination_inputs}, many changes have happened between the 2015 and 2021 combinations in terms of the fitted datasets, theory calculations, and fitting methodologies, this agreement confirms that the PDF4LHC15 uncertainty estimate was realistic.

\begin{figure}[!t]
\centering
\includegraphics[width=0.49\textwidth]{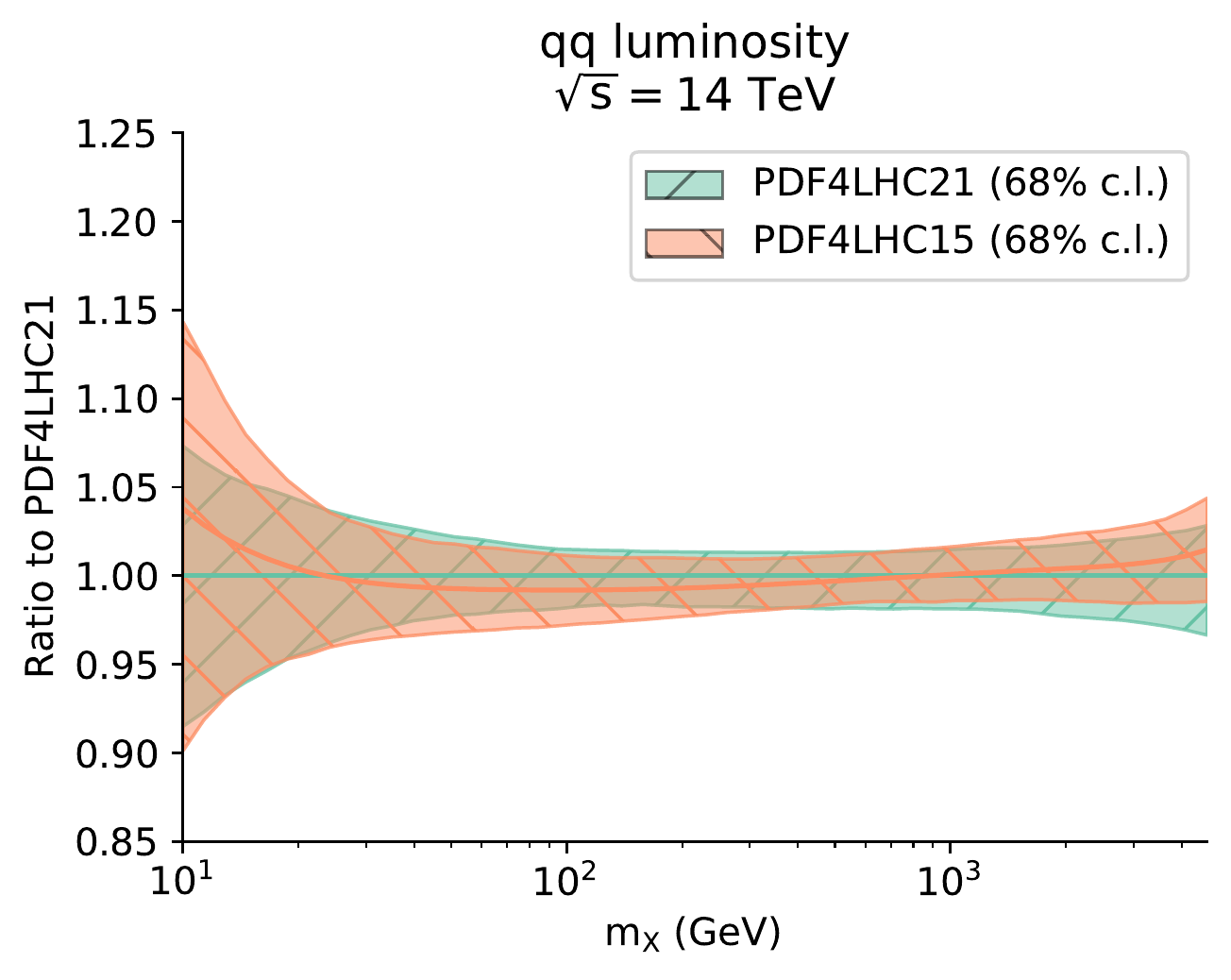}
\includegraphics[width=0.49\textwidth]{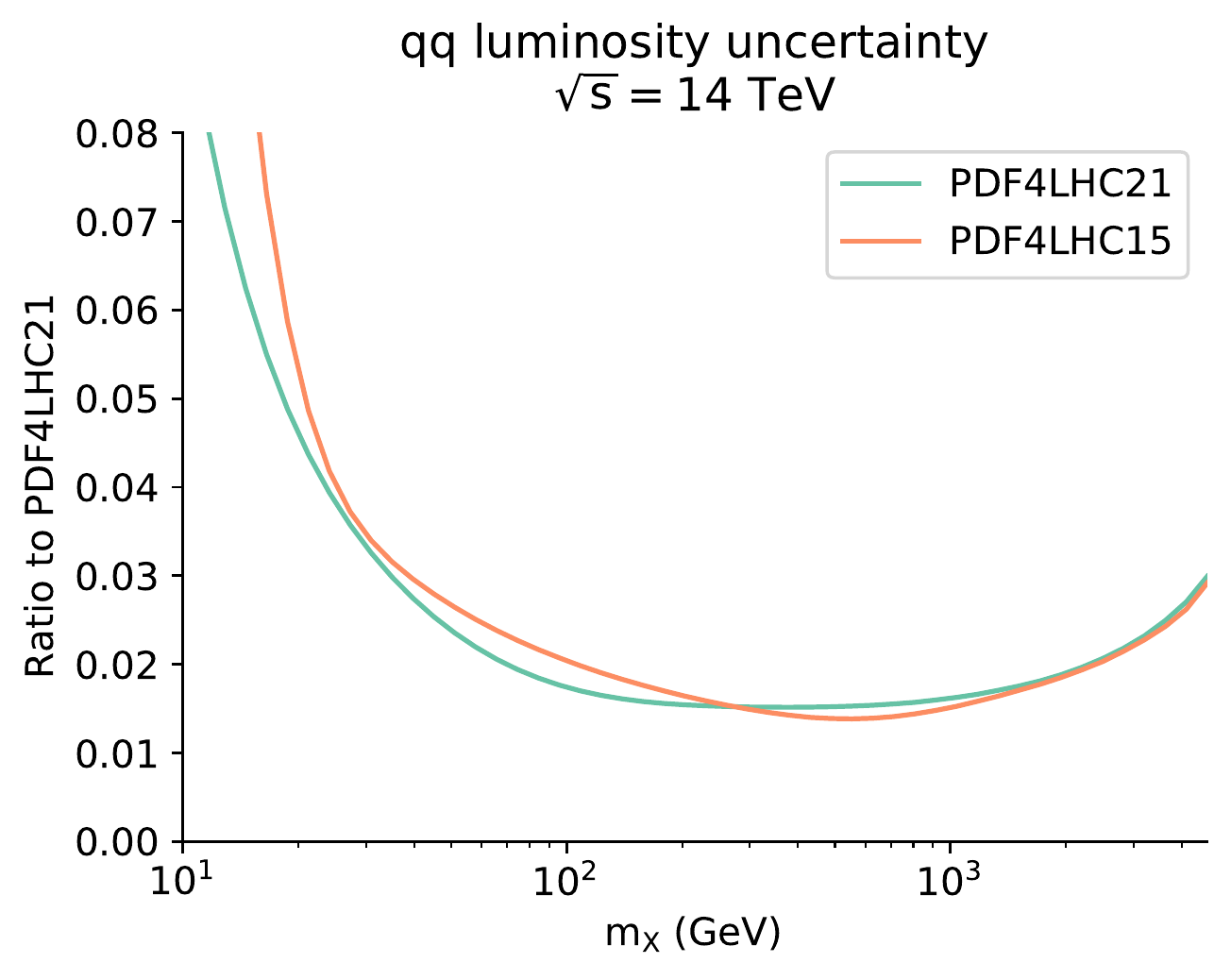}\\
\includegraphics[width=0.49\textwidth]{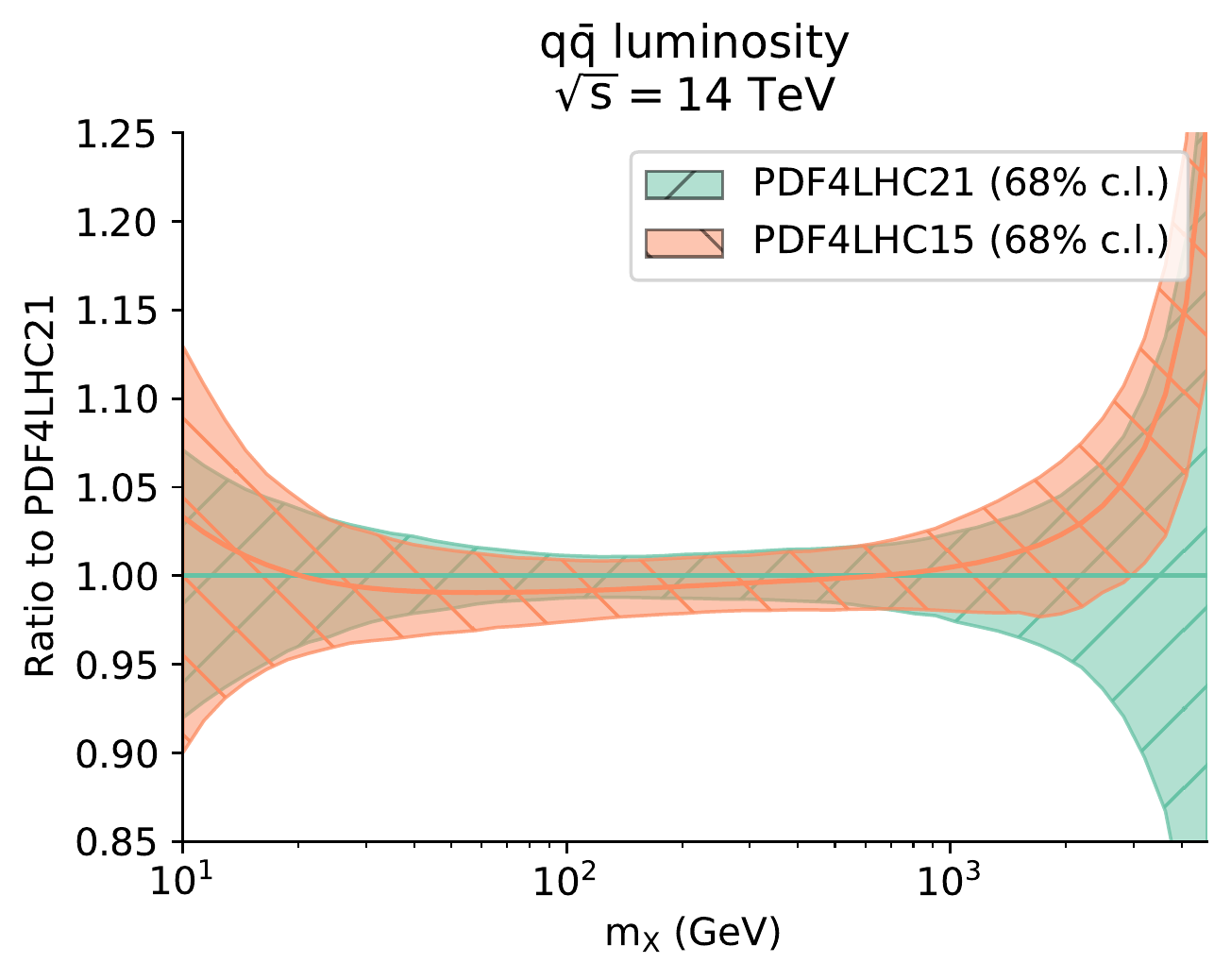}
\includegraphics[width=0.49\textwidth]{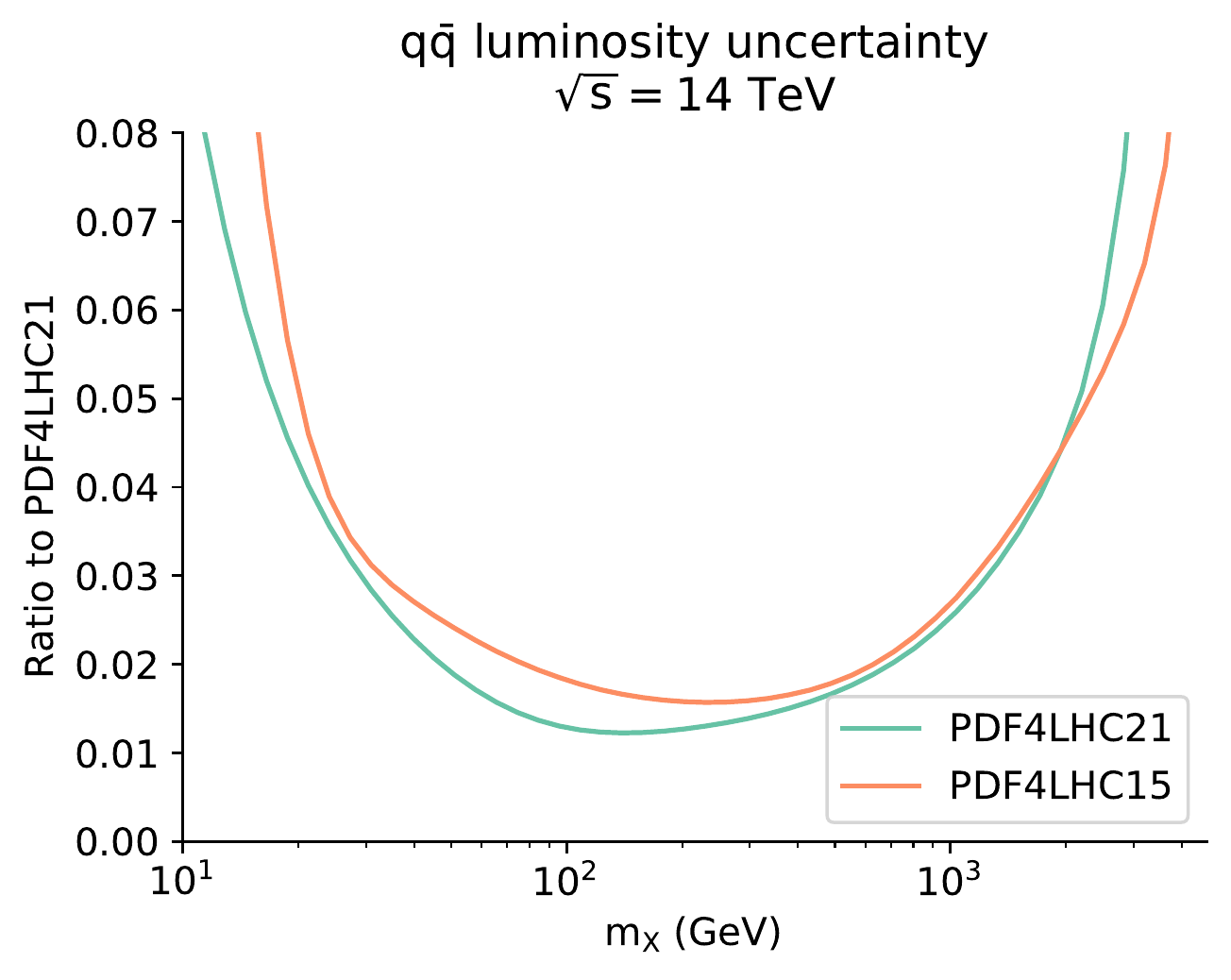}\\
\includegraphics[width=0.49\textwidth]{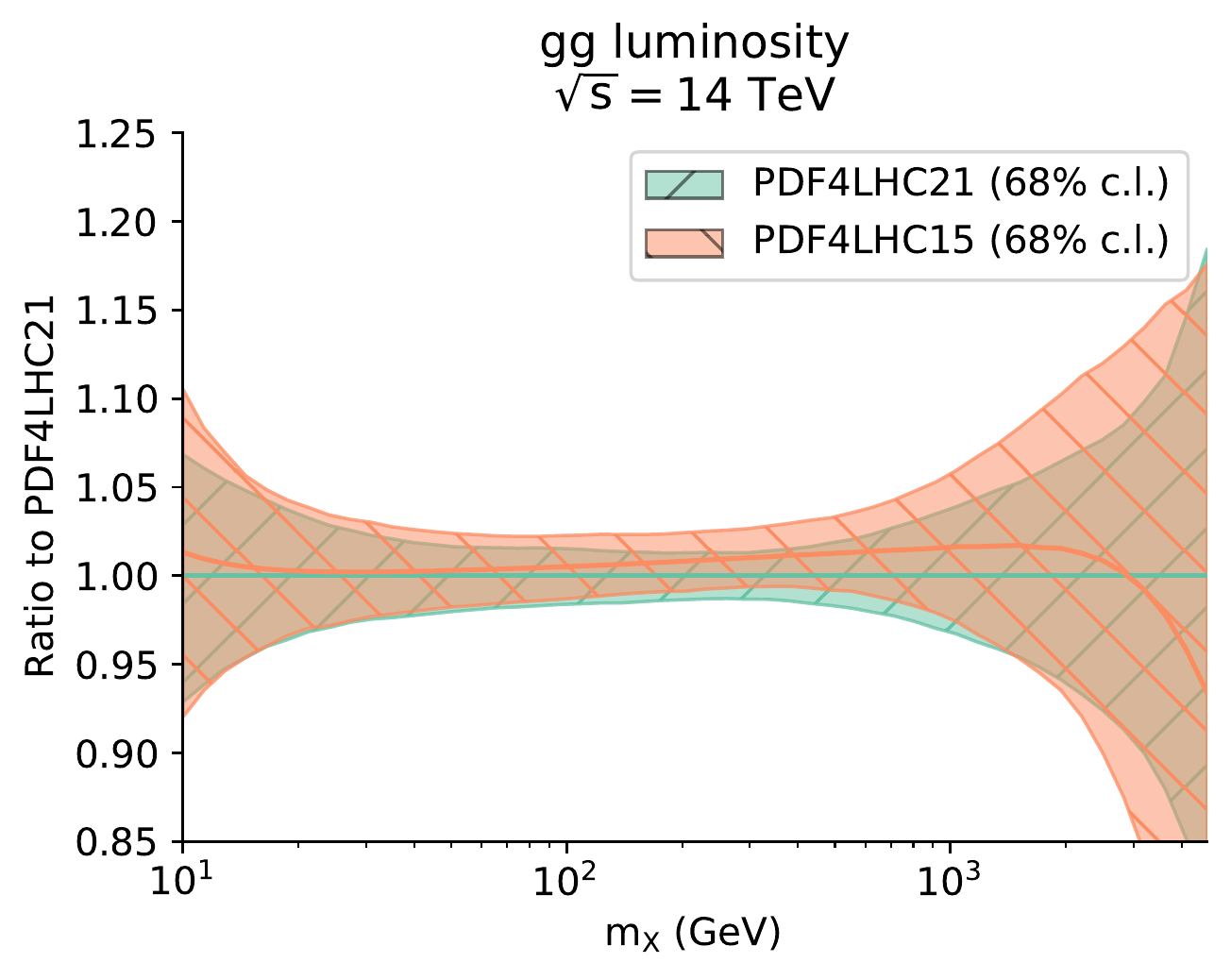}
\includegraphics[width=0.49\textwidth]{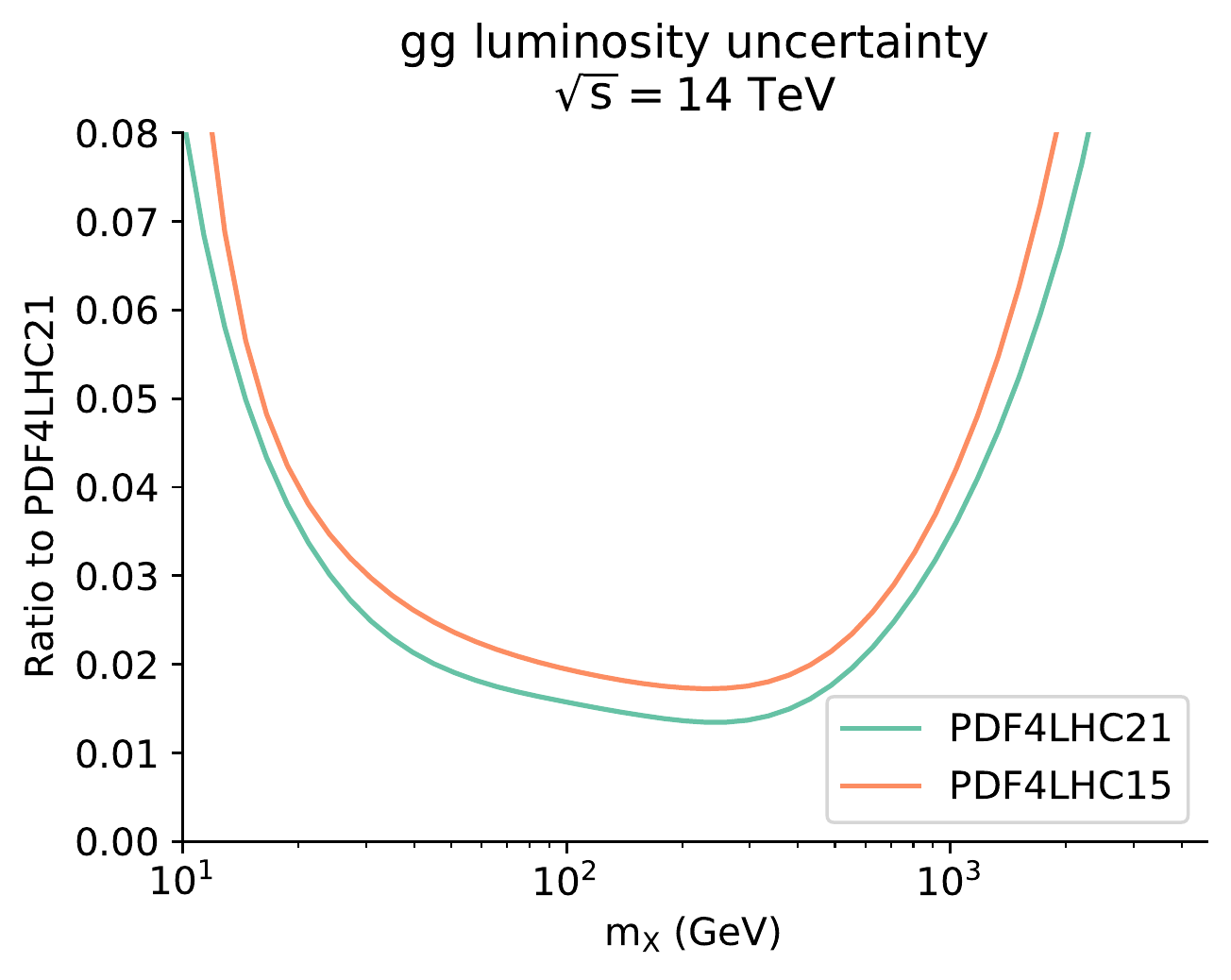}\\
\caption{\small Comparison of the partonic luminosities at $\sqrt{s}=14$ TeV between the baseline PDF4LHC15 and PDF4LHC21 ensembles. Results are shown for the quark-quark, quark-antiquark, and gluon-gluon luminosities as a function of the final state invariant mass $m_X$, and normalised to the central value of the PDF4LHC21 prediction. The right panels display the corresponding 68\% CL relative uncertainties.
}
\label{fig:pdf4lhc21_vs_pdf4lhc15}
\end{figure}

The main differences between PDF4LHC15 and PDF4LHC21 appear to be localised, as expected from the discussion in Sect.~\ref{subsec:inputs_comparison}, to the large-$x$ gluon, the strange quark, and the charm quark. The latter is explained by \NNprime now adopting the fitted charm framework, whereas in PDF4LHC15 the three input sets were all based on a perturbative charm PDF. In the case of the total strangeness, the new combination results in a larger strangeness PDF for $x\gsim 10^{-3}$, as in the input MSHT20 and \NNprime ensembles.
For the gluon, we find that PDF4LHC21 is suppressed in comparison to the previous PDF4LHC15 combination in the region of $0.1 \lsim x \lsim 0.4$.
As shown in Fig.~\ref{fig:pdf4lhc21_vs_pdf4lhc15_errors}, the error band is also larger for the $d$ quark in PDF4LHC21. For $d$ quark PDFs, both \NNprime and MSHT20 yield larger error bands than \CTprime for $x$ around $0.1$, cf. Fig.~\ref{fig:pdf4lhc21_vs_inputs_uncs}, while the error bands of both $\bar{d}$ and $s$ PDFs are larger in \CTprime. The sources of these differences, especially for the gluon and strangeness PDFs, were examined with complementary methods summarized in Apps.~\ref{app:specifics} and \ref{app:l2_sensitivity}.

The PDF uncertainties in general are similar between the new and the previous combination, with some differences worth pointing out. On the one hand, reduction of some PDF uncertainties in the PDF4LHC21 combination is observed, namely for the gluon in the full range of $x$ and for the up and down quarks for $x\lsim 10^{-3}$. On the other hand, for other PDF flavours, the uncertainties are increased in comparison to the PDF4LHC15 baseline. This occurs where the agreement between the three constituent sets has worsened in PDF4LHC21, or more flexible parametric forms have been introduced. One can observe this effect, in particular for the strangeness and charm PDFs for $x\gsim 10^{-2}$ as well as for the down quark PDF in the region $10^{-3}\lsim x \lsim 0.1$.

More relevant in the context of applications to LHC processes is the corresponding comparison of the parton luminosities displayed in Fig.~\ref{fig:pdf4lhc21_vs_pdf4lhc15}
for $\sqrt{s}=14$ TeV. Results are shown for the quark-quark, quark-antiquark, and gluon-gluon luminosities as a function of the final state invariant mass $m_X$, and are normalised to the central values of the PDF4LHC21 prediction. 
The shown range of $m_X$ corresponds to the kinematic coverage attained by present and future LHC measurements.

The central values of the PDF4LHC15 luminosities are always contained within the 68\% CL error band of PDF4LHC21. 
Particularly stable is the  behaviour of the gluon-gluon and gluon-quark luminosities
in the region with $m_X \lsim 1$ TeV, where the central values differ by less than 1\%. Similar considerations apply to the quark-quark luminosity. This stability has direct implications for LHC processes driven by these luminosity combinations, such as Higgs production in gluon fusion, as will be further discussed in Sect.~\ref{sec:phenomenology}. Perhaps the main differences between the two combinations can be observed for the quark-antiquark luminosity, where PDF4LHC21 overshoots PDF4LHC15 for $m_X\sim 100$ GeV, relevant for inclusive $W$ and $Z$ boson production, and then undershoots it for $m_X\gsim 1$ TeV, relevant for heavy BSM resonance production. Nevertheless, in all cases the shift in central values is contained within the 68\% CL uncertainties of the new combination.

Concerning the PDF uncertainties of the partonic luminosities, for the gluon-gluon case
the PDF4LHC21 expectation is systematically more precise in comparison to PDF4LHC15 for the whole kinematic range considered. This implies that predictions for LHC processes driven by the $gg$ luminosity, from charm and top quark pair production to Higgs production in gluon fusion, will benefit from the reduced uncertainties of the PDF4LHC21 combination. A similar improvement is observed for the quark-quark (quark-antiquark) PDF luminosity in the region $m_X\lsim 500~{\rm GeV}$~(1 TeV), relevant i.e. for on-peak electroweak gauge boson production. The only case in which the uncertainties associated to the luminosities based on PDF4LHC21 are markedly larger than those of its predecessor is for the quark-antiquark luminosity in the high-mass region, $m_X\gsim 2$ TeV.

\subsection{The large-$x$ behaviour of PDF4LHC21}
\label{sec:largex_pdf4lhc21}

At the end of Sect.~\ref{sec:hessian_reduction}, we pointed out that the vanishing behaviour of PDF4LHC21 sea PDFs on the verge of the experimentally constrained region at $x>0.4$ demands special attention to obtain physically acceptable predictions for a class of cross sections and their uncertainty ranges in very forward or ultra-heavy final-state production. Toward this goal, the combination should address the non-positivity of a fraction of the baseline Monte Carlo replicas in the large-$x$ (large-mass) extrapolation region. 
In this subsection, we assess this behaviour in more detail. For both the {\sc \small PDF4LHC21\_40} and  {\sc \small PDF4LHC21\_mc} ensembles, we develop  practical recommendations here, and then state them as step-by-step prescriptions in Sect.~\ref{subsec:uncer_prescr},
to obtain non-negative central predictions and uncertainty ranges under such circumstances.

\paragraph{The baseline PDF4LHC21 replicas at large $x$.}
Inspection of the probability distribution associated with PDF4LHC21 reveals that a fraction of its $N_{\rm rep}=900$  replicas may become negative in the large-$x$ extrapolation
region, where experimental constraints are very limited or non-existent.
There are two explanations for this behaviour.
To begin with, the native MC replicas of \NNprime satisfy the positivity of
a certain set of physical observables, but the PDFs themselves are not enforced
to be positive. In such unconstrained regions, the positivity of the PDFs themselves, rather than of the observable cross sections, is a delicate theoretical issue \cite{Candido:2020yat,Collins:2021vke} that is still under investigation.
Hence in the large-$x$ region
some baseline PDF parametrizations may become negative (within uncertainties).
The second reason is that, even for positive-definite Hessian native sets, an MC conversion can result in partially negative replicas in regions with scarce data.  
A log-normal sampling prescription to avoid such negative replicas in positive-definite Hessian sets was proposed in~\cite{Hou:2016sho} and used to produce the $N_{\rm rep}=300$ \CTprime replicas of  the present analysis. 
This prescription was not applied to the MSHT20 replicas, as will be explored below.
The question is then to which extent the negative PDF replicas affect
LHC cross-sections, and how to deal with this situation with an improved error evaluation prescription.

\begin{figure}[!t]
\centering
\includegraphics[height=0.7\textheight]{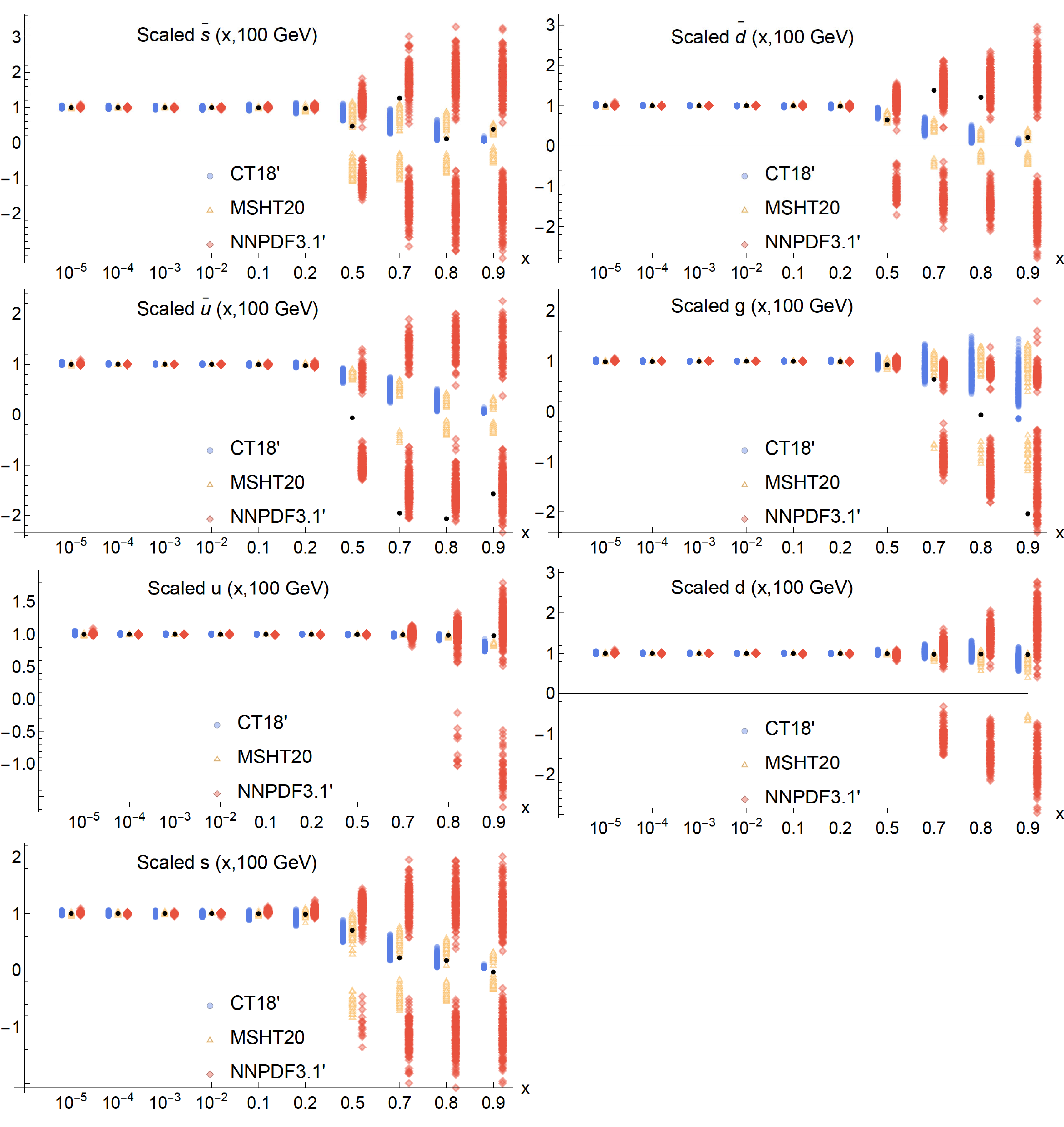}
\vspace{0.2cm}
\caption{\small Distributions of the $N_{\rm rep}=900$ replicas from \CTprime, MSHT20, and \NNprime
  ($N_{\rm rep}=300$ from each group) that constitute the PDF4LHC21 combination
  at $Q=100$ GeV as a function of $x$.
  The PDFs $f$ are plotted as $|f/f_0|^{r}\cdot \mbox{sign}(f)$, where $f_0$ is the positive-definite central PDF of the Hessian
  {\sc \small PDF4LHC21\_40} representation described in Sect.~\ref{sec:hessian_reduction} and $r=0.2$, in
order to better highlight the large-$x$ behaviour of the PDF replicas. The black points are explained in the text.}
\label{fig:pdf4lhc21_900_prior_scaled}
\end{figure}

In order to highlight the large-$x$ behaviour of the PDF4LHC21 replicas,
Fig.~\ref{fig:pdf4lhc21_900_prior_scaled} displays
the distributions of the $N_{\rm rep}=900$ replicas from \CTprime, MSHT20, and \NNprime
  ($N_{\rm rep}=300$ from each group) that constitute the PDF4LHC21 combination
  at $Q=100$ GeV.
  The PDFs $f$ are plotted as $|f/f_0|^{r}\cdot \mbox{sign}(f)$ versus $x$, where $f_0$ is the 
  positive-definite central PDF of the Hessian
  {\sc \small PDF4LHC21\_40} representation, and $r=0.2$.
   Notice that the differences at the largest $x$ are magnified with this plotting format. 
  A good agreement among all three $N_{\rm rep}=300$ replica sets is achieved overall,
  as already mentioned in Sects.~\ref{sec:combination_inputs} and~\ref{sec:generation_mc_replicas}.
  The uncertainties at low-$x$ values show a very good compatibility of the sets.
  However, larger spans are noticed at larger $x$ values, where little to no data are available. While most of the Hessian-originated replicas are positive in the whole $x$ range (\CTprime in blue and MSHT20 in yellow), the \NNprime replicas (in red) span negative values of the PDF space, reflecting the large uncertainties shown in Figs.~\ref{fig:pdf4lhc21_vs_pdf4lhc15_pdfs} and~\ref{fig:pdf4lhc21_vs_pdf4lhc15}.
  Even a fraction of the sea and gluon replica PDFs generated from the 
  MSHT20 Hessian fit is weakly negative in this large-$x$ region, as mentioned above. 
  
\begin{figure}[t]
\centering
\includegraphics[width=0.49\textwidth]{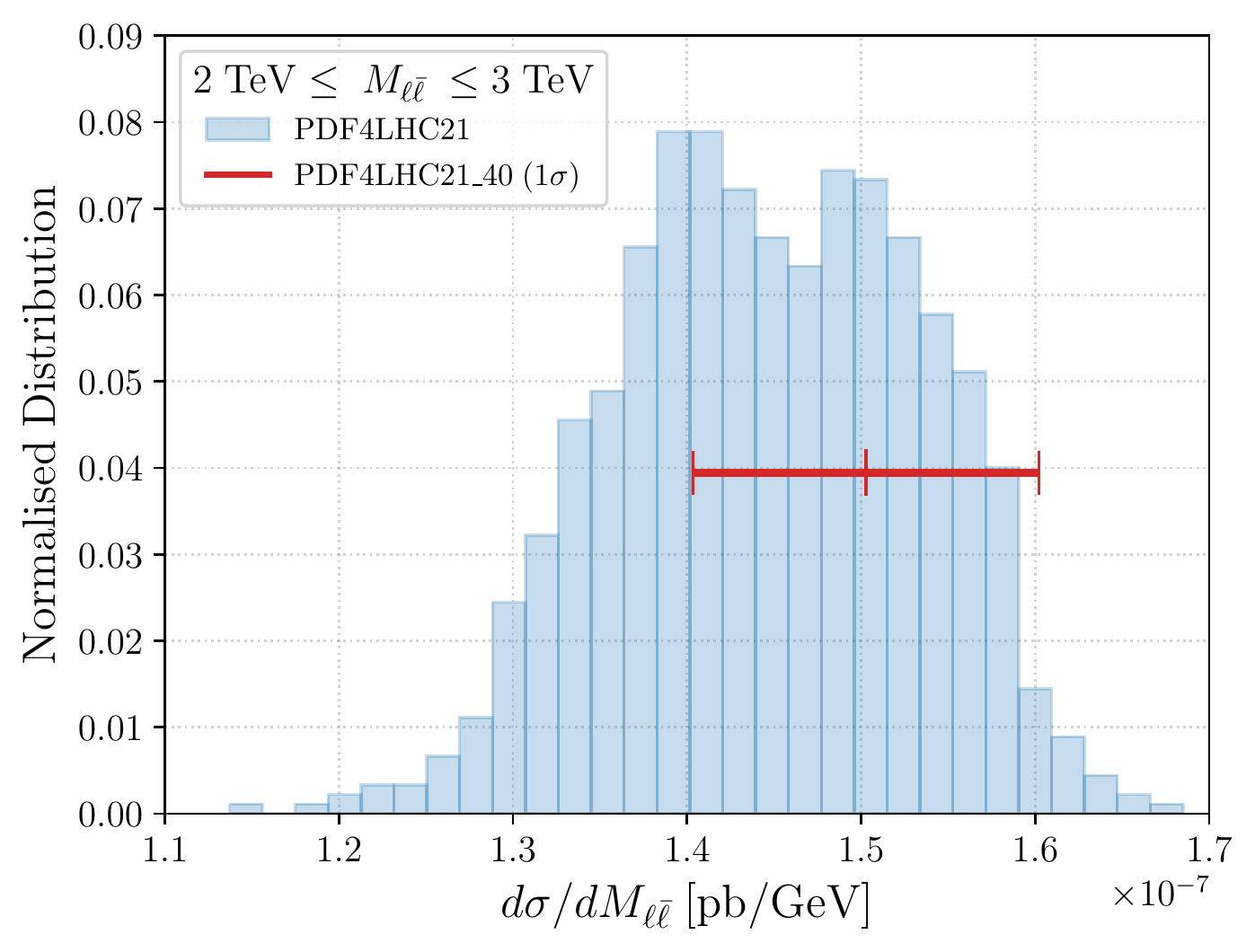}
\includegraphics[width=0.49\textwidth]{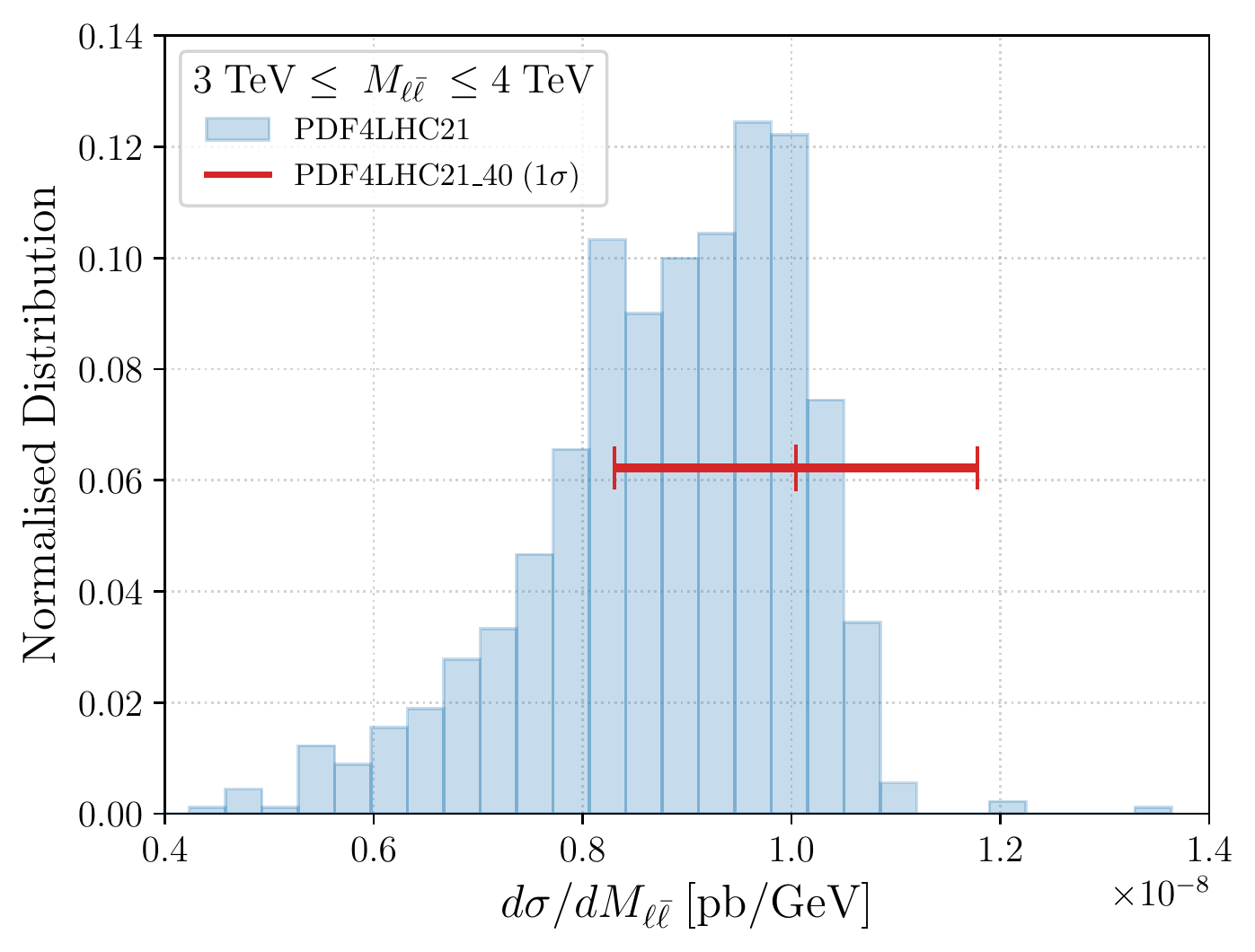}
\includegraphics[width=0.49\textwidth]{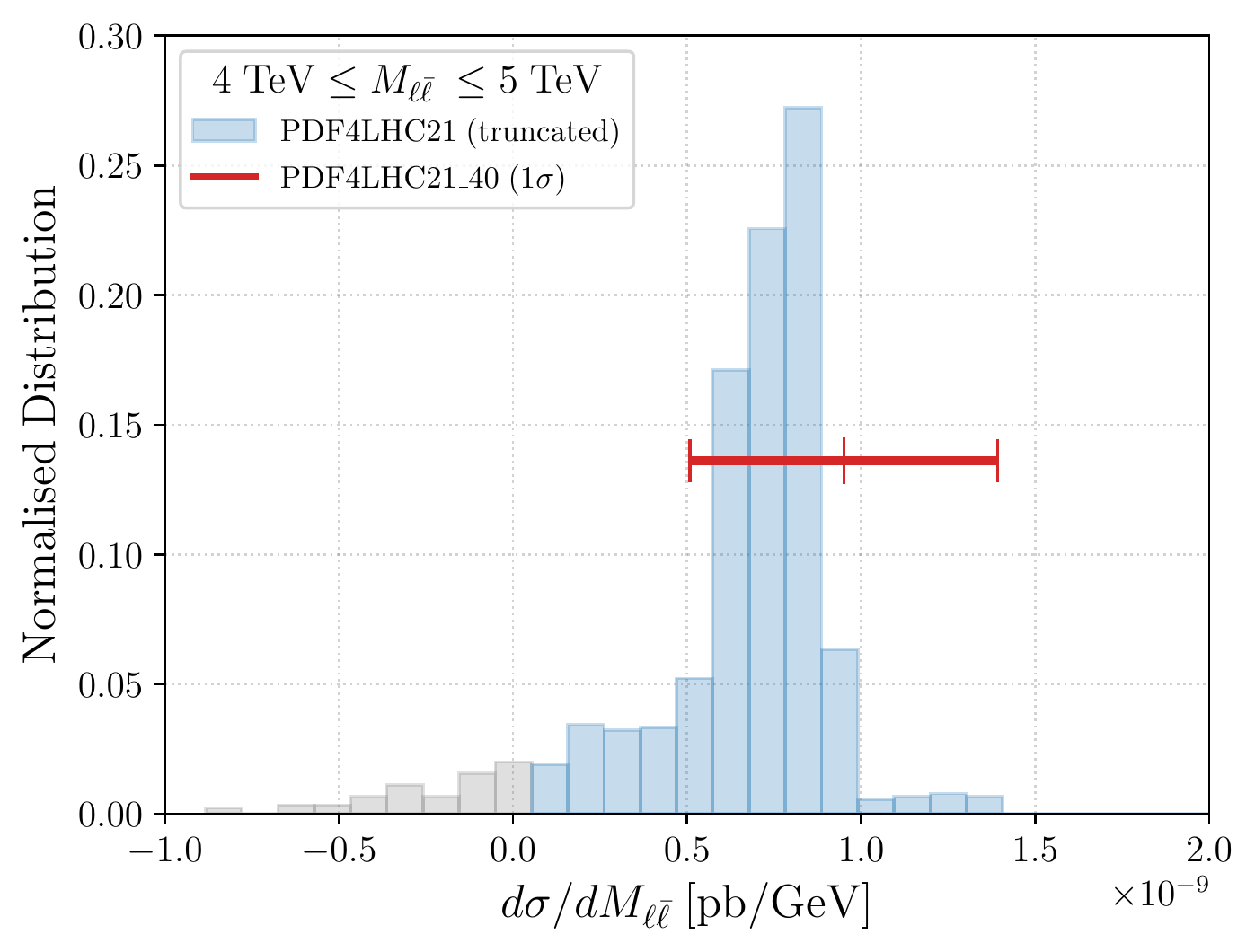}
\includegraphics[width=0.49\textwidth]{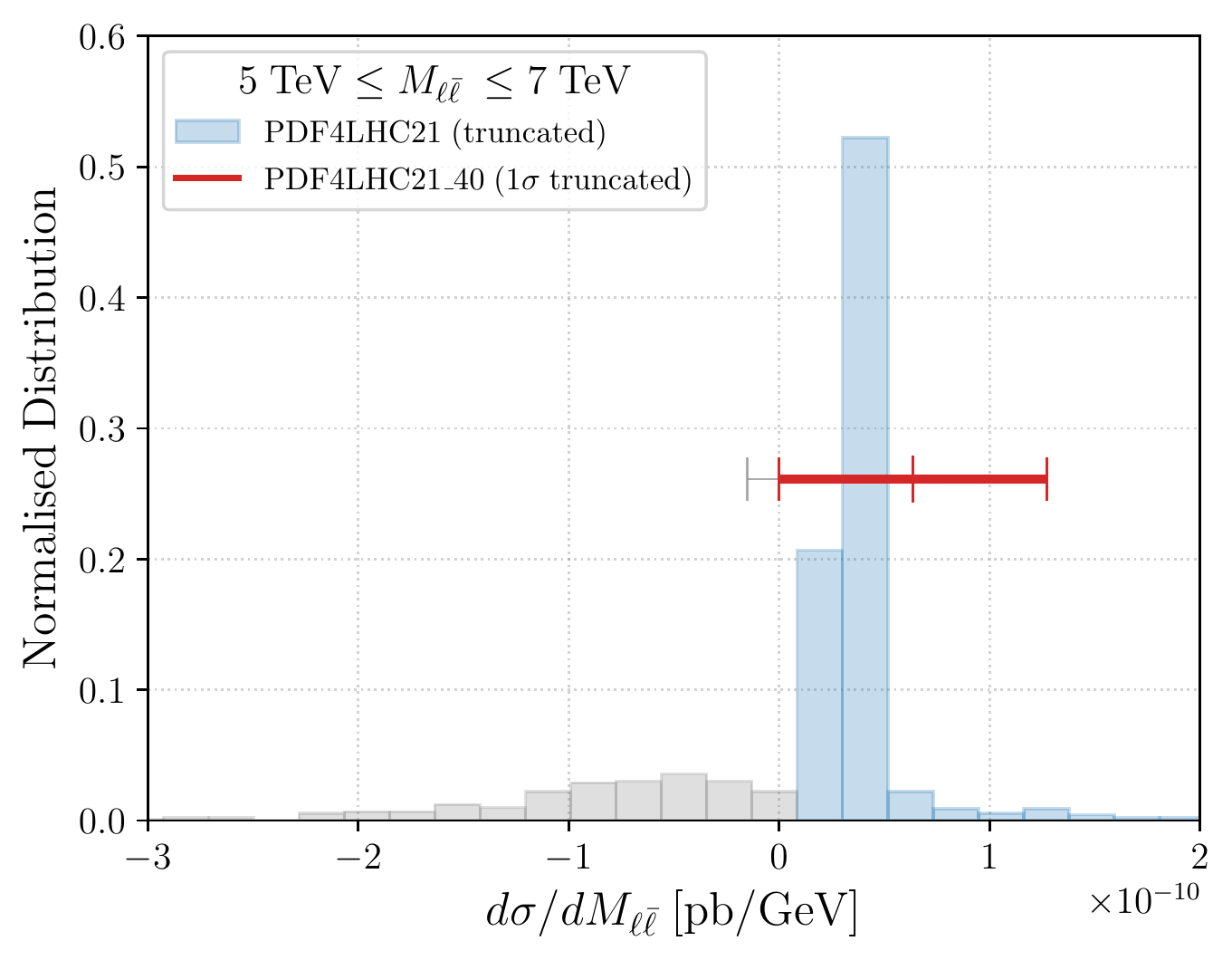}
\caption{\small Normalised invariant mass distributions of the $N_{\rm rep}=900$ Monte Carlo replicas in four invariant mass intervals of the Drell-Yan process at $\sqrt{s}=14~\mathrm{TeV}$. Light-blue bins are for positive cross sections. Replicas with negative cross sections (in greyed bins) have their cross-sections set to zero when evaluating MC uncertainties by following the positivity prescription from Sect.~\ref{sec:uncer_prescr_MC}. The red error bars indicate 68\% CL intervals obtained with the {\sc \small PDF4LHC21\_40} Hessian set. In the fourth $M_{\ell \bar \ell}$ interval, the lower boundary of the Hessian uncertainty is cut off at $d\sigma/dM_{\ell \bar\ell}=0$ in accordance with the positivity prescription for the Hessian case in Sect.~\ref{sec:uncer_prescr_Hessian}.
}
\label{fig:histo_hmdy_nc}
\end{figure}

In the Hessian representations
of PDF4LHC21 described in Sect.~\ref{sec:hessian_reduction},
one provides to the user a {\it central} PDF and eigenvector error sets. It has been customary, although not mandatory, to choose the central PDF to be the mean of all 900 replicas. Other obvious alternatives include choosing the median or mode of 900 replicas when non-Gaussian features are present. 
However, this mean replica is not automatically positive-definite. Furthermore, in the specific realisation adopted for the 40-member Hessian set, one must also define an internal positive-definite {\it reference} PDF for constructing the eigenvector sets, cf. Sect.~\ref{sec:hessian_reduction}. This reference PDF is constructed in App.~\ref{app:tools} and employed as $f_0 > 0$ in Fig.~\ref{fig:pdf4lhc21_900_prior_scaled}  to visualise the large-$x$ differences by plotting $|f/f_0|^{r}\cdot \mbox{sign}(f)$. Now, the black points in Fig.~\ref{fig:pdf4lhc21_900_prior_scaled} correspond to $f$ equal to the mean replica of the baseline ensemble. One can see from the black points that the baseline mean and Hessian reference coincide ($f/f_0=1$) away from the extreme large-$x$ region. The figure also confirms that the input mean PDFs can be negative ($f/f_0 < 0$ for the respective black points) at the largest $x$, especially for $\bar u$ at $x\gsim 0.5$ and $g$ at $x\gsim 0.8$. 

In the default Hessian ensemble, {\sc \small PDF4LHC21\_40}, we choose the central PDF to coincide with the so-described reference PDF. Such central {\sc \small PDF4LHC21\_40} replica coincides with the baseline mean replica in the well-constrained regions, is compatible with the mean replica within the uncertainties everywhere, and is non-negative at large $x$.
This approach naturally leads to positive-definite central cross sections, while still allowing the PDF uncertainties to cover zero cross sections when warranted. In addition to this recommended ensemble, we also provide a 40-member set {\sc \small PDF4LHC21\_40\_nopos} for targeted usage, in which the central PDF coincides with the (generally sign-indefinite) mean replica of the baseline ensemble, while the reference PDFs remain positive-definite internally to reproduce well the uncertainty bands. 

The Monte Carlo
representation, {\sc \small PDF4LHC21\_mc}, is not modified by this procedure. A suitable prescription 
to deal with negative cross-sections in the Monte-Carlo approach is provided in Eqns.~(\ref{eq:presc_pos_1}) and (\ref{eq:presc_pos_last}).

\paragraph{Implications for the LHC high-mass searches.}
In order to illustrate to which extent the large-$x$ behaviour of PDF4LHC21 may
affect LHC phenomenology at highest invariant masses, we consider a representative process of production of high-mass Drell-Yan lepton pairs at $\sqrt{s}=14~\mathrm{TeV}$.
We compute its cross sections using \texttt{MadGraph5$\_$aMC@NLO}~\cite{Alwall:2014hca, Frederix:2018nkq} interfaced with
\texttt{PineAPPL}~\cite{Carrazza:2020gss, christopher_schwan_2022_5846421}, with realistic
acceptance and selection cuts described in Sect.~\ref{sec:phenomenology}.
Figure~\ref{fig:histo_hmdy_nc} displays the distribution of the {\sc \small PDF4LHC21} set ($N_{\rm rep}=900$ Monte Carlo replicas)
for four intervals of high-mass Drell-Yan production with dilepton invariant masses between $M_{\ell \bar{\ell}}= 2 $ TeV and $M_{\ell \bar{\ell}}= 7 $ TeV.
The number of $M_{\ell \bar \ell}$ bins is the same in each plot, and the height of the bin indicates the fraction of replicas that fall into it. The bins with positive (negative) cross sections are indicated with the light-blue (grey) colors. 

For the first two invariant mass bins covering the $M_{\ell \bar{\ell}} \lsim 4$ TeV region,
all $N_{\rm rep}=900$ replicas of PDF4LHC21 lead to positive cross-sections.
In the higher-mass bins, a small fraction of replicas becomes negative,
around 4\%~(10\%) for $4~{\rm TeV} \le M_{\ell \bar{\ell}} \le 5~{\rm TeV}$
($5~{\rm TeV} \le M_{\ell \bar{\ell}} \le 7~{\rm TeV}$).
We note that this kinematic region is at the upper end of the LHC reach, and no measurements
of Standard Model cross-sections are available there.
Furthermore, even for these extreme kinematics, the bulk of the distribution, including
the median, remains positive. Nevertheless, such negative contributions may slightly reduce the net-positive total cross sections integrated over the whole $M_{\ell \bar{\ell}}$ range.
Altogether, negativity of some PDF4LHC21 replicas
in the extreme kinematic regions does not prevent the reliable use of PDF4LHC21 for LHC phenomenology, although
it requires a minor modification of the error evaluation prescription, as will be given in Sect.~\ref{sec:uncer_prescr_MC}. Essentially the effect of this prescription will be to set all negative cross-sections to zero and thereby move all the grey bins in Fig.~\ref{sec:uncer_prescr_MC} into the zero cross-section bin. This leaves the median unaffected (provided it is originally positive) but shifts the lower limit of the 68\% CL interval to zero in the cases where it extends to negative values.

In the same figure, we show the central 68\% CL intervals obtained with the Hessian set {\sc\small PDF4LHC21\_40}. They have been evaluated following the Hessian master formula, Eq.~(\ref{eq:masterformula_hessian}), 
and reproduce well the corresponding intervals of the baseline ensemble. If an uncertainty interval evaluated with the Hessian master formula extends to the negative cross sections, as seen in the bottom right plot, the positivity condition in Sect.~\ref{sec:uncer_prescr_Hessian} boils down to simply setting the lower limit of the interval to zero.
%

\begin{table}[t]
  \centering
  \small
  \renewcommand{\arraystretch}{1.7}
  \begin{tabular}{|c|c|c|c|c|}
    \toprule
    \multirow{2}{*}{Interval}  &    \multicolumn{2}{c|}{Without positivity prescription}           &
    \multicolumn{2}{c|}{With MC positivity prescription}\\
     &  median   &  68\% CL interval & median   &  68\% CL interval \\
    \midrule
    $2~{\rm TeV}\le M_{\ell \bar{\ell}} \le 3~{\rm TeV}$  &  $1.45\cdot 10^{-7}$
    & $\lc 1.36\cdot 10^{-7},  1.54\cdot 10^{-7}\rc $  &  $1.45\cdot 10^{-7}$
    & $\lc 1.36\cdot 10^{-7},  1.54\cdot 10^{-7}\rc$\\
    \midrule
    $3~{\rm TeV}\le M_{\ell \bar{\ell}} \le 4~{\rm TeV}$  &  $9.02\cdot 10^{-9}$ 
    & $\lc  7.65\cdot 10^{-9} , 1.00\cdot 10^{-8}    \rc $ &  $9.02\cdot 10^{-9}$ 
    &  $\lc  7.65\cdot 10^{-9} , 1.00\cdot 10^{-8}    \rc $\\
     \midrule
    $4~{\rm TeV}\le M_{\ell \bar{\ell}} \le 5~{\rm TeV}$  &  $7.26\cdot 10^{-10}$  
    & $\lc  3.85\cdot 10^{-10} , 8.47\cdot 10^{-10}    \rc $  &   $7.26\cdot 10^{-10}$  
     & $\lc  3.85\cdot 10^{-10} , 8.47\cdot 10^{-10}    \rc $ \\
      \midrule
    $5~{\rm TeV}\le M_{\ell \bar{\ell}} \le 7~{\rm TeV}$  &   $3.34\cdot 10^{-11}$  
    & $\lc  -3.68\cdot 10^{-11} , 4.19\cdot 10^{-11}    \rc $  &   $3.34\cdot 10^{-11}$  
    & $\lc  0 , 4.19\cdot 10^{-11}   \rc$\\
    \bottomrule
  \end{tabular}
  \vspace{0.3cm}
  \caption{\small High-mass Drell-Yan cross-sections
    (in pb/GeV)
    for the four dilepton invariant mass intervals shown in Fig.~\ref{fig:histo_hmdy_nc}, obtained using the baseline PDF4LHC21 replicas from the shown histograms.
    For each interval, we indicate the 
    median and the 68\% CL interval computed either with all 900 cross sections or after selecting the positive-definite cross sections according to the prescription in Sect.~\ref{sec:uncer_prescr_MC}. 
     \label{tab:positivity}
  }
\end{table}

In the companion Table~\ref{tab:positivity}, we list the median cross sections and 68\% CL intervals obtained for the MC replicas in Fig.~\ref{fig:histo_hmdy_nc} without and with the MC positivity prescription.
One observes how for all $M_{\ell \bar{\ell}}$ bins considered the median value of the prediction is unaffected by the positivity prescription.
The 68\% CL ranges are also unaffected by the positivity prescription for  $M_{\ell \bar{\ell}}\le 5$ TeV. For the highest-mass bin with $5~{\rm TeV}\le M_{\ell \bar{\ell}} \le 7~{\rm TeV}$, the net effect of the positivity prescription on the 68\% CL interval is to set the lower limit to zero, without affecting the upper limit.

\section{Implications for LHC phenomenology}
\label{sec:phenomenology} 

In this section we present a first assessment
of the implications of the PDF4LHC21 combination for LHC
phenomenology.
All the results presented here have been computed for proton-proton collision 
at $\sqrt{s}=14~\mathrm{TeV}$.
The calculations are done with a combination of several codes, including \texttt{APPLgrid}~\cite{Carli:2010rw}, \texttt{MCFM}~\cite{Campbell:2019dru},
\texttt{Top++}~\cite{Czakon:2011xx}, \texttt{ggHiggs}~\cite{Ahmed:2016otz}, and
\texttt{MadGraph5$\_$aMC@NLO}~\cite{Alwall:2014hca, Frederix:2018nkq} interfaced with
\texttt{PineAPPL}~\cite{Carrazza:2020gss, christopher_schwan_2022_5846421}. For completeness, we present results both for inclusive and 
differential cross-sections.
The main aim of the studies in this section
is to, on the one hand, demonstrate that the compressed
MC and the Hessian representations of PDF4LHC21 accurately reproduce
the predictions from the baseline combination, and on the other hand,
to highlight the similarities and differences between
PDF4LHC21, its predecessor PDF4LHC15, and various individual
PDF sets.

\subsection{Inclusive cross-sections}
\label{sec:incl_xsec}

\begin{figure}[t]
\centering
\includegraphics[width=0.45\textwidth]{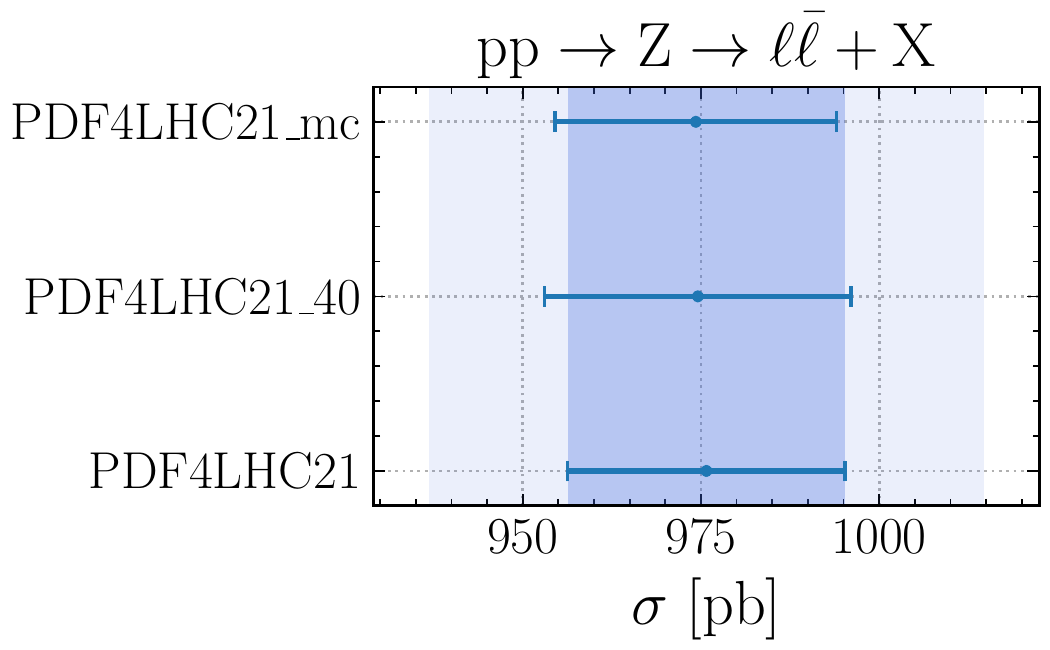}
\includegraphics[width=0.45\textwidth]{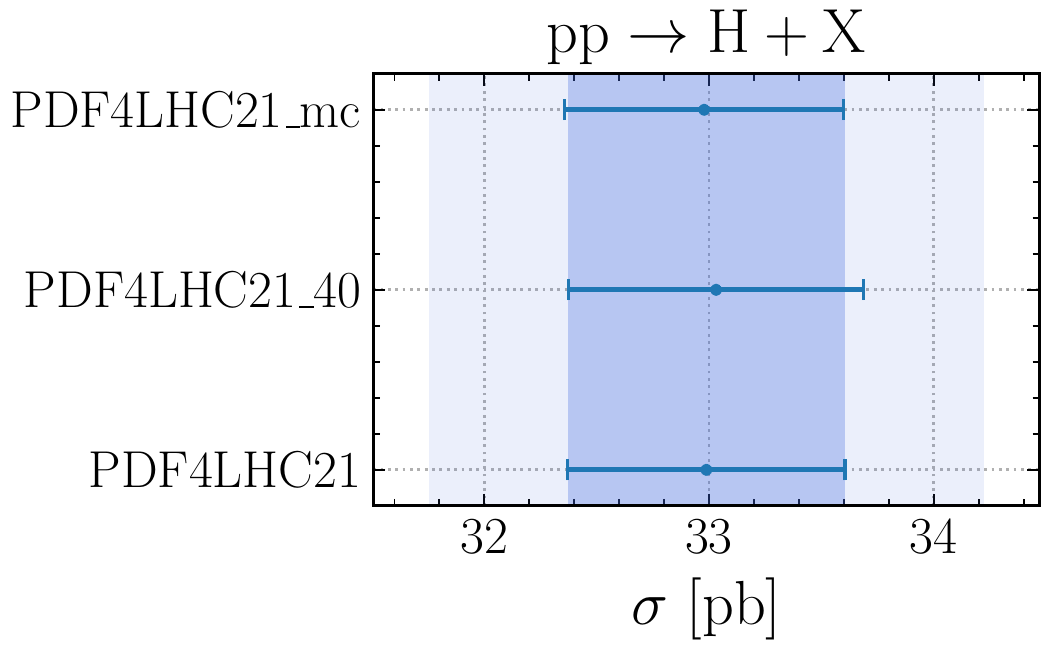}
\includegraphics[width=0.45\textwidth]{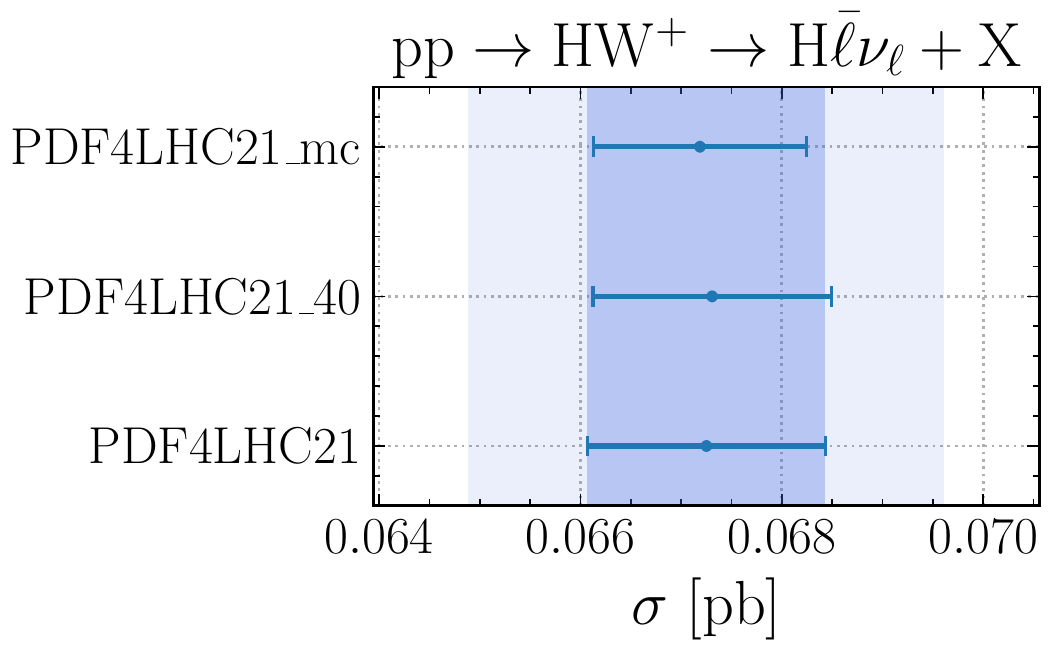}
\includegraphics[width=0.45\textwidth]{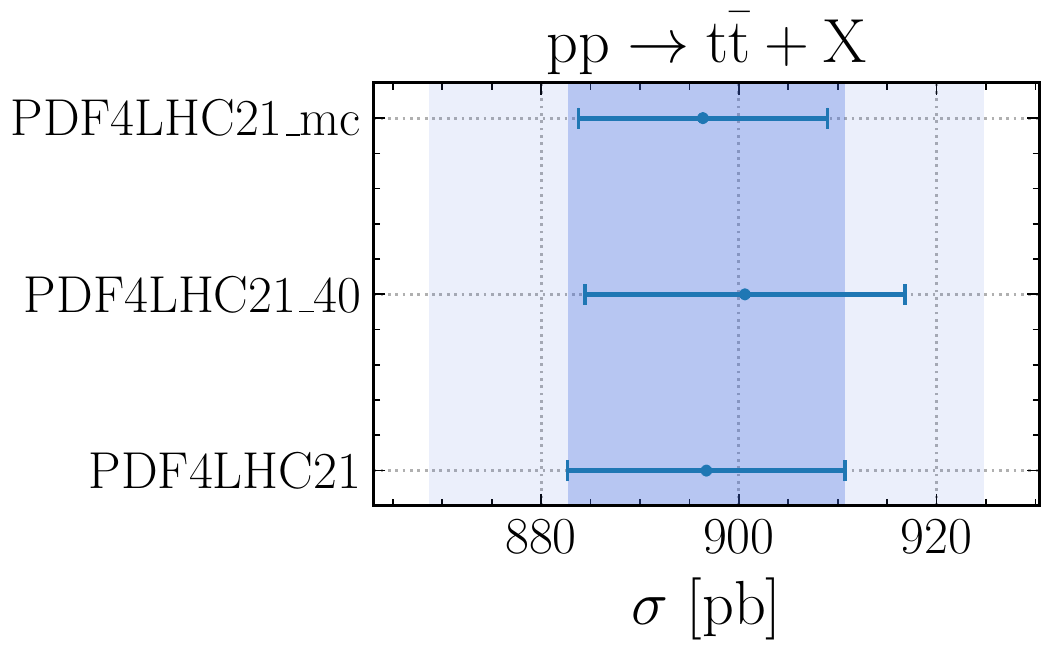}
\includegraphics[width=0.45\textwidth]{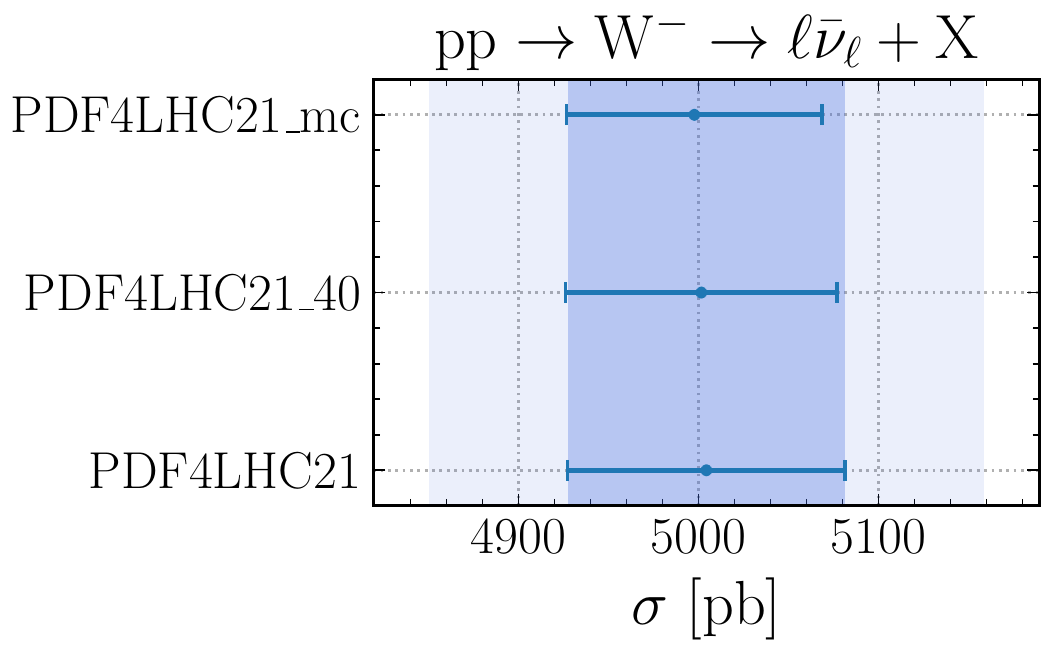}
\includegraphics[width=0.45\textwidth]{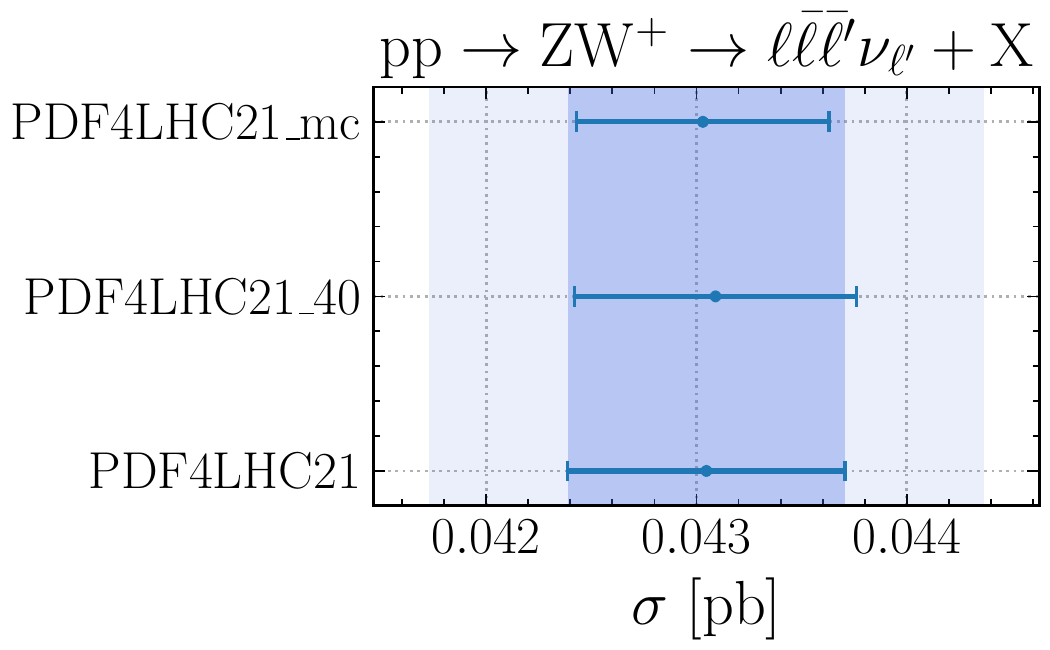}
\caption{\small The fiducial cross-sections for representative LHC processes at
  $\sqrt{s}=14~\mathrm{TeV}$,
  comparing the predictions based on PDF4LHC21 ($N_{\rm rep}=900$) with
  the corresponding Monte Carlo compressed set ($N_{\rm rep}=100$) and the
  Hessian reduction ($N_{\rm eig}=40$) using the META-PDF approach. The darker 
  (lighter) band indicates the 68\% CL (95\% CL) uncertainties associated to 
  the original PDF4LHC21 combination.
}
\label{fig:pheno-reduction}
\end{figure}

To begin with, Fig.~\ref{fig:pheno-reduction} displays predictions for
fiducial cross-sections for representative LHC processes,
comparing the predictions based on the $N_{\rm rep}=900$ Monte Carlo replicas  ({\sc \small PDF4LHC21}) with
those based on the compressed MC {\sc \small PDF4LHC21\_mc}
and reduced Hessian {\sc \small PDF4LHC21\_40} sets.
We display results for single and double gauge boson production, inclusive
top quark pair production, and Higgs production in gluon fusion and in association
with a $W^+$ boson.
\texttt{MadGraph5$\_$aMC@NLO}~\cite{Alwall:2014hca, Frederix:2018nkq} interfaced with
\texttt{PineAPPL}~\cite{Carrazza:2020gss, christopher_schwan_2022_5846421}
has been used to evaluate these fiducial cross-sections 
using next-to-leading (NLO) order theory both in the QCD and electroweak couplings.
In all cases, realistic selection and acceptance cuts on the final state particles
have been applied, 
see~\cite{Ball:2021leu} for the settings of the various calculations.
The darker (lighter) band indicates the 68\% CL (95\% CL)
uncertainties associated to the original PDF4LHC21 combination.

Inspection of Fig.~\ref{fig:pheno-reduction} indicates that
overall the cross sections evaluated with the MC compressed
and Hessian reduced versions of the PDF4LHC21 set agree among each other
as well as with the original set.
The stretched  central PDF of the Hessian variant affects the central value of the cross sections that are sensitive to those PDF flavours that underwent changes at large $x$ to ensure PDF positivity.
For this Hessian set,
cross sections that are sensitive, {\it i.e.}, to the gluon PDF exhibit a slight variation
in their central values, still contained well within the 68\% CL uncertainties associated to the original PDF4LHC21 combination.

As discussed in Sect.~\ref{sec:pdf4lhc21}, the optimal number of
Monte Carlo replicas for the compressed set and that of Hessian eigenvectors
were determined from comparing predictions based on a range of values
of $N_{\rm rep}$ and $N_{\rm eig}$ to those of the PDF4LHC21 baseline, both
in terms of PDFs and of LHC cross-sections.
Analogous plots like Fig.~\ref{fig:pheno-reduction} based on
different values of $N_{\rm rep}$ and $N_{\rm eig}$ are presented in App.~\ref{app:tools}.

\begin{figure}[!t]
\centering
\includegraphics[width=0.48\textwidth]{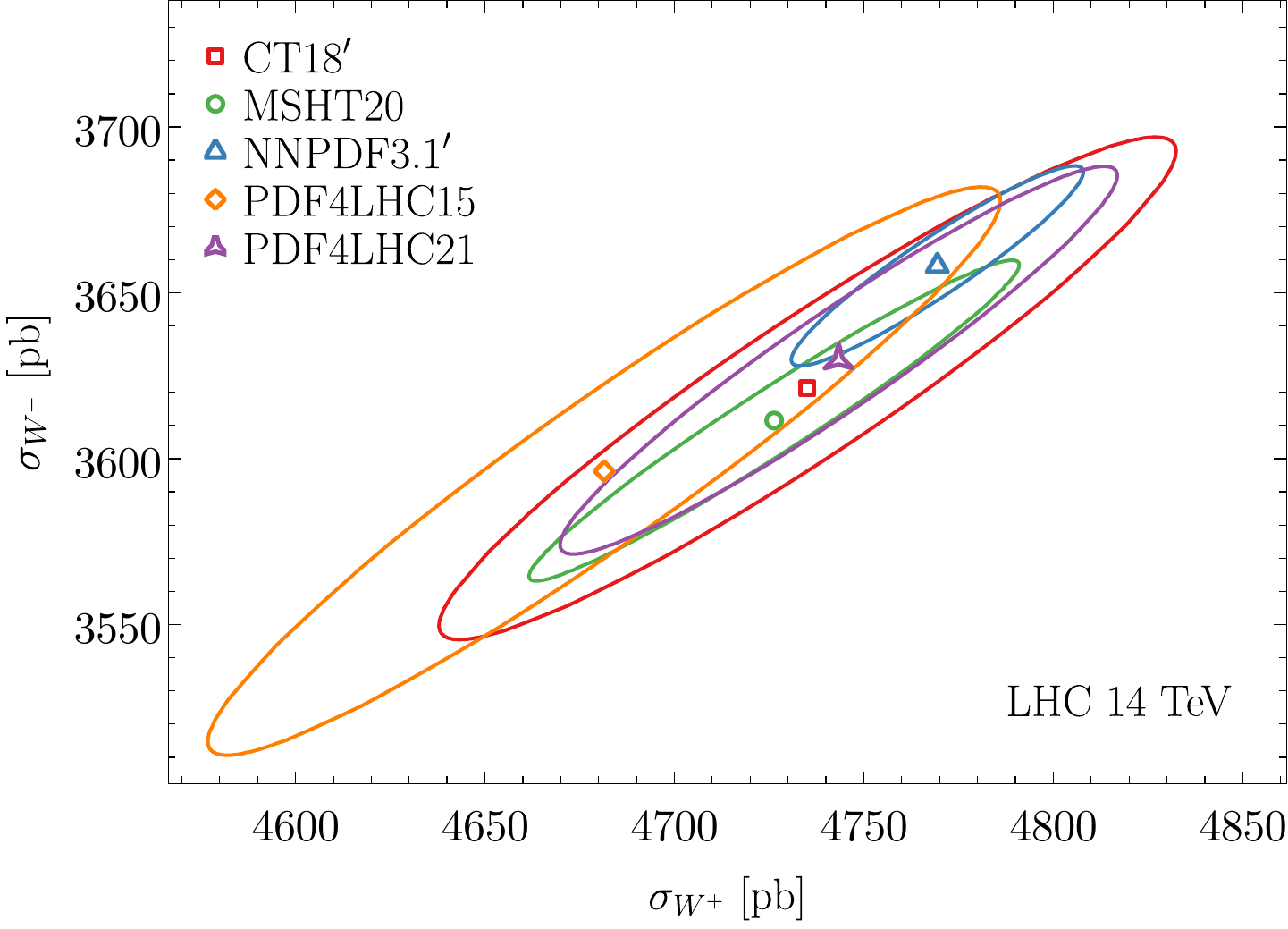}
\includegraphics[width=0.48\textwidth]{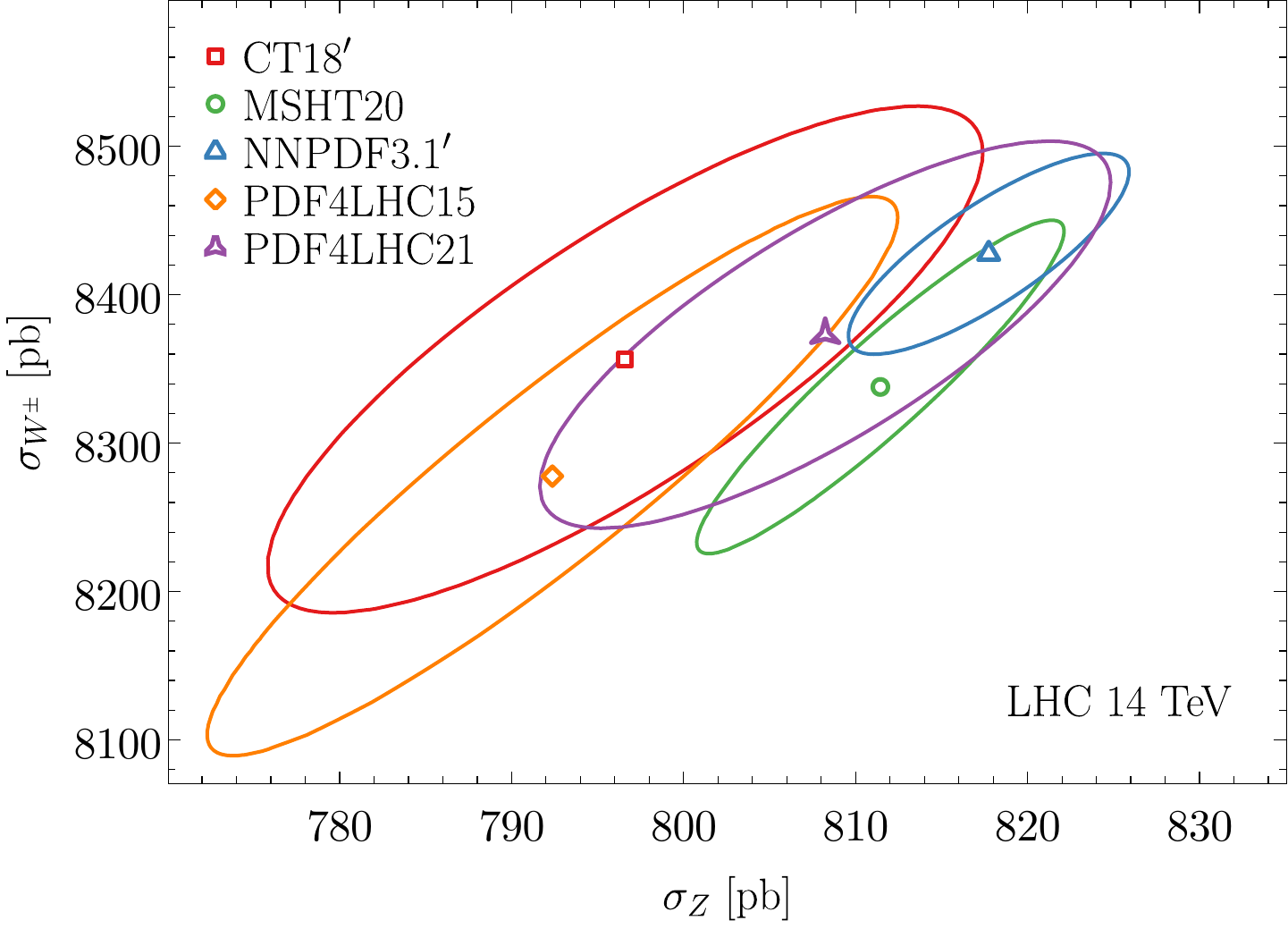}
\includegraphics[width=0.48\textwidth]{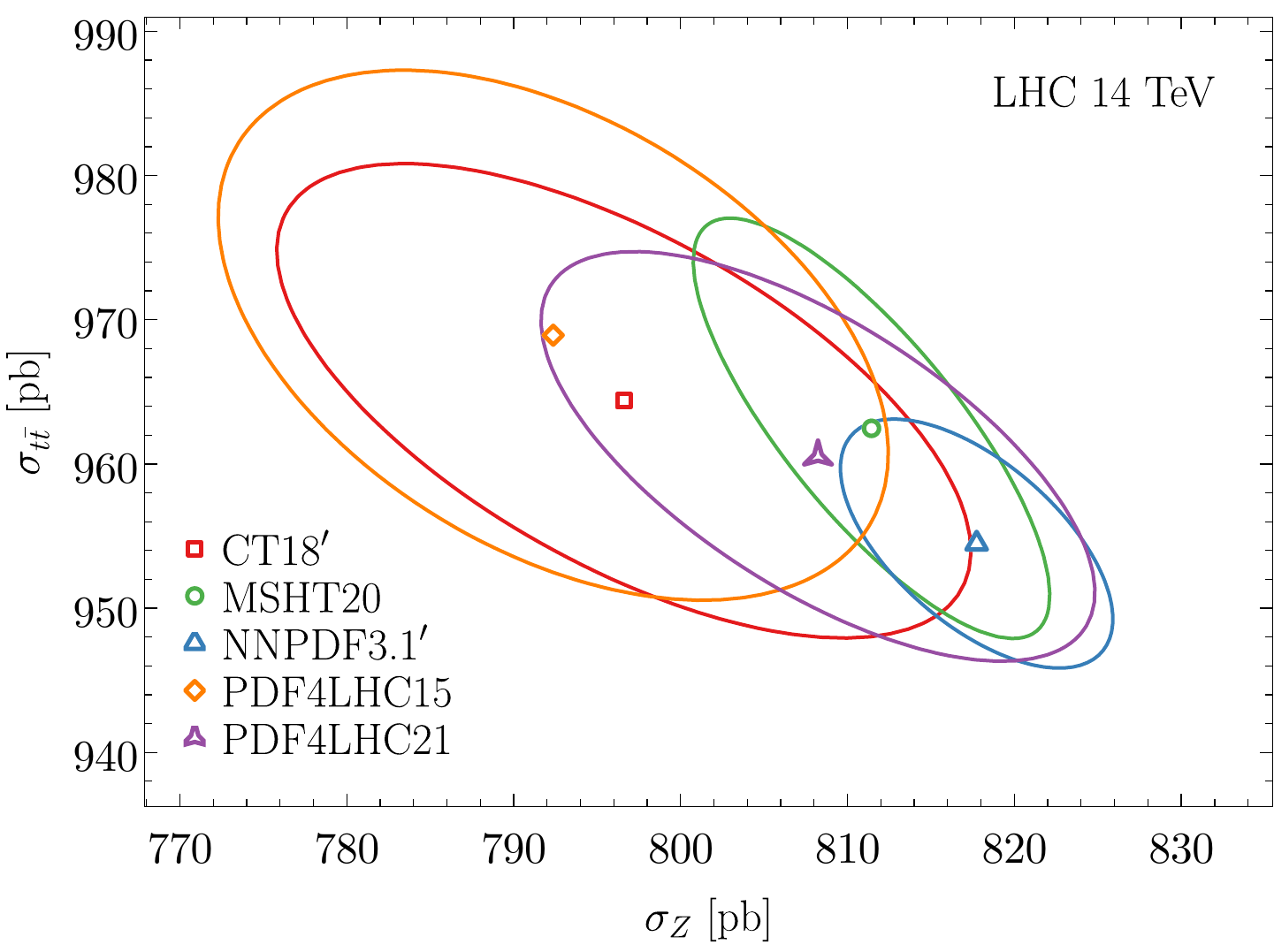}
\includegraphics[width=0.48\textwidth]{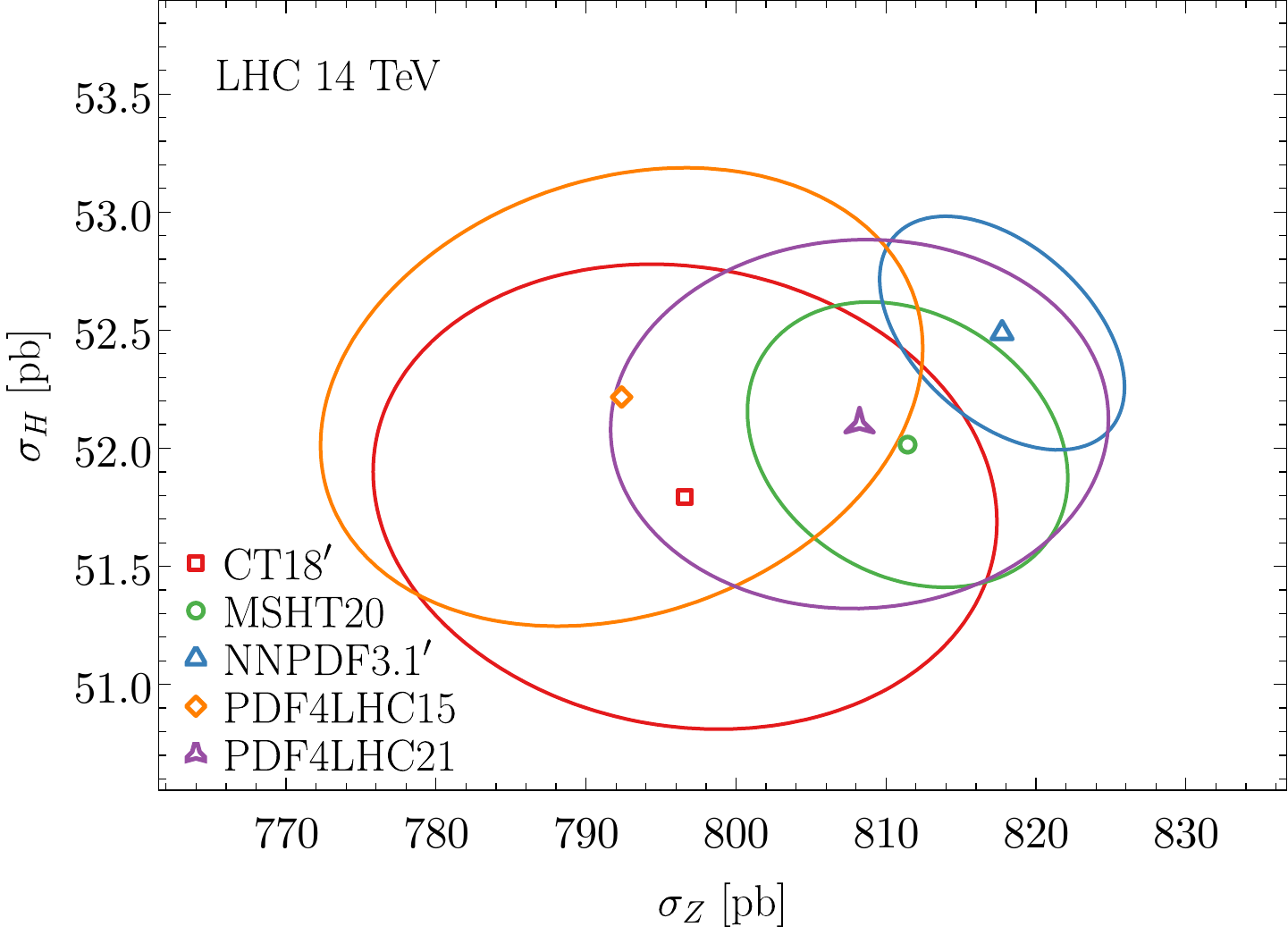}
\includegraphics[width=0.48\textwidth]{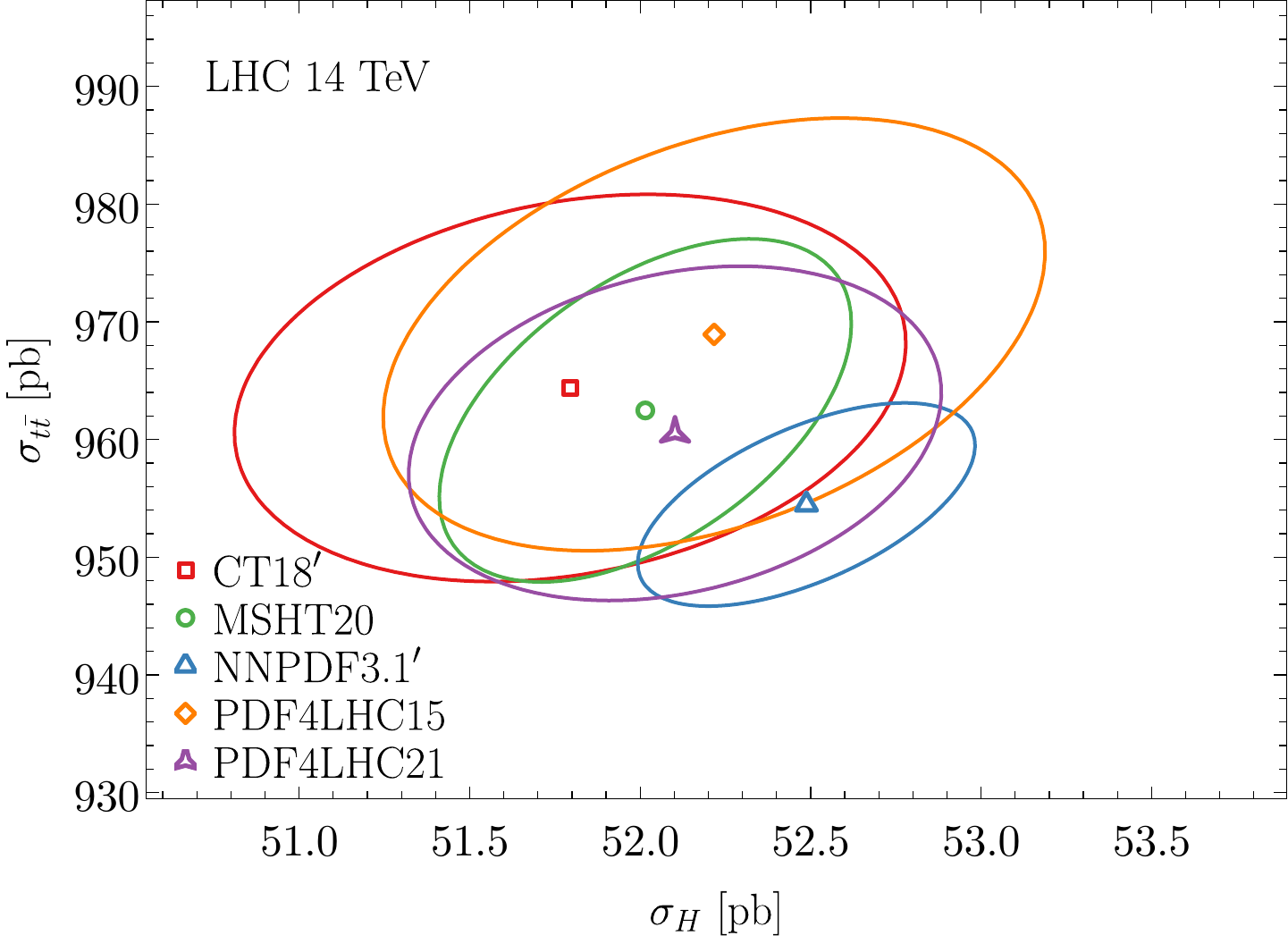}
\includegraphics[width=0.48\textwidth]{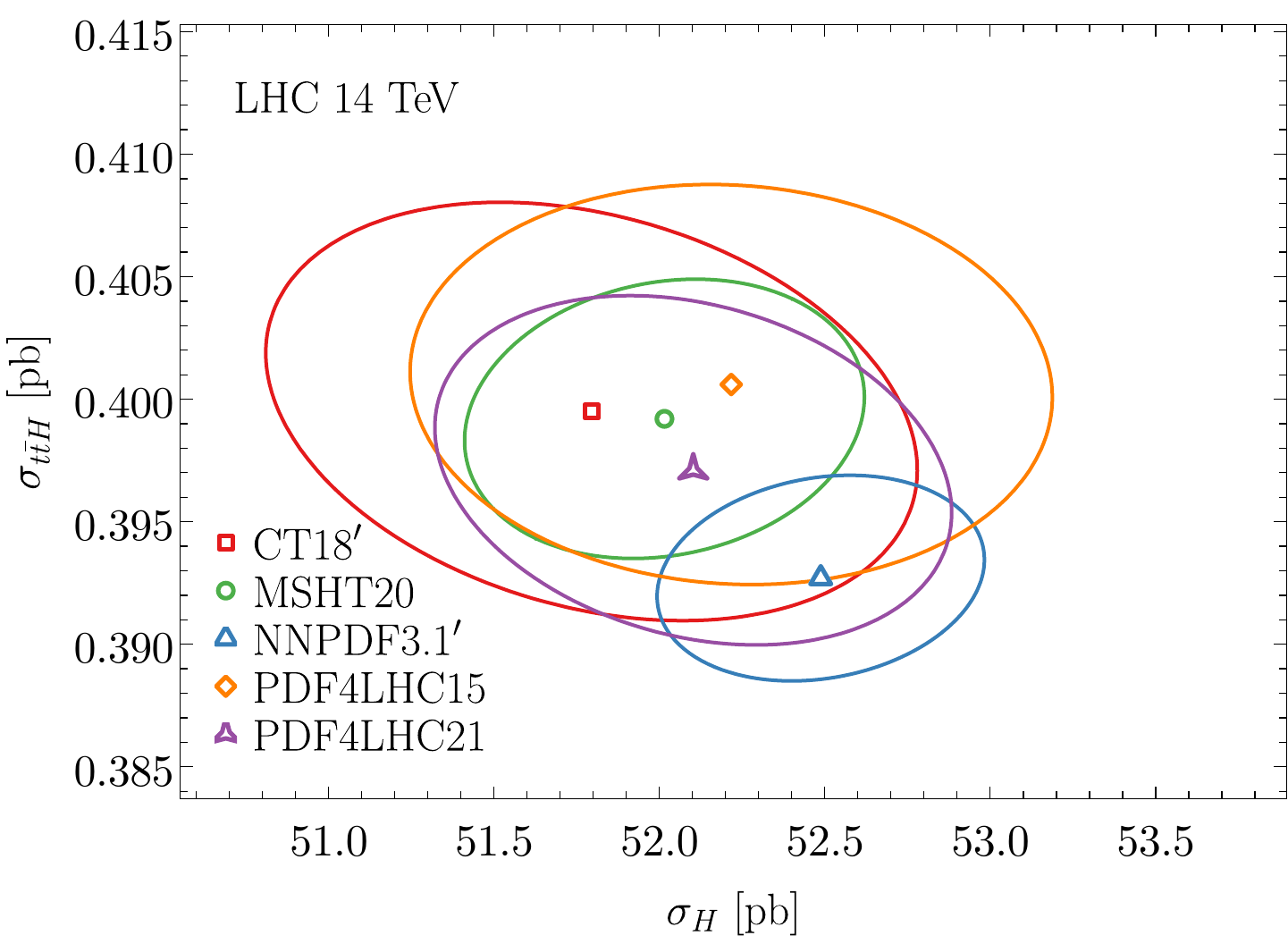}
\caption{The $1\sigma$ ellipses for pairs of inclusive cross sections  among $W^\pm,Z,t\bar{t},H,t\bar{t}H$ production at the LHC at $\sqrt{s}=14~\mathrm{TeV}$.
  The $W^\pm/Z$ cross sections are defined in the ATLAS 13 TeV fiducial volume~\cite{ATLAS:2016fij}, while others correspond to the full phase space. 
  See text for details of the theory calculations.
}
\label{fig:incl_xsec_pdfsets}
\end{figure}

\begin{figure}[!t]
\centering
\includegraphics[width=0.48\textwidth]{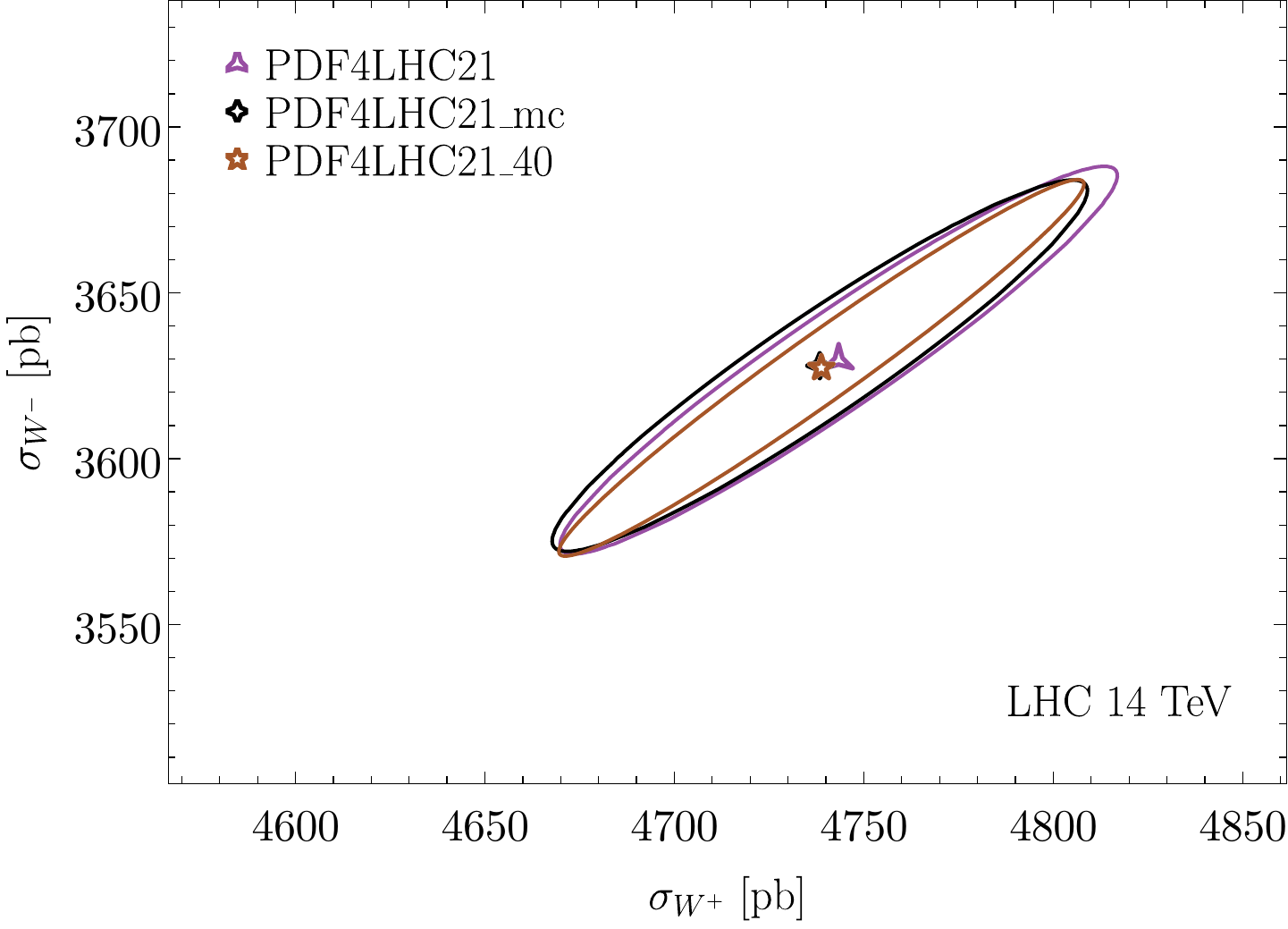}
\includegraphics[width=0.48\textwidth]{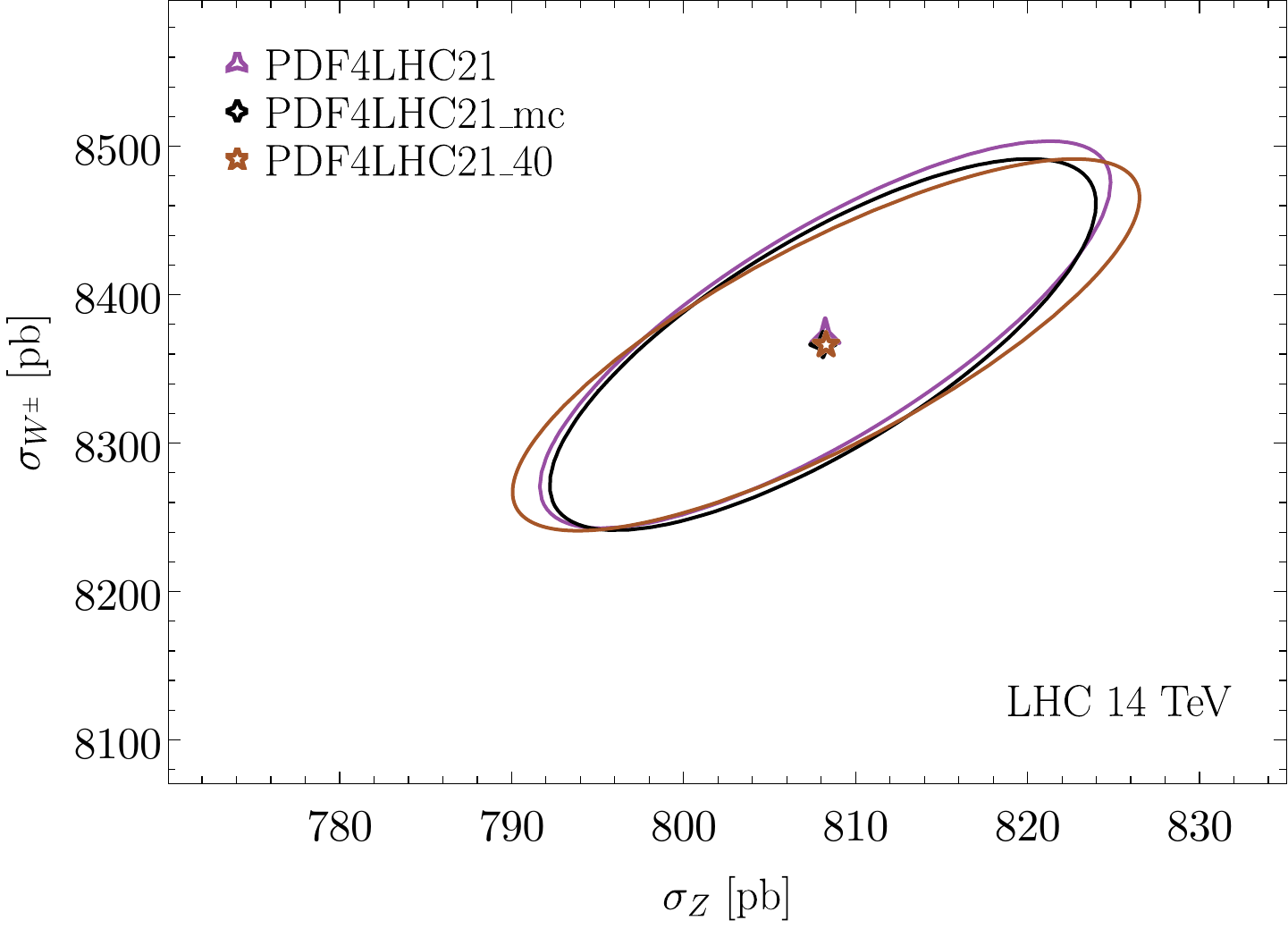}
\includegraphics[width=0.48\textwidth]{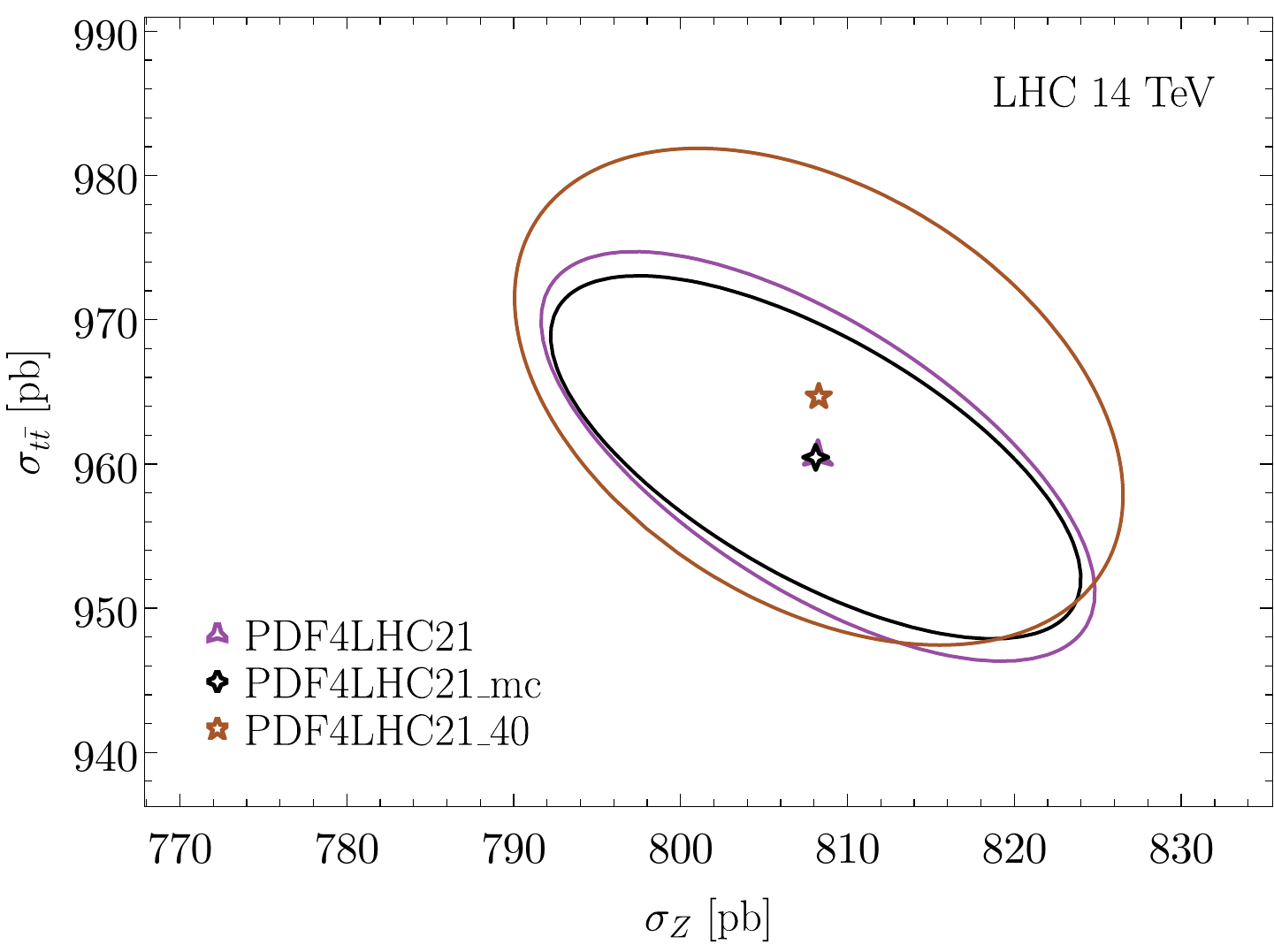}
\includegraphics[width=0.48\textwidth]{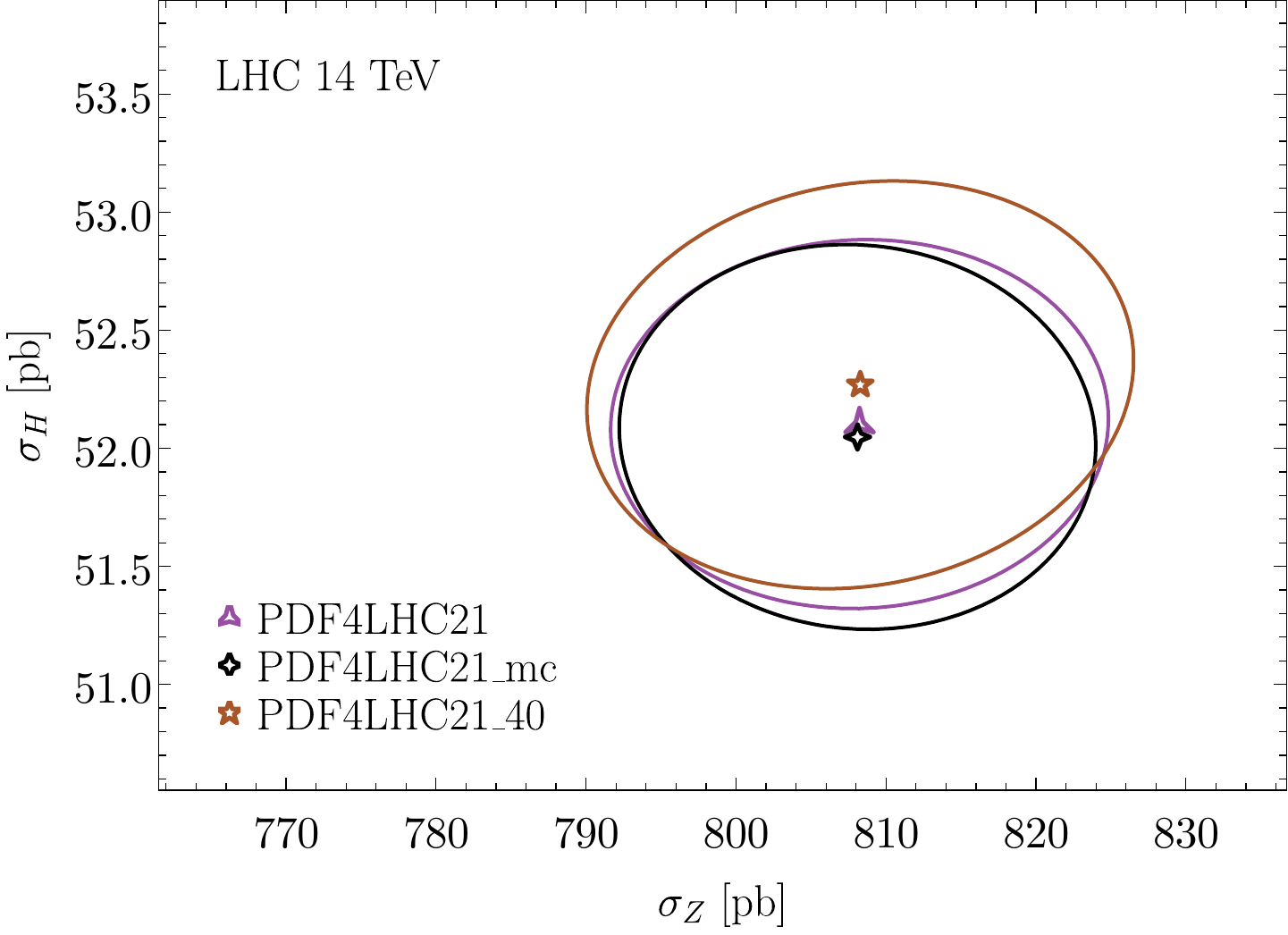}
\includegraphics[width=0.48\textwidth]{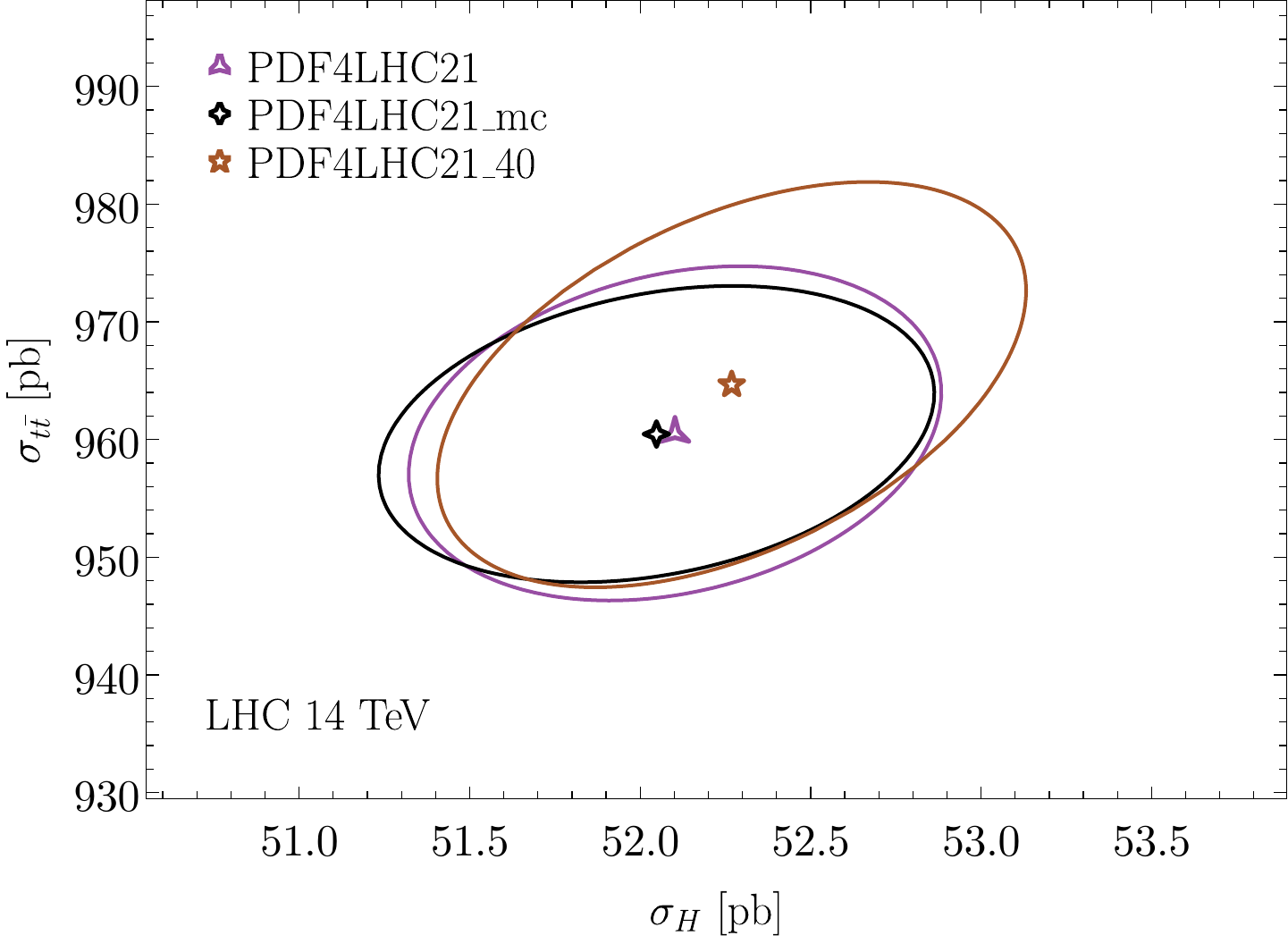}
\includegraphics[width=0.48\textwidth]{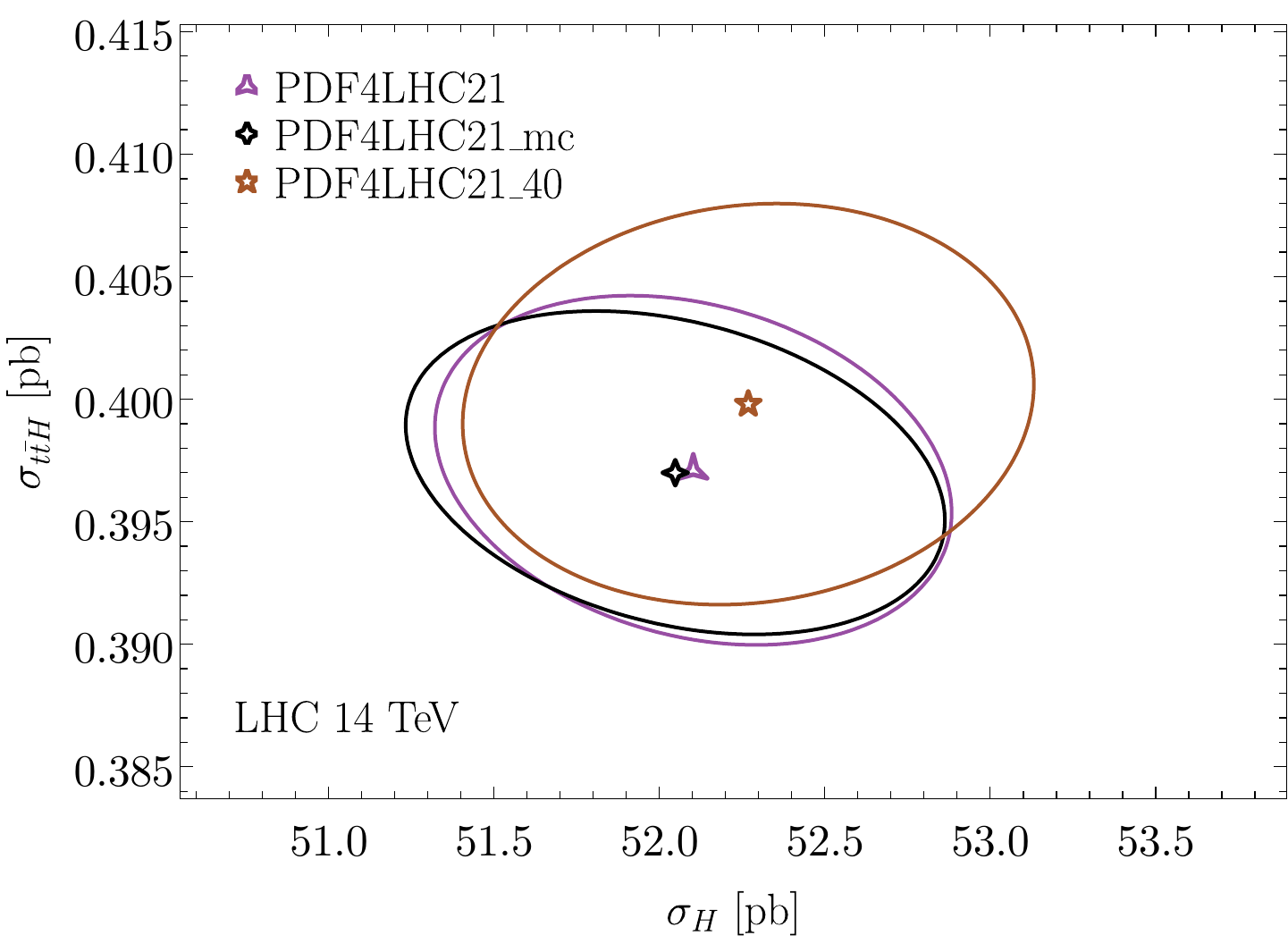}
\caption{Same as Fig.~\ref{fig:incl_xsec_pdfsets} now comparing
  the predictions from the {\sc \small PDF4LHC21} ($N_{\rm rep}=900$) set with
  those of its Monte Carlo compressed  ({\sc \small PDF4LHC21\_mc})
and Hessian reduced  ({\sc \small PDF4LHC21\_40}) representations.
}
\label{fig:incl_xsec_PDF4LHC21}
\end{figure}

In addition to examining the predictions  for individual LHC
cross sections, it is also instructive to study the $1\sigma$ error ellipses highlighting
the correlations between different LHC processes.
This provides a complementary view of the PDF uncertainties themselves, as well as of the correlations inherent in the PDFs, both for the individual global PDF sets and
for the PDF4LHC15~\cite{Butterworth:2015oua} and
PDF4LHC21 combinations.
Several representative LHC processes are shown in Figs.~\ref{fig:incl_xsec_pdfsets} and~\ref{fig:incl_xsec_PDF4LHC21} displaying the
 $1\sigma$ ellipses for pairs of inclusive cross sections  among $W^\pm,Z,t\bar{t},H,t\bar{t}H$ production at the LHC at $\sqrt{s}=14~\mathrm{TeV}$.
  The predictions from PDF4LHC21 are compared to the three constituents sets
  (\CTprime~\cite{Hou:2019efy}, MSHT20~\cite{Bailey:2020ooq}, \NNprime~\cite{NNPDF:2017mvq}) as well as PDF4LHC15 in Fig.~\ref{fig:incl_xsec_pdfsets}, and to the MC compressed and Hessian reduced versions in Fig.~\ref{fig:incl_xsec_PDF4LHC21}, with uncertainties re-scaled to a common $1\sigma$ prescription.
  The $W^\pm/Z$ cross sections are calculated with \texttt{APPLgrid}~\cite{Carli:2010rw} at NLO together with \texttt{MCFM}~\cite{Campbell:2019dru} NNLO $K$-factors. The $t\bar{t}$ process
  is evaluated with \texttt{Top++}~\cite{Czakon:2011xx} at NNLO and with threshold logarithms resummed at NNLL. The inclusive Higgs production predictions are computed with \texttt{ggHiggs}~\cite{Ahmed:2016otz} at N3LO and $t\bar{t}H$ is obtained with \texttt{MadGraph5$\_$aMC@NLO}~\cite{Alwall:2014hca, Frederix:2018nkq} interfaced with \texttt{PineAPPL}~\cite{Carrazza:2020gss, christopher_schwan_2022_5846421} at NLO.
  The $W^\pm/Z$ cross sections are defined in the ATLAS 13 TeV fiducial volume~\cite{ATLAS:2016fij}, while predictions for the other process are inclusive and do not
  include acceptance or selection cuts.

From Fig.~\ref{fig:incl_xsec_pdfsets} 
one can observe how
the $W^+$ cross section is very correlated with the $W^-$ cross section, as shown on the top left panel.
The three global PDF predictions are consistent with each other, and the PDF4LHC21 combination is consistent with that of the PDF4LHC15 combination. The $Z$ boson cross section is shown versus the $W^{\pm}$ (top right). Here, there is more of a spread, with the \NNprime and MSHT20 ellipses touching the edges of \CTprime.
The central PDF4LHC21 $W$ and $Z$ cross sections are higher than those from PDF4LHC15, with the central values being within each other’s $1\sigma$  error ellipse.
Both the Higgs cross section (through $gg$ fusion) and the $t\bar{t}$ cross section depend on the gluon distribution, but in different $x$ ranges, which are mildly correlated, as shown in the lower left panel of Fig.~\ref{fig:incl_xsec_pdfsets}.
This leads to a reduced correlation than observed for $W/Z$ cross sections, as shown in the upper panels. \NNprime predicts a higher Higgs boson cross section then either \CTprime or MSHT20, as observed in the one-dimensional plots. A similar  level of de-correlation can be seen in the the lower right panel, for the $t\bar{t}H$ cross section versus the $H$ cross sections.
The $t\bar{t}$ cross section is anti-correlated with the $Z$ boson cross section as shown middle left. A similar level of anti-correlation is observed for all PDFs sets.
Lastly, the $Z$ boson versus Higgs boson cross section predictions are plotted on the middle right. 
Again, we see the MSHT20 gives predictions in between those of \CTprime and \NNprime, and the PDF4LHC21 average these three groups.
In addition, both the \CTprime and \NNprime central predictions are outside the MSHT20 error ellipse, 
reflecting the various choices made in the three fits.  

One important message from Fig.~\ref{fig:incl_xsec_pdfsets} is that, for
all observables considered here, the predictions of PDF4LHC21 overlap with
those of its predecessor PDF4LHC15, with the central values of the latter
being always included within the $1\sigma$ ellipse of the former.
Furthermore, one can also observe how in general the area covered by the
PDF4LHC21 $1\sigma$ ellipse is smaller than that covered by the
PDF4LHC15 predictions.
Hence, we further confirm the findings at the PDF and partonic luminosity
level presented in Sect.~\ref{subsec:comp_pdf4lhc15} that the new
combination is consistent with the previous one, and that in addition
it typically has associated an improvement in precision concerning
LHC cross-sections.

Next, Fig.~\ref{fig:incl_xsec_PDF4LHC21} presents a comparison
of the predictions for the same LHC cross-sections as in Fig.~\ref{fig:incl_xsec_pdfsets}
now comparing  the predictions from the PDF4LHC21  ($N_{\rm rep}=900$) with
  those of its Monte Carlo compressed  ($N_{\rm rep}=100$)
  and Hessian reduced  ($N_{\rm eig}=40$) representations.
  We can observe how the $N_{\rm rep}=100$ compressed Monte Carlo set provides
  an accurate representation of the $1\sigma$ ellipses of the  PDF4LHC21 baseline,
  hence also reproducing the correlation between processes.
 The Hessian  {\sc \small PDF4LHC21\_40} set provides a good representation of the baseline uncertainties  and correlations, too. 
The fulfilment of positivity at the low scale of the combination for the central PDFs of the Hessian set is reflected through slightly higher central predictions (at most $0.5\%$) for Higgs, $t\bar{t}$, and $t\bar{t}H$ productions as compared to the baseline predictions. The latter set exhibits negative average PDFs, notably for the gluon at very large $x$ as well as for $\bar{u}, \bar{s}$ and $d_{\mbox{\tiny V}}$ in the valence region. A dedicated discussion on the phenomenological implications of positivity at large $x$ is presented in Secs.~\ref{sec:hessian_reduction} and~\ref{sec:largex_pdf4lhc21}. 

\subsection{Differential distributions}
\label{sec:differential_processes}

\begin{figure}[!t]
\centering
\includegraphics[width=0.48\textwidth]{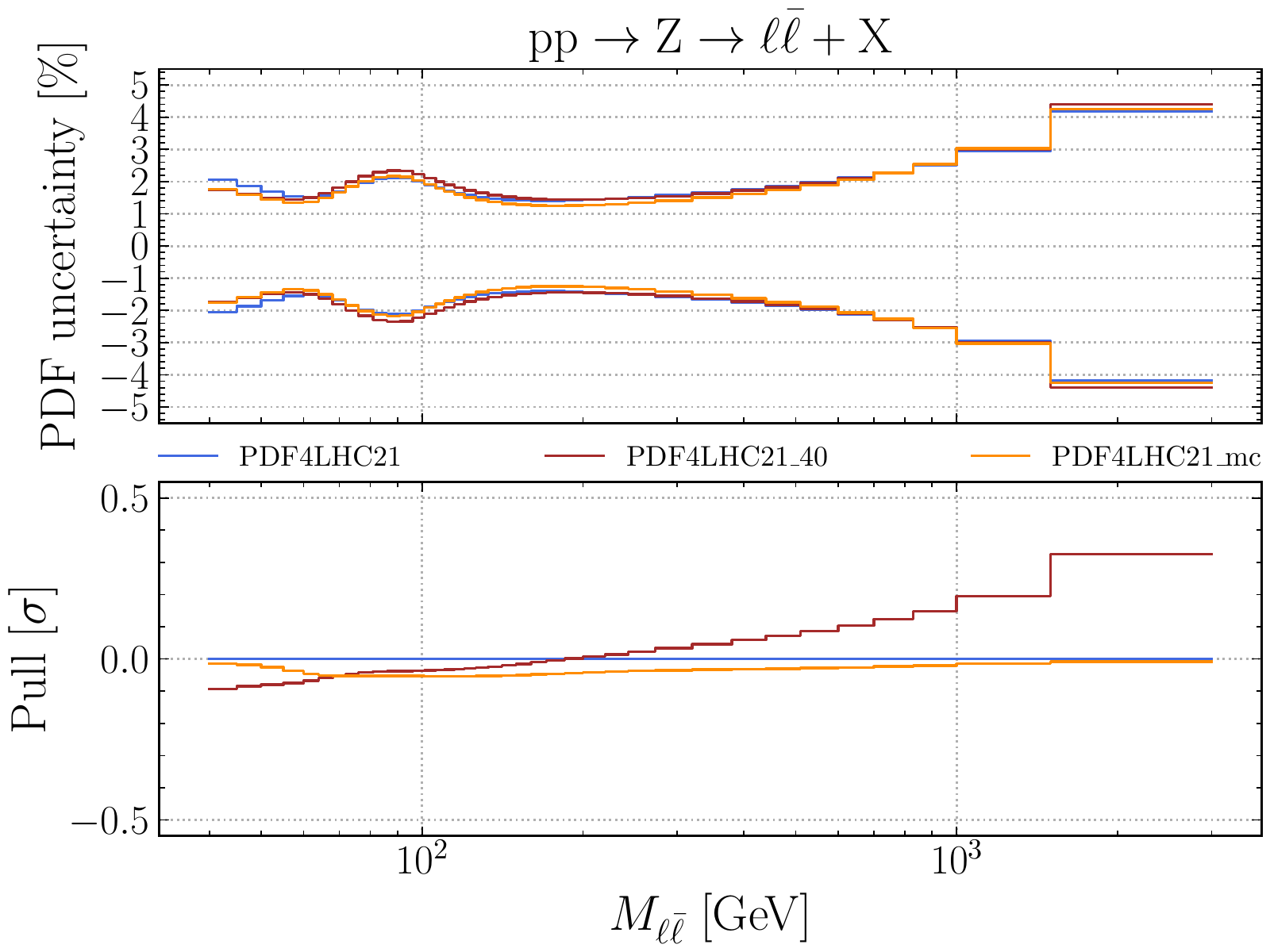}
\includegraphics[width=0.48\textwidth]{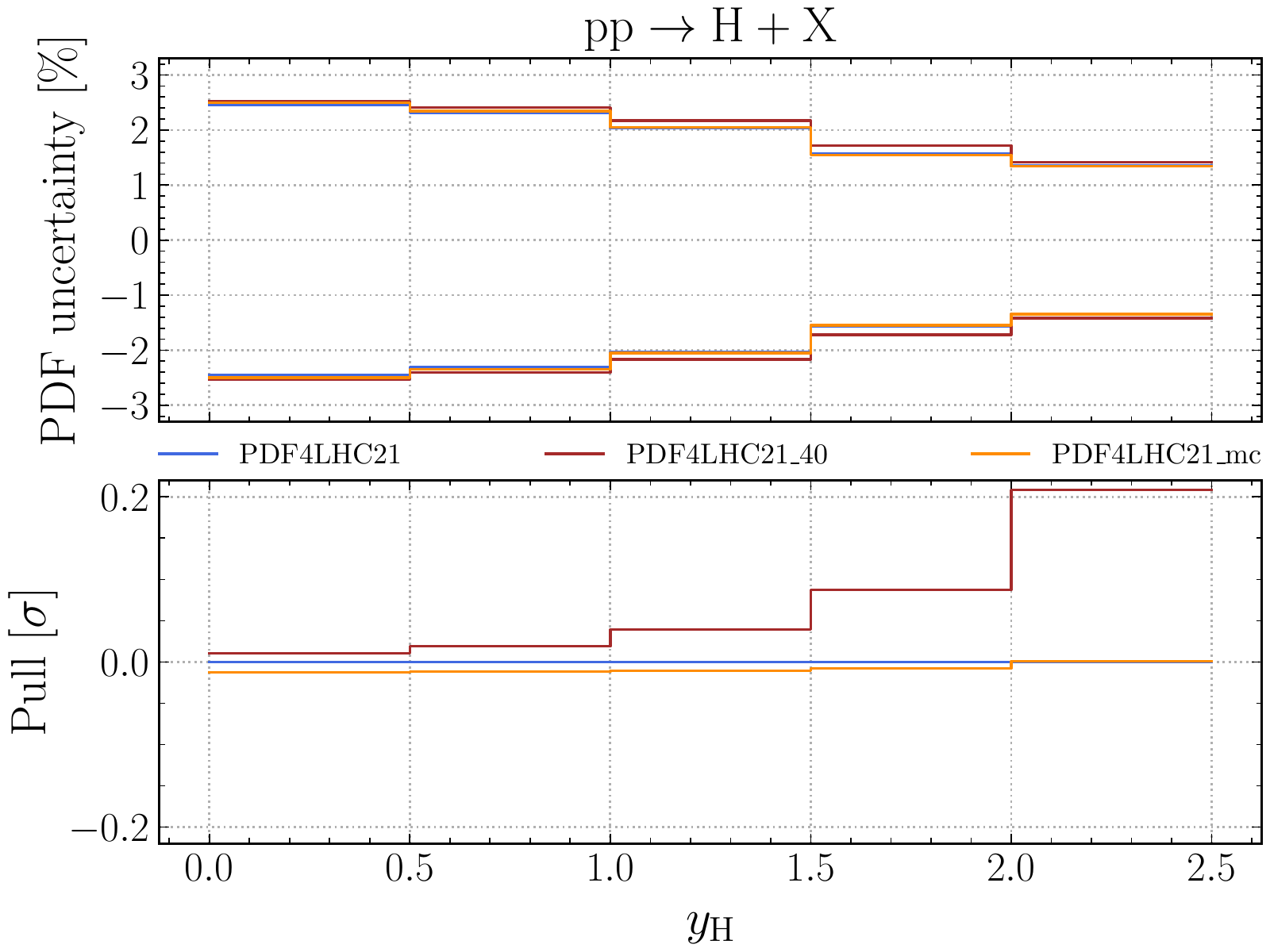}
\includegraphics[width=0.48\textwidth]{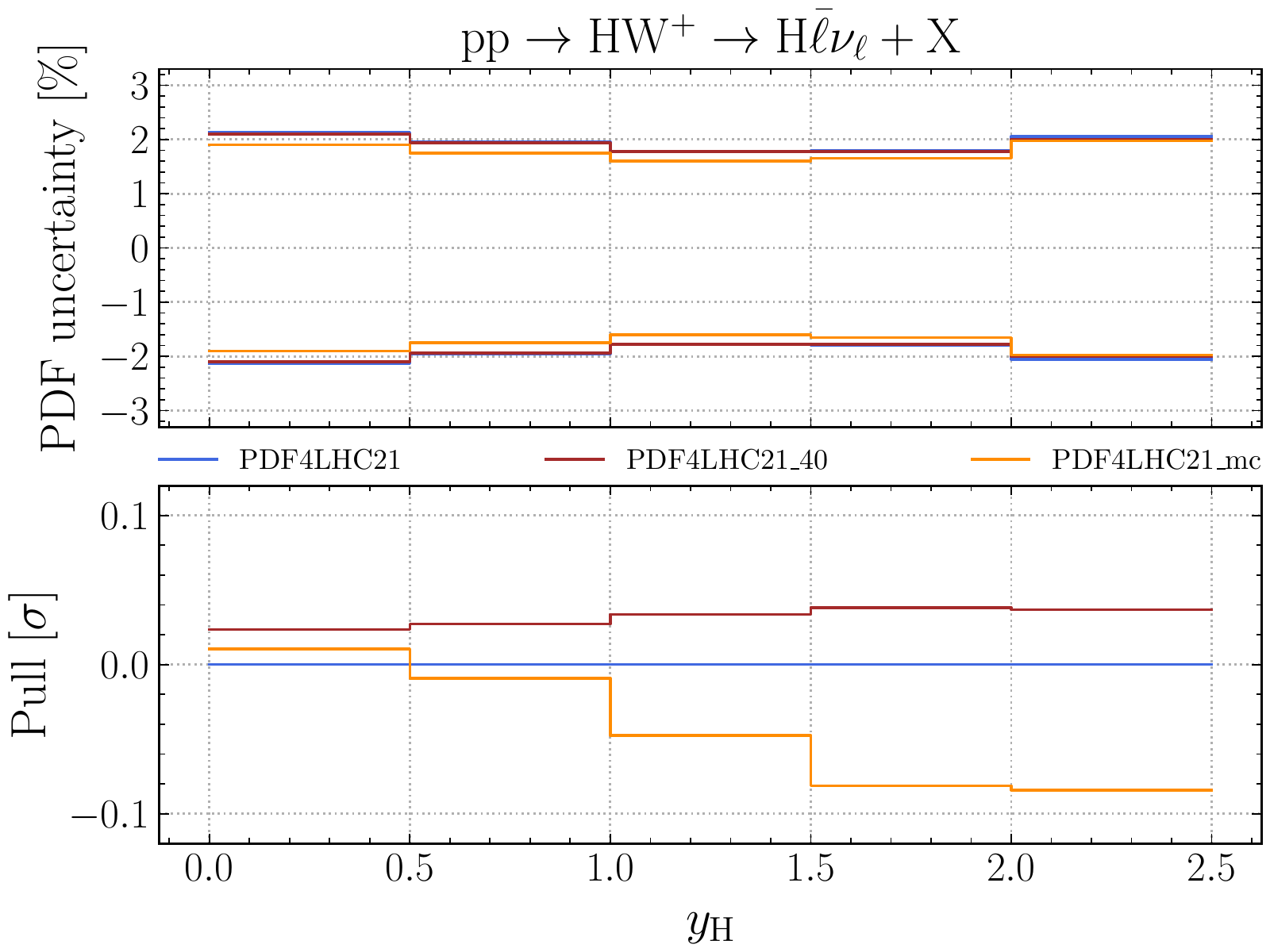}
\includegraphics[width=0.48\textwidth]{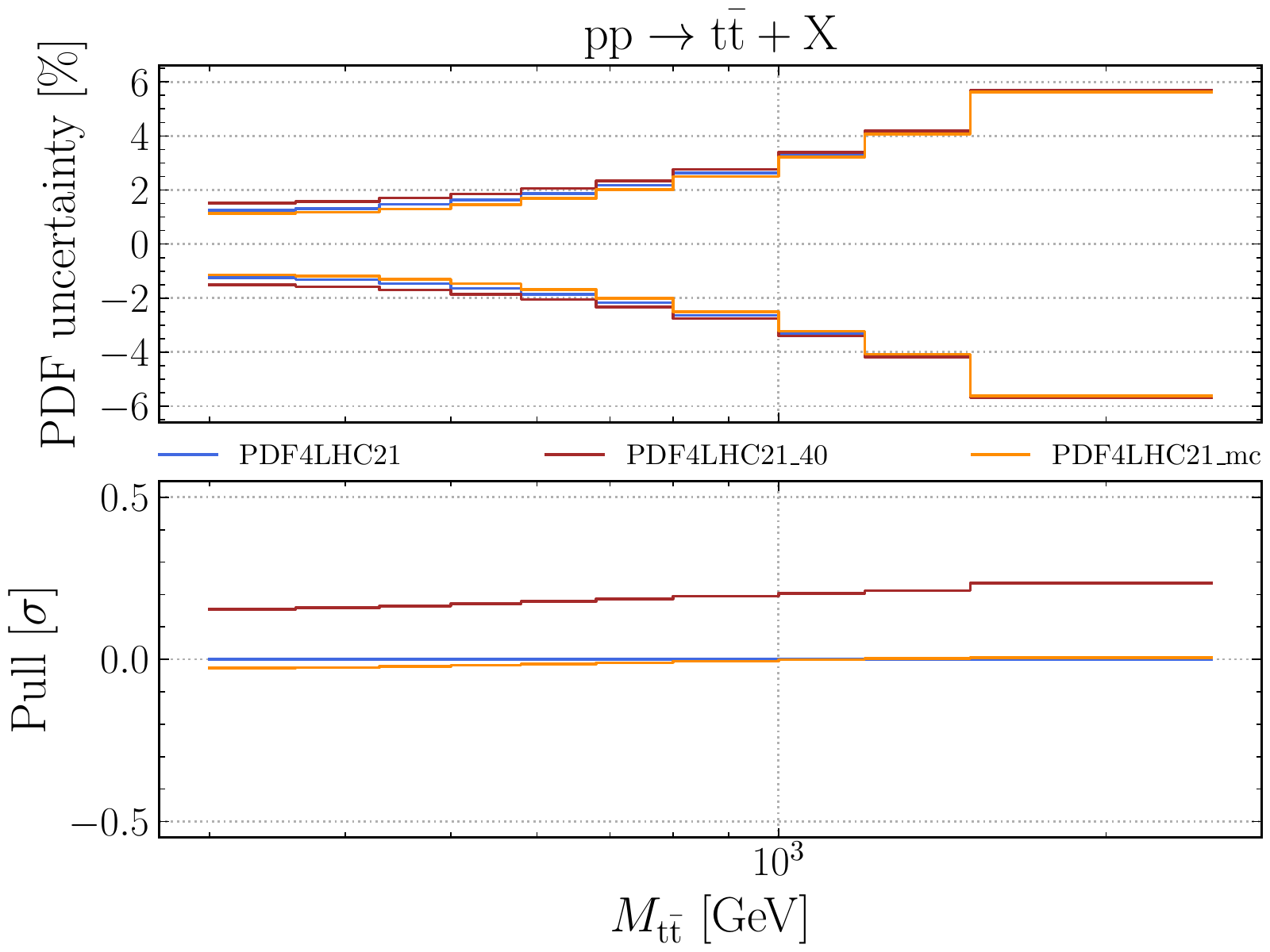}
\includegraphics[width=0.48\textwidth]{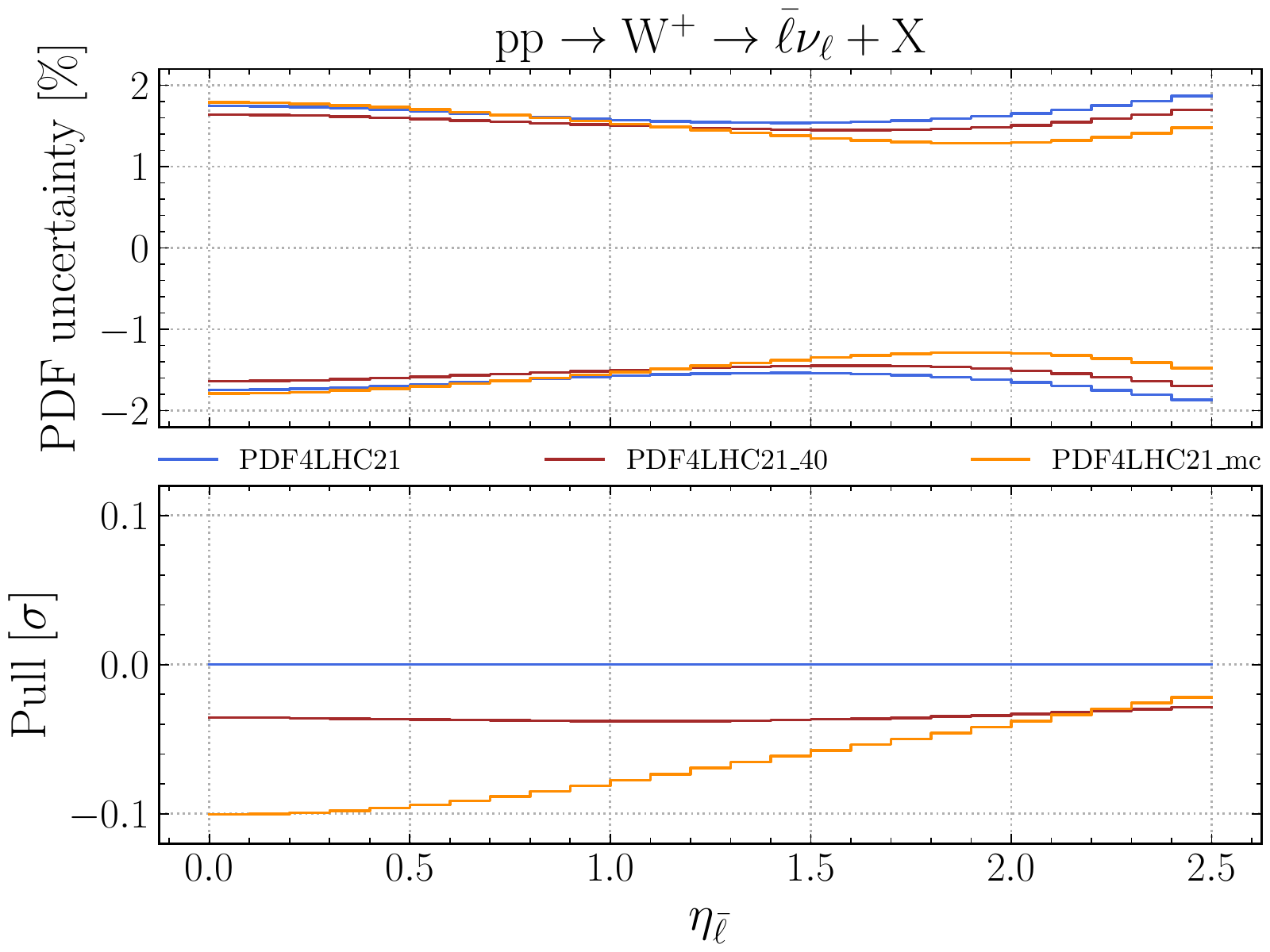}
\includegraphics[width=0.48\textwidth]{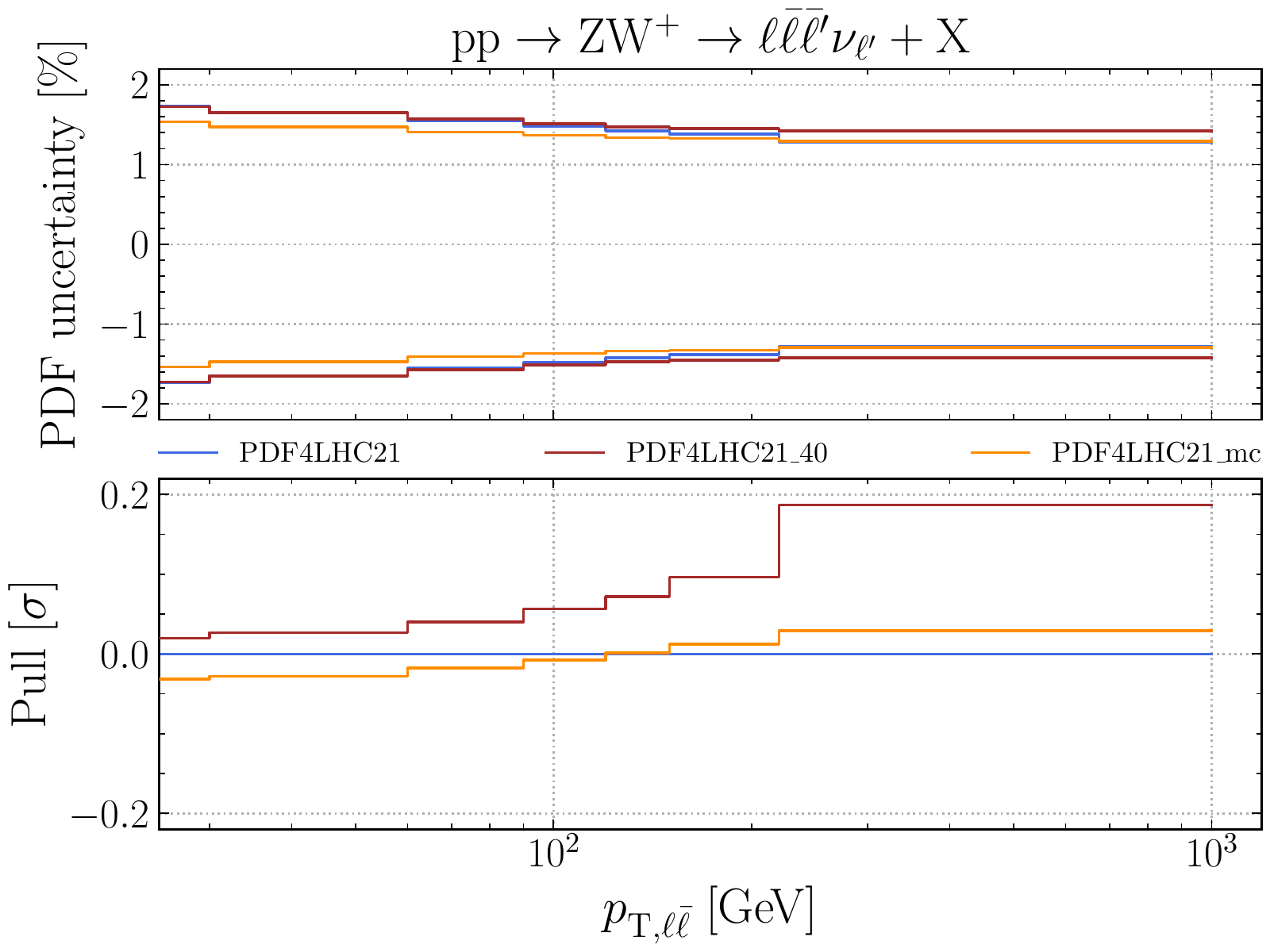}
\caption{\small Same as Fig.~\ref{fig:pheno-reduction} now with the comparison
  presented at the level of differential distributions.
  For each process, the top panels display the relative percentage PDF uncertainty (normalised
  to the corresponding central value), while
  the bottom panels show the pull of the central values with respect
  to the {\sc \small PDF4LHC21} set in units of the PDF uncertainty, as defined in
  Eq.~(\ref{eq:pulldef_xsec}).
  The calculational settings and selection cuts are the same as those adopted in
  Fig.~\ref{fig:pheno-reduction}  and the integral over the differential
  measurements reproduces those fiducial cross-sections.
}
\label{fig:pheno-reduction-differential}
\end{figure}

Following this discussion focused on predictions at the inclusive cross-section level, we move now to consider
the case of differential distributions.
Fig.~\ref{fig:pheno-reduction-differential} displays a similar comparison as in
Fig.~\ref{fig:pheno-reduction} now at the level of differential distributions, where
 the calculational settings and selection cuts adopted are the same as those adopted in
  Fig.~\ref{fig:pheno-reduction}  and the integral over the differential
  measurements reproduces the fiducial cross-sections reported there.
  For each process in Fig.~\ref{fig:pheno-reduction-differential}, the top panels display the relative percentage PDF uncertainty (normalised
  to the corresponding central value), while
  the bottom panels show the pull of the central values with respect
  to the PDF4LHC21 baseline in units of the PDF uncertainty,  defined as
    \be
\label{eq:pulldef_xsec}
P\lp  \sigma_{2,i}, \sigma_{1,i}\rp\equiv \frac{ \sigma^{(0)}_{2,i} -\sigma^{(0)}_{1,i} }{
  \sqrt{ \lp  \delta \sigma_{2,i}\rp^2+\lp  \delta \sigma_{1,i}\rp^2 }} \, , \qquad i=1,\ldots,n_{\rm bin} \, ,
\ee
where  $\sigma^{(0)}_{1,i}$ and $\sigma^{(0)}_{2,i}$ are the central values of the
theory predictions in the $i$-th bin and $\delta \sigma_{1,i}$, $\delta \sigma_{2,i}$ are
the corresponding PDF uncertainties.
For the calculation of these pulls we adopt the PDF4LHC21 $N_{\rm rep}=900$ set
as reference, and hence for this set 
$P=0$ by construction for all bins.
The pull Eq.~(\ref{eq:pulldef_xsec}) quantifies the shift in the central values
between two PDF sets weighted by their combined PDF uncertainties.

From Fig.~\ref{fig:pheno-reduction-differential} one can observe, from inspection of the upper
panels for each process, how the relative uncertainties of the {\sc \small PDF4LHC21} set are well reproduced
by both the compressed MC and the reduced Hessian sets.
Consistent with the results reported at the cross-section level in Sect.~\ref{sec:incl_xsec},
the pull is very small, always below $P\le 0.1$ for the compressed MC set.
The pulls $P$ can be larger
and positive for the $N_{\rm eig}=40$ Hessian set, a consequence of the difference in the central value of the sets that are compared,
and is illustrated in Fig.~\ref{fig-app:META40_CMC100}. 
A non-zero pull is then actually expected for processes
sensitive to the gluon PDF as well as to the quark PDFs at large invariant masses and/or
forward rapidities.
Since positivity of central PDFs is implemented at
the initial combination scale of $Q=2$~GeV,  these shifts are not necessarily restricted to the extrapolation region at higher scales
and can be present (though reasonably small),  e.g. for top quark pair production in the threshold region.

\section{Usage and deliverables}
\label{sec:usage}

In this section we list the PDF4LHC21 combined sets that
are released to the {\sc\small LHAPDF} library~\cite{Buckley:2014ana}. 
Additional variants of PDF4LHC21 considered in
this study are also made available in
the {\sc\small LHAPDF} format through the PDF4LHC
Working Group website as indicated below.
Among the several versions of the combined sets that we make available, we recommend two sets, {\sc \small PDF4LHC21\_40} and {\sc \small PDF4LHC21\_mc}, for typical applications. 
We also provide here explicit formulae for the computation of PDF and $\alpha_s$ uncertainties with the recommended PDF4LHC21 ensembles.

We start by providing recommendations to decide on when to use the PDF4LHC21 sets as opposed to the individual PDF sets from the groups. 
Our previous recommendation issued in 2015 had pointed out the versatility of tasks and  applications involving the PDFs in LHC analyses. As a result, here we state broad guidelines, rather than strict instructions, on using PDF4LHC21 depending on the nature of the application.

\subsection{When to use PDF4LHC21? }
\label{sec:WhenToUsePDF4LHC21}

As discussed in Sect.~\ref{sec:pdf4lhc21}, the two recommended representations of PDF4LHC21, namely 
{\sc \small PDF4LHC21\_40} and {\sc \small PDF4LHC21\_mc}, provide a  reasonably accurate and practical representation of the original distribution of $N_{\rm rep}=900$
replicas from three input global fits.
With these sets, a vast range of LHC applications is covered. %
Both compressed sets have their own advantages and
disadvantages in specific contexts, as stipulated below.
The guidelines on using the PDF4LHC21 sets under the given circumstances largely follow the rationales in the 2015 PDF4LHC recommendation.
In light of the community experience of using the PDF4LHC15 sets over seven years, here we state simplified guidelines that distinguish between three broad usage cases: precision SM studies, other types of experimental analyses (including BSM searches), and theoretical computations. 
With the provided delivery options, we expect that any PDF user will be able to choose the strategy (or a combination of strategies) that is best suited for their needs.

We now state the recommendations for each of three usage cases. 
\begin{enumerate}

\item {\bf Comparisons between data and theory for Standard Model measurements}

{\bf Recommendation:} Use
  {individual PDF sets} from different groups, and, in particular, as many modern
  PDF sets as possible.

{\bf Rationale:} Measurements
  such as jet production,  vector-boson single and pair production, or top-quark pair production, have the potential to constrain
PDFs, and this is best utilised and illustrated by comparing with several individual sets. 
Furthermore, the high precision of some of these measurements requires a detailed quantification of PDF uncertainties. 

As a rule of thumb, {any measurement that potentially can be included
in PDF fits} falls in this category.

The same recommendation applies to the
{extraction of precision SM parameters} that are correlated with PDFs,
  such as the strong coupling $\alpha_s(m_Z^2)$,
  weak-mixing angle $\sin\theta_w$, the $W$ mass, and
  the top quark mass $m_t$, as well as to interpretations
  of LHC data in the Effective Field Theory framework.

\item {\bf Searches for Beyond the Standard Model phenomena, measurements of SM observables of lower precision}

{\bf Recommendation:} Use the  {\sc\small PDF4LHC21\_40} or {\sc\small PDF4LHC21\_mc} sets.

{\bf Rationale}: The combined sets substantially reduce the computational burden in such applications, as they provide an acceptable estimate of the central values and 68\% CL uncertainties that agree with the three recent NNLO global sets, CT18, MSHT20, and NNPDF3.1.1.
BSM searches, in particular for {new massive particles in the TeV
  scale}, often require the knowledge of PDFs in regions where available experimental constraints are limited, notably close to the production threshold where $x \rightarrow 1$. Such predictions can be obtained with the PDF4LHC21 sets. 
  
   Note that predictions of some latest PDF ensembles that are not included in the combination, such as CT18Z or NNPDF4.0, may fall outside the 68\% CL uncertainty bands of the PDF4LHC21 sets. 
   As predictions from such PDF sets tend to lie within a few (say, two) units of the 68\% CL PDF4LHC21 uncertainty, the latter still serves as an informative estimate. When the outcome of the physics analysis strongly depends on the magnitudes of the PDF central value and uncertainty, we recommend to compute the predictions and uncertainties based on such alternative ensembles, in addition to the PDF4LHC21 ensembles. 
  
The primary difference between the {\sc\small PDF4LHC21\_40} or {\sc\small PDF4LHC21\_mc}  is in their 
representation, namely Hessian and Monte Carlo
respectively.

{\bf Hessian representation}: The $N_{\rm eig}=40$ {\sc\small PDF4LHC21\_40} Hessian set is recommended when high computational speed is desirable and a detailed knowledge of the underlying probability distribution is unnecessary.
This ensemble renders the central values and 68\% probability intervals evaluated using the master formulae presented in Sect.~\ref{subsec:uncer_prescr} below.

In addition, because of the positivity requirement imposed on these PDFs, this ensemble renders positive central cross sections in the $x\to 1$ limit, which is achieved by allowing some deviation (within the errors) from the baseline PDFs when the latter go negative in this limit. 

For example, {acceptances, efficiencies or extrapolation factors} often result only in a small correction to  the measured cross-section. Thus one gains in the computational speed with a smaller number of error sets by using the {\sc\small PDF4LHC21\_40} ensemble, while retaining the targeted accuracy of computation for such quantities. 
  
{\bf Monte Carlo representation}: The $N_{\rm rep}=100$ Monte Carlo ensemble {\sc \small PDF4LHC21\_mc} can be used for the same calculations as {\sc\small PDF4LHC21\_40}, in particular,  to predict the 68\% probability intervals. In addition, {\sc \small PDF4LHC21\_mc} reproduces the baseline probability distribution and is suitable for PDF reweighting~\cite{Ball:2010gb,Ball:2011gg} and studies of non-Gaussian aspects of the baseline PDFs. The {\sc \small PDF4LHC21\_mc} central PDFs are the same as the baseline PDF4LHC21 ones. 
Hence, the baseline central PDFs are reproduced, while 
some replicas can be negative in the very large $x$ region
and require the positivity prescription laid out in
Sect.~\ref{subsec:uncer_prescr}.

To further elaborate on the complementarity of the two combined ensembles, we note that the {baseline} combination of PDF replicas is likely to be non-Gaussian in the large-$x$ regions, as illustrated e.g. in Fig.~\ref{fig:histo_hmdy_nc}.
The use of the {\sc\small PDF4LHC21\_mc} set seems appropriate for reproducing such non-Gaussian features. 
On the other hand, in that same large-$x$ region, PDFs can become partially negative. 
This situation, illustrated and explored in Sects.~\ref{sec:hessian_reduction} and~\ref{sec:largex_pdf4lhc21}, can be tackled
either by using the {\sc\small PDF4LHC21\_40} set, whose central value is positive-definite by construction,
or by means of the positivity prescription
for Monte Carlo PDF sets described below.

\item {\bf Theoretical computations.} 

{\bf Recommendation:} Both the {\sc\small PDF4LHC21\_40} or {\sc\small PDF4LHC21\_mc} ensembles are recommended for theoretical calculations that do not require a systematic exploration of the differences between the PDF sets obtained by individual groups.

{\bf Rationale:} The PDF4LHC21 combination includes information from three recent global PDF analyses and combine the input PDF uncertainties before the theoretical calculation is done. The PDF4LHC21 uncertainty estimate is moderately conservative and encloses the central predictions from the three groups. 
It is a consensus prediction for acceptable PDFs, while the PDFs of individual groups may lie outside of this consensus region or have smaller or larger individual uncertainties, as discussed above. 
For example, one can use the PDF4LHC21 predictions to illustrate 
the PDF dependence in new {higher-order QCD calculations} or estimate exclusion limits in {theoretical projections} for proposed BSM searches. 

As in the previous application cases, the user can select the format of the error PDF ensemble, Hessian or Monte Carlo, and choose a PDF set without or with the additional positivity
constraint, depending on the needs of their
specific analysis.

\end{enumerate}

We conclude by noting that the listed three categories of applications are not mutually exclusive. 
Different aspects of a given application may belong to either category. 
Since the two representations are based on the {same 
underlying PDF combination}, one can consistently employ both options in different parts of the same calculation
whenever it is convenient. 
Sect.~6 of the 2015 PDF4LHC working group recommendation~\cite{Butterworth:2015oua} includes an example of such an analysis with the mixed delivery options. 
We also remind the reader that, if a large discrepancy between an experimental observation and theoretical calculation based on a PDF4LHC21 ensemble is observed, it may be essential to explore a wider range of individual PDF sets in the comparison to the data, including the most recent PDF ensembles that are not included in the 2021 combination.

\subsection{PDF4LHC21 sets released}

In Table~\ref{tab:pdf4lhc21} we list  the combined PDF4LHC21 sets
presented in this work that are made available from the {\sc\small LHAPDF6} interface.
Additional variants of PDF4LHC21 considered in
this study are also made available in
the {\sc\small LHAPDF} format through the PDF4LHC
Working Group website:
\begin{center}
    \url{https://www.hep.ucl.ac.uk/pdf4lhc/}
\end{center}
All these sets are based on NNLO QCD theory and adopt a variable-flavour-number
scheme with either $n_f^{\rm max}=5$ or $n_f^{\rm max}=4$ as the maximum number of active flavours.
In the latter case, the PDF4LHC21 grids coincide with those of $n_f^{\rm max}=5$ for $Q < m_b$
while for $Q \ge m_b$ they are evolved with 4 active flavours.
Recall that in the {\sc\small LHAPDF6} grids there is always a zeroth member,
so that the total number of PDF members in a given set
is always $N_{\rm mem}+1$.

The grids listed in  Table~\ref{tab:pdf4lhc21} can be used from $Q=m_c=1.41$ GeV up to $Q=100$ TeV
and from $x_{\rm min}=10^{-7}$ to $x_{\rm max}=1$, while outside of this region
one relies on the default {\sc\small LHAPDF6} extrapolation.
Note that the experimental
constraints considered in the PDF4LHC21 combination extend
down to $x\simeq 10^{-5}$, and hence for
smaller values of $x$ the predictions
from each group, and thus for their combination,
depend on methodological choices associated to the treatment of the small-$x$ extrapolation region. 

\begin{table}[tb]
  \centering
  \small
   \renewcommand{\arraystretch}{1.45}
\begin{tabularx}{\textwidth}{lC{1.8cm}C{1.3cm}C{2.5cm}C{1.2cm}C{4.8cm}}
  \toprule
      {\sc\small LHAPDF6} grid name  & Pert. order  &  $n_f^{\rm max}$ & {\tt ErrorType}  & $N_{\rm mem}$   & $\alpha_s(m_Z^2)$ \\\midrule
               {\sc \small PDF4LHC21\_mc}  & NNLO  & 5 &  {\tt replicas}  &   100  & 0.118 \\ \midrule
          {\sc \small PDF4LHC21\_40}  & NNLO  & 5 & {\tt symmhessian}  &   40  & 0.118 \\\midrule
            \multirow{3}{*}{\sc \small PDF4LHC21\_mc\_pdfas}  & \multirow{3}{*}{NNLO} & \multirow{3}{*}{5}  &  \multirow{3}{*}{{\tt replicas+as}}  &   \multirow{3}{*}{102}  & mem 0:100 $\to$ 0.118  \\
          &  &    &  &  & mem 101 $\to$ 0.117  \\
          &  &    &  &  & mem 102 $\to$ 0.119  \\\midrule
          \multirow{3}{*}{\sc \small PDF4LHC21\_40\_pdfas}  & \multirow{3}{*}{NNLO}  & \multirow{3}{*}{5} &   \multirow{3}{*}{{\tt symmhessian+as}}  &   \multirow{3}{*}{42}  & mem 0:40 $\to$ 0.118  \\
          &  &    &  &  & mem 41 $\to$ 0.117  \\
          &  &   & &    & mem 42 $\to$ 0.119  \\\midrule
                 {\sc \small PDF4LHC21\_mc\_nf4}  & NNLO  & 4 &  {\tt replicas}  &   100  & 0.118 \\ \midrule
          {\sc \small PDF4LHC21\_40\_nf4}  & NNLO  & 4 & {\tt symmhessian}  &   40  & 0.118 \\\midrule
            \multirow{3}{*}{\sc \small PDF4LHC21\_mc\_pdfas\_nf4}  & \multirow{3}{*}{NNLO} & \multirow{3}{*}{4}  &  \multirow{3}{*}{{\tt replicas+as}}  &   \multirow{3}{*}{102}  & mem 0:100 $\to$ 0.118  \\
          &  &    &  &  & mem 101 $\to$ 0.117  \\
          &  &    &  &  & mem 102 $\to$ 0.119  \\\midrule
          \multirow{3}{*}{\sc \small PDF4LHC21\_40\_pdfas\_nf4}  & \multirow{3}{*}{NNLO}  & \multirow{3}{*}{4} &   \multirow{3}{*}{{\tt symmhessian+as}}  &   \multirow{3}{*}{42}  & mem 0:40 $\to$ 0.118  \\
          &  &    &  &  & mem 41 $\to$ 0.117  \\
              &  &   & &    & mem 42 $\to$ 0.119  \\
           \bottomrule
\end{tabularx}
\vspace{0.3cm}
\caption{\small Summary of the combined PDF4LHC21 sets
  presented in this work and made available from  {\sc\small LHAPDF6}.
  All these sets are based on NNLO QCD theory and a variable-flavour-number
  scheme with either $n_f^{\rm max}=5$ or $n_f^{\rm max}=4$ as maximum number of active flavours.
  Recall that in the {\sc\small LHAPDF6} grids there is always a zeroth member,
  so that the total number of PDF members in a given set
  is always $N_{\rm mem}+1$.
  See text for usage instructions. \label{tab:pdf4lhc21}}
  \end{table}

\subsection{Uncertainty prescriptions}
\label{subsec:uncer_prescr}
The uncertainty prescriptions 
associated to both the Hessian and MC representations
of PDF4LHC21 follow from the previous PDF4LHC15 study~\cite{Butterworth:2015oua}. 
We here repeat these master formulae for 
completeness, and also extend them to the specific case of negative cross sections.

\subsubsection{Monte Carlo representation \label{sec:uncer_prescr_MC}}

Let us denote by $\mathcal{F} [ \{
  q^{(k)} \}]$ a physical observable computed
with the $k$-th replica of this Monte Carlo set.
The  $X\%$ CL can be estimated as the
difference between the $\frac12(100-X)$ and 
$\frac12(100+X)$
percentiles. This can in turn be computed from the replica sample as follows:

  \begin{enumerate}

  \item Order the $N_{\rm rep}$ predictions for
    the cross-section $\mathcal{F}$ from the smallest
    to the largest:
    \be
    \mathcal{F}_1 \le \mathcal{F}_2 \le \ldots \le
    \mathcal{F}_{N_{\rm rep}-1} \le \mathcal{F}_{N_{\rm rep}} \, .
    \label{eq:presc_pos_1}
    \ee

  \item Now remove $(100-X)\%$ of replicas
    ($\frac{1}{2}(1-\frac{X}{100})\times N_{\rm rep}$ replicas on each side)
    with the highest and with the lowest values of the
    cross-section.

  \item The resulting interval defines the $X\%$ CL
    interval for the cross section $\mathcal{F}$. For example, a 68\%
    CL interval (corresponding to $1\sigma$ for a Gaussian) is obtained by
    keeping the central 68\% replicas, i.e. 
    \be
    \lc \mathcal{F}_{0.16N_{\rm rep}}, \mathcal{F}_{0.84N_{\rm rep}}  \rc
    \ee
      so that a 68\% CL PDF uncertainty on the cross-section
    is given by
    \be\label{eq:68cl}
    \sigma_{\mathcal{F}} = \frac{ \mathcal{F}_{0.84N_{\rm rep}}
      - \mathcal{F}_{0.16N_{\rm rep}} }{2}
    \ee
    Specifically,  for $N_{\rm rep}=100$, 
    the PDF error computed this way is
    \be
    \sigma_{\mathcal{F}} = \frac{ \mathcal{F}_{84}
      - \mathcal{F}_{16} }{2}
      \label{eq:presc_pos_last}
    \ee
    where the indices number cross-section replicas which
    have been ordered in ascending value.
  \end{enumerate}

\paragraph{Positivity.} 
From Fig.~\ref{fig:histo_hmdy_nc} one observes that the {\sc \small PDF4LHC21\_mc} set
can 
lead to a small fraction 
of negative cross-sections
in the deep extrapolation regions sensitive to large-$x$ PDFs.
These can be dealt with by means of a simple modification of the
prescription described above.
The prescription for handling cases in which $\mathcal{F} [ \{
    q^{(k)} \}]<0$ for some MC replica $k$, while physical constraints would
impose  $\mathcal{F}\ge 0$ (in particular if  $\mathcal{F}$ is an
observable cross-section),
is simply
\begin{itemize}
  \item whenever  $\mathcal{F} [ \{
    q^{(k)} \}]<0$, set $\mathcal{F} [ \{
    q^{(k)} \}]=0$.
    \end{itemize}
      Note that this prescription must be applied at the level
      of the specific observable, say a bin in a differential
      distribution, never at the level of PDF replica itself.
      Afterwards confidence level intervals are evaluated
      as customary for Monte Carlo distributions, described above. 
  When applying this positivity prescription, we recommend to use the
  median as the central prediction and the 68\% confidence
  level in Eq.~(\ref{eq:68cl}) (after application of the positivity prescription) as the uncertainty band.
  In the extreme case when the median is zero, this means that only an upper bound can
  be provided on the observable under consideration.

\subsubsection{Hessian representation \label{sec:uncer_prescr_Hessian}}

Now  $\mathcal{F} [ \{
  q^{(k)} \}]$ denotes a physical observable computed
with the $k$-th eigenvector direction of a Hessian 
PDF set. 
The master formula for the propagation of the PDF uncertainties is given by
\begin{equation}
    \delta^{\mbox{\scriptsize{pdf}}}{\cal F} =\sqrt{\sum_{k=1}^{N_{\rm eig}}\left({\cal F}^{(k)}-{\cal F}^{(0)}\right)^2}
    \label{eq:masterformula_hessian}
\end{equation}
and renders the $68\%$ CL uncertainty according to the symmetric Hessian prescription (with one error set per eigenvector direction) as adopted e.g. in the PDF4LHC15 Hessian and the ABM PDFs~\cite{Alekhin:2011sk}.

\paragraph{Positivity.} 
Due to the Gaussian assumption behind the Hessian sets, the nominal PDF uncertainty can extend to negative cross sections at some confidence level.
In this case, we simply truncate the uncertainty interval at zero. 
An example of this prescription can be viewed in Fig.~\ref{fig:histo_hmdy_nc}.

\subsubsection{Combined PDF+$\alpha_s$ uncertainties}

The PDF4LHC21 combination is based on the
following central value and uncertainty of $\alpha_s(m_Z^2)$,
\begin{equation}
\label{eq:alphaserrPDG}
\alpha_s(m_Z^2)=0.1180 \pm 0.0010 \, ,
\end{equation}
at the 68\% CL.
This choice is consistent with the current PDG average~\cite{ParticleDataGroup:2020ssz}.
As indicated in Table~\ref{tab:pdf4lhc21}, we also provide bundled sets
for the PDF4LHC21 Monte Carlo and Hessian variants which include the central fits
for variations of the strong coupling constant $\alpha_s(m_Z^2)$.
The provided range of $\alpha_s(m_Z^2)=0.118\pm 0.001$ can be used to evaluate
the $1\sigma$ combined PDF+$\alpha_s$ while
keeping track of the underlying cross-correlations in a consistent manner.

The computation of the combined
PDF+$\alpha_s$ uncertainties in PDF4LHC21 is based
on the following prescription.
It is recommended that PDF+$\alpha_s$ uncertainties are
determined by first computing the PDF uncertainty for the central
$\alpha_s$ value, following the corresponding MC or Hessian prescription described above,
then computing the predictions for the upper and lower values
of $\alpha_s$, consistently using the corresponding PDF sets, and then
adding these results in quadrature~\cite{Lai:2010nw,Demartin:2010er}.
Specifically, for a given cross-section $\sigma$, the
$\alpha_s$ uncertainty can be computed, using the same value of $\alpha_s(m_Z^2)$ in 
the partonic cross-sections and in the PDFs, as:
\begin{equation}
\label{eq:aserr}
\delta^{\alpha_s} \sigma = \frac{\sigma(\alpha_s=0.119)-\sigma(\alpha_s=0.117)}{2} \, ,
\end{equation}
corresponding to an uncertainty $\delta\alpha_s=0.001$ at the 68\% confidence
level.
Note that 
Eq.~(\ref{eq:aserr}) is to be computed with central values
of the corresponding PDF4LHC21 sets.
The combined PDF+$\alpha_s$ uncertainty is then computed as
\begin{equation}
\label{eq:pdfaserr}
\delta^{\rm PDF+\alpha_s}\sigma = \sqrt{
  (\delta^{\rm pdf}\sigma )^2
  + (\delta^{\alpha_s}\sigma )^2} \, .
\end{equation}
Note that, for the PDFs in the Hessian representation, this means that the two extra PDFs provided with $\alpha_s=0.117,0.119$ can be effectively regarded as an extra pair of eigenvectors, with the uncertainty $\delta^{\alpha_s}$ computed according to Eq.~(\ref{eq:aserr}). However, for the PDFs provided in the Monte Carlo representation, they are 
not simply two extra replicas -- the PDF uncertainty must be calculated first, and then the 
formula in Eq.~(\ref{eq:pdfaserr}) be applied.

If it is necessary, the result for any other value of
$\delta\alpha_s$,
as compared to the baseline Eq.~(\ref{eq:alphaserrPDG}),
can be obtained
from a trivial rescaling of Eq.~(\ref{eq:aserr}) assuming linear
error propagation.
That is, if we assume a different value for the uncertainty in $\alpha_s$,
\begin{equation}
\tilde{\delta}\alpha_s = r \cdot \delta\alpha_s\, , \qquad
\delta \alpha_s=0.001 \, ,
\end{equation}
then the combined PDF+$\alpha_s$ uncertainty Eq.~(\ref{eq:pdfaserr})
needs to be modified as follows
\begin{equation}
\label{eq:pdfaserr2}
\delta^{\rm PDF+\alpha_s}\sigma = \sqrt{
  (\delta^{\rm pdf}\sigma )^2
  + ( r\cdot \delta^{\alpha_s}\sigma )^2 
} \, .
\end{equation}
which differs from Eq.~(\ref{eq:pdfaserr})
by the rescaling factor $r$ in the second term.

\section{Summary and outlook}
\label{sec:conclusions}

In this final section, we summarise the main findings
of the benchmarking and combination efforts that
have culminated in PDF4LHC21.
We also present a brief outlook about possible future directions
in the benchmarking and combination of global PDFs, and their implications
for the upcoming runs of the LHC.

\paragraph{Summary.}
There are several key take-away messages that may be derived from the outcome of the present study.
The first one is that, after the three groups had carried out the extensive benchmarking of the theory calculations and implementations of experimental datasets, the major differences between the full fits were well understood. 
For example, for the inclusive $Z$ cross section in Fig.~\ref{fig:incl_xsec_pdfsets}, the main remaining difference is due to the fact that the high precision ATLAS 7 TeV $W$, $Z$ data are not included in CT18, by choice.
This is an example of the type of choices that have to be made in PDF fits in situations where a data set provides useful information but may conflict with the other data sets included in the fit. 
All three global PDF fits observe relatively similar types of conflicts, although the corresponding interpretation 
and treatment differs in general for each group.
Nevertheless, even after this benchmarking analysis, there remain mild differences between the three global fits both in terms of central values and of uncertainties. 

The combined uncertainties of PDf4LHC21 reflect both the specific PDF uncertainties of each group and, in some cases, offsets in the corresponding central values.
Some of these differences have decreased since PDF4LHC15, and a few have increased. 
The input PDF uncertainties depend on the choice  of the tolerance (or effective tolerance) regarding what constitutes a good fit, and on different responses to any tensions that may exist in the data sets. 
In addition, both CT18 and MSHT20 have loosened their parametrisations compared to the previous generation of the PDFs, which can also lead to increased uncertainties for specific PDFs, such as the $d$-quark. 
Each collaboration has developed tools to examine these considerations in detail. The differences, in general, represent valid physics choices. The differences in the central values can be due to the selection of the fitted data sets and the variable sensitivity of the global fits to the common data sets in the input fits.

The PDF4LHC21 combination is found to be in good agreement with PDF4LHC15.  
Despite the large number of changes that the three constituent sets have undergone subsequently to the previous combination, PDF4LHC21 not only agrees
within uncertainties with PDF4LHC15 in the whole kinematic range relevant
for the LHC, it also benefits from a modest reduction in the critical PDF uncertainty in the gluon sector. 
Namely, the uncertainty on the gluon distribution and the related $gg$ PDF luminosity went down in the region crucial for Higgs boson production through gluon-gluon fusion. Six years ago, the three global PDFs that went into PDF4LHC15 (CT14, MMHT14 and NNPDF3.0) had gluon distributions (and $gg$ PDF luminosities) that agreed very well both in terms of the central values and uncertainties. 
The PDF4LHC15 gluon uncertainty band thus had a value very similar to each of the individual gluon PDFs. This was viewed at the time as plausibly a fortuitous accident, and indeed the agreement is not as close in PDF4LHC21. Nevertheless, the gluon uncertainty bands for the individual PDFs have decreased, resulting in a small reduction in the corresponding uncertainty for PDF4LHC21.

The differences between PDF4LHC21 and PDF4LHC15 are summarized in 
Figs.~\ref{fig:pdf4lhc21_vs_pdf4lhc15_pdfs} and~\ref{fig:pdf4lhc21_vs_pdf4lhc15_errors}. 
When the uncertainties have mildly increased compared to the PDF4LHC15, it is sometimes due to the improved
fitting methodology, such as the more flexible down quark parametrisation in MSHT20 and the fitted charm PDF in NNPDF3.1. 
In the case of strangeness, the underlying cause is the inclusion of the high precision ATLAS $W$ and $Z$  production data at 7 TeV in MSHT20 and NNPDF3.1$^\prime$, while these data are not included  in  CT18$^\prime$ as discussed in Sect.~\ref{subsec:ct18}.  

One of the first steps in the benchmarking exercise was to choose a  reduced common set of data, in which differences due to data selection are minimised. Good agreement was observed in most cases between the resultant CT18 and MSHT20 reduced PDF uncertainties, with the NNPDF3.1 reduced PDFs exhibiting somewhat smaller uncertainties in general, as, for example, for the gluon distribution at low $x$.
The use of the full experimental data sets leads to a reduction of the PDF uncertainties, with MSHT20 showing the largest decrease (with respect to the results with the reduced data set), and CT18 showing the least decrease.
A number of studies were carried out to further understand these differences, for instance related
to the choice of tolerance in the Hessian sets.
A detailed report is beyond the scope of this publication and will be presented in the future.  
Continued improvements in PDFs for the LHC will depend not only on the implementation of new data sets as they become available, but also a continuation of benchmarking
and methodological studies such as those described here.

These considerations, combined with the user-friendliness of the provided Hessian and Monte Carlo
representations, make PDF4LHC21 a suitable replacement for PDF4LHC15 in providing a conservative
estimate of PDF uncertainties associated to LHC processes.
This said, we emphasise that the availability of  PDF4LHC21 does not preclude the use of individual sets, on
the contrary,
a detailed understanding of the reasons for discrepancies between LHC measurements and theory predictions
necessarily requires inspection of the predictions associated to individual PDF sets and not only of their combination.
 Sect.~\ref{sec:usage} offers general guidelines to assist users in deciding when computations with the PDF4LHC21 combined sets are warranted and lead to substantial simplifications as compared to using the individual input PDF sets. 
 This section also provides the formulae required
 to evaluate the PDF and QCD coupling uncertainties with the Hessian and Monte Carlo PDF4LHC21 sets.

\paragraph{Outlook.}
There are several directions in which these studies could be expanded in future work.
To begin with, one could extend the analyses of reduced PDF fits discussed in Sect.~\ref{sec:benchmarking} and App.~\ref{app:specifics} and \ref{app:l2_sensitivity}
by adding more data to the common dataset.
An even wider ``reduced'' dataset would further highlight which  differences between the three PDF ensembles and PDF sets are traced back to the underlying methodological choices.
In the longer term, eventually the PDF4LHC21 combination will have to be updated once new releases
from the various PDF fitting collaborations are presented.
Furthermore, at some point 
this combination may have to consider 
additional correlations and uncertainties, such as missing higher-order uncertainties, which may be strongly correlated between the groups.

The ultimate hope of these benchmarking and combination efforts
is that progress in experimental measurements and higher-order theory calculations,
combined with better understanding of the different methodological choices adopted
by each group, will result in future PDFs with enhanced precision and accuracy for LHC phenomenology, suitable to match the requirements
of the High-Luminosity LHC era.
\section*{Acknowledgements}
%
We are grateful to the members of the PDF4LHC community for
many useful discussions in the course of the last two years.
We also thank Christopher Schwan for assistance with the computation of the high invariant mass Drell-Yan.
T.~C. and R.~S.~T. thank the Science and Technology Facilities Council (STFC) for support via grant awards ST/P000274/1 and ST/T000856/1.
L.~H.~L. thanks STFC for support via grant awards ST/L000377/1 and ST/T000864/1. R.D.B. and E.R.N. thank the STFC for support by the grant awards ST/P000630/1 and ST/T000600/1.
P.~M.~N. is partially supported
by the U.S. Department of Energy under Grant
No. DE-SC0010129.
J.~R. is partially supported by the Dutch Research Council (NWO).
The research of T.~R. has been partially supported by an ASDI grant
of The Netherlands eScience Center.
The work of CPY is partially supported by the U.S.~National Science Foundation under Grant No.~PHY-2013791. CPY is also grateful for the support from the Wu-Ki Tung endowed chair in particle physics.
A.~C. is supported by UNAM Grant No. DGAPA-PAPIIT IN111222 and CONACyT -- Ciencia de Frontera 2019 No.~51244 (FORDECYT-PRONACES).
The work of K.X. is supported by U.S. Department of Energy under grant No. DE-SC0007914, U.S. National Science Foundation
under Grant No. PHY-2112829, and in part by the PITT PACC. 
A.~C-S. acknowledges financial support
from the Leverhulme Trust. 
The work of T.~J.~Hobbs was supported
by Fermi Research Alliance, LLC under Contract No.~DE-AC02-07CH11359 with the U.S.~Department of Energy, Office of Science, Office of High Energy Physics.
At earlier stages, support at SMU was provided by the U.S.~Department of
Energy under Grant No.~DE-SC0010129 as well as by a JLab EIC Center Fellowship.
SF is supported by the European Research Council under the European Union’s Horizon 2020 research
and innovation Programme (grant agreement n.740006).
%

\appendix

\section{PDF compression and reduction techniques}
\label{app:tools}

In this appendix we review the Monte Carlo compression and Hessian reduction 
techniques that are applied to the PDF4LHC21 combination in order to produce
the {\sc\small LHAPDF} grids that are publicly released.
We also justify the choices for the number of MC replicas and of Hessian eigenvectors for the 
delivered sets, by means of a series of comparisons at the level of both PDFs
and of LHC cross-sections.

\subsection{Compression of Monte Carlo replicas}
\label{subsec:mccompression}

The aim of compressing a Monte Carlo PDF set is to extract a subset of the replicas 
that most faithfully reproduces the statistical properties of the original distribution.
The underpinning idea behind such a compression is to find a subset such that the statistical distance 
between the original and the compressed set is minimal in order to ensure that the loss of information 
is as negligible as possible. As mentioned before, a compression
methodology has to rely on two main ingredients: a proper definition of a distance metric
that quantifies the distinguishability between the original and the compressed distribution,
and an appropriate minimisation algorithm that explores the space of possible combinations
of PDF replicas which leads to such 
minima.

\begin{figure}[!t]
	\centering
	\includegraphics[width=.55\linewidth]{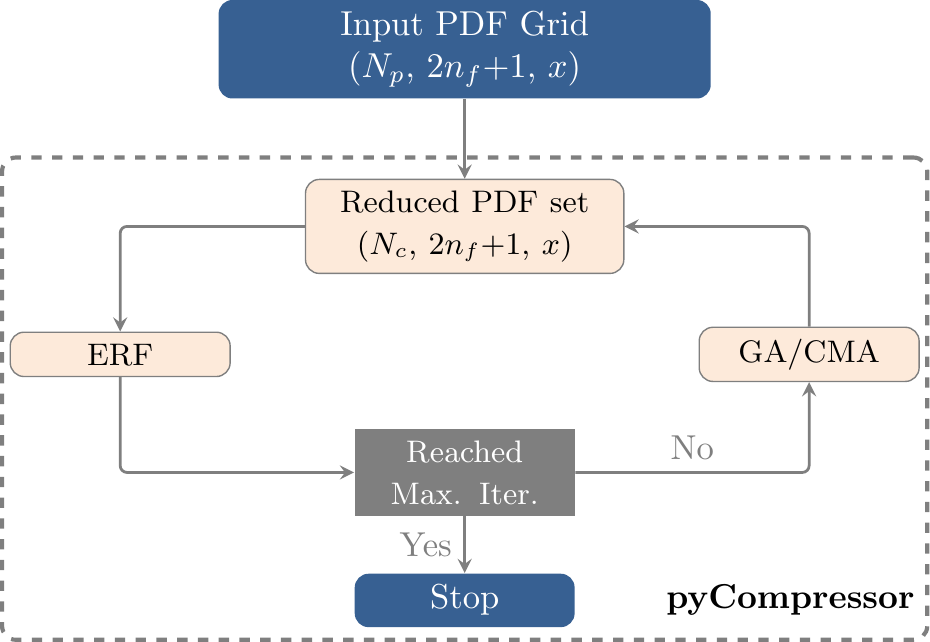}
	\caption{Flowchart describing the compression methodology. The input is a Monte
		Carlo PDF with $N_p$ replicas evaluated as a multidimensional grid at an 
		scale of $Q = 2~\mathrm{GeV}$. The output is a reduced Monte Carlo replica PDF set
		with size $N_c \ll N_p$ that follows the {\sc\small LHAPDF} format.}
	\label{fig-app:compressor}
\end{figure}

In order for a compressed set to faithfully reproduce the underlying probability distribution of
a given input PDF, the definition of the distance metric has to take into account both lower and higher
moments. It was originally proposed in~\cite{Carrazza:2015hva} that a suitable figure 
to measure the difference between two parton density functions is the error function,
\begin{equation}
	\mathrm{ERF}(Q) = \frac{1}{N_\mathrm{EST}} \sum_k \frac{1}{N_k} \sum_{i}
	\left( \frac{C^k (x_i, Q) - P^k (x_i, Q) }{P^k (x_i, Q)} \right)^2 ,
	\label{eq:erf}
\end{equation}
where $k$ runs over a set of statistical estimators with $N_k$ the associated normalisation
factor, $P^k (x_i, Q)$ is the value of the estimator $k$ computed at a scale $Q$ and at a
given point $i$ in the $x$-grid for the original set, and $C^k (x_i, Q)$ is the corresponding
value of the same estimator but for the compressed distribution. Finally, $N_\mathrm{EST}$
denotes the total number of statistical estimators involved in the minimisation. As
statistical estimators, we include in~Eq.~\ref{eq:erf} lower moments (such as mean and
standard deviation), and standardised moments (such as skewness and kurtosis).
In addition,
in order to preserve higher-moments and PDF-induced correlations, which are crucial for
phenomenological studies, we include the Kolmogorov-Smirnov and the correlation between
PDF flavours. It is important to emphasise that for each statistical estimator, the normalisation
factor has to be defined properly in order to compensate for the various orders of magnitude
in different regions of $(x,Q)$-space that mainly arise in higher moments. For a detailed 
description of the various statistical estimators mentioned above with the explicit definition 
of their normalisation factors, we refer the reader to~\cite{Carrazza:2015hva}.

Once the set of statistical estimators and the target size of the compressed set are defined,
the compression algorithm can perform the minimisation procedure by searching for the
combination of PDF replicas that leads to the lowest value of the error function defined
in Eq.~\ref{eq:erf}. There exists various minimisation algorithms to perform the selection
of replicas entering the compressed set, however, due to the discrete nature of the compression 
problem, it is suitable to use Evolution-based minimisation algorithms such as the 
Covariance Matrix Adaptation (CMA) strategy or the Genetic Algorithm (GA). From a practical 
point of view, both minimisation algorithms are equally good and choosing one or the other will 
yield the same result. It is indeed important to emphasise that no matter the choice of minimisation, 
there is no risk of overfitting since the absolute minimum always exists.

In Fig.~\ref{fig-app:compressor} we show a diagrammatic representation of the compression 
workflow. The compression goes as follows. First, we compute a grid of the input PDF for all 
replicas and flavours at every point of the $x$-grid.  The $x$-grid is constructed such that the 
points are only distributed in the region where experimental data are available. In addition, we
choose $x$ to be composed of points that are logarithmically spaced between $\left[
10^{-5}, 10^{-1}\right]$ and linearly spaced in $\left[10^{-1}, 0.9\right]$. Then we 
choose a random subset from the input PDF for a given target size in order to compute the value 
of the error function; this process is repeated for a certain number of times until the maximum
number of iterations is reached--making sure that the minimiser selects a better sample at every 
iteration. The methodology described above was first implemented in a C$++$ code~\cite{Carrazza:2015hva}
and used in the previous PDF4LHC combination~\cite{Butterworth:2015oua}. This was later 
re-implemented and improved in~\cite{Carrazza:2021hny} using state-of-the-art python development.

\begin{figure}[!t]
	\centering
	\includegraphics[width=0.32\textwidth]{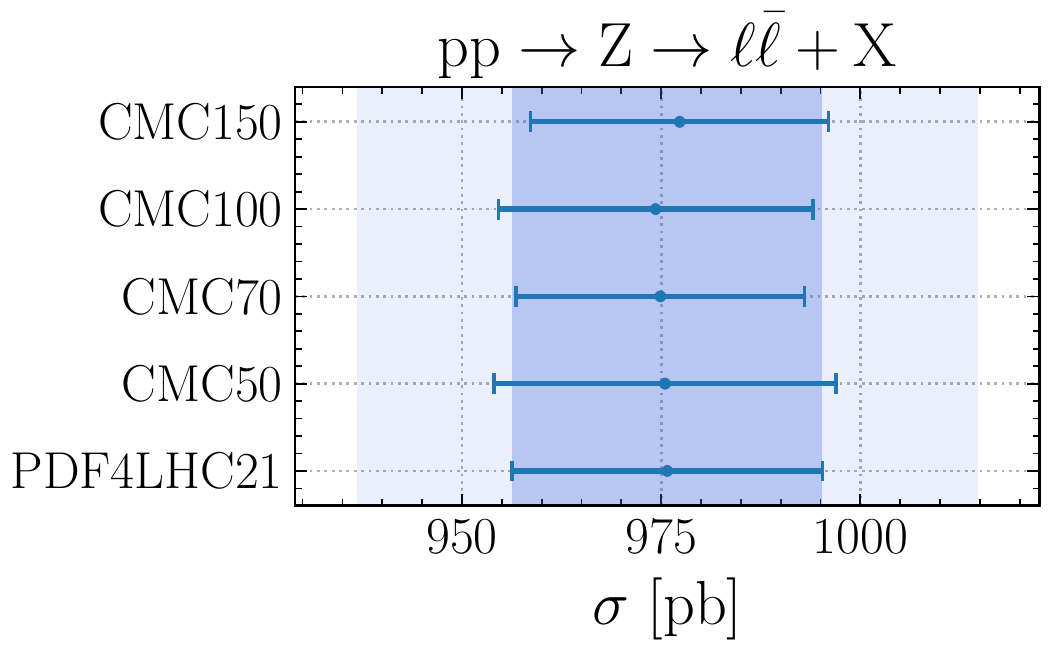}
	\includegraphics[width=0.32\textwidth]{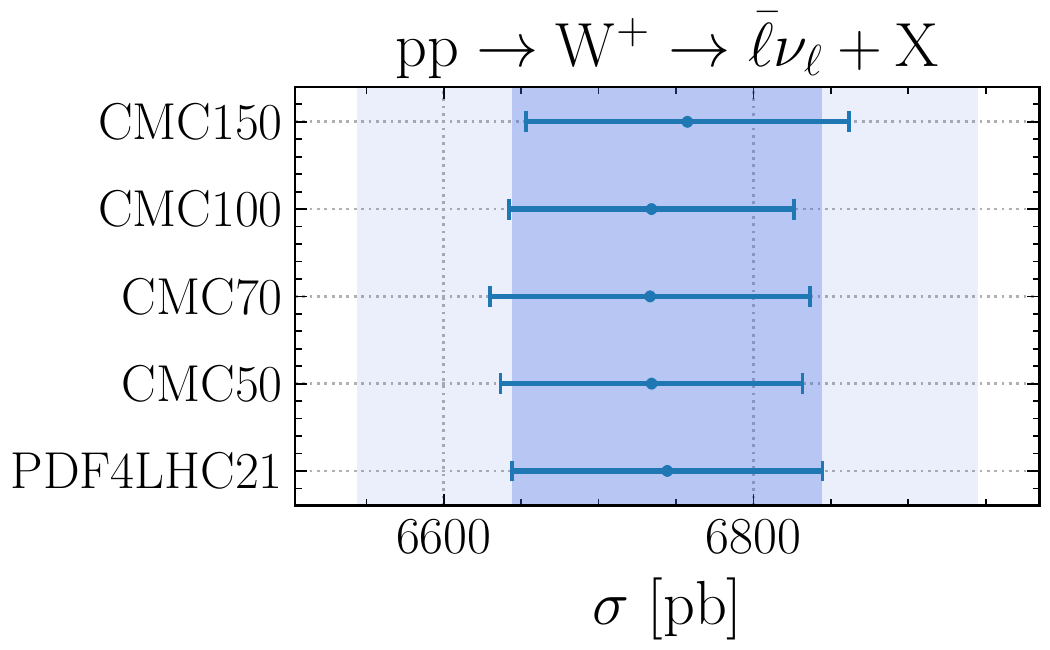}
	\includegraphics[width=0.32\textwidth]{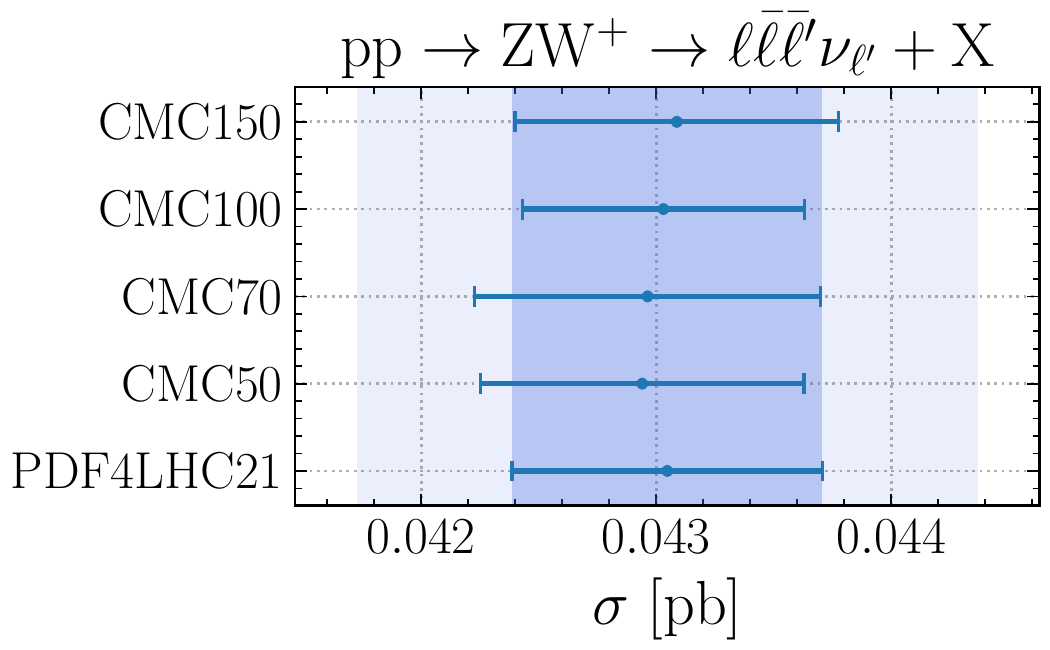}
	\includegraphics[width=0.32\textwidth]{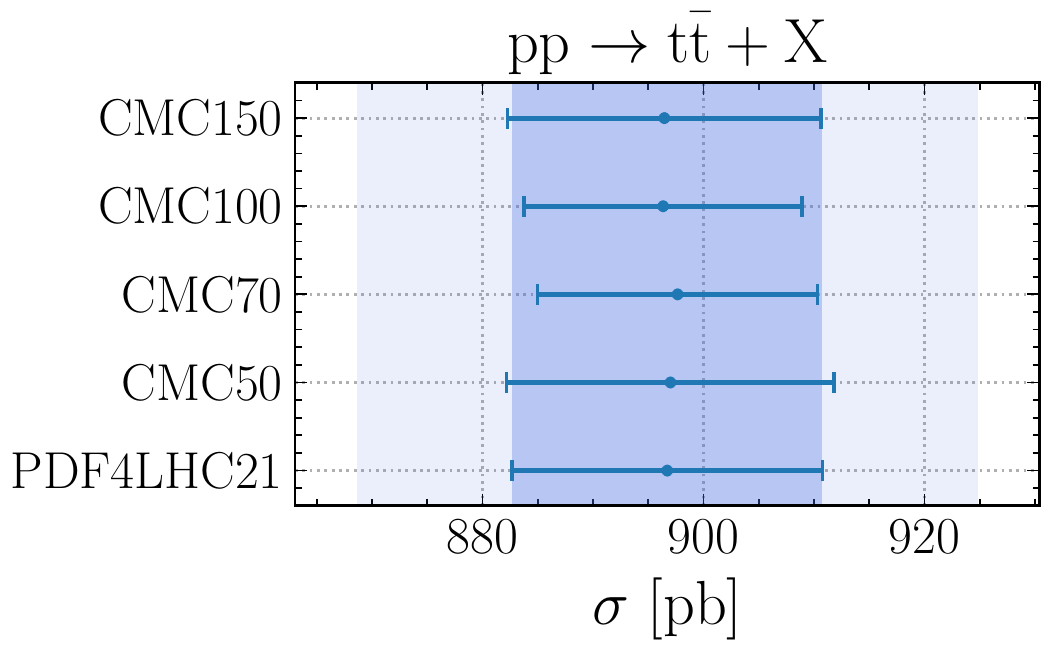}
	\includegraphics[width=0.32\textwidth]{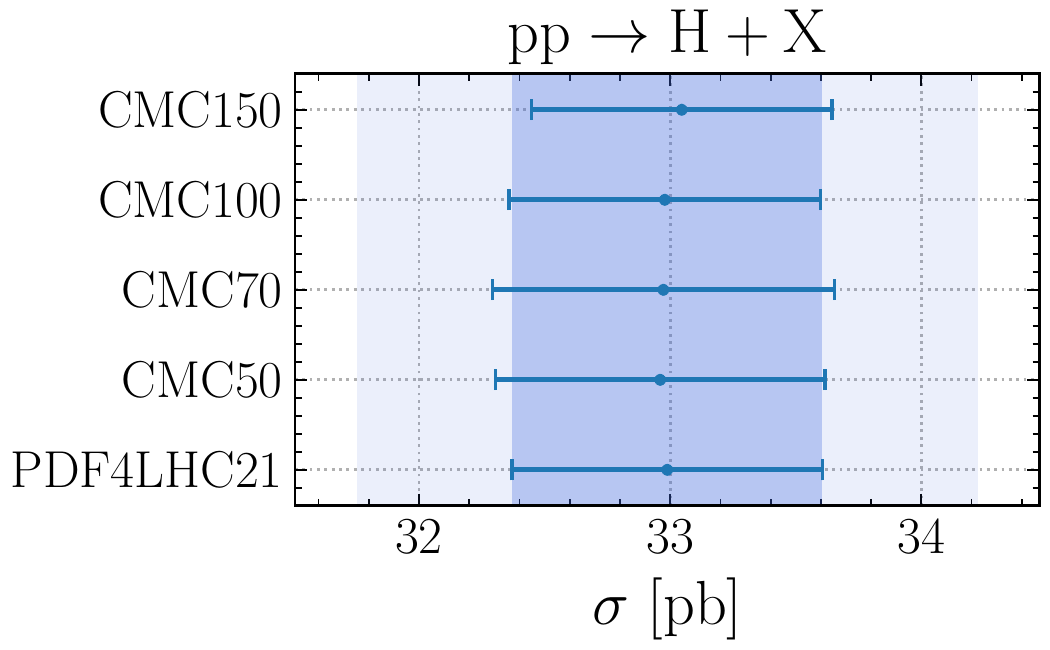}
	\includegraphics[width=0.32\textwidth]{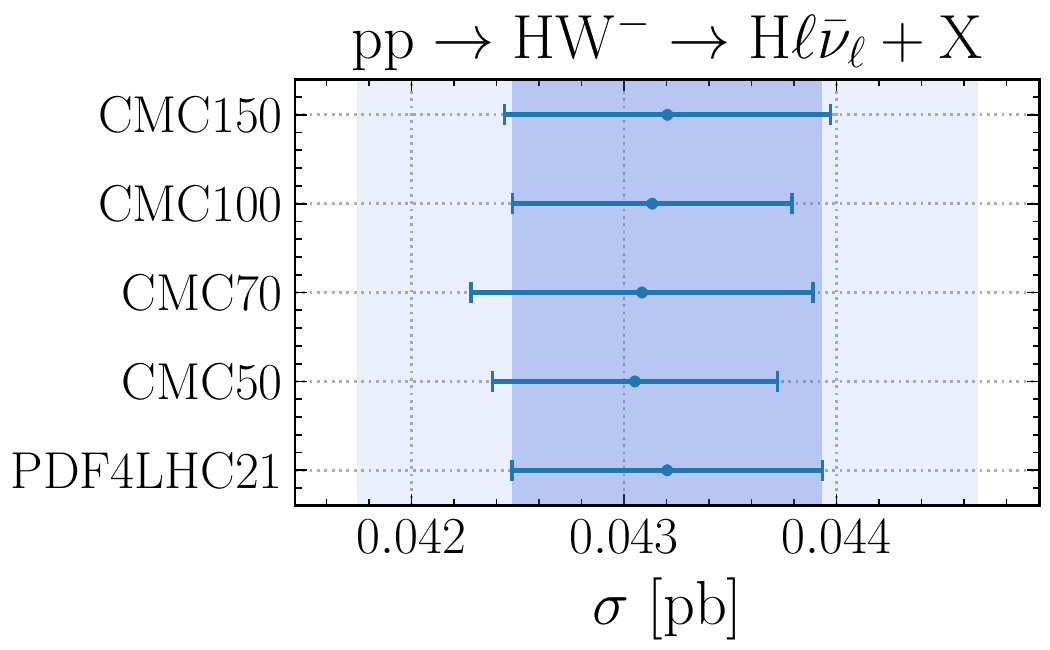}
	\caption{\small Same as Fig.~\ref{fig:pheno-reduction}, now comparing the PDF4LHC21 baseline
		($N_{\rm rep}=900$) with  various Monte Carlo compressed sets 
		(with $N_{\rm rep}=50, 70, 100, 150$ replicas respectively).
	}
	\label{fig:pheno-cmc-reduction}
\end{figure}

\begin{figure}[!b]
	\centering
	\includegraphics[width=.45\linewidth]{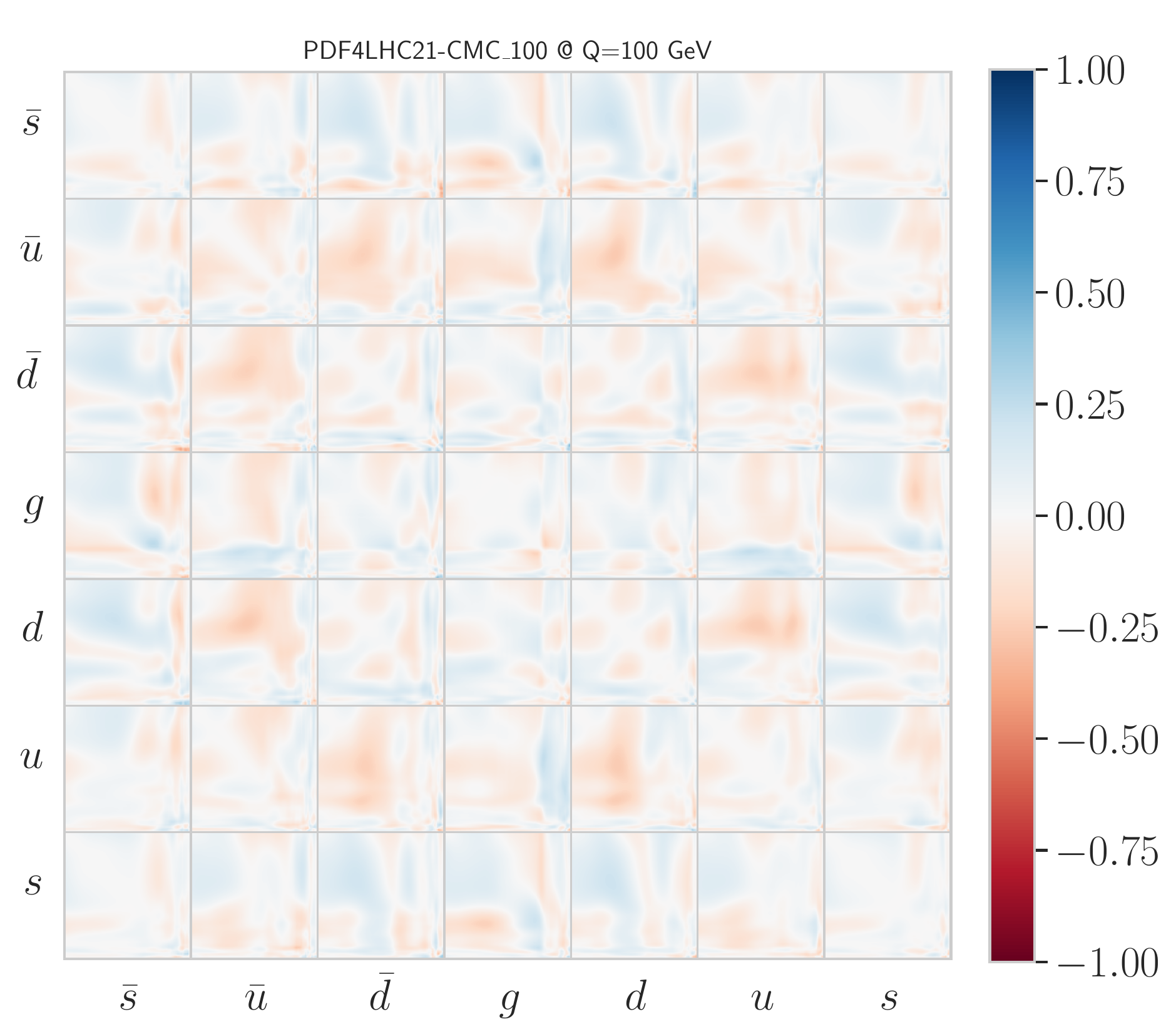}
	\includegraphics[width=.45\linewidth]{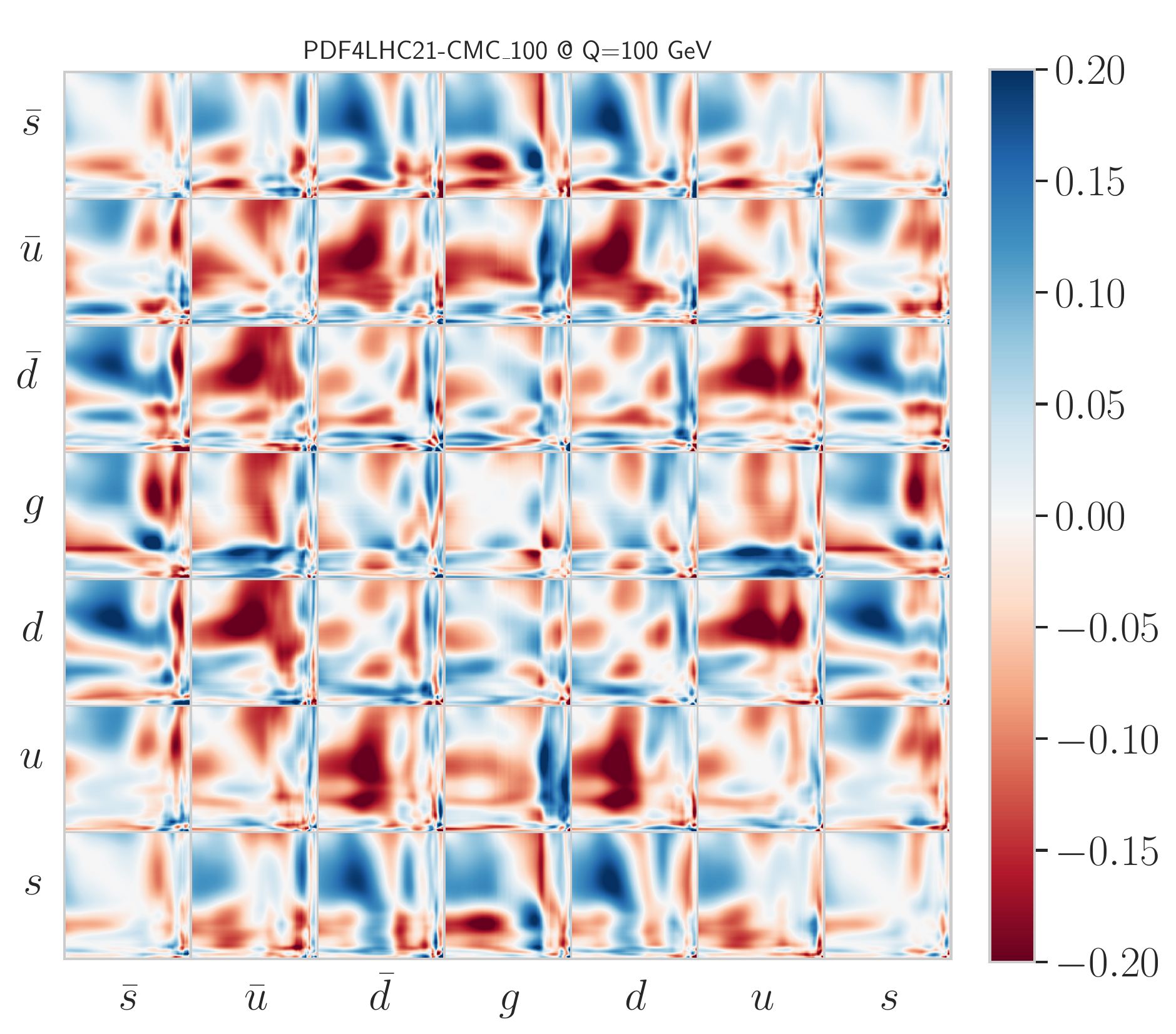}
	\caption{Absolute differences between the PDF correlation matrices of the PDF4LHC21 and the compressed 
	  set ($N_{\rm rep}=100$) in the complete range (left) and in a zoomed range (right panel).
        We show the $7{\times}7$ matrix for all possible combinations of PDF flavours computed in the 
	region of $x$ ranging from $x_\mathrm{min}=10^{-5}$ and $x_\mathrm{max}=0.9$ at fixed 
	$Q=100~\mathrm{GeV}$.}
	\label{fig-app:cmc-corr}
\end{figure}

\begin{figure}[!b]
\centering
\includegraphics[width=0.48\textwidth]{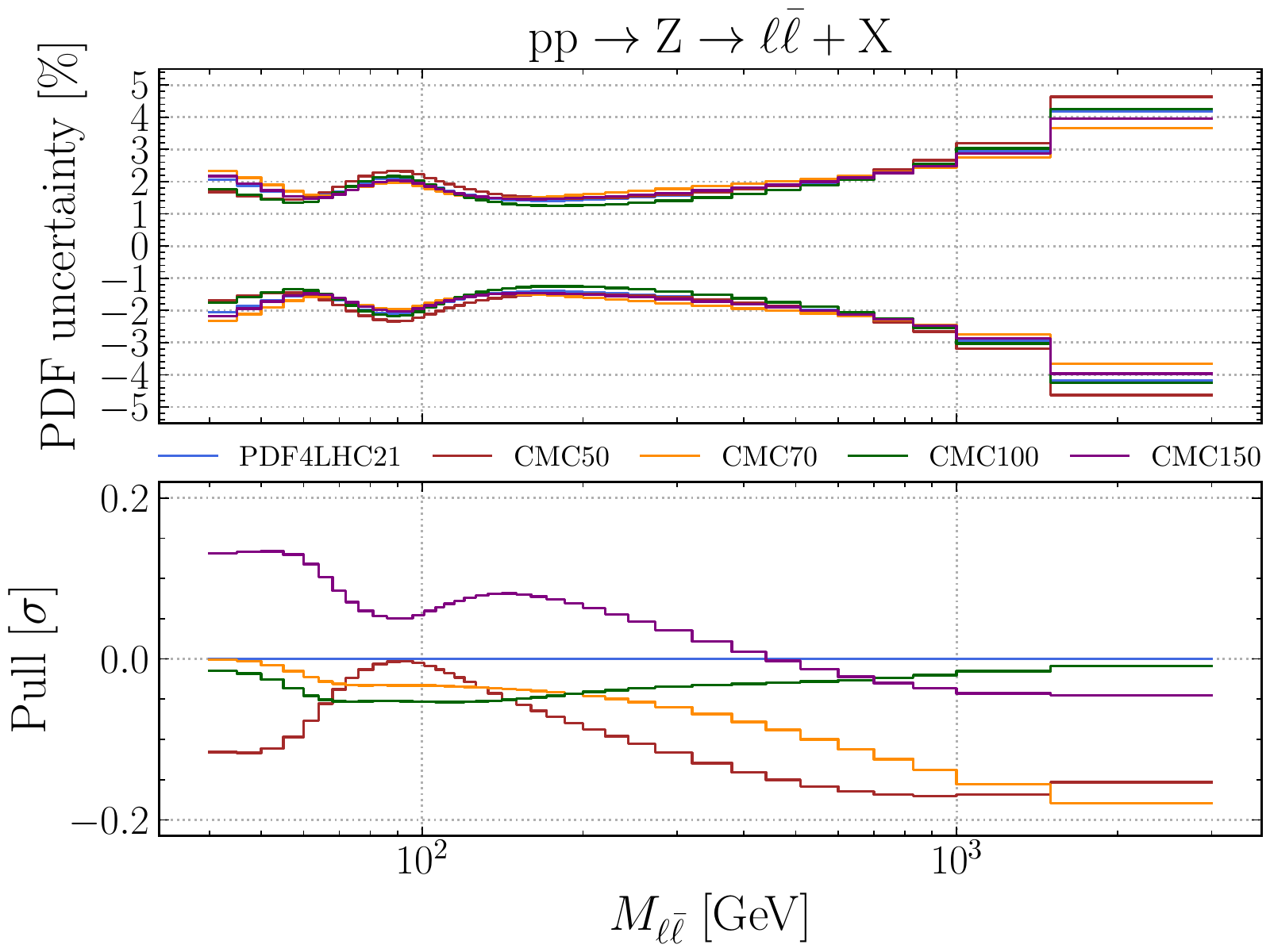}
\includegraphics[width=0.48\textwidth]{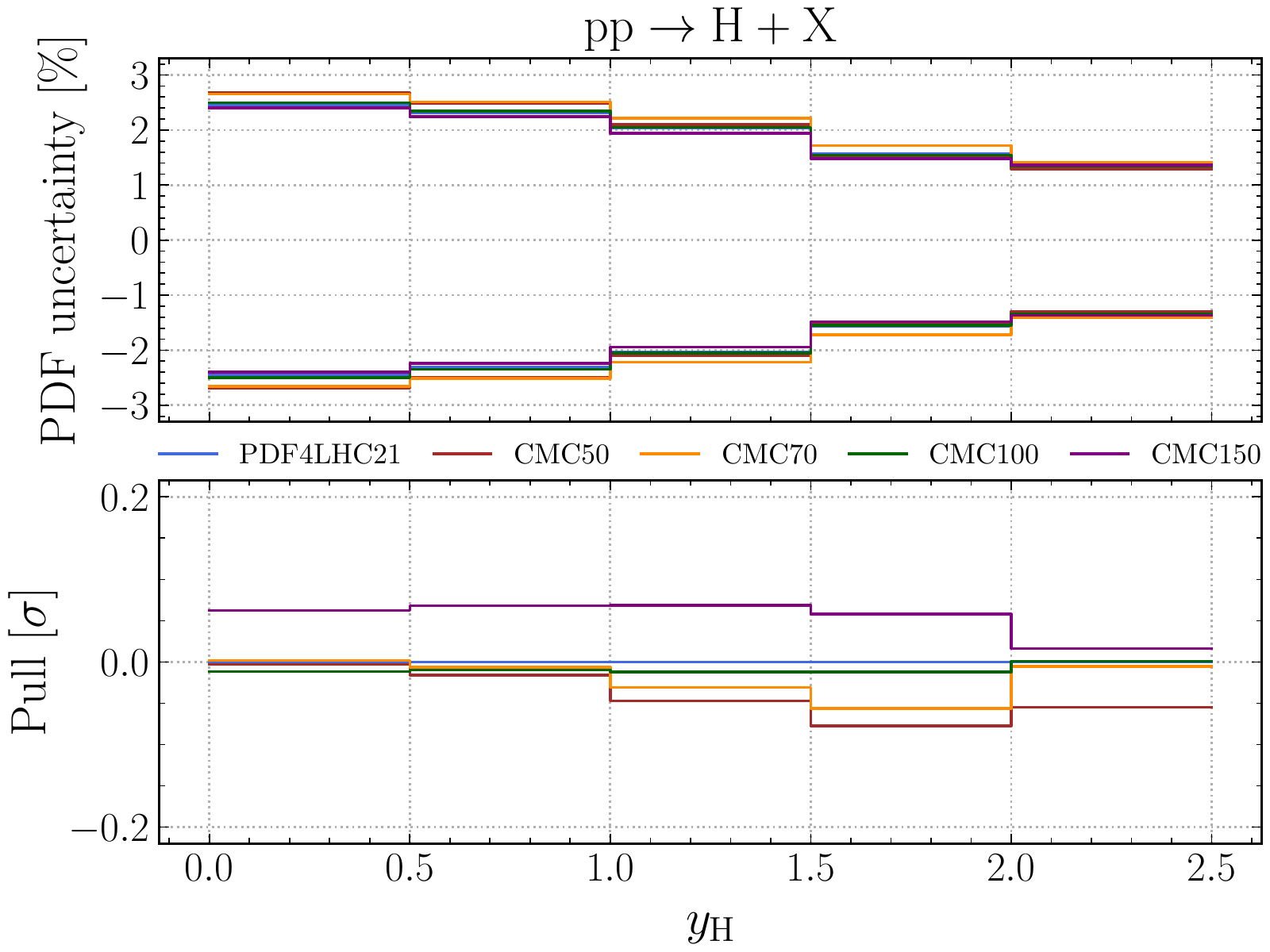}
\includegraphics[width=0.48\textwidth]{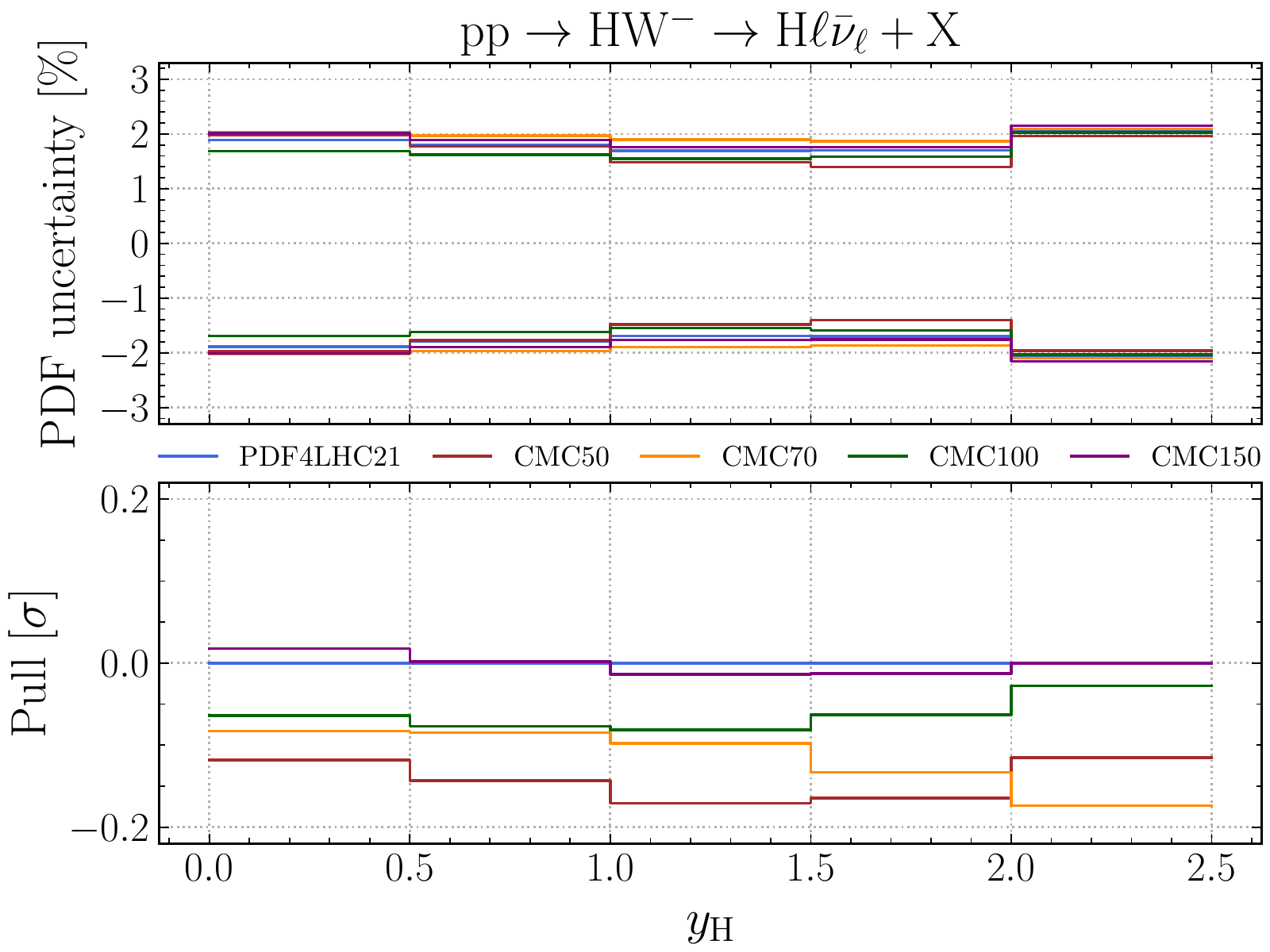}
\includegraphics[width=0.48\textwidth]{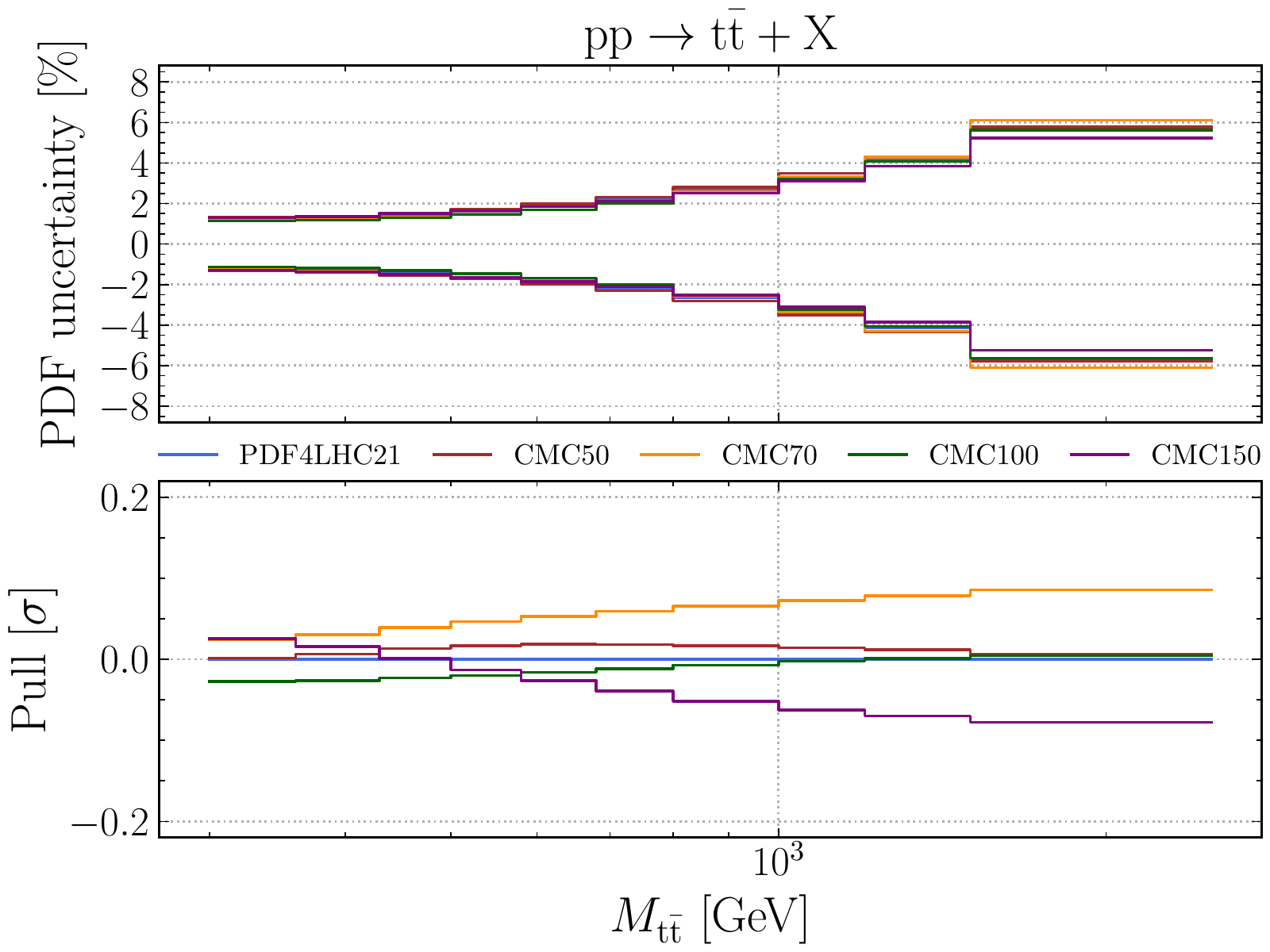}
\includegraphics[width=0.48\textwidth]{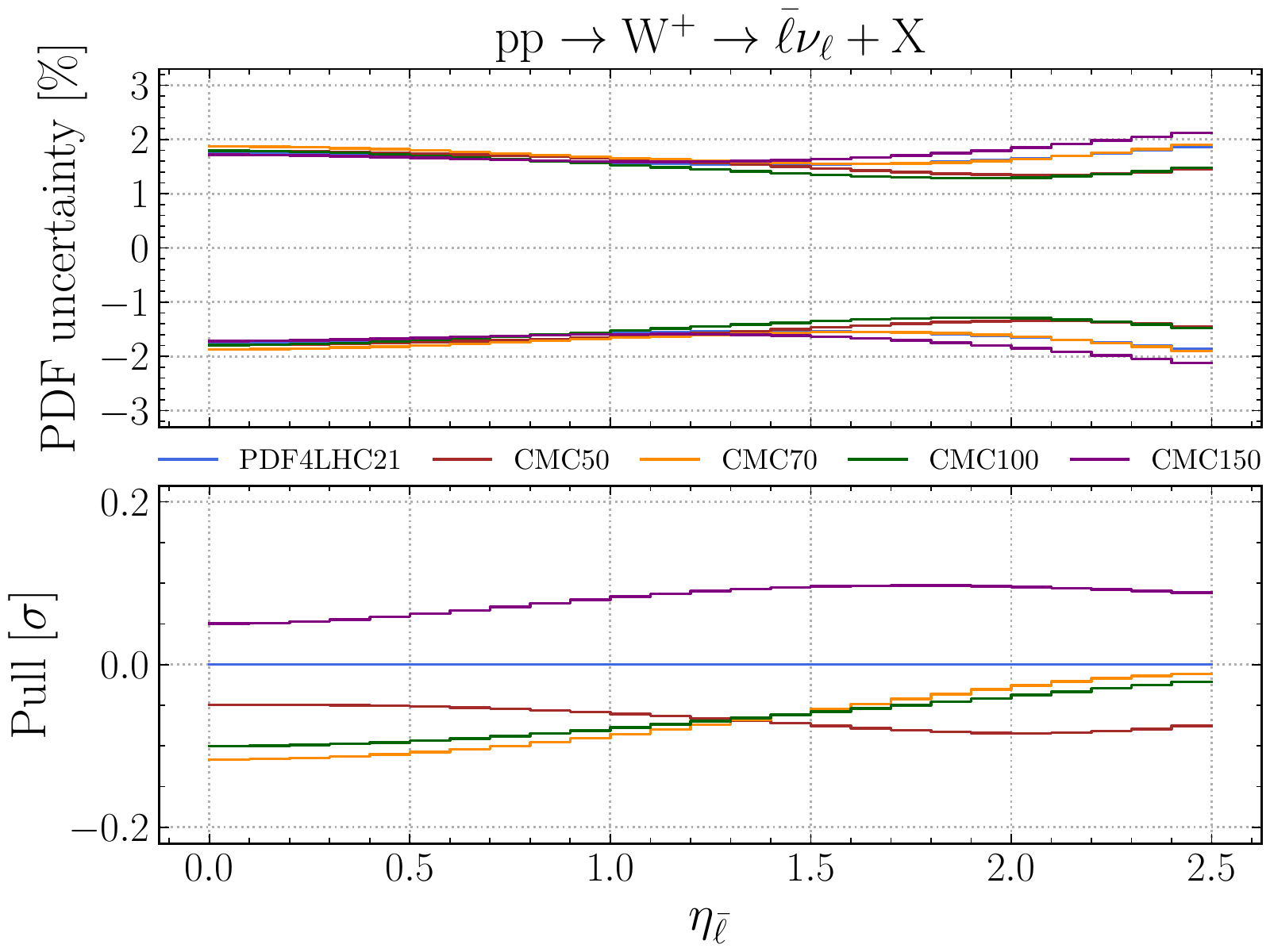}
\includegraphics[width=0.48\textwidth]{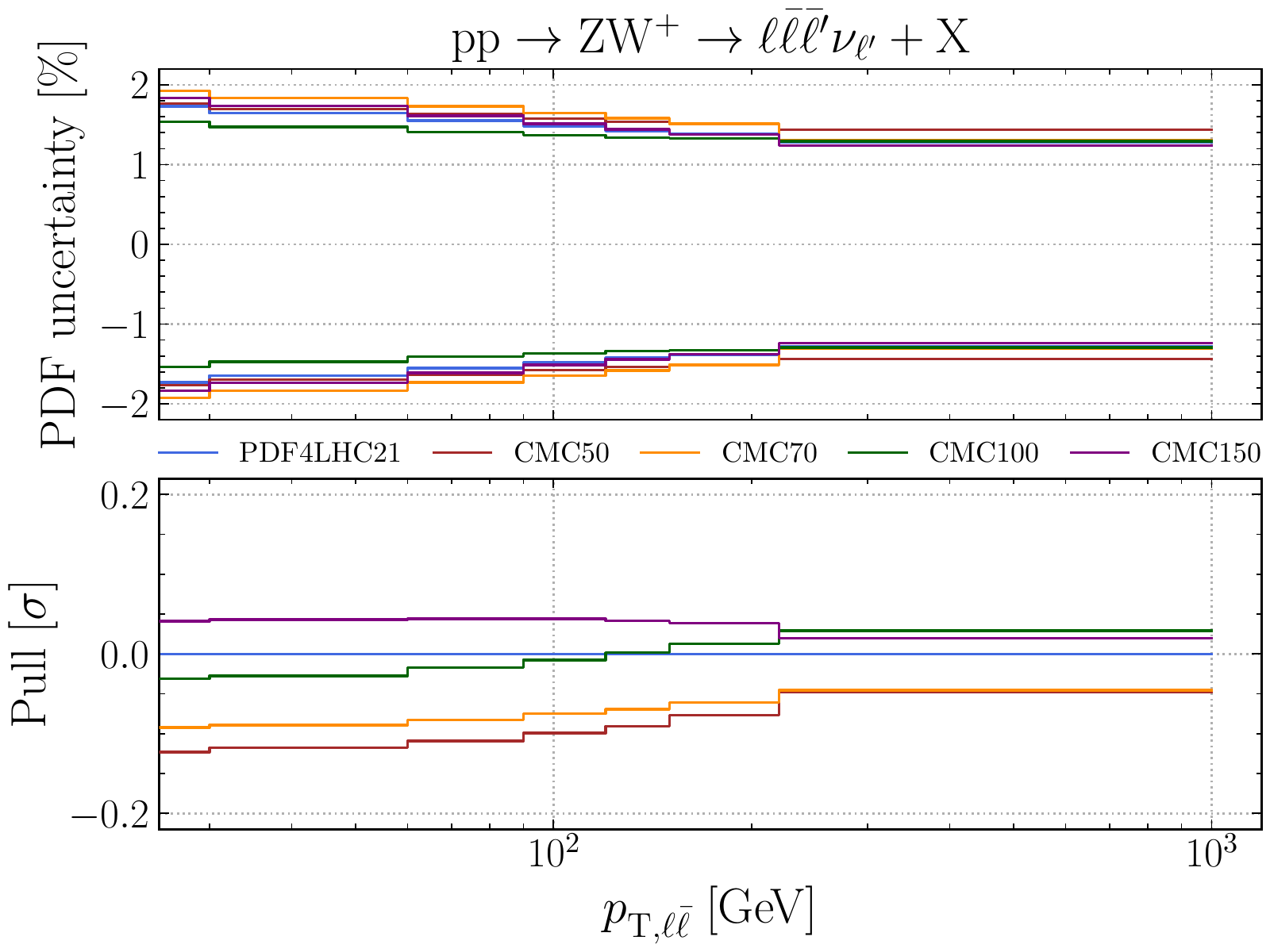}
\caption{\small Same as Fig.~\ref{fig:pheno-reduction-differential} but comparing PDF4LHC21 
    ($N_{\rm rep}=900$) with the various Monte Carlo compressed sets 
    ($N_{\rm rep}=50, 70, 100, 150$).
}
\label{fig:pheno-cmc-reduction-differential}
\end{figure}

In order to decide on the  number of Monte Carlo replicas  to be delivered, we compare at the cross-section 
level various sizes of the compressed set. In Fig.~\ref{fig:pheno-cmc-reduction} we plot the integrated
cross-section for the same processes presented in Fig.~\ref{fig:pheno-reduction}. For each process, we 
compare the PDF4LHC21 set ($N_{\rm rep} = 900$) against various compressed sets with $N_{\rm rep}=50, 70, 100$ 
and $150$ replicas. The same plots are shown in Fig.~\ref{fig:pheno-cmc-reduction-differential} but at the
differential level. In general, we observe good agreement between PDF4LHC21  and the compressed sets
at the $1\sigma$ level. Whilst slight differences can be seen between $N_{\rm rep}=50,70$ and $100$ replicas,
the compressed sets with $100$ and $150$ replicas are almost indistinguishable. This is more apparent when
looking at the differential distributions.
These results suggest that a compressed set with $N_{\rm rep}=100$
replicas sufficiently reproduces the underlying statistical properties of the baseline PDF4LHC21. For completeness 
we show in Fig.~\ref{fig-app:cmc-corr} the difference in the correlation matrix between the original and the delivered 
compressed set ($N_{\rm rep} = 100$) for all combinations of PDF flavours. In order to compute the correlation, we 
considered light flavour PDFs (three quarks and three antiquarks and the gluon), sampled in the region
where experimental data are available at $Q=100~\mathrm{GeV}$. The plot on the left shows the
difference in the range $\left[-1,1\right]$ while the one on the right zooms in the range 
$\left[-0.2,0.2\right]$. The results show very good agreement between the original and the compressed
set with differences never exceeding $0.2$ in modulus, which is in accordance with the expected
range that the correlation coefficients should vary.

The results presented in Figs.~\ref{fig:pheno-cmc-reduction} and~\ref{fig:pheno-cmc-reduction-differential}
are consistent with the corresponding message obtained from comparisons at the level of PDFs
and partonic luminosities.
Hence, we conclude that a compressed set with $N_{\rm rep}=100$ replicas provides an accurate
and faithful representation of the PDF4LHC21 distributions, and is henceforth adopted
by default in this work.

\begin{figure}[!b]
	\centering
	\includegraphics[width=.4\linewidth]{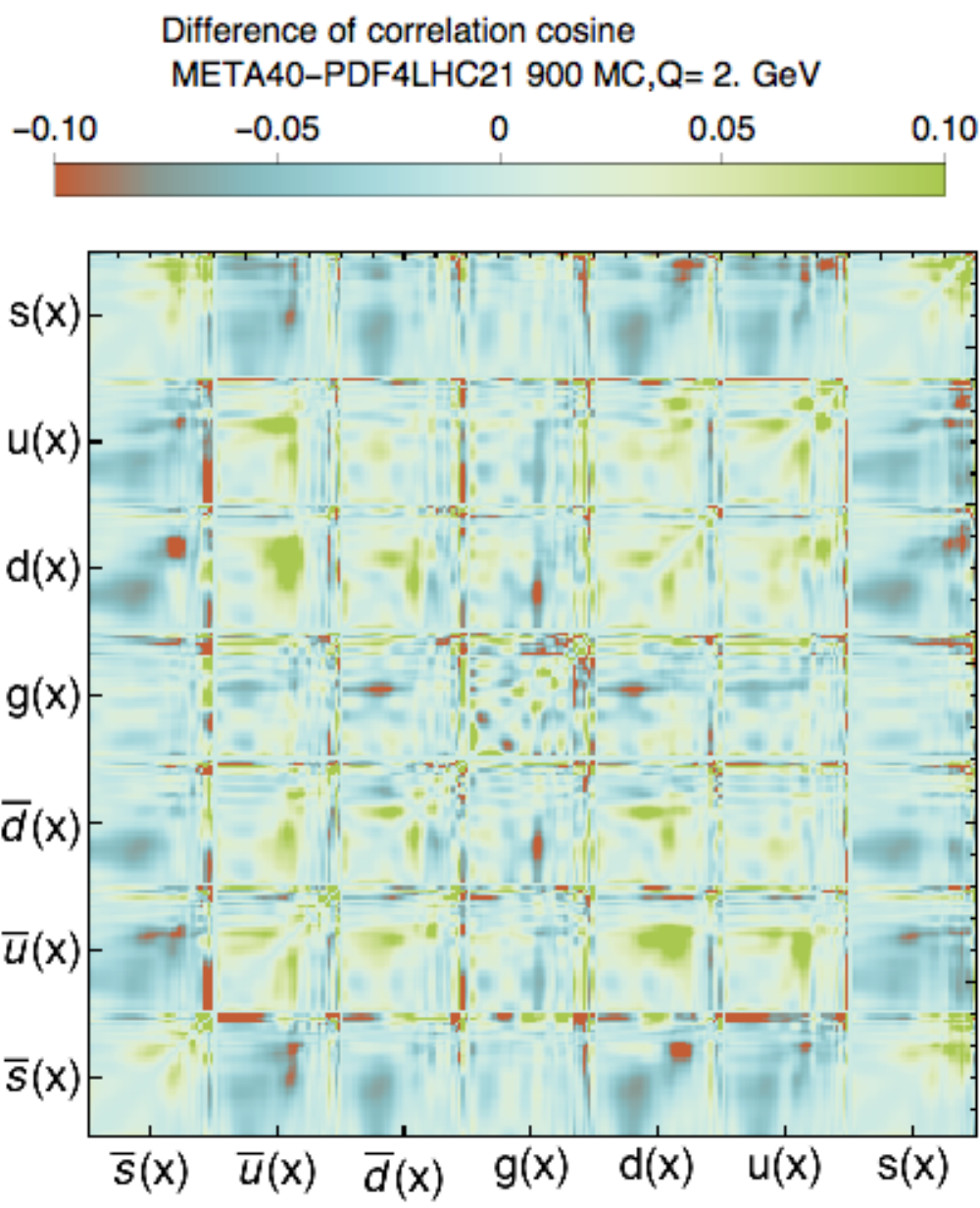}
	\caption{Difference of correlation cosine between the reduced Hessian set, META40, and the original 900 MC set for PDF4LHC21, at $Q=2$GeV.}
	\label{fig-app:META40_900_corr}
\end{figure}

\subsection{Hessian reduction methods}
\label{subsec:HessianConv}

In this section we give concise descriptions of the
two techniques that have been proposed to convert a MC set into a Hessian set, {\sc \small MC2Hessian} and META PDFs. We then compare the outputs obtained using both methods and validate the choice of META PDFs with $N_{\rm eig}=40$ as the output of the combination exercise. While the majority of these PDF sets
are not the released output of the PDF4LHC21 analysis, some of these sets will be available on the PDF4LHC website. We give a brief description of such sets.

\paragraph{META PDFs.}
The META analysis~\cite{Gao:2013bia} results in a {\it meta-parametrisation} of the MC replicas, using Bernstein polynomials. After the META analysis, all input replicas will correspond to the same parametric form, with different parameters of the Bernstein polynomials -- the set of parameters and the degree of the polynomials define the {\it meta-parametrisation}. A fit is performed for each MC replica, starting from the reference PDF $f_{\alpha}^{(0)}$, which is taken to be the mean of the MC set, with $\alpha$ the flavour. 
\begin{equation}
    \Phi_{\alpha}^{(k)}(x;{a})=f_{\alpha}^{(0)}(x, Q_0)\left( 1+\sum_{i=1}^{n+2} a_{\alpha,i}^{(k)} b_i[\phi_S(x, x_S)]\right)
\end{equation}
where the basis function $b_i$ are
\begin{equation}
b_i(x)=\left\{\ln x, \ln (1-x), {\cal B}_{n, i-2}(x) \right\}
\end{equation}
with $n$ the degree of the polynomial. In this analysis, $n=14$ for all flavours except $\bar{d}$, for which we have $n_{\bar{d}}=12$. The scale at which the combination was made is $Q_0=2$ GeV.

\begin{figure}[!t]
	\centering
	\includegraphics[width=1\linewidth]{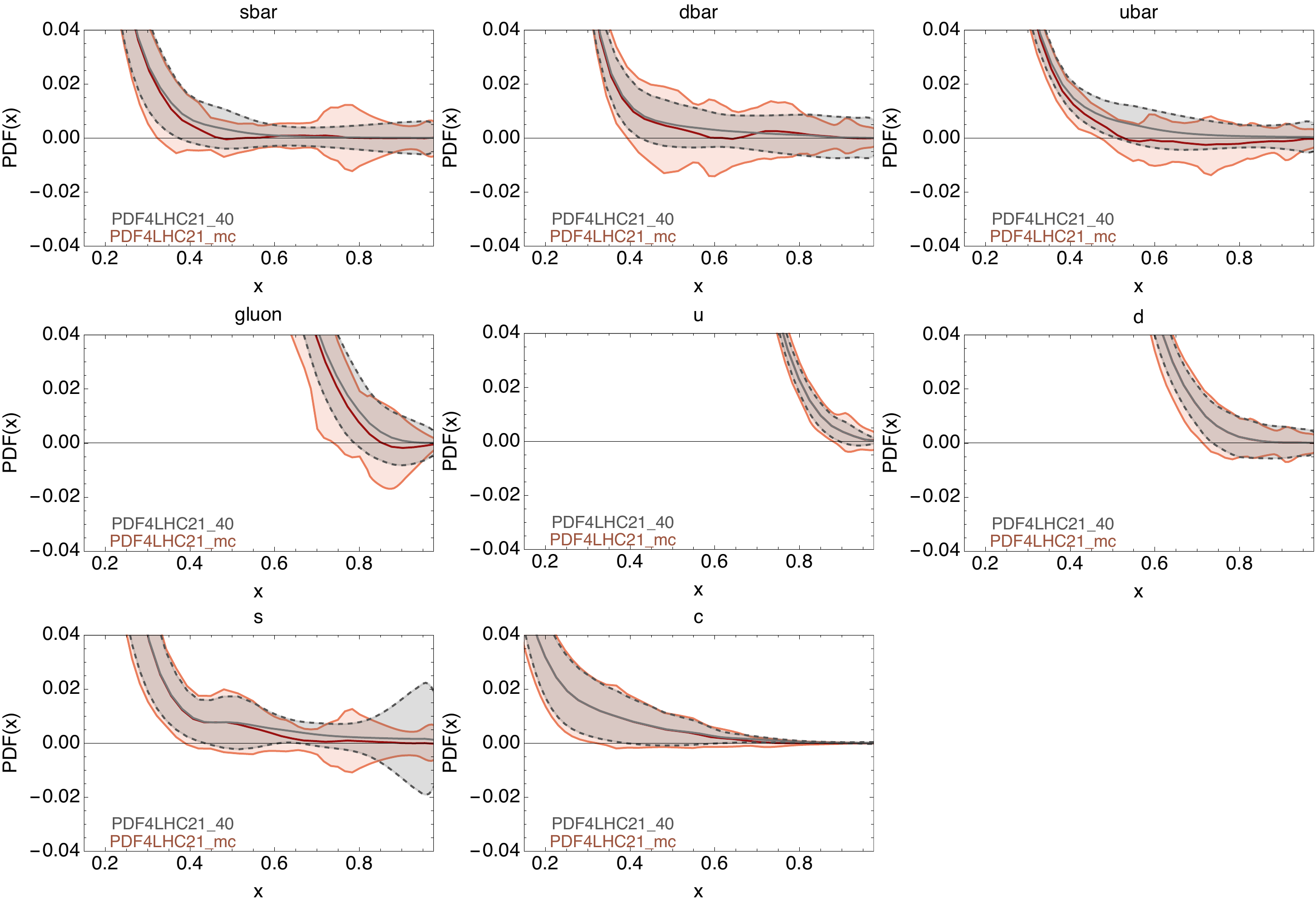}
	\caption{Large-$x$ behavior of the {\sc \small PDF4LHC21\_40} and {\sc \small PDF4LHC21\_mc} sets at $Q=2$ GeV. The central PDF  of the baseline  PDF4LHC21 set is represented in dark red, while the stretched central/reference PDF (see text) of the {\sc \small PDF4LHC21\_40} is shown in gray. The $68\%$ error bands for  {\sc \small PDF4LHC21\_40} and {\sc \small PDF4LHC21\_mc} are shown, respectively, in gray and orange colors.}
	\label{fig-app:META40_CMC100}
\end{figure}

The META-PDF technique was used for the PDF4LHC15 combination, resulting in a 30-member eigenvector set called {\sc \small PDF4LHC15\_nnlo\_30}. With respect to the 2015 version of META PDFs, the reduction proposed here uses a new feature, also discussed in Sect.~\ref{sec:largex_pdf4lhc21}.  For those PDF flavors for which the central PDF of the baseline ensemble becomes negative at large $x$ (typically where there are no data), it is possible to ensure the positivity of the reference PDF for estimation of the Hessian uncertainties by stretching  the $x$-axis towards $x=1$, starting from a value $x_S$ according to some function $\phi_S(x, x_S)$.\footnote{Specifically, we choose $\phi_S(x,x_S)=\left(x^{6}+(1-x_S^6)/x_S^6\right)^{-1/6}$, where $0.465 \leq x_S \leq 1$, depending on the PDF flavor.}
The behaviour of the reference replica is unchanged for $x<x_S$, hence implying minimal physical modifications in the fitted region. The stretching is required for the sea quark and gluon PDFs, while the valence PDFs are positive at the scale of combination except at $x>0.85$, as illustrated in Fig.~\ref{fig-app:META40_CMC100}. By construction, the stretched reference PDFs fulfil the valence and the momentum sum rules within less than $0.05\%$. 

The set of ($8\times 16+1\times 14=142$) parameters defining the meta-parametrisation is obtained by minimising the following function,
\begin{equation}
E\left[f_{\alpha}^{(0)}(x, Q_0), \Phi_{\alpha}^{(k)}(x;{a})\right]
=\sum_{x\in \mbox{\tiny grid}}\, \left(\frac{\ln f_{\alpha}^{(0)}(x_n, Q_0)-\ln \Phi_{\alpha}^{(k)}(x_n;{a})}{\delta(\ln  f_{\alpha}^{(0)}(x_n, Q_0))}\right)^2
\label{eq-app:ErrMeta}
\end{equation}
on a grid of $x$ values running from $[3\times 10^{-5}, 0.95]$. The denominator of Eq.~(\ref{eq-app:ErrMeta}) contains the symmetric PDF uncertainty of the reference PDF $f_{\alpha}^{(0)}$. 
The covariance matrix in this parameter space is evaluated and diagonalised, so as to obtain the eigenvalues $\lambda_l$ corresponding to the eigenvector directions associated with parameters of the {\it meta-parametrisation}. To build the reduced Hessian META set, the eigenvector directions corresponding to the largest (less constrained) eigenvalues are selected. Those eigenvector directions are the ones dominating the uncertainties. In the present analysis, the number of eigenvectors was set to $N_{eig}=40$ after studying the reductions to $N_{eig}=30$ and $50$. This choice gives the best compromise between precision and manageability of the Hessian set. The difference of the correlation cosine computed with the META set for $N_{eig}=40$ and with the $N_{rep}=900$ {\sc \small PDF4LHC21} set, presented in Fig.~\ref{fig-app:META40_900_corr}, shows that the META-PDFs faithfully reproduce the original set.

The new feature of the stretching function allows for a resulting uncertainty that is larger and more compatible with both the {\sc \small PDF4LHC21} and {\sc \small PDF4LHC21\_mc} sets at large $x$, as compared to the PDF4LHC15 exercise. The uncertainty bands for these sets are plotted for all flavors in Fig.~\ref{fig-app:META40_CMC100}. The META combination is carried out the factorization scale $Q=2 $ GeV. The combined PDFs are then evolved at NNLO forward to scales up to $Q>10000$ GeV and backward to $Q=1.4$ GeV using the {\sc \small HOPPET 1.2.0} code \cite{Salam:2008qg}.
\begin{figure}[!t]
	\centering
	\includegraphics[width=1\linewidth]{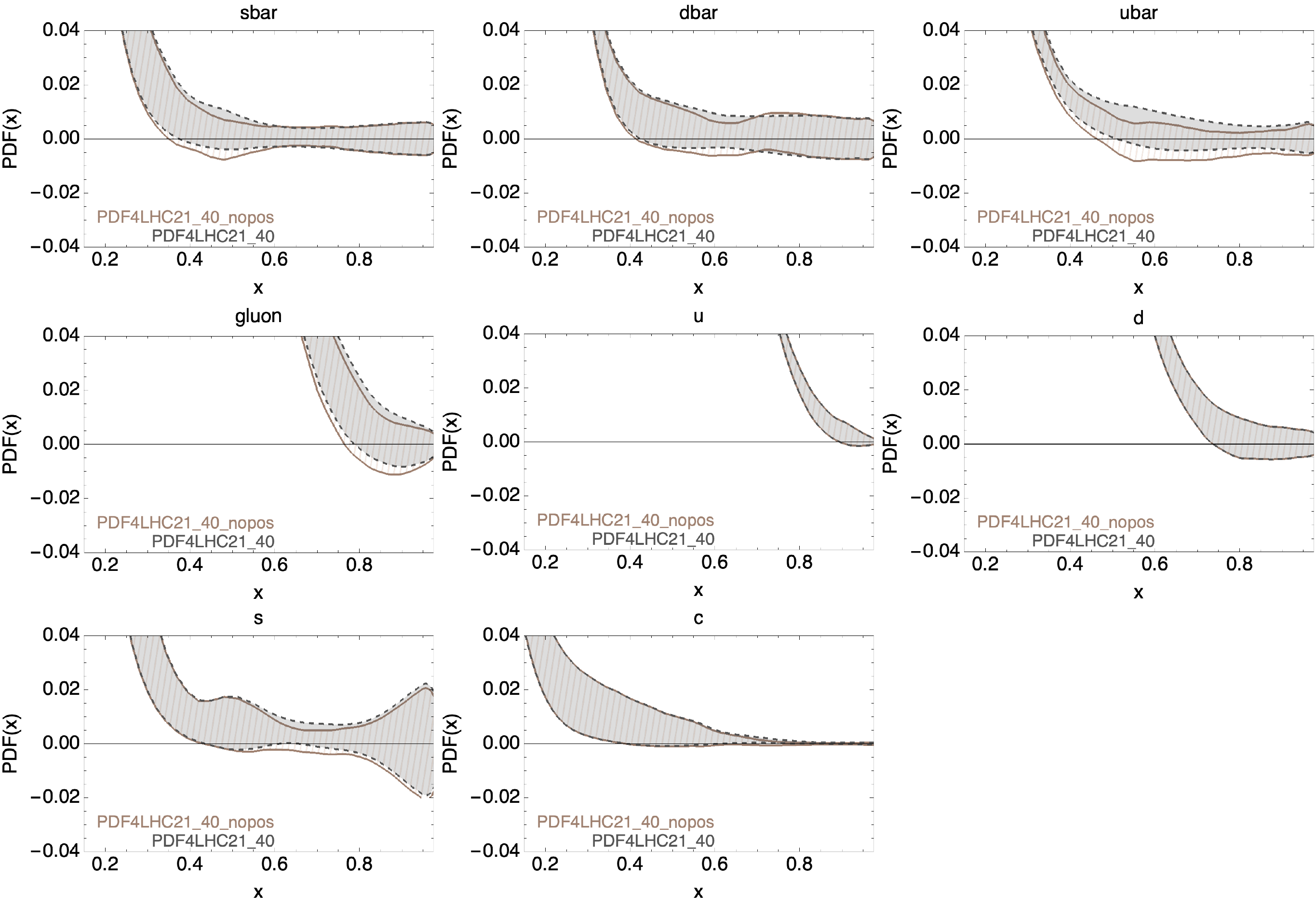}
	\caption{Same format as in Fig.~\ref{fig-app:META40_CMC100}. The $68\%$ error band for  {\sc \small PDF4LHC21\_40} and {\sc \small PDF4LHC21\_40\_nopos} are shown, respectively, in gray (solid fill) and light brown (oblique hatching).}
	\label{fig-app:META40_1_2}
\end{figure}
The Hessian set {\sc\small PDF4LHC21\_40} is built on positive-definite central values for the distributions, as explained above. However, we propose a second set, called {\sc\small PDF4LHC21\_40\_nopos} for which the central value has been shifted to correspond to the average value of the original {\sc\small PDF4LHC21} Monte Carlo set. The two Hessian sets are compared on Fig.~\ref{fig-app:META40_1_2}.

\paragraph{MC2Hessian.}
The second Hessian reduction method that we have considered in this work is
the {\sc\small mc2hessian} algorithm presented in~\cite{Carrazza:2015aoa,Carrazza:2016htc}.
Its basic idea is to use the MC replicas of the prior themselves
to construct a Hessian representation with the replicas' linear expansion basis,
and then to determine the numerical coefficients of the expansion to ensure
that the mean, variance, and correlations of the prior distribution are
reproduced.
Specifically, we use the variant of {\sc\small mc2Hessian} based
on the combination of Principal Component Analysis (PCA) and
Singular Value Decomposition (SVD) to assemble the covariance
matrix of this Hessian representation in the space of MC replicas
of the prior.
While this representation contains in principle a large number of eigenvectors
$N_{\rm eig}$, most of these carry a small weight and can be safely neglected.
In practice, one adds eigenvectors until the size of the differences between the new Hessian
and the original MC representations becomes comparable to the accuracy of the Gaussian
approximation for the PDF flavours and regions of $x$ of interest.

\begin{figure}[p]
	\centering
	\includegraphics[width=1\linewidth]{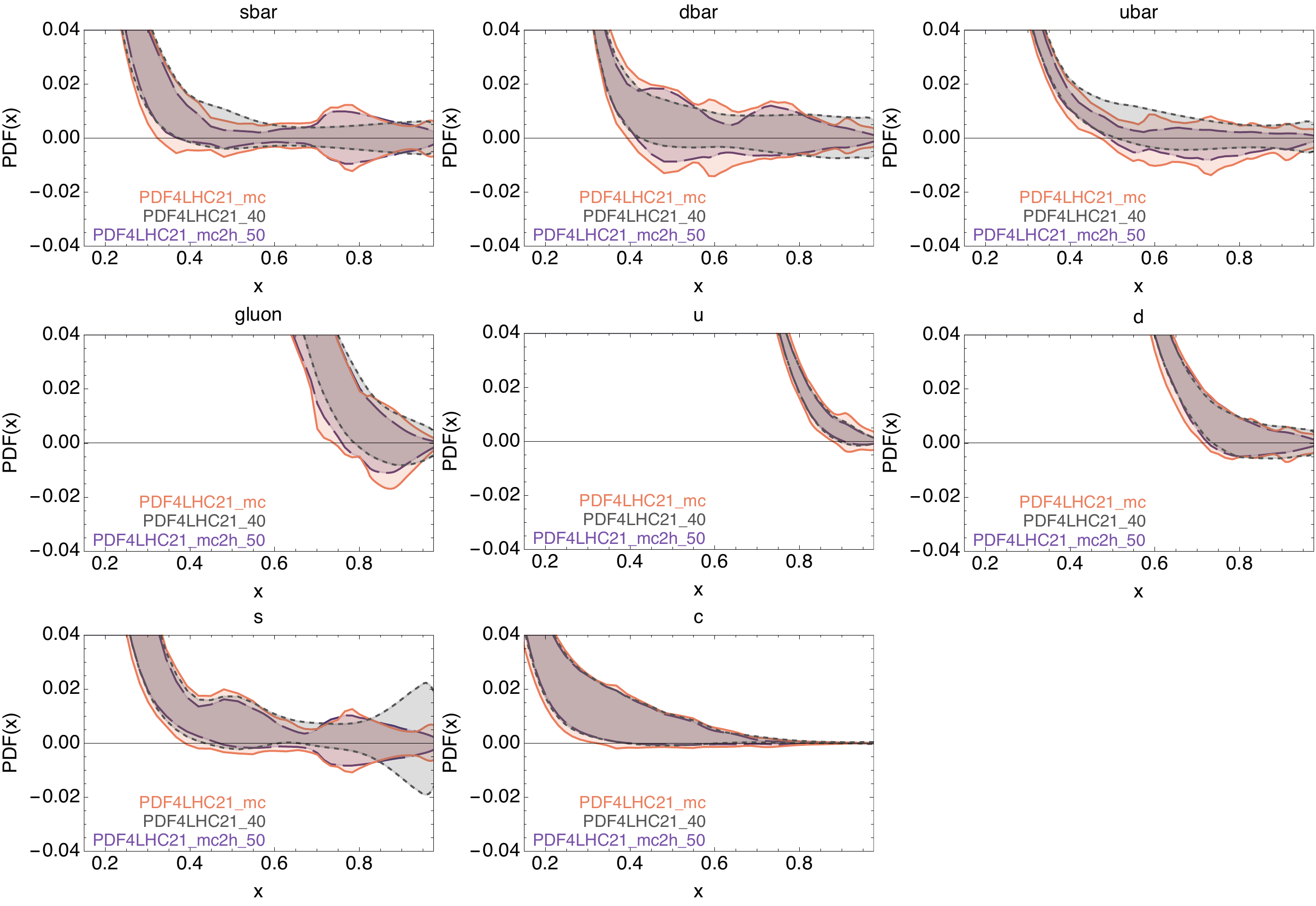}
	\caption{Large-$x$ behavior of the absolute values for {\sc \small MC2Hessian} set with $N_{\rm eig}=50$ compared to {\rm PDF4LHC21\_40} and {\rm PDF4LHC21\_mc} at $Q=2$GeV. The $68\%$ error band are shown, respectively, in purple (long dashed), gray (dashed) and orange (solid).}
	\label{fig-app:META40_MC2H50}
\end{figure}

\begin{figure}[p]
	\centering
	\includegraphics[width=1\linewidth]{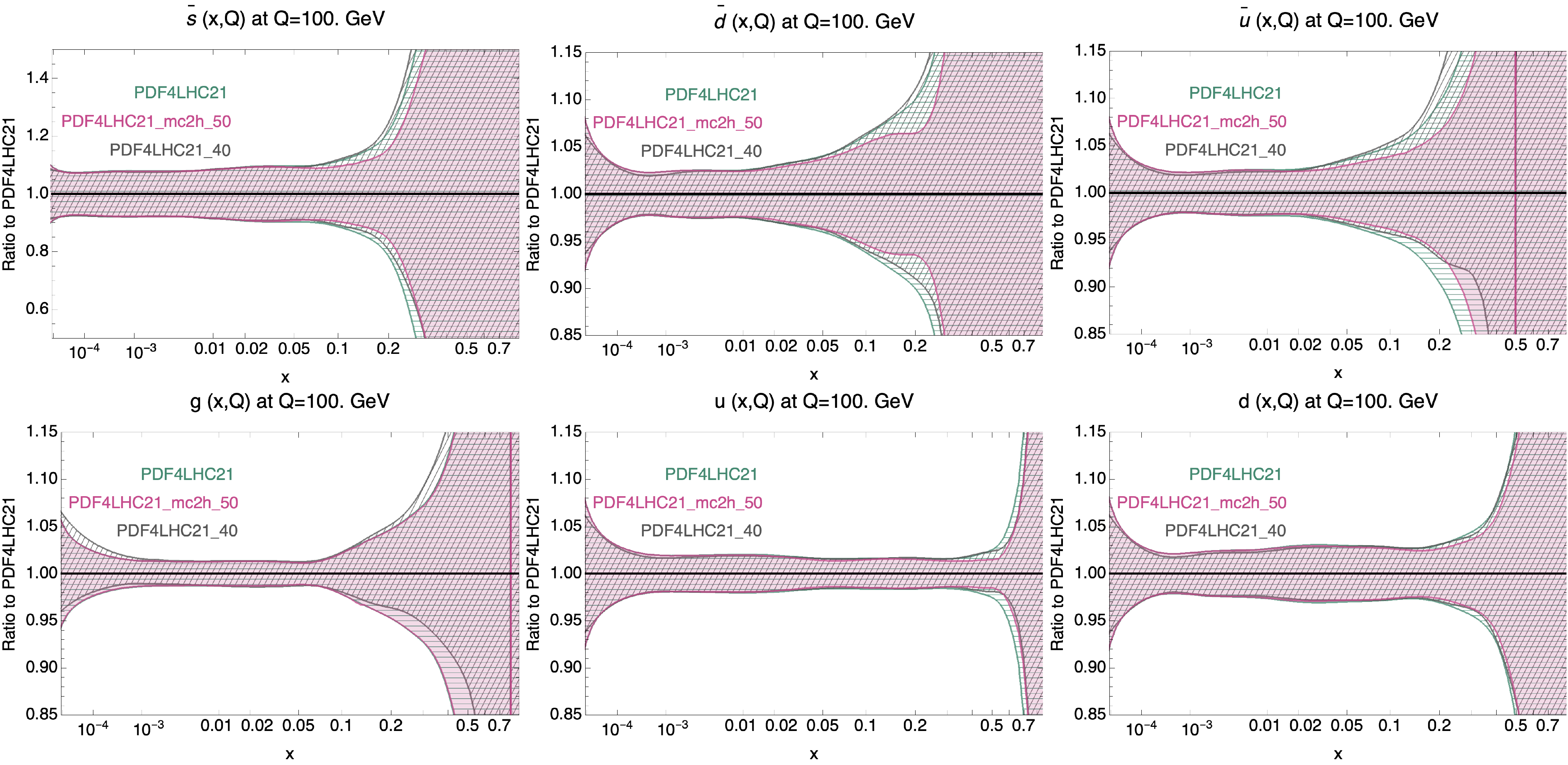}
	\caption{Small-$x$ behavior of  the {\sc \small MC2Hessian} set with $N_{\rm eig}=50$ compared to {\sc \small PDF4LHC21\_40} and the $N_{\rm rep}=900$ {\sc \small PDF4LHC21} sets. The ratio to the average of the latter is shown at $Q=100$ GeV. The $68\%$ error band are shown, respectively, in pink (solid fill), gray (oblique hatching) and cyan (horizontal hatching).
	}
	\label{fig-app:META40_MC2H50_smallx}
\end{figure}

\paragraph{Choice of the Hessian set.}
We will now compare the various outputs of both Hessian reduction techniques
when applied to the {\sc\small PDF4LHC21} set with $N_{\rm rep}=900$ Monte Carlo replicas.
It is found that the two methods exhibit an equally satisfactory performance provided
the number of eigenvectors is large enough, around $N_{\rm eig}=100$.
However, it is convenient to attempt to reduce the number of eigenvectors as much as possible,
since this lowers the computational requirements of analyses based on PDF4LHC21.
It is then found that the latest {\sc\small META-PDF} approach with $N_{\rm eig}=40$ eigenvectors
provides a reasonable description of the prior distribution, somewhat improving
that of {\sc\small mc2hessian} for the same number of eigenvectors, especially
in the large-$x$ region.
We hence adopt the {\sc\small META-PDF} Hessian reduction of PDF4LHC21
with $N_{\rm eig}=40$ eigenvectors as our preferred choice in this work. It is labelled {\sc \small PDF4LHC21\_40}.
We note that this outcome (similar performance of {\sc\small META-PDF} and
{\sc\small mc2hessian} for large $N_{\rm eig}$, with somewhat improved performance
of {\sc\small META-PDF} for smaller  $N_{\rm eig}$) was also obtained
in the context of the PDF4LHC15 combination. 
The large-$x$ behavior of {\sc\small mc2hessian} for $N_{\rm eig}=50$ (labelled {\sc \small PDF4LHC21\_mc2h\_50}) is compared with  {\sc \small PDF4LHC21\_40} at 2 GeV in Fig.~\ref{fig-app:META40_MC2H50}. It is observed that the set obtained through {\sc\small META-PDF} reproduces {\rm PDF4LHC21\_mc} very well at large $x$. A similar plot, Fig.~\ref{fig-app:META40_MC2H50_smallx}, focuses on the ratio to the average PDF of the {\sc \small PDF4LHC21} $N_{\rm rep}=900$ set, comparing the prior PDF4LHC21 to the Hessian sets  {\sc \small PDF4LHC21\_40} and {\sc \small PDF4LHC21\_mc2h\_50}: This shows that the output Hessian set {\sc \small PDF4LHC21\_40} faithfully reproduces the uncertainties of the original set at 100 GeV as well.


\clearpage\newpage\section{Interplay between PDF4LHC21 and NNPDF4.0}
\label{app:nnpdf40}

As mentioned in the introduction, the release of the NNPDF4.0 global analysis~\cite{NNPDF:2021uiq,Ball:2021leu} took place when the PDF4LHC21 combination presented in this paper was already at an advanced stage. The NNPDF4.0 set is therefore not a part of the
present combination. One may however wonder how NNPDF4.0 compares with \NNprime and PDF4LHC21 and also how the PDF4LHC21 combination would have changed had NNPDF4.0 been used in its construction instead of \NNprime. In order to address this question, at least qualitatively, in this appendix we present a comparison between \NNprime, NNPDF4.0, and PDF4LHC21. It is beyond the scope of this appendix to review the many improvements that have been implemented in NNPDF4.0 in terms of experimental data, theoretical calculations, and fitting methodology. For this, we refer the reader to~\cite{Ball:2021leu}.

\begin{figure}[!t]
\centering
\includegraphics[width=0.49\textwidth]{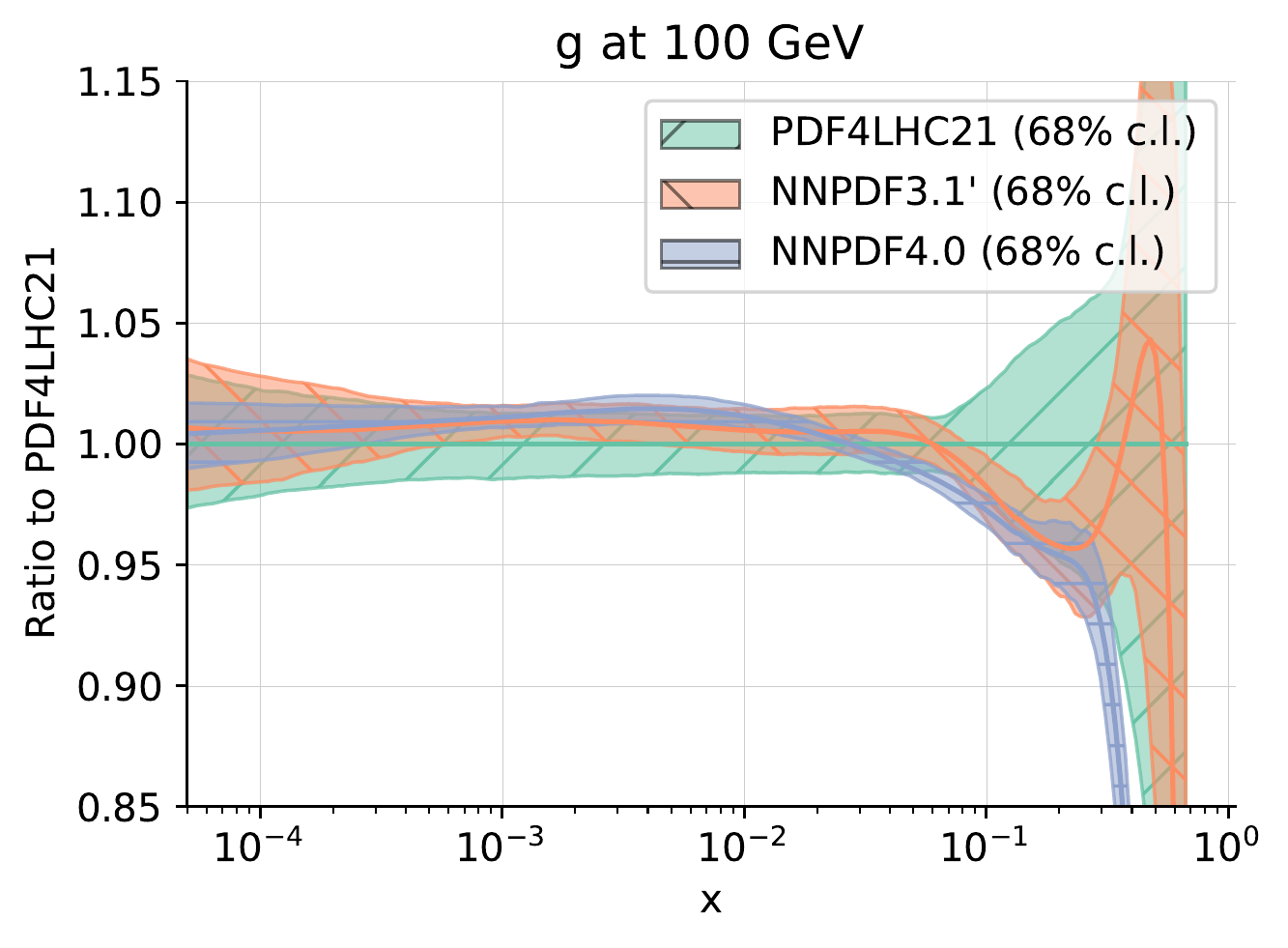}
\includegraphics[width=0.49\textwidth]{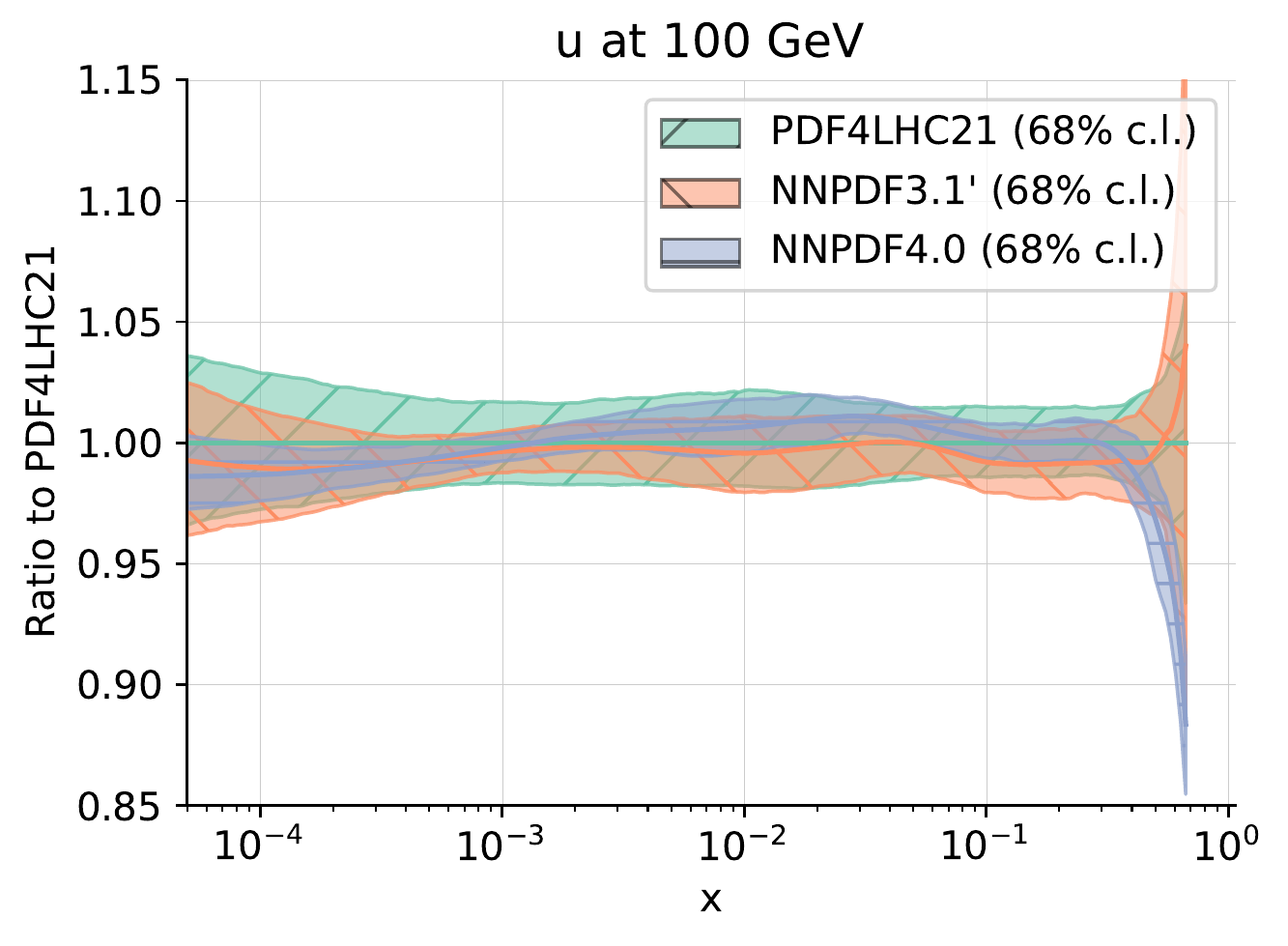}\\
\includegraphics[width=0.49\textwidth]{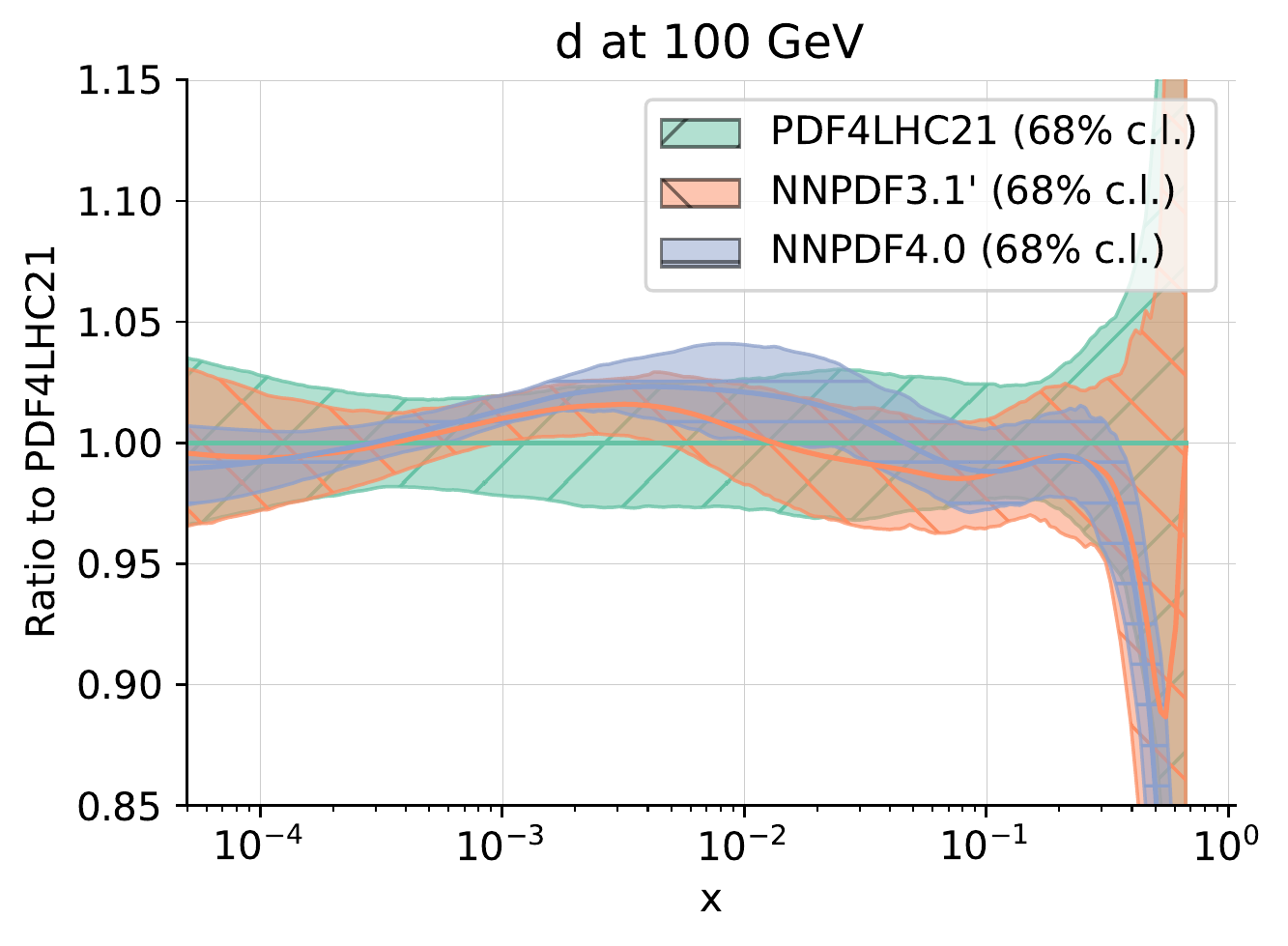}
\includegraphics[width=0.49\textwidth]{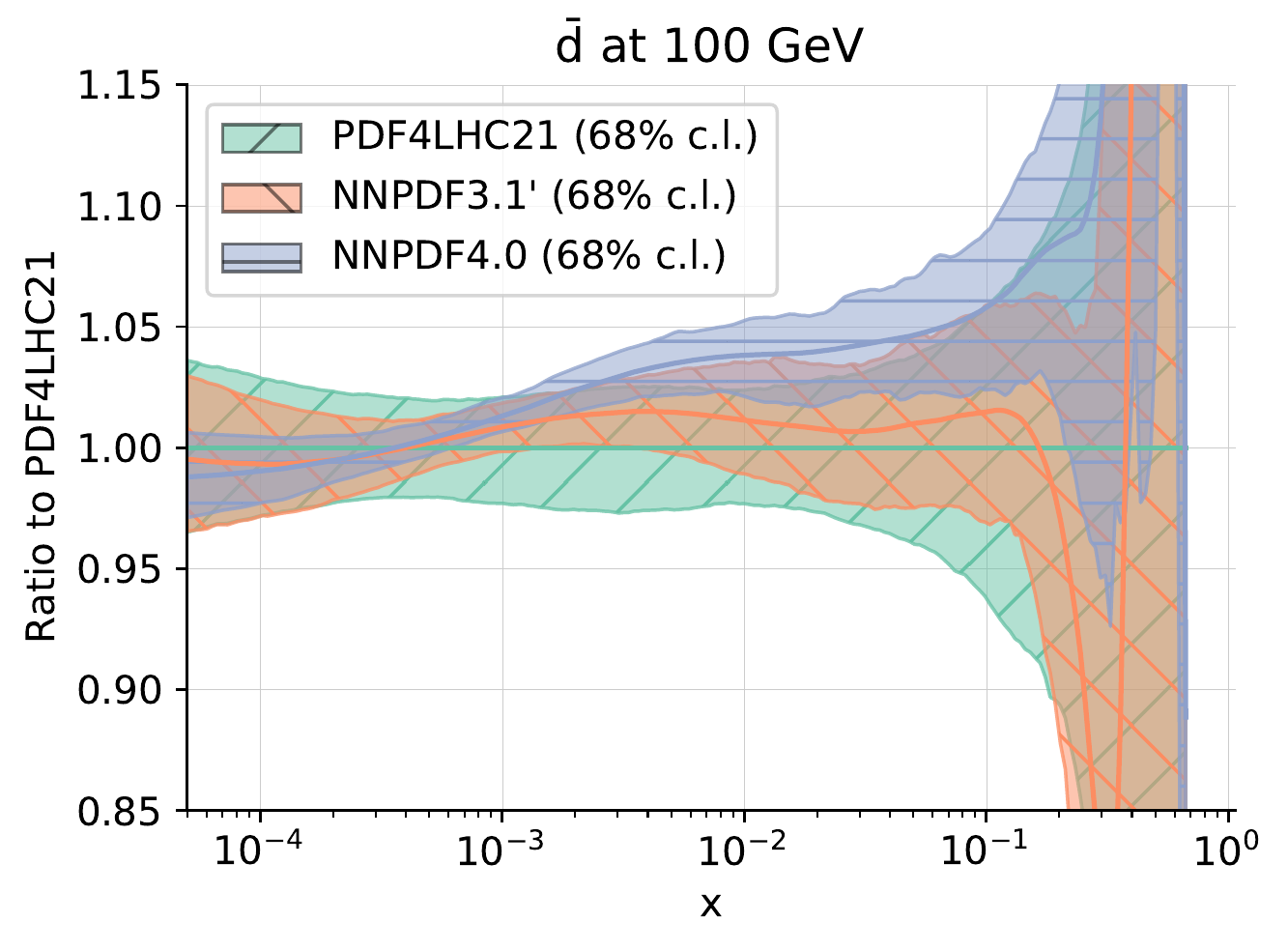}\\
\includegraphics[width=0.49\textwidth]{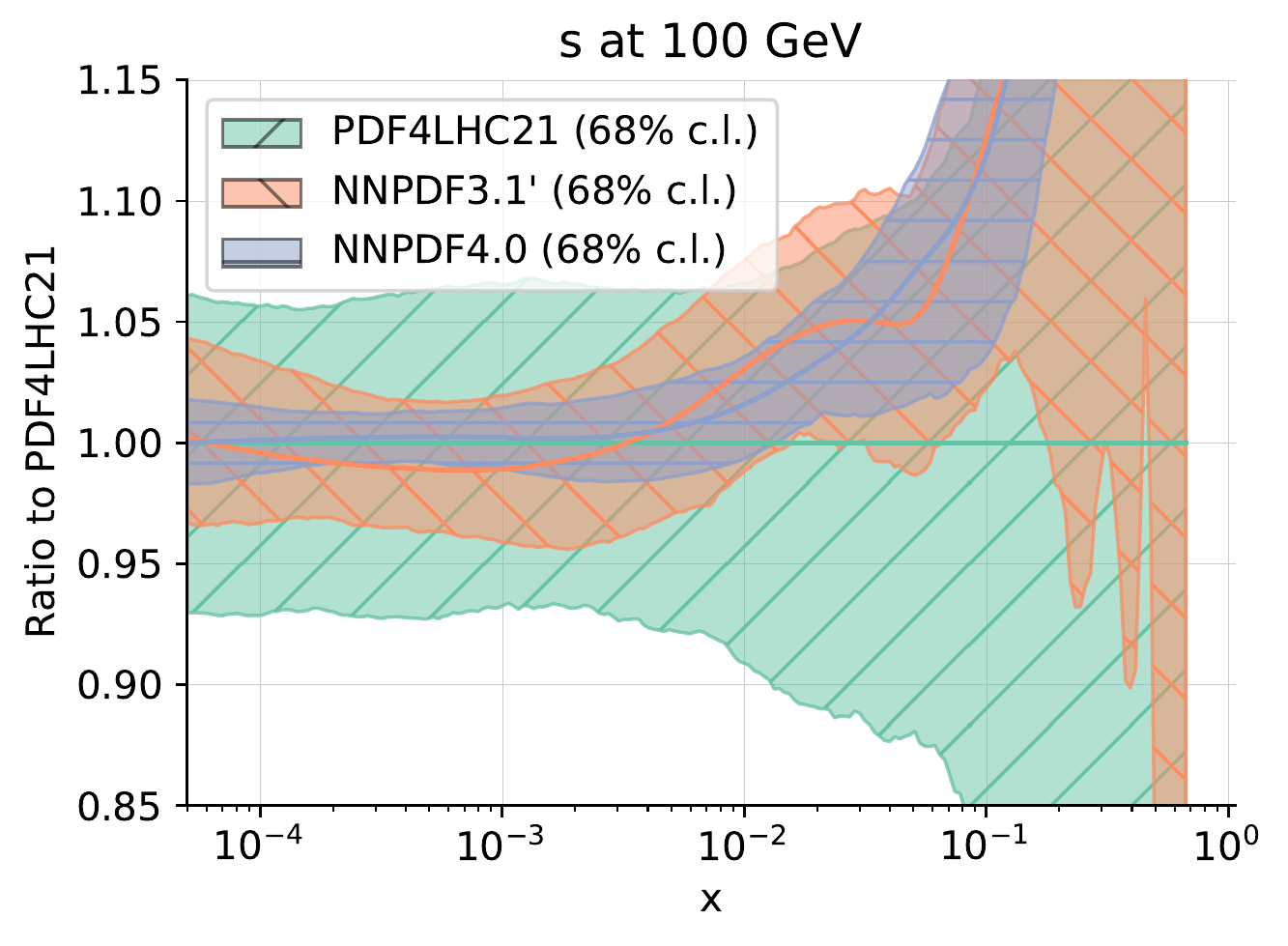}
\includegraphics[width=0.49\textwidth]{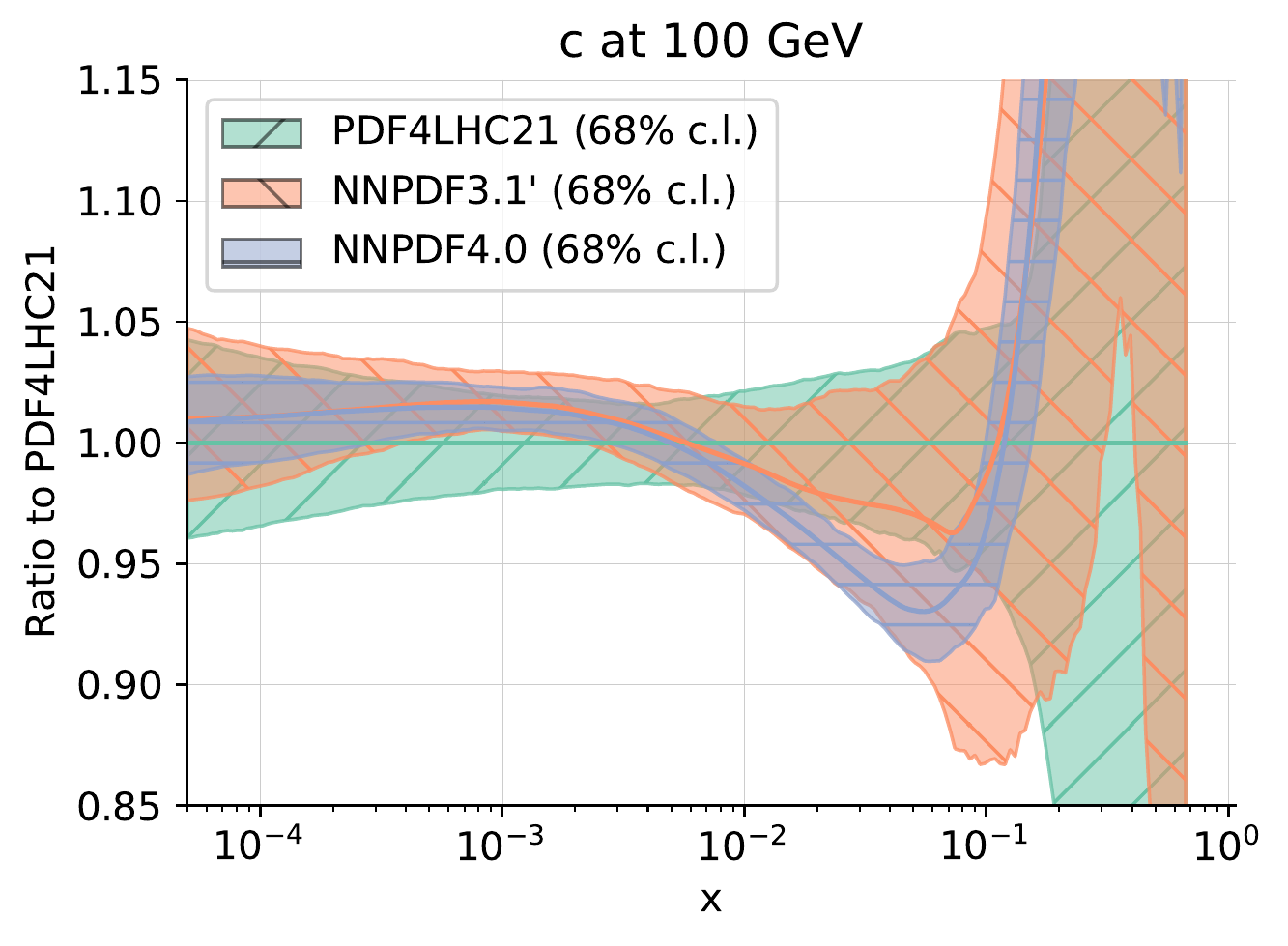}\\
\caption{Comparison of the PDF4LHC21 combination with the \NNprime and NNPDF4.0 global fits. We show results for the gluon and the up, down, anti-down, strange, and charm quark PDFs at $Q=100$~GeV normalised to the central value of PDF4LHC21.
\label{fig:pdf4lhc21_vs_nnpdf40_pdfs}}
\end{figure}

\begin{figure}[!t]
\centering
\includegraphics[width=0.49\textwidth]{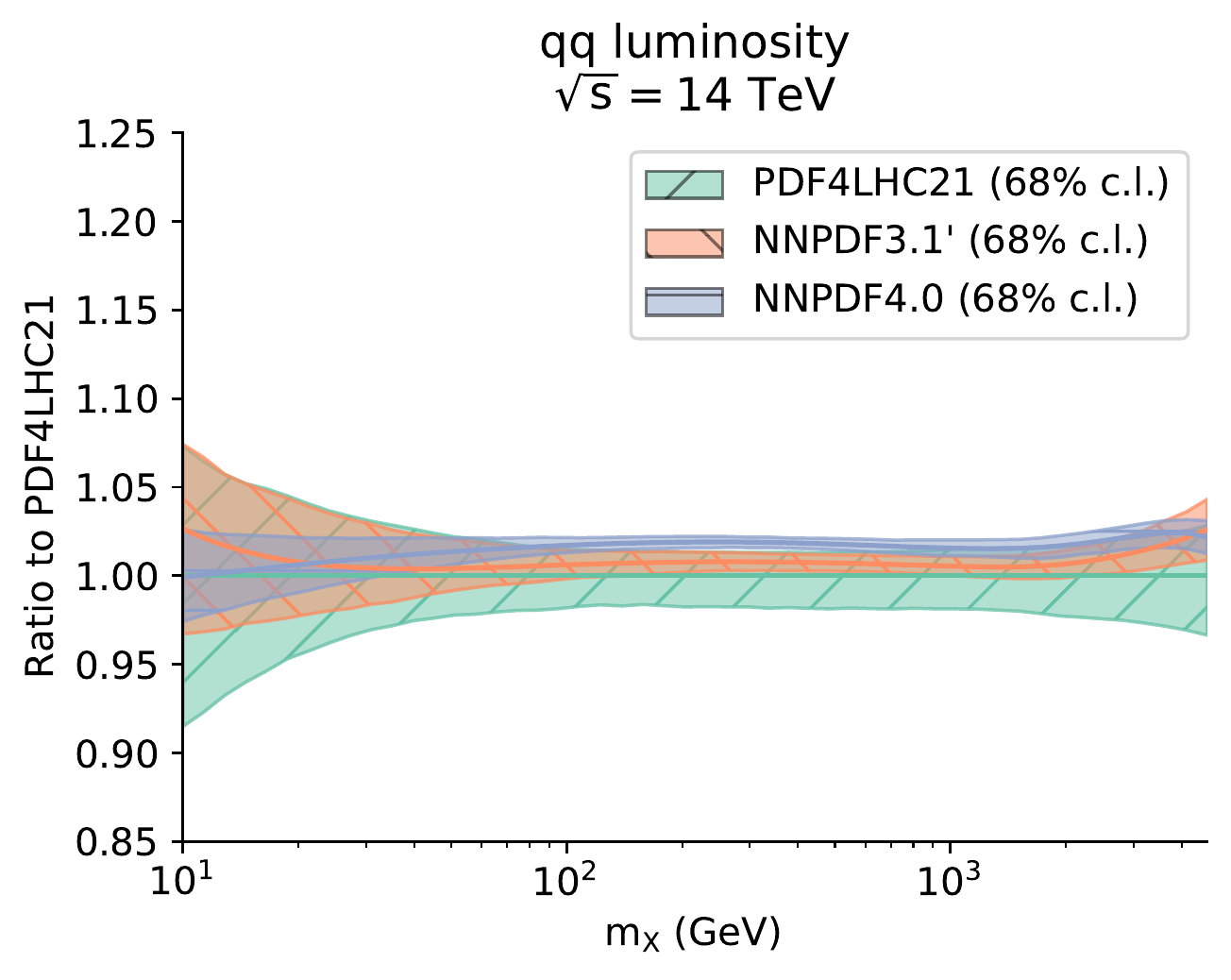}
\includegraphics[width=0.49\textwidth]{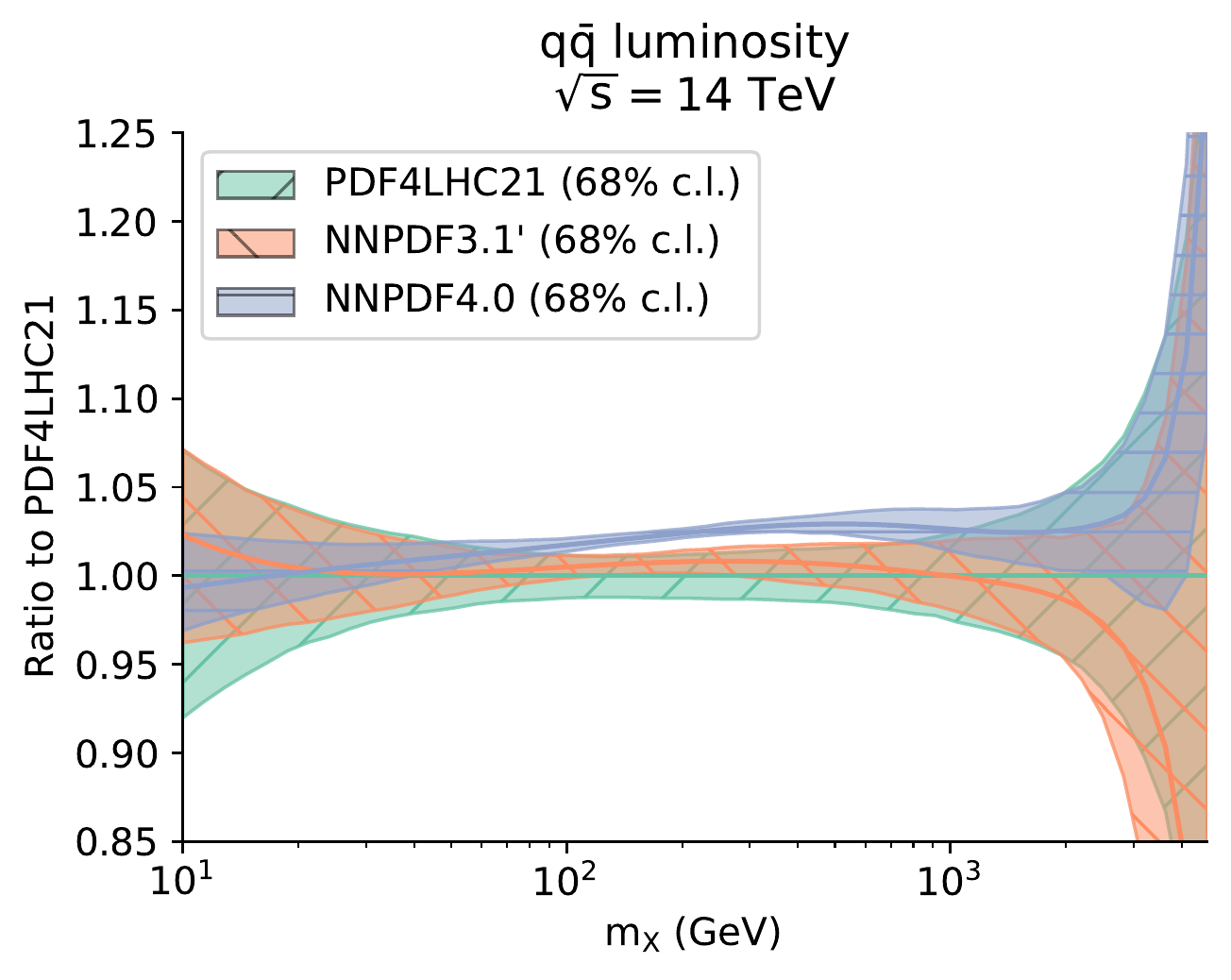}\\
\includegraphics[width=0.49\textwidth]{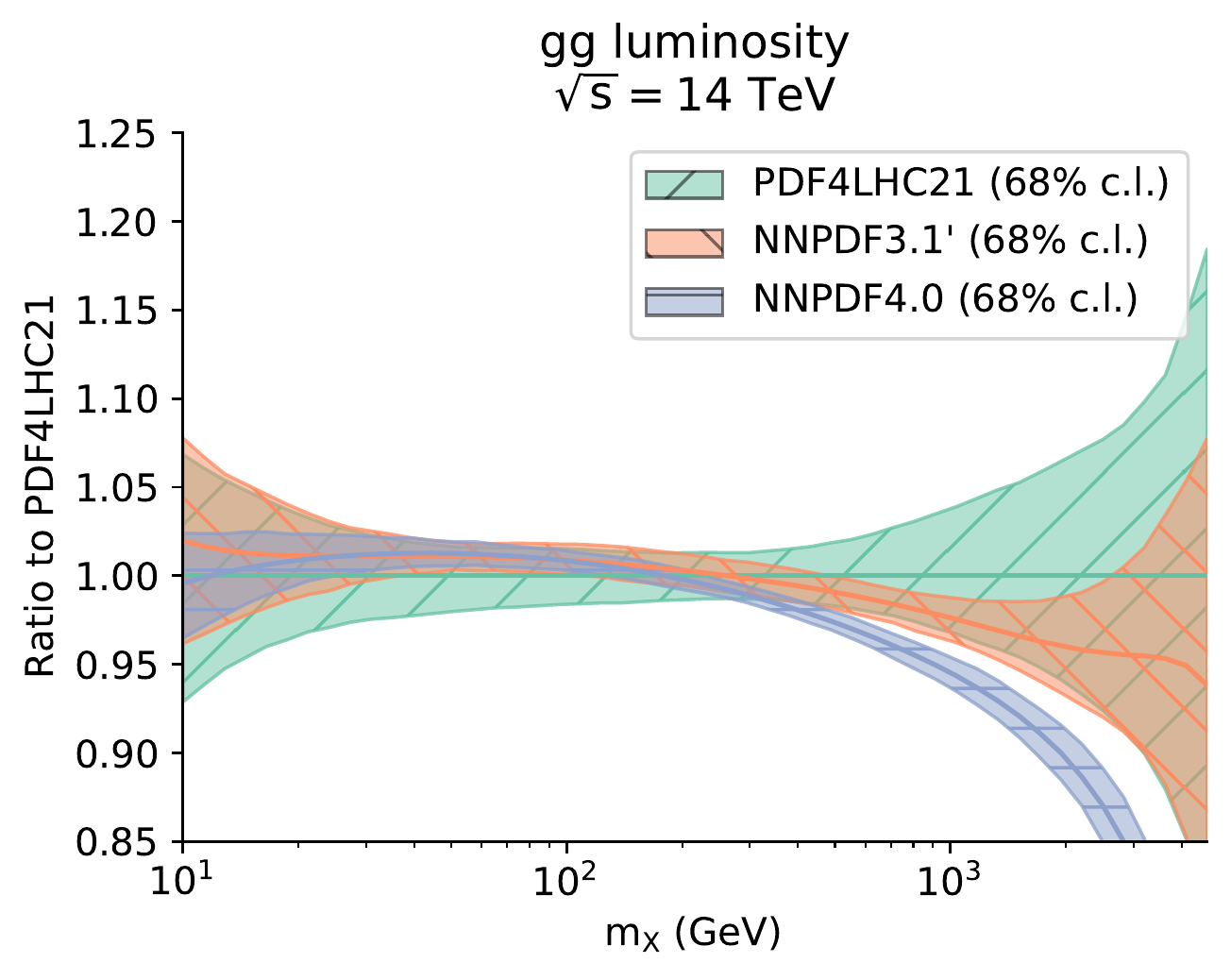}
\includegraphics[width=0.49\textwidth]{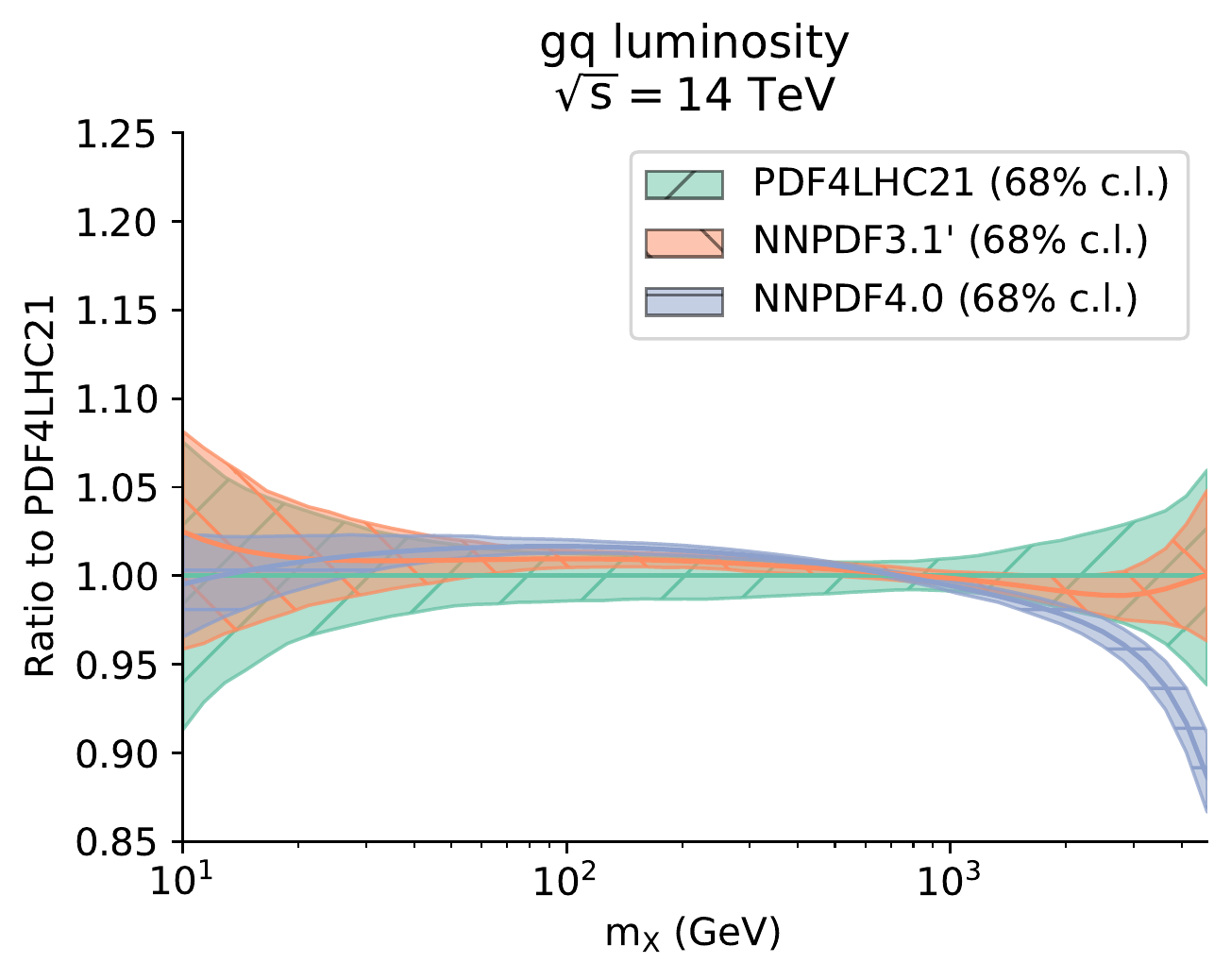}\\
\caption{Same as Fig.~\ref{fig:pdf4lhc21_vs_nnpdf40_pdfs} for partonic luminosities as a function of the invariant mass at $\sqrt{s}=14$~TeV.}
\label{fig:pdf4lhc21_vs_nnpdf40_lumis}
\end{figure}

Figure~\ref{fig:pdf4lhc21_vs_nnpdf40_pdfs} displays a comparison between the PDF4LHC21 combination (with $N_{\rm rep}=900$ replicas) and the \NNprime and NNPDF4.0 global fits. We show results for the gluon and the up, down, anti-down, strange, and charm quark PDFs at $Q=100$ GeV normalised to the central value of the PDF4LHC21 combination. A similar comparison for partonic luminosities as a function of the invariant mass at $\sqrt{s}=14$~TeV is displayed in Fig.~\ref{fig:pdf4lhc21_vs_nnpdf40_lumis}.

Inspection of Figs.~\ref{fig:pdf4lhc21_vs_nnpdf40_pdfs} and~\ref{fig:pdf4lhc21_vs_nnpdf40_lumis} reveals the following features. Concerning PDFs, the central values of NNPDF4.0 are typically contained within  in the \NNprime 68\% CL bands. Deviations slightly larger than the \NNprime 68\% CL band are however seen for the gluon PDF around $x\sim 0.005$ and $x\gtrsim 0.07$, and for the antidown quark PDF around $x\gtrsim 0.01$. The NNPDF4.0 PDF uncertainties are generally smaller than the \NNprime uncertainties. The reason for this is a combination of methodological improvements and the extension of the data set included in NNPDF4.0, as discussed at length in~\cite{Ball:2021leu}. Concerning partonic luminosities, differences are somewhat more apparent: for the quark-antiquark luminosity, where the NNPDF4.0 expectation is enhanced by about 2-3\% for invariant mass values between 100~GeV and 1~TeV in comparison to \NNprime; and for the gluon-gluon luminosity, where the NNPDF4.0 expectation is suppressed by about 5-7\% for invariant mass values larger than 1~TeV in comparison to \NNprime. Other kinematic regions show otherwise a good consistency between NNPDF4.0 and \NNprime for all of the partonic luminosities.

Because of these differences, inclusion of NNPDF4.0 (in lieu of \NNprime) in the PDF4LHC21 combination would have possibly modified some of the features of the combination itself. The central value would have likely been shifted slightly in places. Notably, the gluon PDF would have been suppressed at $x\gtrsim 0.3$, and the $\bar d$ quark PDF would have been enhanced at $x\gtrsim 0.01$. The size of these shifts would, however, have been encompassed by the 68\% CL band of the PDF4LHC21 combination presented. The uncertainties of the combination would have also possibly become wider in these kinematic regions and for the PDFs where the spread across the constituent PDF sets has increased, most notably for the gluon PDF at large $x$. We finally note that NNPDF4.0 enforces positivity of PDFs, see Sect.~3.1.3 in~\cite{Ball:2021leu}. Its inclusion as one of the constituent PDF sets of the combination would have therefore reduced the number of replicas that may have become negative in the large-$x$ region. The strategy outlined in Sect.~\ref{sec:hessian_reduction} and the potential issues discussed in Sect.~\ref{sec:largex_pdf4lhc21} would possibly have needed to be reconsidered in light of this.

\section{Dedicated studies: strangeness and the large-$x$ gluon}
\label{app:specifics}

The results of Sect.~\ref{sec:benchmarking}
demonstrated how  the reduced fits of the three groups are in good agreement in both the PDFs and
in the fit qualities for most datasets.
Nonetheless, we highlighted some differences which we will now scrutinise in this appendix alongside other areas of interest to the global PDF fits. 
The starting point of these investigations are the reduced PDF fits themselves, which
provide a well-understood boundary condition to assess the impact of either changing
some theory settings or adding a new dataset to the fit.
While several studies were carried out in the course of this benchmarking exercise,
here we present results for two of those with particular phenomenological relevance:
the strange PDF in the light of the dimuon production data, and the high-$x$ gluon PDF and resolving the apparent discrepancies observed when fitting top data in the three PDF groups as well the interplay with data on inclusive jet production. 

\subsection{Strangeness and dimuon production}
\label{sec:strangeness}

\begin{figure}[!b]
\centering
\includegraphics[width=0.49\textwidth,trim= 0.1cm 0.1cm 0.1cm 0.2cm,clip]{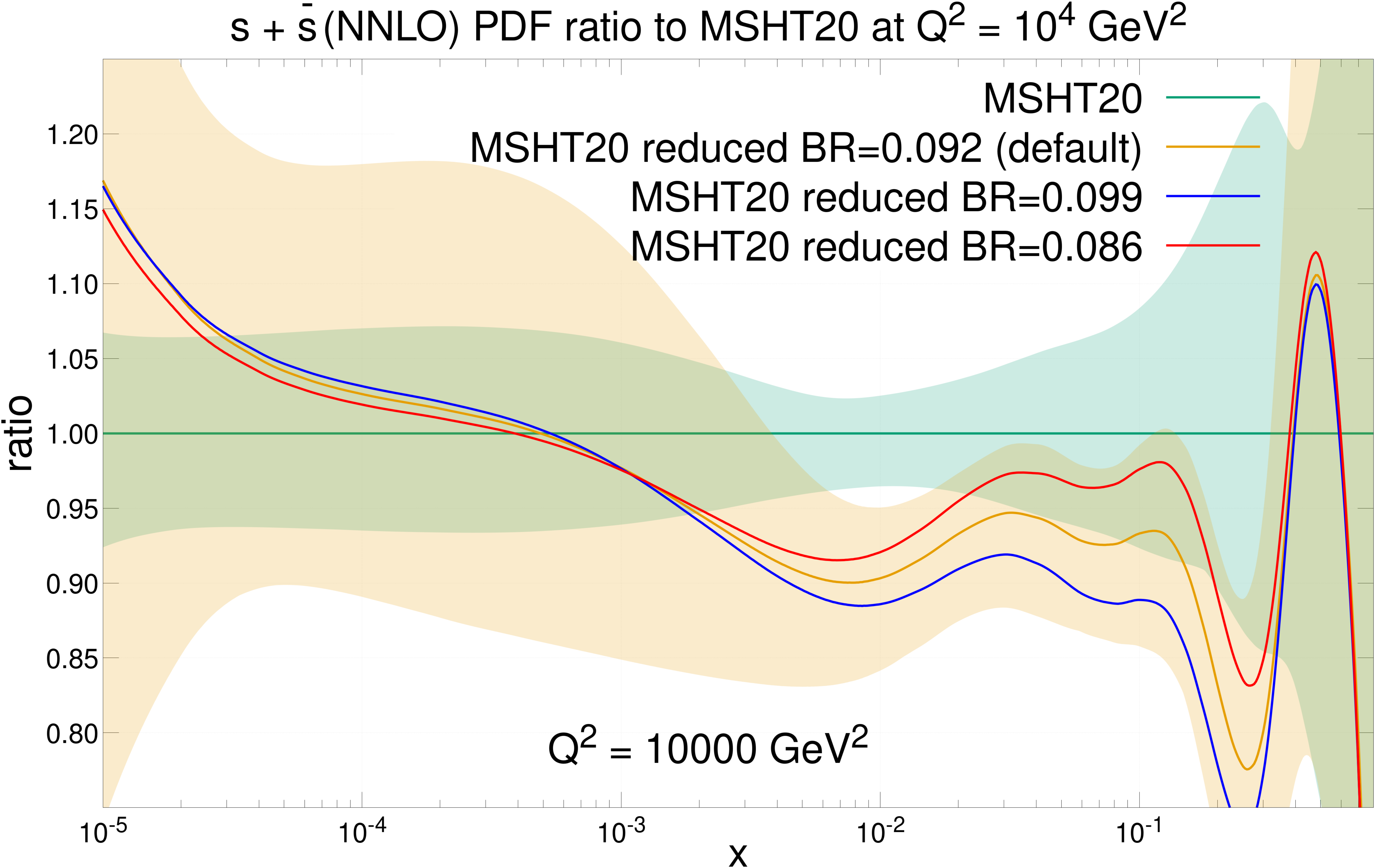}
\caption{The impact that varying the average muon branching ratio (BR) of $D$-mesons
  $B (D \rightarrow \mu)$ on the calculation of the NuTeV dimuon cross-sections has
  on the total strangeness of the MSHT reduced fit, relative to the MSHT20 global fit baseline.
  As the value of BR is decreased from 0.092 (the default in MSHT20) to 0.086 (close
  to the latest PDG average) or increased up to the NuTeV value of 0.099, the strange PDF can increase/decrease by up to $\simeq 5\%$ in
  the region $x\gsim 0.01$ covered by NuTeV.
}
\label{fig:MSHTstrangeness_dimuonBR}
\end{figure}

One of the main differences emerging both in the PDFs and the dataset $\chi^2$ for the reduced fits
reported in Sect.~\ref{sec:benchmarking} was related to the total
strangeness PDF, $s+\bar{s}$ (recall that we set the strangeness asymmetry to zero at the input scale).
The main datasets driving strangeness in the both the reduced fits and the global fits are the NuTeV dimuon data and the ATLAS 7~TeV $W,Z$ (2016) data~\cite{Ball:2009mk,Hou:2019efy,Bailey:2020ooq,NNPDF:2017mvq,Aaboud:2016btc,Thorne:2017aoa,Faura:2020oom}, with the former preferring reduced strangeness at intermediate $x$ and the latter favoring enhanced strangeness in this same region.
However, it is worth emphasising that this different pull does not necessarily imply a tension between these two types of processes, for example the study of~\cite{Faura:2020oom} demonstrates how a satisfactory description of all the strangeness-sensitive datasets in the global fit, including the NOMAD cross-sections~\cite{NOMAD:2013hbk}, can be achieved; a consistent fit was also obtained in \cite{Bailey:2020ooq}.
That said, the NuTeV observable is complex, requiring treatment of the non-isoscalar nature of the iron target, acceptance corrections, and knowledge of the charm hadrons to muons branching ratio~\cite{Mason:2006qa}, $\mathrm{BR}(c \rightarrow \mu)$, and hence it is useful to further study how the outcome of the fit
varies with respect to choices that are in several cases different among the three groups.

While the non-isoscalar nature of the iron target and the acceptance corrections
associated to the NuTeV cross-sections are
treated consistently between the three groups, this is not the case for
branching ratio (BR) of charm hadrons into muons.
NNPDF adopts the PDG value and uncertainty of $0.087\pm 0.005$ \cite{ParticleDataGroup:2020ssz},
CT takes the value of 0.099 used by NuTeV itself \cite{NuTeV:2007uwm} with a normalisation uncertainty,
and MSHT take a value of $0.092\pm 0.01$ following from direct measurements of the same
process~\cite{Harland-Lang:2014zoa,Bolton:1997pq}, which avoids a potential circularity in its determination and use, and further allows the central value to vary with a penalty within this range.

However, it is well known that the total strangeness obtained in the NuTeV data region $10^{-2} \lesssim x \lesssim 0.4$ is anti-correlated with the value of branching ratio, as it maps from the total strangeness to the NuTeV predictions in the fit.
Therefore a smaller branching ratio allows one to fit the same NuTeV data whilst retaining larger strangeness and vice versa.
The impact at the level of the strangeness PDF
of taking the different branching ratios of the three groups within the MSHT reduced fit is shown in Fig.~\ref{fig:MSHTstrangeness_dimuonBR}, where an increase/decrease in the total strangeness relative to the default value of approximately 5\% is observed as the branching ratio is altered.
Upon removal of this difference by determining the reduced fits of all three groups with the same fixed branching ratio of 0.092, the NNPDF strangeness reduces whilst the CT strangeness increases as expected, bringing the reduced fits into closer agreement. Given the lack of constraints in a reduced fit, this improved strangeness agreement also enables a reduction in differences in the flavour decomposition.

After adopting a common value of the muon branching ratio BR, agreement among the strangeness PDFs
of the reduced fits displayed in Fig.~\ref{fig:reducedvsglobal} improved markedly as compared with reduced fits in which each group utilise their default BR value.
This provides a good example of how the reduced fits based on very similar data and theory
provide an excellent test baseline to assess the impact on the fit of parameters or choices
that are different between the three groups.

\subsection{The large-$x$ gluon from top and inclusive jets data}
\label{sec:highxgluon}

Another area of considerable interest in both the reduced fits and global fits is the high-$x$ gluon.
It was found in the comparison of the reduced fits in Fig.~\ref{fig:reducedvsglobal} that the gluon PDF was consistent between the three reduced fits, albeit with notable differences in the low-$x$ uncertainties.
However, this agreement in central values was in contrast with the global fit result, see Fig.~\ref{fig:global_fits_comparison}, where notable differences are present depending on the datasets included, their treatment and relative weights, and the situation is further complicated by various issues in fitting many of these datasets.
Clearly, the choice of datasets adopted by each group in their global fits impacts
the behaviour of the gluon PDF in the large-$x$ region.
Therefore, this is an obvious target that would benefit from benchmarking  using the reduced fits.
By extending these reduced fits, we would hence like to better understand any issues and differences
associated to the choice of dataset in the determination of the large-$x$ gluon.

One can consider three main dataset types which play a role in this region, all largely coming from the LHC in the past few years: these are jet data, top data and $Z$ $p_T$ data.
However, there are many issues with these datasets including difficulties fitting all bins, possible tensions within and between dataset types, and issues of correlated systematics, both within
a single distribution and among different distributions of the same measurement~\cite{Bailey:2019yze,Harland-Lang:2017ytb,Thorne:2019mpt,Boughezal:2017nla,AbdulKhalek:2020jut,Amat:2019upj,Hou:2019gfw,Hou:2019efy,Kovarik:2019xvh,Aad:2015mbv,ATLAS:2018owm,Amoroso:2020lgh,Czakon:2016olj,Czakon:2013tha}.
One of the datasets which has significant issues in its inclusion in the global fits and has demonstrated notable differences between the three groups is the ATLAS 8~TeV $t\bar{t}$ lepton + jets data~\cite{Aad:2015mbv}, for which several kinematic distributions have been measured.
This dataset has been studied by all three groups~\cite{Bailey:2020ooq,Hou:2019efy,Bailey:2019yze,Amat:2019upj,Hou:2019gfw,Kadir:2020yml,Czakon:2016olj,Harland-Lang:2017ytb,Thorne:2019mpt,AbdulKhalek:2020jut} and is presented in terms of one-dimensional distributions in four different variables: $m_{tt}, y_{t}, y_{tt}$ and $p_t^T$, with statistical and systematic correlations provided to enable all four distributions to be fitted together.

However, several groups, including CT and MSHT, have reported difficulties in fitting all four distributions simultaneously with the baseline correlation model provided
by ATLAS, leading to a poor overall data vs theory agreement with $\chi^2/N_{\rm pt} \gtrsim 5$.
Moreover, both groups and also ATLAS themselves found issues fitting either of the rapidity distributions even individually \cite{ATLAS:2018owm,Amoroso:2020lgh}. On the other hand, NNPDF3.0 were able to fit all four distributions individually~\cite{Czakon:2016olj}.
However, the statistical  correlations between the distributions were not available at the time of the NNPDF3.0-based analysis, hence they could not be fit simultaneously.

In order to further analyse the sources of the issues associated
to the PDF interpretation of this dataset,
here we therefore begin by adding it to our reduced fit baseline
presented in the previous section.
First we check the theory predictions and data for fixed PDFs (again, we use PDF4LHC15 NNLO
as the common PDF set), this verifies that the data implementation and theory settings
are in agreement between the three groups if the input PDFs are the same.
Moreover, at this stage all three groups observe the same pattern in the individual distributions' $\chi^2$, shown in the first three rows of Table~\ref{tab:ttbar_chisqs}, with all groups unable to obtain a good fit to the rapidity distributions.
This establishes that the  implementation of the data and NNLO theory
calculation for this dataset is the same or very similar among the three groups.

After fitting however, differences emerge, as shown in the remaining rows of Table~\ref{tab:ttbar_chisqs}. First of all, we have tried adding all four distributions individually without any correlations (whether statistical or systematic) between the distributions, a scenario
which is denoted the ``uncorrelated case''.
This ``uncorrelated'' fit therefore is not how this data should be added to the fit, in fact the data is effectively counted four times as the distributions are treated independently. However, including the data in this way enables the fit qualities for the individual distributions to be evaluated, which offers useful information for understanding the differences.
Whilst CT and MSHT retain the same behaviour as before fitting, NNPDF is now able to fit the rapidity distributions well, albeit at the expense of a worsening of the fit quality to the $p_t^T$
distribution.

\begin{table}[!t]
\small
  \renewcommand\arraystretch{1.5} 
  \centering
  \begin{tabular}{|>{\centering\arraybackslash}m{6.7cm}l|C{1.3cm}|C{1.3cm}|C{1.3cm}|C{1.3cm}|C{1.3cm}|}
   \hline
	Distribution/${N_{\rm pt}}$ &  & $p_t^T/8$ & $y_t/5$ & $y_{tt}/5$ & $m_{tt}/7$ & Total/25 \\ \hline
\multirow{3}{5.4cm}{\centering PDF4LHC15 input (before fitting)} & MSHT & 3.0 & 10.6 & 17.6 & 4.3 & 35.5  \\ 
													  & CT & 3.1 & 10.1 & 15.3 & 4.2 & 32.7 \\ 
													  & NNPDF & 3.4 & 9.5 & 16.2 & 4.2 & 33.3 \\ \hline
\multirow{3}{5.4cm}{\centering After fitting (uncorrelated case) } & MSHT & 3.8 & 8.4 & 12.5 & 6.4 & 31.2  \\ 
													  & CT & 3.4 & 12.9 & 17.3 & 6.1 & 39.7 \\ 
													  & NNPDF & 7.2 & 3.9 & 5.1 & 2.5 & 18.7 \\ \hline
\multirow{2}{5.4cm}{\centering After fitting (correlated case) }    & MSHT & --- & --- & --- & --- & 130.6  \\ 
													  & NNPDF & --- & --- & --- & --- & 122.7 \\ \hline
MSHT20 default (partial decorrelation)                  & MSHT & --- & --- & --- & --- & 35.3 \\
\hline
  \end{tabular}
  \vspace{0.3cm}
  \caption{The values of the $\chi^2$ obtained
    for the ATLAS 8~TeV $t\bar{t}$ lepton + jets dataset distributions when added to the
    reduced fit presented in Sect.~\ref{sec:benchmarking}.
    We show results before fitting (with the common input of the PDF4LHC15 PDFs), after fitting but including all 4 distributions in an ``uncorrelated'' manner, and after fitting maintaining all statistical and systematic correlations between the distributions as provided by the ATLAS
    Collaboration.
    We note that when fitting using either full or partial correlations, only the $\chi^2$
    evaluated over the four distributions is meaningful.
    In the CT case, a fully correlated fit of four distributions has not been performed.
  }
\label{tab:ttbar_chisqs}                                
\end{table}

On the other hand, once all statistical and systematic correlations between the four distributions are retained, following the correlation model provided by ATLAS and denoted as the ``correlated case'', then both MSHT and NNPDF observe similar behaviour. In this case neither the MSHT nor NNPDF reduced fits are able to fit the four distributions together, reflecting the behaviour seen in the global fits.
Finally, in the last row of Table~\ref{tab:ttbar_chisqs} we show that by partially decorrelating the parton shower systematic between and within distributions in the same way as in the MSHT20 default global PDF fit~\cite{Bailey:2020ooq,Bailey:2019yze} then a reasonable fit quality (for the four distributions as a whole rather than individually) can be recovered.

There are two potential explanations for these differences in the quality of the fits to the rapidity distributions individually (i.e. in the ``uncorrelated'' fits) observed by CT and MSHT, on the one hand, and NNPDF, on the other.
The first is a methodological difference between the PDF fitting groups. NNPDF being based on a neural network, must divide each dataset into training and validation to prevent over-fitting, this usually necessitates a 50:50 split of the data.
However, such a split is unfeasible for small datasets, and so all data are placed into the training. This approach potentially increases effective statistical weights of smaller datasets, of which the ATLAS $t\bar{t}$ rapidity distributions are an example, each being composed of only five points.
This would perhaps also partly explain some of the differences noticed in the comparison of the reduced fit dataset by dataset $\chi^2/N_{\rm pt}$ in Table~\ref{tab:reducedfit_chisqs_fit}, where the E866 Drell-Yan ratio data (15 points) and CMS 7~TeV electron asymmetry data (11 points) were fit considerably better in NNPDF than in the CT and MSHT reduced fits.

To investigate this possibility further, a MSHT reduced fit was performed in which these datasets were double-weighted to attempt to approximately mimic any up-weighting.
The result is shown in Table~\ref{tab:smalldatasets_doubleweighting} and reveals a notable improvement in the fit to the ATLAS $t\bar{t}$ data, specifically the rapidity distributions, which are now able to be fit well. This then matches the qualitative pattern seen in NNPDF, suggesting that perhaps the up-weighting (enhanced statistical weights) plays a role, at least in the context of the reduced fits.
In addition, whilst the quality of the fit to the E866 Drell-Yan ratio data is not improved, the fit quality of the CMS 7~TeV electron asymmetry data improves markedly, from $\sim 1.3$ per point to 0.9 per point, more consistent with the $\sim~0.75$ per point seen in the NNPDF reduced fit. Therefore, some of the differences between the reduced fits in Table~\ref{tab:reducedfit_chisqs_fit} appear to be explained.

\begin{table}[t]
\small
  \renewcommand\arraystretch{1.4} 
  \centering
    \begin{tabular}{C{3cm}|C{3cm}|c|c|>{\centering\arraybackslash}m{2.8cm}}
   \toprule
	\multicolumn{2}{c|}{  Dataset~($N_{\rm pt}$)} & MSHT uncorrelated & NNPDF uncorrelated & MSHT uncorrelated double weight \\ \midrule 
	\multicolumn{2}{c|}{ Total} & 2314.1 & 2731.4 & 2313.3  \\ \midrule 
	\multicolumn{2}{c|}{ $\chi^2/N_{\rm pt}$} & 1.15 & 1.20 & 1.15  \\ \midrule
	\multicolumn{2}{c|}{ E866 $\sigma_{pd}/(2\sigma_{pp})$ (15)} & 9.5 & 5.2 & 9.2 \\ \midrule 
	\multicolumn{2}{c|}{ CMS 7~TeV electron $A_{ch}$ (11)} & 14.2 & 8.2 & 10.2 \\ \midrule 
	\multirow{5}{2.5cm}{\centering ATLAS 8~TeV $t\bar{t}$ lepton + jet} & $p_t^T$ (8) & 3.8 & 7.2 & 4.2 \\ 
	 & $y_t$ (5) & 8.4 & 4.3 & 5.8 \\ 
	 & $y_{tt}$ (5) & 12.5 & 5.7 & 7.4 \\ 
	 & $m_{tt}$ (7) & 6.4 & 2.4 & 6.5 \\ \cline{2-5}
	 & total (25) & 31.2 & 19.6 & 23.9 \\ \bottomrule
    \end{tabular}
    \vspace{0.2cm}
    \caption{The values of the  $\chi^2$ in the variants of
      the NNPDF3.1 and MSHT20 reduced fits including the ATLAS 8~TeV $t\bar{t}$ lepton + jets dataset added with all 4 distributions included in the ``uncorrelated'' prescription.
      The final column shows the effect of double weighting the smaller datasets present, namely the E866 $\sigma_{pd}/(2\sigma_{pp})$ ratio, CMS 7~TeV electron $A_{\rm ch}$, and the ATLAS 8~TeV $t\bar{t}$ lepton + jet datasets. Note that for simplicity only a subset of the measurements
      that entered the reduced fits are listed here. See text for more details.}
      \label{tab:smalldatasets_doubleweighting}                                
\end{table}

 Nonetheless, in order for such differences to potentially play a role in the observed differences in fit qualities, it suggests that they may be changing the balance of pulls between different datasets on the PDFs. Therefore possible tensions between datasets in this high $x$ gluon region are a second possible explanation. In order to investigate this, reduced fits were run in which the LHC jet dataset included was varied from the default CMS 8~TeV jets data. The effects of this upon the MSHT reduced fit are shown in Table~\ref{tab:highxgluonjets}. The first column shows the default reduced fit (variant with this top data added), with the rapidity distributions poorly fit in the presence of the CMS 8~TeV jet data. This may indicate a potential tension between the ATLAS 8~TeV $t\bar{t}$ lepton + jets data and the CMS 8~TeV jet data, indeed it is known that the CMS 8~TeV jet dataset prefers a larger gluon at high $x$ \cite{Bailey:2020ooq,Hou:2019efy,AbdulKhalek:2020jut}, whilst the $t\bar{t}$ rapidity distributions favour a lower gluon in this region \cite{Kadir:2020yml}. The second column shows the effect of removing this CMS 8~TeV jet dataset from the reduced fit and thereby having no LHC jet data constraining this high $x$ gluon region. There is consequently more freedom for the reduced fits to fit the $t\bar{t}$ data, albeit with constraints still  provided by non-LHC datasets at high $x$ such as by the BCDMS structure function data. Upon removal of this CMS 8~TeV jet data it is clear that the fit quality to the rapidity distributions improves, with both the $y_{t}$ and $y_{tt}$ distributions now adequately fit at the expense of only a minor worsening in the fit quality for the $p_t^T$ and $m_{tt}$ distributions. This supports the suggestion of a tension between the the ATLAS 8~TeV $t\bar{t}$ lepton + jets data (and particularly its rapidity distributions) and the CMS 8~TeV inclusive jets data and this is also broadly consistent with the pulls observed in the full global fits.

To further verify this, the CMS 7~TeV jets and ATLAS 7~TeV jets data were each added to the reduced fits in place of the CMS 8~TeV jets data. This is shown in the remaining two columns of Table~\ref{tab:highxgluonjets}. These datasets both favour a reduced gluon in the region of interest for the $t\bar{t}$ and so are more consistent with the rapidity distributions, which are then observed to have reasonable fit qualities with $\chi^2/N_{\rm pt} \sim 1$. This clearly demonstrates the effects of tensions in the high $x$ gluon region on the fit quality of the ATLAS 8~TeV $t\bar{t}$ lepton + jets data. Moreover such tensions are also present within the jet datasets themselves, with the CMS 8~TeV jets $\chi^2$ worsening upon inclusion of the CMS 7~TeV or ATLAS 7~TeV jet data and vice versa (not shown). This provides a potential answer to long-standing questions about the different behaviour seen for this $t\bar{t}$ data by the different global fitting groups, with each group investigating this data originally in the context of different baseline fits, and so with different jet datasets and statistical weights of data involved.

\begin{table}[t]
\small
  \renewcommand\arraystretch{1.4} 
  \centering
  \begin{tabular}{C{3cm}|C{3cm}|>{\centering\arraybackslash}m{2.5cm}|>{\centering\arraybackslash}m{2.2cm}|>{\centering\arraybackslash}m{2.2cm}|>{\centering\arraybackslash}m{2.2	cm}}
   \toprule
	\multicolumn{2}{c|}{Dataset~($N_{\rm pt}$)} & MSHT reduced default - CMS8j & MSHT reduced no LHC jets & MSHT reduced CMS7j only & MSHT reduced ATLAS7j only \\ \midrule 
	\multicolumn{2}{c|}{Total~$\chi^2/N_{\rm pt}$} & 1.15 & 1.12 & 1.11 & 1.17   \\ \midrule
	\multicolumn{2}{c|}{CMS 8~TeV jets (174)} & 243.6 & - & - & -    \\ \midrule
	\multicolumn{2}{c|}{CMS 7~TeV jets (158)} & - & - & 156.4 & -    \\ \midrule
	\multicolumn{2}{c|}{ATLAS 7~TeV jets (140)} & - & - & - & 210.4    \\ \midrule
	\multirow{5}{2cm}{\centering ATLAS 8~TeV $t\bar{t}$ lepton + jet} & $p_t^T$ (8) & 3.8 & 4.5 & 4.0 & 4.6 \\
	& $y_t$ (5) & 8.4 & 5.2 & 6.4 & 5.5 \\ 
	& $y_{tt}$ (5) & 12.5 & 6.6 & 7.2 & 5.2 \\ 
	& $m_{tt}$ (8) & 6.4 & 7.4 & 6.4 & 6.4 \\ \cline{2-6}
	& total (25) & 31.2 & 23.8 & 24.0 & 21.6  \\ \bottomrule
  \end{tabular}
  \vspace{0.2cm}
  \caption{The values of the $\chi^2$ in the MSHT reduced fit where in
    addition to the  dataset of Table~\ref{tab:datasets}   and of
    the ATLAS 8~TeV $t\bar{t}$ lepton + jets data (with all four distributions
    fitted simultaneously with the uncorrelated prescription) one
    includes one by one different inclusive jet production datasets:
    the CMS 7 TeV and 8 TeV and the ATLAS 7 TeV measurements. 
As in Table~\ref{tab:smalldatasets_doubleweighting},   only a subset of the measurements
that entered the reduced fits are indicated here.
This comparison is relevant to ascertain the interplay between inclusive
jet and top quark pair production measurements in pinning down the large-$x$ gluon.
  }
\label{tab:highxgluonjets}                                
\end{table}

To summarise the main take-away lessons from this study, we have verified that
the three groups adopt essentially equivalent implementations of the experimental
data and the theory calculations for the ATLAS 8~TeV $t\bar{t}$ lepton + jets distributions,
with very similar $\chi^2$ values obtained when the same fixed input PDFs are utilised to compute the fit qualities.
It is found that the treatment of the correlations among the four distributions has a significant impact on whether or not one can satisfactorily
describe this dataset in a PDF fit. Finally, even in the case where the correlations are dropped the fit quality and the actual impact of the data depends on which other gluon-sensitive
measurements are being added to the fit, such as inclusive jet data, and hence may differ
in the three global fits since each of them adopts a different baseline dataset. This implies that the most complete possible inclusion of data providing constraints on the high-$x$ gluon is ideal in order to limit the sensitivity to more restricted data choices. In the PDF4LHC combination this is effectively achieved by the fact that the input PDF sets include a wider variety of data constraints than each does individually.

\section{\protect{$L_2$} sensitivity studies}
\label{app:l2_sensitivity}

One of the tools developed by CT to understand the role of each experimental data set within a global fit is the $L_2$ sensitivity \cite{Hobbs:2019gob}. 
 The $L_2$ sensitivity, $S_{f, L2}(E)$, for each experiment, $E$, can be computed using the Hessian PDFs as
\begin{equation}
S_{f, L2}(E) = \vec{\nabla} \chi^2_E \cdot \frac{ \vec{\nabla} f } { |\vec{\nabla} f| }
             = \Delta \chi^2_E\, \cos \varphi (f, \chi^2_E)\ ,
\label{eq:L2}
\end{equation}
which yields the variation of the log-likelihood function $\chi^2_E$ due to a unit-length
displacement of the fitted PDF parameters away from the global minimum $\vec{a}_0$ of
$\chi^2(\vec{a})$  in the direction of $\vec{\nabla}f$. $S_{f, L2}(E)$ therefore quantifies the impact that variations of PDFs
at fixed $x$ and $Q$ have upon the description of fitted data sets.
The $L_2$ sensitivity reflects the $\chi^2$ variation, $\Delta \chi^2$, that is obtained when the fitted PDFs, for given values of $(x, Q^2)$, are shifted upward by $1 \sigma$ of their $68\%$-level uncertainty, Eq.~(\ref{eq:L2}). The patterns in the resulting plots summarise the data-driven pulls on the PDFs, as embodied by the relative improvement ($\Delta \chi^2\! <\! 0$) or worsening ($\Delta \chi^2\! >\! 0$) in the description of each data set due to the $1 \sigma$ upward shift in the PDFs. 

The $L_2$ sensitivity approach has been initially developed for Hessian eigenvector sets. Work is in progress on understanding how the sensitivity method can be extended to Monte-Carlo PDFs. 
The $L_2$ sensitivity plots compiled this way were extensively investigated as part of the benchmarking exercise, together with the other considerations detailed in Sec.~\ref{sec:benchmarking} and App.~\ref{app:specifics}.
It was particularly insightful to compare $L_2$ sensitivities from the CT and MSHT full and reduced fits, using the same global tolerance prescription of $T^2=10$ (i.e., with the Hessian eigenvector PDFs obtained to satisfy $\Delta \chi^2_{\mbox{\tiny global}}=10$) to define the PDF uncertainties for both groups. In this case, differences in the $L_2$ sensitivity patterns in the compared fits directly reflect the pulls of experimental data sets and not the differences in the definitions of the PDF uncertainties. 

The $L_2$ sensitivity method has an advantage in that it estimates the pulls of the experiments on the PDFs with all experiments included. 
Since these pulls change when some experiments are removed, the $L_2$ sensitivities complement the study described in App.~\ref{sec:highxgluon}, in which the individual data sets were added to the fit, or taken out, one at a time. 

This method provides a fast approximation to the LM scanning technique repeated at many $x$ values at once, cf.~Sec.~\ref{subsec:ct18}, and is particularly helpful in identifying the experiments with the strongest pulls in some $x$ regions. 
A comparative study of the CT and CJ PDFs \cite{Accardi:2021ysh} demonstrated that the $L_2$ sensitivities provide an easy-to-implement metric for apples-to-apples comparisons of the fits by different groups. In this publication, for the first time we compared the $L_2$ sensitivities in the CT and MSHT reduced fits. We found both similarities and differences, indicating that even in the reduced fits the methodologies of the two fits are not exactly identical. 

While the full discussion of these findings will be postponed for a forthcoming dedicated publication, in a number of cases, the $L_2$ sensitivity studies done by the CT group supported the ultimate choice of the data sets for the reduced fit. The $L_2$ sensitivities, analogously to the LM scans, quantify the strength of the pulls on the PDFs from the data sets across the entire $(x,\ Q^2)$ plane. They also effectively reveal situations when the pulls between the data sets are mutually contradictory. With this in mind, the $L_2$ sensitivity charts \cite{CT18L2Website} prepared for the CT18 analysis helped to arrive at the minimal collection of the experiments for the reduced fits that provide strong constraints to obtain a convergent CT fit and are mutually consistent to avoid complications with the PDF uncertainties. 

For example, including only the E866 $pd/pp$ cross-section ratios in the reduced fits was favored on these grounds to constrain the $\bar d - \bar u$ combination, as opposed to also fitting the absolute E866 $pp$ cross sections that constrain similar sea (anti)quark combinations as BCDMS and prefer a somewhat different trend from BCDMS.
Similarly, some inconsistencies between the jet data sets, observed in the earlier CT studies \cite{Kovarik:2019xvh, Hou:2019efy} and by other groups \cite{Harland-Lang:2017ytb,Bailey:2020ooq,AbdulKhalek:2020jut,Ball:2021leu}, supported the inclusion of only a single set of LHC Run-1 inclusive jet data within the reduced benchmarking exercises.   
This latter observation complements the dedicated studies of inclusive jet data appearing in App.~\ref{app:specifics}. 

\begin{figure*}[tb]
\centering
\includegraphics[width=0.48\textwidth]{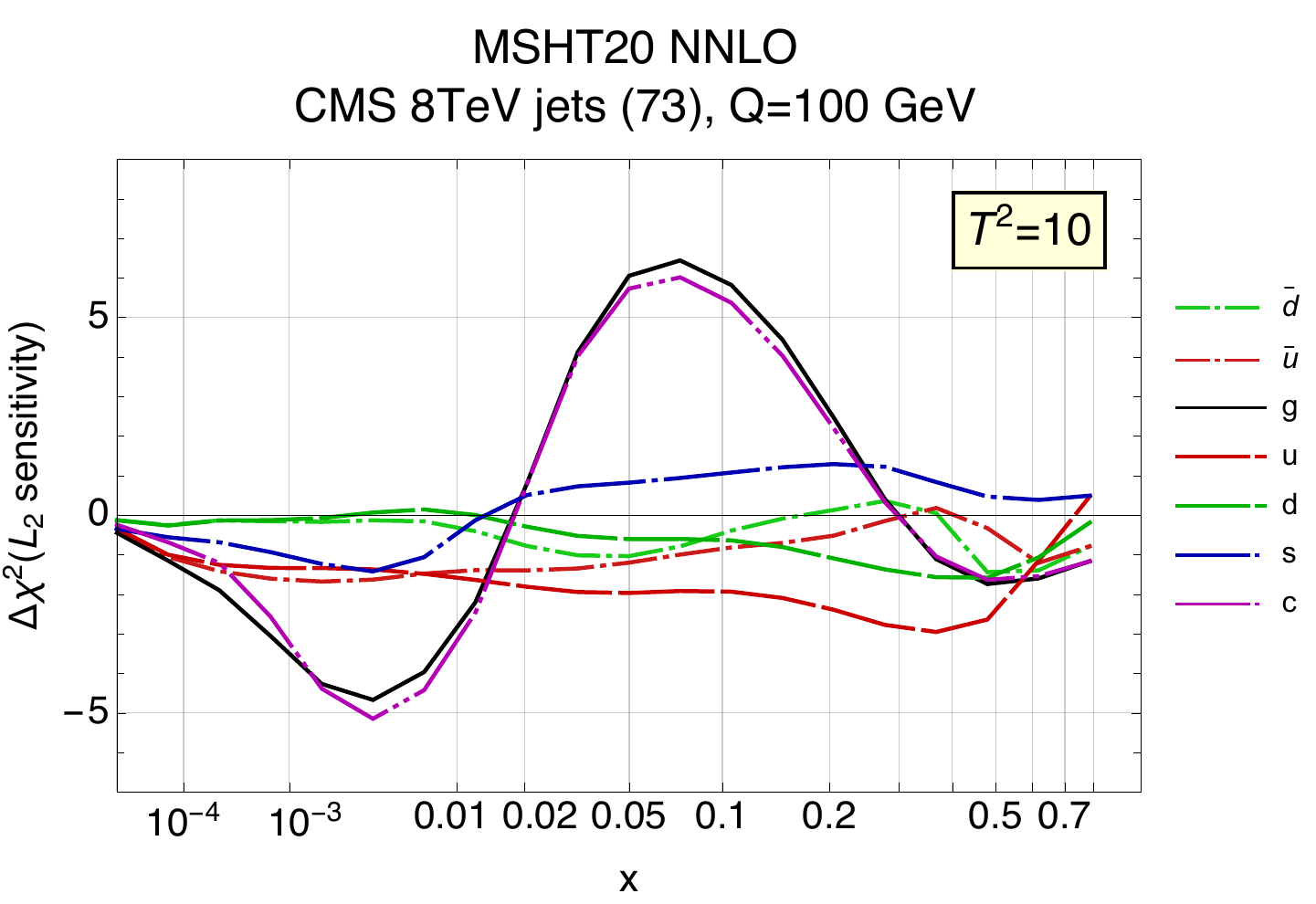}\quad
\includegraphics[width=0.48\textwidth]{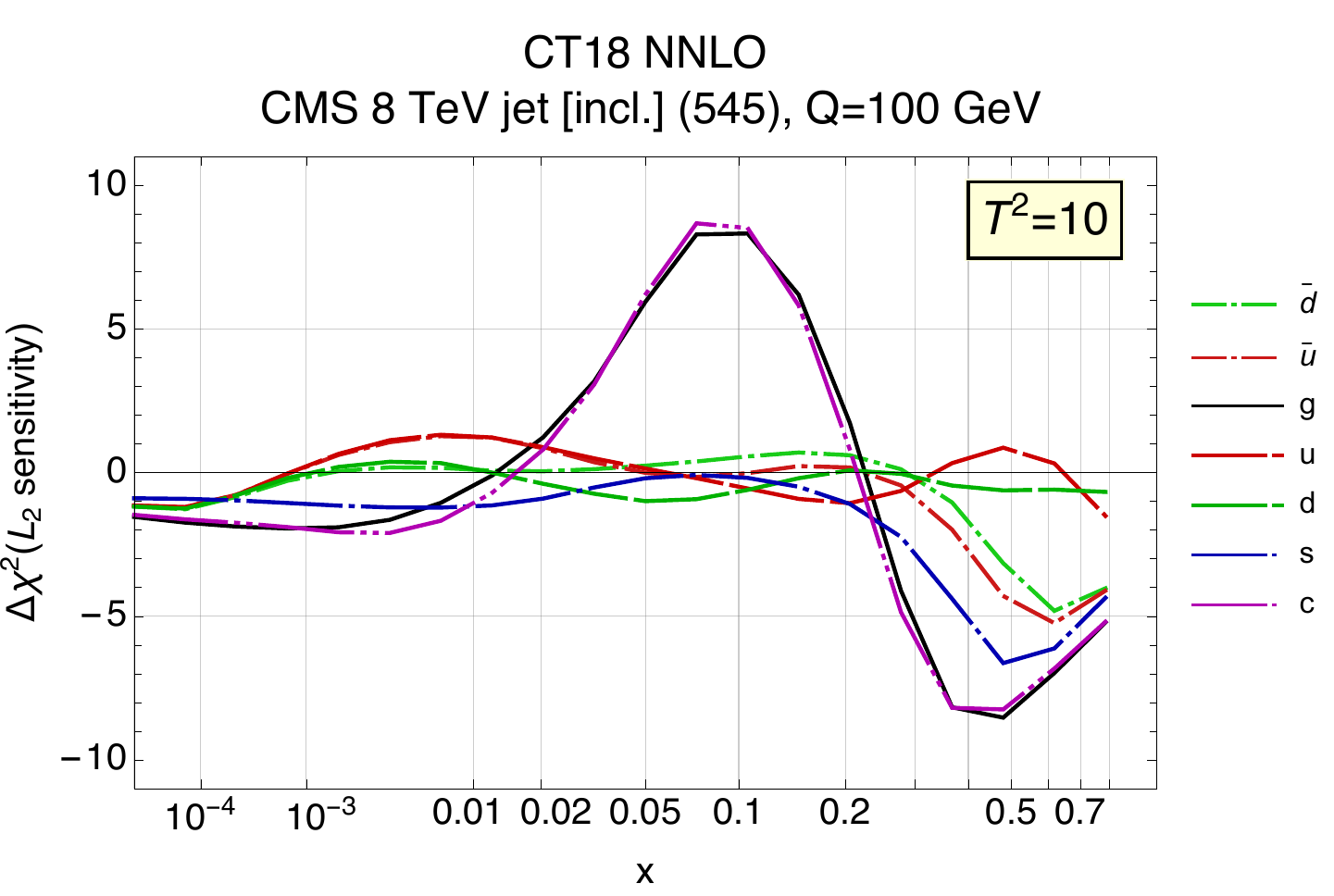}
\\
\includegraphics[width=0.48\textwidth]{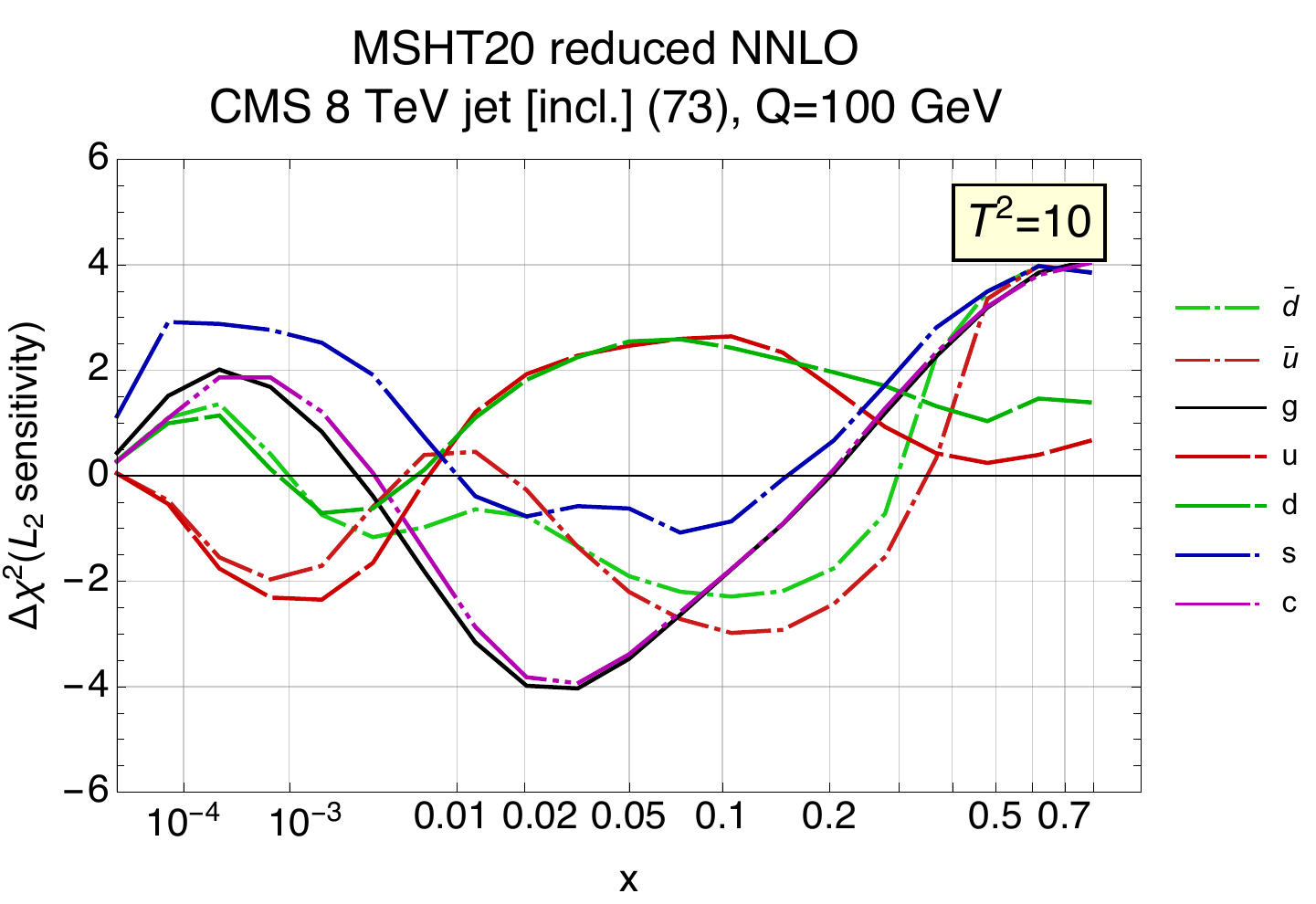}\quad
\includegraphics[width=0.48\textwidth]{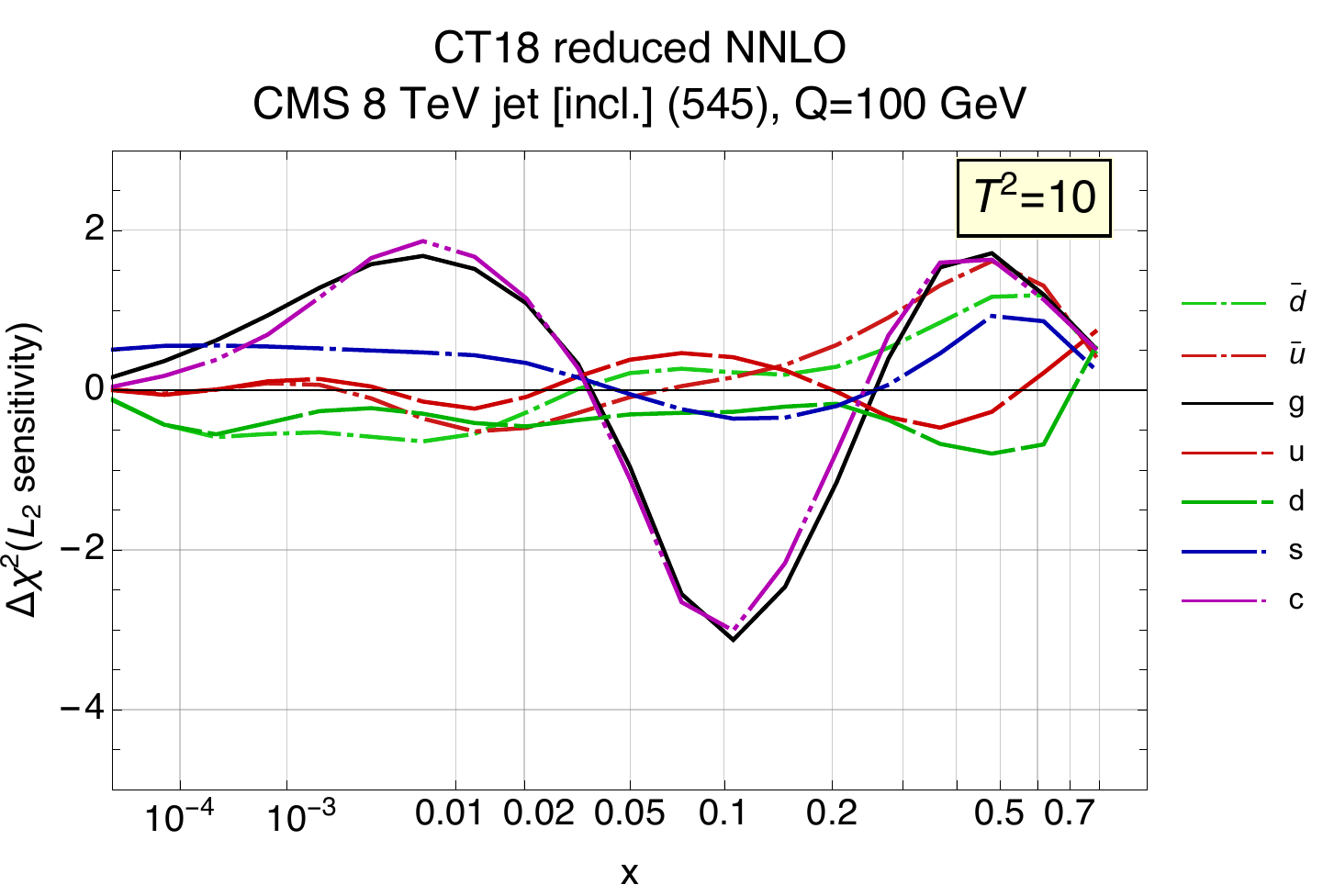}
\caption{The $L_2$ sensitivities at $Q\!=\!100$ GeV of the CMS 8 TeV inclusive jet data to the PDFs obtained with global tolerance $T^2=10$.  Left: MSHT20red NNLO, right: CT18red NNLO.}
\label{fig-app:compar_bench_CMS8TeV}
\end{figure*}

To illustrate the information that can be accessed this way, Figure~\ref{fig-app:compar_bench_CMS8TeV}
shows the pulls on the PDFs of the indicated flavours obtained for the CMS 8 TeV inclusive jet data at $Q\! =\! 100$ GeV via the $L_2$ method. We stress that all eigenvector sets for these $L_2$ studies are generated assuming a global tolerance of $T^2\!=\!10$, such that the uncertainties in this case do not exactly correspond to those shown in Sec.~\ref{sec:benchmarking}. The subfigures in the upper (lower) row are for the full (reduced) MSHT20 and CT18 fits, respectively.

Figure~\ref{fig-app:compar_bench_CMS8TeV} offers a number of insights:
\begin{enumerate}
    \item For this experiment, the patterns of the pulls are similar between the two groups.
    
    \item In the full fits, the CMS 8 TeV jet data impose the strongest pulls (indicated by $\Delta \chi^2 > 5$) to mostly reduce the gluon and charm PDFs at $x=0.02-0.2$, in accord with the kinematics of the process. There are compensating upward pulls on the gluon at other $x$ values either due to the CMS  data themselves or because of the momentum sum rule. The pulls on the (anti)quark PDFs are relatively weak.
    \item For the reduced fits, both groups share a similar pattern for the $L_2$ sensitivity to the gluon (and charm) PDF at large $x\! \sim\! 0.5$. Here a larger gluon is disfavored, as seen by the positive variation of $\chi^2$ in this region of $x$. Intriguingly, this behaviour in the $L_2$ sensitivities is the reverse of that seen for the full fits with MSHT20 and CT18, as in those  the CMS 8 TeV jet data have the tendency of to enhance the gluon at large $x$. The compensating upward pulls on the gluon in the reduced fits are most pronounced at $x=0.005-0.2$ for the reduced MSHT and at $x=0.04-0.2$ for the reduced CT.  
    \item The maximal excursions of $\Delta \chi^2$ reduce from 5-7 in the full fits to 2-4 in the reduced fits. This reduction indicates improved agreement of the CMS jet and other experiments in the reduced fits, as the $L_2$ sensitivities tend to cancel when summed over all experiments. The reduction in the magnitude of the pulls within the bottom row of Fig.~\ref{fig-app:compar_bench_CMS8TeV} thus indicates a sensible interplay among the data sets, as mentioned in App.~\ref{sec:highxgluon}. 

\end{enumerate}

\begin{figure}[tb]
\centering
\includegraphics[width=0.48\textwidth]{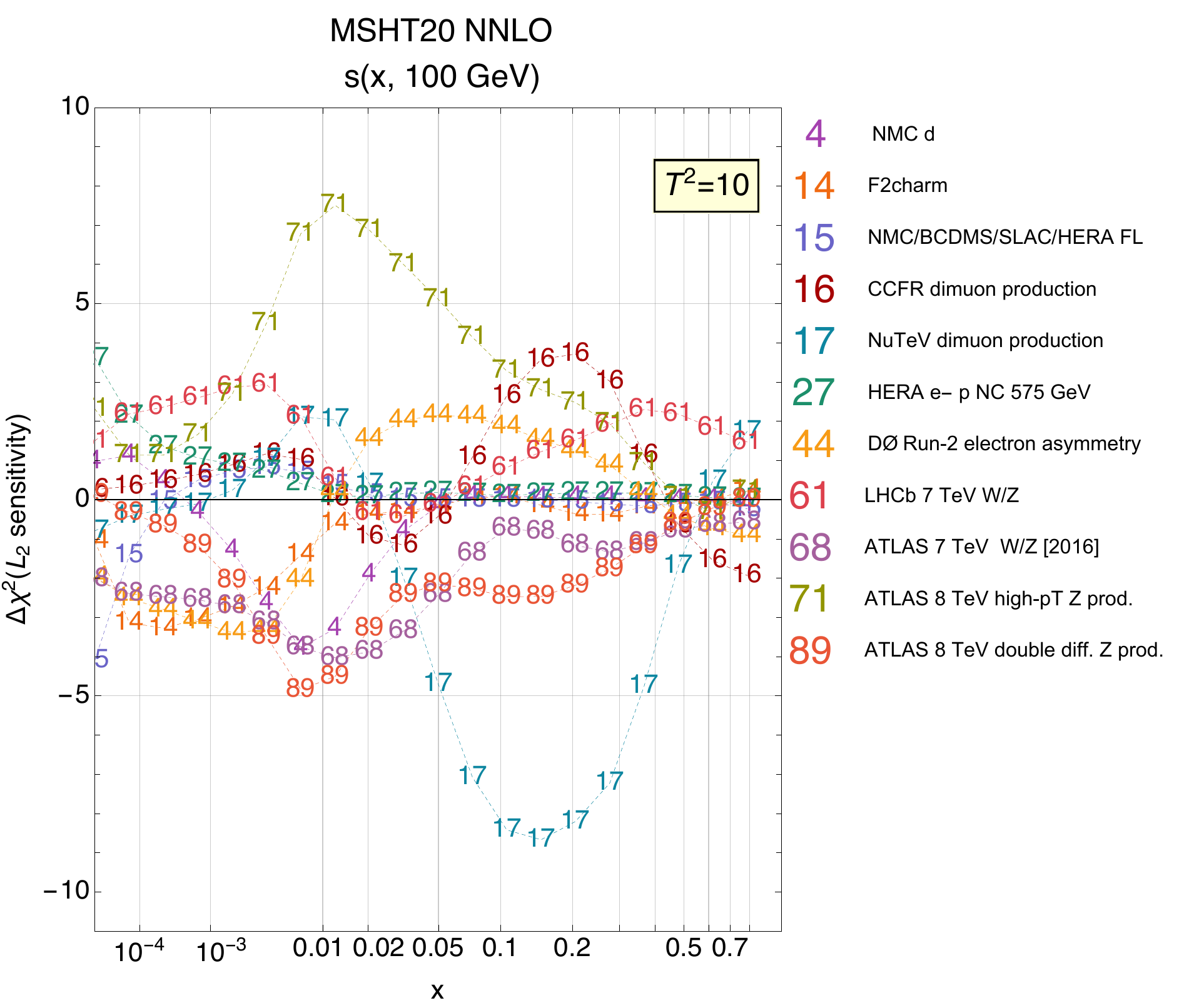}
\includegraphics[width=0.48\textwidth]{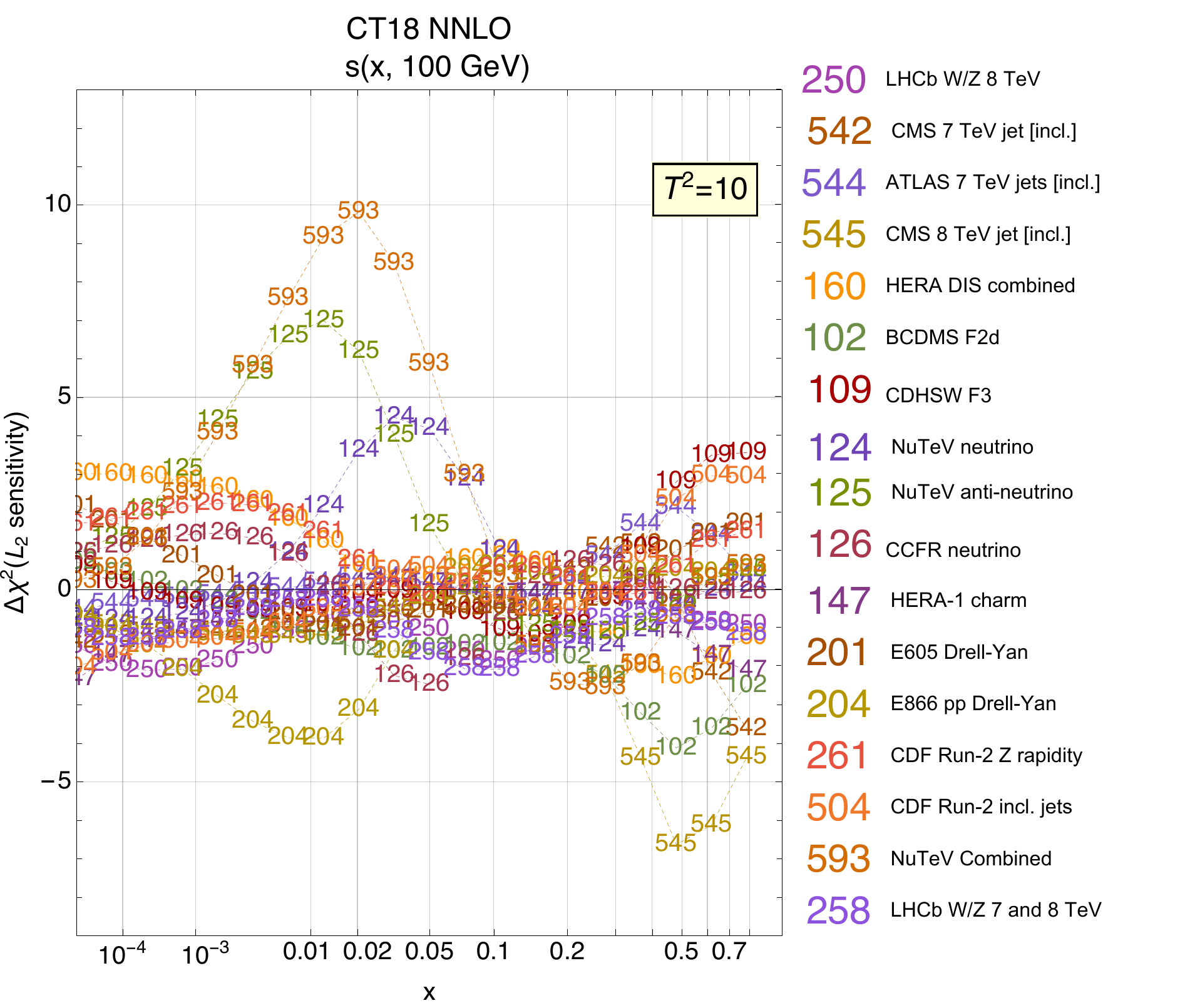}\\
\includegraphics[width=0.48\textwidth]{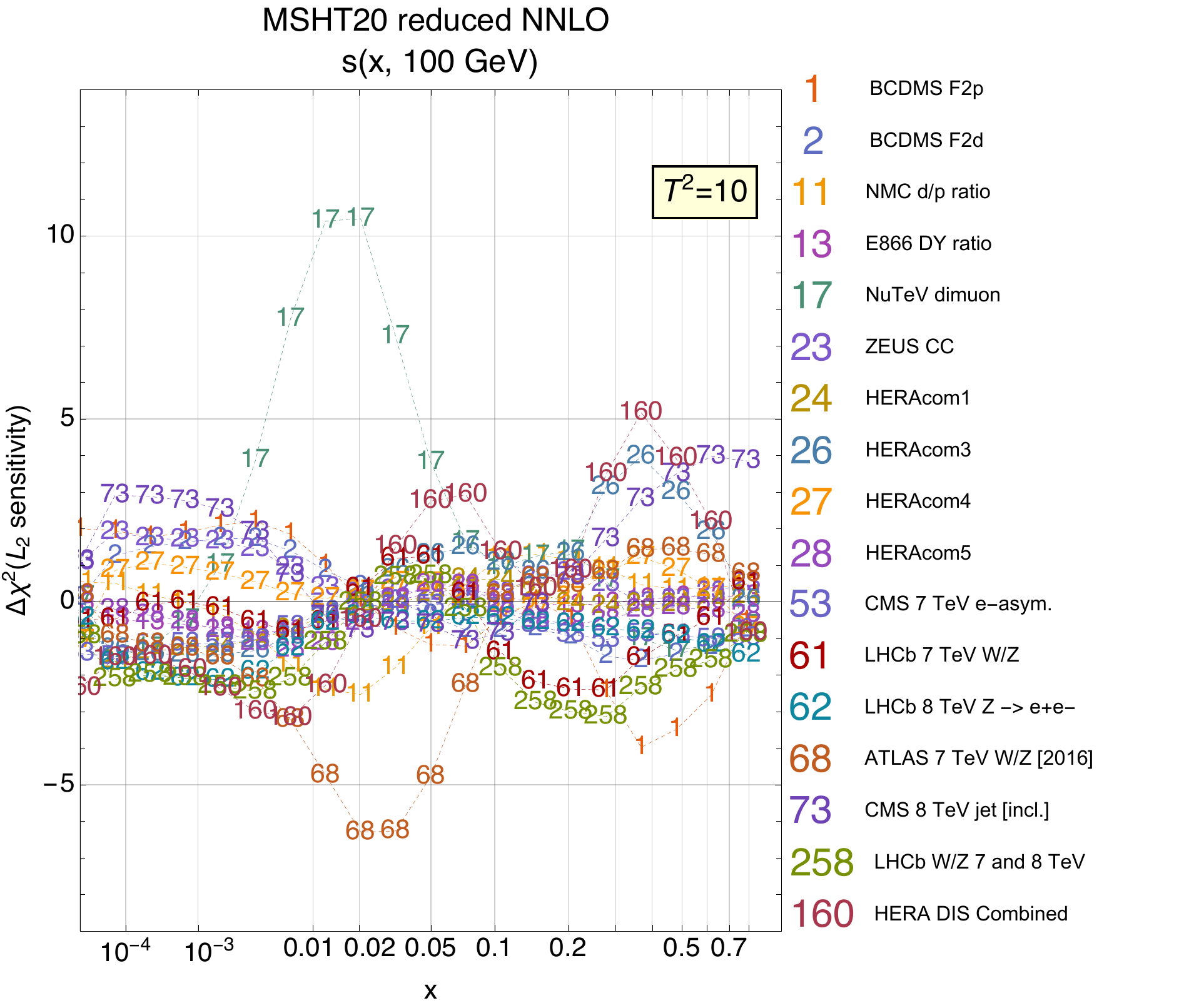}
\includegraphics[width=0.48\textwidth]{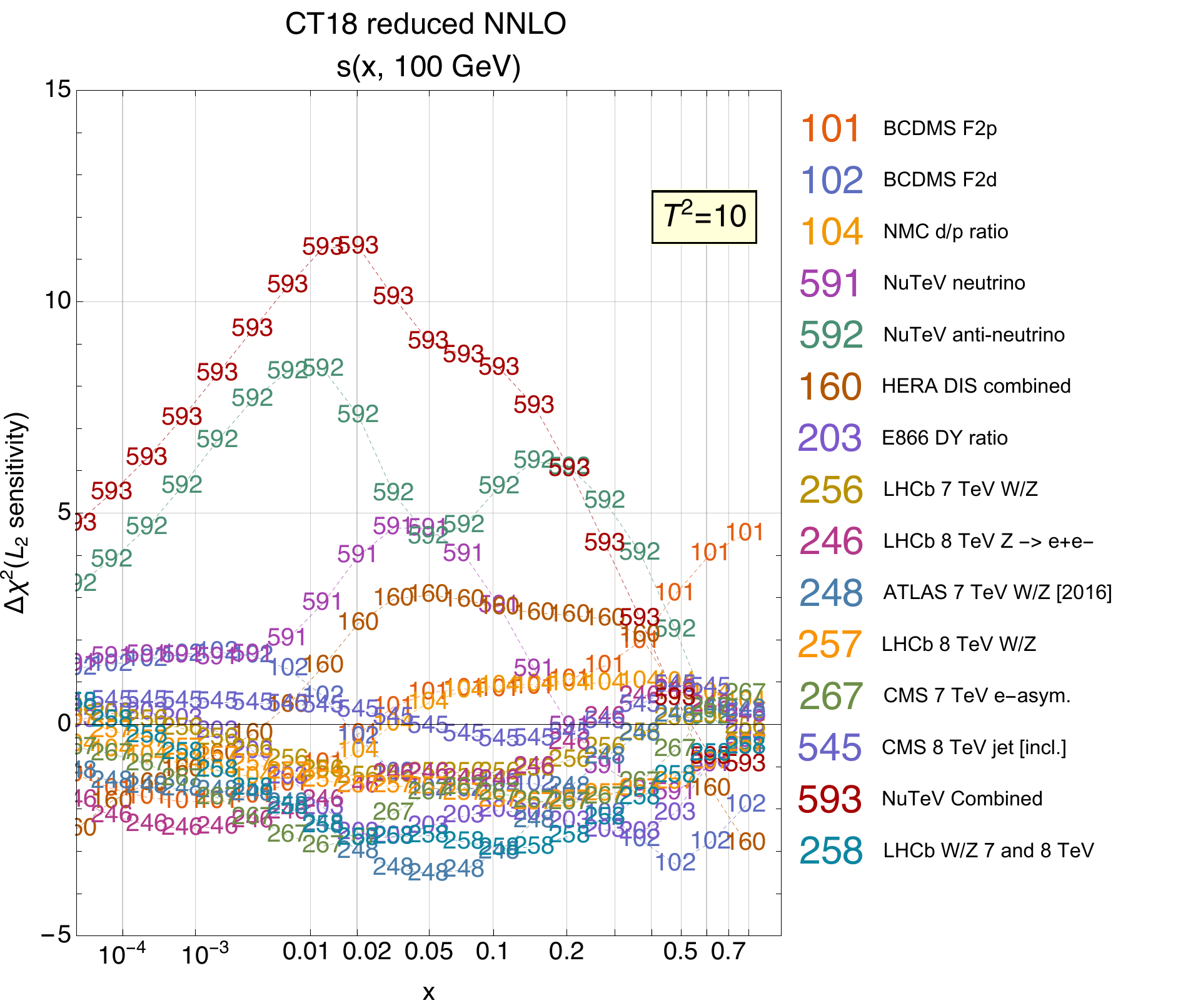}
\caption{The $L_2$ sensitivities of the reduced fit data sets with $T^2\!=\!10$ to the strange PDF at $Q\!=\!100$ GeV. The numerical IDs follow the conventions for the fitted data sets adopted in the respective fits. Different IDs for the same experiment correspond to the different selection of data points. For example,  the combined NuTeV data are labeled as sets $\#17$ and $\#593$ for MSHT and CT, respectively, while CT also shows individual NuTeV sets 124 and 125 for neutrino and anti-neutrino CC DIS scattering, respectively.  Top row: results for the full fits of MSHT20 NNLO (left) and CT18 NNLO (right). Bottom row: the corresponding reduced-fit results.}
\label{fig-app:compar_bench_strange}
\end{figure}

An instructive example of a comparison that identifies the most constraining experiments for a given PDF flavour is the pattern of the $L_2$ sensitivities to strangeness at $Q=100$ GeV shown in Fig.~\ref{fig-app:compar_bench_strange} for the full and reduced fits. Here, the experiments with the largest excursions are the most sensitive. A marked difference in the comparison of the two Hessian sets, CT and MSHT, is evident in the sensitivities of the NuTeV dimuon data to the strange PDF, as can be seen in the top row of Fig.~\ref{fig-app:compar_bench_strange}. 
With the common charm-to-muon branching ratio and selection of data points for the reduced sets, the pattern of the $L_2$ sensitivity shows a better agreement in the bottom row of Fig.~\ref{fig-app:compar_bench_strange}. The combined NuTeV data dominate the pulls around $0.01<x<0.05$, favouring a smaller strangeness. The downward pulls of the NuTeV data sets in this $x$ region compete against the upward ones from the ATLAS 7 TeV $W/Z$ production data and, in the case of the full MSHT fit, ATLAS 8 TeV double-differential $Z$ production. High-$p_T$ $Z$ production exerts a strong downward pull on the full MSHT fit that is absent in the full CT18 fit, although it should be noted that the two groups include different amounts of this data and differ subtly in their treatments of it \cite{Hou:2019efy,Bailey:2020ooq}, this may be relevant to the different pulls observed. 
This study of the $L_2$ sensitivity complements the discussion in App.~\ref{sec:strangeness}.

Investigations of the $L_2$ sensitivity were extended to the parton-parton luminosities. Between the two Hessian fits, CT and MSHT, qualitative agreement was obtained in reduced fits, for instance, for $L_{gg}$. At larger invariant masses, $M_X > 100$ GeV, constraints from the HERA, 8 TeV CMS jet data, and 8 TeV LHCb $Z\to e^+e^-$ data are most prominent.
Analogously, the pulls of the NuTeV data on the parton-parton luminosities ({\it e.g.}, on $L_{gg}$) and their interplay with other fitted experiments as revealed by the $L_2$-sensitivity method formed part of the motivation to investigate the treatment of the neutrino-induced dimuon production data, including choices of branching ratios and QCD theory accuracy.

Additional figures for the $L_2$ sensitivity study presented here are available at
\begin{center}
    \url{http://www.physics.smu.edu/devel/xjing/pdf4lhc21/L2sens/index2.html} \, .
\end{center}

\clearpage

\bibliographystyle{utphys}
\bibliography{pdf4lhc21}

\end{document}